\documentclass[aps,a4paper,showkeys,nofootinbib,longbibliography,notitlepage,twocolumn]{revtex4-2}
\usepackage[utf8]{inputenc}
\usepackage{filecontents}
\usepackage{natbib}
\usepackage{amsmath,amssymb,bm,amsthm}
\usepackage{subfigure}
\usepackage{graphicx}
\usepackage{dcolumn}
\usepackage{bm}
\usepackage[mathlines]{lineno}
\usepackage{booktabs}
\usepackage{color}
\newcounter{one}
\usepackage{url}

\usepackage{textcase}
\usepackage{mathrsfs}

\usepackage{colortbl}
\usepackage{tabularx}
\usepackage{verbatim}
\usepackage{multirow}

\usepackage[T1]{fontenc} 
\usepackage{lmodern}
\usepackage{bbm}
\usepackage[utf8]{inputenc}
\usepackage{amsfonts}
\usepackage{array}

\usepackage{bbm}
\usepackage{enumitem}
\usepackage{umoline}
\usepackage[usenames,svgnames]{xcolor}
\usepackage{natbib}
\usepackage[hyperindex,breaklinks]{hyperref}
\hypersetup{
     colorlinks=true,       		
     linkcolor=Navy,          	
     citecolor=Navy,            
     filecolor=Navy,      		
     urlcolor=Navy,           	
    runcolor=cyan,
 }
\usepackage{graphicx}
\usepackage{amsfonts}
\usepackage{amssymb}
\usepackage{amsmath}
\usepackage{bbm}
\usepackage{enumitem}

\usepackage{xcolor}

\newcommand{\bra}[1]{\langle #1 |}
\newcommand{\ket}[1]{| #1 \rangle}

\newcommand{\tr}[0]{ {\rm tr}}

\newcommand{\Oorderof}{\mathcal{O}}
\newcommand{\orderof}[1]{\Oorderof(#1)} 
\newcommand{\orderofsp}[1]{\Omega(#1)}

\newcommand{\dist}{d}

\newcommand{\co}{{\rm c}}

\newcommand{\poly}{{\rm poly}}
\newcommand{\polylog}{{\rm polylog}}

\def\beq{\begin{equation}}
\def\eeq{\end{equation}}
\def\nbeq{\begin{equation*}}
\def\neeq{\end{equation*}}
\def\<{\langle}
\def\>{\rangle}

\def\tr{{\rm tr}}

\newcommand{\bal}[2]{#1[#2]}

\newcommand{\nb}{\hat{n}}

\newcommand{\br}[1]{\left( #1 \right)}
\newcommand{\brr}[1]{\left[ #1 \right]}

 \newcommand{\norm}[1]{\left \|  #1 \right \|}

\newcommand{\ave}[1]{\left \langle #1 \right \rangle}

\usepackage{mathtools}
\def\multiset#1#2{\ensuremath{\left(\kern-.3em\left(\genfrac{}{}{0pt}{}{#1}{#2}\right)\kern-.3em\right)}}

\begin{document}


\title{Effective light cone and digital quantum simulation of interacting bosons}

\author{Tomotaka Kuwahara$^{1,2,3}$}
\email{E-mail: tomotaka.kuwahara@riken.jp
} 
\author{Tan Van Vu$^{1}$}

\author{Keiji Saito$^{4}$}

\affiliation{$^{1}$ 
Analytical quantum complexity RIKEN Hakubi Research Team, RIKEN Center for Quantum Computing (RQC), Wako, Saitama 351-0198, Japan
}
\affiliation{$^{2}$ 
RIKEN Cluster for Pioneering Research (CPR), Wako, Saitama 351-0198, Japan
}
\affiliation{$^{3}$
PRESTO, Japan Science and Technology (JST), Kawaguchi, Saitama 332-0012, Japan}

\affiliation{$^{4}$ 
Department of Physics, Kyoto University, Kyoto 606-8502, Japan}

\begin{abstract}

The speed limit of information propagation is one of the most fundamental features in non-equilibrium physics. The region of information propagation by finite-time dynamics is approximately restricted inside the effective light cone that is formulated by the Lieb-Robinson bound. To date, extensive studies have been conducted to identify the shape of effective light cones in most experimentally relevant many-body systems. However, the Lieb-Robinson bound in the interacting boson systems, one of the most ubiquitous quantum systems in nature, has remained a critical open problem for a long time. This study reveals a tight effective light cone to limit the information propagation in interacting bosons, where the shape of the effective light cone depends on the spatial dimension. To achieve it, we prove that the speed for bosons to clump together is finite, which in turn leads to the error guarantee of the boson number truncation at each site. Furthermore, we applied the method to provide a provably efficient algorithm for simulating the interacting boson systems. The results of this study settle the notoriously challenging problem and provide the foundation for elucidating the complexity of many-body boson systems. 


\end{abstract}

\maketitle

%
%

\section{Introduction}
Causality is a fundamental principle in physics and imposes the strict prohibition of information propagation outside light cones.
The non-relativistic analog of causality was established by Lieb and Robinson~\cite{ref:LR-bound72}, who proved the existence of the effective light cone. The amount of information outside the light cone decays exponentially with the distance. 
The recent experimental developments have allowed one to directly observe such effective light cones in various experimental setups~\cite{cheneau2012light,richerme2014non,jurcevic2014quasiparticle}.
The Lieb-Robinson bound provides a fundamental and universal speed limit (that is, the Lieb-Robinson velocity) for non-equilibrium structures in real-time evolutions. 
Furthermore, the Lieb-Robinson bound also offers critical insights into the steady states and spectral properties of the systems using the Fourier transformation. 
In the past decades, the Lieb-Robinson bound has found diverse applications in interdisciplinary fields, such as 
the area law of entanglement~\cite{ref:Hastings-AL07,PhysRevLett.111.170501}, 
quasi-adiabatic continuation~\cite{PhysRevB.72.045141},
fluctuation theorem for pure quantum states~\cite{PhysRevLett.119.100601}, 
clustering theorems for correlation functions~\cite{ref:Hastings2006-ExpDec,ref:Nachtergaele2006-LR,PhysRevX.12.021022},
tensor-network based classical simulation of many-body systems~\cite{PhysRevLett.97.157202,PRXQuantum.2.040331}, 
optimal circuit complexity of quantum dynamics~\cite{8555119},
sample complexity of quantum Hamiltonian learning~\cite{Anshu_2021}, and
quantum information scrambling~\cite{PhysRevLett.117.091602}.
Owing to these crucial applications, the Lieb-Robinson bound has become a central topic in the field of quantum many-body physics.  


Lieb and Robinson argued that the speed of information propagation is finitely bounded; that is, the effective light cone is linear with time. One might have a naive expectation that this is true in generic quantum many-body systems. 
However, to justify such an intuition, we must assume the following conditions: (a) the interactions are short-range, and (b) the strength of interactions is finitely bound.
Understanding the breakdown of the two above-mentioned conditions is inevitable for comprehensively describing the information propagation in all experimentally relevant quantum many-body systems. 
The breakdown of condition (a) should be easy to imagine.
Under long-range interactions, the information propagates immediately to an
arbitrarily distant point, causing one to intuitively assume that the effective light cone may no longer be linear~\cite{PhysRevLett.111.260401}. 
Nevertheless, if the interaction decays polynomially with distance, it indicates the existence of a non-trivial effective light cone that depends on the decay rate of the interaction strength. 
The Lieb-Robinson bound for long-range interacting systems has been unraveled significantly in the past decade~\cite{PhysRevLett.114.157201,chen2019finite,PhysRevX.10.031010,PhysRevX.11.031016,PhysRevLett.126.030604,PhysRevLett.127.160401,PhysRevA.104.062420}.

Conversely, the influence of the breakdown of the condition (b) has still been elusive. 
Considering the Lieb-Robinson velocity is roughly proportional to the interaction strength~\cite{ref:Nachtergaele2006-LR,PhysRevLett.97.050401}, 
we can no longer obtain any meaningful effective light cone without condition (b). 
Unfortunately, such quantum systems typically appear in quantum many-body physics because the representative examples include quantum boson Hamiltonians, which describe the atomic, molecular, and optical systems.
In the absence of boson-boson interactions, one can derive the Lieb-Robinson bound with a linear light cone~\cite{cramer2008locality,Nachtergaele2009}. 
In contrast, boson-boson interactions exponentially accelerate transmitting information signals~\cite{PhysRevLett.102.240501}.
In quantum boson systems on a lattice, 
an arbitrary number of bosons can gather at one location, and the on-site energy can become arbitrarily large, resulting in unlimited Lieb-Robinson velocity.  
However, to date, there is no general established method to avoid the unboundedness of the local energy. 
When analyzing the systems, we must truncate the boson number at each site up to a finite number. 
Although practical simulations often adopt this heuristic prescription, the obtained results are always associated with some uncontrolled uncertainty. 
Therefore, the most pressing question is what can happen if we consider the dynamics in unconditional ways. 
The elucidation is crucial in the digital quantum simulation of boson systems with an efficiency guarantee.

As mentioned above, general boson systems inherently cause information propagation with unlimited speed, forcing one to restrict themselves to specific classes of interacting boson systems. 
The most important class is the Bose-Hubbard model, which is a minimal model comprising essential physics for cold atoms in optical lattices (see Refs.~\cite{PhysRevLett.111.230404,PhysRevLett.115.130401,Tong2022} for other boson models). 
In recent studies, cold atom setups have attracted significant attention as a promising platform for programmable quantum simulators~\cite{doi:10.1126/science.1229957,doi:10.1126/science.aal3837,Yang2020,PRXQuantum.2.017003,Ebadi2021}.
Thus far, various researchers have explored this model in theoretical~\cite{PhysRevLett.98.180601,L_uchli_2008,PhysRevLett.100.030602,PhysRevLett.101.063001,PhysRevA.85.053625,PhysRevA.89.031602}
and experimental ways~\cite{Bakr547,cheneau2012light,Baier201}.  
Considering the Lieb-Robinson bound in the Bose-Hubbard type model, 
we must treat the following primary targets separately: i) transport of boson particles~\cite{PhysRevA.84.032309,PhysRevLett.128.150602} and ii) information propagation~\cite{PRXQuantum.1.010303,PhysRevLett.127.070403,PhysRevX.12.021039,Faupin2022}.
The former characterizes the migration speed of boson particles, whereas the latter captures the propagation of all information.
Relevant to the first issue i), Schuch, Harrison, Osborne, and Eisert brought the first breakthrough~\cite{PhysRevA.84.032309} by considering the diffusion of the initially concentrated bosons in the vacuum and ensured that the bosons have a finite propagation speed.
The generalization of the result has been a challenging problem for over a decade. 
Recently, the initial setup has been relaxed to general states while assuming a macroscopic number of boson transport~\cite{PhysRevLett.128.150602}. 
On the second issue ii), Ref.~\cite{PRXQuantum.1.010303} derived the Lieb-Robinson velocity that was proportional to the square root of the total number of bosons. 
Therefore, the result provides a qualitatively better bound, whereas the velocity is still infinitely large in the thermodynamic limit.
Assuming the initial state is steady and has a small number of bosons in each site, it has been proved that the effective light cone is linear with time~\cite{PhysRevLett.127.070403,PhysRevX.12.021039}.
Although these studies have advanced the understanding of the speed limit of Bose-Hubbard-type models, 
the results' application ranges are limited to specific setups, such as the steady initial state (see Ref.~\cite{Faupin2022} for another example).
Until now, we are far from the long-sought goal of characterizing the optimal forms of the effective light cones for the speed of i) and ii) under the condition that arbitrary time-dependent tunings of the Hamiltonian are allowed.

\begin{figure}[tt]
\centering
\includegraphics[clip, width=88mm]{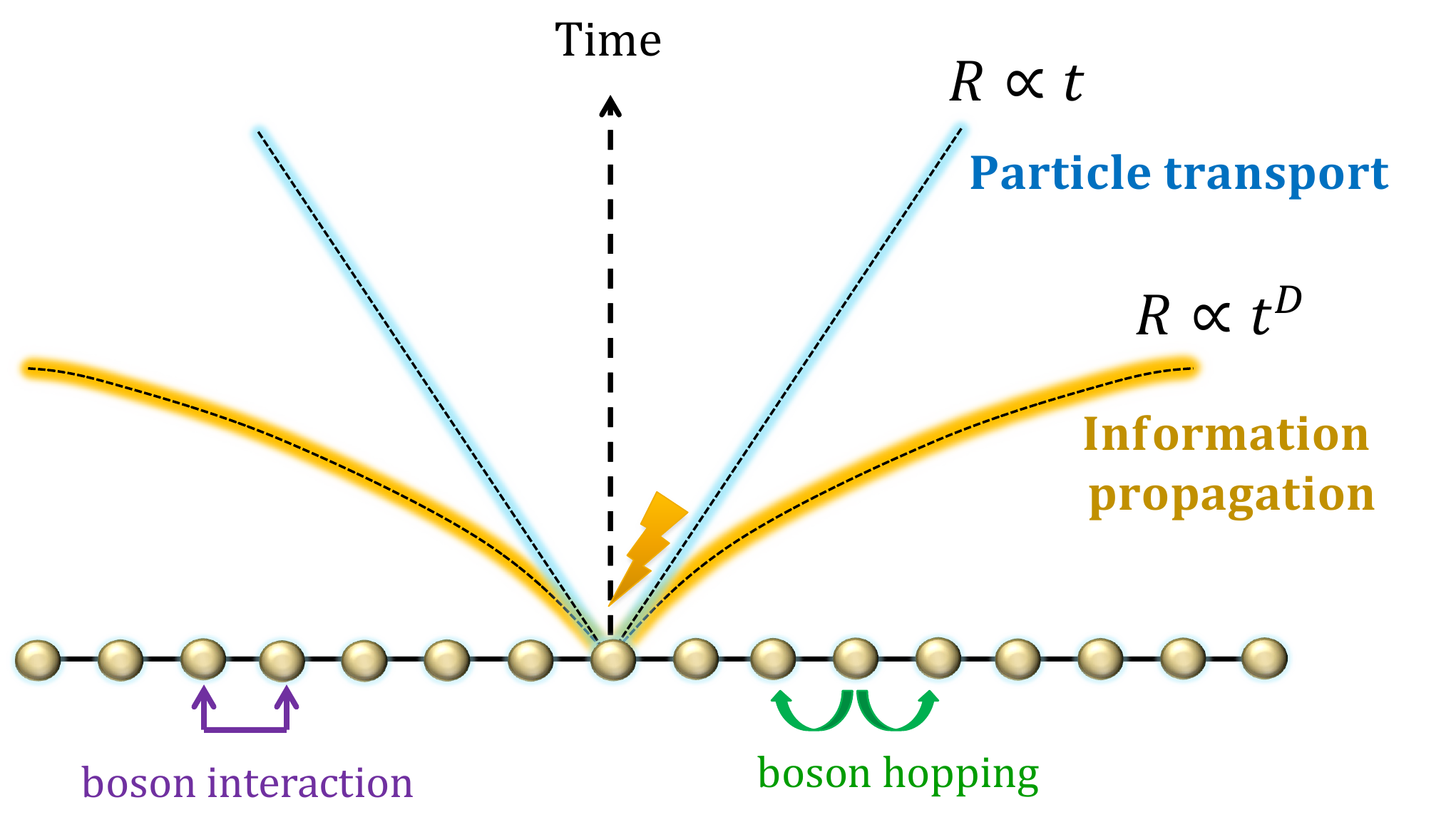}
\caption{Illustration of the effective light cones. 
Herein, we describe the interacting bosons by the Bose-Hubbard type Hamiltonian~\eqref{bose_hubbard_ham}.
We first consider how fast boson particles move to distant regions, as shown in Fig.~\ref{fig:Fig2}. 
The light cone for the boson particle transport is proved to be almost linear up to logarithmic corrections (denoted by the blue shaded line), as shown in Result~1.
Conversely, if we consider the propagation of the full information (see also Fig.~\ref{fig:Fig3}), 
the speed can be much faster than the particle transport. 
The effective light cone is proved to be polynomial with time, and the exponent is equal to the space dimension $D$ (denoted by the orange shaded line), where the mathematical form of the Lieb-Robinson bound is given in Result~2.
We can explicitly construct a protocol to achieve the light cone using dynamics with time-dependent Bose-Hubbard type Hamiltonians (see Fig.~\ref{fig:Fig4}).
}
\label{fig:Fig1}
\end{figure}

In this article, we overcome various difficulties and solve the problem in general setups. We treat arbitrary time-dependent Bose-Hubbard-type Hamiltonians in arbitrary dimensions starting from a non-steady initial state. 
Such a setup is most natural in physics and crucial in estimating the gate complexity of digital quantum simulation of interacting boson systems. 
Figure~\ref{fig:Fig1} summarizes the main results, providing qualitatively optimal effective light cones for both the transport of boson particles and information propagation. 
As a critical difference between bosons and fermions (or spin models), we have clarified that the acceleration of information propagation can occur in high dimensions.
Furthermore, as a practical application, we develop a gate complexity for efficiency-guaranteed digital quantum simulations of interacting bosons based on the Haah-Hastings-Kothari-Low (HHKL) algorithm~\cite{8555119}.

\section{Results}

\subsection*{Speed limit on boson transport}

We consider a quantum system on a $D$-dimensional lattice (graph) with $\Lambda$ set for all sites. 
For an arbitrary subset $X\subseteq \Lambda$, we denote the number of sites in $X$ by $|X|$, that is, the system size is expressed as $|\Lambda|$. 
We define $b_i$ and $b_i^\dagger$ as the bosonic annihilation and creation operators at the site $i\in\Lambda$, respectively.
We focus on the Bose-Hubbard type Hamiltonian in the form of
\begin{align}
\label{bose_hubbard_ham}
&H= \sum_{\langle i, j \rangle} J_{i,j} (b_i b_j^\dagger +{\rm h.c.} ) + V 
\end{align}
with $|J_{i,j}| \le \bar{J}$ and $V:= f\br{ \{\nb_i\}_{i\in \Lambda}}$, 
where $\sum_{\langle i, j \rangle}$ is the summation for all pairs of the adjacent sites $\{i,j\}$ on the lattice and 
$f\br{ \{\nb_i\}_{i\in \Lambda}}$ is an appropriate function of the boson number operators $\{\nb_i\}_{i\in \Lambda}$ with $\nb_i=b_i^\dagger b_i$.
The constraints on the function $f\br{ \{\nb_i\}_{i\in \Lambda}}$ depend on the specific problems under consideration.
These constraints are explicitly detailed in the statements of our main Results 1--3. 
In Result~1, there are no restrictions on $f(\{\hat{n}_i\}_{i\in \Lambda})$; in other words, arbitrary long-range boson-boson couplings are allowed. Result~2 requires a finite interaction length, but no additional constraints. In Result~3, alongside a finite interaction length, we assume that the form of the function is polynomial.
Furthermore, similar to the Lieb-Robinson bound in spin/fermion systems, all our results are applicable to Hamiltonians with arbitrary time dependences.
Although any time-dependences are allowed for Results~1 and 2, we need an additional condition on the norm of the derivative as in Ineq.~\eqref{def_g'_upp} for the proof of Result~3.

We denote the subset Hamiltonian supported on $X\subset \Lambda$ by $H_X$, which picks up all the interactions included in $X$. 
The time evolution of an operator $O$ by the Hamiltonian $H_X$ is expressed as $O(H_X,t):=e^{iH_Xt} O e^{-iH_Xt}$. 
In particular, we denote $O(H,t)$ by $O(t)$ for simplicity.

 \begin{figure}[tt]
\centering
\includegraphics[clip, scale=0.5]{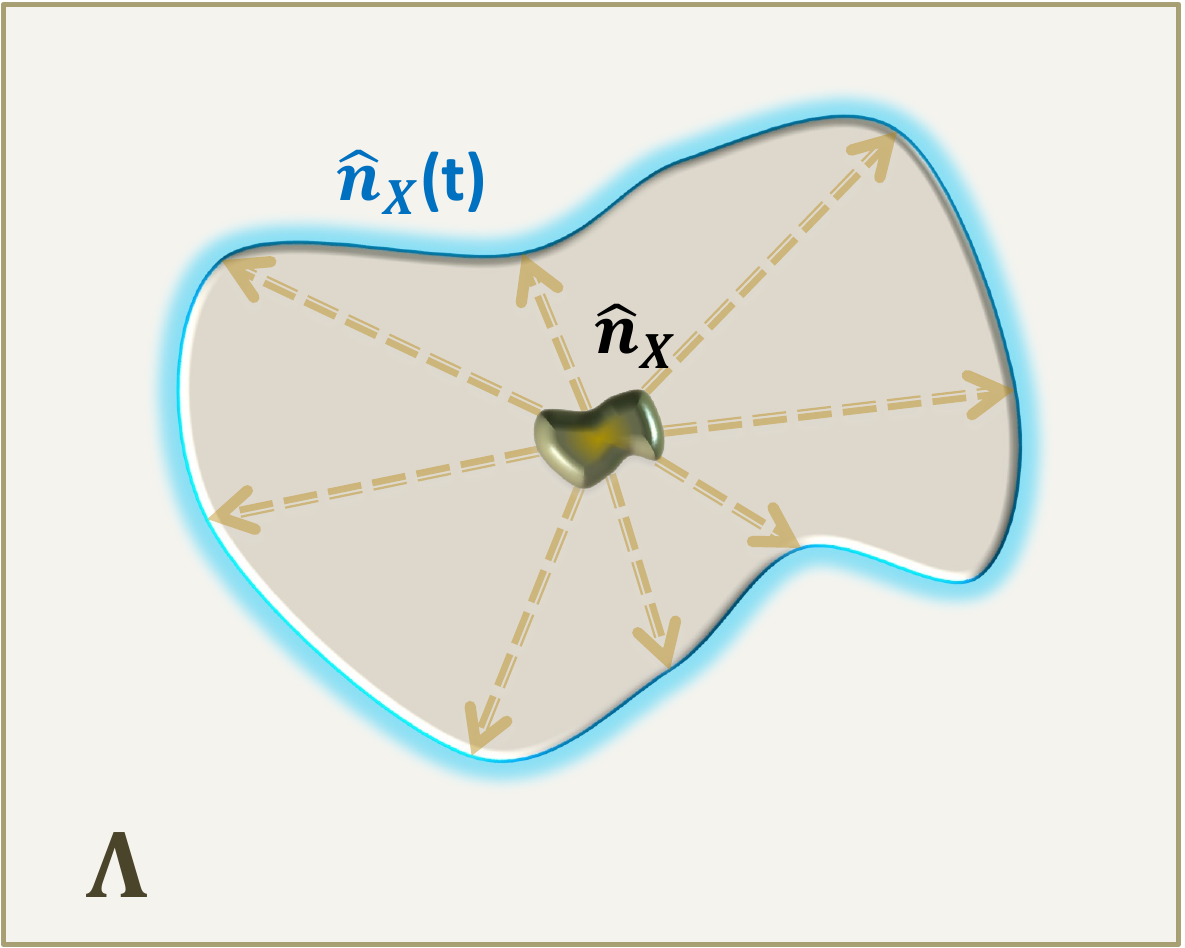}
\caption{Setup of the boson particle transport. 
We consider a boson number operator $\nb_X$ on a region $X$, where $\Lambda$ is the total system.
After time $t$, the bosons initially concentrated on $X$ spread outside the region at a certain speed.  
Assuming there are no bosons outside $X$ at the initial time (i.e., $t=0$), 
the boson particle transport is known to have a finite speed~\cite{PhysRevA.84.032309}; that is, it is approximately restricted in a region $X[R]$ with $R\approx t$ (enclosed by the blue shaded line).
Result~1 generalizes the result to arbitrary initial states up to a logarithmic correction. 
The result plays a crucial role in truncating the boson numbers at each site after time evolution while guaranteeing the desired precision.
}
\label{fig:Fig2}
\end{figure}

We first focus on how fast the bosons spread from a region $X\subset \Lambda$ to the outside (see also Fig.~\ref{fig:Fig2}) by adopting the notation of the extended subset $\bal{X}{r}$ by length $r$ as
\begin{align}
\bal{X}{r}:= \{i\in \Lambda| \dist_{i,X} \le r \} , \label{main_def:bal_X_r}
\end{align}
where $\dist_{i,X}$ is the distance between the subset $X$ and the site $i$.
When $X$ is given by one site (i.e., $X=\{i\}$), $i[r]$ simply denotes the ball region centered at the site $i$. 
We here consider the time evolution of the boson number operator $\nb_X:= \sum_{i\in X}\nb_i$.
We prove that the higher order moment for boson number operator $\brr{\nb_X(t)}^s$ 
is upper-bounded by $\nb^s_{X[R]}$ with an exponentially decaying error with $R$, i.e., $\brr{\nb_X(t)}^s \preceq \nb^s_{X[R]} + e^{-\orderofsp{R/t}}$.  
Throughout this study, we use the notation $O_1 \preceq O_2$ that means $\tr \brr{\sigma(O_2-O_1)} \ge 0$ for an arbitrary quantum state $\sigma$.
Our first result is roughly described by the following statement: 

{\it Result~1.} Let us consider arbitrary boson-boson interactions $V$ in Eq.~\eqref{bose_hubbard_ham} without any assumptions on the function $f$.
For $R\ge c_0 t\log t$, the time-evolution $\nb_X(t)$ satisfies the operator inequality of 
\begin{align}
[\nb_X(t)]^s \preceq \brr{ \nb_{X[R]} + \delta \nb_{X[R]} +c_2 ts }^s, \notag 
\end{align}
where $ \delta \nb_{X[R]} =e^{-c_1R/t} \sum_{j\in \Lambda} e^{-c'_1\dist_{j,X[R]}}\nb_j $ and $\{c_0,c_1,c_1',c_2\}$ are the constants of $\orderof{1}$. 
The operator $ \delta \nb_{X[R]} $ is as small as $e^{-\orderofsp{R/t}}$ if there are not many bosons around the region $X[R]$.
We can apply this theorem to a wide range of setups.
Interestingly, it holds for systems with arbitrary long-range boson-boson interactions, such as the Coulomb interaction. 
Moreover, we can also apply the theorem for imaginary time evolution $\nb_X(it)= e^{-t H} \nb_X e^{tH}$.

From the theorem, we can see that the speed of the boson transport from one region to another is almost constant; at most, it grows logarithmically with time. 
This theorem gives a complete generalization of the result in Ref.~\cite{PhysRevA.84.032309}, 
which discusses the boson transport for initial states that all the bosons are concentrated in a particular region. 
If an initial state has a finite number of bosons at each site,
the probability distribution for the number of bosons still decays exponentially after a time evolution. Herein, the decay form is determined by Result~1.
The estimation provides critical information for simulating the quantum boson systems with guaranteed precision (see Result~3).

Lastly, we notice that Result~1 does not imply that the operator $\nb_X(t)$ is approximated onto region $X[R]$; that is, we cannot ensure 
$\nb_X(t) \approx \nb_X(H_{X[R]},t)=e^{iH_{X[R]}t} O e^{-iH_{X[R]}t}$ as in~\eqref{LR_tightest}. 
For example, if we consider a phase operator $e^{i\nb_X}$, 
the influence can propagate acceleratingly (see below for an explicit example).

\subsection*{Lieb-Robinson bound}

 \begin{figure}[tt]
\centering
\includegraphics[clip, scale=0.45]{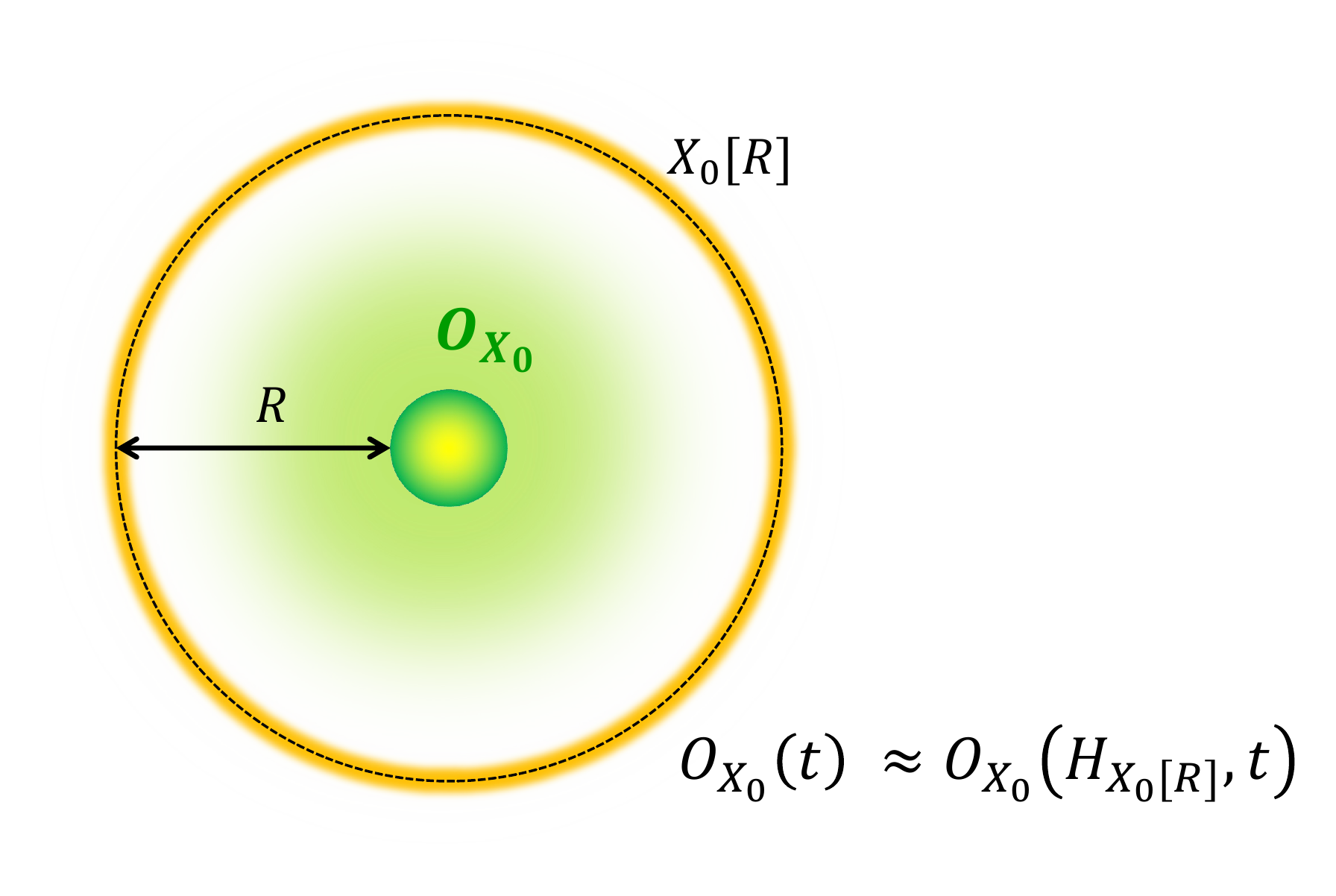}
\caption{Setup of information propagation. 
We consider an operator $O_{X_0}$ supported on the subset $X_0$ and approximate the time-evolved operator $O_{X_0}(t)$ 
onto the extended region $X_0[R]$ (enclosed by the orange shaded line).
Additionally, we assume that the boson number distribution at each site is sub-exponentially suppressed for an initial state, as in~\eqref{low-boson density condition:cond}.
Then, as long as $R\ge t^D \polylog(t)$, the approximation error between $O_{X_0}(t)$ by $O(H_{X_0[R]},t)$ decays sub-exponentially with $R$. 
The effective light cone for the information propagation grows polynomially with time as $R \approx t^D$. 
}
\label{fig:Fig3}
\end{figure}

We next consider the approximation of the time-evolved operator $O_{X_0}(t)$ by $O(H_{X_0[R]},t)$ using the subset Hamiltonian $H_{X_0[R]}$ (see Fig.~\ref{fig:Fig3}), where $O_{X_0}$ is an arbitrary operator. 
Regarding the Holevo capacity, the approximation error characterizes all the information that propagates outside the region $X_0[R]$~\cite{PhysRevLett.97.050401}.
In the following, as a natural setup, we consider an arbitrary initial state with low-boson density condition:
\begin{align}
\tr(\rho_0 \nb_i^s) \le \frac{1}{e} \br{\frac{b_0}{e} s^\kappa }^s ,
\label{low-boson density condition:cond}
\end{align}
where $b_0$ and $\kappa$ ($\ge1$) are the constants of $\orderof{1}$. 
From this condition, the boson number distribution at each site decays (sub)-exponentially at the initial time. 
The simplest example is the Mott state, where a finite fixed number of bosons sit on each site. 
We emphasize that without assuming any conditions for the boson number, there is no speed limit for general information propagation. 
More precisely, the speed of information propagation is directly proportional to the number of bosons at local sites~\cite{PhysRevA.85.053625}. This underscores why the low-boson-density condition is the minimal assumption required to establish a meaningful Lieb-Robinson bound.
This point also makes a clear difference between the information propagation and the particle transport, in which no conditions are imposed for initial states in Result~1.

%
%

Under the condition~\eqref{low-boson density condition:cond}, we can prove the following statement:

{\it Result~2.} 
Let us assume that the range of boson-boson interactions $V$ is finite in Eq.~\eqref{bose_hubbard_ham}.
Then, for an arbitrary operator $O_{X_0}$ ($\norm{O_{X_0}}=1$), the time evolution $O_{X_0}(t)$ is well approximated by using the subset Hamiltonian on $X_0[R]$ with the error of
\begin{align}
&\norm{\brr{ O_{X_0}(t) - O_{X_0}(H_{X_0[R]}, t) } \rho_0}_1 \le e^{-C(R/t^D)^{\frac{1}{\kappa D}}} ,
\label{LR_tightest}
\end{align}
for $R\ge t^D \polylog(t)$, where $\norm{\cdot}_1$ is the trace norm and 
$C$ is an $\orderof{1}$ constant. 

 \begin{figure*}[tt]
\centering
{\includegraphics[clip, scale=0.38]{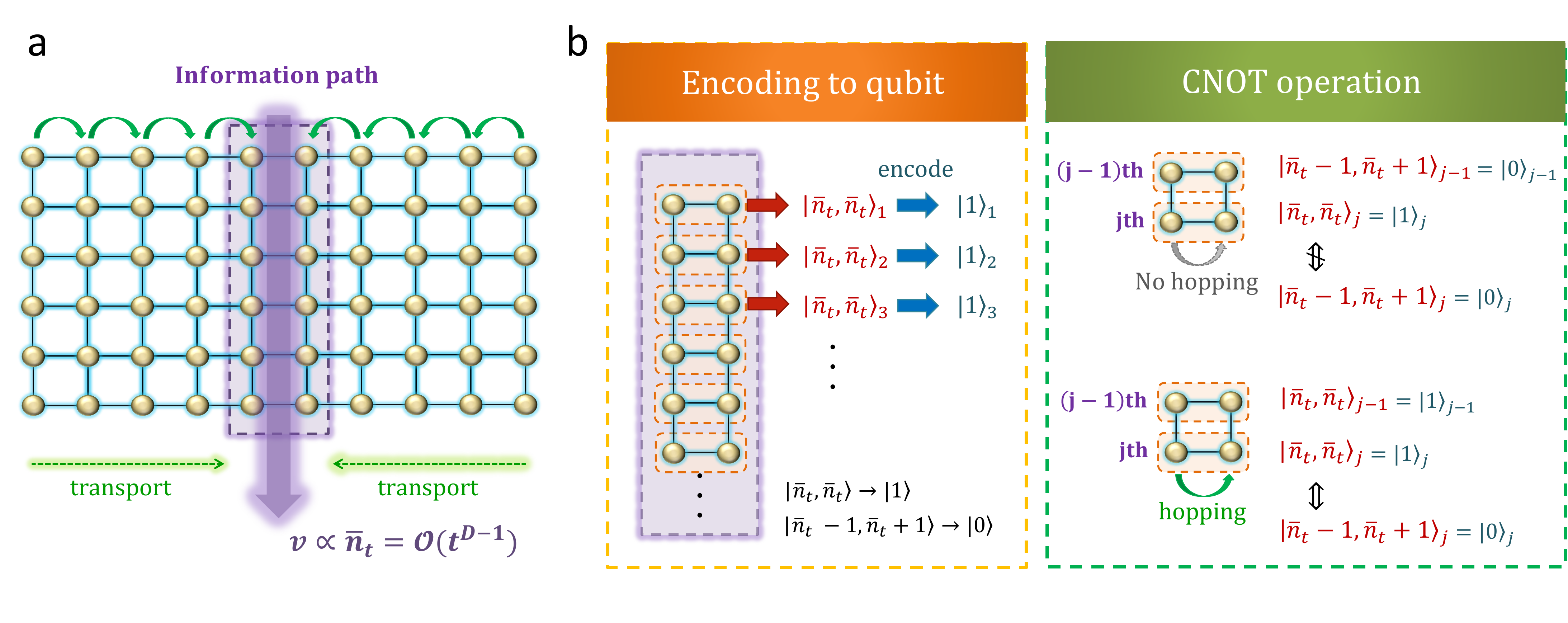}}
\caption{Outline of the proposed protocol to achieve the fastest information transfer by interacting with boson systems.
(a) Schematic picture of information transfer. In the initial state, one boson sits at each site. 
In the first half, we transport the boson particles to a particular one-dimensional region called the information path. 
The path is now given by a ladder (denoted by the purple-shaded region). 
Considering the speed of the boson transport is finite, 
we accumulate the bosons in the region within a distance of $(\bar{J}t)$ from the path. 
Therefore, the number of bosons $\bar{n}_t$ at each site on the path is proportional to $(\bar{J}t)^{D-1}$ (in the picture $D=2$). 
Afterward, we switch off the boson hopping and isolate the information path so that the bosons cannot escape from the path.
(b) Encoding qubits and CNOT operation. The speed of the information propagation is proportional to $\bar{n}_t$. 
By encoding the qubit on the $j$th row as $\ket{\bar{n}_t,\bar{n}_t}_j \to \ket{1}_j$ and $\ket{\bar{n}_t-1,\bar{n}_t+1}_j \to \ket{0}_j$, 
we can implement the CNOT operation by appropriately choosing boson-boson interactions. 
If the state on the $(j-1)$th row is given by $\ket{0}_{j-1}$, there is no boson hopping between the two sites on the $j$th row.
In contrast, if the state on the $(j-1)$th row is given by $\ket{1}_{j-1}$, 
there exists a boson hopping. Herein, the hopping rate of one boson is amplified by $\bar{n}_t$ times, 
and hence, the transition time from $\ket{0}_{j}$ to $\ket{1}_{j}$ is proportional to $1/\bar{n}_t$. 
This enables us to implement the CNOT operation in the time of $1/\bar{n}_t$.}
\label{fig:Fig4}
\end{figure*}

The bound~\eqref{LR_tightest} tells us that any operator spreading is approximated by a local operator on $X_0[R]$ with the accuracy of the right-hand side. It rigorously bounds any information propagation. 
A crucial observation here is that the propagation speed accelerates depending on the spatial dimensions, which is a stark difference from fermionic lattice systems. 

The accelerated information propagation can be interpreted physically as follows. According to Ref.~\cite{PhysRevA.85.053625}, the velocity of information propagation is directly proportional to the number of bosons at local sites in the presence of boson-boson interactions. Consequently, if dynamic processes cause an increase in boson concentrations within specific one-dimensional regions, the boson density in that 1D region will rise over time. As a result, information propagation on this 1D ``information path'' experiences acceleration. This phenomenon is specific to high-dimensional systems. In one-dimensional systems, if bosons concentrate in a specific region, the surrounding areas exhibit sparse boson densities, preventing persistent acceleration.


\subsection*{Optimality of the effective light cone}

As discussed in Result~2, the bound on the speed of information propagation is proportional to $t^{D-1}$.
Herein, we show that the obtained upper bound is qualitatively tight by explicitly developing time evolution to achieve the bound.
As the initial state, we consider the Mott state with only one boson at each site. 
The protocol comprises the following two steps.
\begin{enumerate}
\item{} First, we set the path of the information propagation. We transport the bosons such that they are collected on the path. Here, the path is given by the one-dimensional ladder [see Fig.~\ref{fig:Fig4} (a)]. 
\item{} Then, we encode the qubits on the ladder and realize the CNOT gate by using two-body interactions and boson hopping [see Fig.~\ref{fig:Fig4} (b)]. 
\end{enumerate}
We here consider the first step.
Let us consider two nearest neighbor sites $i_1$ and $i_2$ where the state is given by $\ket{N}_{i_1} \otimes \ket{1}_{i_2}$.
We denote the Fock state on the site $i$ by $\ket{m}_i$ with $m$, the boson number.  
Our task is to move the bosons on the site $i_1$ to the site $i_2$, that is, $$\ket{N}_{i_1} \otimes \ket{1}_{i_2} \to \ket{0}_{i_1} \otimes \ket{N+1}_{i_2}.$$
Such a transformation is realized by combining the free-boson Hamiltonian and Bose-Hubbard Hamiltonian, 
and the necessary time is inversely proportional to the hopping amplitude of the bosons $\bar{J}$ (see the Method section). 
Therefore, based on the time evolution of $t/2$, we can concentrate the bosons in a region within a distance of $(\bar{J}t)$ from the boson path.
The boson number at one site on the boson path is now proportional to $(\bar{J}t)^{D-1}$. 
Herein, we denote the quantum state on the boson path by $\bigotimes_{j=1}^L \ket{\bar{n}_t,\bar{n}_t}_j$ with $\bar{n}_t \propto (\bar{J} t)^{D-1}$. 

In the second step, we encode the quantum states $\ket{\bar{n}_t,\bar{n}_t}_j$ and $\ket{\bar{n}_t-1,\bar{n}_t+1}_j$ on the $j$th row by $\ket{1}_j$ and $\ket{0}_j$, respectively, to prove that the time required to implement the controlled-NOT (CNOT) gate is at most $\orderof{t^{-D+1}}$. 
By using appropriate two-body interactions between the $(j-1)$th row and $j$th row, 
no boson hopping is observed on the $j$th row when the $(j-1)$th row is given by $\ket{\bar{n}_t,\bar{n}_t}_j$; 
however, there exists boson hopping on the $j$th row when the $(j-1)$th row is given by $\ket{\bar{n}_t-1,\bar{n}_t+1}_j$.
One can control the boson-boson interactions such that only the hopping between $\ket{\bar{n}_t,\bar{n}_t}_j\leftrightarrow \ket{\bar{n}_t-1,\bar{n}_t+1}_j $ occurs. 
We thus realize the CNOT operation on the information path (see the Method section), 
where the sufficient time required to realize it is proportional to $1/(\bar{J}\bar{n}_t) \propto1/(\bar{J}^D t^{D-1})$.
Therefore, in half of the total time $t/2$, we can implement $(\bar{J}t)^D$ pieces of the CNOT gates, 
which allows us to propagate the information from one site to another as long as the distance between these sites is smaller than $(\bar{J}t)^D$.


One can demonstrate that the number of CNOT operations is directly linked to the distance of the operator spread, following the discussion in Ref.~\cite{PhysRevLett.97.050401}. To illustrate this, consider two types of operations: flipping or non-flipping the endmost qubit on the information path at the time $t/2$, where the boson concentration has been over. 
By flipping the endmost qubit to $\ket{1}$, $m$-sequential CNOT operations transform the state to $\ket{1}^{\otimes m} \ket{0}^{\otimes \ell-m}$, while without flipping, the state remains unchanged~$\ket{0}^{\otimes \ell}$. Here, $\ell$ represents the total length of the information path. By encoding classical information as flipping ($=0$) or non-flipping ($=1$) of the endmost qubit, one can transmit 1 bit of information through the sequence of CNOT operations. The connection between Holevo capacity and operator spreading~\cite{PhysRevLett.97.050401} implies that this process necessarily induces the operator spreading of the flipping unitary to the endmost qubit in the Heisenberg picture. Thus, $\bar{J}^Dt^{D}$ CNOT operations during the time $t$ achieves the Lieb-Robinson velocity of $\bar{J}^Dt^{D-1}$. This accelerating information propagation must be clearly distinguished from boson transport with a constant velocity, as in Result~1.

We finally discuss a comparison with the mechanism proposed by Eisert and Gross~\cite{PhysRevLett.102.240501}. In their model, the Hamiltonian effectively amplifies the hopping of bosons, with hopping amplitudes directly proportional to their positions. In contrast, our mechanism relies on dynamically adjusting the local boson numbers. The crucial aspect is that, in the presence of boson-boson interactions, the speed of information propagation is enhanced by the local boson numbers.

\subsection*{Provably efficient digital quantum simulation}

In the final application, we consider the quantum simulation of time evolution to estimate the sufficient number of quantum gates that implement the bosonic time evolution $e^{-iHt}$ acting on an initial state $\rho_0$.
In detail, we prove the following statement:

{\it Result~3.} 
Let us assume that the boson-boson interactions in $V=f\br{ \{\nb_i\}_{i\in \Lambda}}$ are finite in length, and
the function $f\br{ \{\nb_i\}_{i\in \Lambda}}$ is given by a polynomial with limited degrees and coefficients.
For an arbitrary initial state $\rho_0$ with the condition~\eqref{low-boson density condition:cond}, 
the number of elementary quantum gates for implementing $e^{-iHt} \rho_0 e^{iHt}$ up to an error $\epsilon$ is at most 
\begin{align}
|\Lambda| t^{D+1} \polylog(|\Lambda| t/\epsilon) ,
\label{simulation_circuit_complexity}
\end{align}
with the depth of the circuit $t^{D+1} \polylog(|\Lambda|t/\epsilon)$, where the error is given in terms of the trace norm. 
Note that $|\Lambda|$ is the number of sites in the total system.

We extend the Haah-Hastings-Kothari-Low (HHKL) algorithm~\cite{8555119} to the interacting bosons. 
Before going to the algorithm, we truncate the boson number at each site up to $\bar{q}$. 
We define $\bar{\Pi}_{\bar{q}}$ as the projection onto the eigenspace of the boson number operators $\{\nb_i\}_{i\in \Lambda}$ with eigenvalues smaller than or equal to $\bar{q}$.  
Then, we consider the time evolution by the effective Hamiltonian $\bar{\Pi}_{\bar{q}}H\bar{\Pi}_{\bar{q}}$. 
In the following, we denote the projected operator $\bar{\Pi}_{\bar{q}}O\bar{\Pi}_{\bar{q}}$ by $\tilde{O}$ for simplicity. 
Assuming low-boson density~\eqref{low-boson density condition:cond}, 
Result~1 gives the upper bound of the approximation error of
\begin{align}
\norm{\rho_0(t) - \rho_0(\tilde{H},t) }_1 \le |\Lambda| e^{-c_3 \brr{\bar{q}/(t\log t)^D}^{1/\kappa}} . 
\label{approximation_error_effective_Ham}
\end{align}
Therefore, to achieve the error of $\epsilon$, we need to choose as $\bar{q}=(t\log t)^D \log^{\kappa} (|\Lambda|/\epsilon) = t^D \polylog(|\Lambda| t/\epsilon)$. 

In the algorithm, we first adopt the interaction picture of the time evolution:
\begin{align}
\label{decomposition_of_effective_time_evo}
e^{-i\tilde{H}t} = e^{-i\tilde{V}t} \mathcal{T} e^{-i\int_0^t \tilde{H}_0(\tilde{V},x) dx} .
\end{align}
First, the time evolution by $\tilde{V}$ is decomposed to $\orderof{|\Lambda|}$ pieces of the local time evolution, considering the interaction terms in $V$ commute with each other. 
Second, we implement the time evolution for the time-dependent Hamiltonian $\tilde{H}_0(\tilde{V},x)$. 
The Hamiltonian $\tilde{H}_0(\tilde{V},x)$ contains interaction terms like $e^{i\tilde{V}x} \tilde{b}_i \tilde{b}_j^\dagger e^{-i\tilde{V}x}$, 
which has a bounded norm by $\orderof{\bar{q}}$ and is described by an $\orderof{1}$ sparse matrix. 
Herein, we say that an operator $O$ is $d$ sparse if it has at most $d$ nonzero entries in any row or column. 
Moreover, the norm of the derivative $\norm{d (e^{i\tilde{V}x} \tilde{b}_i \tilde{b}_j^\dagger e^{-i\tilde{V}x})/dx}$ is upper-bounded by $\poly(\bar{q})$ 
owing to the assumption for $f\br{ \{\nb_i\}_{i\in \Lambda}}$.

Therefore, the problem is equivalent to implementing the time evolution of the Hamiltonian with the following three properties: 
i) the norms of the local interaction terms are upper-bounded by $\orderof{\bar{q}}$, 
ii) the interaction terms are described by $\orderof{\bar{1}}$ sparse matrices, and
iii) the time derivative of the local interactions has a norm of $\poly(\bar{q})$ at most. 
Such cases can be treated by simply combining the previous works in Refs.~\cite{8555119} and \cite{berry2017exponential}. 
As shown in the Method section, the total number of elementary circuits is at most $(|\Lambda| t\bar{q}) \polylog (|\Lambda| t\bar{q}/\epsilon)$, which reduces to~\eqref{simulation_circuit_complexity} 
by applying $\bar{q}= t^D \polylog(|\Lambda| t/\epsilon)$. 

\section*{Discussion}

Our study clarified the qualitatively tight Lieb-Robinson light cone for systems with the Bose-Hubbard type Hamiltonian.
Still, we have various improvements to implement in future works.
First, the obtained bounds incorporate logarithmic corrections, but there is a possibility of their removal through a refinement of the current analyses. 
Currently, it does not seem to be a straightforward problem to remove them using our existing techniques.
A possible starting point to achieve this is to consider the difference between average values as $\tr \brr{ \br{O_{X_0}(t)-O_{X_0}(H_{X_0[R]}, t)} \rho_0}$ instead of the trace norm $\norm{ \br{O_{X_0}(t)-O_{X_0}(H_{X_0[R]}, t)} \rho_0}_1$. 
While this quantity may not capture the propagation of total information, a significantly stronger bound can be proven in one-dimensional systems, where the light cone form strictly follows a linear form with time~\cite{PhysRevX.12.021039}. 
Through the refinement and combination of existing techniques, there is a possibility of eliminating the logarithmic corrections in our current bound in future studies.
Second, although acceleration of the information propagation is possible, 
there are particular cases where the linear light cone is rigorously proved~\cite{PhysRevLett.127.070403,PhysRevX.12.021039}.
Therefore, by appropriately avoiding the acceleration mechanism depicted in our protocol, 
we can possibly establish a broader class that retains the linear light cone.  
Third, it is an interesting open question to seek the possibility of improving the current gate complexity $|\Lambda| t^{D+1} \polylog(|\Lambda| t/\epsilon) $ and clarify the optimal gate complexity to simulate the quantum dynamics.
We hope a more simplified technique may appear for implementing these improvements, considering the current proof techniques are rather complicated.

Other directions include generalizing the current results beyond the Bose-Hubbard type Hamiltonian~\eqref{bose_hubbard_ham}. 
As a straightforward extension, it is intriguing to investigate under what conditions boson-boson interactions like $b_{i_1}b_{i_2}b^\dagger_{i_3}b^\dagger_{i_4}$ can lead to information propagation with a limited speed. 
In this scenario, relying solely on the low-boson-density condition proves insufficient for regulating the speed of information propagation (cf. Ref.~\cite{vanvu2023optimal}).
Still, such an extension has practical importance. 
For example, when applying the quasi-adiabatic continuation technique~\cite{PhysRevB.72.045141} to the Bose-Hubbard type models, 
we must derive the Lieb-Robinson bound for the quasi-adiabatic continuation operator, which is no longer given by the form of Eq.~\eqref{bose_hubbard_ham}.
We expect that our newer techniques will be helpful in developing interacting boson systems where the effective light cone is at most polynomial with time.

Finally, it is intriguing to experimentally or numerically observe the supersonic propagation of quantum signals using a mechanism similar to that illustrated in Figure~\ref{fig:Fig4}. In our protocol, we employed highly artificial boson-boson interactions. Then, a significant open problem is whether acceleration can occur even under time-independent Hamiltonians. We anticipate that the boson transport in the initial step can be achieved by reversing the time evolution (see Supplementary Note 11).
As for the second step in the protocol, Ref.~\cite{PhysRevA.85.053625} has already noted that the group velocity of the propagation front of correlations is proportional to the boson number at each site. Hence, we believe that the proposed acceleration mechanism can be realized within the current experimental setups.

\section*{Methods}

\subsection*{Outline of the proof for Result~1}

Throughout the proof, we denote $\mathcal{O}(1)$ as an arbitrary finite combination of the fundamental parameters, which are detailed in Supplementary Table 1.

\begin{figure}[tt]
\centering
\includegraphics[clip, scale=0.4]{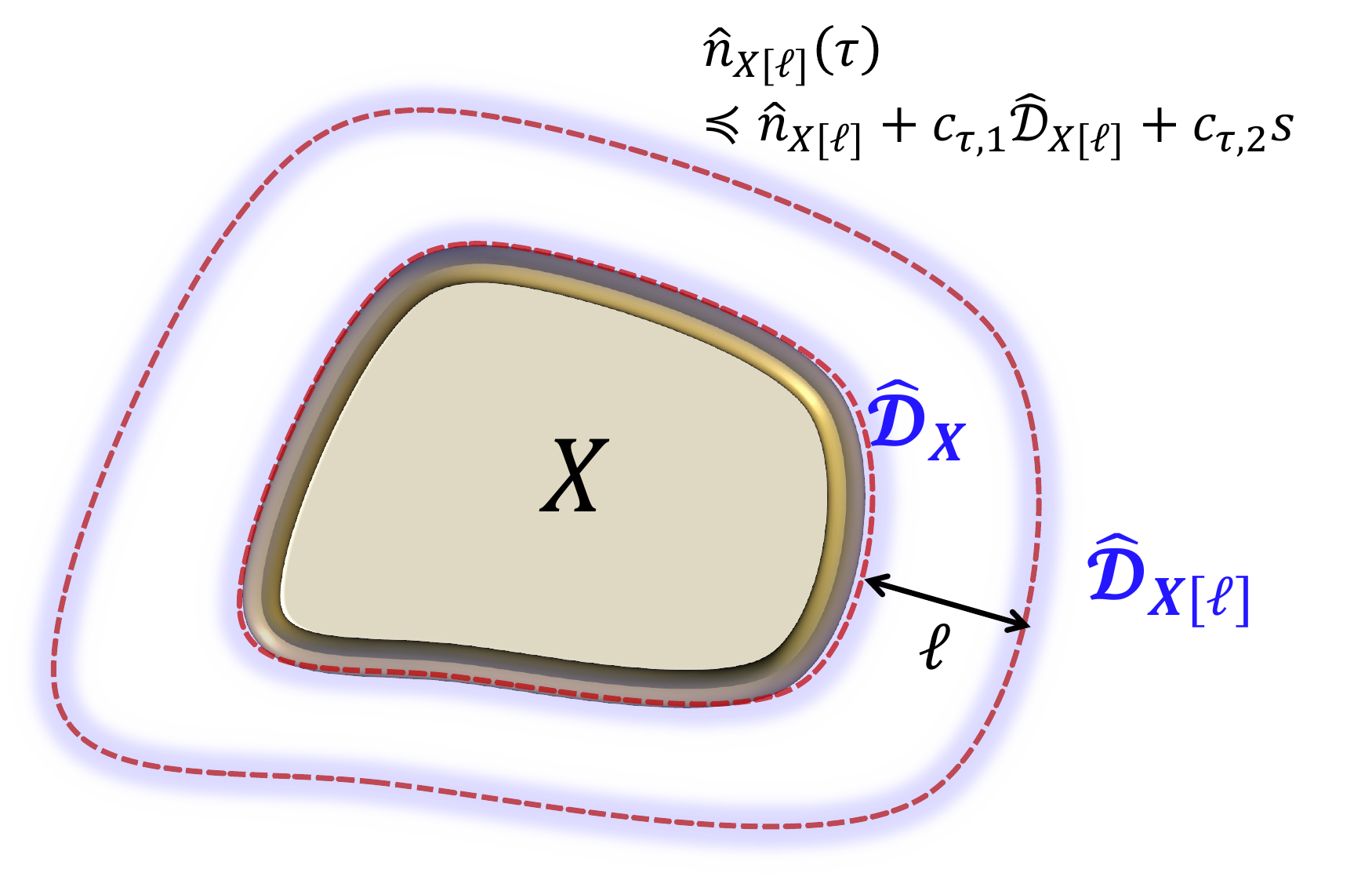}
\caption{Time evolution of boson number operator.
A short-time evolution of $\nb_X(\tau)$ is bounded from above by $\nb_X + c_{\tau,1} \hat{\mathcal{D}}_X + c_{\tau,2}$. 
The influence of $\hat{\mathcal{D}}_X$ exponentially decays with the distance from the surface region $\partial X$ (denoted by the blue shaded region).
The contribution of $\hat{\mathcal{D}}_X$ may be fatal in some classes of the initial states. 
If all bosons concentrate on $\partial X$ in an initial state $\rho$, $\tr(\rho \hat{\mathcal{D}}_X)$ can be as large as $\tr(\rho \nb_X)$.
However, by considering the time evolution of $\nb_{X[\ell]}(\tau)$,
the contribution by the bosons on $\partial X$ is exponentially small with $\ell$ in the operator $\hat{\mathcal{D}}_{X[\ell]}$.
This motivates us to consider the minimization problem~\eqref{optimization_ell} for all choices of $X[r]$ ($0\le r \le \ell$) 
to derive the upper bound~\eqref{optimization_ell_solved}.}
\label{fig:Fig6}
\end{figure}

First, we present several key ideas to prove Result~1 by considering $[\nb_X(t)]^s$ with $s=1$ for simplicity.
By refining the technique in Ref.~\cite{PhysRevA.84.032309}, we begin with the following statement (Supplementary Subtheorem~1):
\begin{align}
\label{ineq_schuch}
\nb_X(\tau) \preceq \nb_X + c_{\tau,1} \hat{\mathcal{D}}_X + c_{\tau,2},
\end{align}
with $\hat{\mathcal{D}}_X := \sum_{i\in \partial X} \sum_{j\in \Lambda} e^{-\dist_{i,j}} \nb_j $, 
where $c_{\tau,1}$ and $c_{\tau,2}$ are the constants that grow exponentially with $\tau$, that is, $c_{\tau,1},c_{\tau,2}=e^{\orderof{\tau}}$. 
We define $\partial X$ as the surface region of the subset $X$, that is, $\partial X = \{i\in X| \dist_{i,X^\co}=1\}.$
From the definition, the operator $\hat{\mathcal{D}}_X$ is roughly given by the boson number operator around the surface region of $X$ with an exponential tail (see Fig.~\ref{fig:Fig6}).
In the inequality~\eqref{ineq_schuch}, the coefficients $c_{\tau,1},c_{\tau,2}$ grow exponentially with $\tau$, and hence the inequality becomes meaningless for large $\tau$, 
which has been the main bottleneck in~\cite{PhysRevA.84.032309}.

The key technique to overcome the above-mentioned difficulty is the connection of the short-time evolution $e^{-iH\tau}$ with $\tau$ a constant of $\orderof{1}$, 
which has played an important role in the previous works~\cite{Kuwahara_2016_njp,PhysRevLett.126.030604,PhysRevLett.127.070403}.
We first refine the upper bound~\eqref{ineq_schuch} to  
\begin{align}
\nb_X(\tau) \preceq \nb_{X[\ell]} + e^{-\orderofsp{\ell}}+ c_{\tau,2} .
\label{upp_extended_X}
\end{align}
If the above inequality holds, 
by iteratively connecting the short-time evolution $(t/\tau)$ times, we obtain 
\begin{align}
\label{upp_extended_X_connect}
\nb_X(t) \preceq \nb_{X[(t/\tau)\ell]} + \frac{t}{\tau} \br{e^{-\orderofsp{\ell}} +c_{\tau,2}} ,  
\end{align}
which yields the desired inequality in Result~1 for $s=1$ by choosing $R=(t/\tau)\ell$ (or $\ell=\tau R/t$). 

Then, we aim to derive the bound~\eqref{upp_extended_X} using the inequality~\eqref{ineq_schuch}.
However, the derivation is not straightforward because the inequality $\nb_X  + c_{\tau,1} \hat{\mathcal{D}}_X + c_{\tau,2} \preceq \nb_{X[\ell]} + e^{-\orderofsp{\ell}}+ c_{\tau,2}$ does not hold in general. 
For example, let us consider a quantum state $\rho$ such that all bosons concentrate on $\partial X$ (see Fig.~\ref{fig:Fig6}). 
Then, we have 
\begin{align}
\tr(\rho\hat{\mathcal{D}}_X) \propto \tr\br{\rho \nb_X}  ,
\label{rho_mathcal_D_X}
\end{align}
which makes $\tr[\rho (\nb_X + c_{\tau,1} \hat{\mathcal{D}}_X)] =  [1+ \Omega(1)]  \tr\br{\rho \nb_X}$.
Therefore, connecting the time evolution $(t/\tau)$ times yields an exponential term $[1+ \Omega(1)]^{t/\tau}$. 
To avoid such exponential growth, we first upper-bound $\nb_X(\tau) \preceq \nb_{X[\ell]}(\tau)$ for $\forall \tau>0$ as a trivial bound, where $\ell$ ($\ge 0$) can be arbitrarily chosen.
Then, we use the inequality~\eqref{ineq_schuch} to obtain
\begin{align}
\nb_X(\tau) \preceq \nb_{X[\ell]}(\tau) \preceq \nb_{X[\ell]} + c_{\tau,1} \hat{\mathcal{D}}_{X[\ell]} + c_{\tau,2}. \notag 
\end{align}
The point here is that $\hat{\mathcal{D}}_{X[\ell]}$ is exponentially localized around the surface of $X[\ell]$. 
Using the above operator inequality, we can resolve the drawback in Eq.~\eqref{rho_mathcal_D_X} that originates from the concentration around the boundary $\partial X$. 
Here, for the quantum state $\rho$ which has the boson concentration on the region $\partial X$, we have
\begin{align}
\tr(\rho\hat{\mathcal{D}}_{X[\ell]}) \approx e^{-\orderofsp{\ell}} \tr\br{\rho \nb_X}  \notag 
\end{align}
instead of Eq.~\eqref{rho_mathcal_D_X}.
Therefore, the contribution from the operator $\hat{\mathcal{D}}_{X[\ell]}$ is exponentially small with the length $\ell$.

From the above discussion, to derive a meaningful upper bound for short-time evolution, we need to consider 
\begin{align}
\label{optimization_ell}
\min_{r :0\le r \le \ell}\tr\brr{\rho \br{\nb_{X[r]} + c_{\tau,1} \hat{\mathcal{D}}_{X[r]} + c_{\tau,2}}} 
\end{align}
for an arbitrary quantum state $\rho$, which also gives an upper bound of $\tr [\rho\nb_{X}(\tau)]$.
We cannot solve the optimization problem~\eqref{optimization_ell} in general but ensure the existence of $r\in [0,\ell]$ that satisfies the following inequality (see Supplementary Lemma~13 and Supplementary Proposition~16):
\begin{align}
\label{optimization_ell_solved}
\nb_X(\tau) \preceq \nb_{X[\ell]} + c_{\tau,1} \delta_{\ell} (\nb_{X[\ell]}+ \hat{\mathscr{D}}_{X[\ell]} )+ c_{\tau,2} ,
\end{align}
where $\delta_{\ell}$ decays exponentially with $\ell$, i.e., $\delta_{\ell}=e^{-\orderofsp{\ell}}$, and we define 
$\hat{\mathscr{D}}_{X[\ell]}=\sum_{j\in X[\ell]^\co} e^{-3\dist_{j,X[\ell]}/4} \nb_j $. 
The inequality~\eqref{optimization_ell_solved} is given in the form of the desired inequality~\eqref{upp_extended_X}, 
which also yields the upper bound~\eqref{upp_extended_X_connect} (Supplementary Theorem~1).
More precisely, the iterative use of the inequality~\eqref{optimization_ell_solved} yields an additional coefficient $(1+c_{\tau,1} \delta_{\ell})^{t/\tau}$ to the first term $\nb_{X[(t/\tau)\ell]} $ in~\eqref{upp_extended_X_connect}. We need the condition $R\ge c_0 t\log t$ (or $\ell\propto \log(t)$) to ensure $(1+c_{\tau,1} \delta_{\ell})^{t/\tau} \lesssim 1+c_{\tau,1}t \delta_{\ell} $.
Therefore, we prove the main inequality in Result~1.

For a general $s$th moment, we apply similar analyses to the case of $s=1$. 
As a remark, we cannot simply obtain $[\nb_X(\tau)]^s \preceq (\nb_X + c_{\tau,1} \hat{\mathcal{D}}_X + c_{\tau,2})^s$ from the inequality~\eqref{ineq_schuch} considering 
$O_1\preceq O_2$ does not imply $O_1^s\preceq O_2^s$.
Instead, we obtained the following modified upper bound: 
\begin{align}
\label{ineq_schuch_s}
[\nb_X(\tau)]^s \preceq \br{\nb_X + c_{\tau,1} \hat{\mathcal{D}}_X + c_{\tau,2}s }^s . 
\end{align}
Then, we consider a similar procedure to the optimization problem~\eqref{optimization_ell} 
and obtain an analogous inequality to~\eqref{optimization_ell_solved} (Supplementary Proposition~18). 
This allows us to connect the short-time evolution to derive Result~1 for general $s$.

\subsection*{Non-acceleration of Boson Transport}

\begin{figure}[tt]
    \centering
    \includegraphics[clip, scale=0.25]{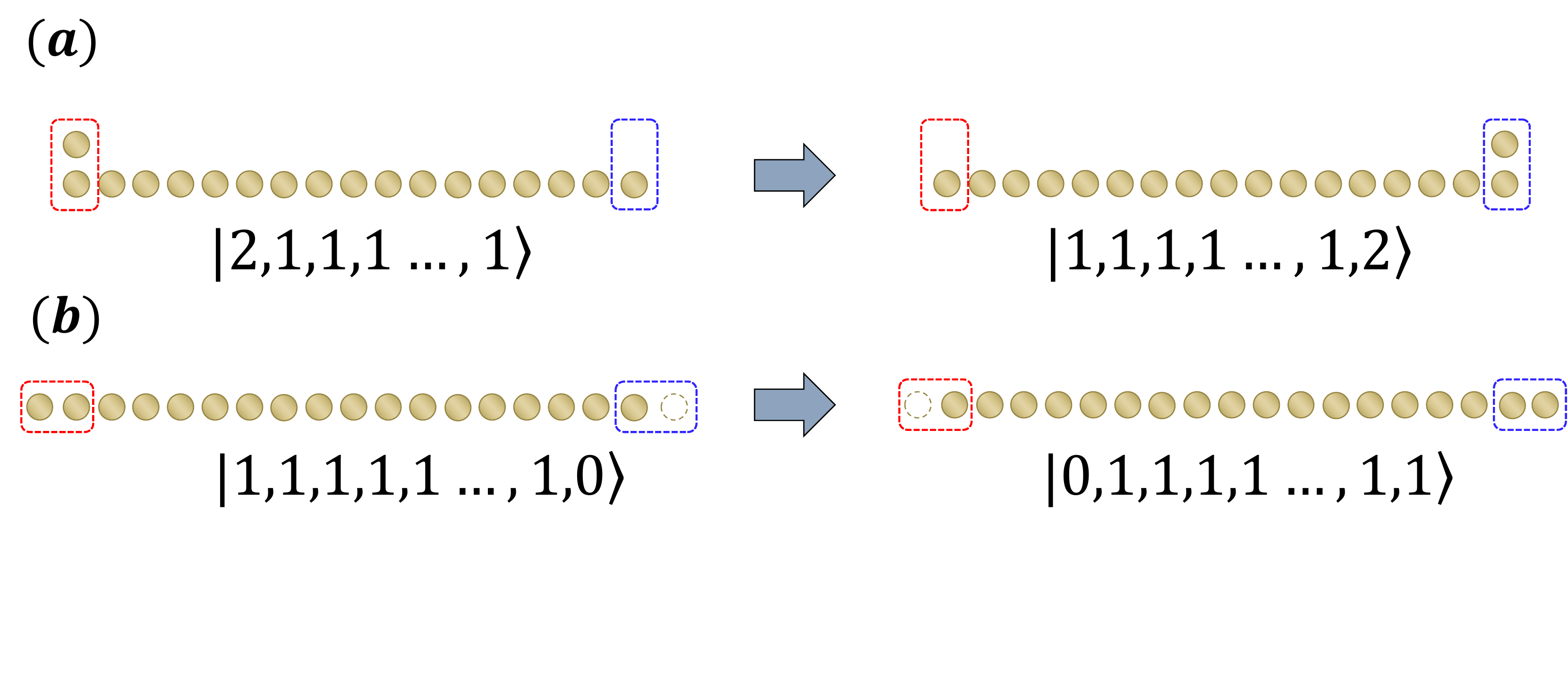}
    \caption{Transformation from $b^\dagger_1 \ket{{\rm M}_1}$ to $b^\dagger_n \ket{{\rm M}_1}$ (a). This process is equivalent to one hopping from left to right of all bosons (b), where the left-end two sites and the right-end two sites are merged into one site, respectively. Then, this process takes time of $\orderof{1}$ for arbitrarily long 1D chains.}
    \label{Fig::pseudo_particle_transport}
\end{figure}

Here, we demonstrate that the protocol in Fig.~\ref{fig:Fig4} cannot induce the acceleration of boson transport. While this protocol enables the transformation
\begin{align}
\label{pseudo_particle_transport}
(b^\dagger_i)^m \ket{{\rm M}_1} \to (b^\dagger_j)^m \ket{{\rm M}_1}
\end{align}
as long as $\dist_{i,j} \lesssim t^D$ (where $\ket{{\rm M}_1}$ is the Mott state with one boson at each site), this process does not imply genuine particle transport due to the indistinguishability of bosons. 
In the first place, even without the Fig.~\ref{fig:Fig4} protocol, the transformation~\eqref{pseudo_particle_transport} can be achieved in a constant time for arbitrary distances (see Fig.~\ref{Fig::pseudo_particle_transport}). To characterize particle transport, it is essential to ensure that the increased bosons indeed originate from a distant region.
This can be achieved in the following cases:
\begin{enumerate}
    \item If $\tr \br { \rho(t) \nb_X} > \tr \br { \rho \nb_{X[R]}}$, we can ensure that a part of the increase in boson number comes from the region $X[R]^\co$, achieving particle transport over a distance $R$.
    \item By making target bosons distinguishable from others (e.g., bosons with the spin degree of freedom), particle transport can be clearly defined.
\end{enumerate}
The first case is addressed in Result~1, where we establish a finite speed. In the second case, we also prove the finite speed of transport by slightly generalizing Result~1. In this case, the Hamiltonian should be generalized to
\begin{align}
    &H=\sum_{\sigma} \sum_{\langle i, j \rangle} J_{i,j} (b_{i,\sigma} b_{j,\sigma}^\dagger +{\rm h.c.} ) +  f\br{ \{\nb_{i,\sigma}\}_{i\in \Lambda, \sigma}} . \notag
\end{align}
Then, the same operator inequality as in Result~1 holds for $\nb_{X,\sigma}(t)$ for corresponding spin degrees $\sigma$.

In the context of this discussion, a more phenomenological explanation to ensure the finite velocity of boson transport is through the particle current. 
The particle current operator $\hat{J}_{i,i+1}$ between the sites $i$ and $i+1$ is defined as~\cite{PhysRevB.55.11029}
\begin{align}
    \hat{J}_{i,i+1} := J (i b_ib_{i+1}^\dagger + \text{{h.c.}}),
\end{align}
where we consider a one-dimensional system for simplicity, and the free Hamiltonian is $H_0=\sum_{i} J (b_ib_{i+1}^\dagger + \text{{h.c.}})$. Usually, the current is defined as the product of particle density and velocity, giving the speed of particle velocity as
\begin{align}
    v_{\text{{transport}}} \sim \norm{\frac{\hat{J}_{i,i+1}}{\nb_i+\nb_{i+1}}} \le 2J,
\end{align}
where we use the operator inequality of $|\hat{J}_{i,i+1}| \preceq 2J ( \nb_i+\nb_{i+1})$ from $|b_ib_{i+1}^\dagger| \preceq \nb_i+\nb_{i+1}$ (see Supplementary Equation 459). Although it is non-trivial to derive our Result~1 only from this discussion, it provides a simple picture of why the speed of particle transport has a finite speed.

\subsection*{Outline of the proof for Result~2: Simpler but looser bound}

Herein, we show how to derive the Lieb-Robinson bound with the effective light cone of $R\propto t^D$ for information propagation. 
The number of bosons created by the operator $O_{X_0}$, say $q_0$, is assumed to be an $\orderof{1}$ constant for simplicity.  
In Supplementary Notes~7, 8, 9, and 10, we treat generic $q_0$, and the obtained Lieb-Robinson bounds depend on $q_0$ (see Supplementary~Theorems~2,3, and 4).
Before going to the tight Lieb-Robinson bound, 
we show the derivation of a looser light cone by using the truncation of the boson number as in Ref.~\cite{PhysRevLett.127.070403}, which gives $R\propto t^{D+1}$ (Supplementary Theorem~2). 
By applying Result~1 with $X=\{i\}$, the time evolution of the boson number operator $\nb_i$ is roughly upper-bounded by the boson number on the ball region 
$i[\ell_t]$, that is, $\nb_{i[\ell_t]}$, where $\ell_t =\orderof{t\log t}$, and we ignored the non-leading terms.
Therefore, if an initial state $\rho_0$ has a finite number of bosons at each site, 
the upper bound of the $s$th moment after a time evolution can be given as 
\begin{align}
\tr\brr{\rho_0(t) \nb_i^s} \lesssim \tr\brr{\rho_0 \nb_{i[\ell_t]}^s} \propto \br{\ell_t^D s^\kappa }^s ,
\end{align}
where we use the condition~\eqref{low-boson density condition:cond} in the second inequality.  
The above inequality characterizes the boson concentration by the time evolution to ensure that the probability distribution of the boson number decays subexponentially 
\begin{align}
\label{upper_bound_prob_dist}
\tr\brr{\rho_0(t) \Pi_{i,\ge x} } \lesssim e^{-\br{x/\ell_t^D}^{1/\kappa}} ,
\end{align}
where $\Pi_{i,\ge x}$ is the projection onto the eigenspace of $\nb_i$ with the eigenvalues larger than or equal to $x$.
Therefore, we expect that the boson number at each site can be truncated up to $\orderof{\ell_t^D}=t^D \polylog(t)$ with guaranteed efficiency.

When deriving the Lieb-Robinson bound, we adopt the projection $\bar{\Pi}_{L,\bar{q}}$ ($L\subseteq \Lambda$) such that
\begin{align}
\label{def:bar_Pi_L_bar_q}
\bar{\Pi}_{L,\bar{q}} := \prod_{i\in L} \Pi_{i,\le \bar{q}} .
\end{align}
This truncates the boson number at each site in the region $L$ up to $\bar{q}$. 
Therefore, the Hamiltonian $\bar{\Pi}_{L,\bar{q}}H\bar{\Pi}_{L,\bar{q}}$ has a finitely bounded energy in the region $L$ under the projection.
The problem is whether we can approximate the exact dynamics $e^{-iHt}$ by using the effective Hamiltonian as $e^{-i\bar{\Pi}_{L,\bar{q}}H\bar{\Pi}_{L,\bar{q}}t}$.
Generally, the error between them is not upper-bounded unless we impose some restrictions on the initial state $\rho_0$.
Under the condition~\eqref{low-boson density condition:cond} of the low-boson density, 
the inequality~\eqref{upper_bound_prob_dist} indicates that the dynamics may be well-approximated by $\bar{\Pi}_{L,\bar{q}}H\bar{\Pi}_{L,\bar{q}}$ as long as 
$\bar{q}\gg \ell_t^D$. 
Indeed, we can prove the following error bound similar to~\eqref{approximation_error_effective_Ham} (Supplementary Proposition~30):
\begin{align}
\label{error_estimation_effective}
&\norm{\br{O_{X_0}(t) - O_{X_0}(\bar{\Pi}_{L,\bar{q}}H\bar{\Pi}_{L,\bar{q}},t) } \rho_0}_1 \notag \\
&\le |L| e^{-c_3 \brr{\bar{q}/(t\log t)^D}^{1/\kappa}} . 
\end{align}
Following the analyses in Ref.~\cite{PhysRevLett.127.070403}, we only have to truncate the boson number in the region $X_0[R]$, that is, $L=X_0[R]$, to estimate the error $\norm{\br{O_{X_0}(t) -  O_{X_0}(H_{X[R]},t) } \rho_0}_1$. 

For the effective Hamiltonian $\bar{\Pi}_{L,\bar{q}}H\bar{\Pi}_{L,\bar{q}}$, the Lieb-Robinson velocity is proportional to $\bar{q}$, and hence, if $\bar{q}t \lesssim R$, we can ensure that the time-evolved operator $O_{X_0}(\bar{\Pi}_{L,\bar{q}}H\bar{\Pi}_{L,\bar{q}},t)$ is well-approximated in the region $X_0[R]$ (Supplementary Lemma~35). 
Therefore, by choosing $\bar{q}\propto R/t$ (see Supplementary Equation 527 for the explicit choice) in~\eqref{error_estimation_effective}, 
the Lieb-Robinson bound is derived as follows:  
\begin{align}
\label{loose_LR_bound}
&\norm{\br{O_{X_0}(t) - O_{X_0}(H_{X[R]},t) } \rho_0}_1 \notag \\
&\le \exp\brr{-c \br{\frac{R}{t(t\log t)^D}}^{1/\kappa} + \log(|X_0[R]|)} . 
\end{align}
This gives the effective light cone in the form of $R = t^{D+1} \polylog(t)$.

\subsection*{Outline of the proof for Result~2: Optimal light cone} 

 \begin{figure}[tt]
\centering
\includegraphics[clip, scale=0.55]{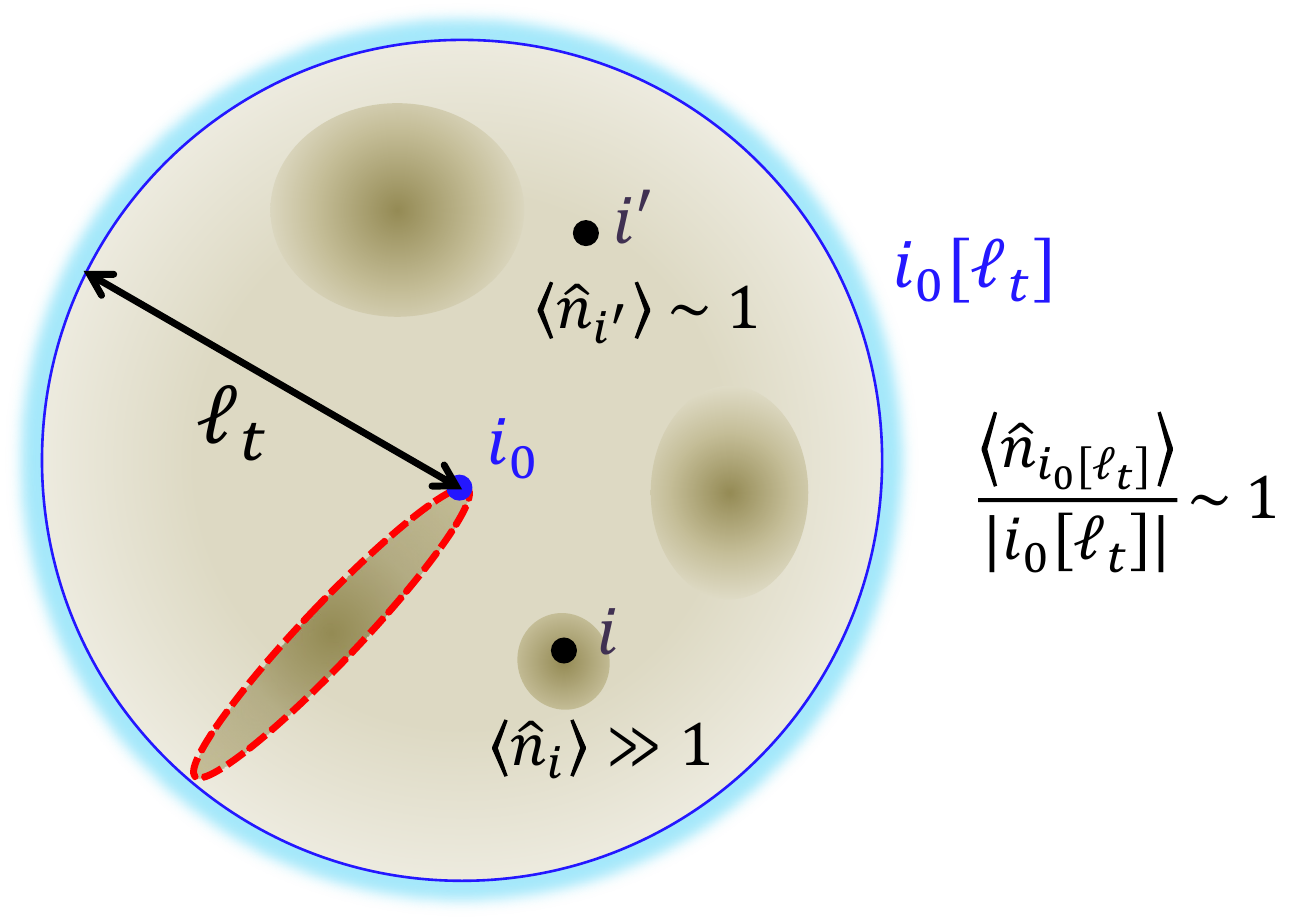}
\caption{Average of the boson number on a region $i_0[\ell_t]$ with $\ell_t=\orderof{t\log (t)}$. Even if bosons can concentrate on a few sites, the average number of bosons on one site is upper-bounded by a constant, as in~\eqref{ave_boson_number_upp}.
The local energy associated with a site is roughly proportional to the boson number on the site; hence, the average of the local energy is finitely bounded. 
However, if bosons concentrate onto a one-dimensional region (enclosed by the red dashed line), they induce an acceleration of information propagation (also see Fig.~\ref{fig:Fig4}). 
}

\label{fig:Ave_boson}
\end{figure}

To refine the bound~\eqref{loose_LR_bound}, we must utilize the fact that the boson number at each site cannot be as large as $\orderof{t^D}$ simultaneously (see Fig.~\ref{fig:Ave_boson}).
From Result~1, after time evolution, the boson number operator $\nb_{i_0[\ell_t]}$ in the ball region $i_0[\ell_t]$ is roughly upper-bounded by that in the extended ball region $i_0[2\ell_t]$ with $\ell_t =\orderof{t\log t}$. We thus obtain 
\begin{align}
\label{ave_boson_number_upp}
\frac{\nb_{i_0[\ell_t]}(t)}{|i_0[\ell_t]|} \lesssim \frac{\nb_{i_0[2\ell_t]}}{|i_0[\ell_t]|} = \frac{\nb_{i_0[2\ell_t]}}{|i_0[2\ell_t]|} \cdot \frac{|i_0[2\ell_t]|}{|i_0[\ell_t]|}.
\end{align}
Considering $|i_0[2\ell_t]|/|i_0[\ell_t]|$ is upper-bounded by the $\orderof{1}$ constant, the average boson number in the region $i_0[\ell_t]$ is still constant as long as 
the initial state satisfies $\ave{\nb_{i_0[2\ell_t]}}/|i_0[2\ell_t]| =\orderof{1}$.
We can ensure that the average local energy is upper-bounded by a constant value from the upper bound on the average number of bosons.
This inspires a feeling of hope to derive a constant Lieb-Robinson velocity.
Unfortunately, such an intuition does not hold, considering bosons clump together to make an information path with high boson density, as shown in Fig.~\ref{fig:Fig4}.
In such a path, up to $\orderof{\ell_t^{D-1}},$ bosons sit on the sites simultaneously. 
Therefore, our task is to prove that the fastest information propagation occurs when bosons clump onto a one-dimensional region.

To address the aforementioned point, we need to consider cases where the interaction strengths in a Hamiltonian depend on the locations. In the standard Lieb-Robinson bound, the Lieb-Robinson velocity is proportional to the maximum local energy~\cite{ref:Hastings2006-ExpDec,ref:Nachtergaele2006-LR}. However, this estimation is insufficient when deriving the bosonic Lieb-Robinson bound, as the local energy depends on the boson number at the local site and can be as large as $\mathcal{O}(t^D)$; our current goal is to derive the Lieb-Robinson velocity as $t^{D-1}$.

For this purpose, we consider a general Hamiltonian in the form of
$H=\sum_{Z\subset \Lambda} h_Z$
with the additional constraint:
\begin{align}
\label{cond_average_interaction}
\frac{1}{m}\sum_{j=1}^m \norm{h_{Z_j}} \le \frac{\bar{g}_0}{m} + \bar{g}_1 ,
\end{align}
where $\{h_{Z_j}\}_{j=1}^m$ are arbitrary interaction terms acting on subsets $\{Z_j\}_{j=1}^m$, respectively. Roughly speaking, the parameter $\bar{g}_0$ corresponds to the maximum local energy on one site, and $\bar{g}_1$ is the average local energy on one site.
Under the above condition, each interaction term $h_Z$ has an upper bound of $\bar{g}_0 + \bar{g}_1$. Thus, the standard Lieb-Robinson bound gives a Lieb-Robinson velocity of $\mathcal{O}(\bar{g}_0 + \bar{g}_1)$, which can be unfavorable if $\bar{g}_0$ is large. Through refined analyses, we can prove that the improved Lieb-Robinson velocity depends on the distance as $\mathcal{O}(\bar{g}_0/R) + \mathcal{O}(\bar{g}_1)$, eventually becoming $\mathcal{O}(\bar{g}_1)$ for sufficiently large $R$ (Supplementary Lemma~42).
To apply this technique to the boson systems, we may come up with an idea to perform site-dependent boson number truncation 
instead of the uniform truncation $\bar{\Pi}_{L,\bar{q}}$ in Eq.~\eqref{def:bar_Pi_L_bar_q}. For example, we consider a projection as 
\begin{align}
\bar{\Pi}_{L,\bold{q}} := \prod_{i\in L} \Pi_{i,\le q_i} ,\quad \bold{q}:=\{q_i\}_{i\in L} .
\end{align}
By the projection, the effective Hamiltonian $\bar{\Pi}_{L,\bold{q}}H\bar{\Pi}_{L,\bold{q}}$ satisfies a similar condition to~\eqref{cond_average_interaction}.

The primary challenge arises from the inability to obtain an accurate approximation for dynamics using the effective Hamiltonian $\bar{\Pi}_{L,\bold{q}}H\bar{\Pi}_{L,\bold{q}}$ for a specific choice of $\bold{q}$. This challenge is rooted in the superposition of quantum states with diverse boson configurations. For instance, consider a quantum state $\ket{\psi}$ represented as the superposition of two states, $\ket{\psi_1}$ and $\ket{\psi_2}$, where $\bar{\Pi}_{L,\bold{q}_1}\ket{\psi_1} = \ket{\psi_1}$ and $\bar{\Pi}_{L,\bold{q}_2}\ket{\psi_2} = \ket{\psi_2}$. Then, time evolution with the effective Hamiltonian $\bar{\Pi}_{L,\bold{q}_1}H\bar{\Pi}_{L,\bold{q}_1}$ provides a reliable approximation for $e^{-iHt}\ket{\psi_1}$ but not for $e^{-iHt}\ket{\psi_2}$. Conversely, time evolution with $\bar{\Pi}_{L,\bold{q}_2}H\bar{\Pi}_{L,\bold{q}_2}$ gives a good approximation for $e^{-iHt}\ket{\psi_2}$ but not for $e^{-iHt}\ket{\psi_1}$. Thus, a specific boson number truncation using $\bar{\Pi}_{L,\bold{q}}$ cannot be applied uniformly to all superposed states. It is necessary to consider different boson number truncations depending on the boson configuration of the superposed states, as discussed in Supplementary Note 8 A.

To resolve the problem, we utilize the connection of short-time unitary evolution~\cite{Kuwahara_2016_njp,PhysRevLett.126.030604,PhysRevLett.127.070403}.
Let $\tau$ be a unit of time that is appropriately chosen afterward.
If we can obtain the approximation error of
\begin{align}
\label{norm_shot_time_0}
\norm{ \brr{ O_{X}(\tau) - O_X(H_{X[\ell]},\tau) } \rho_0(t_1) }_1 
\end{align} 
for arbitrary $O_X$ with $X\subseteq X_0[R]$ and $t_1\le t$, 
we can connect the approximation to obtain the desired error bound $\norm{ \brr{ O_{X_0}(t) - O_{X_0}(H_{X[R]},t) } \rho_0(t) }_1$ (see Supplementary Equation 764).
For sufficiently small $\tau$, the time-evolved state approximately preserves the initial boson distribution. 

To estimate the norm~\eqref{norm_shot_time_0}, we consider a set of projection $\{\mathcal{P}_s\}_{s=1}^M$ such that $\sum_{s=1}^M \mathcal{P}_s=1$, each of which constraints the boson number on the sites.
By using them, we upper-bound the norm~\eqref{norm_shot_time_0} by
 \begin{align}
\label{norm_shot_time_1}
\sum_{s=1}^M \norm{ \brr{ O_{X}(\tau) - O_X(H_{X[\ell]},\tau) } \mathcal{P}_s} \cdot \norm{\mathcal{P}_s\rho_0(t_1) }_1 .
\end{align}
Therefore, although the state $\rho_0(t_1)$ includes various boson number configurations, we can separately treat them.
Because the summation~\eqref{norm_shot_time_1} increases with the number of projections $M$, we need to select a minimal set of $\{\mathcal{P}_s\}_{s=1}^M$ to achieve our goal.
The choice of the projections is rather technical (see Supplementary Note 9 B).

Now, the short-time evolution does not drastically change the original boson number distribution. Hence, 
we perform boson number truncation $\bar{\Pi}_{L,\bold{q}}$ with $q_i$ roughly determined based on the initial boson number around the site $i$. 

In conclusion, we can derive the following upper bound (see Supplementary Proposition~45):
\begin{align}
\label{norm_shot_time_2}
&\norm{ \brr{ O_{X}(\tau) - O_X(H_{X[\ell]},\tau) } \rho_0(t_1) }_1 \notag \\
&\lesssim e^{-(Q/\ell^D)^{1/\kappa}} + \br{\frac{\tau Q\log(q)}{\ell^2}}^{\ell} ,
\end{align} 
where $Q$ is an arbitrary control parameter. 
By choosing $Q$ and $\tau$ appropriately and connecting the short-time evolution, we can prove the main statement~\eqref{LR_tightest} (see 
Supplementary Equations 769, 773, and 774 of Supplementary Note 9 B).
As a final remark, in the case of one-dimensional systems, we cannot utilize the original unitary connection technique~\cite{Kuwahara_2016_njp,PhysRevLett.126.030604,PhysRevLett.127.070403} and have to utilize a refined version (see Supplementary Note 10).

\subsection*{Realization of the CNOT operation}

In the protocol to achieve the information propagation in Fig.~\ref{fig:Fig4},
we need to implement the following two operations that involve two and four sites, respectively.
\begin{align}
\ket{N,1} \leftrightarrow \ket{0,N+1} , \label{first_unitary}
\end{align}
and
\begin{align}
&\ket{\bar{n}_t,\bar{n}_t}\otimes \ket{\bar{n}_t,\bar{n}_t} \leftrightarrow \ket{\bar{n}_t,\bar{n}_t}\otimes \ket{\bar{n}_t-1,\bar{n}_t+1}  , \notag \\
&\ket{\bar{n}_t-1,\bar{n}_t+1}\otimes \ket{\bar{n}_t,\bar{n}_t} \to \ket{\bar{n}_t-1,\bar{n}_t+1}\otimes \ket{\bar{n}_t,\bar{n}_t}, \notag \\
&\ket{\bar{n}_t-1,\bar{n}_t+1}\otimes \ket{\bar{n}_t-1,\bar{n}_t+1}   \notag \\
&\to \ket{\bar{n}_t-1,\bar{n}_t+1}\otimes \ket{\bar{n}_t-1,\bar{n}_t+1} ,
\label{second_unitary}
\end{align}
where we denote the product state $\ket{1} \otimes \ket{N}$ by $\ket{1,N}$ for simplicity. 
We also label the four sites as $1$, $2$, $3$ and $4$.

To achieve the operation~\eqref{first_unitary}, we first transform $\ket{N,1} \to \ket{1,N}$, which is achieved by the free boson Hamiltonian, that is, 
$H_0=J (b_1^\dagger b_2 +{ \rm h.c.} )$ ($J\le \bar{J}$). 
Second, to transform $\ket{1,N} \rightarrow \ket{0,N+1}$, we use the Bose-Hubbard Hamiltonian as 
\begin{align}
H=H_0 + h \nb_2 - U \nb_2^2 ,\quad h=(2N+1)U.
\end{align}
By letting $V= h \nb_2 - U \nb_2^2$, we get $\bra{1,N}V\ket{1,N}=\bra{0,N+1}V\ket{0,N+1}=UN(N+1)$ and 
$\bra{j,N+1-j}V\ket{j,N+1-j}\le UN(N+1) - 2U$ for $\forall j \in[2,N+1]$.
In the limit of $U\to \infty$, the time evolution $e^{-iHt}\ket{1, N}$ is described by the superposition of the two states of 
$\ket{1,N}$ and $\ket{0,N+1}$. Therefore, we achieve the transformation $\ket{1,N} \rightarrow \ket{0,N+1}$ within a time proportional to $J^{-1}$.

The second operation~\eqref{second_unitary} is constructed by the following Hamiltonian
\begin{align}
H=H_0 + h (\nb_2-\nb_1) \nb_3 + U (\nb_3\nb_4 +  \nb_4 -\bar{n}_t)  ,
\end{align}
where we choose $h$ to be infinitely large. 
Owing to the term $h (\nb_2-\nb_1) \nb_3$ ($h\to \infty$), the hopping between the site $3$ and $4$ cannot occur unless the number of bosons on sites 1 and 2 are equal. 
Therefore, we achieve the second and the third operations in~\eqref{second_unitary}.
Next, we denote $V=h (\nb_2-\nb_1) \nb_3 + U (\nb_3\nb_4 +  \nb_4 -\bar{n}_t)$. 
For an arbitrary state as $\ket{\bar{n}_t,\bar{n}_t}\otimes \ket{\bar{n}_t-j,\bar{n}_t+j}$, the eigenvalue of $V$ is given by
\begin{align}
&\bra{\bar{n}_t,\bar{n}_t,\bar{n}_t-j,\bar{n}_t+j}V \ket{\bar{n}_t,\bar{n}_t,\bar{n}_t-j,\bar{n}_t+j} \notag \\
=&U\br{\bar{n}_t^2-j^2 + j}.
\end{align}
Then, the eigenvalue has the same value only for $j=0$ and $j=1$, whereas the other eigenvalues are separated from each other by a width larger than or equal to $2U$. 
Therefore, by letting $U\to \infty$, the first operation~\eqref{second_unitary} can be realized by following the same process as described for~\eqref{first_unitary}.

\subsection*{Gate complexity for quantum simulation}

We here derive the gate complexity to simulate the time evolution by $\tilde{H}_0(\tilde{V},x)$ in Eq.~\eqref{decomposition_of_effective_time_evo}.
The technique herein is similar to the one in Ref.~\cite{maskara2019complexity}, 
which analyzes the quantum simulation for the Bose-Hubbard model with a sufficiently small boson density.
Under the decomposition of Eq.~\eqref{decomposition_of_effective_time_evo}, we must consider the class of time-dependent Hamiltonians as 
\begin{align}
H_t=\sum_{Z\subset \Lambda} h_{t,Z} ,
\end{align}
where each interaction term $\{h_{t,Z}\}_{Z\subset \Lambda}$ is given by the form of $e^{i\tilde{V}t} \tilde{b}_i \tilde{b}_j^\dagger e^{-i\tilde{V}t}$.  
Here, $h_{t,Z}$ satisfies 
\begin{align}
\label{def_g_upp}
\max_{i\in \Lambda}\sum_{Z:Z\ni i} \norm{h_{t,Z}} \le  g =  \orderof{\bar{q}} ,
\end{align}
and 
\begin{align}
\label{def_g'_upp}
\norm{\frac{dh_{t,Z}}{dt}} \le g' = \poly(\bar{q}) , 
\end{align}
where $\bar{q}$ is defined by the boson number truncation as in~\eqref{approximation_error_effective_Ham}. 
Additionally, the local Hilbert space on one site has a dimension of $\bar{q}+1$, whereas the matrix representing $h_{t,Z}$ is $d$ sparse matrix with $d=\orderof{1}$;
that is, it has at most $d$ nonzero elements in any row or column.

 \begin{figure}[tt]
\centering
\includegraphics[clip, scale=0.24]{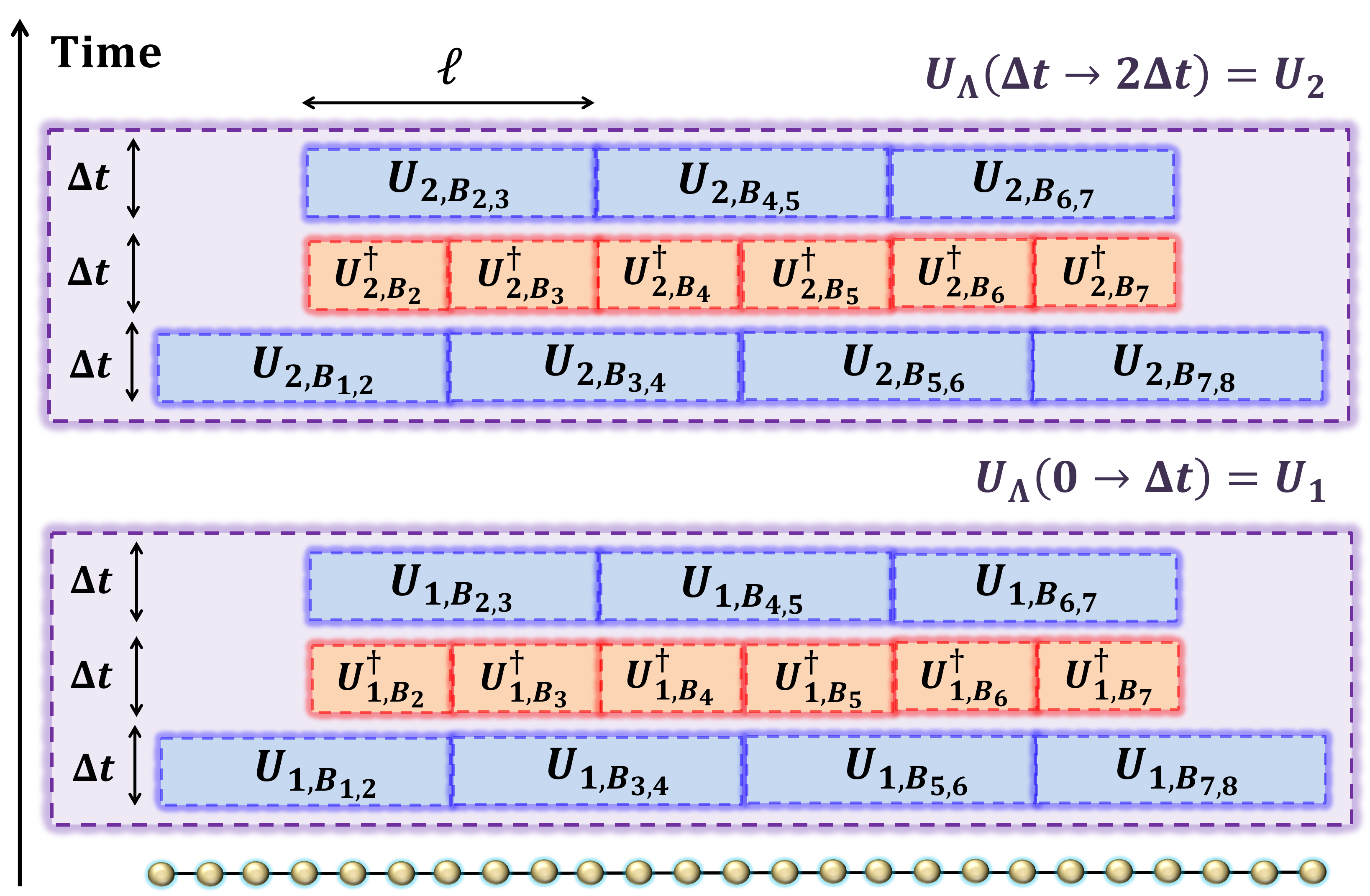}
\caption{Schematic of the HHKL decomposition~\eqref{HHKL_decomposition} in a one-dimensional case.
Each unitary operator $U_{j,B_{s,s+1}}$ is given by the time evolution of the subset Hamiltonian $H_{B_{s,s+1}}(t)$ from the time $(j-1)\Delta t$ to $j\Delta t$.
The error of the decomposition depends on the block size, given as $e^{-\mu \ell +  v g \Delta t}$. 
The local energy of the Hamiltonian $H(t)$ is characterized by the constant $g=\orderof{\bar{q}}$ as in Eq.~\eqref{def_g_upp}. 
When estimating the gate complexity, we choose $\Delta t \propto 1/g = \orderof{1/\bar{q}}$ and $\ell =\log(|\Lambda| t \bar{q}/\epsilon)$.
}
\label{fig:HHKL}
\end{figure}

Now, we consider the subset Hamiltonian $H_{L}(t)$ on an arbitrary subset $L$, defined as 
\begin{align}
H_{t,L}=\sum_{Z\subset L} h_{t,Z}. 
\end{align}
Then, we consider the gate complexity to simulate the dynamics $U_L(0\to \tau) :=\mathcal{T} e^{-i\int_0^\tau H_{t,L} dt}$, 
which has been thoroughly investigated~\cite{Berry2007,berry2017exponential}.
The Hilbert space on the subset $L$ has dimensions of $(\bar{q}+1)^{|L|}$, and hence, the number of 
qubits to represent the Hilbert space is given by $|L| \log_2 (\bar{q}+1)$. 
Additionally, $H_{t,L}$ is given by an $\orderof{d_L}$ sparse matrix, where $d_L=d \times \orderof{|L|}$.  
Therefore, by employing Theorem~2.1 in Ref.~\cite{berry2017exponential}, the gate complexity for simulating $U_L(0\to \tau)$ up to an error $\epsilon$ is upper-bounded by
\begin{align}
\frac{\tilde{\tau} \log(\tilde{\tau}/\epsilon) \log[(\tilde{\tau}+\tilde{\tau}')/\epsilon]}{\log\log(\tilde{\tau}/\epsilon)} |L| \log_2 (\bar{q}+1)
\end{align}
with $\tilde{\tau} := d_L^2g|L|\tau$ and $\tilde{\tau}' := d_L^2g'|L|\tau$. 
By using the inequalities~\eqref{def_g_upp} and~\eqref{def_g'_upp} and $d_L=\orderof{|L|}$, the above quantity reduces to the form of 
\begin{align}
\label{gate_complex_H_L}
\tau \bar{q} |L|^4  \log^2(\tau\bar{q} |L| /\epsilon) \log(\bar{q}).
\end{align}

In the following, we consider the Haah-Hastings-Kothari-Low algorithm~\cite{8555119} to the time evolution of the total system $\Lambda$, that is, 
$U_\Lambda(0\to t)$ by splitting the total time $t$ into $m_0:=t/\Delta t$ pieces and choosing $\Delta t$ as $\orderof{1/\bar{q}}$.
Then, we decompose the total system into blocks $\{B_s\}_{s=1}^{\bar{n}}$, i.e., $\Lambda=\bigcup_{s=1}^{\bar{n}}B_s$, where each block has the size of $\ell$ (see Fig.~\ref{fig:HHKL}).
We then approximate 
\begin{align}
\label{HHKL_decomposition}
U_\Lambda(0\to t)=\mathcal{T} e^{-i\int_0^tH_{x} dx} \approx U_1 U_2 \cdots U_{m_0} ,
\end{align}
where $U_j$ is an approximation for the dynamics from $(j-1)\Delta t$ to $j\Delta t$ as follows:
\begin{align}
U_j= \prod_{s:{\rm odd} } U_{j,B_{s,s+1}}   \prod_{s=2}^{\bar{n}-1} U^\dagger_{j,B_s}  \prod_{s:{\rm even}} U_{j,B_{s,s+1}} .
\end{align}
Herein, we define $B_{s,s+1}:=B_s \cup B_{s+1}$ and define $U_{j,L}$ ($L\subseteq \Lambda$) as 
\begin{align}
U_{j,L} := \mathcal{T} e^{-i\int_{(j-1)\Delta t}^{j\Delta t}H_{x,L} dx}. 
\end{align}
From Ref.~\cite{8555119}, the approximation error of the decomposition~\eqref{HHKL_decomposition} is given as 
\begin{align}
\label{HHKL_decomposition_error}
\norm{U_\Lambda(0\to t) - U_1 U_2 \cdots U_{m_0}} \lesssim  |\Lambda| \frac{t}{\Delta t} e^{-\mu \ell +  v g \Delta t}  ,
\end{align}
where $\mu$ and $v$ are the constants of $\orderof{1}$. 
Due to $\Delta t=\orderof{1/\bar{q}}$  and $g=\orderof{\bar{q}}$ from~\eqref{def_g_upp}, we have $g \Delta t = \orderof{1}$; therefore, 
by choosing $\ell=\log(|\Lambda| t /\epsilon)$, we ensure that the error~\eqref{HHKL_decomposition_error} is smaller than $\epsilon$. 

We now have all the ingredients to estimate the gate complexity. 
From the estimation~\eqref{gate_complex_H_L}, each unitary operator $U_{j,B_s}$ was implemented with a gate complexity of 
\begin{align}
\label{gate_complexity_U_j_B_s}
&|B_s|^4  \log^2(|B_s|  /\epsilon) \log(\bar{q}) ,  
\end{align}
where we use $\Delta t \bar{q}=\orderof{1}$. 
The number of the unitary operators of $U_{j,B_s}$ is proportional to 
\begin{align}
\label{number_of_U_j_B_s}
\frac{|\Lambda|}{|B_s|}\cdot \frac{t}{\Delta t} .
\end{align}
Therefore, by combining the estimations of~\eqref{gate_complexity_U_j_B_s} and \eqref{number_of_U_j_B_s}, the gate complexity implements 
$\{U_{j,B_s}\}_{j,s}$ is given by
 \begin{align}
&\frac{t\ell^{3D}|\Lambda|}{\Delta t}  \log^2(\ell^D /\epsilon) \log(\bar{q}) 
= |\Lambda| t \bar{q} \cdot \polylog(|\Lambda| t \bar{q}/\epsilon) , \notag
\end{align}
where we use $\Delta t = \orderof{1/\bar{q}}$ and $|B_s|=\orderof{\ell^D}$.
We obtained the same estimation when implementing $\{U_{j,B_{s,s+1}}\}_{j,s}$. 
Therefore, we obtain the desired gate complexity to implement the unitary operator $U_\Lambda(0\to t)$.

\section*{Data availability}
Data sharing does not apply to this paper, as no datasets were generated or analyzed
during the current study.

\def\bibsection{\section*{References}}

\providecommand{\noopsort}[1]{}\providecommand{\singleletter}[1]{#1}%
%


\section*{acknowledgments}

{~}\\
T. K. acknowledges Hakubi projects of RIKEN and was supported by Japan Society for the Promotion of Science KAKENHI (Grant No. 18K13475) and Japan Science and Technology
Agency Precursory Research for Embryonic Science and Technology (Grant No. JPMJPR2116). K. S. was supported by JSPS Grants-in-Aid for Scientific Research (No. JP16H02211 and No. JP19H05603).

\section*{Author contributions}

{~}\\
T.K, T.V.V., and K.S. contributed to the conception of the work, the analysis and interpretation, and the preparation and revision of the manuscript.

\section*{Competing Interests}
{~}\\
The authors declare no competing interests.

{~}\\


\providecommand{\noopsort}[1]{}\providecommand{\singleletter}[1]{#1}%
\begin{thebibliography}{56}%
\makeatletter
\providecommand \@ifxundefined [1]{%
 \@ifx{#1\undefined}
}%
\providecommand \@ifnum [1]{%
 \ifnum #1\expandafter \@firstoftwo
 \else \expandafter \@secondoftwo
 \fi
}%
\providecommand \@ifx [1]{%
 \ifx #1\expandafter \@firstoftwo
 \else \expandafter \@secondoftwo
 \fi
}%
\providecommand \natexlab [1]{#1}%
\providecommand \enquote  [1]{``#1''}%
\providecommand \bibnamefont  [1]{#1}%
\providecommand \bibfnamefont [1]{#1}%
\providecommand \citenamefont [1]{#1}%
\providecommand \href@noop [0]{\@secondoftwo}%
\providecommand \href [0]{\begingroup \@sanitize@url \@href}%
\providecommand \@href[1]{\@@startlink{#1}\@@href}%
\providecommand \@@href[1]{\endgroup#1\@@endlink}%
\providecommand \@sanitize@url [0]{\catcode `\\12\catcode `\$12\catcode
  `\&12\catcode `\#12\catcode `\^12\catcode `\_12\catcode `\%12\relax}%
\providecommand \@@startlink[1]{}%
\providecommand \@@endlink[0]{}%
\providecommand \url  [0]{\begingroup\@sanitize@url \@url }%
\providecommand \@url [1]{\endgroup\@href {#1}{\urlprefix }}%
\providecommand \urlprefix  [0]{URL }%
\providecommand \Eprint [0]{\href }%
\providecommand \doibase [0]{https://doi.org/}%
\providecommand \selectlanguage [0]{\@gobble}%
\providecommand \bibinfo  [0]{\@secondoftwo}%
\providecommand \bibfield  [0]{\@secondoftwo}%
\providecommand \translation [1]{[#1]}%
\providecommand \BibitemOpen [0]{}%
\providecommand \bibitemStop [0]{}%
\providecommand \bibitemNoStop [0]{.\EOS\space}%
\providecommand \EOS [0]{\spacefactor3000\relax}%
\providecommand \BibitemShut  [1]{\csname bibitem#1\endcsname}%
\let\auto@bib@innerbib\@empty
\bibitem [{\citenamefont {Lieb}\ and\ \citenamefont
  {Robinson}(1972)}]{ref:LR-bound72}%
  \BibitemOpen
  \bibfield  {author} {\bibinfo {author} {\bibfnamefont {E.~H.}\ \bibnamefont
  {Lieb}}\ and\ \bibinfo {author} {\bibfnamefont {D.~W.}\ \bibnamefont
  {Robinson}},\ }\bibfield  {title} {\bibinfo {title} {{\it The finite group
  velocity of quantum spin systems}},\ }\href
  {https://doi.org/10.1007/BF01645779} {\bibfield  {journal} {\bibinfo
  {journal} {Communications in Mathematical Physics}\ }\textbf {\bibinfo
  {volume} {28}},\ \bibinfo {pages} {251} (\bibinfo {year} {1972})}\BibitemShut
  {NoStop}%
\bibitem [{\citenamefont {Cheneau}\ \emph {et~al.}(2012)\citenamefont
  {Cheneau}, \citenamefont {Barmettler}, \citenamefont {Poletti}, \citenamefont
  {Endres}, \citenamefont {Schau{\ss}}, \citenamefont {Fukuhara}, \citenamefont
  {Gross}, \citenamefont {Bloch}, \citenamefont {Kollath},\ and\ \citenamefont
  {Kuhr}}]{cheneau2012light}%
  \BibitemOpen
  \bibfield  {author} {\bibinfo {author} {\bibfnamefont {M.}~\bibnamefont
  {Cheneau}}, \bibinfo {author} {\bibfnamefont {P.}~\bibnamefont {Barmettler}},
  \bibinfo {author} {\bibfnamefont {D.}~\bibnamefont {Poletti}}, \bibinfo
  {author} {\bibfnamefont {M.}~\bibnamefont {Endres}}, \bibinfo {author}
  {\bibfnamefont {P.}~\bibnamefont {Schau{\ss}}}, \bibinfo {author}
  {\bibfnamefont {T.}~\bibnamefont {Fukuhara}}, \bibinfo {author}
  {\bibfnamefont {C.}~\bibnamefont {Gross}}, \bibinfo {author} {\bibfnamefont
  {I.}~\bibnamefont {Bloch}}, \bibinfo {author} {\bibfnamefont
  {C.}~\bibnamefont {Kollath}},\ and\ \bibinfo {author} {\bibfnamefont
  {S.}~\bibnamefont {Kuhr}},\ }\bibfield  {title} {\bibinfo {title} {{\it
  Light-cone-like spreading of correlations in a quantum many-body system}},\
  }\href {https://doi.org/10.1038/nature10748} {\bibfield  {journal} {\bibinfo
  {journal} {Nature}\ }\textbf {\bibinfo {volume} {481}},\ \bibinfo {pages}
  {484} (\bibinfo {year} {2012})}\BibitemShut {NoStop}%
\bibitem [{\citenamefont {Richerme}\ \emph {et~al.}(2014)\citenamefont
  {Richerme}, \citenamefont {Gong}, \citenamefont {Lee}, \citenamefont {Senko},
  \citenamefont {Smith}, \citenamefont {Foss-Feig}, \citenamefont {Michalakis},
  \citenamefont {Gorshkov},\ and\ \citenamefont {Monroe}}]{richerme2014non}%
  \BibitemOpen
  \bibfield  {author} {\bibinfo {author} {\bibfnamefont {P.}~\bibnamefont
  {Richerme}}, \bibinfo {author} {\bibfnamefont {Z.-X.}\ \bibnamefont {Gong}},
  \bibinfo {author} {\bibfnamefont {A.}~\bibnamefont {Lee}}, \bibinfo {author}
  {\bibfnamefont {C.}~\bibnamefont {Senko}}, \bibinfo {author} {\bibfnamefont
  {J.}~\bibnamefont {Smith}}, \bibinfo {author} {\bibfnamefont
  {M.}~\bibnamefont {Foss-Feig}}, \bibinfo {author} {\bibfnamefont
  {S.}~\bibnamefont {Michalakis}}, \bibinfo {author} {\bibfnamefont {A.~V.}\
  \bibnamefont {Gorshkov}},\ and\ \bibinfo {author} {\bibfnamefont
  {C.}~\bibnamefont {Monroe}},\ }\bibfield  {title} {\bibinfo {title} {{\it
  Non-local propagation of correlations in quantum systems with long-range
  interactions}},\ }\href {https://doi.org/10.1038/nature13450} {\bibfield
  {journal} {\bibinfo  {journal} {Nature}\ }\textbf {\bibinfo {volume} {511}},\
  \bibinfo {pages} {198} (\bibinfo {year} {2014})}\BibitemShut {NoStop}%
\bibitem [{\citenamefont {Jurcevic}\ \emph {et~al.}(2014)\citenamefont
  {Jurcevic}, \citenamefont {Lanyon}, \citenamefont {Hauke}, \citenamefont
  {Hempel}, \citenamefont {Zoller}, \citenamefont {Blatt},\ and\ \citenamefont
  {Roos}}]{jurcevic2014quasiparticle}%
  \BibitemOpen
  \bibfield  {author} {\bibinfo {author} {\bibfnamefont {P.}~\bibnamefont
  {Jurcevic}}, \bibinfo {author} {\bibfnamefont {B.~P.}\ \bibnamefont
  {Lanyon}}, \bibinfo {author} {\bibfnamefont {P.}~\bibnamefont {Hauke}},
  \bibinfo {author} {\bibfnamefont {C.}~\bibnamefont {Hempel}}, \bibinfo
  {author} {\bibfnamefont {P.}~\bibnamefont {Zoller}}, \bibinfo {author}
  {\bibfnamefont {R.}~\bibnamefont {Blatt}},\ and\ \bibinfo {author}
  {\bibfnamefont {C.~F.}\ \bibnamefont {Roos}},\ }\bibfield  {title} {\bibinfo
  {title} {{\it Quasiparticle engineering and entanglement propagation in a
  quantum many-body system}},\ }\href {https://doi.org/10.1038/nature13461}
  {\bibfield  {journal} {\bibinfo  {journal} {Nature}\ }\textbf {\bibinfo
  {volume} {511}},\ \bibinfo {pages} {202} (\bibinfo {year}
  {2014})}\BibitemShut {NoStop}%
\bibitem [{\citenamefont {Hastings}(2007)}]{ref:Hastings-AL07}%
  \BibitemOpen
  \bibfield  {author} {\bibinfo {author} {\bibfnamefont {M.~B.}\ \bibnamefont
  {Hastings}},\ }\bibfield  {title} {\bibinfo {title} {{\it An area law for
  one-dimensional quantum systems}},\ }\href
  {http://stacks.iop.org/1742-5468/2007/i=08/a=P08024} {\bibfield  {journal}
  {\bibinfo  {journal} {Journal of Statistical Mechanics: Theory and
  Experiment}\ }\textbf {\bibinfo {volume} {2007}},\ \bibinfo {pages} {P08024}
  (\bibinfo {year} {2007})},\ \Eprint {https://arxiv.org/abs/arXiv:0705.2024}
  {arXiv:0705.2024} \BibitemShut {NoStop}%
\bibitem [{\citenamefont {Van~Acoleyen}\ \emph {et~al.}(2013)\citenamefont
  {Van~Acoleyen}, \citenamefont {Mari\"en},\ and\ \citenamefont
  {Verstraete}}]{PhysRevLett.111.170501}%
  \BibitemOpen
  \bibfield  {author} {\bibinfo {author} {\bibfnamefont {K.}~\bibnamefont
  {Van~Acoleyen}}, \bibinfo {author} {\bibfnamefont {M.}~\bibnamefont
  {Mari\"en}},\ and\ \bibinfo {author} {\bibfnamefont {F.}~\bibnamefont
  {Verstraete}},\ }\bibfield  {title} {\bibinfo {title} {{\it Entanglement
  Rates and Area Laws}},\ }\href
  {https://doi.org/10.1103/PhysRevLett.111.170501} {\bibfield  {journal}
  {\bibinfo  {journal} {Phys. Rev. Lett.}\ }\textbf {\bibinfo {volume} {111}},\
  \bibinfo {pages} {170501} (\bibinfo {year} {2013})}\BibitemShut {NoStop}%
\bibitem [{\citenamefont {Hastings}\ and\ \citenamefont
  {Wen}(2005)}]{PhysRevB.72.045141}%
  \BibitemOpen
  \bibfield  {author} {\bibinfo {author} {\bibfnamefont {M.~B.}\ \bibnamefont
  {Hastings}}\ and\ \bibinfo {author} {\bibfnamefont {X.-G.}\ \bibnamefont
  {Wen}},\ }\bibfield  {title} {\bibinfo {title} {{\it Quasiadiabatic
  continuation of quantum states: The stability of topological ground-state
  degeneracy and emergent gauge invariance}},\ }\href
  {https://doi.org/10.1103/PhysRevB.72.045141} {\bibfield  {journal} {\bibinfo
  {journal} {Phys. Rev. B}\ }\textbf {\bibinfo {volume} {72}},\ \bibinfo
  {pages} {045141} (\bibinfo {year} {2005})}\BibitemShut {NoStop}%
\bibitem [{\citenamefont {Iyoda}\ \emph {et~al.}(2017)\citenamefont {Iyoda},
  \citenamefont {Kaneko},\ and\ \citenamefont
  {Sagawa}}]{PhysRevLett.119.100601}%
  \BibitemOpen
  \bibfield  {author} {\bibinfo {author} {\bibfnamefont {E.}~\bibnamefont
  {Iyoda}}, \bibinfo {author} {\bibfnamefont {K.}~\bibnamefont {Kaneko}},\ and\
  \bibinfo {author} {\bibfnamefont {T.}~\bibnamefont {Sagawa}},\ }\bibfield
  {title} {\bibinfo {title} {{\it Fluctuation Theorem for Many-Body Pure
  Quantum States}},\ }\href {https://doi.org/10.1103/PhysRevLett.119.100601}
  {\bibfield  {journal} {\bibinfo  {journal} {Phys. Rev. Lett.}\ }\textbf
  {\bibinfo {volume} {119}},\ \bibinfo {pages} {100601} (\bibinfo {year}
  {2017})}\BibitemShut {NoStop}%
\bibitem [{\citenamefont {Hastings}\ and\ \citenamefont
  {Koma}(2006)}]{ref:Hastings2006-ExpDec}%
  \BibitemOpen
  \bibfield  {author} {\bibinfo {author} {\bibfnamefont {M.~B.}\ \bibnamefont
  {Hastings}}\ and\ \bibinfo {author} {\bibfnamefont {T.}~\bibnamefont
  {Koma}},\ }\bibfield  {title} {\bibinfo {title} {{\it Spectral Gap and
  Exponential Decay of Correlations}},\ }\href
  {https://doi.org/10.1007/s00220-006-0030-4} {\bibfield  {journal} {\bibinfo
  {journal} {Communications in Mathematical Physics}\ }\textbf {\bibinfo
  {volume} {265}},\ \bibinfo {pages} {781} (\bibinfo {year}
  {2006})}\BibitemShut {NoStop}%
\bibitem [{\citenamefont {Nachtergaele}\ and\ \citenamefont
  {Sims}(2006)}]{ref:Nachtergaele2006-LR}%
  \BibitemOpen
  \bibfield  {author} {\bibinfo {author} {\bibfnamefont {B.}~\bibnamefont
  {Nachtergaele}}\ and\ \bibinfo {author} {\bibfnamefont {R.}~\bibnamefont
  {Sims}},\ }\bibfield  {title} {\bibinfo {title} {{\it Lieb-Robinson Bounds
  and the Exponential Clustering Theorem}},\ }\href
  {https://doi.org/10.1007/s00220-006-1556-1} {\bibfield  {journal} {\bibinfo
  {journal} {Communications in Mathematical Physics}\ }\textbf {\bibinfo
  {volume} {265}},\ \bibinfo {pages} {119} (\bibinfo {year}
  {2006})}\BibitemShut {NoStop}%
\bibitem [{\citenamefont {Kuwahara}\ and\ \citenamefont
  {Saito}(2022)}]{PhysRevX.12.021022}%
  \BibitemOpen
  \bibfield  {author} {\bibinfo {author} {\bibfnamefont {T.}~\bibnamefont
  {Kuwahara}}\ and\ \bibinfo {author} {\bibfnamefont {K.}~\bibnamefont
  {Saito}},\ }\bibfield  {title} {\bibinfo {title} {{\it Exponential Clustering
  of Bipartite Quantum Entanglement at Arbitrary Temperatures}},\ }\href
  {https://doi.org/10.1103/PhysRevX.12.021022} {\bibfield  {journal} {\bibinfo
  {journal} {Phys. Rev. X}\ }\textbf {\bibinfo {volume} {12}},\ \bibinfo
  {pages} {021022} (\bibinfo {year} {2022})}\BibitemShut {NoStop}%
\bibitem [{\citenamefont {Osborne}(2006)}]{PhysRevLett.97.157202}%
  \BibitemOpen
  \bibfield  {author} {\bibinfo {author} {\bibfnamefont {T.~J.}\ \bibnamefont
  {Osborne}},\ }\bibfield  {title} {\bibinfo {title} {{\it Efficient
  Approximation of the Dynamics of One-Dimensional Quantum Spin Systems}},\
  }\href {https://doi.org/10.1103/PhysRevLett.97.157202} {\bibfield  {journal}
  {\bibinfo  {journal} {Phys. Rev. Lett.}\ }\textbf {\bibinfo {volume} {97}},\
  \bibinfo {pages} {157202} (\bibinfo {year} {2006})}\BibitemShut {NoStop}%
\bibitem [{\citenamefont {Alhambra}\ and\ \citenamefont
  {Cirac}(2021)}]{PRXQuantum.2.040331}%
  \BibitemOpen
  \bibfield  {author} {\bibinfo {author} {\bibfnamefont {A.~M.}\ \bibnamefont
  {Alhambra}}\ and\ \bibinfo {author} {\bibfnamefont {J.~I.}\ \bibnamefont
  {Cirac}},\ }\bibfield  {title} {\bibinfo {title} {{\it Locally Accurate
  Tensor Networks for Thermal States and Time Evolution}},\ }\href
  {https://doi.org/10.1103/PRXQuantum.2.040331} {\bibfield  {journal} {\bibinfo
   {journal} {PRX Quantum}\ }\textbf {\bibinfo {volume} {2}},\ \bibinfo {pages}
  {040331} (\bibinfo {year} {2021})}\BibitemShut {NoStop}%
\bibitem [{\citenamefont {Haah}\ \emph {et~al.}(2018)\citenamefont {Haah},
  \citenamefont {Hastings}, \citenamefont {Kothari},\ and\ \citenamefont
  {Low}}]{8555119}%
  \BibitemOpen
  \bibfield  {author} {\bibinfo {author} {\bibfnamefont {J.}~\bibnamefont
  {Haah}}, \bibinfo {author} {\bibfnamefont {M.}~\bibnamefont {Hastings}},
  \bibinfo {author} {\bibfnamefont {R.}~\bibnamefont {Kothari}},\ and\ \bibinfo
  {author} {\bibfnamefont {G.~H.}\ \bibnamefont {Low}},\ }\bibfield  {title}
  {\bibinfo {title} {{\it Quantum Algorithm for Simulating Real Time Evolution
  of Lattice Hamiltonians}},\ }in\ \href
  {https://doi.org/10.1109/FOCS.2018.00041} {\emph {\bibinfo {booktitle} {2018
  IEEE 59th Annual Symposium on Foundations of Computer Science (FOCS)}}}\
  (\bibinfo {year} {2018})\ pp.\ \bibinfo {pages} {350--360}\BibitemShut
  {NoStop}%
\bibitem [{\citenamefont {Anshu}\ \emph {et~al.}(2021)\citenamefont {Anshu},
  \citenamefont {Arunachalam}, \citenamefont {Kuwahara},\ and\ \citenamefont
  {Soleimanifar}}]{Anshu_2021}%
  \BibitemOpen
  \bibfield  {author} {\bibinfo {author} {\bibfnamefont {A.}~\bibnamefont
  {Anshu}}, \bibinfo {author} {\bibfnamefont {S.}~\bibnamefont {Arunachalam}},
  \bibinfo {author} {\bibfnamefont {T.}~\bibnamefont {Kuwahara}},\ and\
  \bibinfo {author} {\bibfnamefont {M.}~\bibnamefont {Soleimanifar}},\
  }\bibfield  {title} {\bibinfo {title} {{\it Sample-efficient learning of
  interacting quantum systems}},\ }\href
  {https://doi.org/10.1038/s41567-021-01232-0} {\bibfield  {journal} {\bibinfo
  {journal} {Nature Physics}\ }\textbf {\bibinfo {volume} {17}},\ \bibinfo
  {pages} {931} (\bibinfo {year} {2021})}\BibitemShut {NoStop}%
\bibitem [{\citenamefont {Roberts}\ and\ \citenamefont
  {Swingle}(2016)}]{PhysRevLett.117.091602}%
  \BibitemOpen
  \bibfield  {author} {\bibinfo {author} {\bibfnamefont {D.~A.}\ \bibnamefont
  {Roberts}}\ and\ \bibinfo {author} {\bibfnamefont {B.}~\bibnamefont
  {Swingle}},\ }\bibfield  {title} {\bibinfo {title} {{\it Lieb-Robinson Bound
  and the Butterfly Effect in Quantum Field Theories}},\ }\href
  {https://doi.org/10.1103/PhysRevLett.117.091602} {\bibfield  {journal}
  {\bibinfo  {journal} {Phys. Rev. Lett.}\ }\textbf {\bibinfo {volume} {117}},\
  \bibinfo {pages} {091602} (\bibinfo {year} {2016})}\BibitemShut {NoStop}%
\bibitem [{\citenamefont {Eisert}\ \emph {et~al.}(2013)\citenamefont {Eisert},
  \citenamefont {van~den Worm}, \citenamefont {Manmana},\ and\ \citenamefont
  {Kastner}}]{PhysRevLett.111.260401}%
  \BibitemOpen
  \bibfield  {author} {\bibinfo {author} {\bibfnamefont {J.}~\bibnamefont
  {Eisert}}, \bibinfo {author} {\bibfnamefont {M.}~\bibnamefont {van~den
  Worm}}, \bibinfo {author} {\bibfnamefont {S.~R.}\ \bibnamefont {Manmana}},\
  and\ \bibinfo {author} {\bibfnamefont {M.}~\bibnamefont {Kastner}},\
  }\bibfield  {title} {\bibinfo {title} {{\it Breakdown of Quasilocality in
  Long-Range Quantum Lattice Models}},\ }\href
  {https://doi.org/10.1103/PhysRevLett.111.260401} {\bibfield  {journal}
  {\bibinfo  {journal} {Phys. Rev. Lett.}\ }\textbf {\bibinfo {volume} {111}},\
  \bibinfo {pages} {260401} (\bibinfo {year} {2013})}\BibitemShut {NoStop}%
\bibitem [{\citenamefont {Foss-Feig}\ \emph {et~al.}(2015)\citenamefont
  {Foss-Feig}, \citenamefont {Gong}, \citenamefont {Clark},\ and\ \citenamefont
  {Gorshkov}}]{PhysRevLett.114.157201}%
  \BibitemOpen
  \bibfield  {author} {\bibinfo {author} {\bibfnamefont {M.}~\bibnamefont
  {Foss-Feig}}, \bibinfo {author} {\bibfnamefont {Z.-X.}\ \bibnamefont {Gong}},
  \bibinfo {author} {\bibfnamefont {C.~W.}\ \bibnamefont {Clark}},\ and\
  \bibinfo {author} {\bibfnamefont {A.~V.}\ \bibnamefont {Gorshkov}},\
  }\bibfield  {title} {\bibinfo {title} {{\it Nearly Linear Light Cones in
  Long-Range Interacting Quantum Systems}},\ }\href
  {https://doi.org/10.1103/PhysRevLett.114.157201} {\bibfield  {journal}
  {\bibinfo  {journal} {Phys. Rev. Lett.}\ }\textbf {\bibinfo {volume} {114}},\
  \bibinfo {pages} {157201} (\bibinfo {year} {2015})}\BibitemShut {NoStop}%
\bibitem [{\citenamefont {Chen}\ and\ \citenamefont
  {Lucas}(2019)}]{chen2019finite}%
  \BibitemOpen
  \bibfield  {author} {\bibinfo {author} {\bibfnamefont {C.-F.}\ \bibnamefont
  {Chen}}\ and\ \bibinfo {author} {\bibfnamefont {A.}~\bibnamefont {Lucas}},\
  }\bibfield  {title} {\bibinfo {title} {{\it Finite Speed of Quantum
  Scrambling with Long Range Interactions}},\ }\href
  {https://doi.org/10.1103/PhysRevLett.123.250605} {\bibfield  {journal}
  {\bibinfo  {journal} {Phys. Rev. Lett.}\ }\textbf {\bibinfo {volume} {123}},\
  \bibinfo {pages} {250605} (\bibinfo {year} {2019})}\BibitemShut {NoStop}%
\bibitem [{\citenamefont {Kuwahara}\ and\ \citenamefont
  {Saito}(2020)}]{PhysRevX.10.031010}%
  \BibitemOpen
  \bibfield  {author} {\bibinfo {author} {\bibfnamefont {T.}~\bibnamefont
  {Kuwahara}}\ and\ \bibinfo {author} {\bibfnamefont {K.}~\bibnamefont
  {Saito}},\ }\bibfield  {title} {\bibinfo {title} {{\it Strictly Linear Light
  Cones in Long-Range Interacting Systems of Arbitrary Dimensions}},\ }\href
  {https://doi.org/10.1103/PhysRevX.10.031010} {\bibfield  {journal} {\bibinfo
  {journal} {Phys. Rev. X}\ }\textbf {\bibinfo {volume} {10}},\ \bibinfo
  {pages} {031010} (\bibinfo {year} {2020})}\BibitemShut {NoStop}%
\bibitem [{\citenamefont {Tran}\ \emph
  {et~al.}(2021{\natexlab{a}})\citenamefont {Tran}, \citenamefont {Guo},
  \citenamefont {Deshpande}, \citenamefont {Lucas},\ and\ \citenamefont
  {Gorshkov}}]{PhysRevX.11.031016}%
  \BibitemOpen
  \bibfield  {author} {\bibinfo {author} {\bibfnamefont {M.~C.}\ \bibnamefont
  {Tran}}, \bibinfo {author} {\bibfnamefont {A.~Y.}\ \bibnamefont {Guo}},
  \bibinfo {author} {\bibfnamefont {A.}~\bibnamefont {Deshpande}}, \bibinfo
  {author} {\bibfnamefont {A.}~\bibnamefont {Lucas}},\ and\ \bibinfo {author}
  {\bibfnamefont {A.~V.}\ \bibnamefont {Gorshkov}},\ }\bibfield  {title}
  {\bibinfo {title} {{\it Optimal State Transfer and Entanglement Generation in
  Power-Law Interacting Systems}},\ }\href
  {https://doi.org/10.1103/PhysRevX.11.031016} {\bibfield  {journal} {\bibinfo
  {journal} {Phys. Rev. X}\ }\textbf {\bibinfo {volume} {11}},\ \bibinfo
  {pages} {031016} (\bibinfo {year} {2021}{\natexlab{a}})}\BibitemShut
  {NoStop}%
\bibitem [{\citenamefont {Kuwahara}\ and\ \citenamefont
  {Saito}(2021{\natexlab{a}})}]{PhysRevLett.126.030604}%
  \BibitemOpen
  \bibfield  {author} {\bibinfo {author} {\bibfnamefont {T.}~\bibnamefont
  {Kuwahara}}\ and\ \bibinfo {author} {\bibfnamefont {K.}~\bibnamefont
  {Saito}},\ }\bibfield  {title} {\bibinfo {title} {{\it Absence of Fast
  Scrambling in Thermodynamically Stable Long-Range Interacting Systems}},\
  }\href {https://doi.org/10.1103/PhysRevLett.126.030604} {\bibfield  {journal}
  {\bibinfo  {journal} {Phys. Rev. Lett.}\ }\textbf {\bibinfo {volume} {126}},\
  \bibinfo {pages} {030604} (\bibinfo {year} {2021}{\natexlab{a}})}\BibitemShut
  {NoStop}%
\bibitem [{\citenamefont {Tran}\ \emph
  {et~al.}(2021{\natexlab{b}})\citenamefont {Tran}, \citenamefont {Guo},
  \citenamefont {Baldwin}, \citenamefont {Ehrenberg}, \citenamefont
  {Gorshkov},\ and\ \citenamefont {Lucas}}]{PhysRevLett.127.160401}%
  \BibitemOpen
  \bibfield  {author} {\bibinfo {author} {\bibfnamefont {M.~C.}\ \bibnamefont
  {Tran}}, \bibinfo {author} {\bibfnamefont {A.~Y.}\ \bibnamefont {Guo}},
  \bibinfo {author} {\bibfnamefont {C.~L.}\ \bibnamefont {Baldwin}}, \bibinfo
  {author} {\bibfnamefont {A.}~\bibnamefont {Ehrenberg}}, \bibinfo {author}
  {\bibfnamefont {A.~V.}\ \bibnamefont {Gorshkov}},\ and\ \bibinfo {author}
  {\bibfnamefont {A.}~\bibnamefont {Lucas}},\ }\bibfield  {title} {\bibinfo
  {title} {{\it Lieb-Robinson Light Cone for Power-Law Interactions}},\ }\href
  {https://doi.org/10.1103/PhysRevLett.127.160401} {\bibfield  {journal}
  {\bibinfo  {journal} {Phys. Rev. Lett.}\ }\textbf {\bibinfo {volume} {127}},\
  \bibinfo {pages} {160401} (\bibinfo {year} {2021}{\natexlab{b}})}\BibitemShut
  {NoStop}%
\bibitem [{\citenamefont {Chen}\ and\ \citenamefont
  {Lucas}(2021)}]{PhysRevA.104.062420}%
  \BibitemOpen
  \bibfield  {author} {\bibinfo {author} {\bibfnamefont {C.-F.}\ \bibnamefont
  {Chen}}\ and\ \bibinfo {author} {\bibfnamefont {A.}~\bibnamefont {Lucas}},\
  }\bibfield  {title} {\bibinfo {title} {{\it Optimal Frobenius light cone in
  spin chains with power-law interactions}},\ }\href
  {https://doi.org/10.1103/PhysRevA.104.062420} {\bibfield  {journal} {\bibinfo
   {journal} {Phys. Rev. A}\ }\textbf {\bibinfo {volume} {104}},\ \bibinfo
  {pages} {062420} (\bibinfo {year} {2021})}\BibitemShut {NoStop}%
\bibitem [{\citenamefont {Bravyi}\ \emph {et~al.}(2006)\citenamefont {Bravyi},
  \citenamefont {Hastings},\ and\ \citenamefont
  {Verstraete}}]{PhysRevLett.97.050401}%
  \BibitemOpen
  \bibfield  {author} {\bibinfo {author} {\bibfnamefont {S.}~\bibnamefont
  {Bravyi}}, \bibinfo {author} {\bibfnamefont {M.~B.}\ \bibnamefont
  {Hastings}},\ and\ \bibinfo {author} {\bibfnamefont {F.}~\bibnamefont
  {Verstraete}},\ }\bibfield  {title} {\bibinfo {title} {{\it Lieb-Robinson
  Bounds and the Generation of Correlations and Topological Quantum Order}},\
  }\href {https://doi.org/10.1103/PhysRevLett.97.050401} {\bibfield  {journal}
  {\bibinfo  {journal} {Phys. Rev. Lett.}\ }\textbf {\bibinfo {volume} {97}},\
  \bibinfo {pages} {050401} (\bibinfo {year} {2006})}\BibitemShut {NoStop}%
\bibitem [{\citenamefont {Cramer}\ \emph
  {et~al.}(2008{\natexlab{a}})\citenamefont {Cramer}, \citenamefont
  {Serafini},\ and\ \citenamefont {Eisert}}]{cramer2008locality}%
  \BibitemOpen
  \bibfield  {author} {\bibinfo {author} {\bibfnamefont {M.}~\bibnamefont
  {Cramer}}, \bibinfo {author} {\bibfnamefont {A.}~\bibnamefont {Serafini}},\
  and\ \bibinfo {author} {\bibfnamefont {J.}~\bibnamefont {Eisert}},\
  }\href@noop {} {\bibinfo {title} {{\it Locality of dynamics in general
  harmonic quantum systems}}} (\bibinfo {year} {2008}{\natexlab{a}}),\ \Eprint
  {https://arxiv.org/abs/0803.0890} {arXiv:0803.0890 [quant-ph]} \BibitemShut
  {NoStop}%
\bibitem [{\citenamefont {Nachtergaele}\ \emph {et~al.}(2009)\citenamefont
  {Nachtergaele}, \citenamefont {Raz}, \citenamefont {Schlein},\ and\
  \citenamefont {Sims}}]{Nachtergaele2009}%
  \BibitemOpen
  \bibfield  {author} {\bibinfo {author} {\bibfnamefont {B.}~\bibnamefont
  {Nachtergaele}}, \bibinfo {author} {\bibfnamefont {H.}~\bibnamefont {Raz}},
  \bibinfo {author} {\bibfnamefont {B.}~\bibnamefont {Schlein}},\ and\ \bibinfo
  {author} {\bibfnamefont {R.}~\bibnamefont {Sims}},\ }\bibfield  {title}
  {\bibinfo {title} {{\it Lieb-Robinson Bounds for Harmonic and Anharmonic
  Lattice Systems}},\ }\href {https://doi.org/10.1007/s00220-008-0630-2}
  {\bibfield  {journal} {\bibinfo  {journal} {Communications in Mathematical
  Physics}\ }\textbf {\bibinfo {volume} {286}},\ \bibinfo {pages} {1073}
  (\bibinfo {year} {2009})}\BibitemShut {NoStop}%
\bibitem [{\citenamefont {Eisert}\ and\ \citenamefont
  {Gross}(2009)}]{PhysRevLett.102.240501}%
  \BibitemOpen
  \bibfield  {author} {\bibinfo {author} {\bibfnamefont {J.}~\bibnamefont
  {Eisert}}\ and\ \bibinfo {author} {\bibfnamefont {D.}~\bibnamefont {Gross}},\
  }\bibfield  {title} {\bibinfo {title} {{\it Supersonic Quantum
  Communication}},\ }\href {https://doi.org/10.1103/PhysRevLett.102.240501}
  {\bibfield  {journal} {\bibinfo  {journal} {Phys. Rev. Lett.}\ }\textbf
  {\bibinfo {volume} {102}},\ \bibinfo {pages} {240501} (\bibinfo {year}
  {2009})}\BibitemShut {NoStop}%
\bibitem [{\citenamefont {J\"unemann}\ \emph {et~al.}(2013)\citenamefont
  {J\"unemann}, \citenamefont {Cadarso}, \citenamefont {P\'erez-Garc\'{\i}a},
  \citenamefont {Bermudez},\ and\ \citenamefont
  {Garc\'{\i}a-Ripoll}}]{PhysRevLett.111.230404}%
  \BibitemOpen
  \bibfield  {author} {\bibinfo {author} {\bibfnamefont {J.}~\bibnamefont
  {J\"unemann}}, \bibinfo {author} {\bibfnamefont {A.}~\bibnamefont {Cadarso}},
  \bibinfo {author} {\bibfnamefont {D.}~\bibnamefont {P\'erez-Garc\'{\i}a}},
  \bibinfo {author} {\bibfnamefont {A.}~\bibnamefont {Bermudez}},\ and\
  \bibinfo {author} {\bibfnamefont {J.~J.}\ \bibnamefont
  {Garc\'{\i}a-Ripoll}},\ }\bibfield  {title} {\bibinfo {title} {{\it
  Lieb-Robinson Bounds for Spin-Boson Lattice Models and Trapped Ions}},\
  }\href {https://doi.org/10.1103/PhysRevLett.111.230404} {\bibfield  {journal}
  {\bibinfo  {journal} {Phys. Rev. Lett.}\ }\textbf {\bibinfo {volume} {111}},\
  \bibinfo {pages} {230404} (\bibinfo {year} {2013})}\BibitemShut {NoStop}%
\bibitem [{\citenamefont {Woods}\ \emph {et~al.}(2015)\citenamefont {Woods},
  \citenamefont {Cramer},\ and\ \citenamefont
  {Plenio}}]{PhysRevLett.115.130401}%
  \BibitemOpen
  \bibfield  {author} {\bibinfo {author} {\bibfnamefont {M.~P.}\ \bibnamefont
  {Woods}}, \bibinfo {author} {\bibfnamefont {M.}~\bibnamefont {Cramer}},\ and\
  \bibinfo {author} {\bibfnamefont {M.~B.}\ \bibnamefont {Plenio}},\ }\bibfield
   {title} {\bibinfo {title} {{\it Simulating Bosonic Baths with Error Bars}},\
  }\href {https://doi.org/10.1103/PhysRevLett.115.130401} {\bibfield  {journal}
  {\bibinfo  {journal} {Phys. Rev. Lett.}\ }\textbf {\bibinfo {volume} {115}},\
  \bibinfo {pages} {130401} (\bibinfo {year} {2015})}\BibitemShut {NoStop}%
\bibitem [{\citenamefont {Tong}\ \emph {et~al.}(2022)\citenamefont {Tong},
  \citenamefont {Albert}, \citenamefont {McClean}, \citenamefont {Preskill},\
  and\ \citenamefont {Su}}]{Tong2022}%
  \BibitemOpen
  \bibfield  {author} {\bibinfo {author} {\bibfnamefont {Y.}~\bibnamefont
  {Tong}}, \bibinfo {author} {\bibfnamefont {V.~V.}\ \bibnamefont {Albert}},
  \bibinfo {author} {\bibfnamefont {J.~R.}\ \bibnamefont {McClean}}, \bibinfo
  {author} {\bibfnamefont {J.}~\bibnamefont {Preskill}},\ and\ \bibinfo
  {author} {\bibfnamefont {Y.}~\bibnamefont {Su}},\ }\bibfield  {title}
  {\bibinfo {title} {{\it Provably accurate simulation of gauge theories and
  bosonic systems}},\ }\href {https://doi.org/10.22331/q-2022-09-22-816}
  {\bibfield  {journal} {\bibinfo  {journal} {Quantum}\ }\textbf {\bibinfo
  {volume} {6}},\ \bibinfo {pages} {816} (\bibinfo {year} {2022})}\BibitemShut
  {NoStop}%
\bibitem [{\citenamefont {Childs}\ \emph {et~al.}(2013)\citenamefont {Childs},
  \citenamefont {Gosset},\ and\ \citenamefont
  {Webb}}]{doi:10.1126/science.1229957}%
  \BibitemOpen
  \bibfield  {author} {\bibinfo {author} {\bibfnamefont {A.~M.}\ \bibnamefont
  {Childs}}, \bibinfo {author} {\bibfnamefont {D.}~\bibnamefont {Gosset}},\
  and\ \bibinfo {author} {\bibfnamefont {Z.}~\bibnamefont {Webb}},\ }\bibfield
  {title} {\bibinfo {title} {{\it Universal Computation by Multiparticle
  Quantum Walk}},\ }\href {https://doi.org/10.1126/science.1229957} {\bibfield
  {journal} {\bibinfo  {journal} {Science}\ }\textbf {\bibinfo {volume}
  {339}},\ \bibinfo {pages} {791} (\bibinfo {year} {2013})}\BibitemShut
  {NoStop}%
\bibitem [{\citenamefont {Gross}\ and\ \citenamefont
  {Bloch}(2017)}]{doi:10.1126/science.aal3837}%
  \BibitemOpen
  \bibfield  {author} {\bibinfo {author} {\bibfnamefont {C.}~\bibnamefont
  {Gross}}\ and\ \bibinfo {author} {\bibfnamefont {I.}~\bibnamefont {Bloch}},\
  }\bibfield  {title} {\bibinfo {title} {{\it Quantum simulations with
  ultracold atoms in optical lattices}},\ }\href
  {https://doi.org/10.1126/science.aal3837} {\bibfield  {journal} {\bibinfo
  {journal} {Science}\ }\textbf {\bibinfo {volume} {357}},\ \bibinfo {pages}
  {995} (\bibinfo {year} {2017})}\BibitemShut {NoStop}%
\bibitem [{\citenamefont {Yang}\ \emph {et~al.}(2020)\citenamefont {Yang},
  \citenamefont {Sun}, \citenamefont {Ott}, \citenamefont {Wang}, \citenamefont
  {Zache}, \citenamefont {Halimeh}, \citenamefont {Yuan}, \citenamefont
  {Hauke},\ and\ \citenamefont {Pan}}]{Yang2020}%
  \BibitemOpen
  \bibfield  {author} {\bibinfo {author} {\bibfnamefont {B.}~\bibnamefont
  {Yang}}, \bibinfo {author} {\bibfnamefont {H.}~\bibnamefont {Sun}}, \bibinfo
  {author} {\bibfnamefont {R.}~\bibnamefont {Ott}}, \bibinfo {author}
  {\bibfnamefont {H.-Y.}\ \bibnamefont {Wang}}, \bibinfo {author}
  {\bibfnamefont {T.~V.}\ \bibnamefont {Zache}}, \bibinfo {author}
  {\bibfnamefont {J.~C.}\ \bibnamefont {Halimeh}}, \bibinfo {author}
  {\bibfnamefont {Z.-S.}\ \bibnamefont {Yuan}}, \bibinfo {author}
  {\bibfnamefont {P.}~\bibnamefont {Hauke}},\ and\ \bibinfo {author}
  {\bibfnamefont {J.-W.}\ \bibnamefont {Pan}},\ }\bibfield  {title} {\bibinfo
  {title} {{\it Observation of gauge invariance in a 71-site Bose--Hubbard
  quantum simulator}},\ }\href {https://doi.org/10.1038/s41586-020-2910-8}
  {\bibfield  {journal} {\bibinfo  {journal} {Nature}\ }\textbf {\bibinfo
  {volume} {587}},\ \bibinfo {pages} {392} (\bibinfo {year}
  {2020})}\BibitemShut {NoStop}%
\bibitem [{\citenamefont {Altman}\ \emph {et~al.}(2021)\citenamefont {Altman},
  \citenamefont {Brown}, \citenamefont {Carleo}, \citenamefont {Carr},
  \citenamefont {Demler}, \citenamefont {Chin}, \citenamefont {DeMarco},
  \citenamefont {Economou}, \citenamefont {Eriksson}, \citenamefont {Fu},
  \citenamefont {Greiner}, \citenamefont {Hazzard}, \citenamefont {Hulet},
  \citenamefont {Koll\'ar}, \citenamefont {Lev}, \citenamefont {Lukin},
  \citenamefont {Ma}, \citenamefont {Mi}, \citenamefont {Misra}, \citenamefont
  {Monroe}, \citenamefont {Murch}, \citenamefont {Nazario}, \citenamefont {Ni},
  \citenamefont {Potter}, \citenamefont {Roushan}, \citenamefont {Saffman},
  \citenamefont {Schleier-Smith}, \citenamefont {Siddiqi}, \citenamefont
  {Simmonds}, \citenamefont {Singh}, \citenamefont {Spielman}, \citenamefont
  {Temme}, \citenamefont {Weiss}, \citenamefont {Vu\ifmmode \check{c}\else
  \v{c}\fi{}kovi\ifmmode~\acute{c}\else \'{c}\fi{}}, \citenamefont
  {Vuleti\ifmmode~\acute{c}\else \'{c}\fi{}}, \citenamefont {Ye},\ and\
  \citenamefont {Zwierlein}}]{PRXQuantum.2.017003}%
  \BibitemOpen
  \bibfield  {author} {\bibinfo {author} {\bibfnamefont {E.}~\bibnamefont
  {Altman}}, \bibinfo {author} {\bibfnamefont {K.~R.}\ \bibnamefont {Brown}},
  \bibinfo {author} {\bibfnamefont {G.}~\bibnamefont {Carleo}}, \bibinfo
  {author} {\bibfnamefont {L.~D.}\ \bibnamefont {Carr}}, \bibinfo {author}
  {\bibfnamefont {E.}~\bibnamefont {Demler}}, \bibinfo {author} {\bibfnamefont
  {C.}~\bibnamefont {Chin}}, \bibinfo {author} {\bibfnamefont {B.}~\bibnamefont
  {DeMarco}}, \bibinfo {author} {\bibfnamefont {S.~E.}\ \bibnamefont
  {Economou}}, \bibinfo {author} {\bibfnamefont {M.~A.}\ \bibnamefont
  {Eriksson}}, \bibinfo {author} {\bibfnamefont {K.-M.~C.}\ \bibnamefont {Fu}},
  \bibinfo {author} {\bibfnamefont {M.}~\bibnamefont {Greiner}}, \bibinfo
  {author} {\bibfnamefont {K.~R.}\ \bibnamefont {Hazzard}}, \bibinfo {author}
  {\bibfnamefont {R.~G.}\ \bibnamefont {Hulet}}, \bibinfo {author}
  {\bibfnamefont {A.~J.}\ \bibnamefont {Koll\'ar}}, \bibinfo {author}
  {\bibfnamefont {B.~L.}\ \bibnamefont {Lev}}, \bibinfo {author} {\bibfnamefont
  {M.~D.}\ \bibnamefont {Lukin}}, \bibinfo {author} {\bibfnamefont
  {R.}~\bibnamefont {Ma}}, \bibinfo {author} {\bibfnamefont {X.}~\bibnamefont
  {Mi}}, \bibinfo {author} {\bibfnamefont {S.}~\bibnamefont {Misra}}, \bibinfo
  {author} {\bibfnamefont {C.}~\bibnamefont {Monroe}}, \bibinfo {author}
  {\bibfnamefont {K.}~\bibnamefont {Murch}}, \bibinfo {author} {\bibfnamefont
  {Z.}~\bibnamefont {Nazario}}, \bibinfo {author} {\bibfnamefont {K.-K.}\
  \bibnamefont {Ni}}, \bibinfo {author} {\bibfnamefont {A.~C.}\ \bibnamefont
  {Potter}}, \bibinfo {author} {\bibfnamefont {P.}~\bibnamefont {Roushan}},
  \bibinfo {author} {\bibfnamefont {M.}~\bibnamefont {Saffman}}, \bibinfo
  {author} {\bibfnamefont {M.}~\bibnamefont {Schleier-Smith}}, \bibinfo
  {author} {\bibfnamefont {I.}~\bibnamefont {Siddiqi}}, \bibinfo {author}
  {\bibfnamefont {R.}~\bibnamefont {Simmonds}}, \bibinfo {author}
  {\bibfnamefont {M.}~\bibnamefont {Singh}}, \bibinfo {author} {\bibfnamefont
  {I.}~\bibnamefont {Spielman}}, \bibinfo {author} {\bibfnamefont
  {K.}~\bibnamefont {Temme}}, \bibinfo {author} {\bibfnamefont {D.~S.}\
  \bibnamefont {Weiss}}, \bibinfo {author} {\bibfnamefont {J.}~\bibnamefont
  {Vu\ifmmode \check{c}\else \v{c}\fi{}kovi\ifmmode~\acute{c}\else
  \'{c}\fi{}}}, \bibinfo {author} {\bibfnamefont {V.}~\bibnamefont
  {Vuleti\ifmmode~\acute{c}\else \'{c}\fi{}}}, \bibinfo {author} {\bibfnamefont
  {J.}~\bibnamefont {Ye}},\ and\ \bibinfo {author} {\bibfnamefont
  {M.}~\bibnamefont {Zwierlein}},\ }\bibfield  {title} {\bibinfo {title} {{\it
  Quantum Simulators: Architectures and Opportunities}},\ }\href
  {https://doi.org/10.1103/PRXQuantum.2.017003} {\bibfield  {journal} {\bibinfo
   {journal} {PRX Quantum}\ }\textbf {\bibinfo {volume} {2}},\ \bibinfo {pages}
  {017003} (\bibinfo {year} {2021})}\BibitemShut {NoStop}%
\bibitem [{\citenamefont {Ebadi}\ \emph {et~al.}(2021)\citenamefont {Ebadi},
  \citenamefont {Wang}, \citenamefont {Levine}, \citenamefont {Keesling},
  \citenamefont {Semeghini}, \citenamefont {Omran}, \citenamefont {Bluvstein},
  \citenamefont {Samajdar}, \citenamefont {Pichler}, \citenamefont {Ho},
  \citenamefont {Choi}, \citenamefont {Sachdev}, \citenamefont {Greiner},
  \citenamefont {Vuleti{\'{c}}},\ and\ \citenamefont {Lukin}}]{Ebadi2021}%
  \BibitemOpen
  \bibfield  {author} {\bibinfo {author} {\bibfnamefont {S.}~\bibnamefont
  {Ebadi}}, \bibinfo {author} {\bibfnamefont {T.~T.}\ \bibnamefont {Wang}},
  \bibinfo {author} {\bibfnamefont {H.}~\bibnamefont {Levine}}, \bibinfo
  {author} {\bibfnamefont {A.}~\bibnamefont {Keesling}}, \bibinfo {author}
  {\bibfnamefont {G.}~\bibnamefont {Semeghini}}, \bibinfo {author}
  {\bibfnamefont {A.}~\bibnamefont {Omran}}, \bibinfo {author} {\bibfnamefont
  {D.}~\bibnamefont {Bluvstein}}, \bibinfo {author} {\bibfnamefont
  {R.}~\bibnamefont {Samajdar}}, \bibinfo {author} {\bibfnamefont
  {H.}~\bibnamefont {Pichler}}, \bibinfo {author} {\bibfnamefont {W.~W.}\
  \bibnamefont {Ho}}, \bibinfo {author} {\bibfnamefont {S.}~\bibnamefont
  {Choi}}, \bibinfo {author} {\bibfnamefont {S.}~\bibnamefont {Sachdev}},
  \bibinfo {author} {\bibfnamefont {M.}~\bibnamefont {Greiner}}, \bibinfo
  {author} {\bibfnamefont {V.}~\bibnamefont {Vuleti{\'{c}}}},\ and\ \bibinfo
  {author} {\bibfnamefont {M.~D.}\ \bibnamefont {Lukin}},\ }\bibfield  {title}
  {\bibinfo {title} {{\it Quantum phases of matter on a 256-atom programmable
  quantum simulator}},\ }\href {https://doi.org/10.1038/s41586-021-03582-4}
  {\bibfield  {journal} {\bibinfo  {journal} {Nature}\ }\textbf {\bibinfo
  {volume} {595}},\ \bibinfo {pages} {227} (\bibinfo {year}
  {2021})}\BibitemShut {NoStop}%
\bibitem [{\citenamefont {Kollath}\ \emph {et~al.}(2007)\citenamefont
  {Kollath}, \citenamefont {L\"auchli},\ and\ \citenamefont
  {Altman}}]{PhysRevLett.98.180601}%
  \BibitemOpen
  \bibfield  {author} {\bibinfo {author} {\bibfnamefont {C.}~\bibnamefont
  {Kollath}}, \bibinfo {author} {\bibfnamefont {A.~M.}\ \bibnamefont
  {L\"auchli}},\ and\ \bibinfo {author} {\bibfnamefont {E.}~\bibnamefont
  {Altman}},\ }\bibfield  {title} {\bibinfo {title} {{\it Quench Dynamics and
  Nonequilibrium Phase Diagram of the Bose-Hubbard Model}},\ }\href
  {https://doi.org/10.1103/PhysRevLett.98.180601} {\bibfield  {journal}
  {\bibinfo  {journal} {Phys. Rev. Lett.}\ }\textbf {\bibinfo {volume} {98}},\
  \bibinfo {pages} {180601} (\bibinfo {year} {2007})}\BibitemShut {NoStop}%
\bibitem [{\citenamefont {Läuchli}\ and\ \citenamefont
  {Kollath}(2008)}]{L_uchli_2008}%
  \BibitemOpen
  \bibfield  {author} {\bibinfo {author} {\bibfnamefont {A.~M.}\ \bibnamefont
  {Läuchli}}\ and\ \bibinfo {author} {\bibfnamefont {C.}~\bibnamefont
  {Kollath}},\ }\bibfield  {title} {\bibinfo {title} {{\it Spreading of
  correlations and entanglement after a quench in the one-dimensional
  Bose{\textendash}Hubbard model}},\ }\href
  {https://doi.org/10.1088/1742-5468/2008/05/p05018} {\bibfield  {journal}
  {\bibinfo  {journal} {Journal of Statistical Mechanics: Theory and
  Experiment}\ }\textbf {\bibinfo {volume} {2008}},\ \bibinfo {pages} {P05018}
  (\bibinfo {year} {2008})}\BibitemShut {NoStop}%
\bibitem [{\citenamefont {Cramer}\ \emph
  {et~al.}(2008{\natexlab{b}})\citenamefont {Cramer}, \citenamefont {Dawson},
  \citenamefont {Eisert},\ and\ \citenamefont
  {Osborne}}]{PhysRevLett.100.030602}%
  \BibitemOpen
  \bibfield  {author} {\bibinfo {author} {\bibfnamefont {M.}~\bibnamefont
  {Cramer}}, \bibinfo {author} {\bibfnamefont {C.~M.}\ \bibnamefont {Dawson}},
  \bibinfo {author} {\bibfnamefont {J.}~\bibnamefont {Eisert}},\ and\ \bibinfo
  {author} {\bibfnamefont {T.~J.}\ \bibnamefont {Osborne}},\ }\bibfield
  {title} {\bibinfo {title} {{\it Exact Relaxation in a Class of Nonequilibrium
  Quantum Lattice Systems}},\ }\href
  {https://doi.org/10.1103/PhysRevLett.100.030602} {\bibfield  {journal}
  {\bibinfo  {journal} {Phys. Rev. Lett.}\ }\textbf {\bibinfo {volume} {100}},\
  \bibinfo {pages} {030602} (\bibinfo {year} {2008}{\natexlab{b}})}\BibitemShut
  {NoStop}%
\bibitem [{\citenamefont {Cramer}\ \emph
  {et~al.}(2008{\natexlab{c}})\citenamefont {Cramer}, \citenamefont {Flesch},
  \citenamefont {McCulloch}, \citenamefont {Schollw\"ock},\ and\ \citenamefont
  {Eisert}}]{PhysRevLett.101.063001}%
  \BibitemOpen
  \bibfield  {author} {\bibinfo {author} {\bibfnamefont {M.}~\bibnamefont
  {Cramer}}, \bibinfo {author} {\bibfnamefont {A.}~\bibnamefont {Flesch}},
  \bibinfo {author} {\bibfnamefont {I.~P.}\ \bibnamefont {McCulloch}}, \bibinfo
  {author} {\bibfnamefont {U.}~\bibnamefont {Schollw\"ock}},\ and\ \bibinfo
  {author} {\bibfnamefont {J.}~\bibnamefont {Eisert}},\ }\bibfield  {title}
  {\bibinfo {title} {{\it Exploring Local Quantum Many-Body Relaxation by Atoms
  in Optical Superlattices}},\ }\href
  {https://doi.org/10.1103/PhysRevLett.101.063001} {\bibfield  {journal}
  {\bibinfo  {journal} {Phys. Rev. Lett.}\ }\textbf {\bibinfo {volume} {101}},\
  \bibinfo {pages} {063001} (\bibinfo {year} {2008}{\natexlab{c}})}\BibitemShut
  {NoStop}%
\bibitem [{\citenamefont {Barmettler}\ \emph {et~al.}(2012)\citenamefont
  {Barmettler}, \citenamefont {Poletti}, \citenamefont {Cheneau},\ and\
  \citenamefont {Kollath}}]{PhysRevA.85.053625}%
  \BibitemOpen
  \bibfield  {author} {\bibinfo {author} {\bibfnamefont {P.}~\bibnamefont
  {Barmettler}}, \bibinfo {author} {\bibfnamefont {D.}~\bibnamefont {Poletti}},
  \bibinfo {author} {\bibfnamefont {M.}~\bibnamefont {Cheneau}},\ and\ \bibinfo
  {author} {\bibfnamefont {C.}~\bibnamefont {Kollath}},\ }\bibfield  {title}
  {\bibinfo {title} {{\it Propagation front of correlations in an interacting
  Bose gas}},\ }\href {https://doi.org/10.1103/PhysRevA.85.053625} {\bibfield
  {journal} {\bibinfo  {journal} {Phys. Rev. A}\ }\textbf {\bibinfo {volume}
  {85}},\ \bibinfo {pages} {053625} (\bibinfo {year} {2012})}\BibitemShut
  {NoStop}%
\bibitem [{\citenamefont {Carleo}\ \emph {et~al.}(2014)\citenamefont {Carleo},
  \citenamefont {Becca}, \citenamefont {Sanchez-Palencia}, \citenamefont
  {Sorella},\ and\ \citenamefont {Fabrizio}}]{PhysRevA.89.031602}%
  \BibitemOpen
  \bibfield  {author} {\bibinfo {author} {\bibfnamefont {G.}~\bibnamefont
  {Carleo}}, \bibinfo {author} {\bibfnamefont {F.}~\bibnamefont {Becca}},
  \bibinfo {author} {\bibfnamefont {L.}~\bibnamefont {Sanchez-Palencia}},
  \bibinfo {author} {\bibfnamefont {S.}~\bibnamefont {Sorella}},\ and\ \bibinfo
  {author} {\bibfnamefont {M.}~\bibnamefont {Fabrizio}},\ }\bibfield  {title}
  {\bibinfo {title} {{\it Light-cone effect and supersonic correlations in one-
  and two-dimensional bosonic superfluids}},\ }\href
  {https://doi.org/10.1103/PhysRevA.89.031602} {\bibfield  {journal} {\bibinfo
  {journal} {Phys. Rev. A}\ }\textbf {\bibinfo {volume} {89}},\ \bibinfo
  {pages} {031602} (\bibinfo {year} {2014})}\BibitemShut {NoStop}%
\bibitem [{\citenamefont {Bakr}\ \emph {et~al.}(2010)\citenamefont {Bakr},
  \citenamefont {Peng}, \citenamefont {Tai}, \citenamefont {Ma}, \citenamefont
  {Simon}, \citenamefont {Gillen}, \citenamefont {F{\"o}lling}, \citenamefont
  {Pollet},\ and\ \citenamefont {Greiner}}]{Bakr547}%
  \BibitemOpen
  \bibfield  {author} {\bibinfo {author} {\bibfnamefont {W.~S.}\ \bibnamefont
  {Bakr}}, \bibinfo {author} {\bibfnamefont {A.}~\bibnamefont {Peng}}, \bibinfo
  {author} {\bibfnamefont {M.~E.}\ \bibnamefont {Tai}}, \bibinfo {author}
  {\bibfnamefont {R.}~\bibnamefont {Ma}}, \bibinfo {author} {\bibfnamefont
  {J.}~\bibnamefont {Simon}}, \bibinfo {author} {\bibfnamefont {J.~I.}\
  \bibnamefont {Gillen}}, \bibinfo {author} {\bibfnamefont {S.}~\bibnamefont
  {F{\"o}lling}}, \bibinfo {author} {\bibfnamefont {L.}~\bibnamefont
  {Pollet}},\ and\ \bibinfo {author} {\bibfnamefont {M.}~\bibnamefont
  {Greiner}},\ }\bibfield  {title} {\bibinfo {title} {{\it Probing the
  Superfluid{\textendash}to{\textendash}Mott Insulator Transition at the
  Single-Atom Level}},\ }\href {https://doi.org/10.1126/science.1192368}
  {\bibfield  {journal} {\bibinfo  {journal} {Science}\ }\textbf {\bibinfo
  {volume} {329}},\ \bibinfo {pages} {547} (\bibinfo {year}
  {2010})}\BibitemShut {NoStop}%
\bibitem [{\citenamefont {Baier}\ \emph {et~al.}(2016)\citenamefont {Baier},
  \citenamefont {Mark}, \citenamefont {Petter}, \citenamefont {Aikawa},
  \citenamefont {Chomaz}, \citenamefont {Cai}, \citenamefont {Baranov},
  \citenamefont {Zoller},\ and\ \citenamefont {Ferlaino}}]{Baier201}%
  \BibitemOpen
  \bibfield  {author} {\bibinfo {author} {\bibfnamefont {S.}~\bibnamefont
  {Baier}}, \bibinfo {author} {\bibfnamefont {M.~J.}\ \bibnamefont {Mark}},
  \bibinfo {author} {\bibfnamefont {D.}~\bibnamefont {Petter}}, \bibinfo
  {author} {\bibfnamefont {K.}~\bibnamefont {Aikawa}}, \bibinfo {author}
  {\bibfnamefont {L.}~\bibnamefont {Chomaz}}, \bibinfo {author} {\bibfnamefont
  {Z.}~\bibnamefont {Cai}}, \bibinfo {author} {\bibfnamefont {M.}~\bibnamefont
  {Baranov}}, \bibinfo {author} {\bibfnamefont {P.}~\bibnamefont {Zoller}},\
  and\ \bibinfo {author} {\bibfnamefont {F.}~\bibnamefont {Ferlaino}},\
  }\bibfield  {title} {\bibinfo {title} {{\it Extended Bose-Hubbard models with
  ultracold magnetic atoms}},\ }\href {https://doi.org/10.1126/science.aac9812}
  {\bibfield  {journal} {\bibinfo  {journal} {Science}\ }\textbf {\bibinfo
  {volume} {352}},\ \bibinfo {pages} {201} (\bibinfo {year}
  {2016})}\BibitemShut {NoStop}%
\bibitem [{\citenamefont {Schuch}\ \emph {et~al.}(2011)\citenamefont {Schuch},
  \citenamefont {Harrison}, \citenamefont {Osborne},\ and\ \citenamefont
  {Eisert}}]{PhysRevA.84.032309}%
  \BibitemOpen
  \bibfield  {author} {\bibinfo {author} {\bibfnamefont {N.}~\bibnamefont
  {Schuch}}, \bibinfo {author} {\bibfnamefont {S.~K.}\ \bibnamefont
  {Harrison}}, \bibinfo {author} {\bibfnamefont {T.~J.}\ \bibnamefont
  {Osborne}},\ and\ \bibinfo {author} {\bibfnamefont {J.}~\bibnamefont
  {Eisert}},\ }\bibfield  {title} {\bibinfo {title} {{\it Information
  propagation for interacting-particle systems}},\ }\href
  {https://doi.org/10.1103/PhysRevA.84.032309} {\bibfield  {journal} {\bibinfo
  {journal} {Phys. Rev. A}\ }\textbf {\bibinfo {volume} {84}},\ \bibinfo
  {pages} {032309} (\bibinfo {year} {2011})}\BibitemShut {NoStop}%
\bibitem [{\citenamefont {Faupin}\ \emph
  {et~al.}(2022{\natexlab{a}})\citenamefont {Faupin}, \citenamefont {Lemm},\
  and\ \citenamefont {Sigal}}]{PhysRevLett.128.150602}%
  \BibitemOpen
  \bibfield  {author} {\bibinfo {author} {\bibfnamefont {J.}~\bibnamefont
  {Faupin}}, \bibinfo {author} {\bibfnamefont {M.}~\bibnamefont {Lemm}},\ and\
  \bibinfo {author} {\bibfnamefont {I.~M.}\ \bibnamefont {Sigal}},\ }\bibfield
  {title} {\bibinfo {title} {{\it Maximal Speed for Macroscopic Particle
  Transport in the Bose-Hubbard Model}},\ }\href
  {https://doi.org/10.1103/PhysRevLett.128.150602} {\bibfield  {journal}
  {\bibinfo  {journal} {Phys. Rev. Lett.}\ }\textbf {\bibinfo {volume} {128}},\
  \bibinfo {pages} {150602} (\bibinfo {year} {2022}{\natexlab{a}})}\BibitemShut
  {NoStop}%
\bibitem [{\citenamefont {Wang}\ and\ \citenamefont
  {Hazzard}(2020)}]{PRXQuantum.1.010303}%
  \BibitemOpen
  \bibfield  {author} {\bibinfo {author} {\bibfnamefont {Z.}~\bibnamefont
  {Wang}}\ and\ \bibinfo {author} {\bibfnamefont {K.~R.}\ \bibnamefont
  {Hazzard}},\ }\bibfield  {title} {\bibinfo {title} {{\it Tightening the
  Lieb-Robinson Bound in Locally Interacting Systems}},\ }\href
  {https://doi.org/10.1103/PRXQuantum.1.010303} {\bibfield  {journal} {\bibinfo
   {journal} {PRX Quantum}\ }\textbf {\bibinfo {volume} {1}},\ \bibinfo {pages}
  {010303} (\bibinfo {year} {2020})}\BibitemShut {NoStop}%
\bibitem [{\citenamefont {Kuwahara}\ and\ \citenamefont
  {Saito}(2021{\natexlab{b}})}]{PhysRevLett.127.070403}%
  \BibitemOpen
  \bibfield  {author} {\bibinfo {author} {\bibfnamefont {T.}~\bibnamefont
  {Kuwahara}}\ and\ \bibinfo {author} {\bibfnamefont {K.}~\bibnamefont
  {Saito}},\ }\bibfield  {title} {\bibinfo {title} {{\it Lieb-Robinson Bound
  and Almost-Linear Light Cone in Interacting Boson Systems}},\ }\href
  {https://doi.org/10.1103/PhysRevLett.127.070403} {\bibfield  {journal}
  {\bibinfo  {journal} {Phys. Rev. Lett.}\ }\textbf {\bibinfo {volume} {127}},\
  \bibinfo {pages} {070403} (\bibinfo {year} {2021}{\natexlab{b}})}\BibitemShut
  {NoStop}%
\bibitem [{\citenamefont {Yin}\ and\ \citenamefont
  {Lucas}(2022)}]{PhysRevX.12.021039}%
  \BibitemOpen
  \bibfield  {author} {\bibinfo {author} {\bibfnamefont {C.}~\bibnamefont
  {Yin}}\ and\ \bibinfo {author} {\bibfnamefont {A.}~\bibnamefont {Lucas}},\
  }\bibfield  {title} {\bibinfo {title} {{\it Finite Speed of Quantum
  Information in Models of Interacting Bosons at Finite Density}},\ }\href
  {https://doi.org/10.1103/PhysRevX.12.021039} {\bibfield  {journal} {\bibinfo
  {journal} {Phys. Rev. X}\ }\textbf {\bibinfo {volume} {12}},\ \bibinfo
  {pages} {021039} (\bibinfo {year} {2022})}\BibitemShut {NoStop}%
\bibitem [{\citenamefont {Faupin}\ \emph
  {et~al.}(2022{\natexlab{b}})\citenamefont {Faupin}, \citenamefont {Lemm},\
  and\ \citenamefont {Sigal}}]{Faupin2022}%
  \BibitemOpen
  \bibfield  {author} {\bibinfo {author} {\bibfnamefont {J.}~\bibnamefont
  {Faupin}}, \bibinfo {author} {\bibfnamefont {M.}~\bibnamefont {Lemm}},\ and\
  \bibinfo {author} {\bibfnamefont {I.~M.}\ \bibnamefont {Sigal}},\ }\bibfield
  {title} {\bibinfo {title} {{\it On Lieb--Robinson Bounds for the
  Bose--Hubbard Model}},\ }\bibfield  {journal} {\bibinfo  {journal}
  {Communications in Mathematical Physics}\ }\href
  {https://doi.org/10.1007/s00220-022-04416-8} {10.1007/s00220-022-04416-8}
  (\bibinfo {year} {2022}{\natexlab{b}})\BibitemShut {NoStop}%
\bibitem [{\citenamefont {Berry}\ \emph {et~al.}(2017)\citenamefont {Berry},
  \citenamefont {Childs}, \citenamefont {Cleve}, \citenamefont {Kothari},\ and\
  \citenamefont {Somma}}]{berry2017exponential}%
  \BibitemOpen
  \bibfield  {author} {\bibinfo {author} {\bibfnamefont {D.~W.}\ \bibnamefont
  {Berry}}, \bibinfo {author} {\bibfnamefont {A.~M.}\ \bibnamefont {Childs}},
  \bibinfo {author} {\bibfnamefont {R.}~\bibnamefont {Cleve}}, \bibinfo
  {author} {\bibfnamefont {R.}~\bibnamefont {Kothari}},\ and\ \bibinfo {author}
  {\bibfnamefont {R.~D.}\ \bibnamefont {Somma}},\ }\bibfield  {title} {\bibinfo
  {title} {{\it Exponential improvement in precision for simulating sparse
  Hamiltonians}},\ }in\ \href {https://doi.org/10.1017/fms.2017.2} {\emph
  {\bibinfo {booktitle} {Forum of Mathematics, Sigma}}},\ Vol.~\bibinfo
  {volume} {5}\ (\bibinfo {organization} {Cambridge University Press},\
  \bibinfo {year} {2017})\ p.~\bibinfo {pages} {e8}\BibitemShut {NoStop}%
\bibitem [{\citenamefont {Vu}\ \emph {et~al.}(2023)\citenamefont {Vu},
  \citenamefont {Kuwahara},\ and\ \citenamefont {Saito}}]{vanvu2023optimal}%
  \BibitemOpen
  \bibfield  {author} {\bibinfo {author} {\bibfnamefont {T.~V.}\ \bibnamefont
  {Vu}}, \bibinfo {author} {\bibfnamefont {T.}~\bibnamefont {Kuwahara}},\ and\
  \bibinfo {author} {\bibfnamefont {K.}~\bibnamefont {Saito}},\ }\href@noop {}
  {\bibinfo {title} {Optimal form of light cones for bosonic transport in
  long-range systems}} (\bibinfo {year} {2023}),\ \Eprint
  {https://arxiv.org/abs/2307.01059} {arXiv:2307.01059 [quant-ph]} \BibitemShut
  {NoStop}%
\bibitem [{\citenamefont {Kuwahara}(2016)}]{Kuwahara_2016_njp}%
  \BibitemOpen
  \bibfield  {author} {\bibinfo {author} {\bibfnamefont {T.}~\bibnamefont
  {Kuwahara}},\ }\bibfield  {title} {\bibinfo {title} {{\it Exponential bound
  on information spreading induced by quantum many-body dynamics with
  long-range interactions}},\ }\href
  {https://doi.org/10.1088/1367-2630/18/5/053034} {\bibfield  {journal}
  {\bibinfo  {journal} {New Journal of Physics}\ }\textbf {\bibinfo {volume}
  {18}},\ \bibinfo {pages} {053034} (\bibinfo {year} {2016})}\BibitemShut
  {NoStop}%
\bibitem [{\citenamefont {Zotos}\ \emph {et~al.}(1997)\citenamefont {Zotos},
  \citenamefont {Naef},\ and\ \citenamefont {Prelovsek}}]{PhysRevB.55.11029}%
  \BibitemOpen
  \bibfield  {author} {\bibinfo {author} {\bibfnamefont {X.}~\bibnamefont
  {Zotos}}, \bibinfo {author} {\bibfnamefont {F.}~\bibnamefont {Naef}},\ and\
  \bibinfo {author} {\bibfnamefont {P.}~\bibnamefont {Prelovsek}},\ }\bibfield
  {title} {\bibinfo {title} {Transport and conservation laws},\ }\href
  {https://doi.org/10.1103/PhysRevB.55.11029} {\bibfield  {journal} {\bibinfo
  {journal} {Phys. Rev. B}\ }\textbf {\bibinfo {volume} {55}},\ \bibinfo
  {pages} {11029} (\bibinfo {year} {1997})}\BibitemShut {NoStop}%
\bibitem [{\citenamefont {Maskara}\ \emph {et~al.}(2022)\citenamefont
  {Maskara}, \citenamefont {Deshpande}, \citenamefont {Ehrenberg},
  \citenamefont {Tran}, \citenamefont {Fefferman},\ and\ \citenamefont
  {Gorshkov}}]{maskara2019complexity}%
  \BibitemOpen
  \bibfield  {author} {\bibinfo {author} {\bibfnamefont {N.}~\bibnamefont
  {Maskara}}, \bibinfo {author} {\bibfnamefont {A.}~\bibnamefont {Deshpande}},
  \bibinfo {author} {\bibfnamefont {A.}~\bibnamefont {Ehrenberg}}, \bibinfo
  {author} {\bibfnamefont {M.~C.}\ \bibnamefont {Tran}}, \bibinfo {author}
  {\bibfnamefont {B.}~\bibnamefont {Fefferman}},\ and\ \bibinfo {author}
  {\bibfnamefont {A.~V.}\ \bibnamefont {Gorshkov}},\ }\bibfield  {title}
  {\bibinfo {title} {{\it Complexity Phase Diagram for Interacting and
  Long-Range Bosonic Hamiltonians}},\ }\href
  {https://doi.org/10.1103/PhysRevLett.129.150604} {\bibfield  {journal}
  {\bibinfo  {journal} {Phys. Rev. Lett.}\ }\textbf {\bibinfo {volume} {129}},\
  \bibinfo {pages} {150604} (\bibinfo {year} {2022})}\BibitemShut {NoStop}%
\bibitem [{\citenamefont {Berry}\ \emph {et~al.}(2007)\citenamefont {Berry},
  \citenamefont {Ahokas}, \citenamefont {Cleve},\ and\ \citenamefont
  {Sanders}}]{Berry2007}%
  \BibitemOpen
  \bibfield  {author} {\bibinfo {author} {\bibfnamefont {D.~W.}\ \bibnamefont
  {Berry}}, \bibinfo {author} {\bibfnamefont {G.}~\bibnamefont {Ahokas}},
  \bibinfo {author} {\bibfnamefont {R.}~\bibnamefont {Cleve}},\ and\ \bibinfo
  {author} {\bibfnamefont {B.~C.}\ \bibnamefont {Sanders}},\ }\bibfield
  {title} {\bibinfo {title} {{\it Efficient Quantum Algorithms for Simulating
  Sparse Hamiltonians}},\ }\href {https://doi.org/10.1007/s00220-006-0150-x}
  {\bibfield  {journal} {\bibinfo  {journal} {Communications in Mathematical
  Physics}\ }\textbf {\bibinfo {volume} {270}},\ \bibinfo {pages} {359}
  (\bibinfo {year} {2007})}\BibitemShut {NoStop}%
\end{thebibliography}%


\providecommand{\noopsort}[1]{}\providecommand{\singleletter}[1]{#1}%
\begin{thebibliography}{11}%
\makeatletter
\providecommand \@ifxundefined [1]{%
 \@ifx{#1\undefined}
}%
\providecommand \@ifnum [1]{%
 \ifnum #1\expandafter \@firstoftwo
 \else \expandafter \@secondoftwo
 \fi
}%
\providecommand \@ifx [1]{%
 \ifx #1\expandafter \@firstoftwo
 \else \expandafter \@secondoftwo
 \fi
}%
\providecommand \natexlab [1]{#1}%
\providecommand \enquote  [1]{``#1''}%
\providecommand \bibnamefont  [1]{#1}%
\providecommand \bibfnamefont [1]{#1}%
\providecommand \citenamefont [1]{#1}%
\providecommand \href@noop [0]{\@secondoftwo}%
\providecommand \href [0]{\begingroup \@sanitize@url \@href}%
\providecommand \@href[1]{\@@startlink{#1}\@@href}%
\providecommand \@@href[1]{\endgroup#1\@@endlink}%
\providecommand \@sanitize@url [0]{\catcode `\\12\catcode `\$12\catcode
  `\&12\catcode `\#12\catcode `\^12\catcode `\_12\catcode `\%12\relax}%
\providecommand \@@startlink[1]{}%
\providecommand \@@endlink[0]{}%
\providecommand \url  [0]{\begingroup\@sanitize@url \@url }%
\providecommand \@url [1]{\endgroup\@href {#1}{\urlprefix }}%
\providecommand \urlprefix  [0]{URL }%
\providecommand \Eprint [0]{\href }%
\providecommand \doibase [0]{https://doi.org/}%
\providecommand \selectlanguage [0]{\@gobble}%
\providecommand \bibinfo  [0]{\@secondoftwo}%
\providecommand \bibfield  [0]{\@secondoftwo}%
\providecommand \translation [1]{[#1]}%
\providecommand \BibitemOpen [0]{}%
\providecommand \bibitemStop [0]{}%
\providecommand \bibitemNoStop [0]{.\EOS\space}%
\providecommand \EOS [0]{\spacefactor3000\relax}%
\providecommand \BibitemShut  [1]{\csname bibitem#1\endcsname}%
\let\auto@bib@innerbib\@empty
\bibitem [{\citenamefont {Kuwahara}\ and\ \citenamefont
  {Saito}(2021{\natexlab{a}})}]{PhysRevLett.127.070403}%
  \BibitemOpen
  \bibfield  {author} {\bibinfo {author} {\bibfnamefont {T.}~\bibnamefont
  {Kuwahara}}\ and\ \bibinfo {author} {\bibfnamefont {K.}~\bibnamefont
  {Saito}},\ }\bibfield  {title} {\bibinfo {title} {{\it Lieb-Robinson Bound
  and Almost-Linear Light Cone in Interacting Boson Systems}},\ }\href
  {https://doi.org/10.1103/PhysRevLett.127.070403} {\bibfield  {journal}
  {\bibinfo  {journal} {Phys. Rev. Lett.}\ }\textbf {\bibinfo {volume} {127}},\
  \bibinfo {pages} {070403} (\bibinfo {year} {2021}{\natexlab{a}})}\BibitemShut
  {NoStop}%
\bibitem [{\citenamefont {Schuch}\ \emph {et~al.}(2011)\citenamefont {Schuch},
  \citenamefont {Harrison}, \citenamefont {Osborne},\ and\ \citenamefont
  {Eisert}}]{PhysRevA.84.032309}%
  \BibitemOpen
  \bibfield  {author} {\bibinfo {author} {\bibfnamefont {N.}~\bibnamefont
  {Schuch}}, \bibinfo {author} {\bibfnamefont {S.~K.}\ \bibnamefont
  {Harrison}}, \bibinfo {author} {\bibfnamefont {T.~J.}\ \bibnamefont
  {Osborne}},\ and\ \bibinfo {author} {\bibfnamefont {J.}~\bibnamefont
  {Eisert}},\ }\bibfield  {title} {\bibinfo {title} {{\it Information
  propagation for interacting-particle systems}},\ }\href
  {https://doi.org/10.1103/PhysRevA.84.032309} {\bibfield  {journal} {\bibinfo
  {journal} {Phys. Rev. A}\ }\textbf {\bibinfo {volume} {84}},\ \bibinfo
  {pages} {032309} (\bibinfo {year} {2011})}\BibitemShut {NoStop}%
\bibitem [{\citenamefont {Finner}(1992)}]{Finner1992}%
  \BibitemOpen
  \bibfield  {author} {\bibinfo {author} {\bibfnamefont {H.}~\bibnamefont
  {Finner}},\ }\bibfield  {title} {\bibinfo {title} {{\it A Generalization of
  Holder's Inequality and Some Probability Inequalities}},\ }\href
  {http://www.jstor.org/stable/2244734} {\bibfield  {journal} {\bibinfo
  {journal} {The Annals of Probability}\ }\textbf {\bibinfo {volume} {20}},\
  \bibinfo {pages} {1893} (\bibinfo {year} {1992})},\ \bibinfo {note} {full
  publication date: Oct., 1992}\BibitemShut {NoStop}%
\bibitem [{\citenamefont {Sutter}(2018)}]{sutter2018approximate}%
  \BibitemOpen
  \bibfield  {author} {\bibinfo {author} {\bibfnamefont {D.}~\bibnamefont
  {Sutter}},\ }\bibfield  {title} {\bibinfo {title} {{\it Approximate quantum
  Markov chains}},\ }\href@noop {} {\bibfield  {journal} {\bibinfo  {journal}
  {arXiv preprint arXiv:1802.05477}\ } (\bibinfo {year} {2018})},\ \Eprint
  {https://arxiv.org/abs/arXiv:1802.05477} {arXiv:1802.05477} \BibitemShut
  {NoStop}%
\bibitem [{\citenamefont {Kuwahara}(2015)}]{Phd_kuwahara}%
  \BibitemOpen
  \bibfield  {author} {\bibinfo {author} {\bibfnamefont {T.}~\bibnamefont
  {Kuwahara}},\ }\bibfield  {title} {\bibinfo {title} {{\it Fundamental
  inequalities in quantum many-body systems}},\ }\href
  {http://hatano-lab.iis.u-tokyo.ac.jp/thesis/dron2014/thesis_kuwahara.pdf}
  {\bibfield  {journal} {\bibinfo  {journal} {PhD thesis, University of Tokyo}\
  } (\bibinfo {year} {2015})}\BibitemShut {NoStop}%
\bibitem [{\citenamefont {Hastings}\ and\ \citenamefont
  {Koma}(2006)}]{ref:Hastings2006-ExpDec}%
  \BibitemOpen
  \bibfield  {author} {\bibinfo {author} {\bibfnamefont {M.~B.}\ \bibnamefont
  {Hastings}}\ and\ \bibinfo {author} {\bibfnamefont {T.}~\bibnamefont
  {Koma}},\ }\bibfield  {title} {\bibinfo {title} {{\it Spectral Gap and
  Exponential Decay of Correlations}},\ }\href
  {https://doi.org/10.1007/s00220-006-0030-4} {\bibfield  {journal} {\bibinfo
  {journal} {Communications in Mathematical Physics}\ }\textbf {\bibinfo
  {volume} {265}},\ \bibinfo {pages} {781} (\bibinfo {year}
  {2006})}\BibitemShut {NoStop}%
\bibitem [{\citenamefont {Nachtergaele}\ \emph {et~al.}(2006)\citenamefont
  {Nachtergaele}, \citenamefont {Ogata},\ and\ \citenamefont
  {Sims}}]{Nachtergaele2006}%
  \BibitemOpen
  \bibfield  {author} {\bibinfo {author} {\bibfnamefont {B.}~\bibnamefont
  {Nachtergaele}}, \bibinfo {author} {\bibfnamefont {Y.}~\bibnamefont
  {Ogata}},\ and\ \bibinfo {author} {\bibfnamefont {R.}~\bibnamefont {Sims}},\
  }\bibfield  {title} {\bibinfo {title} {{\it Propagation of Correlations in
  Quantum Lattice Systems}},\ }\href
  {https://doi.org/10.1007/s10955-006-9143-6} {\bibfield  {journal} {\bibinfo
  {journal} {Journal of Statistical Physics}\ }\textbf {\bibinfo {volume}
  {124}},\ \bibinfo {pages} {1} (\bibinfo {year} {2006})}\BibitemShut {NoStop}%
\bibitem [{\citenamefont {Kuwahara}(2016)}]{Kuwahara_2016_njp}%
  \BibitemOpen
  \bibfield  {author} {\bibinfo {author} {\bibfnamefont {T.}~\bibnamefont
  {Kuwahara}},\ }\bibfield  {title} {\bibinfo {title} {{\it Exponential bound
  on information spreading induced by quantum many-body dynamics with
  long-range interactions}},\ }\href
  {https://doi.org/10.1088/1367-2630/18/5/053034} {\bibfield  {journal}
  {\bibinfo  {journal} {New Journal of Physics}\ }\textbf {\bibinfo {volume}
  {18}},\ \bibinfo {pages} {053034} (\bibinfo {year} {2016})}\BibitemShut
  {NoStop}%
\bibitem [{\citenamefont {Kuwahara}\ and\ \citenamefont
  {Saito}(2021{\natexlab{b}})}]{PhysRevLett.126.030604}%
  \BibitemOpen
  \bibfield  {author} {\bibinfo {author} {\bibfnamefont {T.}~\bibnamefont
  {Kuwahara}}\ and\ \bibinfo {author} {\bibfnamefont {K.}~\bibnamefont
  {Saito}},\ }\bibfield  {title} {\bibinfo {title} {{\it Absence of Fast
  Scrambling in Thermodynamically Stable Long-Range Interacting Systems}},\
  }\href {https://doi.org/10.1103/PhysRevLett.126.030604} {\bibfield  {journal}
  {\bibinfo  {journal} {Phys. Rev. Lett.}\ }\textbf {\bibinfo {volume} {126}},\
  \bibinfo {pages} {030604} (\bibinfo {year} {2021}{\natexlab{b}})}\BibitemShut
  {NoStop}%
\bibitem [{\citenamefont {Yin}\ and\ \citenamefont
  {Lucas}(2022)}]{PhysRevX.12.021039}%
  \BibitemOpen
  \bibfield  {author} {\bibinfo {author} {\bibfnamefont {C.}~\bibnamefont
  {Yin}}\ and\ \bibinfo {author} {\bibfnamefont {A.}~\bibnamefont {Lucas}},\
  }\bibfield  {title} {\bibinfo {title} {{\it Finite Speed of Quantum
  Information in Models of Interacting Bosons at Finite Density}},\ }\href
  {https://doi.org/10.1103/PhysRevX.12.021039} {\bibfield  {journal} {\bibinfo
  {journal} {Phys. Rev. X}\ }\textbf {\bibinfo {volume} {12}},\ \bibinfo
  {pages} {021039} (\bibinfo {year} {2022})}\BibitemShut {NoStop}%
\bibitem [{\citenamefont {Barmettler}\ \emph {et~al.}(2012)\citenamefont
  {Barmettler}, \citenamefont {Poletti}, \citenamefont {Cheneau},\ and\
  \citenamefont {Kollath}}]{PhysRevA.85.053625}%
  \BibitemOpen
  \bibfield  {author} {\bibinfo {author} {\bibfnamefont {P.}~\bibnamefont
  {Barmettler}}, \bibinfo {author} {\bibfnamefont {D.}~\bibnamefont {Poletti}},
  \bibinfo {author} {\bibfnamefont {M.}~\bibnamefont {Cheneau}},\ and\ \bibinfo
  {author} {\bibfnamefont {C.}~\bibnamefont {Kollath}},\ }\bibfield  {title}
  {\bibinfo {title} {{\it Propagation front of correlations in an interacting
  Bose gas}},\ }\href {https://doi.org/10.1103/PhysRevA.85.053625} {\bibfield
  {journal} {\bibinfo  {journal} {Phys. Rev. A}\ }\textbf {\bibinfo {volume}
  {85}},\ \bibinfo {pages} {053625} (\bibinfo {year} {2012})}\BibitemShut
  {NoStop}%
\end{thebibliography}%
\end{document}



\title{Supplementary Information for  ``Effective light cone and digital quantum simulation of interacting bosons''}

\author{Tomotaka Kuwahara$^{1,2,3}$}

\author{Tan Van Vu$^{1}$}

\author{Keiji Saito$^{4}$}

\affiliation{$^{1}$ 
Analytical quantum complexity RIKEN Hakubi Research Team, RIKEN Center for Quantum Computing (RQC), Wako, Saitama 351-0198, Japan
}
\affiliation{$^{2}$ 
RIKEN Cluster for Pioneering Research (CPR), Wako, Saitama 351-0198, Japan
}
\affiliation{$^{3}$
PRESTO, Japan Science and Technology (JST), Kawaguchi, Saitama 332-0012, Japan}

\affiliation{$^{4}$ 
Department of Physics, Kyoto University, Kyoto 606-8502, Japan}

\maketitle

\tableofcontents





\renewcommand\thefootnote{*\arabic{footnote}}

\clearpage
\newpage


\setcounter{equation}{0}
\setcounter{section}{0}

\setcounter{tocdepth}{1}

\renewcommand{\figurename}{Supplementary Figure}

\renewcommand{\tablename}{Supplementary Table}
\renewcommand{\thetable}{\arabic{table}}

\renewcommand{\thesection}{Supplementary Note~\arabic{section}}

\section{Set up}
We first show the detailed setup, which partially overlaps with the one in the main text.
We consider a quantum system on a $D$ dimensional lattice (graph),
where the bosons interact with each other. 
The unbounded number of bosons can sit on each site; hence, the local Hilbert dimension is infinitely large.
We denote $\Lambda$ by the set of total sites and for an arbitrary partial set $X\subseteq \Lambda$, we represent the cardinality (the number of sites contained in $X$) by $|X|$.
We also denote the complementary subset and the surface subset of $X$ by $X^\co := \Lambda\setminus X$ and $\partial X:=\{ i\in X| \dist_{i,X^\co}=1\}$, respectively.

For arbitrary subsets $X, Y \subseteq \Lambda$, we define $\dist_{X,Y}$ as the shortest path length on the graph that connects $X$ and $Y$; that is, if $X\cap Y \neq \emptyset$, $\dist_{X,Y}=0$. 
When $X$ is composed of only one element (i.e., $X=\{i\}$), we denote $\dist_{\{i\},Y}$ by $\dist_{i,Y}$ for the simplicity.
We also define $\diam(X)$ as follows: 
\begin{align}
\diam(X):  =1+ \max_{i,j\in X} (\dist_{i,j}).
\end{align}
For a subset $X\subseteq \Lambda$, we define the extended subset $\bal{X}{r}$ as
\begin{align}
\bal{X}{r}:= \{i\in \Lambda| \dist_{X,i} \le r \} , \label{def:bal_X_r}
\end{align}
where $\bal{X}{0}=X$ and $r$ is an arbitrary positive number (i.e., $r\in \mathbb{R}^+$).

 \begin{figure}[tt]
\centering
\includegraphics[clip, scale=0.5]{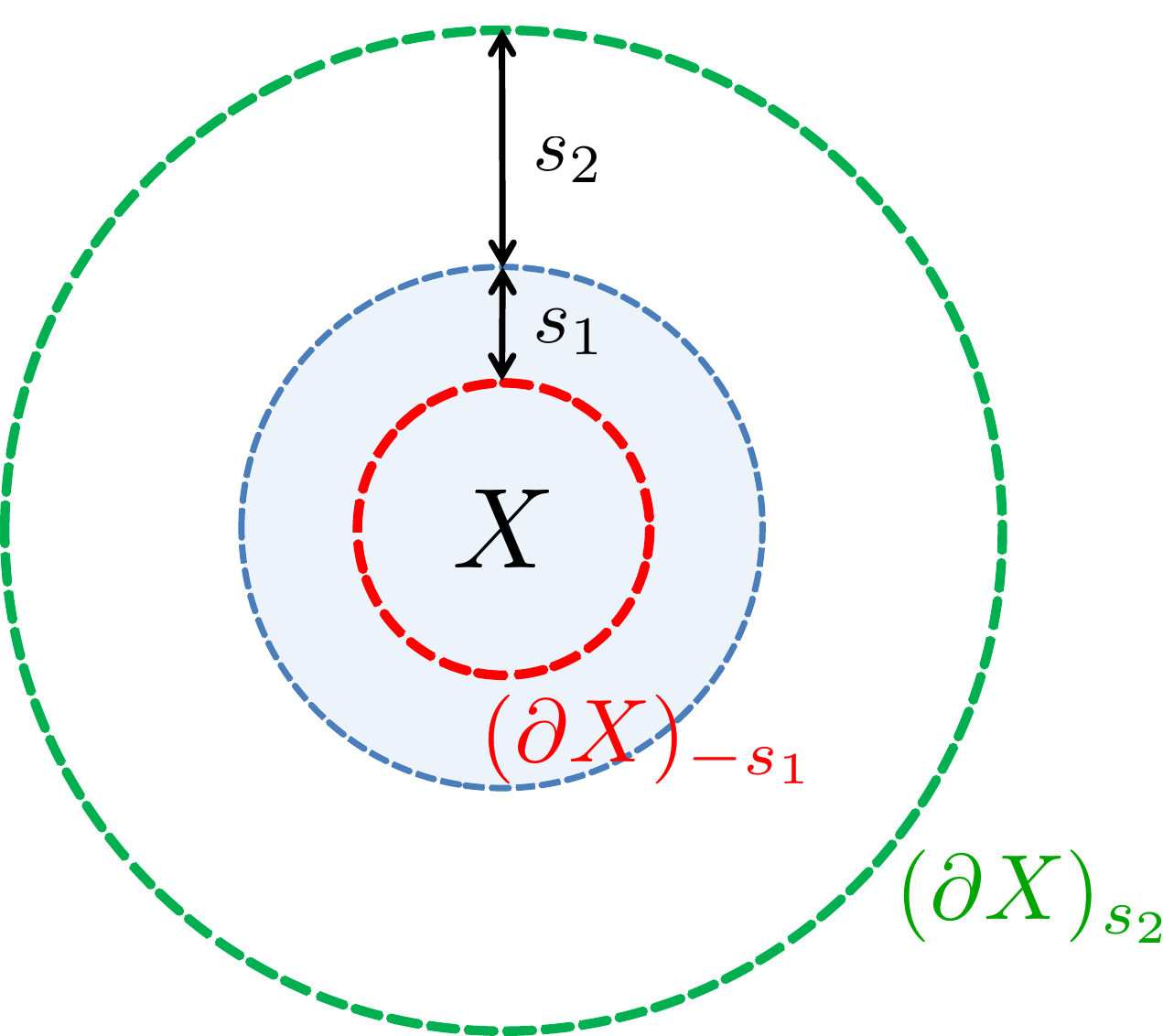}
\caption{Schematic picture of the definition of $(\partial X)_s$ for positive and negative $s$. 
}
\label{supp_fig_partial_X_s}
\end{figure}

Moreover, we define $(\partial X)_s$ as follows (see Fig.~\ref{supp_fig_partial_X_s}): 
\begin{align}
\label{supp_notation_partial_X_s}
(\partial X)_s:= 
\begin{cases}
\{i\in X| \dist_{i,\partial X} =|s|\} &\for s\le 0, \\
\{i\in X^\co | \dist_{i,\partial X} =s\} &\for s> 0, 
\end{cases}
\end{align}
where $(\partial X)_0= \partial X$ and we have 
\begin{align}
X = \bigcup_{s=-\infty}^0 (\partial X)_{s} ,\quad \Lambda= \bigcup_{s=-\infty}^\infty  (\partial X)_s .
\end{align}
For a subset $X\subseteq \Lambda$, we define the extended subset $\bal{X}{r}$ as
\begin{align}
\bal{X}{r}:= \bigcup_{s=-\infty}^r (\partial X)_{s}, \label{supp_def:bal_X_r}
\end{align}
where $\bal{X}{0}=X$ and $r$ is an arbitrary integer.
In particular, for $r>0$, $\bal{X}{r}$ is an extended region from $X$:
\begin{align}
\bal{X}{r}= \{i\in \Lambda| \dist_{X,i} \le r \} .
\end{align}

We introduce a geometric parameter $\gamma$ 
which is determined only by the lattice structure.
We define $\gamma \ge 1$ as a lattice constant which gives 
\begin{align}
\label{supp_parameter_gamma_X}
\abs{ X[\ell] \setminus X } \le \gamma \ell^{D} |\partial X| -1 ,\quad  \abs{ \partial (X[\ell]) } \le \gamma \ell^{D-1} |X| .
\end{align}
for an arbitrary $X\subseteq \Lambda$ and $\ell \ge 1$. 
By choosing $X=\{i\}$, we have 
\begin{align}
\label{supp_parameter_gamma_X_i}
\abs{ i[\ell] } \le \gamma \ell^{D},\quad \abs{ \partial (i[\ell]) } \le \gamma \ell^{D-1} .
\end{align}

We often use the notation of $\Theta(x_1,x_2,\ldots,x_s)$ that means a function with respect to $\{x_1,x_2,\ldots,x_s\}$ as follows:
\begin{align}
\label{Theta_notation_def}
\Theta(x_1,x_2,\ldots,x_s) = \sum_{\sigma_1,\sigma_2,\ldots,\sigma_s=0,1} c_{\sigma_1,\sigma_2,\ldots,\sigma_s} x_1^{\sigma_1}x_2^{\sigma_2} \cdots x_s^{\sigma_s} \quad (0<c_{\sigma_1,\sigma_2,\ldots,\sigma_s}<\infty )  ,
\end{align}
where the constants $\{c_{\sigma_1,\sigma_2,\ldots,\sigma_s}\}$ only depend on the fundamental parameters listed in Table~\ref{tab:fund_para}. 
Also, we use the notation of $\tO(x)$ in the following sense:
\begin{align}
\tO(x)= \mathcal{O} \brr{x \cdot {\rm polylog} (x)} . 
\end{align}

\begin{table*}[tt]
 \caption{Fundamental parameters in our statements}
  \label{tab:fund_para} 
\begin{ruledtabular}
\begin{tabular}{lr}
\textrm{\textbf{Definition}}&\textrm{\textbf{Parameters}} 
\\
\colrule
Spatial dimension    
&  $D$ \\
Structural constant [see Ineq.~\eqref{supp_parameter_gamma_X}]    
& $\gamma$  \\
 Hopping amplitudes [see Eq.~\eqref{def:Ham}] 
&    $\bar{J}$ \\
Maximum interaction length (see Def.~\ref{def_Short-range interactions_0})    
&  $k$   \\
Amplitudes of the boson-boson interactions (see Def.~\ref{def_Finite coupling constants_0})   
& $g$, $\vH$\\
Parameters in the low-boson-density condition (see Assumption~\ref{only_the_assumption_initial})   
& $b_\rho$, $\kappa$
\end{tabular}
\end{ruledtabular}
\end{table*}

Throughout the paper, we mainly consider three types of norms:  i) the operator norm $\norm{\cdot}$, ii) the trace norm $\norm{\cdot}_1$ and 
iii) the Frobenius norm $\norm{\cdot}_F$.
For an arbitrary operator $O$, the operator norm $\norm{O}$ means the maximum eigenvalue of $\sqrt{O^\dagger O}$. 
Also, the trace norm $\norm{O}_1$ is defined as $\tr\br{\sqrt{O^\dagger O}}$.
Finally, the Frobenius norm $\norm{O}_F$ is defined as $\sqrt{\tr\br{O^\dagger O}}$.

\begin{figure}[tt]
\centering
\includegraphics[clip, scale=0.4]{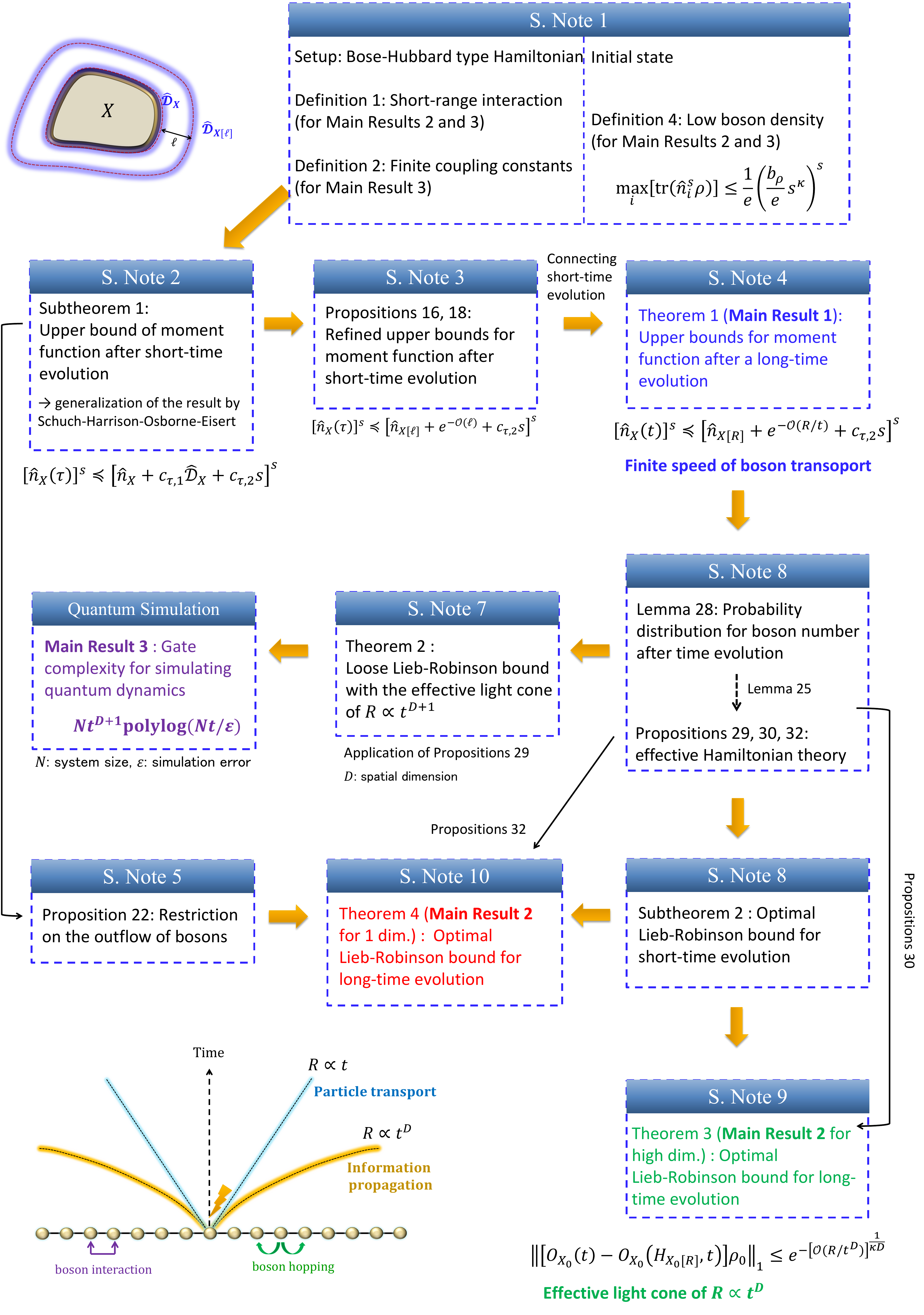}
\caption{Proof outline.} 
\label{supp_fig_overview.pdf}
\end{figure}

The overview of this supplementary information is summarized in Fig.~\ref{supp_fig_overview.pdf}. 

\clearpage

\clearpage

\subsection{Often used inequalities}

As convenient inequalities, we prove the following lemma:
\begin{lemma} \label{Lemm_Often used inequalities}
Let $D$ be an arbitrary positive integer. Then, for arbitrary positive parameters $c,\Phi,\xi,\kappa$ and $s$, we have 
\begin{align}
\label{summation_s_D_e_a_m}
&\sum_{m=1}^\infty m^D e^{-c m} \le e^{-c} +e^{-c} 2^D D! \brr{\max(1,1/c)}^{D+1} ,    \\
&\sum_{j\in \Lambda}  e^{-c\dist_{i,j}} \le \lambda_c  , \quad \lambda_c\coloneqq 1+c^{-D-1}e^c\gamma D! , \label{oftern_used_inequality_lambda_c} \\
& \sum_{i\in L[\ell]^\co} \exp\brr{-\br{ \frac{q e^{\dist_{i,L[\ell]}/\xi}}{\Phi}}^{1/\kappa} }   \le    \gamma |L|   2^{2D+2} D!  \max(\ell^D,1) e^{- \brr{q/\Phi }^{1/\kappa}/(\xi\kappa)} 
   \for q\ge (\xi \kappa)^\kappa \Phi , 
\label{ineq:summation_set_S_L_xi_N} \\
&\sum_{x=1}^\infty (x-1)^s e^{-(x/\Phi)^{1/\kappa}}\le \Phi^{s+1} \kappa \Gamma [\kappa (s+1)] ,\label{summation_for_exp_x^s}\\
& x^s e^{-(x/\Phi)^{1/\kappa}}\le  2^s  (2^\kappa \Phi)^{s+1} \kappa \Gamma [\kappa (s+1)]   e^{-(x/\Phi)^{1/\kappa}/2} \for \kappa\ge1 ,\quad \Phi\ge 1,
\label{x^m_e^-x/Phi_upper_bound}
\end{align}
where $\Gamma(\cdot)$ is the gamma function. 
\end{lemma}

\subsubsection{Proof of Lemma~\ref{Lemm_Often used inequalities}}

The proof of the inequality~\eqref{summation_s_D_e_a_m} is given as follows.
First of all, for $D\ge 3$, we have 
\begin{align}
\label{summation_s_D_e_a_m_derive}
\sum_{m=1}^\infty m^D e^{-c m} = e^{-c} + \sum_{m=2}^\infty m^D e^{-c m} 
&\le  e^{-c} +e^{-c} \int_0^\infty (z+2)^D e^{-cz} dz \notag \\
&\le e^{-c} +e^{-c} 2^D D!\max(1,1/c)^{D+1} . 
\end{align}
For $D=1,2$, we use 
 \begin{align}
\sum_{m=1}^\infty m^D e^{-c m} =\br{-\frac{d}{dc}}^D \sum_{m=1}^\infty e^{-c m}= \br{-\frac{d}{dc}}^D \frac{1}{e^c-1} .
\end{align}
We have 
 \begin{align}
\br{-\frac{d}{dc}} \frac{1}{e^c-1} = \frac{e^c}{(e^c-1)^2} , \quad\br{-\frac{d}{dc}}^2 \frac{1}{e^c-1} =\frac{2e^{2c}}{(e^c-1)^3} -  \frac{e^c}{(e^c-1)^2},
\end{align}
each of which has $ e^{-c} +e^{-c} 2^D D!\max(1,1/c)^{D+1}$ as the upper bound.

Also, we can prove the inequality~\eqref{oftern_used_inequality_lambda_c} for an arbitrary site $i\in \Lambda$:
\begin{align}
\sum_{j\in \Lambda}  e^{-c\dist_{i,j}} &=1+ \sum_{x=1}^\infty \sum_{j: \dist_{i,j}=x} e^{-cx} 
 \le 1 + \gamma  \int_0^\infty x^D e^{c(1-x)}dx = 1+ c^{-D-1}e^c \gamma D! =\lambda_c,
\end{align}
where we use $\{j\in \Lambda: \dist_{i,j}=x\} = \partial i[x]$ and $|\partial i[x]|\le \gamma x^D$ from the inequality~\eqref{supp_parameter_gamma_X_i}.

We here let $q_i=q e^{\dist_{i,L[\ell]}/\xi}$. 
We then upper-bound the LHS of~\eqref{ineq:summation_set_S_L_xi_N} as 
\begin{align}
\label{proof_lem:probability_q_q_Set_S_3}
 \sum_{i\in L[\ell]^\co} e^{- (q_i/\Phi)^{1/\kappa} }
 &\le  \sum_{r=1}^\infty \sum_{i\in \partial (L[\ell+r])}  e^{- \brr{q e^{r/\xi} /\Phi }^{1/\kappa}}   \notag \\
 &\le   \sum_{r=1}^\infty \abs{L[\ell+r]} e^{- e^{r/(\xi\kappa)}  \brr{q/\Phi  }^{1/\kappa}} \notag \\
 &\le  \gamma|L|  \sum_{r=1}^\infty (\ell+r)^D e^{- r/(\xi\kappa) \cdot \brr{q/\Phi  }^{1/\kappa}}   \notag \\
& \le  \gamma2^D |L|  \sum_{r=1}^\infty \br{\ell^D  +r^D}  e^{- r/(\xi\kappa) \cdot \brr{q/\Phi  }^{1/\kappa}} ,
\end{align} 
where we use $e^{r/(\xi \kappa)} \ge  r/(\xi\kappa)$ for $r>0$ and $(x+y)^D\le 2^D (x^D+y^D)$ for $x>0$ and $y>0$.
By using the inequality~\eqref{summation_s_D_e_a_m}, we can obtain 
\begin{align}
\label{proof_lem:probability_q_q_Set_S_4}
\sum_{r=1}^\infty \br{\ell^D  +r^D}  e^{- r/(\xi\kappa) \cdot \brr{q/\Phi}^{1/\kappa}} 
&\le e^{- \brr{q/\Phi }^{1/\kappa}/(\xi\kappa)} 
\br{2\ell^D + 1+ 2^D D! } \notag \\
& \le  2^{D+2} D!  \max(\ell^D,1) e^{- \brr{q/\Phi }^{1/\kappa}/(\xi\kappa)} ,
\end{align} 
where we use the condition~$q\ge (\xi \kappa)^\kappa \Phi$, which implies $\brr{q/\Phi}^{1/\kappa}/(\xi\kappa) \ge 1$, in the first inequality.
By combining the inequalities~\eqref{proof_lem:probability_q_q_Set_S_3} and \eqref{proof_lem:probability_q_q_Set_S_4}, we prove the main inequality~\eqref{ineq:summation_set_S_L_xi_N}.

The inequality~\eqref{summation_for_exp_x^s} is immediately derived as follows:
\begin{align}
\sum_{x=1}^\infty (x-1)^s e^{-(x/\Phi)^{1/\kappa}} \le \int_{0}^\infty x^s   e^{-(x/\Phi)^{1/\kappa}} dx 
&= \Phi^{s+1} \kappa \Gamma [\kappa (s+1)] ,
\end{align}
where $\Gamma(\cdot)$ is the gamma function. 
Using the above inequality, we can also prove the inequality~\eqref{x^m_e^-x/Phi_upper_bound}.
In the case of $x\ge2$, we have $x^s \le 2^s (x-1)^s$, and hence,  
\begin{align}
&x^s e^{-(x/\Phi)^{1/\kappa}} \le 2^s(x-1)^s e^{-(x/\Phi)^{1/\kappa}} 
\le 2^s e^{-(x/\Phi)^{1/\kappa}/2} \sum_{x=1}^\infty (x-1)^s e^{-(x/\Phi)^{1/\kappa}/2} \notag \\
&\le  2^s  (2^\kappa \Phi)^{s+1} \kappa \Gamma [\kappa (s+1)]   e^{-(x/\Phi)^{1/\kappa}/2}.
\end{align}
From the conditions $\Phi\ge 1$ and $\kappa\ge 1$, the above upper bound trivially holds for $x\le 2$. 
This completes the proof of Lemma~\ref{Lemm_Often used inequalities}. $\square$

\subsection{Boson operators}

We define $b_i$ and $b_i^\dagger$ as the annihilation and the creation operators of boson, respectively.
We also define $\nb_i$ as the number operator of bosons on a site $i \in \Lambda$, namely $\nb_i\coloneqq b_i^\dagger b_i$.
We denote the boson number on a subset $X$ by $\nb_X$, namely 
\begin{align}
\nb_X = \sum_{i\in X} \nb_i .
\end{align}
For an arbitrary subset $X\subseteq \Lambda$, we define $\Pi_{X,N}$ as the projection onto the eigenspace of $\nb_X$ with the eigenvalue $N$:
\begin{align}
\label{def_Pi_X_q}
\nb_X \Pi_{X,N} = N \Pi_{X,N}.
\end{align}
In the case where $X$ includes only one site (i.e., $X=\{i\}$), we denote $\Pi_{\{i\},N}$ by $\Pi_{i,N}$ for simplicity. 
We also define $\Pi_{X, \ge N}$ as $\sum_{N'=N}^\infty \Pi_{X,N'}$.

\subsection{Bose-Hubbard type Hamiltonian}

We consider a Hamiltonian in the form of
\begin{align}
&H= H_0+ V= \sum_{Z\subseteq \Lambda} h_Z , \notag \\
&H_0\coloneqq  \sum_{\langle i, j \rangle} J_{i,j} (b_i b_j^\dagger +{\rm h.c.} ), \quad  |J_{i,j}| \le \bar{J} \notag\\
& V\coloneqq f\br{ \{\nb_i\}_{i\in \Lambda}} = \sum_{Z\subseteq \Lambda} v_Z\label{def:Ham}
\end{align}
where $\sum_{\langle i, j \rangle}$ means the summation for all the pairs of the adjacent sites $\{i,j\}$ on the lattice, 
the interaction terms $h_Z$ or $v_Z$ are supported on the subset $Z\subseteq \Lambda$,
and $f\br{ \{\nb_i\}_{i\in \Lambda}}$ is an arbitrary function for the boson number operators $\{\nb_i\}_{i\in \Lambda}$.
For example, $f\br{ \{\nb_i\}_{i\in \Lambda}}$ includes interactions like $\nb_i^4$, $\nb_i^2e^{\nb_j}$, $e^{\nb_i^2\nb_j^3}$ and so on.
Throughout the proof, we consider the time-independent Hamiltonian for simplicity, but all the following results can be generalized to time-dependent Hamiltonians. 

For an arbitrary subset $X \subseteq \Lambda$, we define the subset Hamiltonians $H_{0,X}$, $V_X$ and $H_X$ as follows:
\begin{align}
&H_X= H_{0,X}+ V_X , \notag \\
&H_{0,X}\coloneqq  \sum_{i,j \in X}  J_{i,j} (b_i b_j^\dagger +{\rm h.c.} ), \quad   V_X\coloneqq   \sum_{Z \subseteq X} v_Z .
 \label{def:Ham_subset}
\end{align}
Note that they are supported on the subset $X$.
Also, we define the boundary hopping terms on the region $X$ as follows:
\begin{align}
\partial h_{X}\coloneqq  H_0-H_{0,X} - H_{0,X^\co}=\sum_{i \in \partial X} \sum_{j\in X^\co: \dist_{i,j}=1} J_{i,j} (b_i b_j^\dagger +{\rm h.c.} ) .
 \label{def:Ham_surface}
\end{align}

\subsubsection{Assumptions on the Hamiltonian}

\begin{table*}[tt]
 \caption{Assumptions for the main results}
  \label{tab:necessary_assumption} 
\begin{ruledtabular}
\begin{tabular}{lr}
\textrm{\textbf{Restrictions on $\pmb{f\br{ \{\nb_i\}_{i\in \Lambda}}}$}}  &\textrm{\textbf{Obtained theorems}} 
\\
\colrule
(i) No assumptions
& Result~1 in the main manuscript (Theorem~\ref{main_theorem0_boson_concentration}) \\
(ii) Short-range interactions (Def.~\ref{def_Short-range interactions_0}) &   Result~2 in the main manuscript (Theorems~\ref{main_theorem_looser_Lieb_Robinson}, \ref{main_thm:boson_Lieb-Robinson_main} and \ref{main_thm:boson_Lieb-Robinson_main_1D}) \\
$\quad \ \And$ Low boson density (Def.~\ref{only_the_assumption_initial}) & \\
(iii) Short-range interactions (Def.~\ref{def_Short-range interactions_0}) & Result~3 in the main manuscript \\
$\quad \ \ \And$ Finite coupling constants (Def.~\ref{def_Finite coupling constants_0}) & \\
$\quad \ \ \And$ Low boson density (Def.~\ref{only_the_assumption_initial})
&
\end{tabular}
\end{ruledtabular}
\end{table*}

We can derive several of our main results for the Hamiltonian without any assumptions on $f\br{ \{\nb_i\}_{i\in \Lambda}}$.
In Table~\ref{tab:necessary_assumption}, we summarize what types of Hamiltonians are assumed in deriving the main theorems. 
For example, in the derivation of Result~1 (or Theorem~\ref{main_theorem0_boson_concentration}), $f\br{ \{\nb_i\}_{i\in \Lambda}}$ can include even the macroscopic interactions like $\nb_1\nb_2\nb_3\cdots \nb_{|\Lambda|}$. 
We thus apply our results to arbitrary long-range interacting systems, e.g., the coulomb interaction, the dipole-dipole interaction, and so on.

However, in considering the Lieb-Robinson bound as shown in Theorems~\ref{main_theorem_looser_Lieb_Robinson}, \ref{main_thm:boson_Lieb-Robinson_main} and \ref{main_thm:boson_Lieb-Robinson_main_1D}, we need an additional assumption on the form of $f\br{ \{\nb_i\}_{i\in \Lambda}}$, i.e., short-rangeness of the interactions:
\begin{definition}[Short-range interactions] \label{def_Short-range interactions_0}
If all the boson-boson interactions in $V$ are finite range with interaction length as most $k$, i.e., 
\begin{align}
V\coloneqq f\br{ \{\nb_i\}_{i\in \Lambda}}= \sum_{Z \subset \Lambda: \diam(Z) \le k} \norm{v_Z} ,   \label{def:Ham_short} 
\end{align}
we call that the Hamiltonian is short-range. 
Here, $v_Z $ is still an arbitrary function of $\{\nb_i\}_{i\in Z}$; that is, it may be infinitely large. 
\end{definition}

\noindent 
The simplest example of the short-range Hamiltonian is the Bose-Hubbard model: 
\begin{align}
H= \sum_{\langle i, j \rangle}J (b_i b_j^\dagger +{\rm h.c.} ) + \frac{U}{2} \sum_{i\in \Lambda} \nb_i(\nb_i-1)
-\mu\sum_{i\in \Lambda} \nb_i  , \label{def:Ham_BH}
\end{align}
where $U$ and $\mu$ are arbitrary constants.

Furthermore, in considering quantum simulation of the time evolution (Result~3 in the main manuscript), we have to assume the additional condition as follows:
\begin{definition}[Finite coupling constants]\label{def_Finite coupling constants_0}
The function $f\br{ \{\nb_i\}_{i\in \Lambda}}$ is given by a polynomial with limited degrees and coefficients.
That is, the boson-boson couplings satisfy the following inequality:
\begin{align}
\max_{i \in \Lambda} \sum_{Z:Z\ni i} v_Z \preceq g \br{\nb_{i[k]}}^{\vH},  \label{def:Ham_short_upper_bound} 
\end{align}
where $g$ and $\vH$ are $\orderof{1}$ constants. 
\end{definition}

\noindent 
In the case of the Bose-Hubbard Hamiltonian~\eqref{def:Ham_BH}, we have $k=0$, $\vH=2$, and $g=|U|/2+|\mu|$. 
Hence, the finiteness of $g$ implies $U=\orderof{1}$ and $\mu=\orderof{1}$.

%

\subsection{Time evolution}

For an arbitrary operator $O$, we denote the time evolution by an operator $A$ as $O(A,t)$, namely
\begin{align}
O(A,t) \coloneqq  e^{iAt} O e^{-iA t}.
\end{align}
In particular, when $A=H$, we often denote $O(H,t)$ by $O(t)$ for simplicity. 

When the operator $A$ is time-dependent, we utilize the following notations.  
As in Ref.~\cite{PhysRevLett.127.070403}, for an arbitrary time-dependent operator $A_t$, we define as 
\begin{align}
\label{notation_time_dependent_operator}
&U_{A_t,\tau_1\to \tau_2} := \mathcal{T} e^{-i \int_{\tau_1}^{\tau_2} A_t   dt} ,\notag \\
&O(A_t , \tau_1\to \tau_2) := U_{A_t, \tau_1\to \tau_2}^\dagger O U_{A_t, \tau_1\to \tau_2},
\end{align} 
where $\mathcal{T}$ is the time-ordering operator.

\subsection{Definitions on the boson density}

We here define the condition of the low boson density as follows:
\begin{definition}[Low boson density] \label{only_the_assumption_initial}
For a quantum state $\rho$, we say that the state satisfies the low-boson-density condition if 
there exists $\orderof{1}$ constants $b_\rho$ and $\kappa$ such that 
\begin{align}
\label{condition_for_moment_generating}
\max_{i\in \Lambda} \tr  \br{ \nb_i^s \rho} \le  
\frac{1}{e} \br{ \frac{b_\rho}{e}  s^\kappa }^s .
\end{align}
\end{definition}

{\bf Remark.}  
This assumption is adopted in deriving Results~2 and 3 in the main manuscript, 
while Result~1 does not depend on the initial state and we do not need the assumption.

{~} \\  

\noindent
From the assumption, we can also obtain an upper bound for $\nb_X$  
\begin{align}
\label{condition_for_moment_generating_X}
\tr  \br{ \nb_X^s \rho} \le  \brrr{\sum_{i\in X} \brr{\tr \br{ \nb_i^s \rho}}^{1/s} }^s 
\le  \br{\frac{1}{e^{1/s}} \sum_{i\in X} \frac{b_\rho}{e}  s^\kappa} ^s = 
\frac{1}{e}  \br{ \frac{b_\rho}{e}  s^\kappa |X|} ^s .
\end{align}
Therefore, for $X=i[\ell]$, we have 
\begin{align}
\label{condition_for_moment_generating_i_ell}
\tr  \br{ \nb_{i[\ell]}^s \rho} \le 
\frac{1}{e}  \br{ \frac{\gamma b_\rho}{e}  s^\kappa \ell^D}^s .
\end{align}

Also, we often utilize the following set of boson numbers:
\begin{definition}
Let $\bold{N}$ be a set of boson numbers on $\Lambda$, i.e., $\bold{N} =\{N_i\}_{i\in \Lambda}$.
Then, we define the set $\mathcal{S}(L,\xi,N)$ as follows:
\begin{align}
\label{mathcal_S_L_xi_N/def}
\mathcal{S}(L,\xi,N) \coloneqq  \brrr{ \bold{N} | N_i \le N e^{\xi\dist_{i,L}} \for i\in  \Lambda} .
\end{align}
If $\bold{N}\in \mathcal{S}(L,\xi,N)$, the number of bosons cannot be too large around the region $L$. 
Also, we define the projection $\Pi_{\mathcal{S}(L,\xi,N)}$ as 
\begin{align}
\label{mathcal_S_L_xi_N/def_proj}
\Pi_{\mathcal{S}(L,\xi,N)} := \sum_{\bold{N}\in \mathcal{S}(L,\xi,N)}  P_{\bold{N}} , 
\end{align}
where $P_{\bold{N}}=\ket{\bold{N}} \bra{\bold{N}}$ is the projection onto the Mott state $\ket{\bold{N}}$ such that $\nb_i\ket{\bold{N}}=N_i\ket{\bold{N}}$ ($i\in \Lambda$). 
\end{definition}

{\bf Remark.} 
If a quantum state $\rho$ satisfies 
\begin{align}
\Pi_{\mathcal{S}(L,\xi,N)} \rho = \rho ,
\end{align}
we can expand the quantum state $\rho$ by using the Mott state $\ket{\bold{N}}$ with $\bold{N} \in \mathcal{S}(L,\xi,N)$:
\begin{align}
\rho= \sum_{\bold{N}\in \mathcal{S}(L,\xi,N) } P_{\bold{N}}\ \rho  . 
\end{align}

\subsubsection{Basic lemmas on the boson density}

From the condition, we can ensure that the probability for many bosons to be concentrated on one site is (sub)exponentially small.
In the following, for a fixed $p\in \mathbb{N}$ and $\forall \ell \in \mathbb{N}$, we consider the quantity of
\begin{align}
\tr\br{\rho \nb_{i[\ell]}^p \Pi_{i[\ell], \ge x} }   , 
\end{align}
which immediately yields the upper bounds on the probability distribution by choosing $p=0$
\begin{align}
\tr\br{\rho \Pi_{i, \ge x} } \le\tr\br{\rho \Pi_{i[\ell], \ge x} } . 
\end{align}

In the following general lemma, we prove an upper bound on the probability distribution for boson numbers.

\begin{lemma}\label{lem:probability_dist_moment}
Let $\rho$ be an arbitrary quantum state such that 
\begin{align}
\label{assump_rho_moment_nb_thm_pre}
\tr(\rho \nb_{i[\ell]}^s) \le  \frac{1}{e} \brr{ \frac{b_\rho}{e} \br{q_0+ s^\kappa \ell^D}}^s
\end{align}
for a site $i\in \Lambda$ and $\ell\ge 1$, where $b_\rho$ is an appropriate constant.  
Then, the assumption~\eqref{assump_rho_moment_nb_thm_pre} implies the upper bound of the boson number distribution as
\begin{align}
\label{ineq:lem:probability_dist_moment_pre}
\tr\br{\rho \nb_{i[\ell]}^p \Pi_{i[\ell], \ge x} }  \le x^p  e^{- \brr{(x-b_\rho q_0)/(b_\rho \ell^D) }^{1/\kappa}}  
\for x\ge \max(e, b_\rho q_0).
\end{align}
That is, for $p=0$, the distribution decays (sub)exponentially for $x\gtrsim b_\rho q_0 + b_\rho \ell^D$. 
\end{lemma}

{\bf Remark.}
If the assumption~\eqref{assump_rho_moment_nb_thm_pre} holds up to $s=s_0$ with $s_0$ a fixed constant, 
the main inequality reduces to the form of 
\begin{align}
\label{ineq:lem:probability_dist_moment_pre_s0}
\tr\br{\rho \nb_{i[\ell]}^p \Pi_{i[\ell], \ge x} } \le \max\br{x^p e^{-s_0-1} , x^p e^{- \brr{(x-b_\rho q_0)/(b_\rho \bar{\ell}^D) }^{1/\kappa}}  } .
\end{align}
The inequality is immediately proved from the first inequality in \eqref{prob_upper_bound_e^-s_ineq}.

\subsubsection{Proof of Lemma~\ref{lem:probability_dist_moment}}

The proof immediately follows from the Markov inequality:  
\begin{align}
\tr\br{\rho \nb_{i[\ell]}^p \Pi_{i[\ell], \ge x} } = \tr\br{\rho \nb_{i[\ell]}^p \nb_{i[\ell]}^{s-p} \cdot \nb_{i[\ell]}^{-s+p}  \Pi_{i[\ell], \ge x} } 
&\le  \norm{\nb_{i[\ell]}^{-s+p}  \Pi_{i[\ell], \ge x}} \tr\br{\rho \nb_{i[\ell]}^s} \notag \\
&\le  \frac{1}{e} \brr{ \frac{b_\rho}{e} \br{q_0+ s^\kappa \ell^D}}^p \br{\frac{b_\rho}{e} \cdot \frac{q_0+s^\kappa \ell^D }{x}}^{s-p}  ,
\label{lem:probability_dist_moment_proof_start}
\end{align}
where we assume that $s\ge p$. 
Then, by choosing $s$ such that
\begin{align}
\frac{b_\rho}{e} \br{q_0+s^\kappa \ell^D}  \le \frac{x}{e}  
\longrightarrow  s^\kappa \le  \frac{x - b_\rho q_0}{b_\rho \ell^D} 
\longrightarrow  s = \floorf { \br{\frac{x - b_\rho q_0}{b_\rho \ell^D}}^{1/\kappa} },
\end{align}
we reduce the inequality~\eqref{lem:probability_dist_moment_proof_start} to 
\begin{align}
\label{prob_upper_bound_e^-s_ineq}
\tr\br{\rho \nb_{i[\ell]}^p \Pi_{i[\ell], \ge x} } 
\le \frac{1}{e} \br{\frac{x}{e}}^p e^{-s+p}  
&\le\frac{1}{e} \br{\frac{x}{e}}^p  \exp\brr{p+1-  \br{\frac{x-b_\rho q_0}{b_\rho \ell^D}}^{1/\kappa} } \notag \\
&= x^p \exp\brr{-  \br{\frac{x-b_\rho q_0}{b_\rho \ell^D}}^{1/\kappa} } .
\end{align}
In the case of $s<p$ (or $s\le p-1$) and $x\ge e$, the above inequality trivially holds since the first RHS of the above inequality (i.e., $\frac{1}{e} \br{\frac{x}{e}}^p e^{-s+p} $) is larger than $1$. 
We thus prove the inequality~\eqref{ineq:lem:probability_dist_moment_pre}. $\square$

 {~}

\hrulefill{\bf [ End of Proof of Lemma~\ref{lem:probability_dist_moment}] }

{~}

We consider how an operator influences the moment function as another essential lemma.
For example, we are interested in the moment difference between the original state $\rho$ and $O_L^\dagger \rho O_L$, where $O_L$ is supported on $L\subset \Lambda$. 
We here prove the following lemma: 
\begin{lemma}\label{lem:boson_operator_norm}
Let $O_q$ be an arbitrary operator such that 
\begin{align}
\label{O_X_condition/_norm_lemma_operator}
\Pi_{L, > m+q} O_q \Pi_{L, m} , \quad \Pi_{L, < m-q} O_q \Pi_{L, m} =0  \for \forall m>0.
\end{align}
Note that $O_q$ is not necessarily supported on the subset $L$.
Then, we obtain 
\begin{align}
\label{main_ineq_lem:boson_operator_norm}
O_q^\dagger \br{\sum_{i\in \Lambda} \nu_i \nb_i}^m O_q 
\preceq 4^{m}\norm{O_q}^{2} \br{4 \bar{\nu}q^3 + \bar{\nu}\nb_L + \sum_{i\notin L} \nu_i \nb_i }^m  ,
\end{align}
where we define $\bar{\nu}:= \max_{i\in L} (\nu_i)$ with $\nu_i>0$. 
Here, for arbitrary operators $O_1$ and $O_2$, the notation $O_1 \preceq O_2$ means that $O_2-O_1$ is a semi-positive definite operator.   
\end{lemma}

\subsubsection{Proof of Lemma~\ref{lem:boson_operator_norm}}

We here define $\hat{Q}$ as 
\begin{align}
\label{definitiion_of_hat_Q}
\hat{Q} :=  \br{\bar{\nu}\nb_L + \sum_{i\notin L} \nu_i \nb_i}^{1/2} .
\end{align}
Then, from the definition of $\bar{\nu}$, we obtain
\begin{align}
\label{main_ineq_lem:boson_operator_norm_pre_0}
O_q^\dagger \br{\sum_{i\in \Lambda} \nu_i \nb_i}^m O_q 
\preceq O_q^\dagger \hat{Q}^{2m} O_q .  
\end{align}

First of all, by using the condition~\eqref{O_X_condition/_norm_lemma_operator}, we aim to prove 
\begin{align}
\label{INEQ_commutator_norm_hatWQ}
\norm{ [\hat{Q} , O_q]} \le 2q \sqrt{\bar{\nu}q}  \norm{O_q} .
\end{align}
For proof, we first decompose 
\begin{align}
O_q = \sum_{q_1=-q}^{q}  O_q^{(q_1)} , 
\end{align}
where $O_q^{(q_1)}$ satisfies $\Pi_{L, m'} O_q^{(q_1)} \Pi_{L, m} \neq 0$ only for $m'=m+q_1$. 
For arbitrary $q_1 \in [-q,q]$, we obtain 
\begin{align}
\brr{ \hat{Q} , O_q^{(q_1)} } = \brr{\br{\bar{\nu}\nb_L + \sum_{i\notin L} \nu_i \nb_i}^{1/2}  - \br{\bar{\nu} (\nb_L-q_1) + \sum_{i\notin L} \nu_i \nb_i}^{1/2} } O_q^{(q_1)} ,
\end{align}
which yields 
\begin{align}
\label{commute_O_q_q_1}
\norm{\brr{\hat{Q} , O_q^{(q_1)} }} \le  \sqrt{\bar{\nu} |q_1|} \norm{O_q^{(q_1)}} . 
\end{align}
where we use $|\sqrt{x} - \sqrt{x-y}| \le \sqrt{|y|}$.
We then aim to prove 
\begin{align}
\norm{O_q^{(q_1)}}\le  \norm{O_q} .
\end{align}
From the definition, we have $[ O_q^{(q_1)\dagger} O_q^{(q_1)} , \nb_L]=0$, and hence the eigenstate of $O_q^{(q_1)\dagger} O_q^{(q_1)}$ is expressed as $\ket{\psi_m}$ 
such that $\nb_L\ket{\psi_m}=m\ket{\psi_m}$, where $\ket{\psi_m}$ is an appropriate superposition of the eigenstates of $\nb_L$ with the eigenvalue $m$. 
On the other hand, from the properties of $O_q^{(q_1)}$, we obtain
\begin{align}
O_q^{(q_1)}\ket{\psi_m} =\Pi_{L, m+q_1}  O_q^{(q_1)}\ket{\psi_m}  , \quad O_q\ket{\psi_m} = \sum_{q_1=-q}^{q} \Pi_{L, m+q_1}  O_q^{(q_1)}  \ket{\psi_m} ,
\end{align}
which immediately yields 
\begin{align}
\label{norm_O_q_q_1_eq_O_q}
\norm{O_q^{(q_1)}} =\max_{\ket{\psi_m}} \norm{O_q^{(q_1)}\ket{\psi_m}} \le \max_{\ket{\psi_m}} \norm{ O_q\ket{\psi_m} } \le \norm{O_q} . 
\end{align}
By combining the inequalities~\eqref{commute_O_q_q_1} and \eqref{norm_O_q_q_1_eq_O_q}, 
we have 
\begin{align}
\norm{ [\hat{Q} , O_q]} \le \sum_{q_1=-q}^{q}   \norm{ \brr{\hat{Q} , O_q^{(q_1)}}} \le 
\norm{O_q} \sum_{q_1=-q}^{q} \sqrt{\bar{\nu} |q_1|} \le 2q \sqrt{\bar{\nu} q} \norm{O_q} , 
\end{align}
which gives the desired inequality~\eqref{INEQ_commutator_norm_hatWQ}.

In the following,  by induction on $m$, we aim to prove 
\begin{align}
\label{lem:boson_operator_norm_induction}
\norm{\hat{Q}^m O_q \ket{\psi}} \le 2^{m/2} \norm{O_q} \cdot  \norm{\br{\hat{Q}+2q\sqrt{\bar{\nu}q}}^m  \ket{\psi}} . 
\end{align}
for arbitrary quantum state $\ket{\psi}$ and operator $O_q$ satisfying Eq.~\eqref{O_X_condition/_norm_lemma_operator}. 
For $m=1$, we can immediately prove the inequality~\eqref{lem:boson_operator_norm_induction} as follows:
\begin{align}
\norm{\hat{Q} O_q \ket{\psi}} \le\norm{[\hat{Q},  O_q]  \ket{\psi}} + \norm{ O_q \hat{Q}  \ket{\psi}}  
&\le \norm{O_q} \br{ \norm{2q\sqrt{\bar{\nu}q}  \ket{\psi}} + \norm{ \hat{Q}  \ket{\psi}} } \notag \\
&\le \sqrt{2} \norm{O_q}  \br{ \norm{\br{\hat{Q} +2q\sqrt{\bar{\nu}q} } \ket{\psi}} }  .
\end{align}
In deriving the last inequality above, for arbitrary positive commutative operators $\Phi_1$ and $\Phi_2$, we have used
\begin{align}
\norm{(\Phi_1+\Phi_2)\ket{\psi}}^2 =\bra{\psi} (\Phi_1^2 + 2\Phi_1 \Phi_2 +\Phi_2^2)\ket{\psi}
\ge \bra{\psi} \Phi_1^2\ket{\psi}  +\bra{\psi} \Phi_2^2\ket{\psi} \ge \br{\frac{\sqrt{\bra{\psi} \Phi_1^2\ket{\psi}}  +\sqrt{\bra{\psi} \Phi_2^2\ket{\psi}}}{\sqrt{2}}}^2 ,
\end{align}
which yields
\begin{align}
\label{norm_phi_C_phi+C_ineq}
\norm{\Phi_1\ket{\psi}} + \norm{\Phi_2\ket{\psi}} \le \sqrt{2}\norm{(\Phi_1+\Phi_2)\ket{\psi}}. 
\end{align}

We then prove the case of $m+1$ by assuming Eq.~\eqref{lem:boson_operator_norm_induction} for a fixed $m$. 
We then obtain 
\begin{align}
\label{lem:boson_operator_norm_induction/m+1}
\norm{\hat{Q}^{m+1}  O_q \ket{\psi}} 
&=
\norm{\hat{Q}^{m} \br{ [\hat{Q},O_q] + O_q \hat{Q}} \ket{\psi}}  \le \norm{\hat{Q}^{m} [\hat{Q},O_q]  \ket{\psi}} + \norm{\hat{Q}^{m}O_q \hat{Q} \ket{\psi}} .
\end{align}
First, for $\norm{\hat{Q}^{m} [\hat{Q},O_q]  \ket{\psi}}$ in the above inequality, Eq.~\eqref{lem:boson_operator_norm_induction} gives 
\begin{align}
\label{lem:boson_operator_norm_induction/m+1_1}
\norm{\hat{Q}^{m} [\hat{Q},O_q]  \ket{\psi}} \le 2^{m/2} \norm{[\hat{Q},O_q]}\cdot\norm{\br{\hat{Q}+2q\sqrt{\bar{\nu}q}}^m  \ket{\psi}} 
\le 2^{m/2} \norm{O_q}\cdot\norm{ 2q\sqrt{\bar{\nu}q} \br{\hat{Q}+2q\sqrt{\bar{\nu}q}}^m  \ket{\psi}} ,
\end{align}
where we use the inequality~\eqref{lem:boson_operator_norm_induction} by letting $O_q \to [\hat{Q},O_q]$ in the first inequality and apply~\eqref{INEQ_commutator_norm_hatWQ} to the second inequality.
Note that we can apply the inequality~\eqref{lem:boson_operator_norm_induction} to $[\hat{Q},O_q]$ since it still satisfies the condition~\eqref{O_X_condition/_norm_lemma_operator}.
Second, for $\norm{\hat{Q}^{m}O_q \hat{Q} \ket{\psi}}$ in the inequality~\eqref{lem:boson_operator_norm_induction/m+1}, we have from Eq.~\eqref{lem:boson_operator_norm_induction} 
\begin{align}
\label{lem:boson_operator_norm_induction/m+1_2}
\norm{\hat{Q}^{m}O_q \hat{Q} \ket{\psi}}\le 2^{m/2} \norm{O_q}\cdot \norm{\br{\hat{Q}+2q\sqrt{\bar{\nu}q}}^m   \hat{Q} \ket{\psi}}  .
\end{align}
By applying the inequalities~\eqref{lem:boson_operator_norm_induction/m+1_1} and \eqref{lem:boson_operator_norm_induction/m+1_2} to \eqref{lem:boson_operator_norm_induction/m+1}, we obtain 
\begin{align}
\label{lem:boson_operator_norm_induction/m+1_3}
\norm{\hat{Q}^{m+1}  O_q \ket{\psi}} 
&\le 2^{m/2}  \norm{O_q} \br{ \norm{2q \sqrt{\bar{\nu}q} \br{\hat{Q}+2q\sqrt{\bar{\nu}q}}^m  \ket{\psi}} + \norm{\hat{Q} \br{ \hat{Q}+2q\sqrt{\bar{\nu}q}}^m  \ket{\psi}}  } \notag \\
&\le 2^{(m+1)/2}  \norm{O_q} \cdot  \norm{\br{\hat{Q}+2q\sqrt{\bar{\nu}q}}^{m+1}  \ket{\psi}},
\end{align}
where we use~\eqref{norm_phi_C_phi+C_ineq} in the second inequality. 
We thus prove the inequality~\eqref{lem:boson_operator_norm_induction} for $m+1$.

From the inequality~\eqref{lem:boson_operator_norm_induction}, we obtain 
\begin{align}
\label{lem:boson_operator_norm_induction/4}
\bra{\psi} O_q^\dagger \hat{Q}^{2m} O_q \ket{\psi}=\norm{\hat{Q}^m O_q \ket{\psi}}^2 
\le 2^m \norm{O_q}^2 \cdot  \bra{\psi} \br{\hat{Q}+2q \sqrt{\bar{\nu}q}}^{2m}  \ket{\psi}  
\end{align}
By using the definition~\eqref{definitiion_of_hat_Q} and the inequality of 
\begin{align}
\br{\hat{Q}+2q\sqrt{\bar{\nu}q}}^2 \preceq 2 \br{\hat{Q}^2+4 \bar{\nu}q^3 } ,
\end{align}
we reduce the inequality~\eqref{lem:boson_operator_norm_induction/4} to the form of~\eqref{main_ineq_lem:boson_operator_norm}.
This completes the proof. $\square$

\section{Upper bound on moment function after short-time evolution}
\label{Sec:Proof:density of bosons after time evolution}

\subsection{Statement}

We first consider the $s$th moment function in the form of 
\begin{align}
\label{moment_general_X_1_X_s}
M_{X_1,X_2,\ldots,X_s}(\tau) := \tr\brr{ \nb_{X_1}\nb_{X_2} \cdots \nb_{X_s} \rho(\tau) }  .
\end{align}
As long as $X_j \supseteq X$ ($j=1,2,\ldots,s$), because of 
\begin{align}
\label{multi_paramt_upper_bound}
\nb_{X}^s \preceq   \nb_{X_1}\nb_{X_2} \cdots \nb_{X_s}  , 
\end{align}
we have 
\begin{align}
M_{X}^{(s)}(\tau) \le  M_{X_1,X_2,\ldots,X_s}(\tau)  .
\end{align}
In the following, we aim to upper-bound a short-time evolution of the generalized moment function~\eqref{moment_general_X_1_X_s}. 
The primary reason is to derive Proposition~\ref{prop:upp_Moment_s_th} by resolving the obstacle addressed in Sec.~\ref{sec_sub_Bottleneck for the generalization} below.

In our analyses, the following subtheorem plays a central role: 
\begin{subtheorem} \label{Schuch_boson_extend}
Let $\rho$ be an arbitrary quantum state, and $\{X_j\}_{j=1}^s$ be arbitrary subsets.  
Then, as long as $\tau\le 1/(4\gamma \bar{J})$ we obtain the following upper bound for $M_{X_1,X_2,\ldots,X_s}(\tau)$: 
\begin{align}
\label{main_ineN_Schuch_boson_extend}
M_{X_1,X_2,\ldots,X_s}(\tau) \le  \tr \brr{ \prod_{j=1}^s \br{ \nb_{X_j} + c_{\tau,1} \hat{\mathcal{D}}_{X_j} + c_{\tau,2} s}     \rho } 
\end{align}
with
\begin{align}
\label{Eq:def_tilde_D_X}
\hat{\mathcal{D}}_X:= \sum_{i\in \partial X} \sum_{j\in \Lambda} e^{-\dist_{i,j}} \nb_j 
\end{align}
where $c_{\tau,1}$ and $c_{\tau,2}$ are defined as 
\begin{align}\label{choice of c_t_1_ct_2}
c_{\tau,1}= 40e^{4\gamma \bar{J}\tau}, \quad 
c_{\tau,2}=  e^{4\gamma\bar{J} \tau+1} \br{1+8\gamma\bar{J}\tau +\tau} . 
\end{align}
Notice that they are $\orderof{1}$ constants as long as $\tau=\orderof{1}$. 
\end{subtheorem}

{\bf Remark.} 
This subtheorem is a generalization of the previous work by Schuch, Harrison, Osborne, and Eisert~\cite{PhysRevA.84.032309} that essentially treated the case of $s=1$.  
Because the inequality~\eqref{main_ineN_Schuch_boson_extend} holds for arbitrary quantum states unconditionally, we can also obtain the operator inequality as follows:
\begin{align}
\label{moment_general_X_1_X_s_op_ineq}
\prod_{j=1}^s \nb_{X_j} (\tau) \preceq \prod_{j=1}^s \br{ \nb_{X_j} + c_{\tau,1} \hat{\mathcal{D}}_{X_j} + c_{\tau,2} s} .
\end{align}
In particular, for $s=1$, we have 
\begin{align}
\label{case_s=1_moment/general}
\nb_{X} (\tau) \preceq \nb_{X} + c_{\tau,1} \hat{\mathcal{D}}_{X} + c_{\tau,2}  .
\end{align}

\subsubsection{Effective Hamiltonian}

We here consider a time evolution by the Hamiltonian $H$, but the statement can be extended to a more general class of Hamiltonian as follow:
\begin{align}
T_{\nb}^\dagger H T_{\nb} ,\quad [T_{\nb},\nb_i]=0 \for \forall i\in \Lambda ,\quad \norm{T_{\nb}} \le 1. 
\end{align}
By following the proof of Subtheorem~\ref{Schuch_boson_extend}, we can immediately prove that the same statement as Subtheorem~\ref{Schuch_boson_extend} holds for the Hamiltonian $T^\dagger_{\nb}H T_{\nb}$:
\begin{corol}\label{Schuch_boson_extend_eff_Ham}
For the Hamiltonian $T_{\nb}^\dagger H T_{\nb}$, the time evolution $M_{X_1,X_2,\ldots,X_s}(T_{\nb}^\dagger H T_{\nb},\tau) $ satisfies the same inequality as~\eqref{main_ineN_Schuch_boson_extend}.
The proof is the same as that for Subtheorem~\ref{Schuch_boson_extend}.
\end{corol}

\subsubsection{Imaginary time evolution}

We here consider a real-time evolution as $e^{-iH\tau}$, but the same proof below can be applied to the imaginary time evolution. 
We then immediately obtain the following corollary:
\begin{corol}\label{Schuch_boson_extend_imaginary_time}
For the imaginary time evolution 
$$\rho(i\tau)= e^{\tau H} \rho  e^{-\tau H},$$
and
$$O(i\tau)= e^{-\tau H} O  e^{\tau H}, $$
we obtain
\begin{align}
\label{moment_general_X_1_X_s_op_ineq_imag}
\prod_{j=1}^s \nb_{X_j} (i\tau) \preceq \prod_{j=1}^s \br{ \nb_{X_j} + c_{\tau,1} \hat{\mathcal{D}}_{X_j} + c_{\tau,2} s} .
\end{align}
\end{corol}

\noindent 
Using the upper bound, we can also obtain the same statement as Theorem~\ref{main_theorem0_boson_concentration}.
We hence ensure that during the imaginary time evolution the boson number can be truncated at each site up to $\orderof{\tau^D}$, 
which is helpful in quantum simulation of the imaginary time evolution\footnote{Precisely speaking, we have to estimate the expectation as $$\bra{\psi(i\tau)}\nb_i^s \ket{\psi(i\tau)} :=\bra{\psi} e^{-\tau H}\nb_i^s  e^{-\tau H} \ket{\psi} $$ with $\ket{\psi}$ an appropriate initial state, where we set the energy origin such that 
$\norm{e^{-\tau H} \ket{\psi}}=1$. 
To connect the above expectation to $[\nb_i(i\tau)]^s$, we rewrite as 
$$\bra{\psi(i\tau)}\nb_i^s \ket{\psi(i\tau)} =\bra{\psi(2i\tau)} [\nb_i(i\tau)]^s \ket{\psi} \le \norm{\ket{\psi(2i\tau)}} \sqrt{\bra{\psi} [\nb_i(i\tau)]^{2s} \ket{\psi}}.$$
Thus, if $\norm{\ket{\psi(2i\tau)}}/\norm{\ket{\psi(i\tau)}}\lesssim 1$, we can obtain a meaningful upper bound by applying the inequality~\eqref{moment_general_X_1_X_s_op_ineq_imag}.
}.

\subsection{Proof of Subtheorem~\ref{Schuch_boson_extend}}

In the proof, we consider the case of $X_1=X_2=\cdots = X_s$ for simplicity of the notations.
The choice of the subsets does not influence the proof, and the generalization to generic $\{X_j\}_{j=1}^s$ is straightforward.

As a difficulty, it is not sufficient to analyze only the $1$st moment in deriving the main inequality~\eqref{main_ineN_Schuch_boson_extend}.
Even if we obtain the operator inequality~\eqref{case_s=1_moment/general} for $s=1$, i.e., 
\begin{align}
\nb_{X_1} (\tau)  \preceq  \nb_{X_1} + c_{\tau,1} \hat{\mathcal{D}}_{X_1} + c_{\tau,2},
\quad {\rm and} \quad \nb_{X_2} (\tau) \preceq \nb_{X_2} + c_{\tau,1} \hat{\mathcal{D}}_{X_2} + c_{\tau,2}  ,
\end{align}
we cannot simply connect the inequality to obtain the $2$nd order moment:
\begin{align}
\label{case_s=1_moment/general2}
\nb_{X_1} (\tau) \nb_{X_2} (\tau)  \preceq  \br{ \nb_{X_1} + c_{\tau,1} \hat{\mathcal{D}}_{X_1} + c_{\tau,2}} \br{ \nb_{X_2} + c_{\tau,1} \hat{\mathcal{D}}_{X_2} + c_{\tau,2}}  ,
\end{align}
which is NOT correct. 
Usually, for two operators $A$ and $B$, the operator inequalities $A\preceq A'$ and $B\preceq B'$ do not lead to $AB \preceq A'B'$.  
Therefore, we must develop an analytical technique to treat the $s$th moment.

In the following, we aim to obtain a simple inequality for $M^{(s)}_X(\tau)$ by appropriately upper-bounding the derivatives concerning the time $\tau$.
Our main purpose is to prove the inequality~\eqref{mth_order_derivative_mathcal_D_integral_0} below. 
To make the point clear, we first consider the first differential equation for $M^{(s)}_X(\tau)$:
\begin{align}
\label{sth_moment_der_0}
\frac{d}{d\tau} M^{(s)}_X(\tau) &=-i \tr \left ( \hat{n}_X^s [H, \rho(\tau) ]  \right) =i\tr \left ( [H,\hat{n}_X^s]  \rho(\tau)  \right) .
\end{align}
The form of the Hamiltonian~\eqref{def:Ham} gives 
\begin{align}
\label{Ham_commute_nb_i_s}
 [H,\hat{n}_X^s ] = [H_0,\hat{n}_X^s ] = [\partial h_X ,\hat{n}_X^s ] =\sum_{i \in \partial X}  \sum_{j \in X^\co:\dist_{i,j}=1}J_{i,j} [b_i b_{j}^\dagger +{\rm h.c.} , \hat{n}_X^s ] ,
\end{align}
where we use $[H_{0,X},\hat{n}_X^s]=0$ and $[H_{0,X^\co},\hat{n}_X^s]=0$ [see~\eqref{def:Ham_surface} for the notation of $\partial h_X$].
Note that $[V,\hat{n}_X^s]=0$ because of $[\hat{n}_i,\hat{n}_j]=0$ for $\forall j \in \Lambda$. 
By using $[b_i,\nb_X]=b_i$ for $\forall i\in X$, we have 
\begin{align}
\label{commutator_nb_X_H0}
&[b_ib_j^\dagger,\nb_X^s] = \sum_{s_1=0}^{s-1} \nb_X^{s_1} [b_ib_j^\dagger,\nb_X]\nb_X^{s-s_1-1} = 
\sum_{s_1=0}^{s-1} \nb_X^{s_1} b_ib_j^\dagger \nb_X^{s-s_1-1}, 
\end{align}
which yields 
\begin{align}
\label{derivative_M_S_i_t_0}
\left| \frac{d}{d\tau} M^{(s)}_X(\tau) \right |
&=  \left| \tr \bigl( [H,\nb_X^s] \rho(\tau)\bigr)  \right|  \notag \\
&\le 
 \sum_{s_1=0}^{s-1}
\left| \tr\br{ \nb_X^{s_1} \sum_{i \in \partial X} \sum_{j \in X^\co:\dist_{i,j}=1}\bar{J} \br{b_i b_{j}^\dagger -{\rm h.c.}} \nb_X^{s-s_1-1}  \rho(\tau) } \right| .
\end{align}

We now need to obtain an upper bound for
\begin{align}
\left| \tr \br{  \nb_X^{s_1} b_i b_{j}^\dagger \nb_X^{s-s_1-1}  \rho(\tau) } \right| .
\end{align}
In general, we can prove the following proposition: 
\begin{prop}\label{prop:expectation_b_j_b_i_poly_n_i}
For arbitrary $s_1$, $s_1'$, $s_2$ and $s_2'$, we obtain the upper bound as
\begin{align}
\left| \tr \brr{ f(\hat{\bold{n}})\nb_i^{s_1}\nb_j^{s_2}  b_i b_{j}^\dagger \nb_i^{s'_1}\nb_j^{s'_2}    \rho(\tau) } \right|
 &\le \frac{1}{2}\tr \brr{ f(\hat{\bold{n}})\br{\nb_i^{s_1+s_1'}(\nb_j+1)^{s_2+s_2'}+(\nb_i+1)^{s_1+s_1'}\nb_j^{s_2+s_2'}}  (\nb_i+\nb_j) \rho(\tau) } ,
 \label{b_i_b_j_n_i+_n_j_ineN_general}
\end{align}
where $ f(\hat{\bold{n}})$ is an arbitrary non-negative function with respect to $\{\hat{n}_s\}_{s\neq i,j}$ (i.e., it satisfies $[f(\hat{\bold{n}}), b_i] = [f(\hat{\bold{n}}), b_j] =0$). 
\end{prop}

\subsubsection{Proof of Proposition~\ref{prop:expectation_b_j_b_i_poly_n_i}} \label{sec:Proof of prop:expectation_b_j_b_i_poly_n_i}

Let $\ket{\bold{N}}$ ($\bold{N}=\{N_i\}_{i\in \Lambda}$) be an eigenstate such of the boson number operators, i.e., 
$\nb_i \ket{\bold{N}}=N_i \ket{\bold{N}}$ for $\forall i\in \Lambda$. 
For the proof, by using $b_i\ket{\bold{N}}  =\sqrt{N_i}\ket{N_i-1, N_j ,\bold{N}_{i,j}}$ and $b_j^\dagger\ket{\bold{N}}  =\sqrt{N_j+1}\ket{N_i, N_j+1 ,\bold{N}_{i,j}}$, we first transform
\begin{align}
\label{start_eq/_prop:expectation_b_j_b_i_poly_n_i}
&\tr \brr{ f(\hat{\bold{n}})\nb_i^{s_1}\nb_j^{s_2}  b_i b_{j}^\dagger \nb_i^{s'_1}\nb_j^{s'_2}    \rho(\tau) }  = \tr \brr{ f(\hat{\bold{n}})\nb_i^{s_1}\nb_j^{s_2}  b_i b_{j}^\dagger \nb_i^{s'_1}\nb_j^{s'_2}  \sum_{\bold{N}} P_{\bold{N}}  \rho(\tau) } \quad (P_{\bold{N}} := \ket{\bold{N}}\bra{\bold{N}})  \notag \\
&=\sum_{\bold{N}}f(\bold{N}_{i,j})N_i^{s_1'}N_j^{s_2'}\sqrt{N_i}\sqrt{N_j+1}(N_i-1)^{s_1}(N_j+1)^{s_2} \bra{N_i,N_j,\bold{N}_{i,j}}\rho(\tau)\ket{N_i-1,N_j+1,\bold{N}_{i,j}} , 
\end{align}
where we use a notation of $\bold{N} = \{N_i, N_j, \bold{N}_{i,j}\}$, where $\bold{N}_{i,j}$ includes the boson number in $\Lambda\setminus \{i,j\}$. 
From the Cauchy-Schwartz inequality, we have 
\begin{align} 
\label{Cauchy-Schwartz inequality_apply_ninjnvec}
&\bra{N_i,N_j,\bold{N}_{i,j}}\rho(\tau)\ket{N_i-1,N_j+1,\bold{N}_{i,j}} \notag \\
&\le \frac{1}{2}  \br{\bra{N_i,N_j,\bold{N}_{i,j}}\rho(\tau)\ket{N_i,N_j,\bold{N}_{i,j}}  + \bra{N_i-1,N_j+1,\bold{N}_{i,j}}\rho(\tau)\ket{N_i-1,N_j+1,\bold{N}_{i,j}}  }  .
\end{align} 
By applying the above inequality to Eq.~\eqref{start_eq/_prop:expectation_b_j_b_i_poly_n_i}, we obtain 
\begin{align}
&\abs{\tr \brr{ f(\hat{\bold{n}})\nb_i^{s_1}\nb_j^{s_2}  b_i b_{j}^\dagger \nb_i^{s'_1}\nb_j^{s'_2}    \rho(\tau) } } \notag \\
&\le \frac{1}{2}\sum_{\bold{N}}f(\bold{N}_{i,j})N_i^{s_1'}N_j^{s_2'}\sqrt{N_i}\sqrt{N_j+1}(N_i-1)^{s_1}(N_j+1)^{s_2} \bra{N_i,N_j,\bold{N}_{i,j}}\rho(\tau)\ket{N_i,N_j,\bold{N}_{i,j}} \notag\\
&+\frac{1}{2}\sum_{\bold{N}}f(\bold{N}_{i,j})N_i^{s_1'}N_j^{s_2'}\sqrt{N_i}\sqrt{N_j+1}(N_i-1)^{s_1}(N_j+1)^{s_2} \bra{N_i-1,N_j+1,\bold{N}_{i,j}}\rho(\tau)\ket{N_i-1,N_j+1,\bold{N}_{i,j}} \notag\\
&\le \frac{1}{2}\sum_{\bold{N}}f(\bold{N}_{i,j})N_i^{s_1+s_1'}(N_j+1)^{s_2+s_2'}(N_i+N_j)\bra{N_i,N_j,\bold{N}_{i,j}}\rho(\tau)\ket{N_i,N_j,\bold{N}_{i,j}} \notag\\
&+\frac{1}{2}\sum_{\bold{N}}f(\bold{N}_{i,j})(N_i+1)^{s_1+s_1'}N_j^{s_2+s_2'}(N_i+N_j)\bra{N_i,N_j,\bold{N}_{i,j}}\rho(\tau)\ket{N_i,N_j,\bold{N}_{i,j}}\notag\\
&=\frac{1}{2}\sum_{\bold{N}}f(\bold{N}_{i,j})\br{N_i^{s_1+s_1'}(N_j+1)^{s_2+s_2'}+(N_i+1)^{s_1+s_1'}N_j^{s_2+s_2'}}\br{N_i+N_j}\bra{N_i,N_j,\bold{N}_{i,j}}\rho(\tau)\ket{N_i,N_j,\bold{N}_{i,j}}\\
&=\frac{1}{2}\tr \brr{ f(\hat{\bold{n}})\br{\nb_i^{s_1+s_1'}(\nb_j+1)^{s_2+s_2'}+(\nb_i+1)^{s_1+s_1'}\nb_j^{s_2+s_2'}}  (\nb_i+\nb_j)  \sum_{\bold{N}} P_{\bold{N}}  \rho(\tau) }  \notag\\
&=\frac{1}{2}\tr \brr{ f(\hat{\bold{n}})\br{\nb_i^{s_1+s_1'}(\nb_j+1)^{s_2+s_2'}+(\nb_i+1)^{s_1+s_1'}\nb_j^{s_2+s_2'}}  (\nb_i+\nb_j) \rho(\tau) },
\end{align}
where we use the facts that the operator $ f(\hat{\bold{n}})$ commutes with $b_i$ and $b_j$, $ f(\bold{N}_{i,j})\ge 0$, and $f(\hat{\bold{n}})\ket{\bold{N}} =f(\bold{N}_{i,j})\ket{\bold{N}}$.
Note that $$\max\br{ \sqrt{N_i(N_j+1)}, \sqrt{(N_i+1)N_j}}\le N_i+N_j \for \forall N_i,N_j\in\mathbb{N}.$$ This completes the proof. $\square$

 {~}

\noindent\hrulefill{\bf [ End of Proof of Proposition~\ref{prop:expectation_b_j_b_i_poly_n_i}] }

{~}

By applying Proposition~\ref{prop:expectation_b_j_b_i_poly_n_i} to the inequality~\eqref{derivative_M_S_i_t_0}, we have 
\begin{align}
\left| \frac{d}{d\tau} M^{(s)}_X(\tau) \right | &\le \sum_{s_1=0}^{s-1} \left| \tr\br{ \nb_X^{s_1} \sum_{i \in \partial X} \sum_{j \in X^\co:\dist_{i,j}=1}\bar{J}
\br{ b_i b_{j}^\dagger - {\rm h.c.} }\nb_X^{s-s_1-1}  \rho(\tau) } \right| \notag \\
&\le\sum_{s_1=0}^{s-1}\tr\brr{(\nb_X+1)^{s_1}  \sum_{i \in \partial X} \sum_{j \in X^\co:\dist_{i,j}=1}2\bar{J} (\nb_i+\nb_j) (\nb_X+1)^{s-s_1-1}  \rho(\tau) } \notag \\
&\le2 \sum_{s_1=0}^{s-1}\tr\brr{(\nb_X+1)^{s_1} \sum_{i \in \partial X} \br{  \gamma \bar{J} \nb_i  + \sum_{j :\dist_{i,j}=1} \bar{J} \nb_j } (\nb_X+1)^{s-s_1-1}  \rho(\tau) } ,\label{eq:tmp_bound_MXs}
\end{align} 
where we use $\sum_{j :\dist_{i,j}=1}1 \le |i[1]| \le \gamma$ in the third inequality. 
Note that the parameter $\gamma$ have been defined in Eq.~\eqref{supp_parameter_gamma_X}.  

In the same way, we can also upper-bound the derivative of 
\begin{align}
\left| \frac{d}{d\tau} \tr\brr{(\nb_X+1)^{s_1} \sum_{i \in \partial X} \br{  \gamma \bar{J} \nb_i  + \sum_{j :\dist_{i,j}=1}  \bar{J} \nb_j } (\nb_X+1)^{s-s_1-1}  \rho(\tau) } \right|  
\end{align} 
by using the same analyses and Proposition~\ref{prop:expectation_b_j_b_i_poly_n_i}. 
In the following, we aim to upper-bound the $m$th-order derivative by utilizing the upper bound for $(m-1)$th-order derivative.
However, the notation becomes more and more complicated as the order of derivatives increases.  

To simplify the notation, we define the following super-operator $\mathcal{M}$, which acts on arbitrary $\nb_i$ ($i\in \Lambda$), by 
\begin{equation}\label{def_mathcal_M_mat}
\mathcal{M}(\nb_i)\coloneqq  \gamma\bar{J}\nb_i+\sum_{j: \dist_{i,j}=1} \bar{J} \nb_j \for \forall i \in \Lambda.
\end{equation} 
The super-operator $\mathcal{M}$ is linear in the sense that 
\begin{align}
\mathcal{M}\br{\sum_{i\in\Lambda} \nu_i \nb_i } \coloneqq  \sum_{i\in\Lambda} \nu_i  \mathcal{M}\br{\nb_i }  .
\end{align} 
Recursively, we can define $\mathcal{M}^m$ for general $m\in \mathbb{N}$ by
\begin{align}
\mathcal{M}^0(\nb_i)& \coloneqq \nb_i,\\
\mathcal{M}^m(\nb_i)& \coloneqq \mathcal{M}(\mathcal{M}^{m-1}(\nb_i)),
\end{align} 
which are also linear super-operators.
Using the super-operator $\mathcal{M}$, we define the super-operator $\mathcal{D}^m$, which acts on $\nb_{X}$, by 
\begin{align}
&\mathcal{D}^0(\nb_{X}) \coloneqq \nb_X,\quad  \mathcal{D}^1(\nb_{X}) \coloneqq \sum_{i\in \partial X} \mathcal{M}(\nb_i) ,  \notag \\
&\mathcal{D}^{m+1}(\nb_{X})\coloneqq  \mathcal{M}(\mathcal{D}^{m}(\nb_X))  \for \forall m\ge 1 .
\label{def_D_mathcal_m}
\end{align}

From the definitions~\eqref{def_mathcal_M_mat} and \eqref{def_D_mathcal_m}, by acting the super-operator $\mathcal{D}^m$, the operator $\mathcal{D}^m(\nb_{X})$ is still a linear combination of $\{\nb_i\}_{i\in \Lambda}$. 
For the operator $\mathcal{D}^m(\nb_{X})$, we can prove the following lemma:
\begin{lemma}\label{lem:upp_max_alpha_i_D^m}
Let us define $X_{m}$ as the set of all sites where the number operators $\{\nb_i\}$ appear in $\mathcal{D}^m(\nb_{X})$, i.e., 
\begin{align}
\mathcal{D}^m(\nb_{X})=\sum_{i\in X_m}\alpha_i^{(m)}\nb_i ,
\end{align}
where $\alpha_i^{(m)}>0$. Then, we can upper-bound $\alpha_i^{(m)}$ as 
\begin{align}
\label{upp_alpha/_i_max}
\max_{i\in X_m}\br{\alpha_i^{(m)}}\le (2\gamma \bar J)^m.
\end{align} 
\end{lemma}

\subsubsection{Proof of Lemma~\ref{lem:upp_max_alpha_i_D^m}}
Because of the definition~\eqref{def_mathcal_M_mat} for $\mathcal{M}$, we obtain 
\begin{align}
\mathcal{D}^{m+1}(\nb_{X})=\sum_{i\in X_m}\alpha_i^{(m)} \mathcal{M} (\nb_i)
&=\sum_{i\in X_m}\alpha_i^{(m)}\br{ \gamma\bar{J}\nb_i+\sum_{j: \dist_{i,j}=1} \bar{J} \nb_j}\notag \\
 & \preceq \sum_{i\in X_{m+1}}\br{\alpha_i^{(m)} \gamma\bar{J}  +\sum_{j\in X_m: \dist_{i,j}=1} \alpha_j^{(m)}\bar{J}  }\nb_i  , \notag 
\end{align}
which yields 
\begin{align}
\label{lem:upp_max_alpha_i_D^m_proof}
\max_{i\in X_{m+1}}\brr{\alpha_i^{(m+1)}} 
\le \br{ \gamma\bar{J}+\sum_{j: \dist_{i,j}=1} \bar{J} }\max_{i\in X_m} \brr{\alpha_i^{(m)}} 
\le 2\gamma \bar{J} \max_{i\in X_m} \brr{\alpha_i^{(m)}} ,
\end{align}
where we use $\sum_{j: \dist_{i,j}=1} \bar{J} \le \bar{J} |i[1]| \le \gamma \bar{J}$.
We thus prove the inequality~\eqref{upp_alpha/_i_max} by iteratively using the inequality~\eqref{lem:upp_max_alpha_i_D^m_proof}. 
This completes the proof. $\square$

 {~}

\noindent\hrulefill{\bf [ End of Proof of Lemma~\ref{lem:upp_max_alpha_i_D^m}] }

{~}

By using the notation $\mathcal{D}$, we simplify the inequality~\eqref{eq:tmp_bound_MXs} as follows:
\begin{align}
\left| \frac{d}{d\tau} M^{(s)}_X(\tau) \right | &\le 
2\sum_{s_1=0}^{s-1}\tr\brr{(\nb_X+1)^{s_1} \mathcal{D}(\nb_X) (\nb_X+1)^{s-s_1-1}  \rho(\tau) } . \label{eq:tmp_bound_MXs_notation}
\end{align} 
Also, by using the same analytical techniques, we upper-bound the derivative 
\begin{align}
\left| \frac{d}{d\tau} \sum_{s_1=0}^{s-1}\tr\brr{(\nb_X+1)^{s_1} \mathcal{D}(\nb_X) (\nb_X+1)^{s-s_1-1}  \rho(\tau) }\right |  
\end{align} 
in the form of 
\begin{align}
\label{secon_order_der_upp_norm}
&\left| \frac{d}{d\tau} \sum_{s_1=0}^{s-1}\tr\brr{(\nb_X+1)^{s_1} \mathcal{D}(\nb_X) (\nb_X+1)^{s-s_1-1}  \rho(\tau) }\right |   \notag \\
\le& 4\sum_{0\le s_1 \le s_2 \le s-2}\tr\brr{(\nb_X+2)^{s_1} [ \mathcal{D}(\nb_X) + 2\gamma \bar{J} ] (\nb_X+2)^{s_2-s_1}[ \mathcal{D}(\nb_X) + 2\gamma \bar{J} ]  (\nb_X+2)^{s-s_2-2}  \rho(\tau) } \notag \\
&+ 2\sum_{0\le s_1 \le s-1} \tr\brr{(\nb_X+2)^{s_1} \mathcal{D}^2(\nb_X)   (\nb_X+2)^{s-s_1-1}  \rho(\tau) } .
\end{align} 
Here, by changing as $\nb_i\to \nb_i+1$ ($i\in X$) in using Proposition~\ref{prop:expectation_b_j_b_i_poly_n_i}, we need to perform the replacement of
\begin{align}
\mathcal{D}(\nb_X) \to \mathcal{D}(\nb_X) + 2\gamma \bar{J} 
\end{align} 
from the inequality~\eqref{upp_alpha/_i_max} in Lemma~\ref{lem:upp_max_alpha_i_D^m}.

The challenging task here is to further simplify the upper bounds like \eqref{secon_order_der_upp_norm} in considering the higher order derivatives.
As a key technique, we prove the following proposition:
\begin{prop} \label{prop:simplify_high_der}
Let us define
\begin{align}
\label{def_tilde_mathcal_D_nb_X_s}
\tilde{\mathcal{K}}^m(\nb_X^s)&\coloneqq\sum_{m_1+\cdots+m_s=m}\frac{m!}{m_1!\cdots m_s!}\brr{\mathcal{D}^{m_1}(\nb_X)+m(2\gamma\bar J)^{m_1}}\cdots \brr{\mathcal{D}^{m_s}(\nb_X)+m(2\gamma\bar J)^{m_s}},\\
M_X^{(s,m)}(\tau)&\coloneqq \tr\brr{\tilde{\mathcal{K}}^m(\nb_X^s)\rho(\tau)}.
\end{align}
It is evident that $\tilde{\mathcal{K}}^0(\nb_X^s)=\nb_X^s$, and hence $M_X^{(s,0)}(\tau)=\tr\br{\nb_X^s\rho(\tau)}$.
Then, we can prove that
\begin{equation}
\left|\frac{d}{d\tau}M_X^{(s,m)}(\tau) \right|\le 2M_X^{(s,m+1)}(\tau) \for \forall m\ge 0.
\label{main_eq_prop:simplify_high_der}
\end{equation}
\end{prop}

{\bf Remark.} In considering $M_{X_1,X_2,\ldots,X_s}(\tau)$ in Eq.~\eqref{moment_general_X_1_X_s}, 
we need to replace the inequality~\eqref{main_eq_prop:simplify_high_der} as follows:
\begin{equation}
\left|\frac{d}{d\tau}M_{X_1,X_2 ,\ldots, X_s}^{(s,m)}(\tau) \right|\le 2M_{X_1,X_2 ,\ldots, X_s}^{(s,m+1)}(\tau) \for \forall m\ge 0.
\label{main_eq_prop:simplify_high_der_gene}
\end{equation}
with  
\begin{align}
\label{def_tilde_mathcal_D_nb_X_s_gene}
\tilde{\mathcal{K}}^m(\nb_{X_1} \nb_{X_2} \cdots \nb_{X_s} )&
\coloneqq\sum_{m_1+\cdots+m_s=m}\frac{m!}{m_1!\cdots m_s!}\brr{\mathcal{D}^{m_1}(\nb_{X_1})+m(2\gamma\bar J)^{m_1}}\cdots \brr{\mathcal{D}^{m_s}(\nb_{X_s})+m(2\gamma\bar J)^{m_s}},\\
M_{X_1,X_2 ,\ldots, X_s}^{(s,m)}(\tau)&\coloneqq \tr\brr{\tilde{\mathcal{K}}^m(\nb_{X_1} \nb_{X_2} \cdots \nb_{X_s} ) \rho(\tau)}.
\end{align}
We notice that the proof for the inequality~\eqref{main_eq_prop:simplify_high_der_gene} is exactly same as that for~\eqref{main_eq_prop:simplify_high_der}. 

\subsubsection{Proof of Proposition~\ref{prop:simplify_high_der}}

To derive the upper bound~\eqref{main_eq_prop:simplify_high_der}, we start from the differential equation for $M_X^{(s,m)}(\tau)$:
\begin{align}
\label{sth_moment_der}
\frac{d}{d\tau} M^{(s,m)}_X(\tau) &=-i \tr \left ( \sum_{m_1+\cdots+m_s=m}\frac{m!}{m_1!\cdots m_s!}
\br{\mathcal{D}^{m_1}(\nb_X)+m(2\gamma\bar J)^{m_1}}\cdots
\br{\mathcal{D}^{m_s}(\nb_X)+m(2\gamma\bar J)^{m_s}}  [H, \rho(\tau) ]  \right) \notag \\
& =i\tr \left (\sum_{m_1+\cdots+m_s=m}\frac{m!}{m_1!\cdots m_s!}\left[H,\br{\mathcal{D}^{m_1}(\nb_X)+m(2\gamma\bar J)^{m_1}}\cdots
\br{\mathcal{D}^{m_s}(\nb_X)+m(2\gamma\bar J)^{m_s}}\right] \rho(\tau)  \right) \notag\\
&= i\sum_{m_1+\cdots+m_s=m}\frac{m!}{m_1!\cdots m_s!}\sum_{k=1}^s\tr \left (\mathcal{D}_{<k}^{(m)} \left[H,\mathcal{D}^{m_k}(\nb_{X})\right] \mathcal{D}_{>k}^{(m)} \rho(\tau)  \right),
\end{align}
where we have defined 
\begin{align}
\mathcal{D}_{<k}^{(m)}\coloneqq\prod_{l<k} [\mathcal{D}^{m_l}(\nb_{X})+m(2\gamma\bar J)^{m_l}]  , \quad 
\mathcal{D}_{>k}^{(m)}\coloneqq\prod_{l>k} [\mathcal{D}^{m_l}(\nb_{X})+m(2\gamma\bar J)^{m_l}]. 
\label{def_mathcal_D_<>k_m}
\end{align}

The form of the Hamiltonian~\eqref{def:Ham} gives 
\begin{align}
\label{Ham_commute_nb_i_s_2}
[H,\mathcal{D}^{m_k}(\nb_{X}) ] = [H_0,\mathcal{D}^{m_k}(\nb_{X})],
\end{align}
where we have used $[V,\mathcal{D}^{m_k}(\nb_{X})]=0$ because of $[\hat{n}_i,\hat{n}_j]=0$ for $\forall i,j \in \Lambda$. 
To estimate the upper bound of the norm of~\eqref{Ham_commute_nb_i_s_2}, we start from the case of $m_k=0$, i.e., $[H_0,\mathcal{D}^{m_k}(\nb_{X})]= [H_0,\mathcal{D}^{0}(\nb_{X})]= [H_0,\nb_{X}]$.
In this case, we have already obtained Eqs.~\eqref{Ham_commute_nb_i_s} and \eqref{commutator_nb_X_H0}, which give 
\begin{align}\label{eq:HD_com_tmp1}
[H_0,\mathcal{D}^{0}(\nb_{X})] &= 
\sum_{i \in \partial X}  \sum_{j \in X^\co:\dist_{i,j}=1}J_{i,j}\br{b_i b_{j}^\dagger - b_i^\dagger b_{j}}. 
\end{align}
We thus obtain 
\begin{align}
&\left| \tr \left (\mathcal{D}_{<k}^{(m)} \left[H,\mathcal{D}^{m_k}(\nb_{X})\right] \mathcal{D}_{>k}^{(m)} \rho(\tau)  \right)  \right|  
\le \sum_{i \in \partial X} \sum_{j \in X^\co :\dist_{i,j}=1} \left| J_{i,j} \tr\br{\mathcal{D}_{<k}^{(m)} \br{b_i b_{j}^\dagger - b_i^\dagger b_{j}}  \mathcal{D}_{>k}^{(m)} \rho(\tau) } \right| \notag \\
&\le  \sum_{i \in \partial X} \sum_{j \in X^\co :\dist_{i,j}=1}  \bar{J} 
\br{ \left| \tr\br{\mathcal{D}_{<k}^{(m)}  b_i b_{j}^\dagger  \mathcal{D}_{>k}^{(m)} \rho(\tau) } \right| + \left| \tr\br{\mathcal{D}_{<k}^{(m)}  b_i^\dagger  b_{j} \mathcal{D}_{>k}^{(m)} \rho(\tau) } \right|}.
\label{derivative_M_S_i_t_mk=0}
\end{align}
We here need to obtain an upper bound for
\begin{align}
\left| \tr\br{\mathcal{D}_{<k}^{(m)}  b_i b_{j}^\dagger  \mathcal{D}_{>k}^{(m)} \rho(\tau) } \right| \AND \left| \tr\br{\mathcal{D}_{<k}^{(m)}  b_i^\dagger  b_{j} \mathcal{D}_{>k}^{(m)} \rho(\tau) } \right|.
\end{align}
For the purpose, we apply the proposition \ref{prop:expectation_b_j_b_i_poly_n_i}.
Here, because of the inequality~\eqref{upp_alpha/_i_max} in Lemma~\ref{lem:upp_max_alpha_i_D^m}, when one of $\{\nb_i\}_{i\in X_{m_l}}$ increases as $\nb_i\to \nb_i+ 1$, 
the operator $\mathcal{D}^{m_l}(\nb_{X})$ is replaced by 
\begin{align}
\mathcal{D}^{m_l}(\nb_{X}) \to \mathcal{D}^{m_l}(\nb_{X}) + (2\gamma \bar{J})^{m_l} ,
\end{align}
which leads to the replacement of
\begin{align}
\mathcal{D}_{<k}^{(m)} \to \mathcal{D}_{<k}^{(m+1)} ,\quad \mathcal{D}_{>k}^{(m)} \to \mathcal{D}_{>k}^{(m+1)}  
\end{align}
from the definitions~\eqref{def_mathcal_D_<>k_m}. 
We thus apply the inequality~\eqref{b_i_b_j_n_i+_n_j_ineN_general} to $\left| \tr\br{\mathcal{D}_{<k}^{(m)}  b_i b_{j}^\dagger  \mathcal{D}_{>k}^{(m)} \rho(\tau) } \right| $ and obtain 
\begin{align}
\label{b_i_b_j_prop_expec_apply}
\left| \tr\br{\mathcal{D}_{<k}^{(m)}  b_i b_{j}^\dagger  \mathcal{D}_{>k}^{(m)} \rho(\tau) } \right| 
\le \tr\br{\mathcal{D}_{<k}^{(m+1)}  (\nb_i+\nb_j)  \mathcal{D}_{>k}^{(m+1)} \rho(\tau) }  .
\end{align}
We can obtain the same upper bound for $\left| \tr\br{\mathcal{D}_{<k}^{(m)}  b_i^\dagger  b_{j} \mathcal{D}_{>k}^{(m)} \rho(\tau) } \right|$. 

By combining the inequality~\eqref{b_i_b_j_prop_expec_apply} with~\eqref{derivative_M_S_i_t_mk=0}, we obtain
\begin{align}
\left| \tr \left (\mathcal{D}_{<k}^{(m)} \left[H,\mathcal{D}^{0}(\nb_{X})\right] \mathcal{D}_{>k}^{(m)} \rho(\tau)  \right)  \right|  
&\le  2 \tr\br{\mathcal{D}_{<k}^{(m+1)}\sum_{i \in \partial X}   \sum_{j \in X^\co :\dist_{i,j}=1} \bar{J} (\nb_i+\nb_j)  \mathcal{D}_{>k}^{(m+1)} \rho(\tau) } \notag\\
&\le  2  \tr\br{\mathcal{D}_{<k}^{(m+1)} \sum_{i \in \partial X}  \br{\gamma \bar{J}\nb_i+  \sum_{j \in\Lambda :\dist_{i,j}=1}\bar{J}  \nb_j}  \mathcal{D}_{>k}^{(m+1)} \rho(\tau) }  \notag \\
&= 2 \tr\br{\mathcal{D}_{<k}^{(m+1)} \mathcal{D}(\nb_X) \mathcal{D}_{>k}^{(m+1)} \rho(\tau) }  ,
 \label{eq:tmp_bound_MXs_mk=0}
\end{align} 
where we use the definitions~\eqref{def_mathcal_M_mat} and \eqref{def_D_mathcal_m} to obtain $\mathcal{D}(\nb_X)=\sum_{i \in \partial X}  \br{\gamma \bar{J}\nb_i+  \sum_{j \in\Lambda :\dist_{i,j}=1}\bar{J}  \nb_j} $.

For general $m_k$ that is larger than $0$, we use the same analyses to estimate the upper bound of the norm of~\eqref{Ham_commute_nb_i_s_2}. 
By expanding as $\mathcal{D}^{m_k}(\nb_{X})=  \sum_{i \in X_{m_k}}\alpha_i^{(m_k)} \nb_i$, we have 
\begin{align}
\label{Ham_commute_nb_i_s_2_m_k}
 [H,\mathcal{D}^{m_k}(\nb_{X})] = \sum_{i \in X_{m_k}}\alpha_i^{(m_k)} \sum_{j \in\Lambda :\dist_{i,j}=1}   J_{i,j}\br{b_i b_{j}^\dagger - b_i^\dagger b_{j}}.
\end{align}
in the same way as the derivation of Eq.~\eqref{eq:HD_com_tmp1}.
We therefore obtain a similar inequality to~\eqref{derivative_M_S_i_t_mk=0} as follows:
\begin{align}
\label{derivative_M_S_i_t}
\left| \tr \left (\mathcal{D}_{<k}^{(m)} \left[H,\mathcal{D}^{m_k}(\nb_{X})\right] \mathcal{D}_{>k}^{(m)} \rho(\tau)  \right)  \right|  
&\le  2  \bar{J} \sum_{i \in X_{m_k}}\sum_{j \in\Lambda :\dist_{i,j}=1} \alpha_i^{(m_k)} \left| \tr\br{\mathcal{D}_{<k}^{(m)} b_i b_{j}^\dagger   \mathcal{D}_{>k}^{(m)} \rho(\tau) } \right|  \notag \\
&\le 2 \sum_{i \in X_{m_k}} \alpha_i^{(m_k)} \tr\brr{\mathcal{D}_{<k}^{(m+1)}  \br{\gamma \bar{J}\nb_i+  \sum_{j \in\Lambda :\dist_{i,j}=1}\bar{J}  \nb_j} \mathcal{D}_{>k}^{(m+1)} \rho(\tau) }  ,
\end{align}
where in the last inequality we use the inequality~\eqref{b_i_b_j_prop_expec_apply}. 
Then, from the definitions~\eqref{def_mathcal_M_mat} and \eqref{def_D_mathcal_m}, we have 
\begin{equation}
\sum_{i \in X_{m_k}}\alpha_i^{(m_k)}\br{\gamma\bar{J}\nb_i+ \sum_{j \in\Lambda :\dist_{i,j}=1}   \bar{J} \nb_j} 
=\sum_{i \in X_{m_k}}\alpha_i^{(m_k)}\mathcal{M} (\nb_i) = 
\mathcal{M} (\mathcal{D}^{m_k}(\nb_X))  =  \mathcal{D}^{m_k+1}(\nb_X), \label{eq:ineq_tmp}
\end{equation}
which reduces the inequality~\eqref{derivative_M_S_i_t} to 
\begin{align}
\label{derivative_M_S_i_t//2}
\left| \tr \left (\mathcal{D}_{<k}^{(m+1)} \left[H,\mathcal{D}^{m_k}(\nb_{X})\right] \mathcal{D}_{>k}^{(m+1)} \rho(\tau)  \right)  \right|  
&\le 2  \tr\brr{\mathcal{D}_{<k}^{(m+1)}  \mathcal{D}^{m_k+1}(\nb_X) \mathcal{D}_{>k}^{(m+1)} \rho(\tau) }   \for m_k >0 .
\end{align}

By combining the inequalities~\eqref{eq:tmp_bound_MXs_mk=0} and \eqref{derivative_M_S_i_t//2}, we obtain 
\begin{align}
\label{derivative_M_S_i_t//2_general_m}
\left| \tr \left (\mathcal{D}_{<k}^{(m+1)} \left[H,\mathcal{D}^{m_k}(\nb_{X})\right] \mathcal{D}_{>k}^{(m+1)} \rho(\tau)  \right)  \right|  
&\le 2\tr\brr{\mathcal{D}_{<k}^{(m+1)}  \mathcal{D}^{m_k+1}(\nb_X) \mathcal{D}_{>k}^{(m+1)} \rho(\tau) }   
\end{align}
for arbitrary $m_k\ge 0$. Hence, from Eq.~\eqref{sth_moment_der}, we arrive at the desired inequality~\eqref{main_eq_prop:simplify_high_der}: 
\begin{align}
\left| \frac{d}{d\tau} M^{(s,m)}_X(\tau) \right |
&\le  \sum_{m_1+\cdots+m_s=m}\frac{m!}{m_1!\cdots m_s!}\sum_{k=1}^s
\left| \tr \left (\mathcal{D}_{<k}^{(m+1)} \left[H,\mathcal{D}^{m_k}(\nb_{X})\right] \mathcal{D}_{>k}^{(m+1)} \rho(\tau)  \right)  \right|  
\notag \\
&\le 2 \sum_{m_1+\cdots+m_s=m}\frac{m!}{m_1!\cdots m_s!}\sum_{k=1}^s  \tr\br{\mathcal{D}_{<k}^{(m+1)} \mathcal{D}^{m_k+1}(\nb_X) \mathcal{D}_{>k}^{(m+1)} \rho(\tau) }  \notag \\
&\le 2\tr \brr{ \tilde{\mathcal{K}}^{m+1}(\nb_X^s)\rho(\tau)}  = 2M_X^{(s,m+1)}(\tau),\label{eq:tmp_bound_MXs2}
\end{align} 
where in the last inequality, we use the following inequality:
\begin{align}
\label{derivative_M_S_i_t//2_general_m_fin}
&\sum_{m_1+\cdots+m_s=m}\frac{m!}{m_1!\cdots m_s!}\sum_{k=1}^s  \tr\br{\mathcal{D}_{<k}^{(m+1)} \mathcal{D}^{m_k+1}(\nb_X) \mathcal{D}_{>k}^{(m+1)} \rho(\tau) } \notag \\
&\le \sum_{m_1+\cdots+m_s=m}\frac{m!}{m_1!\cdots m_s!}\sum_{k=1}^s  \tr\br{\mathcal{D}_{<k}^{(m+1)} \brr{\mathcal{D}^{m_k+1}(\nb_X)+(m+1)(2\gamma\bar J)^{m_k+1}} \mathcal{D}_{>k}^{(m+1)} \rho(\tau) } \notag \\
&=\sum_{k=1}^s  \sum_{\substack{m_1+\cdots+m_s=m+1 \\ m_k \ge 1}}\frac{m!}{m_1!\cdots m_{k-1}! (m_k-1)!m_{k+1}!\cdots m_s!}
\tr\br{\prod_{i}\brr{\mathcal{D}^{m_i}(\nb_X)+(m+1)(2\gamma\bar J)^{m_i}}\rho(\tau)} \notag \\
&=\sum_{m_1+\cdots+m_s=m+1} \br{\sum_{k=1}^s\frac{m_k}{m+1}} \frac{(m+1)!}{m_1!\cdots m_s!}
\tr\br{\prod_{i}\brr{\mathcal{D}^{m_i}(\nb_X)+(m+1)(2\gamma\bar J)^{m_i}}\rho(\tau)} \notag \\
&= \sum_{m_1+\cdots+m_s=m+1}\frac{(m+1)!}{m_1!\cdots m_s!}
\tr\br{\prod_{i}\brr{\mathcal{D}^{m_i}(\nb_X)+(m+1)(2\gamma\bar J)^{m_i}}\rho(\tau)} 
= \tr \br{ \tilde{\mathcal{K}}^{m+1}(\nb_X^s)\rho(\tau)} ,
\end{align}
where we use the definition~\eqref{def_tilde_mathcal_D_nb_X_s} in the last equation.
This completes the proof. $\square$

 {~}

\noindent\hrulefill{\bf [ End of Proof of Proposition~\ref{prop:simplify_high_der}] }

{~}

Applying Eq.~\eqref{eq:tmp_bound_MXs2} immediately yields
\begin{align}
\label{m_derivative_mathcal_D_integral}
M_X^{(s,m)}(\tau)&\le M_X^{(s,m)}(0) +  2\int_0^\tau M_X^{(s,m+1)}(\tau) d\tau. 
\end{align}
For $m=0$, we have 
\begin{align}
\label{m_derivative_mathcal_D_integral_m=0}
M_X^{(s)}(\tau) = M_X^{(s,0)}(\tau)&\le M_X^{(s,0)}(0) +  2\int_0^\tau M_X^{(s,1)}(\tau) d\tau. 
\end{align}
In the same way, by applying the inequality~\eqref{m_derivative_mathcal_D_integral} for $m=1$, we have 
\begin{align}
\label{m_derivative_mathcal_D_integral_m=1}
M_X^{(s)}(\tau)&\le M_X^{(s,0)}(0) +  2\int_0^\tau \br{ M_X^{(s,1)}(0) +  2\int_0^{\tau_1} M_X^{(s,2)}(\tau_2) d\tau_2 }  d\tau_1 \notag \\
&= M_X^{(s,0)}(0) +  2\tau M_X^{(s,1)}(0) +  2^2\int_0^\tau\int_0^{\tau_1} M_X^{(s,2)}(\tau_2) d\tau_2   d\tau_1 . 
\end{align}
Then, by iteratively repeating this process, we finally obtain  
\begin{align}
\label{mth_order_derivative_mathcal_D_integral_0}
M_X^{(s)}(\tau)\le \sum_{m=0}^\infty \frac{(2\tau)^m}{m!} M_X^{(s,m)}(0)  = \sum_{m=0}^\infty \frac{(2\tau)^m}{m!}\tr\br{\tilde{\mathcal{K}}^m(\nb_{X}^s)\rho} .
\end{align}

Next, we further treat the operator $\tilde{\mathcal{K}}^m(\nb_X^s)$ in Eq.~\eqref{def_tilde_mathcal_D_nb_X_s}:
\begin{align}
\tilde{\mathcal{K}}^m(\nb_X^s)&=\sum_{m_1+\cdots+m_s=m}\frac{m!}{m_1!\cdots m_s!}[\mathcal{D}^{m_1}(\nb_X)+m(2\gamma\bar J)^{m_1}]\cdots[\mathcal{D}^{m_s}(\nb_X)+m(2\gamma\bar J)^{m_s}] \notag \\
&=\sum_{s_1=0}^s\sum_{m_1+\cdots+m_s=m}\frac{m!}{m_1!\cdots m_s!} \sum_{\{i_1, \ldots,i_{s_1}\}\subset[s]}  \mathcal{D}^{m_{i_1}}(\nb_X)\cdots\mathcal{D}^{m_{i_{s_1}}}(\nb_X)m^{s-s_1} (2\gamma\bar{J})^{m_{j_1} + m_{j_2} + \cdots +m_{j_{s-s_1}}}  ,
\end{align}
where we denote $[s]\setminus \{i_1,i_2, \ldots,i_{s_1}\} = \{j_1,j_2,\ldots, j_{s-s_1}\}$.
Then, we use the following relations  
\begin{align}
 \tau^m m^{s-s_1} \le \frac{d^{s-s_1}}{d\tau^{s-s_1}} \br{\tau^{m+s-s_1}} , \quad  \sum_{m=0}^\infty \sum_{m_1+\cdots+m_s=m}=\sum_{m_1=0}^\infty \sum_{m_2=0}^\infty \cdots \sum_{m_s=0}^\infty ,
\end{align}
which yields the inequality of
\begin{align}
\label{mth_order_derivative_mathcal_D_integral}
&\sum_{m=0}^\infty \frac{(2\tau)^m}{m!} \tilde{\mathcal{K}}^m(\nb_{X}^s) \notag\\
&\preceq \sum_{s_1=0}^s \sum_{\{i_1,i_2, \ldots,i_{s_1}\}\subset[s]}  \frac{d^{s-s_1}}{d\tau^{s-s_1}} \tau^{s-s_1}  
 \sum_{m_1=0}^\infty \frac{(2\tau)^{m_1}}{m_1!} \cdots \sum_{m_s=0}^\infty \frac{(2\tau)^{m_s}}{m_s!} \mathcal{D}^{m_{i_1}}(\nb_X)\cdots\mathcal{D}^{m_{i_{s_1}}}(\nb_X)(2\gamma\bar{J})^{m_{j_1} + \cdots +m_{j_{s-s_1}}} \notag \\
&= \sum_{s_1=0}^s \sum_{\{i_1, \ldots,i_{s_1}\}\subset[s]}  \frac{d^{s-s_1}}{d\tau^{s-s_1}} \tau^{s-s_1}  \br{ e^{4\gamma\bar{J} \tau}}^{s-s_1} 
\br{\sum_{m=0}^\infty \frac{(2\tau)^m}{m!} \mathcal{D}^{m}(\nb_X) }^{s_1} \notag \\
&=  \sum_{s_1=0}^s \sum_{\{i_1, \ldots,i_{s_1}\}\subset[s]}  \frac{d^{s'}}{d\tau^{s'}} \br{\tau^{s'} e^{4\gamma\bar{J} s' \tau} 
 \brr{\nb_X + \tilde{\mathcal{D}}_{\tau}  (\nb_{X}) }^{s_1} },
\end{align}
where we set $s'=s-s_1$ and define $\tilde{\mathcal{D}}_{\tau}(\nb_{X}) $ as 
\begin{align}
\label{def_tilde_mathcal_D_tau_nb_X}
\tilde{\mathcal{D}}_{\tau} (\nb_{X})  := \sum_{m=1}^\infty \frac{(2\tau)^m}{m!} \mathcal{D}^{m}(\nb_X). 
\end{align}
To estimate the upper bound of the derivative in the RHS of~\eqref{mth_order_derivative_mathcal_D_integral}, we separately consider 
\begin{align}
\frac{d^{p}}{d\tau^{p}} \br{\tau^{s'} e^{4\gamma\bar{J} s' \tau} },  \quad \frac{d^{p}}{d\tau^{p}} \brr{\nb_X + \tilde{\mathcal{D}}_{\tau}  (\nb_{X}) }^{s_1}  \for p\le s'.
\end{align}

For the first, we obtain 
\begin{align}
\label{derivative_first_term_dt_dnb}
\frac{d^{p}}{d\tau^{p}} \br{\tau^{s'} e^{4\gamma\bar{J} s' \tau} } 
&=\sum_{p_1=0}^p \binom{p}{p_1} \frac{d^{p_1}}{d\tau^{p_1}} \br{\tau^{s'}} \frac{d^{p-p_1}}{d\tau^{p-p_1}} e^{4\gamma\bar{J} s' \tau} \notag \\
&= \sum_{p_1=0}^p \binom{p}{p_1} s' (s'-1) \cdots (s'-p_1+1)  \tau^{s'-p_1} \br{4\gamma\bar{J} s'}^{p-p_1} e^{4\gamma\bar{J} s' \tau}  \notag \\
&\le  \tau^{s'-p} \sum_{p_1=0}^p \binom{p}{p_1} (s')^{p_1}  \tau^{p-p_1} \br{4\gamma\bar{J} s'}^{p-p_1} e^{4\gamma\bar{J} s' \tau}  \le (s')^{p} \tau^{s'-p} e^{4\gamma\bar{J} s' \tau}  (1+4\gamma\bar{J}\tau )^p .
\end{align}
Second, we consider the derivatives of $\brr{\nb_X + \tilde{\mathcal{D}}_{\tau} (\nb_X)}^{s_1}$. 
For convenience, we here define $\tilde{\mathcal{D}}_\tau^{(p)} (\nb_{X})$ as 
\begin{align}
\tilde{\mathcal{D}}_\tau^{(p)} (\nb_{X})\coloneqq  \frac{d^p}{d\tau^p}  \tilde{\mathcal{D}}_{\tau} (\nb_{X}).
\end{align} 
We then prove the following lemma:
\begin{lemma} \label{lem:der_p_upp}
For an arbitrary integer $p\ge 0$, under the condition of $ \tau\le 1/(4\gamma \bar{J})$, we have 
\begin{align}
\tilde{\mathcal{D}}_\tau^{(p)} (\nb_{X}) \preceq p! (4\gamma \bar{J})^{p}  \bar{\mathcal{D}}_{\tau} (\nb_{X}),
\end{align} 
where $\bar{\mathcal{D}}_{\tau} (\nb_{X}) $ is defined as  
\begin{align}
\label{main_lem:der_p_upp}
\bar{\mathcal{D}}_{\tau} (\nb_{X}) \coloneqq  40e^{4\gamma \bar{J}\tau} \sum_{i \in \partial X}\sum_{j\in\Lambda}  e^{-d_{i,j}} \nb_j .
\end{align} 
Note that by considering the case of $p=0$, we obtain the upper bound of 
\begin{align}
\tilde{\mathcal{D}}_{\tau}(\nb_{X}) =  \tilde{\mathcal{D}}_\tau^{(0)} (\nb_{X}) \le \bar{\mathcal{D}}_{\tau} (\nb_{X}) . 
\end{align} 
\end{lemma}


\subsubsection{Proof of Lemma~\ref{lem:der_p_upp}}

By using the definitions of $\mathcal{D}$ in Eq.~\eqref{def_D_mathcal_m}, we have 
\begin{align}
\tilde{\mathcal{D}}_{\tau} (\nb_{X}) 
:=  \sum_{m=1}^\infty \frac{(2\tau)^{m}}{m!} \mathcal{D}^{m}(\nb_X) 
&= \sum_{i \in \partial X}  \sum_{m=1}^\infty \frac{(2\tau)^{m}}{m!}  \mathcal{M}^{m}(\nb_i ) ,
\end{align}
and hence,
\begin{align}
\label{der_express_1}
\tilde{\mathcal{D}}_\tau^{(p)} (\nb_{X})=\frac{d^p}{d\tau^p}\tilde{\mathcal{D}}_{\tau} (\nb_{X}) 
&\preceq 2^p\sum_{i \in \partial X } \sum_{m=p}^\infty \frac{(2\tau)^{m-p}}{(m-p)!}  \mathcal{M}^{m}(\nb_i ).
\end{align}

Now, let us define $(\mathcal{M}^m)_{i,j} \in \mathbb{R}^+$ ($q\in \mathbb{N}$) by
\begin{align}
\label{qth_power_of_M}
\mathcal{M}^m(\nb_i) = \sum_{j\in \Lambda} (\mathcal{M}^m)_{i,j} \nb_j .
\end{align}
That is, $(\mathcal{M}^m)_{i,j}$ is the coefficient for $\nb_j$ in expanding $\mathcal{M}^m(\nb_i)$.  
Then, the definition~\eqref{def_mathcal_M_mat} implies that $ (\mathcal{M}^m)_{i,j} =0$ as long as $m< \dist_{i,j}$. On the other hand, for $m\ge \dist_{i,j}$, we have 
\begin{align}
\label{qth_power_of_M_upp}
(\mathcal{M}^m)_{i,j}\le (2\gamma \bar{J})^m \for m\ge \dist_{i,j} ,
\end{align}
which immediately results from Lemma~\ref{lem:upp_max_alpha_i_D^m}. 
By applying~\eqref{qth_power_of_M} and \eqref{qth_power_of_M_upp} to Eq.~\eqref{der_express_1}, we have 
\begin{align}
\label{der_express_2}
\tilde{\mathcal{D}}_\tau^{(p)} (\nb_{X})
&\preceq  \sum_{i \in \partial X}\sum_{j\in\Lambda} (\bar{\mathcal{M}})_{i,j} \nb_j,
\end{align}
where the matrix $\bar{\mathcal{M}}$ is defined as 
\begin{align}
\br{\bar{\mathcal{M}}}_{i,j} 
&:=  2^p \sum_{m=p}^\infty \frac{(2\tau)^{m-p}}{(m-p)!}  \br{\mathcal{M}^{m}}_{i,j} 
=2^p \sum_{m=0}^\infty \frac{(2\tau)^{m}}{m!}  \br{\mathcal{M}^{m+p}}_{i,j} \notag \\
&\le  2^p\sum_{m\ge \max\{0,\dist_{i,j}-p\}}^\infty \frac{(2\tau)^{m}}{m!}  (2\gamma \bar{J})^{m+p} \notag \\
&\le 
\begin{cases}
(4\gamma \bar{J})^{d_{i,j}} e^{4\gamma \bar{J}\tau} \tau^{\dist_{i,j}-p} \frac{1}{(\dist_{i,j}-p)!} &\for p\le \dist_{i,j}, \\ 
(4\gamma \bar{J})^{p} e^{4\gamma \bar{J}\tau} &\for p> \dist_{i,j},
\end{cases}
\label{upper_bound_/for_bar_math_M}
\end{align}
where in the last inequality, we have used 
\begin{align}
\sum_{m\ge m_0} \frac{z^m}{m!} = \sum_{m=0}^\infty \frac{z^{m+m_0}}{(m+m_0)!} \le 
\sum_{m=0}^\infty \frac{z^{m+m_0}}{m! \cdot m_0!}= e^z \frac{z^{m_0}}{m_0!}.
\end{align}

Finally, by using the inequality
\begin{align}
\frac{1}{(\dist_{i,j}-p)!} = \frac{p!}{\dist_{i,j}!} \binom{\dist_{i,j}}{p} \le 2^{\dist_{i,j}}\frac{p!}{\dist_{i,j}!}  ,
\end{align}
and the condition
\begin{align}
\tau\le \frac{1}{4\gamma \bar{J}},
\end{align}
we have for $p\le \dist_{i,j}$
\begin{align}
(4\gamma \bar{J})^{\dist_{i,j}} e^{4\gamma \bar{J}\tau} \tau^{\dist_{i,j}-p} \frac{1}{(\dist_{i,j}-p)!} 
\le (4\gamma \bar{J})^{p} e^{4\gamma \bar{J}\tau}  2^{\dist_{i,j}}\frac{p!}{\dist_{i,j}!}  =p! (4\gamma \bar{J})^{p} \cdot e^{4\gamma \bar{J}\tau} \frac{2^{\dist_{i,j}}}{\dist_{i,j}!}  .
\label{upper_bound_/for_bar_math_M_case1}
\end{align}
In the case of $p>d_{i,j}$ in~\eqref{upper_bound_/for_bar_math_M}, the following inequality trivially holds:
\begin{equation}
(4\gamma \bar{J})^{p} e^{4\gamma \bar{J}\tau}\le p! (4\gamma \bar{J})^{p} \cdot e^{4\gamma \bar{J}\tau} \frac{2^{\dist_{i,j}}}{\dist_{i,j}!} .
\label{upper_bound_/for_bar_math_M_case2}
\end{equation}
Thus, by applying the inequalities~\eqref{upper_bound_/for_bar_math_M_case1} and \eqref{upper_bound_/for_bar_math_M_case2} to \eqref{upper_bound_/for_bar_math_M}, we arrive at the inequality of 
\begin{align}
\br{\bar{\mathcal{M}}}_{i,j} \le p! (4\gamma \bar{J})^{p} \cdot e^{4\gamma \bar{J}\tau} \frac{2^{\dist_{i,j}}}{\dist_{i,j}!}.
\end{align}
Applying this inequality to Eq.~\eqref{der_express_2}, we have
\begin{align}
\tilde{\mathcal{D}}_\tau^{(p)} (\nb_{X})
&\preceq  p! (4\gamma \bar{J})^{p}\sum_{i \in \partial X }\sum_{j\in\Lambda} e^{4\gamma \bar{J}\tau} \frac{2^{\dist_{i,j}}}{\dist_{i,j}!} \nb_j,\notag\\
&\preceq p! (4\gamma \bar{J})^{p} \cdot e^{4\gamma \bar{J}\tau} \sum_{i \in \partial X}\sum_{j\in\Lambda}  \frac{2^{\dist_{i,j}}}{\dist_{i,j}!} \nb_j,\notag\\
&\preceq p! (4\gamma \bar{J})^{p} \cdot 40e^{4\gamma \bar{J}\tau}\sum_{i \in \partial X}\sum_{j\in\Lambda} e^{-\dist_{i,j}} \nb_j=: p! (4\gamma \bar{J})^{p}\bar{\mathcal{D}}_{\tau} (\nb_{X})   ,
\end{align}
where in the third inequality, we have applied the following inequality:
\begin{align}
\frac{2^{\dist_{i,j}}}{\dist_{i,j}!}  \le 40e^{-d_{i,j}} .
\end{align} 
This completes the proof. $\square$

{~}

\noindent\hrulefill{\bf [ End of Proof of Lemma~\ref{lem:der_p_upp}] }

{~}

\noindent

By using Lemma~\ref{lem:der_p_upp}, we obtain the upper bound of 
\begin{align}
\nb_X+ \tilde{\mathcal{D}}_\tau^{(p)} (\nb_{X}) \le \nb_X+ p! (4\gamma \bar{J})^p \bar{\mathcal{D}}_{\tau} (\nb_{X}) 
\le p! (4\gamma \bar{J}+1)^p \brr{\nb_X + \bar{\mathcal{D}}_{\tau} (\nb_{X}) } .
\end{align} 
By using the above inequality, we upper-bound the derivatives of $\brr{\nb_X + \tilde{\mathcal{D}}_{\tau} (\nb_X)}^{s_1}$ as follows:
\begin{align}
&\frac{d^{p}}{d\tau^{p}} \brr{\nb_X + \tilde{\mathcal{D}}_{\tau}  (\nb_{X}) }^{s_1} 
\preceq \sum_{\substack{p_1,p_2,\ldots,p_{s_1}  \\  p_1+p_2+\cdots +p_{s_1}=p}} \frac{p!}{p_1!p_2!\cdots p_{s_1}!}
 \prod_{j=1}^{s_1} \brr{\nb_X+ \tilde{\mathcal{D}}_\tau^{(p_j)} (\nb_{X})}\notag \\
&\preceq  \
\sum_{\substack{p_1,p_2,\ldots,p_{s_1} \\  p_1+p_2+\cdots +p_{s_1}=p}} \frac{p!}{p_1!p_2!\cdots p_{s_1}!}
(4\gamma \bar{J}+1)^{p_1+p_2+\cdots +p_{s_1}}  p_1!p_2!\cdots p_{s_1}! 
\brr{\nb_X+ \bar{\mathcal{D}}_{\tau} (\nb_{X})}^{s_1}  \notag \\
 &= (4\gamma \bar{J}+1)^{p} p! \sum_{\substack{p_1,p_2,\ldots,p_{s_1} \\  p_1+p_2+\cdots +p_{s_1}=p}}
 \brr{\nb_X +\bar{\mathcal{D}}_{\tau} (\nb_{X})}^{s_1} \notag \\
 &= (4\gamma \bar{J}+1)^{p} p!   \multiset{s_1}{p}  \brr{\nb_X+\bar{\mathcal{D}}_{\tau} (\nb_{X})}^{s_1}  ,
 \label{derivative_second_term_dt_dnb}
\end{align}
where $\multiset{s_1}{p}$ is the ($p$)-multicombination from a set of $s_1$ elements, which satisfies 
\begin{align}
\multiset{s_1}{p} = \binom{s_1+p-1}{p}.
\end{align}

By combining the inequalities \eqref{derivative_first_term_dt_dnb} and \eqref{derivative_second_term_dt_dnb}, we obtain
\begin{align}
\label{mth_order_derivative_mathcal_D_integral_sd}
& \frac{d^{s'}}{d\tau^{s'}} \br{ \tau^{s'} e^{4\gamma\bar{J} s' \tau} 
 \brr{\nb_X + \tilde{\mathcal{D}}_{\tau}  (\nb_{X}) }^{s_1} } \notag \\
 =& \sum_{p=0}^{s'} \binom{s'}{p} \frac{d^{p}}{d\tau^{p}} \br{ \tau^{s'} e^{4\gamma\bar{J} s' \tau}  }
 \frac{d^{s'-p}}{d\tau^{s'-p}}   \brr{\nb_X + \tilde{\mathcal{D}}_{\tau}  (\nb_{X}) }^{s_1}  \notag \\
\preceq &  \sum_{p=0}^{s'} \binom{s'}{p}  (s')^{p} \tau^{s'-p} e^{4\gamma\bar{J} s' \tau}  (1+4\gamma\bar{J}\tau )^p \cdot 
 (4\gamma \bar{J}+1)^{s'-p} (s'-p)!  \multiset{s_1}{s'-p}    \brr{\nb_X+\bar{\mathcal{D}}_{\tau} (\nb_{X})}^{s_1} \notag \\
\preceq &(s')^{s'}e^{4\gamma\bar{J} s' \tau}  \multiset{s_1}{s'}  \brr{\nb_X+\bar{\mathcal{D}}_{\tau} (\nb_{X})}^{s_1} \sum_{p=0}^{s'} \binom{s'}{p} \br{1+4\gamma\bar{J}\tau }^{p}  [\tau(4\gamma \bar{J}+1)]^{s'-p} \notag \\
=&\multiset{s_1}{s'} (s')^{s'} \brr{e^{4\gamma\bar{J} \tau} \br{1+8\gamma\bar{J}\tau +\tau}}^{s'} \brr{\nb_X+\bar{\mathcal{D}}_{\tau} (\nb_{X})}^{s_1} \notag\\
\preceq &  \brr{e^{4\gamma\bar{J} \tau+1} \br{1+8\gamma\bar{J}\tau +\tau} s }^{s'} \brr{\nb_X+\bar{\mathcal{D}}_{\tau} (\nb_{X})}^{s_1}  ,
\end{align}
where in the second inequality, we use the fact that the multiset $\multiset{s_1}{p}$ monotonically increases with $p$, and in the last inequality, we use $s'=s-s_1$ and 
\begin{align}
\multiset{s_1}{s'} (s')^{s'} \le e^{s'}  \binom{s_1+s'-1}{s'} s'! =e^{s'} \binom{s-1}{s'} s'! =e^{s'} \frac{(s-1)!}{(s-1-s')!}  \le (e s)^{s'}  . 
\end{align}  
Note that $s'! \ge (s'/e)^{s'}$. 
By applying the inequality~\eqref{mth_order_derivative_mathcal_D_integral_sd} to~\eqref{mth_order_derivative_mathcal_D_integral}, we finally obtain 
\begin{align}
\label{mth_order_derivative_mathcal_D_integral_fin}
\sum_{m=0}^\infty \frac{(2\tau)^m}{m!} \tilde{\mathcal{K}}^m(\nb_{X}^s)
&\preceq  \sum_{s_1=0}^s \sum_{\{i_1, \ldots,i_{s_1}\}\subset[s]} \brr{e^{4\gamma\bar{J} \tau+1} \br{1+8\gamma\bar{J}\tau +\tau} s }^{s'} \brr{\nb_X+\bar{\mathcal{D}}_{\tau} (\nb_{X})}^{s_1}     \notag\\ 
&= \sum_{s_1=0}^s \binom{s}{s_1} \brr{e^{4\gamma\bar{J} \tau+1} \br{1+8\gamma\bar{J}\tau +\tau} s }^{s'} \brr{\nb_X+\bar{\mathcal{D}}_{\tau} (\nb_{X})}^{s_1}      \notag \\
&= \brr{\nb_X+40e^{4\gamma \bar{J}\tau} \sum_{i \in \partial X}\sum_{j\in\Lambda}  e^{-d_{i,j}} \nb_j  + e^{4\gamma\bar{J} \tau+1} \br{1+8\gamma\bar{J}\tau +\tau} s }^{s'} ,
\end{align}
where in the last equation, we use the explicit form~\eqref{main_lem:der_p_upp} of $\bar{\mathcal{D}}_{\tau} (\nb_{X})$. 
Hence, by choosing $c_{\tau,1}$ and $c_{\tau,2}$ as in Eq.~\eqref{choice of c_t_1_ct_2}, we obtain the main inequality~\eqref{main_ineN_Schuch_boson_extend}.
This completes the proof. $\square$

\subsection{More general statement}

In some cases, we are interested in the moment function in the form of $|\nb_X - N|^s$ instead of $\nb_X^s$, where $N$ is an arbitrary integer.  
In particular, we indeed utilize it in deriving Proposition~\ref{prop:lower_probability} below, which plays a critical role in proving the optimal Lieb-Robinson bound in one-dimensional systems (see Sec.~\ref{sec:Proof of Lemma_lemma:Probability_upper_bound} for the proof of Lemma~\ref{lemma:Probability_upper_bound}).  
We here consider 
\begin{align}
\label{moment_general_X_1_X_s_more_general}
&M'_{X_1,X_2,\ldots,X_s}(\tau) := \tr\brr{ \nb'_{X_1}\nb'_{X_2} \cdots \nb'_{X_s} \rho(\tau) }  , \notag \\
&\nb'_{X_j} := |\nb_{X_j} - N_j|  \for j=1,2,\ldots, s ,
\end{align}
where $\{N_j\}_{j=1}^s$ are arbitrary positive numbers. 
For $M'_{X_1,X_2,\ldots,X_s}(\tau)$, we can prove the same statement as Subtheorem~\ref{Schuch_boson_extend}.

\begin{prop} \label{Schuch_boson_extend_more_general}
Let $\rho$ be an arbitrary quantum state, and $\{X_j\}_{j=1}^s$ be arbitrary subsets.  
Then, as long as $\tau\le 1/(4\gamma \bar{J})$, we obtain the following upper bound for $M'_{X_1,X_2,\ldots,X_s}(\tau)$: 
\begin{align}
\label{main_ineN_Schuch_boson_extend_general}
M'_{X_1,X_2,\ldots,X_s}(\tau) \le  \tr \brr{ \prod_{j=1}^s \br{ \nb'_{X_j} + c_{\tau,1} \hat{\mathcal{D}}_{X_j} + c_{\tau,2} s}     \rho } ,
\end{align}
where $\hat{\mathcal{D}}_X$, $c_{\tau,1}$ and $c_{\tau,2}$ have been defined in Eqs.~\eqref{Eq:def_tilde_D_X} and \eqref{choice of c_t_1_ct_2}, respectively.  
\end{prop}

\subsubsection{Proof of Proposition~\ref{Schuch_boson_extend_more_general}} 

For the proof, we utilize the same technique. 
Here, we slightly extend the definition of the super-operator $\mathcal{D}$ in Eq.~\eqref{def_D_mathcal_m} as follows:
\begin{align}
&\mathcal{D}^0(\nb'_{X}) \coloneqq \nb'_X,\quad  \mathcal{D}^1(\nb'_{X}) \coloneqq \sum_{i\in \partial X} \mathcal{M}(\nb_i) ,  \notag \\
&\mathcal{D}^{m+1}(\nb_{X})\coloneqq  \mathcal{M}(\mathcal{D}^{m}(\nb_X))  \for \forall m\ge 1
\label{def_D_mathcal_m_change}
\end{align}
Note that we have 
\begin{align}
\label{same_for_nb'_X}
\mathcal{D}^m(\nb'_{X}) \coloneqq \mathcal{D}^m(\nb_{X}) \for m \ge 1 . 
\end{align}
Then, if the same statement as Proposition~\ref{prop:simplify_high_der} is proved, 
we can derive the main inequality~\eqref{main_ineN_Schuch_boson_extend_general} by employing the same analyses as in the proof of Subtheorem~\ref{Schuch_boson_extend}. 
Note that we only replace the inequality~\eqref{mth_order_derivative_mathcal_D_integral} by
\begin{align}
\label{mth_order_derivative_mathcal_D_integral_general}
&\sum_{m=0}^\infty \frac{(2\tau)^m}{m!} \tilde{\mathcal{K}}^m(\nb_{X}^s) 
\preceq  \sum_{s_1=0}^s \sum_{\{i_1, \ldots,i_{s_1}\}\subset[s]}  \frac{d^{s'}}{d\tau^{s'}} \br{\tau^{s'} e^{4\gamma\bar{J} s' \tau} 
 \brr{\nb'_X + \tilde{\mathcal{D}}_{\tau}  (\nb_{X}) }^{s_1} },
\end{align}
where we can adopt the same definition~\eqref{def_tilde_mathcal_D_tau_nb_X} for $\tilde{\mathcal{D}}_{\tau}  (\nb_{X})$ because of Eq.~\eqref{same_for_nb'_X}.

For the proof of Proposition~\ref{prop:simplify_high_der} in the present case, we start from the equation~\eqref{sth_moment_der} as follows:
\begin{align}
\label{sth_moment_der_general}
\frac{d}{d\tau} M'^{(s,m)}_X(\tau) &= i\sum_{m_1+\cdots+m_s=m}\frac{m!}{m_1!\cdots m_s!}\sum_{k=1}^s\tr \left (\mathcal{D}_{<k}^{(m)} \left[H,\mathcal{D}^{m_k}(\nb'_{X})\right] \mathcal{D}_{>k}^{(m)} \rho(\tau)  \right)
\end{align}
with 
\begin{align}
\mathcal{D}_{<k}^{(m)}\coloneqq\prod_{l<k} [\mathcal{D}^{m_l}(\nb'_{X})+m(2\gamma\bar J)^{m_l}]  , \quad 
\mathcal{D}_{>k}^{(m)}\coloneqq\prod_{l>k} [\mathcal{D}^{m_l}(\nb'_{X})+m(2\gamma\bar J)^{m_l}]. 
\label{def_mathcal_D_<>k_m_general}
\end{align}
The equation~\eqref{Ham_commute_nb_i_s_2} gives 
\begin{align}
\label{Ham_commute_nb_i_s_general}
[H,\mathcal{D}^{m_k}(\nb'_{X}) ] = [H_0,\mathcal{D}^{m_k}(\nb'_{X})],
\end{align}

For $m_k= 0$, because of 
\begin{align}
[\nb_X' , H_{0,X} ]= 0, \And  [\nb_X' , H_{0,X^\co} ]= 0, 
\end{align}
we have 
\begin{align}\label{eq:HD_com_tmp1_2}
[H_0,\mathcal{D}^{0}(\nb'_{X})] = [\partial h_X,\nb'_{X}]
&= \sum_{i \in \partial X}  \sum_{j \in X^\co:\dist_{i,j}=1}J_{i,j} [ b_i b_j^\dagger + b_j b_i^\dagger , \nb'_X]  \notag \\
&=\sum_{i \in \partial X}  \sum_{j \in X^\co:\dist_{i,j}=1}J_{i,j} [ b'_{X,i} b_j^\dagger -  b_j (b'_{X,i})^\dagger ] ,
\end{align} 
where we use the notation of~\eqref{def:Ham_surface} and define 
\begin{align}
b'_{X,i} := [b_i, \nb'_X]  = [b_i , |\nb_X - N| ] .
\end{align}
Clearly, for $N=0$ (i.e., $|\nb_X - N| = |\nb_X| = \nb_X$), we have $b'_{X,i}=[b_i, \nb_i] = b_i$. 
By using the notation, we obtain a similar inequality to~\eqref{derivative_M_S_i_t_mk=0}:
\begin{align}
&\left| \tr \left (\mathcal{D}_{<k}^{(m)} \left[H,\mathcal{D}^{m_k}(\nb_{X})\right] \mathcal{D}_{>k}^{(m)} \rho(\tau)  \right)  \right|  \notag \\
&\le  \sum_{i \in \partial X} \sum_{j \in X^\co :\dist_{i,j}=1}  \bar{J} 
\br{ \left| \tr\br{\mathcal{D}_{<k}^{(m)}  b'_{X,i} b_{j}^\dagger  \mathcal{D}_{>k}^{(m)} \rho(\tau) } \right| + \left| \tr\br{\mathcal{D}_{<k}^{(m)} (b'_{X,i})^\dagger   b_{j} \mathcal{D}_{>k}^{(m)} \rho(\tau) } \right|}.
\label{derivative_M_S_i_t_mk=0_general}
\end{align}
Then, the primary problem is to derive upper bounds for 
\begin{align}
\label{upper_bound_aim_mathcal_D_b'_x}
\left| \tr\br{\mathcal{D}_{<k}^{(m)} b'_{X,i} b_{j}^\dagger    \mathcal{D}_{>k}^{(m)} \rho(\tau) } \right| \AND \left| \tr\br{\mathcal{D}_{<k}^{(m)}   (b'_{X,i})^\dagger   b_{j}  \mathcal{D}_{>k}^{(m)} \rho(\tau) } \right|.
\end{align}
We note that the operator $b'_{X,i}$ is a non-local operator supported on the whole region of $X$. 
However, by taking the upper bound, the non-locality disappears as in \eqref{b_i_b_j_prop_expec_apply_general}, where $(b'_{X,i})^\dagger   b_{j}$ is simply upper-bounded by using $\nb_i+\nb_j$.  

We here extend the proposition~\ref{prop:expectation_b_j_b_i_poly_n_i} as follows:
\begin{lemma}\label{lemma:expectation_b_j_b_i_poly_n_i}
Let us assume $i \in X$ and $j \in X^\co$.  
Then, for arbitrary $s_0$, $s_1$, $s_1'$, $s_2$ and $s_2'$, we obtain the upper bound as
\begin{align}
&\left| \tr \brr{ f(\hat{\bold{n}}) (\nb_X')^{s_0} \nb_i^{s_1}\nb_j^{s_2}  b'_{X,i} b_{j}^\dagger (\nb_X')^{s_0'}  \nb_i^{s'_1}\nb_j^{s'_2}    \rho(\tau) } \right| \notag \\
 &\le \frac{1}{2} \brr{ f(\hat{\bold{n}})  (\nb_X'+1)^{s_0+s_0'} \brr{\nb_i^{s_1+s_1'}(\nb_j+1)^{s_2+s_2'} + (\nb_i+1)^{s_1+s_1'}\nb_j^{s_2+s_2'} } (\nb_i+\nb_j) \rho(\tau) },
 \label{ienq_b_i_b_j_n_i+_n_j_ineN_general}
\end{align}
where $ f(\hat{\bold{n}})$ is an arbitrary non-negative function with respect to $\{\hat{n}_s\}_{s\neq i,j}$ or equivalently satisfies $[f(\hat{\bold{n}}), b_i] = [f(\hat{\bold{n}}), b_j] =0$. 
We have the same inequality if we replace $b'_{X,i} b_{j}^\dagger$ with $(b'_{X,i})^\dagger b_{j}$.
\end{lemma}

\subsubsection{Proof of Lemma~\ref{lemma:expectation_b_j_b_i_poly_n_i}} \label{sec:Proof of lemma:expectation_b_j_b_i_poly_n_i}

We use the notation of $\bold{N} = \{N_i, \bold{N}_{i}\}$, where $\bold{N}_{i}$ includes the boson number in $\Lambda\setminus \{i\}$. 
In the same way, we define $\bold{N} = \{N_i, N_j, \bold{N}_{i,j}\}$, where $\bold{N}_{i,j}$ includes the boson number in $\Lambda\setminus \{i,j\}$.  

For the proof, we first consider the explicit form of the operator $b'_{X,i}$:
\begin{align}
&b'_{X,i} := [b_i, \nb'_X]   \notag \\
&=\sum_{N_{X_i}=0}^\infty \sum_{N_i=0}^\infty |N_i + N_{X_i} -N| \cdot [b_i , \ket{N_i,N_{X_i}}\bra{N_i,N_{X_i}}] \notag \\
& =\sum_{N_{X_i}=0}^\infty \sum_{N_i=0}^\infty |N_i + N_{X_i} -N|  \br{ \sqrt{N_i}  \ket{N_i-1,N_{X_i}}\bra{N_i,N_{X_i}} - \sqrt{N_i+1} \ket{N_i,N_{X_i}}\bra{N_i+1,N_{X_i}}   }  \notag \\
&=\sum_{N_{X_i}=0}^\infty \sum_{N_i=0}^\infty  \br{  |N_i + N_{X_i} -N|   -  |N_i -1 + N_{X_i} -N|   }  \sqrt{N_i}  \ket{N_i-1,N_{X_i}}\bra{N_i,N_{X_i}} \notag \\
&=:\sum_{N_{X_i}=0}^\infty \sum_{N_i=0}^\infty  g(N_i,N_{X_i},N) \sqrt{N_i}  \ket{N_i-1,N_{X_i}}\bra{N_i,N_{X_i}} ,
\end{align}
where we define $g(N_i,N_{X_i},N):=  |N_i + N_{X_i} -N|   -  |N_i -1 + N_{X_i} -N| $. 
From the definition, we trivially obtain 
\begin{align}
\label{g_is_smaller_than_1}
| g(N_i,N_{X_i},N)| \le 1 .
\end{align}
We therefore obtain 
\begin{align}
\label{b_X_i_b_X_i_dagger_prop}
&b_{X,i}\ket{\bold{N}}  =g(N_i,N_{X_i},N) \sqrt{N_i}\ket{N_i-1, N_j ,\bold{N}_{i,j}} , \notag \\
& (b_{X,i})^\dagger\ket{\bold{N}}  =g(N_i+1,N_{X_i},N) \sqrt{N_i+1}\ket{N_i+1, N_j ,\bold{N}_{i,j}} .
\end{align}

Because of $i \in X$ and $j \in X^\co$, we have 
$b'_{X,i} b_{j}^\dagger  \ket{\bold{N}} \propto  \ket{N_i-1, N_j+1 ,\bold{N}_{i,j}}$
and
$\nb_X'  b'_{X,i} b_{j}^\dagger  \ket{\bold{N}}= |N_X -1 - N|^{s_0}  b'_{X,i} b_{j}^\dagger  \ket{\bold{N}} $.
Hence, by using Eq.~\eqref{b_X_i_b_X_i_dagger_prop}, we transform 
\begin{align}
\label{start_eq/_prop:expectation_b_j_b_i_poly_n_i___2}
&\tr \brr{  f(\hat{\bold{n}}) (\nb_X')^{s_0} \nb_i^{s_1}\nb_j^{s_2}  b'_{X,i} b_{j}^\dagger \nb_i^{s'_1}\nb_j^{s'_2}    \rho(\tau) }  = \tr \brr{  f(\hat{\bold{n}}) (\nb_X')^{s_0} \nb_i^{s_1}\nb_j^{s_2}  b'_{X,i} b_{j}^\dagger \nb_i^{s'_1}\nb_j^{s'_2}     \sum_{\bold{N}} P_{\bold{N}}  \rho(\tau) }  \notag \\
&=\sum_{\bold{N}}f(\bold{N}_{i,j}) |N_X - N|^{s_0'} N_i^{s_1'}N_j^{s_2'}\sqrt{N_i}\sqrt{N_j+1} |N_X -1 - N|^{s_0}(N_i-1)^{s_1}(N_j+1)^{s_2}  \notag \\
&\quad  \times g(N_i,N_{X_i},N) \bra{N_i,N_j,\bold{N}_{i,j}}\rho(\tau)\ket{N_i-1,N_j+1,\bold{N}_{i,j}}  \notag \\
&\le \sum_{\bold{N}}f(\bold{N}_{i,j})  |N_X- N|^{s_0'} N_i^{s_1'}N_j^{s_2'}\sqrt{N_i}\sqrt{N_j+1}|N_X -1 - N|^{s_0} (N_i-1)^{s_1}(N_j+1)^{s_2}   \notag \\
&\quad \times \frac{1}{2}  \br{\bra{N_i,N_j,\bold{N}_{i,j}}\rho(\tau)\ket{N_i,N_j,\bold{N}_{i,j}}  + \bra{N_i-1,N_j+1,\bold{N}_{i,j}}\rho(\tau)\ket{N_i-1,N_j+1,\bold{N}_{i,j}}  }    ,
\end{align}
where we use the inequalities~\eqref{Cauchy-Schwartz inequality_apply_ninjnvec} and \eqref{g_is_smaller_than_1} in the last inequality.
Then, by using $\sqrt{N_i(N_j+1)}\le N_i+N_j$ and $\sqrt{(N_i+1)N_j}\le N_i+N_j$ for $\forall N_i,N_j\in\mathbb{N}$, we can reduce 
\begin{align}
\label{start_eq/_prop:expectation_b_j_b_i_poly_n_i_1}
&\frac{1}{2}  \sum_{\bold{N}}f(\bold{N}_{i,j}) |N_X - N|^{s_0'}  N_i^{s_1'}N_j^{s_2'}\sqrt{N_i}\sqrt{N_j+1} |N_X -1- N|^{s_0} (N_i-1)^{s_1}(N_j+1)^{s_2} \bra{N_i,N_j,\bold{N}_{i,j}}\rho(\tau)\ket{N_i,N_j,\bold{N}_{i,j}}       \notag \\
&\le \frac{1}{2}  \tr \brr{  f(\hat{\bold{n}}) ( |\nb_X - N| +1)^{s_0+s_0'}  \nb_i^{s_1'}\nb_j^{s_2'}(\nb_i + \nb_j)(\nb_i-1)^{s_1}(\nb_j+1)^{s_2} \rho(\tau) }  \notag \\
&\le \frac{1}{2}  \tr \brr{  f(\hat{\bold{n}}) ( \nb'_X +1)^{s_0+s_0'}  \nb_i^{s_1+s_1'}(\nb_j+1)^{s_2+s_2'}(\nb_i + \nb_j) \rho(\tau) } .
\end{align}
In the same way, we obtain 
\begin{align}
\label{start_eq/_prop:expectation_b_j_b_i_poly_n_i_2}
&\frac{1}{2}  \sum_{\bold{N}}f(\bold{N}_{i,j}) |N_X - N|^{s_0'}  N_i^{s_1'}N_j^{s_2'}\sqrt{N_i}\sqrt{N_j+1}|N_X -1- N|^{s_0} (N_i-1)^{s_1}(N_j+1)^{s_2} \notag \\
&\quad \quad \times\bra{N_i-1,N_j+1,\bold{N}_{i,j}}\rho(\tau)\ket{N_i-1,N_j+1,\bold{N}_{i,j}}     \notag \\
&=\frac{1}{2}  \sum_{\bold{N}}f(\bold{N}_{i,j})  |N_X - N+1|^{s_0'}  (N_i+1)^{s_1'} (N_j-1)^{s_2'}\sqrt{N_i+1}\sqrt{N_j}|N_X - N|^{s_0}  N_i^{s_1}N_j^{s_2} \bra{N_i,N_j,\bold{N}_{i,j}}\rho(\tau)\ket{N_i,N_j,\bold{N}_{i,j}}   \notag \\
&=\frac{1}{2}  \tr \brr{  f(\hat{\bold{n}})  |\nb_X - N+1|^{s_0'}  (\nb_i+1)^{s_1'} (\nb_j-1)^{s_2'}(\nb_i + \nb_j)|\nb_X - N|^{s_0}\nb_i^{s_1} \nb_j^{s_2} \rho(\tau) }  \notag \\
&\le \frac{1}{2}  \tr \brr{  f(\hat{\bold{n}}) ( \nb'_X +1)^{s_0+s_0'}  (\nb_i+1)^{s_1+s_1'}\nb_j^{s_2+s_2'}(\nb_i + \nb_j) \rho(\tau) } .
\end{align}
By combining the inequalities~\eqref{start_eq/_prop:expectation_b_j_b_i_poly_n_i_1} and \eqref{start_eq/_prop:expectation_b_j_b_i_poly_n_i_2}, 
we obtain the desired inequality~\eqref{ienq_b_i_b_j_n_i+_n_j_ineN_general}. 
This completes the proof. $\square$

 {~}

\noindent\hrulefill{\bf [ End of Proof of Lemma~\ref{lemma:expectation_b_j_b_i_poly_n_i}] }

{~}

By using Lemma~\ref{lemma:expectation_b_j_b_i_poly_n_i}, we can upper-bound the quantities~\eqref{upper_bound_aim_mathcal_D_b'_x} in the same way of~\eqref{b_i_b_j_prop_expec_apply}: 
\begin{align}
\label{b_i_b_j_prop_expec_apply_general}
&\left| \tr\br{\mathcal{D}_{<k}^{(m)}  b_{X,i} b_{j}^\dagger  \mathcal{D}_{>k}^{(m)} \rho(\tau) } \right| 
\le \tr\br{\mathcal{D}_{<k}^{(m+1)}  (\nb_i+\nb_j)  \mathcal{D}_{>k}^{(m+1)} \rho(\tau) },   \notag \\
&\left| \tr\br{\mathcal{D}_{<k}^{(m)}  (b_{X,i})^\dagger  b_{j} \mathcal{D}_{>k}^{(m)} \rho(\tau) } \right| 
\le \tr\br{\mathcal{D}_{<k}^{(m+1)}  (\nb_i+\nb_j)  \mathcal{D}_{>k}^{(m+1)} \rho(\tau) } ,  
\end{align}
which yields the same inequality as \eqref{eq:tmp_bound_MXs_mk=0}. 

For general $m_k\ge 1$, we apply the same analyses because of 
\begin{align}
[H,\mathcal{D}^{m_k}(\nb'_{X}) ] = [H_0,\mathcal{D}^{m_k}(\nb_{X})]  \for m_k\ge 1,
\end{align}
and hence we immediately obtain all the inequalities from~\eqref{derivative_M_S_i_t} and~\eqref{derivative_M_S_i_t//2_general_m_fin}. 
We thus derive the same statement as Proposition~\ref{prop:simplify_high_der}.
As mentioned, based on the statement, we can immediately prove our main inequality~\eqref{main_ineN_Schuch_boson_extend_general}.  
This completes the proof. $\square$

\subsection{Lower bound on the moment function}

We have derived the upper bound for the $s$th moment function.
The same analyses can be applied to derive a lower bound. 
We here prove the following corollary:
\begin{corol}\label{corol_lower_bound_moment}
For an arbitrary subset $X$, the moment function $M^{(s)}_X(\tau)$ is lower-bounded by
 \begin{align}
 \label{main_inq;corol_lower_bound_moment}
M^{(s)}_X(\tau) \ge \ave{\nb_X - c_{\tau,1} \hat{\mathcal{D}}_X - c_{\tau,2} }_\rho^s  , 
\end{align}
where $\tau\le 1/(4\bar{J} \gamma)$ and $\ave{\nb_X - c_{\tau,1} \hat{\mathcal{D}}_X - c_{\tau,2} }_\rho\ge 0$ is assumed. 
\end{corol}

\subsubsection{Proof of Corollary~\ref{corol_lower_bound_moment}}

For the proof, it is enough to derive the first order moment because of
\begin{align}
\label{moment_lwoer_relation_trivial}
\tr(\rho(\tau) \nb_X^s  ) \ge [ \tr(\rho(\tau) \nb_X) ]^s 
\end{align}
for $s\ge 1$.  
Here, the inequality~\eqref{mth_order_derivative_mathcal_D_integral_0} is changed to 
\begin{align}
\label{mth_order_derivative_mathcal_D_integral_lower}
M_X^{(s)}(\tau)&\ge M_X^{(s)}(0) - \sum_{m=1}^\infty \frac{(2\tau)^m}{m!}\tr\br{\tilde{\mathcal{K}}^m(\nb_{X}^s)\rho} \notag \\
&= 2M_X^{(s)}(0) -  \sum_{m=0}^\infty \frac{(2\tau)^m}{m!}\tr\br{\tilde{\mathcal{K}}^m(\nb_{X}^s)\rho}  \notag \\
&\ge 2M_X^{(s)}(0) - \tr \brr { \br{ \nb_{X} + c_{\tau,1} \hat{\mathcal{D}}_X + c_{\tau,2} s }^s\rho} ,
\end{align}
where in the last inequality, we use the upper bound~\eqref{main_ineN_Schuch_boson_extend}.  
By considering $s=1$, we prove 
\begin{align}
\nb_X(\tau) \succeq 2 \nb_X - \br{ \nb_{X} + c_{\tau,1} \hat{\mathcal{D}}_X + c_{\tau,2}  } = \nb_X - c_{\tau,1} \hat{\mathcal{D}}_X - c_{\tau,2}  ,
\end{align}
which reduces the inequality~\eqref{moment_lwoer_relation_trivial} to 
\begin{align}
\tr\brr{ \rho \nb_X(\tau)^s } \ge  \brrr{\tr\brr{\rho \nb_X(\tau)  } }^s \ge \ave{\nb_X - c_{\tau,1} \hat{\mathcal{D}}_X - c_{\tau,2} }_\rho^s.
\end{align}
This completes the proof. $\square$

\section{Refined moment bounds after short-time evolution}

\subsection{Strategy for refined bounds}

 \begin{figure}[tt]
\centering
\includegraphics[clip, scale=0.4]{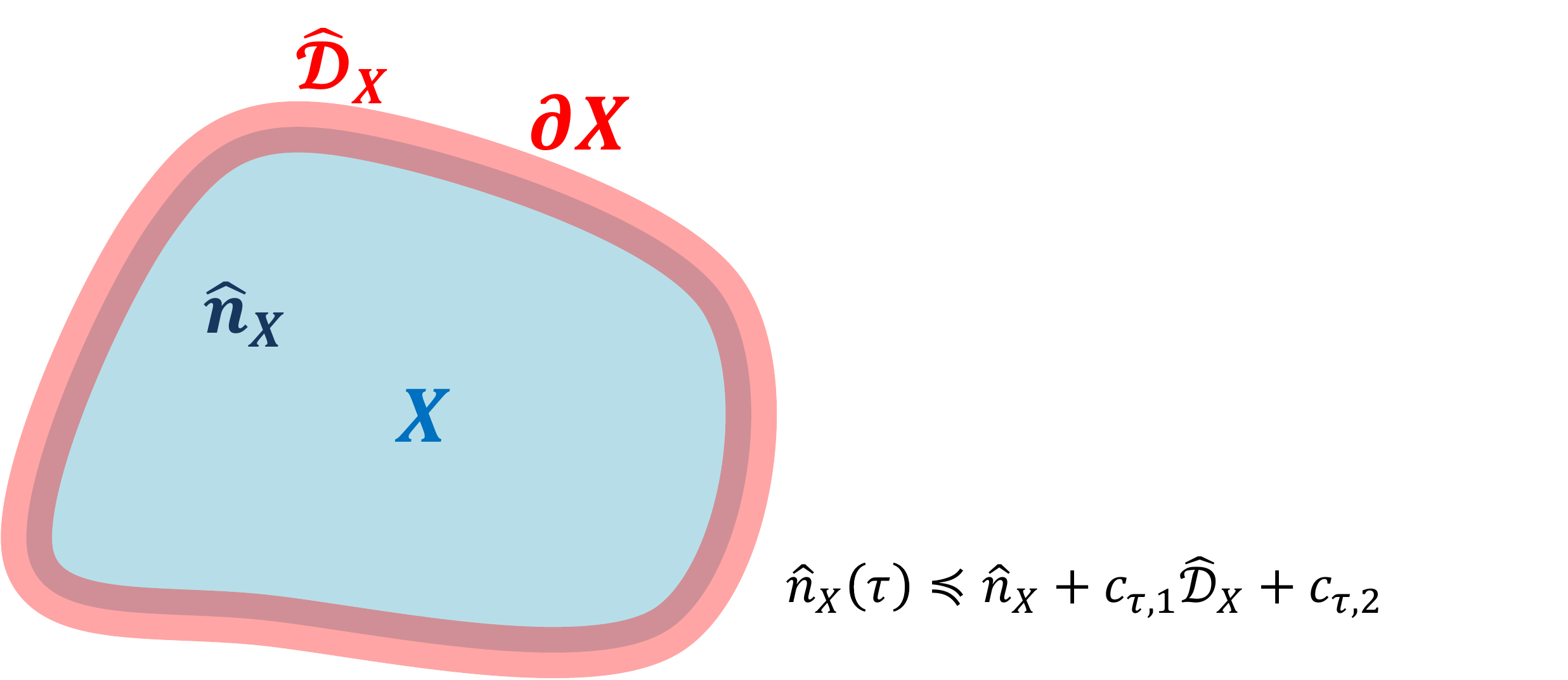}
\caption{In the upper bound~\eqref{assump_start_point0_re1} on the moment function, the operator $\nb_X(\tau)$ is bounded from above by 
$\br{ \nb_{X} + c_{\tau,1} \hat{\mathcal{D}}_{X} + c_{\tau,2} }$. 
Here, the operator $\hat{\mathcal{D}}_{X}$ is roughly approximated on the surface region $\partial X$. 
Hence, we expect that $\nb_{X} + c_{\tau,1} \hat{\mathcal{D}}_{X} + c_{\tau,2} \approx \nb_X \br{1 + |\partial X|/|X| }$. 
To make the expectation rigorous, we need to consider the possibility that most of the bosons concentrate on $\partial X$, which spoils the above discussions.
However, by carefully choosing the subsets $X[r]$ in~\eqref{refined_uppe_rough_boson_00}, we can obtain a refined upper bound as in~\eqref{refined_uppe_rough_boson}. 
}
\label{supp_fig_surface_bulk.pdf}
\end{figure}

We here show a rough description of refining the obtained bound in Subtheorem~\ref{Schuch_boson_extend}.
We have shown a similar explanation in the Method section of the main manuscript (see the subsection entitled ``Outline of the proof for Result 1'').

According to Subtheorem~\ref{Schuch_boson_extend}, we have
\begin{align}
\label{assump_start_point0_re1}
M_X^{(1)} (\tau)= \tr\brr{\nb_X^s \rho(\tau)} 
\le   \tr \brr{  \br{ \nb_{X} + c_{\tau,1} \hat{\mathcal{D}}_{X} + c_{\tau,2} }     \rho }    , 
\quad \hat{\mathcal{D}}_X = \sum_{i\in  \partial X} \sum_{j \in \Lambda} e^{-\dist_{i,j}} \nb_j ,
\end{align}
where we consider $s=1$ for simplicity.
The operator $\hat{\mathcal{D}}_{X}$ is approximately supported around the surface region $\partial X$.
Therefore, we expect that the contribution from $\hat{\mathcal{D}}_{X}$ is ignorable in comparison with that from $\nb_{X} $ as the size of $X$ increases (see Fig.~\ref{supp_fig_surface_bulk.pdf}). 
Note that we have $|\partial X|/|X| \to 0$ as $\diam(X) \to \infty$.  
However, we have a possibility that the bosons concentrate on the surface region, which may imply 
\begin{align}
\tr \brr{\br{\nb_X + c_{\tau,1} \hat{\mathcal{D}}_X }  \rho }  = [1+ \orderof{1}] \tr \br{ \nb_X  \rho }  
\longrightarrow  M_X^{(1)} (\tau) = [1+ \orderof{1}]  M_X^{(1)} (0) . 
\end{align}
In the above case, by connecting the short-time evolution, the moment function $M_X^{(1)}(\tau)$ exponentially increases as $e^{\orderof{t/\tau}}$ for general $t$, which leads to a meaningless upper bound.  

In the following, we start from the following trivial inequality:
 \begin{align}
\label{refined_uppe_rough_boson_00}
M_X^{(1)} (\tau) \le  M_{X[r]}^{(1)} (\tau) =  \tr \brr{  \br{ \nb_{X[r]} + c_{\tau,1} \hat{\mathcal{D}}_{X[r]} + c_{\tau,2} }  \rho }  .
\end{align}
We then prove that by appropriately choosing $r$ in \eqref{refined_uppe_rough_boson_00} ($0\le r\le \ell$), we can upper-bound $M_X^{(1)} (\tau)$ in the form of
\begin{align}
\label{refined_uppe_rough_boson}
M_X^{(1)} (\tau) \le   \tr \brr{ \br{ \nb_{X[\ell]} + c_{\tau,2} s +e^{-\orderofsp{\ell}}}     \rho }    
\end{align}
with $\ell$ a control parameter.  
In Sec.~\ref{sec:Moment upper bound: 1st order}, we consider the case of $s=1$, and afterward, we extend the result to general $s$ in Sec.~\ref{sec:Moment upper bound: higher order}.
     

%

\subsection{Convenient lemmas}

Before going to the proof, we first derive several convenient lemmas for our analyses. 
The main statement starts from Sec.~\ref{sec:Moment upper bound: 1st order} and readers can skip this subsection. 

Let $\{\alpha_j\}_{j=0}^\infty$ be an arbitrary positive sequence. 
For fixed $\{N_j\}_{j=1}^{{\tilde{p}}}$ ($N_j\in \mathbb{N}$, $j\in [1:{\tilde{p}}]$), we here define $\tilde{N}_p$ as follows:
\begin{align}
\label{def_tilde_N_p}
\tilde{N}_p :=   \sum_{j=1}^{{\tilde{p}}} \alpha_{|p-j|} N_j ,
\end{align} 
where $\{\alpha_{j}\}_{j=0}^\infty$ are arbitrary positive numbers.
In practical applications, we consider $\{\alpha_{j}\}_{j=0}^\infty$ that $\alpha_j$ decays exponentially as $j$ increases. 
We then consider the following quantity ($1\le p\le {\tilde{p}}$):
\begin{align}
\tilde{N}_p + N_{1:p}  ,\quad N_{1:p} = \sum_{j=1}^{p} N_j . 
\end{align} 
We, in the following, aim to estimate the minimization of 
\begin{align}
\label{minimization_num_shell}
\min_{p\in [1:{\tilde{p}}]} \br{\tilde{N}_p + N_{1:p} } .
\end{align}
We can prove the following statement, which claims that the above quantity is upper-bounded by $(1+e^{-\orderofsp{\tilde{p}}}) N_{1:\tilde{p}}$:  

\begin{lemma}\label{lemma:minimization_num_shell}
The quantity~\eqref{minimization_num_shell} is upper-bounded as follows:
\begin{align}
\label{ineq:lemma:minimization_num_shell}
\min_{p\in [1:{\tilde{p}}]} \br{ \tilde{N}_p + N_{1:p}  } \le 
\brr{1+ \bar{\alpha}_\infty \br{1+\frac{1}{\bar{\alpha}_\infty + \chi_\alpha}}^{-{\tilde{p}}+1}} N_{1:{\tilde{p}}},
\end{align}
where $\bar{\alpha}_\infty$ and $\chi_\alpha$ are defined by Eqs.~\eqref{def:bar_alpha_infty} and \eqref{definition_chi_alpha}, i.e., 
 \begin{align} 
&\bar{\alpha}_{\infty}:= \sum_{j=-\infty}^{\infty} \alpha_{|j|}, \quad \chi_\alpha := \max_{j \in [0,\infty]}\br{ \sum_{p=1}^{\infty}  \alpha_{p+j}/\alpha_{j} } .
\end{align}
\end{lemma}

{\bf Remark.} The parameters $\bar{\alpha}_\infty$ and $\chi_\alpha$ may be unbounded. 
In particular, when $\alpha_j$ decays polynomially with $j$, the parameter $\chi_{\alpha}$ is infinitely large.
If $\alpha_j$ decays exponentially with $j$, both of the parameters $\bar{\alpha}_\infty$ and $\chi_\alpha$ are $\orderof{1}$ constants.
For example, for $\alpha_j=0$ for $\forall j\ge 1$, we can trivially obtain $\bar{\alpha}_{\infty}= \alpha_0$ and $\chi_\alpha=0$.

\subsubsection{Proof of Lemma~\ref{lemma:minimization_num_shell}} 

For proof, we can restrict ourselves to the case of 
\begin{align}
\label{inequality_assume_1}
\min_{p\in [1:{\tilde{p}}]} \br{\tilde{N}_p + N_{1:p} } \ge  N_{1:{\tilde{p}}} .
\end{align}
Note that in the case where the above inequality does not hold, the main inequality~\eqref{ineq:lemma:minimization_num_shell} trivially holds.
From the inequality~\eqref{inequality_assume_1}, we obtain 
\begin{align}
\label{inequality_assume_reduce}
\tilde{N}_p + N_{1:p}  \ge N_{1:{\tilde{p}}}   \longrightarrow 
\tilde{N}_p \ge N_{p+1:{\tilde{p}}} .
\end{align}
The key tool to analyze the condition~\eqref{inequality_assume_reduce} is the following inequality (see below for the proof):
\begin{align}
\label{key_ineq_for_proof_shell_num_main}
&\tilde{N}_{p+1:{\tilde{p}}} \le 
\bar{\alpha}_{\infty} N_{p+1:{\tilde{p}}} + 
\chi_\alpha \tilde{N}_p  \for \forall p , \notag \\
&\tilde{N}_{p_1:p_2} := \sum_{p\in [p_1,p_2]} \tilde{N}_p .  
\end{align}
By using the above inequality, the condition~\eqref{inequality_assume_reduce} reduces to
\begin{align}
&\tilde{N}_p \ge N_{p+1:{\tilde{p}}} \ge \frac{1}{\bar{\alpha}_{\infty}}\br{\tilde{N}_{p+1:{\tilde{p}}} - \chi_\alpha \tilde{N}_p } \notag \\
&\longrightarrow  \tilde{N}_p \ge \frac{1}{\bar{\alpha}_{\infty} + \chi_\alpha} \tilde{N}_{p+1:{\tilde{p}}} =: \frac{1}{\bar{\alpha}} \tilde{N}_{p+1:{\tilde{p}}}  \for \forall p,
\label{key_ineq_for_proof_shell_num}
\end{align}
where we define $\bar{\alpha}:=\bar{\alpha}_{\infty} + \chi_\alpha$.

By using the inequality~\eqref{key_ineq_for_proof_shell_num}, we first reduce the quantity $\tilde{N}_{1:{\tilde{p}}}$ to
\begin{align}
\tilde{N}_{1:{\tilde{p}}} = \tilde{N}_{1} + \tilde{N}_{2:{\tilde{p}}} \ge \br{1+\frac{1}{\bar{\alpha}}}\tilde{N}_{2:{\tilde{p}}} .
\end{align}
In the same way, we have 
\begin{align}
\tilde{N}_{1:{\tilde{p}}} \ge \br{1+\frac{1}{\bar{\alpha}}}\tilde{N}_{2:{\tilde{p}}} \ge \br{1+\frac{1}{\bar{\alpha}}} 
\br{ \tilde{N}_{2} +\tilde{N}_{3:{\tilde{p}}}} 
\ge  \br{1+\frac{1}{\bar{\alpha}}}^2\tilde{N}_{3:{\tilde{p}}}   .
\end{align}
By repeating the process, we obtain 
\begin{align}
\tilde{N}_{1:{\tilde{p}}} \ge \br{1+\frac{1}{\bar{\alpha}}}^{{\tilde{p}}-1}\tilde{N}_{{\tilde{p}}:{\tilde{p}}} = \br{1+\frac{1}{\bar{\alpha}}}^{{\tilde{p}}-1}\tilde{N}_{{\tilde{p}}}   .
\end{align}
Therefore, we obtain
\begin{align}
\tilde{N}_{{\tilde{p}}}\le \br{1+\frac{1}{\bar{\alpha}}}^{-{\tilde{p}}+1}  \tilde{N}_{1:{\tilde{p}}} \le 
\bar{\alpha}_\infty \br{1+\frac{1}{\bar{\alpha}}}^{-{\tilde{p}}+1}  N_{1:{\tilde{p}}} ,
\end{align}
where we use the definition~\eqref{def:bar_alpha_infty} below to obtain $\tilde{N}_{1:\tilde{p}}  \le \bar{\alpha}_\infty  N_{1:\tilde{p}}$ as follows:
 \begin{align}
\tilde{N}_{1:{\tilde{p}}} = \sum_{p'=1}^{{\tilde{p}}} \tilde{N}_{p'}=
  \sum_{p'=1}^{{\tilde{p}}} \sum_{j=1}^{{\tilde{p}}} \alpha_{|p'-j|} N_j  
  \le  \sum_{p'=-\infty}^{\infty} \alpha_{|p'|}  \sum_{j=1}^{{\tilde{p}}} N_j  = \bar{\alpha}_\infty  N_{1:\tilde{p}}. 
  \label{def:bar_alpha_infty_ineq}
\end{align}
We thus arrive at the inequality of 
\begin{align}
N_{1:{\tilde{p}}}+\tilde{N}_{{\tilde{p}}}\le 
\brr{1+ \bar{\alpha}_\infty \br{1+\frac{1}{\bar{\alpha}}}^{-{\tilde{p}}+1}}  N_{1:{\tilde{p}}} .
\end{align}
The above upper bound is applied to $\min_{p\in [1:{\tilde{p}}]} \br{N_{1:p}+\tilde{N}_p}$, which yields the desired inequality~\eqref{ineq:lemma:minimization_num_shell} by replacing $\bar{\alpha} \to \bar{\alpha}_{\infty} + \chi_\alpha$.
This completes the proof. $\square$

{~}\\

{[\bf Proof of the inequality~\eqref{key_ineq_for_proof_shell_num_main}]}

We here consider the quantity of
\begin{align}
\tilde{N}_{p+1:{\tilde{p}}} = \sum_{p'=p+1}^{{\tilde{p}}} \tilde{N}_{p'} =
  \sum_{p'=p+1}^{{\tilde{p}}} \sum_{j=1}^{{\tilde{p}}} \alpha_{|p'-j|} N_j  .
\end{align}
By defining $\bar{\alpha}_{\infty}$ as 
\begin{align}
\label{def:bar_alpha_infty}
\bar{\alpha}_{\infty}:= \sum_{j=-\infty}^{\infty} \alpha_{|j|} ,
\end{align}
we first obtain
\begin{align}
\label{key_ineq_for_proof_shell_num_proof1}
\tilde{N}_{p+1:{\tilde{p}}} \le 
\bar{\alpha}_{\infty} N_{p+1:{\tilde{p}}} + 
\sum_{j=1}^{p} 
 \sum_{p'=p+1}^{{\tilde{p}}}  \alpha_{|p'-j|} N_j  
&= \bar{\alpha}_{\infty} N_{p+1:{\tilde{p}}} + 
\sum_{j=1}^{p} 
 \sum_{p'=1}^{{\tilde{p}}-p}  \alpha_{p'+p-j} N_j .
\end{align}

We here introduce the parameter $\chi_\alpha$ as 
\begin{align}
\label{definition_chi_alpha}
\chi_\alpha = \max_{j \in [0,\infty]} \br{\sum_{p=1}^{\infty}  \alpha_{p+j}/\alpha_{j}} ,
\end{align}
which trivially satisfies
\begin{align}
 \sum_{p'=1}^{{\tilde{p}}-p}  \alpha_{p'+p-j} \le  \sum_{p'=1}^{\infty}  \alpha_{p'+p-j} \le  \chi_\alpha \alpha_{p-j} \for \forall j \in [0,p] .
\end{align}
We then obtain 
\begin{align}
\label{key_ineq_for_proof_shell_num_proof2}
\sum_{j=1}^{p} 
 \sum_{p'=1}^{{\tilde{p}}-p}  \alpha_{p'+p-j} N_j 
 \le \sum_{j=1}^{p}  \chi_\alpha \alpha_{p-j} N_j \le 
 \chi_\alpha 
 \sum_{j=1}^{{\tilde{p}}}\alpha_{|p-j|} N_{j} = \chi_\alpha \tilde{N}_{p},
\end{align}
where we use the definition~\eqref{def_tilde_N_p} in the last equation.
By combining the inequalities~\eqref{key_ineq_for_proof_shell_num_proof1} and \eqref{key_ineq_for_proof_shell_num_proof2}, 
we arrive at the main inequality~\eqref{key_ineq_for_proof_shell_num_main}. 
This completes the proof. $\square$

 {~}

\hrulefill{\bf [ End of Proof of Lemma~\ref{lemma:minimization_num_shell}] }

{~}

Conversely, we also consider the maximization of 
\begin{align}
\label{minimization_num_shell_2}
\max_{p\in [1:{\tilde{p}}]} \br{-\tilde{N}_p + N_{1:p} }   .
\end{align}
We can prove the following similar statement:  

\begin{lemma}\label{lemma:maximization_num_shell}
The quantity~\eqref{minimization_num_shell_2} is lower-bounded as follows:
\begin{align}
\label{ineq:lemma:minimization_num_shell_2}
\max_{p\in [1:{\tilde{p}}]} \br{-\tilde{N}_p + N_{1:p} }    \ge 
- \bar{\alpha}_\infty \br{1+\frac{1}{\bar{\alpha}_\infty + \chi_\alpha}}^{-{\tilde{p}}+1} N_{1:{\tilde{p}}},
\end{align}
where $\bar{\alpha}_\infty$ and $\chi_\alpha$ have been defined in Eqs.~\eqref{def:bar_alpha_infty} and \eqref{definition_chi_alpha}.
\end{lemma}

\subsubsection{Proof of Lemma~\ref{lemma:maximization_num_shell}} 

The proof is almost the same as the one for Lemma~\ref{lemma:minimization_num_shell}. 
We can restrict ourselves to the case of 
\begin{align}
\label{inequality_assume_1_2}
\max_{p\in [1:{\tilde{p}}]} \br{-\tilde{N}_p + N_{1:p} }   \le  0 .
\end{align}
Otherwise, the inequality~\eqref{ineq:lemma:minimization_num_shell_2} trivially holds.
Then, we obtain 
 \begin{align}
\label{inequality_assume_reduce_2}
-\tilde{N}_p + N_{1:p}  \le 0   \longrightarrow 
\tilde{N}_p \ge N_{1:p} .
\end{align}
We can obtain a similar inequality to the inequality~\eqref{key_ineq_for_proof_shell_num_main}: 
\begin{align}
\label{key_ineq_for_proof_shell_num_main_2}
\tilde{N}_{1:p-1} \le 
\bar{\alpha}_{\infty} N_{1:p} + 
\chi_\alpha \tilde{N}_p  \for \forall p ,
\end{align}
where we show the proof below. 
By using~\eqref{key_ineq_for_proof_shell_num_main_2}, we can reduce the inequality~\eqref{inequality_assume_reduce_2} to 
 \begin{align}
 \label{key_ineq_for_proof_shell_num_3}
\tilde{N}_p \ge N_{1:p} \ge \frac{1}{\bar{\alpha}_{\infty}}\br{\tilde{N}_{1:p-1} - \chi_\alpha \tilde{N}_p } 
\longrightarrow  \tilde{N}_p \ge \frac{1}{\bar{\alpha}_{\infty} + \chi_\alpha} \tilde{N}_{1:p-1} 
\end{align}

By using the inequality~\eqref{key_ineq_for_proof_shell_num_3}, we first reduce the quantity $\tilde{N}_{1:{\tilde{p}}}$ to
\begin{align}
\label{upper_boundtilde_N/1;q}
\tilde{N}_{1:{\tilde{p}}} = \tilde{N}_{{\tilde{p}}} + \tilde{N}_{1:{\tilde{p}}-1} \ge \br{1+\frac{1}{\bar{\alpha}}}\tilde{N}_{1:{\tilde{p}}-1} ,\quad 
\bar{\alpha} := \bar{\alpha}_{\infty} + \chi_\alpha. 
\end{align}
In the same way, we have 
\begin{align}
\tilde{N}_{1:{\tilde{p}}} \ge \br{1+\frac{1}{\bar{\alpha}}}\tilde{N}_{1:{\tilde{p}}-1} \ge \br{1+\frac{1}{\bar{\alpha}}} 
\br{ \tilde{N}_{\tilde{p}-1} +\tilde{N}_{1:{\tilde{p}}-2}} 
\ge  \br{1+\frac{1}{\bar{\alpha}}}^2\tilde{N}_{1:{\tilde{p}}-2}   .
\end{align}
By repeating the process, we obtain 
\begin{align}
\tilde{N}_{1:{\tilde{p}}} \ge \br{1+\frac{1}{\bar{\alpha}}}^{{\tilde{p}}-1}\tilde{N}_{1:1} = \br{1+\frac{1}{\bar{\alpha}}}^{{\tilde{p}}-1}\tilde{N}_{1}   .
\end{align}
Therefore, we arrive at the inequality of
\begin{align}
\tilde{N}_{1}\le \br{1+\frac{1}{\bar{\alpha}}}^{-{\tilde{p}}+1}  \tilde{N}_{1:{\tilde{p}}} \le 
\bar{\alpha}_\infty \br{1+\frac{1}{\bar{\alpha}}}^{-{\tilde{p}}+1}  N_{1:{\tilde{p}}} ,
\end{align}
where the second inequality immediately results from the inequality~\eqref{def:bar_alpha_infty_ineq}. 
We thus obtain the inequality of 
\begin{align}
N_{1:1}-\tilde{N}_1\ge 
N_1-  \bar{\alpha}_\infty \br{1+\frac{1}{\bar{\alpha}}}^{-{\tilde{p}}+1}  N_{1:{\tilde{p}}} .
\end{align}
The above lower bound is also applied to $\max_{p\in [1:{\tilde{p}}]} \br{-\tilde{N}_p + N_{1:p}}$ under the condition of~\eqref{inequality_assume_1_2}, which yields the desired inequality~\eqref{ineq:lemma:minimization_num_shell_2}.
This completes the proof. $\square$

{~} \\

{[\bf Proof of the inequality~\eqref{key_ineq_for_proof_shell_num_main_2}]}

The proof is almost the same as for the inequality~\eqref{key_ineq_for_proof_shell_num_main}. 
We here consider 
\begin{align}
\tilde{N}_{1:p-1} = \sum_{p'=1}^{p-1} \tilde{N}_{p'} =
  \sum_{p'=1}^{p-1} \sum_{j=1}^{{\tilde{p}}} \alpha_{|p'-j|} N_j  .
\end{align}
From the definition of  $\bar{\alpha}_{\infty}:= \sum_{j=-\infty}^{\infty} \alpha_{|j|}$, 
we first obtain
\begin{align}
\label{key_ineq_for_proof_shell_num_proof1_2}
\tilde{N}_{1:p-1} \le 
\bar{\alpha}_{\infty} N_{1:p} + 
\sum_{j=p+1}^{{\tilde{p}}} 
 \sum_{p'=1}^{p-1}  \alpha_{|p'-j|} N_j  
&= \bar{\alpha}_{\infty} N_{1:p} + 
\sum_{j=p+1}^{{\tilde{p}}} 
\sum_{p'=1}^{p-1}  \alpha_{j-p'} N_j    .
\end{align}

By using the parameter $\chi_\alpha$ in Eq.~\eqref{definition_chi_alpha}, i.e.,  
$$
\chi_\alpha = \max_{j \in [0,\infty]} \br{\sum_{p=1}^{\infty}  \alpha_{p+j}/\alpha_{j}} ,
$$
we have 
\begin{align}
 \sum_{p'=1}^{p-1}  \alpha_{j-p'} = \sum_{p'=1}^{p-1}  \alpha_{j-p + p-p'}
 \le  \sum_{p''=1}^{\infty}\alpha_{j-p+p''} \le  \chi_\alpha \alpha_{j-p} \for j \in [p+1,{\tilde{p}}] .
\end{align}
We then obtain 
\begin{align}
\label{key_ineq_for_proof_shell_num_proof2_2}
\sum_{j=p+1}^{{\tilde{p}}} 
\sum_{p'=1}^{p-1}  \alpha_{j-p'} N_j 
 \le \sum_{j=p+1}^{{\tilde{p}}}   \chi_\alpha \alpha_{j-p} N_{j} 
 \le  \chi_\alpha 
 \sum_{j=1}^{{\tilde{p}}}\alpha_{|p-j|} N_{j} = \chi_\alpha \tilde{N}_{p}.
\end{align}
By combining the inequalities~\eqref{key_ineq_for_proof_shell_num_proof1_2} and \eqref{key_ineq_for_proof_shell_num_proof2_2}, 
we arrive at the main inequality~\eqref{key_ineq_for_proof_shell_num_main_2}. 
This completes the proof. $\square$

 {~}

\hrulefill{\bf [ End of Proof of Lemma~\ref{lemma:maximization_num_shell}] }

{~}

Finally, in proving Proposition~\ref{prop:lower_probability}, we need to estimate the minimization of 
\begin{align}
\label{minimization_num_shell_frac}
\min_{p\in [1:{\tilde{p}}]} \frac{\tilde{N}_p}{N_{1:p} } ,
\end{align}
where $\tilde{N}_p :=   \sum_{j=1}^{{\tilde{p}}} \alpha_{|p-j|} N_j $. 
By using similar analyses to the proof of Lemma~\ref{lemma:maximization_num_shell}, we can prove the following statement:
\begin{lemma}\label{lemma:minimization_num_shell_frac}
If there does not exist a parameter $p\in [1:{\tilde{p}}]$ such that 
\begin{align}
\label{ineq:lemma:minimization_num_shell_frac}
\frac{\tilde{N}_p}{N_{1:p} } \le \frac{1}{\Phi}  ,
\end{align}
the quantity $\tilde{N}_1$ is upper-bounded by
\begin{align}
&\tilde{N}_1 \le \br{1+\frac{1}{\Phi \bar{\alpha}_{\infty} + \chi_\alpha}  }^{-{\tilde{p}}+1} \bar{\alpha}_\infty  N_{1:{\tilde{p}}}   ,
\end{align}
where $\bar{\alpha}_\infty$ and $\chi_\alpha$ are defined by Eqs.~\eqref{def:bar_alpha_infty} and \eqref{definition_chi_alpha}.
\end{lemma}

{\bf Remark.} From the lemma, we can ensure that either of the following statements is satisfied:
\begin{align}
\min_{p\in [1:{\tilde{p}}]} \frac{\tilde{N}_p}{N_{1:p} } \le \frac{1}{\Phi}  \Or  \tilde{N}_1  \le e^{-\orderofsp{{\tilde{p}}}} N_{1:{\tilde{p}}} .
\end{align}
The lemma plays a central role in proving Lemma~\ref{Lema:upp_pi_X_ell_m_0_q+1} below (see Case~2 in the proof of  Lemma~\ref{Lema:upp_pi_X_ell_m_0_q+1}, Sec.~\ref{sec:Proof of Lemma_Lema:upp_pi_X_ell_m_0_q+1}).

\subsubsection{Proof of Lemma~\ref{lemma:minimization_num_shell_frac}}

We aim to estimate the upper bound of $\tilde{N}_1$ under the condition of 
\begin{align}
\label{minimization_num_shell_frac_cond}
\frac{\tilde{N}_p}{N_{1:p} } \ge \frac{1}{\Phi} \for \forall p \in [1,\tilde{p}].
\end{align}
By using the inequality~\eqref{key_ineq_for_proof_shell_num_main_2}, we obtain
\begin{align}
&\tilde{N}_p \ge \frac{N_{1:p}}{\Phi} \ge \frac{1}{\Phi}\cdot \frac{\tilde{N}_{1:p-1}- \chi_\alpha \tilde{N}_p }{\bar{\alpha}_{\infty}}  \longrightarrow \tilde{N}_p  \ge \frac{\tilde{N}_{1:p-1}}{\Phi \bar{\alpha}_{\infty} + \chi_\alpha} ,
\end{align}
which yields 
\begin{align}
&\tilde{N}_{1:p} =\tilde{N}_{1:p-1} + \tilde{N}_{p}  \ge \br{1+\frac{1}{\Phi \bar{\alpha}_{\infty} + \chi_\alpha}  } \tilde{N}_{1:p-1} . 
\end{align}
By repeatedly using the above inequality, we reach the inequality of 
\begin{align}
&\tilde{N}_{1:{\tilde{p}}}  \ge \br{1+\frac{1}{\Phi \bar{\alpha}_{\infty} + \chi_\alpha}  }^{{\tilde{p}}-1} \tilde{N}_{1} . 
\end{align}
Hence, by using $\tilde{N}_{1:{\tilde{p}}}\le \bar{\alpha}_\infty N_{1:{\tilde{p}}}$ from the inequality~\eqref{def:bar_alpha_infty_ineq}, we obtain the main inequality:
\begin{align}
&\tilde{N}_1 \le \br{1+\frac{1}{\Phi \bar{\alpha}_{\infty} + \chi_\alpha}  }^{-{\tilde{p}}+1} \bar{\alpha}_\infty  N_{1:{\tilde{p}}}   .
\end{align}
This completes the proof. $\square$

 {~}

\hrulefill{\bf [ End of Proof of Lemma~\ref{lemma:minimization_num_shell_frac}] }

{~}

\subsection{Moment upper bound: 1st order} \label{sec:Moment upper bound: 1st order}

We first consider the 1st order moment, i.e., 
\begin{align}
\label{assump_start_point0_re1_1st}
M_X^{(1)} (\tau)=\ave{\nb_X(\tau)}_{\rho} \le \ave{  \nb_{X} + c_{\tau,1} \hat{\mathcal{D}}_{X} + c_{\tau,2}}_\rho      ,
\end{align}
where we use Subtheorem~\ref{Schuch_boson_extend} in the inequality.
For the above quantity, we can prove the following proposition in the form of the operator inequality, which yields the desired bound as~\eqref{refined_uppe_rough_boson}:
\begin{prop} \label{prop:upp_Moment_1th}
For an arbitrary time of $\tau \le 1/(4\gamma \bar{J})$, we obtain the following operator inequality:
\begin{align}
\label{Ineq:prop:upp_Moment_1th}
\nb_X(\tau)&\preceq 
\br{1+ 5c_{\tau,1} \lambda_{1/2} e^{-f_\tau \ell} } \nb_{X[\ell]} + c_{\tau,1}e^{-3\ell/16} \hat{\mathscr{D}}_{X[\ell]}+ c_{\tau,2} ,
\end{align}
where we can choose the length $\ell$ arbitrarily, and the operator $\hat{\mathscr{D}}_{X[\ell]}$ and the constant $f_\tau$ are defined by using parameter $\lambda_c$ in~\eqref{oftern_used_inequality_lambda_c} as follows:
\begin{align}
\label{def_delta_X_ell_prop}
\hat{\mathscr{D}}_{X[\ell]}\coloneqq   \sum_{j\in X[\ell]^\co} e^{-3\dist_{j,X[\ell]}/4} \nb_j   
\end{align}
and 
\begin{align}
\label{def_f_tau_ell_prop}
f_\tau\coloneqq  \frac{1}{2} \log \br{1+\frac{1}{5c_{\tau,1} \lambda_{1/2} + 2}}.
\end{align}
\end{prop}

 \begin{figure}[tt]
\centering
\includegraphics[clip, scale=0.5]{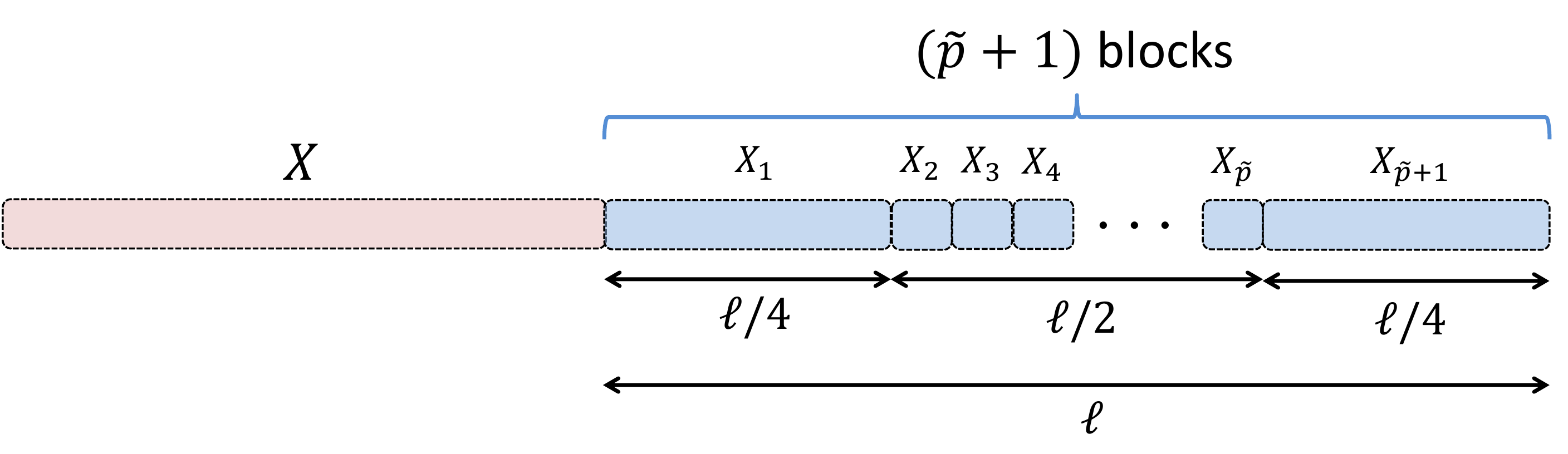}
\caption{We decompose the subset $X[\ell] \setminus X$ to $\tilde{p}+1$ pieces. 
In the picture, a one-dimensional case is considered. 
We define the extended subset of $X$ from $X_1$ to $X_p$ as $X_{0:p} := X \cup X_1 \cup \cdots \cup X_p$. 
Then, we upper-bound the moment function $M^{(1)}_X (\tau)$ by using $M^{(1)}_{X_{0:p}}(\tau)$ with an appropriate choice of $p$.
In the choice, we need to suppress the contribution from the surface region (i.e., the term $\hat{\mathcal{D}}_{X_{0:p}}$).
}
\label{supp_fig_shell_choice.pdf}
\end{figure}

\subsection{Proof of Proposition~\ref{prop:upp_Moment_1th}}

We first define the subsets $\{X_p\}_{p=1}^{{\tilde{p}}+1}$ as follows (see Fig.~\ref{supp_fig_shell_choice.pdf}):
\begin{align}
\label{df_X_1_X_j_X_q+1_0}
&X_0\coloneqq X,\quad X_1 \coloneqq  X[\ell/4] \setminus X , \notag \\
& X_p \coloneqq  X[\ell/4+p-1] \setminus X[\ell/4+p-2]\quad (2\le p\le {\tilde{p}}) , \notag \\
& X_{{\tilde{p}}+1} \coloneqq   X[\ell]\setminus X[3\ell/4],
\end{align} 
where $\tilde{p}=\ell/2 +1$. 
We then consider the first order moment as 
\begin{align}
M_{X_{0:p}}^{(1)} (\tau) = \tr\brr { \nb_{X_{0:p}}(\tau) \rho },\quad X_{0:p} \coloneqq  \bigcup_{j=0}^p X_j  
\end{align} 
for an arbitrary quantum state $\rho$.
Note that because of $X\subseteq X_{0:p}$ (i.e., $\nb_X \preceq \nb_{X_{0:p}}$) we trivially obtain 
\begin{align}
\label{trivial_upp_moment_extend}
M_{X}^{(1)} (\tau)  \le M_{X_{0:p}}^{(1)} (\tau)  
\end{align} 
for arbitrary $p\ge 0$. 

From the inequality~\eqref{assump_start_point0_re1_1st}, we obtain the upper bound of 
\begin{align}
\label{Start_for_the_proof_prop_upp_Moment_sth}
M_{X_{0:p}}^{(1)} (\tau) &\le  \tr\brr{\br{\nb_{X_{0:p}}+ c_{\tau,1} \hat{\mathcal{D}}_{X_{0:p}} + c_{\tau,2}} \rho}   , 
\quad \hat{\mathcal{D}}_{X_{0:p}} = \sum_{i\in  \partial (X_{0:p})} \sum_{j \in \Lambda} e^{-\dist_{i,j}} \nb_j  ,\notag \\
&= \ave{\nb_{X_{0:p}}}_\rho+ c_{\tau,1} \ave{\hat{\mathcal{D}}_{X_{0:p}}}_\rho    + c_{\tau,2}  \notag \\
&= N^{(\rho)}_{X_{0:p}} +  c_{\tau,1}\tilde{N}^{(\rho)}_{X_{0:p}} + c_{\tau,2} ,\quad  \tilde{N}^{(\rho)}_{X_{0:p}} := \sum_{i\in  \partial (X_{0:p})} \sum_{j \in \Lambda} e^{-\dist_{i,j}} N^{(\rho)}_j,
\end{align}
where we denote $\ave{\nb_i}_\rho = N^{(\rho)}_i$ for $\forall i\in \Lambda$. 
In the following, we aim to estimate the best choice of $p$ that minimizes the moment as 
\begin{align}
\label{min_prob_N_rho_X_0:p}
\min_{p\in [1,{\tilde{p}}]} \br{N^{(\rho)}_{X_{0:p}} +  c_{\tau,1}\tilde{N}^{(\rho)}_{X_{0:p}} } ,
\end{align}
from which we obtain an optimal upper bound for $M_{X}^{(1)} (\tau) $ because of the inequality~\eqref{trivial_upp_moment_extend}.

For the estimation of the RHS of the inequality~\eqref{Start_for_the_proof_prop_upp_Moment_sth}, 
we reduce the problem~\eqref{min_prob_N_rho_X_0:p} to the form for which Lemma~\ref{lemma:minimization_num_shell} can be applied. 
For this purpose, we first need to upper-bound $\tilde{N}_{X_{0:p}}$ to make it a tractable form.
We prove the following lemma (see Sec.~\ref{proof_of_lmma:tilde_N_X_p_upper_bound} below for the proof):
\begin{lemma} \label{lmma:tilde_N_X_p_upper_bound}
By using the parameter $\lambda_c$ in Eq.~\eqref{oftern_used_inequality_lambda_c}, we obtain the upper bound of 
\begin{align}
\label{ineq_lmma:tilde_N_X_p_upper_bound}
&\tilde{N}^{(\rho)}_{X_{0:p}}
\le \lambda_{1/2} \sum_{j=1}^{{\tilde{p}}} e^{-|p-j|/2} M_j  +\delta N^{(\rho)}_{X,\ell}  \quad (1\le p\le {\tilde{p}}), \notag \\
&M_j = \begin{cases} 
N^{(\rho)}_{X_j}& \for 1\le j \le {\tilde{p}}-1 , \\
N^{(\rho)}_{X_{\tilde{p}}} +N^{(\rho)}_{X_{{\tilde{p}}+1}} &\for j={\tilde{p}},
\end{cases}
\end{align} 
with
\begin{align}
\label{def_f_tau_ell_prop_re}
\delta N^{(\rho)}_{X,\ell}\coloneqq  \lambda_{1/4}e^{- 3\ell/16} \br{  \sum_{j\in X} N^{(\rho)}_j   + \sum_{j\in X[\ell]^\co }e^{- 3\dist_{j,X[\ell]}/4} N^{(\rho)}_j  }.
\end{align}
\end{lemma}

By applying Lemma~\ref{lmma:tilde_N_X_p_upper_bound} to Eq.~\eqref{min_prob_N_rho_X_0:p}, we reduce the optimization problem to  
\begin{align}
\label{max_N_X_0:p_tilde_N_X_p}
\min_{p\in[1,{\tilde{p}}]} \br{N^{(\rho)}_{X_{0:p}}+ c_{\tau,1} \tilde{N}^{(\rho)}_{X_{0:p}}}
\le N^{(\rho)}_X + c_{\tau,1}\delta N^{(\rho)}_{X,\ell} +  \min_{p\in[1,{\tilde{p}}]} \br{M_{1:p}+ c_{\tau,1} \lambda_{1/2} \sum_{j=1}^{{\tilde{p}}} e^{-|p-j|/2} M_j  } ,
\end{align}
where $M_{1:p}=\sum_{j=1}^p M_j$ and we use the definition of $\{M_j\}_{j=1}^{\tilde{p}}$ to obtain 
$$N^{(\rho)}_{X_{0:p}}= N^{(\rho)}_X +(M_1+M_2 +\cdots+M_p) = N^{(\rho)}_X + M_{1:p} \for p\le \tilde{p}-1.$$
To solve the minimization problem in \eqref{max_N_X_0:p_tilde_N_X_p}, we utilize Lemma~\ref{lemma:minimization_num_shell}, where $\alpha_j$ corresponds to $\alpha_j=c_{\tau,1} \lambda_{1/2} e^{-j/2}$.  
The parameters $\bar{\alpha}_{\infty}$ and $\chi_\alpha$ are estimated as follows:
\begin{align} 
\label{bar_alpha_infty_apply_upp}
\bar{\alpha}_{\infty}=c_{\tau,1} \lambda_{1/2} \sum_{j=-\infty}^{\infty} e^{-|j|/2}= \br{ \frac{2}{1-e^{-1/2}} -1}c_{\tau,1} \lambda_{1/2} = (4.0829\cdots)c_{\tau,1} \lambda_{1/2} \le 5 c_{\tau,1} \lambda_{1/2},
\end{align}
and 
\begin{align}
\label{chi_alpha_infty_apply_upp}
\chi_\alpha = \sum_{p=1}^{\infty} e^{-p/2}=  \frac{e^{-1/2}}{1-e^{-1/2}}  = (1.5414\cdots ) < 2. 
\end{align}
Thus, the inequality~\eqref{ineq:lemma:minimization_num_shell} yields 
\begin{align}
\label{max_N_X_0:p_tilde_N_X_p_Lemma}
\min_{p\in[1,{\tilde{p}}]} \br{M_{1:p}+ c_{\tau,1} \lambda_{1/2} \sum_{j=1}^{{\tilde{p}}} e^{-|p-j|/2} M_j  }
&\le \brr{1+ 5c_{\tau,1} \lambda_{1/2}\br{1+\frac{1}{5c_{\tau,1} \lambda_{1/2} + 2}}^{-{\tilde{p}}+1}} M_{1:{\tilde{p}}} \notag\\
&= \brr{1+ 5c_{\tau,1} \lambda_{1/2}\br{1+\frac{1}{5c_{\tau,1} \lambda_{1/2} + 2}}^{-\ell/2}} ( N^{(\rho)}_{X[\ell]} -N^{(\rho)}_X),
\end{align}
where we have used ${\tilde{p}}=\ell/2+1$ and $M_{1:{\tilde{p}}}=N^{(\rho)}_{X[\ell]} -N^{(\rho)}_X$.
By combining the inequalities~\eqref{max_N_X_0:p_tilde_N_X_p} and \eqref{max_N_X_0:p_tilde_N_X_p_Lemma} with Eq.~\eqref{def_f_tau_ell_prop_re}, we obtain 
\begin{align}
&\min_{p\in[1,{\tilde{p}}]} \br{N^{(\rho)}_{X_{0:p}}+ c_{\tau,1} \tilde{N}^{(\rho)}_{X_{0:p}}}\notag \\
&\le \brr{1+ 5c_{\tau,1} \lambda_{1/2} \br{1+\frac{1}{5c_{\tau,1} \lambda_{1/2} + 2}}^{-\ell/2}} ( N^{(\rho)}_{X[\ell]} -N^{(\rho)}_X)
+ N^{(\rho)}_X+c_{\tau,1} \lambda_{1/4}e^{- 3\ell/16}  \br{ N^{(\rho)}_X   +  \sum_{j\in X[\ell]^\co }e^{- 3\dist_{j,X[\ell]}/4} N^{(\rho)}_j  }   \notag \\
&\le \brr{1+ 5c_{\tau,1} \lambda_{1/2} \br{1+\frac{1}{5c_{\tau,1} \lambda_{1/2} + 2}}^{-\ell/2}} N^{(\rho)}_{X[\ell]}   +c_{\tau,1} \lambda_{1/4}e^{- 3\ell/16}  \sum_{j\in X[\ell]^\co }e^{- 3\dist_{j,X[\ell]}/4} N^{(\rho)}_j  .
\end{align}
By applying the above inequality to~\eqref{Start_for_the_proof_prop_upp_Moment_sth} with \eqref{trivial_upp_moment_extend}, we finally obtain
\begin{align}
M_{X}^{(1)} (\tau)=  \tr\brr { \rho  \nb_X(\tau)}
&\le  
\br{1+ 5c_{\tau,1} \lambda_{1/2} e^{-f_\tau \ell} } N^{(\rho)}_{X[\ell]}   +c_{\tau,1} \lambda_{1/4}e^{- 3\ell/16}  \sum_{j\in X[\ell]^\co }e^{- 3\dist_{j,X[\ell]}/4} N^{(\rho)}_j  + c_{\tau,2} \notag \\
&= \tr \brrr{\rho \brr{ \br{1+ 5c_{\tau,1} \lambda_{1/2} e^{-f_\tau \ell} } \nb_{X[\ell]} + c_{\tau,1}e^{-3\ell/16} \hat{\mathscr{D}}_{X[\ell]}+ c_{\tau,2} }} ,
\end{align} 
where we use the notation of $\hat{\mathscr{D}}_{X[\ell]}$ and $f_\tau$ in Eqs.~\eqref{def_delta_X_ell_prop} and \eqref{def_f_tau_ell_prop}.
Because the above inequality holds for an arbitrary quantum state $\rho$, we prove the desired operator inequality~\eqref{Ineq:prop:upp_Moment_1th}. 
This completes the proof. $\square$

\subsubsection{Proof of Lemma~\ref{lmma:tilde_N_X_p_upper_bound}} \label{proof_of_lmma:tilde_N_X_p_upper_bound}

For arbitrary $i\in  \partial (X_{0:p}) \subseteq X_p$ and $j'\in X_j$, we have $\dist_{i,j'}\ge\dist_{X_p,X_j}\ge |p-j|$ (see Fig.~\ref{supp_fig_shell_choice.pdf}), and hence
\begin{align}
\label{prooflmma:tilde_N_X_p_upper_bound_1}
 \sum_{i\in  \partial (X_{0:p})} \sum_{j' \in X_j} e^{-\dist_{i,j'}} N^{(\rho)}_{j'} 
 \le  \sum_{j' \in X_j}  N^{(\rho)}_{j'} \sum_{i: \dist_{i,j'}\ge |p-j|} e^{-\dist_{i,j'}} 
 &\le  e^{-|p-j|/2} N^{(\rho)}_{X_j} \sum_{i: \dist_{i,j'}\ge 0} e^{-\dist_{i,j'}/2} \notag \\
 &\le  \lambda_{1/2} e^{-|p-j|/2} N^{(\rho)}_{X_j},
\end{align} 
where we apply the equation $\sum_{j' \in X_j}  N^{(\rho)}_{j'}=N^{(\rho)}_{X_j}$ in the second inequality and use~\eqref{oftern_used_inequality_lambda_c} with $c=1/2$ in the last inequality. 
We thus reduce $\tilde{N}^{(\rho)}_{X_{0:p}}$ in~\eqref{Start_for_the_proof_prop_upp_Moment_sth} to the form of 
\begin{align}
\label{reduce_tilde_N_X_p_upp}
\tilde{N}^{(\rho)}_{X_{0:p}} 
&=\sum_{i\in  \partial (X_{0:p})} \sum_{j \in \Lambda} e^{-\dist_{i,j}} N^{(\rho)}_j 
= \sum_{i\in  \partial (X_{0:p})} \br{ \sum_{j \in X_1}+\sum_{j \in X_2}+\cdots \sum_{j \in X_{\tilde{p}+1}} +\sum_{j \in X_{1:\tilde{p}+1}^\co}  } e^{-\dist_{i,j}} N^{(\rho)}_j 
\notag \\
&\le \lambda_{1/2}  \sum_{j=1}^{{\tilde{p}}-1} e^{-|p-j|/2} N^{(\rho)}_{X_j} + \lambda_{1/2}\br{e^{-|p-{\tilde{p}}|/2} N^{(\rho)}_{X_{\tilde{p}}} + e^{-|p-{\tilde{p}}-1|/2} N^{(\rho)}_{X_{{\tilde{p}}+1}}} 
+\sum_{i\in  \partial (X_{0:p})}  \sum_{j\in X_{1:{\tilde{p}}+1}^\co} e^{-\dist_{i,j}} N^{(\rho)}_j   \notag \\
&\le \lambda_{1/2} \sum_{j=1}^{{\tilde{p}}} e^{-|p-j|/2} M_j  +\sum_{i\in  \partial (X_{0:p})}  \sum_{j\in X_{1:{\tilde{p}}+1}^\co} e^{-\dist_{i,j}} N^{(\rho)}_j   \quad (1\le p\le {\tilde{p}}),
\end{align} 
with $M_j=N^{(\rho)}_{X_j}$ for $1\le j \le {\tilde{p}}-1$ and $M_{\tilde{p}} = N^{(\rho)}_{X_{\tilde{p}}} +N^{(\rho)}_{X_{{\tilde{p}}+1}}$, 
where in the last inequality we use $e^{-|p-{\tilde{p}}-1|/2} \le e^{-|p-{\tilde{p}}|/2}$ for $p\le \tilde{p}$.

We then aim to upper-bound the second term of the RHS in \eqref{reduce_tilde_N_X_p_upp}.
From the definitions~\eqref{df_X_1_X_j_X_q+1_0}, we obtain
\begin{align}
\label{lower_bound_dist_ij_X_1_X_0p}
& \dist_{i,j} \ge  \ell/4 \for \forall j\in X ,\quad \forall i\in  \partial (X_{0:p})  \quad (1\le p\le {\tilde{p}}), \\
&\dist_{j,X[\ell]} \le \dist_{i,j} - \ell/4 \for \forall j\in X[\ell]^\co ,\quad \forall i\in  \partial (X_{0:p})  \quad (1\le p\le {\tilde{p}}) .
\end{align}
We recall that $X_{1:{\tilde{p}}+1}= X[\ell] \setminus X$ (i.e., $X_{1:{\tilde{p}}+1}^\co = X[\ell]^\co \cup  X$) from the definition~\eqref{df_X_1_X_j_X_q+1_0}. 
Then, from $e^{-\dist_{i,j}} \le e^{-\dist_{i,j}/4 - 3\ell/16}$ for $j\in X$ and $e^{-\dist_{i,j}} \le e^{-\dist_{i,j}/4 - (3/4) (\dist_{j,X[\ell]}+ \ell/4)}$ for $j\in X[\ell]^\co$, we can decompose as 
\begin{align}
\sum_{i\in  \partial (X_{0:p})}  \sum_{j\in X_{1:{\tilde{p}}+1}^\co} e^{-\dist_{i,j}} N^{(\rho)}_j 
&=\sum_{i\in  \partial (X_{0:p})}  \sum_{j\in X} e^{-\dist_{i,j}/4 - 3\ell/16} N^{(\rho)}_j   
+ \sum_{i\in  \partial (X_{0:p})}  \sum_{j\in X[\ell]^\co }e^{-\dist_{i,j}/4 - (3/4) (\dist_{j,X[\ell]}+ \ell/4)} N^{(\rho)}_j 
\notag  \\
&\le  \lambda_{1/4}e^{- 3\ell/16} \br{  \sum_{j\in X} N^{(\rho)}_j   + \sum_{j\in X[\ell]^\co }e^{- 3\dist_{j,X[\ell]}/4} N^{(\rho)}_j  }  ,
\label{reduce_tilde_N_X_p_upp_2nd_term_fin_r}
\end{align} 
where we use the inequality~\eqref{oftern_used_inequality_lambda_c} to derive 
\begin{align}
\sum_{i\in  \partial (X_{0:p})} e^{-\dist_{i,j}/4 } \le \sum_{i\in  \Lambda} e^{-\dist_{i,j}/4 } \le \lambda_{1/4} \for \forall j\in \Lambda .
\end{align} 
By combining the inequalities~\eqref{reduce_tilde_N_X_p_upp} and \eqref{reduce_tilde_N_X_p_upp_2nd_term_fin_r}, 
we prove the main inequality~\eqref{ineq_lmma:tilde_N_X_p_upper_bound}. 
This completes the proof. $\square$
 
{~}

\noindent\hrulefill{\bf [ End of Proof of Lemma~\ref{lmma:tilde_N_X_p_upper_bound}] }

{~}

\subsection{Moment upper bound: higher order} \label{sec:Moment upper bound: higher order}

\subsubsection{Difficulty for the generalization} \label{sec_sub_Bottleneck for the generalization}

In this section, we generalize Proposition~\ref{prop:upp_Moment_1th} to the moment functions with arbitrary orders.
Here, we have to consider
\begin{align}
\tr\br{\nb_X^s \rho(\tau)} = \tr\br{\nb_X(\tau)^s \rho}  .
\end{align}
The problem is precisely the same as the case for the first order moment if the quantum state is given by the Mott state, i.e., $\rho=\ket{\bold{N}} \bra{\bold{N}}$.
In this case, we obtain 
\begin{align}
\tr\br{\nb_X^s \rho(\tau)} &\le \tr \brr{ (\nb_X + c_{\tau,1} \hat{\mathcal{D}}_X + c_{\tau_2} s)^s  \ket{\bold{N}}\bra{\bold{N}} } 
=\br{  \bra{\bold{N}} \br{\nb_X + c_{\tau,1} \hat{\mathcal{D}}_X + c_{\tau_2} s} \ket{\bold{N}}}^s  .
\end{align}
Thus, we can apply Proposition~\ref{prop:upp_Moment_1th} to $ \bra{\bold{N}} \br{\nb_X + c_{\tau,1} \hat{\mathcal{D}}_X+ c_{\tau_2} s }\ket{\bold{N}}$, which yields an upper bound in the form of 
\begin{align}
\tr\br{\nb_X^s \rho(\tau)} &\le  \brr{\bra{\bold{N}} \br{1+ 5c_{\tau,1} \lambda_{1/2} e^{-f_\tau \ell} } \nb_{X[\ell]} + c_{\tau,1}e^{-3\ell/16} \hat{\mathscr{D}}_{X[\ell]}+ c_{\tau,2} s \ket{\bold{N}}}^s  \notag \\
&=  \bra{\bold{N}} \brr{\br{1+ 5c_{\tau,1} \lambda_{1/2} e^{-f_\tau \ell} } \nb_{X[\ell]} + c_{\tau,1}e^{-3\ell/16} \hat{\mathscr{D}}_{X[\ell]}+ c_{\tau,2} s}^s  \ket{\bold{N}}.
\end{align}
However, in general, we have 
\begin{align}
\ave{\nb_X^s}_\rho \ge  \ave{\nb_X}_\rho^s  ,
\end{align}
and hence, we cannot utilize the same analyses to generic quantum states $\rho$ by extending the analysis for 1st order moment.  

As a remark, we note that the following discussion cannot be applied to extend the results for the Mott states to an arbitrary state $\rho$. 
 To see the point, we consider the decomposition of 
 \begin{align}
\tr(\rho(\tau) \nb_X^s) =\tr(\rho\nb_X(\tau)^s) 
&=  \sum_{\bold{N}} \tr \br{    \rho \ket{\bold{N}} \bra{\bold{N}} \brr{\nb_X(\tau)}^s }  \notag \\
&\le \sum_{\bold{N}}   \norm{ \rho \ket{\bold{N}} \bra{\bold{N}}}_1    \norm{ \brr{\nb_X(\tau)}^s  \ket{\bold{N}} }  \notag \\
&= \sum_{\bold{N}} \sqrt{\bra{\bold{N}}\rho^2 \ket{\bold{N}}} \sqrt{\bra{\bold{N}}[\nb_X(t)]^{2s}\ket{\bold{N}} }   ,
\label{start_small_time_evo_connect_not_work}
\end{align}
where we use the H\"older inequality in the  inequality and $\norm{\cdot}_1$ is the trace norm, i.e., $\norm{O}_1 := \tr \br{\sqrt{O^\dagger O}}$. 
Now, we can apply Proposition~\ref{prop:upp_Moment_1th} to $\bra{\bold{N}}[\nb_X(t)]^{2s}\ket{\bold{N}}$, but we cannot ensure 
 \begin{align}
\sum_{\bold{N}} \sqrt{\bra{\bold{N}}\rho^2 \ket{\bold{N}}}  = \orderof{1} .
\end{align}
Indeed, when $\rho$ is given by a random pure state, i.e., $\rho=\ket{\rm rand}\bra{\rm rand}$, we have 
\begin{align}
\sum_{\bold{N}} \sqrt{\bra{\bold{N}}\rho^2 \ket{\bold{N}}}  =  \sum_{\bold{N}} |\langle \bold{N}   \ket{\rm rand}| = e^{\orderof{|\Lambda|}} , 
\end{align}
which leads the inequality~\eqref{start_small_time_evo_connect_not_work} to a meaningless upper bound. 
This point necessitates us to utilize more refined analyses. 

\subsubsection{Main results}

In considering the first order moment, we have considered the upper bound of
\begin{align}
\tr\br{\nb_X(\tau) \rho} \le \min_{X': X' \supseteq X} \tr\br{\nb_{X'}(\tau) \rho}  .
\end{align}
We have aimed to estimate the minimization in the RHS of the above inequality. 
A simple generalization to high-order moments gives 
\begin{align}
\tr\br{ \nb_X(\tau)^s \rho} \le \min_{X': X' \supseteq X} \tr\br{\nb_{X'}(\tau)^s \rho}  . 
\end{align}
However, it is not enough to control a single parameter $X'$ in deriving the desired bound below. 
In deriving a meaningful bound for higher order moments, the key idea here is to utilize the multi-parameter optimization by using the inequality~\eqref{multi_paramt_upper_bound}: 
\begin{align}
\label{multi_paramter_optimization_moment}
\tr\br{\nb_X(\tau)^s \rho} \le \min_{X_1: X_1 \supseteq X}\min_{X_2: X_2 \supseteq X}\cdots \min_{X_s: X_s \supseteq X}  
\tr\brr{ \nb_{X_1}(\tau)\nb_{X_2}(\tau) \cdots \nb_{X_s} (\tau)\rho }  .
\end{align}
Also, in Subtheorem~\ref{Schuch_boson_extend}, we have already obtained a general upper bound for $M_{X_1,X_2,\ldots,X_s}(\tau)$.

Then, by solving the multi-parameter optimization problem~\eqref{multi_paramter_optimization_moment}, we can prove the following proposition:
\begin{prop} \label{prop:upp_Moment_s_th}
We here adopt the same setup as that in Proposition~\ref{prop:upp_Moment_1th}. 
Then, we obtain
\begin{align}
\label{Ineq:prop:upp_Moment_s_th}
[\nb_X(\tau)]^s&\preceq  \brr{  \br{1+ 5c_{\tau,1} \lambda_{1/2} e^{-f_\tau \ell} } \nb_{X[\ell]} + c_{\tau,1}e^{-3\ell/16} \hat{\mathscr{D}}_{X[\ell]}+ c_{\tau,2} s  }^s ,
\end{align}
where $\hat{\mathscr{D}}_{X[\ell]}$ and $f_\tau$ have been defined in Eqs.~\eqref{def_delta_X_ell_prop} and \eqref{def_f_tau_ell_prop}. 
\end{prop}

\subsubsection{Proof of Proposition~\ref{prop:upp_Moment_s_th}}
 
Without loss of generality, we can assume that $\rho$ is given by a pure state, i.e., $\rho=\ket{\psi}\bra{\psi}$.
We note that $O_1 \preceq O_2$ is derived from $\bra{\psi} O_1\ket{\psi}  \preceq \bra{\psi}O_2 \ket{\psi}$ for all the quantum states. 
Then, by using Subtheorem~\ref{Schuch_boson_extend}, the multi-parameter optimization~\eqref{multi_paramter_optimization_moment} reduces to
\begin{align}
\label{multi_paramter_optimization_moment_psi}
\bra{\psi} \nb_X(\tau)^s \ket{\psi} 
& \le \min_{X_1: X_1 \supseteq X}\min_{X_2: X_2 \supseteq X}\cdots \min_{X_s: X_s \supseteq X} \bra{\psi}\nb_{X_1}(\tau)\nb_{X_2}(\tau) \cdots \nb_{X_s} (\tau) \ket{\psi}    \notag\\
&\le \min_{X_1: X_1 \supseteq X}\min_{X_2: X_2 \supseteq X}\cdots \min_{X_s: X_s \supseteq X}  
\Bra{\psi} \prod_{j=1}^s  \br{  \nb_{X_j} + c_{\tau,1}\hat{\mathcal{D}}_{X_j} + c_{\tau,2}s} \Ket{\psi}  \notag \\
&= \min_{X_1: X_1 \supseteq X}\min_{X_2: X_2 \supseteq X}\cdots \min_{X_s: X_s \supseteq X}  
\Bra{\psi} \prod_{j=1}^s  \nb_{\tau, X_j} \Ket{\psi},
\end{align}
where in the last equation we denote $\nb_{X_j} + c_{\tau,1}\hat{\mathcal{D}}_{X_j} + c_{\tau,2}s$ by $\nb_{\tau, X_j}$. 
We note that each of $\{\nb_{\tau, X_j}\}$ commutes with $\nb_i$ for $\forall i\in \Lambda$. 

We define the following unnormalized quantum states $\ket{\psi_{X_{1:j}}}$ ($1\le j\le s$):
\begin{align}
&\ket{\psi_{X_{1:j-1}}} :=\br{\nb_{\tau, X_1} \nb_{\tau, X_2} \cdots \nb_{\tau, X_{j-1}} \hat{\mathcal{N}}_{\tau,X[\ell]}^{s-j}}^{1/2} \ket{\psi} , \notag \\
& \hat{\mathcal{N}}_{\tau,X[\ell]}:= \br{1+ 5c_{\tau,1} \lambda_{1/2} e^{-f_\tau \ell} } \nb_{X[\ell]} + c_{\tau,1}e^{-3\ell/16} \hat{\mathscr{D}}_{X[\ell]}+ c_{\tau,2} s.
\label{def_nb_uta_X_ell}
\end{align}
We then rewrite the optimization problem in \eqref{multi_paramter_optimization_moment_psi} as 
\begin{align}
\label{multi_paramter_optimization_moment_psi_1step}
&\min_{X_1: X_1 \supseteq X}\min_{X_2: X_2 \supseteq X}\cdots \min_{X_s: X_s \supseteq X}  
\Bra{\psi} \prod_{j=1}^s  \nb_{\tau, X_j} \Ket{\psi} = \min_{X_1: X_1 \supseteq X}\min_{X_2: X_2 \supseteq X}\cdots \min_{X_s: X_s \supseteq X} \bra{\psi_{X_{1:s-1}}} \nb_{\tau, X_s} \ket{\psi_{X_{1:s-1}}} .
\end{align}
We can apply Proposition~\ref{prop:upp_Moment_1th} to $\bra{\psi_{X_{1:s-1}}} \nb_{\tau, X_s} \ket{\psi_{X_{1:s-1}}} $, which yields 
\begin{align}
\min_{X_s: X_s \supseteq X} \bra{\psi_{X_{1:s-1}}} \nb_{\tau, X_s} \ket{\psi_{X_{1:s-1}}} 
\le   \bra{\psi_{X_{1:s-1}}} \hat{\mathcal{N}}_{\tau,X[\ell]} \ket{\psi_{X_{1:s-1}}}  
\end{align}
for arbitrary choices of $\{X_j\}_{j=1}^{s-1}$ in $\ket{\psi_{X_{1:s-1}}}$. We then reduce Eq.~\eqref{multi_paramter_optimization_moment_psi_1step} to 
\begin{align}
\label{multi_paramter_optimization_moment_psi_2step}
&\min_{X_1: X_1 \supseteq X}\min_{X_2: X_2 \supseteq X}\cdots \min_{X_s: X_s \supseteq X}  
\Bra{\psi} \prod_{j=1}^s  \nb_{\tau, X_j} \Ket{\psi} \notag \\
& \le  \min_{X_1: X_1 \supseteq X}\min_{X_2: X_2 \supseteq X}\cdots \min_{X_{s-1}: X_{s-1} \supseteq X}  \bra{\psi_{X_{1:s-1}}} \hat{\mathcal{N}}_{\tau,X[\ell]} \ket{\psi_{X_{1:s-1}}}   \notag \\
&=  \min_{X_1: X_1 \supseteq X}\min_{X_2: X_2 \supseteq X}\cdots \min_{X_{s-1}: X_{s-1} \supseteq X}  \bra{\psi_{X_{1:s-2}}} \nb_{\tau, X_{s-1}}\ket{\psi_{X_{1:s-2}}}.
\end{align}
In the same way, we apply Proposition~\ref{prop:upp_Moment_1th} to $\bra{\psi_{X_{1:s-2}}} \nb_{\tau, X_{s-1}} \ket{\psi_{X_{1:s-2}}}$, which further upper-bounds the RHS of the above inequality as 
\begin{align}
&\min_{X_1: X_1 \supseteq X}\min_{X_2: X_2 \supseteq X}\cdots \min_{X_s: X_s \supseteq X}  
\Bra{\psi} \prod_{j=1}^s  \nb_{\tau, X_j} \Ket{\psi} \notag \\
& \le \min_{X_1: X_1 \supseteq X}\min_{X_2: X_2 \supseteq X}\cdots \min_{X_{s-2}: X_{s-2} \supseteq X}  \bra{\psi_{X_{1:s-2}}} \hat{\mathcal{N}}_{\tau,X[\ell]}\ket{\psi_{X_{1:s-2}}} \notag \\
&=  \min_{X_1: X_1 \supseteq X}\min_{X_2: X_2 \supseteq X}\cdots \min_{X_{s-2}: X_{s-2} \supseteq X}  \bra{\psi_{X_{1:s-3}}} \nb_{\tau, X_{s-2}}\ket{\psi_{X_{1:s-3}}} .
\end{align}
By repeating the same process, we finally obtain 
\begin{align}
&\min_{X_1: X_1 \supseteq X}\min_{X_2: X_2 \supseteq X}\cdots \min_{X_s: X_s \supseteq X}  
\Bra{\psi} \prod_{j=1}^s  \nb_{\tau, X_j} \Ket{\psi} 
 \le \bra{\psi_{X_{1:0}}}\hat{\mathcal{N}}_{\tau,X[\ell]} \ket{\psi_{X_{1:0}}} = \bra{\psi}\hat{\mathcal{N}}_{\tau,X[\ell]}^s  \ket{\psi}  .
\end{align}
By applying the definition~\eqref{def_nb_uta_X_ell} to the above inequality, we prove the main inequality~\eqref{Ineq:prop:upp_Moment_s_th}.
This completes the proof. $\square$

\section{Main theorem on the boson concentration}

\begin{figure}[tt]
\centering
\includegraphics[clip, scale=0.37]{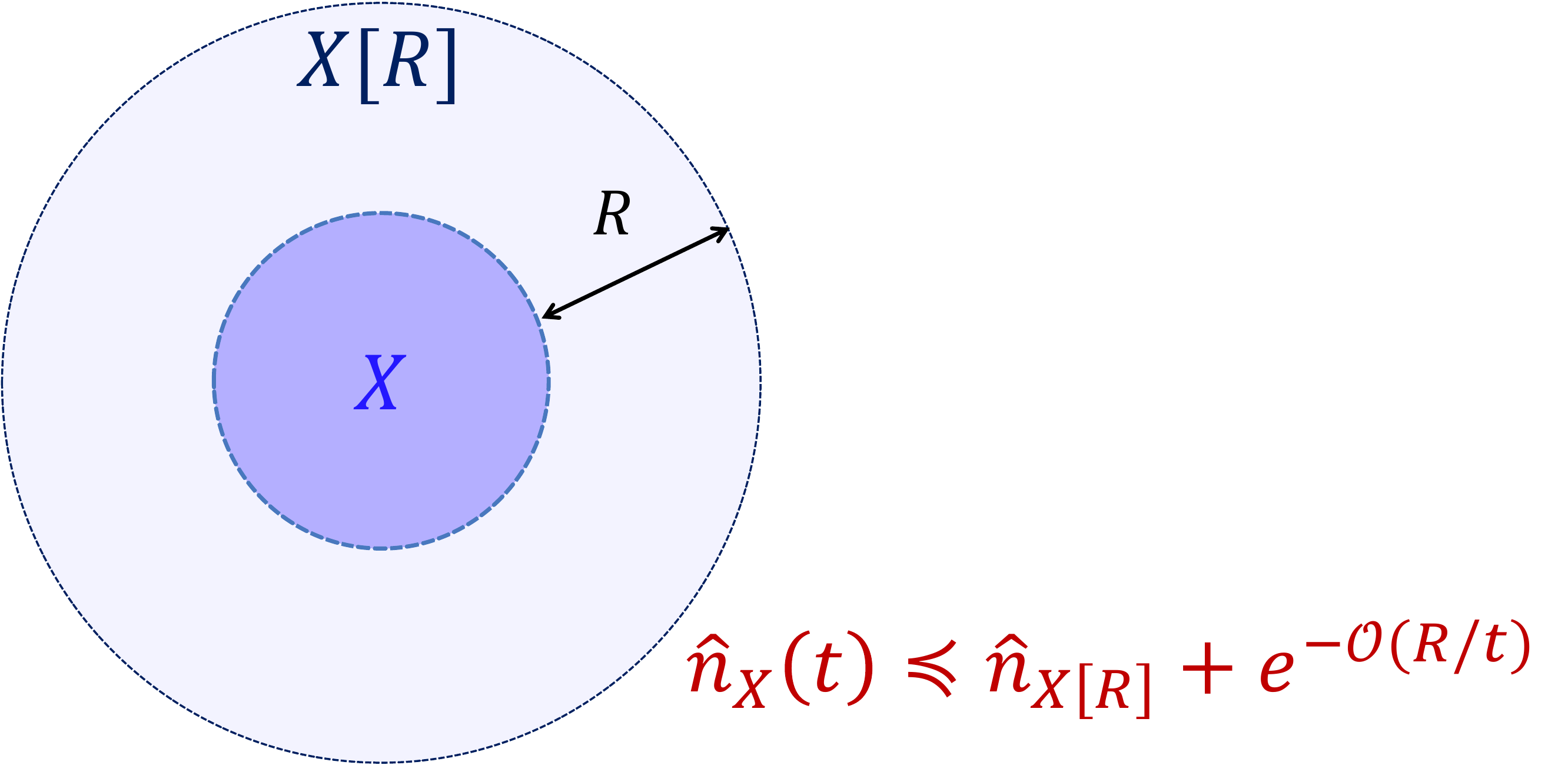}
\caption{Schematic picture of our setup for Theorem~\ref{main_theorem0_boson_concentration}. 
We consider the boson number operator $\nb_X$ in a region $X$.
After a time evolution, $\nb_X(t)$ is upper-bounded by the boson number $\nb_{X[R]}$ on an extended subset $X[R]$, where the error is given by $e^{-\orderofsp{R/t}}$. 
The obtained inequality is given in the form of the operator inequality [see \eqref{main_ineq:main_theorem0_boson_concentration}], and hence it can be applied to an arbitrary initial state. 
}
\label{supp_fig_main_thm_boson_number.pdf}
\end{figure}

In this section, we consider a boson density after a time evolution for an arbitrary time $t$. 
For this purpose, we connect the short-time evolution by iteratively applying Proposition~\ref{prop:upp_Moment_s_th}.
Our goal is to prove that the boson number operator in a region $X$ is upper-bounded by a number operator on an extended region $X[R]$ after a time evolution $e^{-iHt}$, 
where the length $R$ is almost linear with the time $t$ (see Fig.~\ref{supp_fig_main_thm_boson_number.pdf}). 

We here prove the following theorem:
\begin{theorem} \label{main_theorem0_boson_concentration}
Let $\tau$ be the maximum number such that $\tau\le 1/(4\gamma\bar{J})$ and $t/\tau$ becomes integer. 
Also, for $\ell = \floorf{R/(t/\tau)}=\Theta(R/t)$, we assume 
\begin{align}
\label{basic_cond_ell_f_tau}
e^{-f_\tau \ell} \le \frac{\tau}{5c_{\tau,1} \lambda_{1/2} t} 
\longrightarrow  \ell \ge \frac{1}{f_\tau} \log \br{\frac{5c_{\tau,1} \lambda_{1/2} t}{\tau}} ,
\end{align}
where $f_\tau$ has been defined in Eq.~\eqref{def_f_tau_ell_prop}.
We then upper-bound $[\nb_X(t)]^s$ in the form of
\begin{align}
 \label{main_ineq:main_theorem0_boson_concentration}
[\nb_X(t)]^s& \preceq \brr{ \br{1+ \frac{3t}{\tau} \delta_\ell } \nb_{X[R]} +2 c_{\tau,1} e^{-3\ell/16} \hat{\mathscr{D}}_{X[R]}+  \br{\frac{c_{\tau,2}t}{\tau} +\tilde{\delta}_{\ell}}  s }^s 
\end{align}
with $\hat{\mathscr{D}}_{X[R]}=\sum_{j\in X[R]^\co} e^{-3\dist_{j,X[R]}/4} \nb_j
$ as in Eq.~\eqref{def_delta_X_ell_prop} and 
\begin{align}
\delta_\ell := 5c_{\tau,1} \lambda_{1/2} e^{-f_\tau \ell} ,\quad 
\tilde{\delta}_{\ell} := \frac{c_{\tau,2}t}{\tau} \br{ \frac{3t\delta_\ell}{\tau}+ 2 c_{\tau,1} \lambda_{3/4}  e^{-3\ell/16}  |\partial X[R]|}.
\end{align}
\end{theorem}

{\bf Remark.} 
Because of the condition~\eqref{basic_cond_ell_f_tau} and $\ell \propto R/t$, the length $R$ should be as large as 
\begin{align}
R \gtrsim t\log(t).
\end{align}
Then, because of $\delta_\ell =e^{-\Theta(\ell)}$ and $\tilde{\delta}_\ell =e^{-\Theta(\ell)}$, we obtain 
\begin{align}
[\nb_X(t)]^s& \preceq \brr{ \nb_{X[R]} + e^{-\Theta(R/t)} \br{\nb_{X[R]}+\hat{\mathscr{D}}_{X[R]}} + \Theta(st) }^s,
\end{align}
where we use the $\Theta$ notation in Eq.~\eqref{Theta_notation_def}.
Therefore, for sufficiently large $R$, we have $[\nb_X(t)]^s \lesssim \brr{ \nb_{X[R]} + \Theta(st)  }^s $.

\subsection{Proof of Theorem~\ref{main_theorem0_boson_concentration}}

For the proof, we iteratively utilize Proposition~\ref{prop:upp_Moment_s_th}.
We decompose the total time $t$ into $m_0$ pieces:
\begin{align}
m_0 = \frac{t}{\tau} = \orderof{t} ,
\end{align}
where we have chosen the maximum $\tau$ such that $\tau \le 1/(4\gamma \bar{J})$ and $t/\tau$ becomes integer.
In the proof, we consider the case that $\ell= \floorf{R/m_0} = R/m_0$, but the generalization is straightforward. 
For simplicity of notation, we denote $C_0$ by
\begin{align}
\label{Ineq:def_C_0_1}
C_0 := 1+ 5c_{\tau,1} \lambda_{1/2} e^{-f_\tau \ell} = 1+ \delta_\ell , \quad \delta_\ell := 5c_{\tau,1} \lambda_{1/2} e^{-f_\tau \ell} 
\end{align}
We then reduce the inequality~\eqref{Ineq:prop:upp_Moment_s_th} to 
\begin{align}
\label{Ineq:prop:upp_Moment_s_th_C_0}
[\nb_X(\tau)]^s&\preceq  \brr{ C_0 \nb_{X[\ell]} + c_{\tau,1}e^{-3\ell/16} \hat{\mathscr{D}}_{X[\ell]}+ c_{\tau,2} s  }^s 
\end{align}
for arbitrary $X\subseteq \Lambda$ and $\tau\le 1/(4\gamma \bar{J})$.

The condition~\eqref{basic_cond_ell_f_tau} immediately implies the inequality of $\delta_\ell  \le 1/m_0$ from the definition $\delta_\ell := 5c_{\tau,1} \lambda_{1/2} e^{-f_\tau \ell}$.
This yields 
\begin{align}
\label{Ineq:def_C_0_1_cond_basic}
C_0^m = \br{1+ \delta_\ell }^m
&\le 1 + m (e-1) \delta_\ell   \le 1+ 2m \delta_\ell   \for m\le m_0,
\end{align}
where we use $(1+z)^m \le 1+ m (e-1) z$ for $z\le 1/m$. 
Moreover, $\delta_\ell  \le 1/m_0 \le 1$ implies 
\begin{align}
\label{e^-f_tau_ell_inequality}
 e^{-f_\tau \ell} = \br{1+\frac{1}{5c_{\tau,1} \lambda_{1/2} + 2}}^{-\ell/2} \le  \frac{1}{5c_{\tau,1} \lambda_{1/2}}.
\end{align} 
Because $\brr{1+1/\br{5c_{\tau,1} \lambda_{1/2} + 2}}^{1/2} \le (8/7)^{1/2} \le e^{1/14} $ from $c_{\tau,1}\ge 1$ and $\lambda_{1/2}\ge1$, 
we also obtain  
\begin{align}
\label{Ineq:e^-ell/14}
e^{-\ell/14}\le  \frac{1}{5c_{\tau,1} \lambda_{1/2}} .
\end{align} 

For the time-evolved operator $[\nb_X(m\tau)]^s$, we prove the following proposition:
\begin{prop}\label{prop:connect_time_evo_nb_X_m_tau}
For an arbitrary time $m\tau$ ($m\le m_0$), we obtain the operator inequality as 
\begin{align}
[\nb_X(m\tau)]^s \preceq \brr{ \br{1+ 3m \delta_\ell } \nb_{X[m\ell]} +2 c_{\tau,1} e^{-3\ell/16} \hat{\mathscr{D}}_{X[m\ell]}+  \br{c_{\tau,2} m +\tilde{\delta}_{\ell}}  s }^s,
\end{align}
where we define $\tilde{\delta}_{\ell}$ as 
\begin{align}
\tilde{\delta}_{\ell}:=\frac{c_{\tau,2}t}{\tau} \br{ \frac{3t\delta_\ell}{\tau}+ 2 c_{\tau,1} \lambda_{3/4}  e^{-3\ell/16}  |\partial X[R]|} .
\end{align} 
\end{prop}

From Proposition~\ref{prop:connect_time_evo_nb_X_m_tau} with $m=m_0$ (i.e., $m\tau=m_0\tau =t$), we obtain
\begin{align}
[\nb_X(t)]^s &\preceq \brr{ \br{1+ 3m_0 \delta_\ell } \nb_{X[m_0\ell]} +2 c_{\tau,1} e^{-3\ell/16} \hat{\mathscr{D}}_{X[m_0\ell]}+  \br{c_{\tau,2} m_0 +\tilde{\delta}_{\ell}}  s }^s \notag \\
&= \brr{ \br{1+ \frac{3t}{\tau} \delta_\ell } \nb_{X[R]} +2 c_{\tau,1} e^{-3\ell/16} \hat{\mathscr{D}}_{X[R]}+  \br{\frac{c_{\tau,2}t}{\tau} +\tilde{\delta}_{\ell}}  s }^s ,
\end{align}
where we use $m_0\ell = R$ and $m_0=t/\tau$. 
This proves the main inequality~\eqref{main_ineq:main_theorem0_boson_concentration}. $\square$

\subsection{Proof of Proposition~\ref{prop:connect_time_evo_nb_X_m_tau}}

For the proof, we upper-bound $[\nb_X(m\tau)]^s$ in the form of 
\begin{align}
\label{proof_start_boson_concent_n_X_m}
[\nb_X(m\tau)]^s \preceq \brr{ C_{m,0} \nb_{X[m\ell]} + C_{m,1} \hat{\mathscr{D}}_{X[m\ell]}+ C_{m,2} s }^s.
\end{align}
For $m=0$, we trivially obtain the inequality~\eqref{proof_start_boson_concent_n_X_m} by choosing 
\begin{align}
\label{C_0_0,1,2_eq}
C_{0,0}= 1 ,\quad C_{0,1}=C_{0,2} = 0. 
\end{align}
For $m=1$, the parameters are given by  
\begin{align}
C_{1,0}= C_0 ,\quad C_{0,1}=c_{\tau,1}e^{-3\ell/16} ,\quad  C_{0,2} = c_{\tau,2}.  
\end{align}
In the following, we upper-bound the parameters $\{C_{m+1,0},C_{m+1,1},C_{m+1,2}\}$ under the inequality of~\eqref{proof_start_boson_concent_n_X_m} for a fixed $m$. 

We start from the time evolution of 
\begin{align}
[\nb_X(m\tau + \tau)]^s = e^{iH\tau} [\nb_X(m\tau)]^s e^{- iH\tau}
\preceq \brr{ C_{m,0} \nb_{X[m\ell]} (\tau)+ C_{m,1} \hat{\mathscr{D}}_{X[m\ell]}(\tau)+ C_{m,2} s }^s ,
\end{align}
where we use $U^\dagger O_1 U \preceq U^\dagger O_2 U$ for an arbitrary unitary operator $U$ if $O_1  \preceq O_2$.
By using Proposition~\ref{prop:upp_Moment_s_th}, we obtain
\begin{align}
&\brr{ C_{m,0} \nb_{X[m\ell]}(\tau) + C_{m,1} \hat{\mathscr{D}}_{X[m\ell]}(\tau) + C_{m,2} s }^s 
\preceq \Biggl\{ C_{m,0}  \brr{C_0 \nb_{X[m\ell+\ell]} + c_{\tau,1} e^{-3\ell/16} \hat{\mathscr{D}}_{X[m\ell+\ell]}+ c_{\tau,2} s }  \notag \\
&\quad \quad \quad \quad \quad \quad\quad \quad \quad \quad \quad \quad
\quad \quad \quad \quad   \quad \quad+ C_{m,1} \sum_{j\in X[m\ell]^\co} e^{-3\dist_{j,X[m\ell]}/4} \br{ \nb_j  + c_{\tau,1}\hat{\mathcal{D}}_j + c_{\tau,2} s}  + C_{m,2} s 
\Biggr\}^s,
\label{mth_time_evo_upp_C_m_1_2}
\end{align}
where we apply Subtheorem~\ref{Schuch_boson_extend} to $\hat{\mathscr{D}}_{X[m\ell]}(\tau) = \sum_{j\in X[m\ell]^\co} e^{-3\dist_{j,X[m\ell]}/4} \nb_j (\tau)$.
First of all, we consider the coefficient of $s$ in \eqref{mth_time_evo_upp_C_m_1_2}, which is given by
\begin{align}
c_{\tau,2} C_{m,0}  + C_{m,2}+  c_{\tau,2}  C_{m,1} \sum_{j\in X[m\ell]^\co} e^{-3\dist_{j,X[m\ell]}/4}
&\preceq  c_{\tau,2} C_{m,0}  + C_{m,2}+  c_{\tau,2}  C_{m,1} \lambda_{3/4} |\partial X[m\ell]|, 
\label{upp_C_m+1_2}
\end{align}
where the inequality is derived from
\begin{align}
\sum_{j\in X[m\ell]^\co} e^{-3\dist_{j,X[m\ell]}/4} 
\le  \sum_{j\in X[m\ell]^\co} \sum_{i\in \partial X[m\ell]} e^{-3\dist_{i,j}/4} 
\le \sum_{i\in \partial X[m\ell]}  \sum_{j\in \Lambda} e^{-3\dist_{i,j}/4}  
\le \lambda_{3/4} |\partial X[m\ell]|  .
\end{align}

We next consider the following term in the inequality~\eqref{mth_time_evo_upp_C_m_1_2}: 
\begin{align}
 C_{m,1}  \sum_{j\in X[m\ell]^\co} e^{-3\dist_{j,X[m\ell]}/4} \br{ \nb_j  + c_{\tau,1}\hat{\mathcal{D}}_j }   
=&\  C_{m,1}\sum_{j\in X[m\ell]^\co} e^{-3\dist_{j,X[m\ell]}/4}  \br{ \nb_j  + c_{\tau,1}\sum_{j'\in \Lambda} e^{-\dist_{j,j'}} \nb_{j'}     }  \notag \\
\preceq & \ C_{m,1}   \sum_{j\in \Lambda} e^{-3\dist_{j,X[m\ell]}/4} 
\br{ 1 + c_{\tau,1}  \lambda_{1/4}} \nb_j , 
\label{C_m_1_summation_exp_ctau_1_D_j}
\end{align}
where we use the inequality of $\dist_{j,X}+\dist_{j,j'} \ge \dist_{j',X}$ for $\forall X\subset \Lambda$, which yields
\begin{align}
\sum_{j\in \Lambda} e^{-3\dist_{j,X[m\ell]}/4}  \sum_{j'\in \Lambda} e^{-\dist_{j,j'}}\nb_{j'} 
&\le \sum_{j\in \Lambda} \sum_{j'\in \Lambda} e^{-3\dist_{j',X[m\ell]}/4}  e^{-\dist_{j,j'}/4}\nb_{j'}   \le \lambda_{1/4} \sum_{j'\in \Lambda} e^{-3\dist_{j',X[m\ell]}/4} \nb_{j'} .
\end{align}
We further decompose 
\begin{align}
&\sum_{j\in \Lambda} e^{-3\dist_{j,X[m\ell]}/4}  \nb_j  
=   \sum_{j\in X[m\ell+\ell]}  e^{-3\dist_{j,X[m\ell]}/4}  \nb_j  
+  \sum_{j\in X[m\ell+\ell]^\co}    e^{-3\dist_{j,X[m\ell]}/4}  \nb_j  . \label{decomposition_X_m_ell_exp_nb_j}
\end{align}
For the first term in Eq.~\eqref{decomposition_X_m_ell_exp_nb_j}, we immediately obtain 
\begin{align}
 \sum_{j\in X[m\ell+\ell]}  e^{-3\dist_{j,X[m\ell]}/4}  \nb_j    \le 
 \sum_{j\in X[m\ell+\ell]}  \nb_j = \nb_{X[m\ell+\ell]}    .
 \label{decomposition_X_m_ell_exp_nb_j_1st}
\end{align}
For the second term in Eq.~\eqref{decomposition_X_m_ell_exp_nb_j}, by using 
$ \dist_{j,X[m\ell]}=\dist_{j,X[m\ell+\ell]}  + \ell$ for $j\in X[m\ell+\ell]^\co$, we have 
\begin{align}
\sum_{j\in X[m\ell+\ell]^\co}    e^{-3\dist_{j,X[m\ell]}/4}  \nb_j 
=e^{-3\ell/4}  \sum_{j\in X[m\ell+\ell]^\co}    e^{-3\dist_{j,X[m\ell+\ell]}/4}  \nb_j  
=e^{-3\ell/4} \hat{\mathscr{D}}_{X[m\ell+\ell]} .
\label{decomposition_X_m_ell_exp_nb_j_2nd}
\end{align}
Then, by combining Eqs.~\eqref{decomposition_X_m_ell_exp_nb_j},~\eqref{decomposition_X_m_ell_exp_nb_j_1st}, and \eqref{decomposition_X_m_ell_exp_nb_j_2nd} with the inequality~\eqref{C_m_1_summation_exp_ctau_1_D_j}, we obtain
\begin{align}
 C_{m,1}  \sum_{j\in X[m\ell]^\co} e^{-3\dist_{j,X[m\ell]}/4} \br{ \nb_j  + c_{\tau,1}\hat{\mathcal{D}}_j }   
\preceq &\ C_{m,1}   \br{ 1 + c_{\tau,1}\lambda_{1/4}}  \br{ \nb_{X[m\ell+\ell]} +e^{-3\ell/4} \hat{\mathscr{D}}_{X[m\ell+\ell]}  } . 
\label{C_m_1_summation_exp_ctau_1_D_j_fin}
\end{align}

Finally, by applying the inequalities~\eqref{upp_C_m+1_2} and~\eqref{C_m_1_summation_exp_ctau_1_D_j_fin} to~\eqref{mth_time_evo_upp_C_m_1_2}, 
we arrive at the inequality of
\begin{align}
&\brr{ C_{m,0} \nb_{X[m\ell]}(\tau) + C_{m,1} \hat{\mathscr{D}}_{X[m\ell]}(\tau) + C_{m,2} s }^s \notag \\
&\preceq \Biggl\{ \brr{C_0C_{m,0} +C_{m,1} \br{ 1 + c_{\tau,1}\lambda_{1/4}} } \nb_{X[m\ell+\ell]}
+\brr{ c_{\tau,1} e^{-3\ell/16} + C_{m,1} \br{ 1 + c_{\tau,1}\lambda_{1/4}}  e^{-3\ell/4}}    \hat{\mathscr{D}}_{X[m\ell+\ell]} \notag \\
&\quad  +s \br{c_{\tau,2} C_{m,0}  + C_{m,2}+  c_{\tau,2}  C_{m,1} \lambda_{3/4} |\partial X[m\ell]|  } 
\Biggr\}^s. 
\label{mth_time_evo_upp_C_m_1_2_fin}
\end{align}
The above inequality yields
 \begin{align}
&C_{m+1,0} = C_0C_{m,0} +C_{m,1} \br{ 1 + c_{\tau,1}\lambda_{1/4}}, \notag \\
&C_{m+1,1} =c_{\tau,1} e^{-3\ell/16} + C_{m,1} \br{ 1 + c_{\tau,1}\lambda_{1/4}}  e^{-3\ell/4} , \notag \\
&C_{m+1,2} =c_{\tau,2} C_{m,0}  + C_{m,2}+  c_{\tau,2}  C_{m,1} \lambda_{3/4} |\partial X[m\ell]|  .
\label{eq_m+1_C_m012}
\end{align}
For an arbitrary numerical sequence $\{a_m\}$ such that $a_{m+1} = b_1 a_m + b_2$, we have 
\begin{align}
a_m = \frac{b_2}{1-b_1} + b_1^m \br{a_0-\frac{b_2}{1-b_1}  } \for \forall m . \label{sequence_explicit_eq_uation}
\end{align}
We first derive the explicit form of $C_{m,1}$. 
For the second equation in~\eqref{eq_m+1_C_m012}, we use Eq.~\eqref{sequence_explicit_eq_uation} with $C_{0,1}=0$ as in Eq.~\eqref{C_0_0,1,2_eq} to derive
 \begin{align}
C_{m,1} =\brrr{1- \brr{ \br{ 1 + c_{\tau,1}\lambda_{1/4}}  e^{-3\ell/4}}^m} \frac{c_{\tau,1} e^{-3\ell/16}}{1- \br{ 1 + c_{\tau,1}\lambda_{1/4}} e^{-3\ell/4}  }  
\le  2 c_{\tau,1} e^{-3\ell/16} \le 2 c_{\tau,1}e^{-2f_\tau \ell}  ,  
\label{C_m_1upp_bound}
\end{align} 
where we use the inequality~\eqref{Ineq:e^-ell/14} to derive $1- \br{ 1 + c_{\tau,1}\lambda_{1/4}} e^{-3\ell/4}  \ge 1/2$ 
and use $e^{-f_\tau} > e^{-1/14}$ to derive $e^{-3\ell/16} \le e^{-2f_\tau \ell}$.  

We then reduce the first equation in~\eqref{eq_m+1_C_m012} to
 \begin{align}
C_{m+1,0}  \le C_0C_{m,0} +2 c_{\tau,1}e^{-2f_\tau \ell} \br{ 1 + c_{\tau,1}\lambda_{1/4}},
\end{align}
which yields from $C_{0,0}=1$ in Eq.~\eqref{C_0_0,1,2_eq}
 \begin{align}
C_{m,0} &\le C_0^m + \frac{2 c_{\tau,1}e^{-2f_\tau \ell}\br{ 1 + c_{\tau,1}\lambda_{1/4}}}{1-C_0} (1- C_0^m ) \notag \\
&\le  1+2m \brr{ \delta_\ell +  2 c_{\tau,1}e^{-2f_\tau \ell} \br{ 1 + c_{\tau,1}\lambda_{1/4}}} \le  1+ 3m \delta_\ell  .
\label{C_m_0upp_bound}
\end{align}
 In the second inequality above, we use the inequality~\eqref{Ineq:def_C_0_1_cond_basic} and the following upper bound: 
 \begin{align}
\frac{C_0^m-1}{C_0-1} = \sum_{m'=0}^{m-1} C_0^{m'} \le m + 2 \delta_\ell \sum_{m'=0}^{m-1} m'  = m+  \delta_\ell m(m-1) \le 2m ,
\end{align} 
Note that we have $\delta_\ell  \le 1/m_0 \le 1/m$. 
In the third inequality of~\eqref{C_m_0upp_bound}, we use the definition of $\delta_\ell=5c_{\tau,1} \lambda_{1/2} e^{-f_\tau \ell}$ and the inequality~\eqref{e^-f_tau_ell_inequality}.
Finally, by using the inequalities~\eqref{C_m_1upp_bound} and \eqref{C_m_0upp_bound}, we reduce the third equation in~\eqref{eq_m+1_C_m012} to
 \begin{align}
C_{m+1,2} &\le c_{\tau,2}\br{1+ 3m\delta_\ell  } + C_{m,2}+  2 c_{\tau,1}   c_{\tau,2}  e^{-3\ell/16} \lambda_{3/4} |\partial X[m\ell]|   \notag \\
&\le C_{m,2}+  c_{\tau,2} \brr{1+ 3m\delta_\ell+ 2 c_{\tau,1} \lambda_{3/4}  e^{-3\ell/16}  |\partial X[R]| } ,  
\end{align}
where we use $m\le m_0$ and $m\ell \le R$. 
The above inequality immediately gives
 \begin{align}
C_{m,2} \le c_{\tau,2} m +c_{\tau,2} m_0\br{ 3m_0\delta_\ell+ 2 c_{\tau,1} \lambda_{3/4}  e^{-3\ell/16}  |\partial X[R]| }=c_{\tau,2} m +\tilde{\delta}_{\ell} ,
\label{C_m_2upp_bound}
\end{align}
where we use $C_{0,2}=0$. 

By using the inequalities~\eqref{C_m_1upp_bound}, \eqref{C_m_0upp_bound} and \eqref{C_m_2upp_bound}, we reduce the upper bound~\eqref{proof_start_boson_concent_n_X_m} to
\begin{align}
\label{proof_start_boson_concent_n_X_m_fin}
[\nb_X(m\tau)]^s \preceq \brr{ \br{1+ 3m \delta_\ell } \nb_{X[m\ell]} +2 c_{\tau,1} e^{-3\ell/16} \hat{\mathscr{D}}_{X[m\ell]}+  \br{c_{\tau,2} m +\tilde{\delta}_{\ell}}  s }^s.
\end{align}
We thus prove Proposition~\ref{prop:connect_time_evo_nb_X_m_tau}. $\square$


\subsection{Boson density after time evolution for an initial state with low-boson density}

\begin{figure}[tt]
\centering
\includegraphics[clip, scale=0.37]{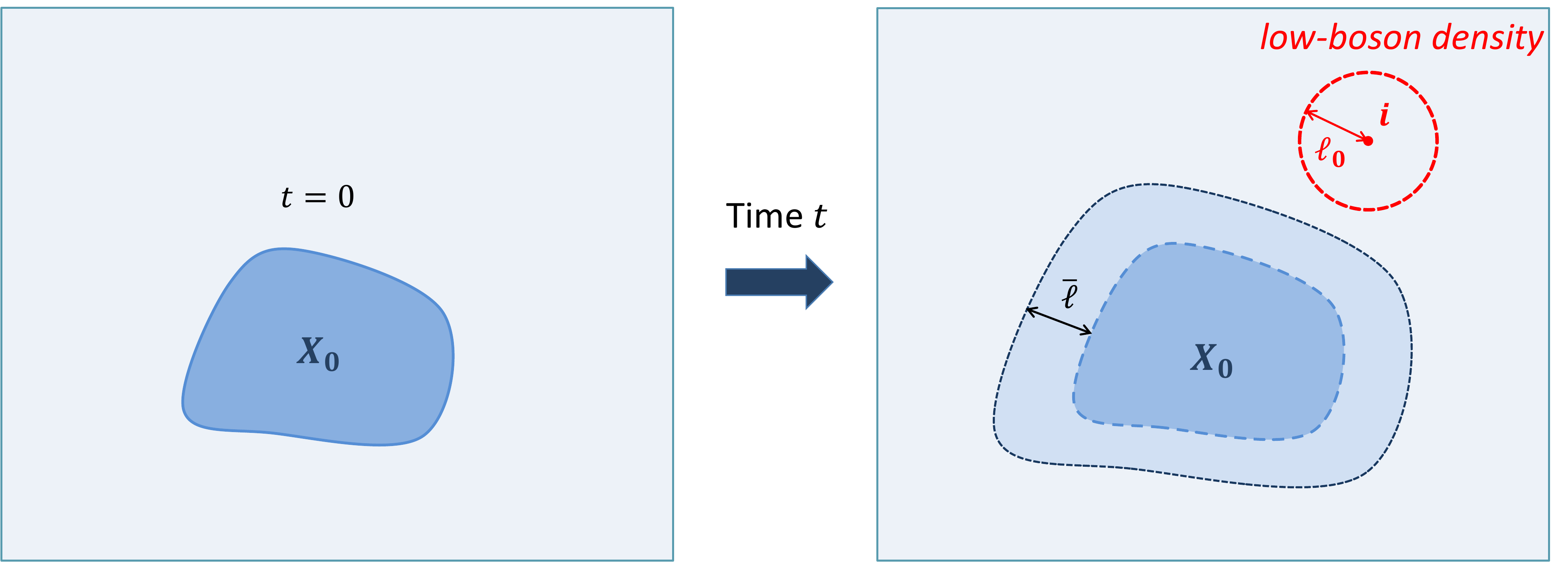}
\caption{Schematic picture of our setup for Proposition~\ref{main_theorem_boson_concentration}. 
We here consider an initial state where the low-boson density condition in Definition~\ref{only_the_assumption_initial} is satisfied outside the region $X_0$. 
After a time evolution, the concentration of the boson density may occur. 
Around the region $X_0$ (or in the extended region $X_0[\bar{\ell}]$), we cannot ensure that the boson density should be suppressed.
On the other hand, in the region $X_0[\bar{\ell}]^\co$, we can prove that 
the number of bosons in a ball region $i[\ell_0]$ obeys the (sub)exponentially decaying probability distribution as long as $\ell_0$ satisfies~\eqref{main_ineq:main_theorem_boson_concentration_condition}.
In our theorem, we prove that the length of $\bar{\ell}$ is roughly given by $t\log(t)$ [see~\eqref{cond_main_theorem_boson_concentration}]. 
}
\label{supp_fig_Setup_boson_concentration.pdf}
\end{figure}

We apply Theorem~\ref{main_theorem0_boson_concentration} to a boson density after a time evolution for an initial state $\rho_0$, which satisfies the following conditions: 
\begin{align}
&\tr\br{\rho_0 \nb_{X_0[r]}^s} \le \frac{1}{e} \brr{ \frac{b_0}{e} \br{q_0+s^\kappa q_1\max(r^D,1) |X_0|}}^s  ,
\label{assump_main_theorem_boson_concentration_inX_0} \\
&\tr\br{\rho_0 \nb_{i[\ell_0]}^s} \le \frac{1}{e} \br{\frac{b_0}{e} s^\kappa \ell_0^D}^s  \for  i[\ell_0] \in X_0^\co, \quad \forall \ell_0 \ge \bar{\ell}_0,
\label{assump_main_theorem_boson_concentration}
\end{align}
where $b_0$ and $\kappa$ ($\ge 1$) are constants of $\orderof{1}$.
In this setup, unlike the condition in Definition~\ref{only_the_assumption_initial}, the boson density can be large only in the region $X_0$. 
Our purpose is to prove that the boson density around the site $i$ is still small after a time evolution if $i$ is sufficiently separated from the region $X$ (see Fig.~\ref{supp_fig_Setup_boson_concentration.pdf}). 

We here prove the following corollary:
\begin{prop} \label{main_theorem_boson_concentration}
Let us consider an initial state $\rho_0$ that satisfies the assumptions~\eqref{assump_main_theorem_boson_concentration_inX_0} and \eqref{assump_main_theorem_boson_concentration}.
Then, under the condition of 
\begin{align}
\label{main_ineq:main_theorem_boson_concentration_condition}
\ell_0\ge \max(\bar{\ell}_0, 2D \bar{\ell}),
\end{align}
we obtain 
\begin{align}
\label{main_ineq:main_theorem_boson_concentration}
&\tr\brr{\rho_0(t) \nb_{i[\ell_0]}^s} \le \frac{1}{e} \br{\frac{2b_0}{e} s^\kappa \ell_0^D}^s   \for i[\ell_0]\in X_0[\bar{\ell}]^\co ,
\end{align}
and 
\begin{align}
\label{main_ineq:main_theorem_boson_concentration_X}
\tr\brr{\rho_0(t) \nb_{X_0[\ell_0]}^s} \le \frac{1}{e} \brr{\frac{2b_0}{e} \br{q_0+ s^\kappa q_1 \ell_0^D |X_0|}}^s ,  
\end{align}
where $\bar{\ell}$ is given by 
\begin{align}
\label{cond_main_theorem_boson_concentration}
\bar{\ell} =  t \log [\Theta( t q_0 q_1 |X_0| )] . 
\end{align}
\end{prop}

{\bf Remark.} 
From the inequality~\eqref{main_ineq:main_theorem_boson_concentration} with $s=1$, roughly speaking, the average boson number in a region $i[\bar{\ell}]$ (i.e., $s=1$) is proportional to $\bar{\ell}^D$, which gives 
$$\frac{1}{|i[\ell_0]|}\tr\brr{\rho_0(t) \nb_{i[\ell_0]}^s} = \orderof{1}.$$  

In particular, if we have no constraints in the region $X_0$, we can simplify Proposition~\ref{main_theorem_boson_concentration}.
Here, we obtain the inequality~\eqref{main_ineq:main_theorem_boson_concentration} for arbitrary $i\in \Lambda$ with
 \begin{align}
\label{cond_main_theorem_boson_concentration_X_0=empty}
\bar{\ell} =  t \log [\Theta(t)] . 
\end{align}

\subsubsection{Proof of Proposition~\ref{main_theorem_boson_concentration}}

Let us consider an arbitrary operator $Q$ such that 
\begin{align}
Q =\nu_X \nb_X  + \sum_{i\in X^\co} \nu_i \nb_i ,
\end{align}
where $X$ is arbitrarily chosen and we assume $\nu_X\ge 0$ and $\nu_i \ge 0$ ($i\in X^\co$).
Then, for arbitrary quantum state $\rho$, we have 
\begin{align}
\tr \br{\rho Q^s }  = \tr\brr{ \rho \br{\nu_X \nb_X  + \sum_{i\in X^\co} \nu_i \nb_i }^s  } 
 \le  \br{\nu_X\ave{\nb_X^s}_\rho^{1/s} + \sum_{i\in \Lambda} \nu_i  \ave{\nb_i^s}_\rho^{1/s} }^s    ,
 \label{rho_Q^s_upp_bound_inq}
\end{align}
where we use the generalized H\"older inequality~\cite{Finner1992} to obtain
\begin{align}
\label{generalized Holder_ineq}
\tr\br{ \rho \nb_{X_1}\nb_{X_2}\cdots \nb_{X_s} } \le \prod_{j=1}^s  \ave{\nb_{X_j}^s}_\rho^{1/s} .
\end{align}
Note that $\ave{\nb_i^s}_\rho$ means $\ave{\nb_i^s}_\rho=\tr\br{\rho \nb_i^s}$.

We here apply the inequality~\eqref{rho_Q^s_upp_bound_inq} to $\tr\brr{\rho_0(t) \nb_X^s}= \tr\brrr{\rho_0[\nb_X(t)]^s}$.
We first use Proposition~\ref{prop:upp_Moment_s_th} to derive 
\begin{align}
\tr\brr{\rho_0(t) \nb_X^s} \le \tr \brrr{ \rho_0 \brr{ \br{1+ \frac{3t}{\tau} \delta_\ell } \nb_{X[R]} +2 c_{\tau,1} e^{-3\ell/16} \hat{\mathscr{D}}_{X[R]}+  \br{\frac{c_{\tau,2}t}{\tau} +\tilde{\delta}_{\ell}}  s }^s  } 
\end{align}
with $\ell = \Theta(R/t)$, 
where we choose $R$ such that the condition~\eqref{basic_cond_ell_f_tau} is satisfied, i.e., $R\gtrsim t\log(t)$. 
The inequality~\eqref{rho_Q^s_upp_bound_inq} reduces the above inequality
\begin{align}
\label{upp_moment_bound_s_1/s}
\brrr{ \tr\brr{\rho_0(t) \nb_X^s} }^{1/s} 
 \le   \br{1+ \frac{3t}{\tau} \delta_\ell } \ave{\nb_{X[R]}^s}_{ \rho_0}^{1/s} +2 c_{\tau,1} e^{-3\ell/16} \ave{\hat{\mathscr{D}}_{X[R]}^s}_{ \rho_0}^{1/s}+  \br{\frac{c_{\tau,2}t}{\tau} +\tilde{\delta}_{\ell}}  s  .
\end{align}

We now consider the case of $X=i[\ell_0]$ such that $i[\ell_0+R] \cap X_0=\emptyset$ and $\ell_0 \ge 2D R$.
From the conditions~\eqref{assump_main_theorem_boson_concentration_inX_0} and \eqref{assump_main_theorem_boson_concentration}, 
we obtain 
\begin{align}
\ave{\nb_{i[\ell_0+R]}^s}_{\rho_0}^{1/s} \le  \frac{1}{e^{1/s}} \frac{b_0}{e} s^\kappa (\ell_0+R)^D,
\end{align}
and 
\begin{align}
 \ave{\hat{\mathscr{D}}_{i[\ell_0+R]}^s}_{ \rho_0}^{1/s}
 & \le \ave{\nb_{X_0}^s}_{\rho_0}^{1/s} + 
\ave{ \br{ \sum_{j\in i[\ell_0+R]^\co \setminus X_0} e^{-3\dist_{j,i[\ell_0+R]}/4} \nb_j}^s}_{\rho_0}^{1/s} \notag \\
&\le \frac{1}{e^{1/s}}\frac{b_0}{e}( q_0+s^\kappa q_1 |X_0| )+\frac{1}{e^{1/s}} \br{ \frac{b_0}{e} s^\kappa \ell_0^D} e^{3/2} \brr{1+2^DD! (4/3)^{D+1}},
\end{align}
where we use $R\le \ell_0$ from $R\le \ell_0/(2D)$ and 
\begin{align}
&\ave{ \br{ \sum_{j\in i[\ell_0+R]^\co \setminus X_0} e^{-3\dist_{j,i[\ell_0+R]}/4} \nb_j}^s}_{\rho_0}^{1/s}
\le \sum_{m=1}^\infty  e^{-3(m-1)\ell_0/4}  \ave{\nb_{i[(m+1)\ell_0+R]\setminus i[m\ell_0+R]}^s}_{\rho_0}^{1/s}  \notag \\
&\le \sum_{m=1}^\infty  e^{-3(m-1)/4}  \ave{\nb_{i[(m+2)\ell_0]}^s}_{\rho_0}^{1/s} 
\le  \frac{1}{e^{1/s}} \br{ \frac{b_0}{e} s^\kappa \ell_0^D} \sum_{m=1}^\infty  e^{-3(m-1)/4}  (m+2)^D \notag \\
&\le \frac{1}{e^{1/s}} \br{ \frac{b_0}{e} s^\kappa \ell_0^D} \sum_{m=1}^\infty  e^{-3(m+2)/4 + 9/4}  (m+2)^D
\le \frac{1}{e^{1/s}} \br{ \frac{b_0}{e} s^\kappa \ell_0^D} e^{3/2} \brr{1+2^DD! (4/3)^{D+1}}.
\end{align}
By using the above inequalities, we have  
\begin{align}
&\brrr{\tr\brr{\rho_0(t) \nb_{i[\ell_0]}^s}}^{1/s}  \notag \\
&\le \frac{1}{e^{1/s}} \br{\frac{b_0}{e} s^\kappa \ell_0^D} \Biggl [ \br{1+ \frac{3t}{\tau} \delta_{\ell}}(1+R/\ell_0)^D  +2 c_{\tau,1} e^{-3\ell/16} 
\br{\frac{q_0s^{-\kappa}+ q_1 |X_0| }{\ell_0^D} +  e^{3/2} \brr{1+2^DD! (4/3)^{D+1}} }\notag \\
&\quad + \frac{e^{1+1/s}}{b_0 \ell_0^D}  \br{\frac{c_{\tau,2}t}{\tau} +\tilde{\delta}_{\ell}}s^{1-\kappa}  \Biggr] .
\end{align}

By using the $\Theta$ notation in Eq.~\eqref{Theta_notation_def}, we can obtain  
\begin{align}
&\br{1+ \frac{3t}{\tau} \delta_{\ell}}(1+R/\ell_0)^D  +2 c_{\tau,1} e^{-3\ell/16} 
\br{\frac{q_0s^{-\kappa}+ q_1 |X_0| }{\ell_0^D} + e^{3/2} \brr{1+2^DD! (4/3)^{D+1}} }  +\frac{e^{1+1/s}}{b_0 \ell_0^D}  \br{\frac{c_{\tau,2}t}{\tau} +\tilde{\delta}_{\ell}}s^{1-\kappa} \notag \\
&\le \br{1+ \Theta(t) e^{-\Theta(R/t)}} (1+R/\ell_0)^D + e^{-\Theta(R/t)} \Theta\br{q_0, q_1 |X_0|} .
\end{align}
Thus, under the assumption of $R/\ell_0 \le 1/(2D)$ (or $\ell_0 \ge 2D R$), we can find a length $R$ such that
\begin{align}
\br{1+ \Theta(t) e^{-\Theta(R/t)}} (1+R/\ell_0)^D + e^{-\Theta(R/t)} \Theta\br{q_0, q_1 |X_0|}  \le 2 ,
\end{align}
which is achieved by choosing $R= \bar{\ell}$ with  
\begin{align}
\label{choice_bar_ell}
\bar{\ell} = t \log [\Theta( t q_0 q_1 |X_0| )] .
\end{align}
We thus prove the first inequality~\eqref{main_ineq:main_theorem_boson_concentration}. 

We then consider the case where $X=X_0[\ell_0]$. 
From the condition~\eqref{assump_main_theorem_boson_concentration_inX_0}, 
we obtain 
\begin{align}
\ave{\nb_{X_0[\ell_0+R]}^s}_{\rho_0}^{1/s} \le \frac{1}{e}\brr{\frac{b_0}{e}\br{q_0+s^\kappa q_1(\ell_0+R)^D |X_0|}},
\end{align}
and 
\begin{align}
 \ave{\hat{\mathscr{D}}_{X_0[\ell_0+R]}^s}_{ \rho_0}^{1/s}
 & \le \ave{ \br{ \sum_{j\in X_0[\ell_0+R]^\co} e^{-3\dist_{j,X_0[\ell_0+R]}/4} \nb_j}^s}_{\rho_0}^{1/s} \notag \\
&\le \frac{1}{e^{1/s}} \brr{\frac{b_0}{e}\br{q_0+s^\kappa q_1\ell_0^D |X_0|}}   e^{3/2} \brr{1+2^DD! (4/3)^{D+1}},
\end{align}
where we use 
\begin{align}
&\ave{ \br{\sum_{j\in X_0[\ell_0+R]^\co}  e^{-3\dist_{j,X_0[\ell_0+R]}/4} \nb_j}^s}_{\rho_0}^{1/s}
\le \sum_{m=1}^\infty  e^{-3(m-1)/4}  \ave{\nb_{X_0[(m+1)\ell_0+R]\setminus X_0[m\ell_0+R]}^s}_{\rho_0}^{1/s}  \notag \\
&\le \sum_{m=1}^\infty  e^{-3(m-1)/4}  \ave{\nb_{X_0[(m+2)\ell_0]}^s}_{\rho_0}^{1/s} 
\le  \sum_{m=1}^\infty  e^{-3(m-1)/4}  \frac{1}{e^{1/s}}\brr{\frac{b_0}{e}\br{q_0+s^\kappa q_1[(m+2)\ell_0]^D |X_0|}}  \notag \\
&\le \frac{1}{e^{1/s}} \brr{\frac{b_0}{e}\br{q_0+s^\kappa q_1\ell_0^D |X_0|}}   \sum_{m=1}^\infty  e^{-3(m+2)/4 + 9/4}  (m+2)^D
\le\frac{1}{e^{1/s}} \brr{\frac{b_0}{e}\br{q_0+s^\kappa q_1\ell_0^D |X_0|}}   e^{3/2} \brr{1+2^DD! (4/3)^{D+1}}. \notag
\end{align}
By applying the above inequalities to~\eqref{upp_moment_bound_s_1/s}, we can derive
\begin{align}
&\brrr{\tr\brr{\rho_0(t) \nb_{X_0[\ell_0]}^s}}^{1/s}  \notag \\
&\le \frac{1}{e}\brr{\frac{b_0}{e}\br{q_0+s^\kappa q_1\ell_0^D |X_0|}} \Biggl [ \br{1+ \frac{3t}{\tau} \delta_{\ell}} (1+R/\ell_0)^D  +2 c_{\tau,1} e^{-3\ell/16} 
\cdot e^{3/2} \brr{1+2^DD! (4/3)^{D+1}}\notag \\
&\quad + \frac{e^{1+1/s}}{b_0 q_1\ell_0^D |X_0| }  \br{\frac{c_{\tau,2}t}{\tau} +\tilde{\delta}_{\ell}}s^{1-\kappa}  \Biggr] .
\end{align}
Therefore, under the assumption of $\ell_0 \ge 2D R$, we can obtain 
\begin{align}
\br{1+ \frac{3t}{\tau} \delta_{\ell}} (1+R/\ell_0)^D  +2 c_{\tau,1} e^{-3\ell/16} 
e^{3/2} \brr{1+2^DD! (4/3)^{D+1}} + \frac{e^{1+1/s}}{b_0 q_1\ell_0^D |X_0| }  \br{\frac{c_{\tau,2}t}{\tau} +\tilde{\delta}_{\ell}}s^{1-\kappa}  \le 2 
\end{align}
 by choosing $R=\bar{\ell}$ as in Eq.~\eqref{choice_bar_ell}. 
We thus prove the second inequality~\eqref{main_ineq:main_theorem_boson_concentration_X}. 
This completes the proof. $\square$

 \section{Restriction on the outflow of bosons} \label{sec:Restriction on the outflow of bosons}

\begin{figure}[tt]
\centering
\includegraphics[clip, scale=0.5]{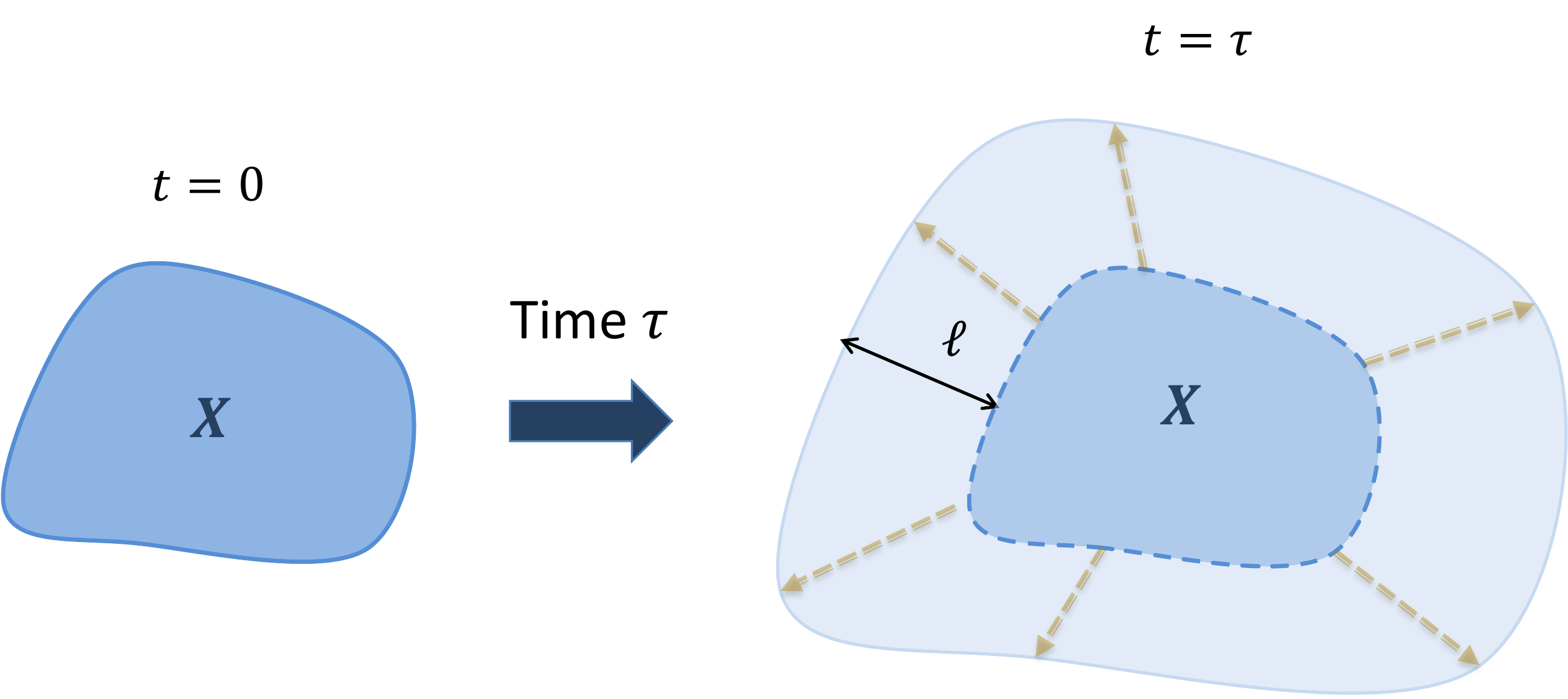}
\caption{Schematic picture of the setup in Sec.~\ref{sec:Restriction on the outflow of bosons}.
We consider an initial state where at least $N$ bosons exist inside the region $X\subset \Lambda$.  
After a time evolution, these bosons may escape outside the region $X$. 
Even though, if the time is short, we expect that the bosons still remain in the region of around $X$.
Mathematically, as the width $\ell$ increases, the probability of observing $\orderof{N}$ bosons in $X[\ell]$ is exponentially close to $1$
(see Proposition~\ref{prop:lower_probability}).
}
\label{supp_fig_prob_lower.pdf}
\end{figure}

In this section, we aim to bound a boson number in a particular region from below.
As in Fig.~\ref{supp_fig_prob_lower.pdf}, we consider a setup that many bosons originally concentrated in a region $X$: 
 \begin{align}
 \label{initial:cond_lemma}
\Pi_{X, \ge N} \rho = \rho,
\end{align}
where $N$ is a fixed number. 
Unlike the setup in Ref.~\cite{PhysRevA.84.032309}, there exist bosons outside the region $X$. 
Then, after a short-time evolution, the bosons are expected to concentrate around the region $X$, i.e.,  
 \begin{align}
 \label{desired_inequality_for_boson_outflow}
\tr \brr{ \Pi_{X[\ell], \le N-\delta N} \rho(\tau) } \approx g(\delta N)
\end{align}
for sufficiently large $\ell$, where $g(x)$ is a rapidly decaying function.  
We need to consider such an upper bound in proving the optimal Lieb-Robinson bound in one-dimensional systems (see Sec.~\ref{sec:Proof of Lemma_lemma:Probability_upper_bound}).

\subsection{Moment lower bound} \label{subsec:Moment lower bound}
First of all, we can prove that in the level of the expectation value. 
We can obtain   
\begin{align}
\bra{\psi_0(\tau)} \nb_{X[\ell]} \ket{\psi_0(\tau)} \gtrsim \bra{\psi_0} \nb_{X} \ket{\psi_0} - e^{-\orderofsp{\ell}}  \ge N - e^{-\orderofsp{\ell}}.
\end{align}
Thus, we ensure that the number of bosons is lower-bounded as long as we consider an extended region $X[\ell]$ with a sufficiently large $\ell$. 
For higher-order moments, we can also derive  
\begin{align}
\label{psi_0_X_ell_moment_lower}
\bra{\psi_0(\tau)} \nb_{X[\ell]}^s \ket{\psi_0(\tau)} \gtrsim \br{ N - e^{-\orderofsp{\ell}} }^s.
\end{align}
Mathematically, we prove the following corollary by using the same analyses as in the proof for Proposition~\ref{prop:upp_Moment_1th}: 
\begin{lemma}\label{corol_lower_bound_moment_refine}
Let us consider the same setup as that in Proposition~\ref{prop:upp_Moment_1th}. 
For an arbitrary subset $X$, the moment function $M^{(1)}_X(\tau)$ is lower-bounded by
 \begin{align}
  \label{lower_bound_moment_sth_1tyh_main}
\nb_X(\tau) \succeq  \nb_{X[-\ell]} - c_{\tau,1}\lambda_{1/4}e^{- 3\ell/16} \hat{\mathscr{D}}_X  - 5c_{\tau,1} \lambda_{1/2}  e^{-f_\tau \ell } \nb_{X} -c_{\tau,2},
\end{align}
where $\hat{\mathscr{D}}_X $ and $f_\tau$ have been defined in Eqs.~\eqref{def_delta_X_ell_prop} and \eqref{def_f_tau_ell_prop}, respectively.
Also, for moment functions $M^{(s)}_X(\tau)$ with arbitrary orders, the inequality~\eqref{main_inq;corol_lower_bound_moment} yields 
 \begin{align}
 \label{lower_bound_moment_sth_main}
M^{(s)}_X(\tau) \ge  \br{\nb_{X[-\ell]} - c_{\tau,1}\lambda_{1/4}e^{- 3\ell/16} \hat{\mathscr{D}}_X  - 5c_{\tau,1} \lambda_{1/2}  e^{-f_\tau \ell } \nb_{X} -c_{\tau,2}}^s  . 
\end{align}
\end{lemma}

{\bf Remark.} To derive the inequality~\eqref{psi_0_X_ell_moment_lower}, 
we consider the moment function of $M^{(s)}_{X[\ell]}(\tau)$ and apply the inequality~\eqref{lower_bound_moment_sth_main}, which yields 
 \begin{align}
 \label{lower_bound_moment_sth_main_re}
M^{(s)}_{X[\ell]} (\tau)
& \ge  \br{\ave{\nb_{X}}_{\rho} - c_{\tau,1}\lambda_{1/4}e^{- 3\ell/16} \ave{\hat{\mathscr{D}}_X}_\rho  - 5c_{\tau,1} \lambda_{1/2}  e^{-f_\tau \ell } \ave{\nb_{X}}_\rho -c_{\tau,2} }^s  \notag \\
& \ge  \br{\ave{\nb_{X}}_{\rho}  -e^{-\orderofsp{\ell}} -c_{\tau,2} }^s  \ge \br{ N - e^{-\orderofsp{\ell}}-c_{\tau,2} }^s,
\end{align}
where we assume $\ave{\nb_{X}}_{\rho}\ge N$ in the last inequality.

\subsubsection{Proof of Lemma~\ref{corol_lower_bound_moment_refine}} 
As shown in Corollary~\ref{corol_lower_bound_moment}, we have $\tr(\rho(\tau) \nb_X^s  ) \ge [ \tr(\rho(\tau) \nb_X) ]^s $ for an arbitrary quantum state $\rho$,
and hence it is enough to consider the 1st order moment in proving the inequality~\eqref{lower_bound_moment_sth_main} for the $s$th order moment.
In the following, we aim to derive a lower bound for the moment function $M_{X}^{(1)} (\tau)=\tr[\rho(\tau) \nb_X ]$ for an arbitrary $\rho$. 

\begin{figure}[tt]
\centering
\includegraphics[clip, scale=0.5]{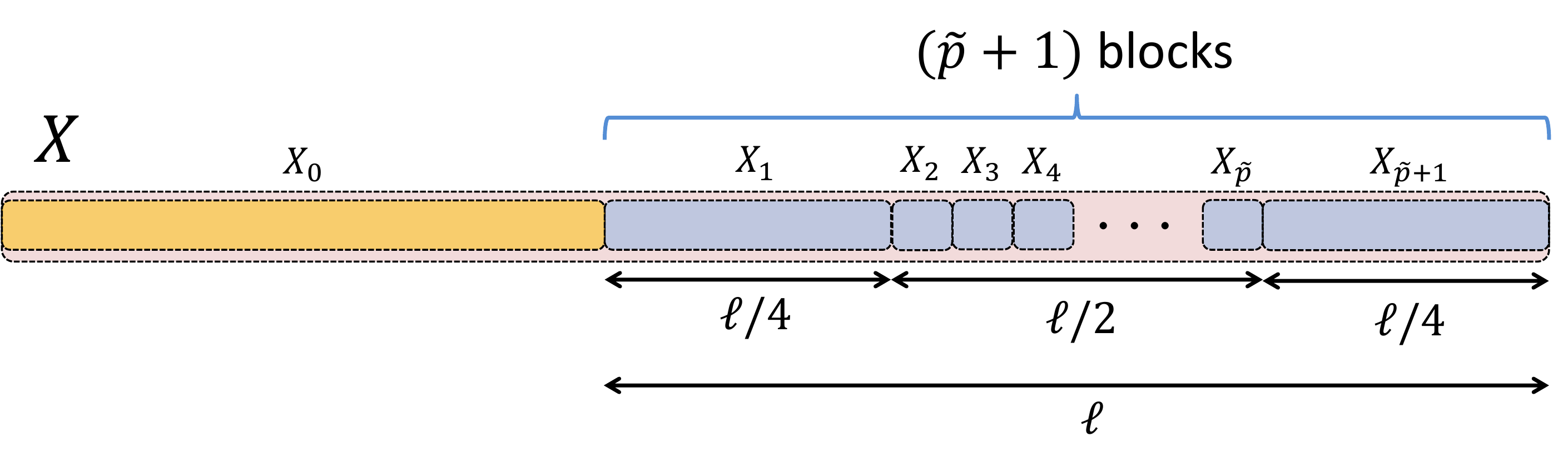}
\caption{Choice of the subsets $\{X_p\}_{p=0}^{\tilde{p} +1}$. 
Unlike Fig.~\ref{supp_fig_shell_choice.pdf}, we define the subsets inside the region $X$. 
We can lower-bound the moment function $M^{(1)}_X (\tau)$
 by using $M^{(1)}_{X_{0:p}}(\tau)$ because of $X_{0:p} \subseteq X$.
Our task is to find the optimal $p \in [1,\tilde{p}]$ which gives the lower bound~\eqref{lower_bound_moment_sth_1tyh_main}.
}
\label{supp_fig_shell_choice_lower_bound.pdf}
\end{figure}

We first define the subsets $\{X_p\}_{p=0}^{{\tilde{p}}+1}$ as follows (see Fig.~\ref{supp_fig_shell_choice_lower_bound.pdf}):
\begin{align}
\label{df_X_1_X_j_X_q+1}
&X_0 \coloneqq  X[-\ell] ,\quad  X_1 \coloneqq  X_0[\ell/4] \setminus X_0 , \notag \\
& X_p \coloneqq  X_0[\ell/4+p-1] \setminus X_0[\ell/4+p-2]\quad (2\le p\le \tilde{p}) , \notag \\
& X_{{\tilde{p}}+1} \coloneqq   X_0[\ell]\setminus X_0[3\ell/4],
\end{align} 
where $\tilde{p}=\ell/2 +1$. 
We then obtain 
\begin{align}
\label{trivial_upp_moment_extend_lower}
M_{X}^{(1)} (\tau)  \ge M_{X_{0:p}}^{(1)} (\tau)  
\end{align} 
for arbitrary $p\ge 0$, where we use $\nb_X \succeq \nb_{X_{0:p}}$ from $X \supseteq X_{0:p}$.

We start from a similar inequality to \eqref{Start_for_the_proof_prop_upp_Moment_sth}:
\begin{align}
\label{Start_for_the_proof_prop_upp_Moment_sth_2_}
M_{X_{0:p}}^{(1)} (\tau) &\ge  N^{(\rho)}_{X_{0:p}} - \tilde{N}^{(\rho)}_{X_{0:p}} - c_{\tau,2}   ,\quad  
\tilde{N}^{(\rho)}_{X_{0:p}} := \sum_{i\in  \partial (X_{0:p})} \sum_{j \in \Lambda} e^{-\dist_{i,j}} N^{(\rho)}_j ,
\end{align}
where we have denoted $\ave{\nb_i}_\rho $ by $N^{(\rho)}_i$ for $\forall i\in \Lambda$. 
Then, we aim to lower-bound  
\begin{align}
\label{lower_bouned_optimization_moment_lower}
\max_{p\in [1,{\tilde{p}}]} \br{N^{(\rho)}_{X_{0:p}} - c_{\tau,1} \tilde{N}^{(\rho)}_{X_{0:p}} } ,
\end{align}
which yields the lower bound for $M_{X}^{(1)} (\tau) $ because of the inequality~\eqref{trivial_upp_moment_extend_lower}.

By using Lemma~\ref{lmma:tilde_N_X_p_upper_bound}, we reduce the maximization problem~\eqref{lower_bouned_optimization_moment_lower} to the form of  
\begin{align}
\label{max_N_X_0:p_tilde_N_X_p_lower}
&\max_{p\in[1,{\tilde{p}}]} \br{N^{(\rho)}_{X_{0:p}}- c_{\tau,1} \tilde{N}^{(\rho)}_{X_{0:p}}}
\ge N^{(\rho)}_{X_0} - c_{\tau,1}\delta N_{X_0,\ell}^{(\rho)} +  \max_{p\in[1,{\tilde{p}}]} \br{M_{1:p}- c_{\tau,1} \lambda_{1/2} \sum_{j=1}^{{\tilde{p}}} e^{-|p-j|/2} M_j  } , \notag \\
&M_j = \begin{cases} 
N^{(\rho)}_{X_j}& \for 1\le j \le {\tilde{p}}-1, \\
N^{(\rho)}_{X_{\tilde{p}}} +N^{(\rho)}_{X_{{\tilde{p}}+1}} &\for j={\tilde{p}},
\end{cases}
\end{align}
where we use $N^{(\rho)}_{X_{0:p}}:= N^{(\rho)}_{X_0} + M_{1:p}$ and 
$\delta N_{X,\ell}^{(\rho)}$ has been defined in Eq.~\eqref{def_f_tau_ell_prop_re}, i.e.,
 \begin{align}
\delta N^{(\rho)}_{X_0,\ell}= \lambda_{1/4}e^{- 3\ell/16}\tr \brr{\rho \br{ \nb_{X_0}   + \hat{\mathscr{D}}_{X_0[\ell]}}}
\end{align}
with $\hat{\mathscr{D}}_{X_0[\ell]}$ given by Eq.~\eqref{def_delta_X_ell_prop}.
Then, by using Lemma~\ref{lemma:maximization_num_shell} with $\bar{\alpha}$ and $\chi_\alpha$ as in Eqs.~\eqref{bar_alpha_infty_apply_upp} and \eqref{chi_alpha_infty_apply_upp}, we obtain  
\begin{align}
\label{max_N_X_0:p_tilde_N_X_p_lower_2nd}
\max_{p\in[1,{\tilde{p}}]} \br{M_{1:p}- c_{\tau,1} \lambda_{1/2} \sum_{j=1}^{{\tilde{p}}} e^{-|p-j|/2} M_j  }
 &\ge - 5c_{\tau,1} \lambda_{1/2} \br{1+\frac{1}{5c_{\tau,1} \lambda_{1/2}  + 2}}^{-{\tilde{p}}+1} M_{1:{\tilde{p}}} \notag \\
 &= - 5c_{\tau,1} \lambda_{1/2} \br{1+\frac{1}{5c_{\tau,1} \lambda_{1/2}  + 2}}^{-\ell/2} \br{N^{(\rho)}_{X} -N^{(\rho)}_{X_0}} ,
\end{align}
where we have used ${\tilde{p}}=\ell/2+1$ and $M_{1:{\tilde{p}}}=N^{(\rho)}_{X} -N^{(\rho)}_{X_0} $.
By combining the inequalities~\eqref{max_N_X_0:p_tilde_N_X_p_lower} and \eqref{max_N_X_0:p_tilde_N_X_p_lower_2nd}, we obtain
\begin{align}
\label{max_N_X_0:p_tilde_N_X_p_lower_fin}
&\max_{p\in[1,{\tilde{p}}]} \br{N^{(\rho)}_{X_{0:p}}- c_{\tau,1} \tilde{N}^{(\rho)}_{X_{0:p}}} \notag \\
&\ge N^{(\rho)}_{X_0} - c_{\tau,1}\lambda_{1/4}e^{- 3\ell/16} \brr{N^{(\rho)}_{X_0} + \tr \br{\rho \hat{\mathscr{D}}_{X_0[\ell]}} } 
 - 5c_{\tau,1} \lambda_{1/2} \br{1+\frac{1}{5c_{\tau,1} \lambda_{1/2}  + 2}}^{-\ell/2} \br{N^{(\rho)}_{X} -N^{(\rho)}_{X_0}}   , \notag \\
&\ge N^{(\rho)}_{X_0} - c_{\tau,1}\lambda_{1/4}e^{- 3\ell/16} \tr \br{\rho \hat{\mathscr{D}}_{X_0[\ell]}} 
 - 5c_{\tau,1} \lambda_{1/2} \br{1+\frac{1}{5c_{\tau,1} \lambda_{1/2}  + 2}}^{-\ell/2}N^{(\rho)}_{X} ,
\end{align}
which yields the main inequality~\eqref{lower_bound_moment_sth_1tyh_main} form $X_0=X[-\ell]$. 
This completes the proof. $\square$

\subsection{Bound on the probability distribution} \label{subsec:Bound on the probability distribution}

In the previous section, we have proved that 
\begin{align}
\label{lower_boson_moment_sth_N+e^-oroderofell}
\bra{\psi_0(\tau)} \nb_{X[\ell]}^s \ket{\psi_0(\tau)} \gtrsim \br{N - e^{-\orderofsp{\ell}}}^s.
\end{align}
Unfortunately, we cannot utilize it to restrict the boson number distribution as in Eq.~\eqref{desired_inequality_for_boson_outflow}. 
For example, if a time-evolved state $\ket{\psi_0(\tau)}$ satisfies 
\begin{align}
\norm{\Pi_{X[\ell], 2N} \ket{\psi_0(\tau)} }^2 = 1/2, \quad \norm{\Pi_{X[\ell] , 0} \ket{\psi_0(\tau)} }^2 = 1/2 ,
\end{align}
the inequality~\eqref{lower_boson_moment_sth_N+e^-oroderofell} holds, but we still have a probability of $\orderof{1}$ that all the bosons flow outside the region $X[\ell]$.  
In this sense, it is not enough to consider only the lower bound on the moment function. 
By refining the analyses, we can prove the following statement:

\begin{prop} \label{prop:lower_probability}
Let $\rho$ be an arbitrary quantum state $\rho$ such that 
\begin{align}
\label{cond_psi_in_derivation:s_2}
\Pi_{\mathcal{S}(X[\ell],1/2, q_\ast)} \rho  = \rho , \quad \Pi_{X, \ge N} \rho = \rho,
\end{align}
where the set $\mathcal{S}(L,\xi,N)$ has been defined in Eq.~\eqref{mathcal_S_L_xi_N/def}\footnote{Roughly speaking, the projection~$\Pi_{\mathcal{S}(X[\ell],1/2, q_\ast)}$ implies that there does not exist too many bosons around $X[\ell]$.}. 
We then obtain 
\begin{align}
\label{main_ineq:prop:lower_probability}
\tr \brr{ \Pi_{X[\ell], \le N-\delta N} \rho(\tau) } \le \frac{2}{1-e^{-1/(8e c_{\tau,2})}}\exp\brr{\frac{-\delta N}{8e c_{\tau,2}} + 3c_\ell' \br{1 + c_\ell'  D }\log^2 \br{q_\ast \gamma |X|} },
\end{align}
where the length $\ell$ is given by
\begin{align}
\label{main_ineq:prop:lower_probability_ell_cond}
\ell = \ceilf{\frac{2}{\tilde{f}_\tau} \log \br{\frac{5c_{\tau,1} \lambda_{1/2} e^{2\tilde{f}_\tau} \gamma (D/\tilde{f}_\tau)^D q_\ast  |X| }{2c_{\tau,2}} }}= c'_\ell \log \br{q_\ast \gamma |X|} ,
\end{align}
with 
\begin{align}
\label{def_f_tau_ell_prop_dash_original}
\tilde{f}_\tau \coloneqq  \frac{1}{2} \log \br{1+\frac{1}{10e c_{\tau,1} \lambda_{1/2}  + 2} }. 
\end{align}
Here, $c_\ell'$ is a constant of $\orderof{1}$.
\end{prop}

{\bf Remark.}
From the proposition, as long as $\delta N \gtrsim \log^2 \br{q_\ast}$, 
we obtain the desired restriction in Eq.~\eqref{desired_inequality_for_boson_outflow}. 
Here, the probability distribution obeys the large deviation bound, i.e.,
\begin{align}
\tr \brr{ \Pi_{X[\ell], \le N-\delta N} \rho(\tau) } = e^{-\Theta(N)}  \for \delta N= \Theta(N).
\end{align}
We will utilize the inequality~\eqref{main_ineq:prop:lower_probability} in deriving the upper bound~\eqref{main_ineq:prop:lower_probability_apply___0} hereafter.

 \begin{figure}[tt]
\centering
\includegraphics[clip, scale=0.5]{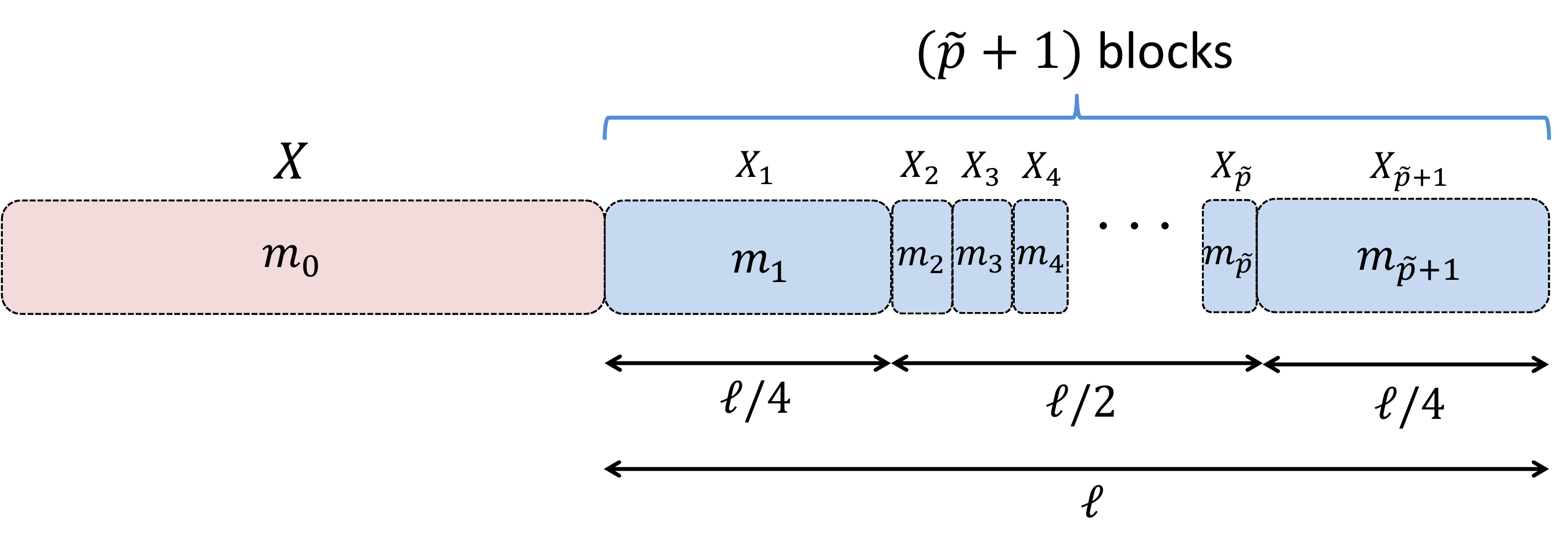}
\caption{We adopt the same notation as in Fig.~\ref{supp_fig_shell_choice.pdf}. We decompose the extended subset $X[\ell]\setminus X$ to $(\tilde{p}+1)$ pieces.
In the spectral decomposition~\eqref{Pi_X_ell_N-delta N_psi_tau_upp}, we assign the boson numbers $\{m_p\}_{p=0}^{\tilde{p}+1}$ in the subsets $\{X_p\}_{p=0}^{\tilde{p}+1}$, respectively. To prove the main inequality~\eqref{main_ineq:prop:lower_probability}, we derive the 
upper bound of \eqref{proof_start_for_boson_move_outside_fin} and aim to identify the optimal $p$ which minimizes the RHS of~\eqref{proof_start_for_boson_move_outside_fin}.
}
\label{supp_fig_shell_choice_out.pdf}
\end{figure}

\subsection{Proof of Proposition~\ref{prop:lower_probability}}

By using the spectral decomposition of $\rho$ as $\rho=\sum_j p_j \ket{\psi_j}\bra{\psi_j}$, we have 
\begin{align}
&\tr \br{ \Pi_{X[\ell], \le N-\delta N} \rho(\tau) } =\sum_{j}p_j \bra{\psi_j(\tau)} \Pi_{X[\ell], \le N-\delta N} \ket{\psi_j(\tau)} 
=\sum_{j}p_j \norm{ \Pi_{X[\ell], \le N-\delta N} \ket{\psi_j(\tau)}}^2  ,\notag \\
&\Pi_{\mathcal{S}(X[\ell],1/2, q_\ast)}\ket{\psi_j} = \ket{\psi_j} \for  \forall j .
\end{align}
Hence, we consider the case that $\rho$ is given by a pure state $\ket{\psi}$ without loss of generality. 
In the following, we choose the length $\ell$ such that 
\begin{align}
\label{delta_X_ell_q_ast_upp_choice}
\delta_{X,\ell,q_\ast}: = e^{-\ell/8} q_\ast |X| \br{ \lambda_{1/2}   + \gamma \lambda_{1/4} \ell^{D-1}e^{-\ell/16}} \le \frac{c_{\tau,2}}{c_{\tau,1}} s ,
\end{align} 
where $\delta_{X,\ell,q_\ast}$ is going to be defined in Eq.~\eqref{delta_X_ell_q_ast_definition}. 

We adopt the same notation for the subsets $\{X_p\}_{p=1}^{{\tilde{p}}+1}$ as in Eq.~\eqref{df_X_1_X_j_X_q+1_0}:
\begin{align}
\label{df_X_1_X_j_X_q+1_re}
&X_0 \coloneqq  X ,\quad  X_1 \coloneqq  X[\ell/4] \setminus X , \notag \\
& X_p \coloneqq  X[\ell/4+p-1] \setminus X[\ell/4+p-2]\quad (2\le p\le {\tilde{p}}) , \notag \\
& X_{{\tilde{p}}+1} \coloneqq   X[\ell]\setminus X[3\ell/4],
\end{align} 
where $\tilde{p}=\ell/2 +1$. 
We here introduce the projection $\Pi_{m_{0:{\tilde{p}}+1}}$ as 
\begin{align}
\label{Pi_m_0:q+1_def}
\Pi_{m_{0:{\tilde{p}}+1}}:= \Pi_{X, m_0} \prod_{j=1}^{{\tilde{p}}+1} \Pi_{X_j, m_j} .
\end{align} 
Note that $\Pi_{X, N}$ has been defined in Eq.~\eqref{def_Pi_X_q}, i.e., $\nb_X\Pi_{X,N}=N\Pi_{X,N}$.
We then obtain 
\begin{align}
\label{Pi_X_ell_N-delta N_psi_tau_upp}
\norm{ \Pi_{X[\ell], \le N-\delta N} \ket{\psi(\tau)}}  
&= \norm{ \sum_{m_0,m_1,\ldots,m_{\tilde{p}+1}} \Pi_{X[\ell], \le N-\delta N} e^{-iHt} \Pi_{m_{0:{\tilde{p}}+1}}\ket{\psi} } \notag \\
&\le\sum_{m_0,m_1,\ldots,m_{\tilde{p}+1}} \norm{  \Pi_{X[\ell], \le N-\delta N} e^{-iHt} \ket{\psi_{m_{0:{\tilde{p}}+1}} }} ,
\end{align}
where we define the unnormalized state $\ket{\psi_{m_{0:{\tilde{p}}+1}}} :=  \Pi_{m_{0:{\tilde{p}}+1}}\ket{\psi}$. 
From the condition~\eqref{cond_psi_in_derivation:s_2}, we have $\Pi_{X, m_0} \ket{\psi}=0$ for $m_0 < N$, and hence in the summation~\eqref{Pi_X_ell_N-delta N_psi_tau_upp} we only have to consider the range of $m_0 \ge N$.

For the estimation of $\norm{  \Pi_{X[\ell], \le N-\delta N} e^{-iHt} \ket{\psi_{m_{0:{\tilde{p}}+1}} }}$, we consider
\begin{align}
\label{proof_start_for_boson_move_outside}
\norm{  \Pi_{X[\ell], \le N-\delta N} e^{-iHt} \ket{\psi_{m_{0:{\tilde{p}}+1}} }} 
&\le  \norm{  \Pi_{X[\ell], \le N-\delta N}  |\nb_{X_{0:p}} - N_p|^{-s/2}} \cdot   \norm{ |\nb_{X_{0:p}}- N_p|^{s/2}  e^{-iHt} \ket{\psi_{m_{0:{\tilde{p}}+1}} }}  
\end{align}
for an arbitrary choice of $p\in[0, \tilde{p}+1]$, 
where we choose $N_p$ appropriately such that $N_p$ is larger than $N$. 
Because of $X[\ell] \supseteq X_{0:p}$ for $\forall p$, we have  $\Pi_{X[\ell], \le N-\delta N}=\Pi_{X[\ell], \le N-\delta N} \Pi_{X_{0:p}, \le N-\delta N}$, which yields
\begin{align}
\label{proof_start_for_boson_move_outside_2}
\norm{  \Pi_{X[\ell], \le N-\delta N} |\nb_{X_{0:p}} - N_p|^{-s/2} }  \le  \frac{1}{|N_p - N+\delta N|^{s/2}}   ,
\end{align}
where we use $N_p\ge N$ to obtain the inequality.
Also, by using Proposition~\ref{Schuch_boson_extend_more_general}, we obtain an upper bound of  
\begin{align}
\label{proof_start_for_boson_move_outside_3}
 \norm{ |\nb_{X_{0:p}}- N_p|^{s/2}  e^{-iHt} \ket{\psi_{m_{0:{\tilde{p}}+1}} }}^2
&= \bra{\psi_{m_{0:{\tilde{p}}+1}}} e^{iHt}  |\nb_{X_{0:p}}- N_p|^{s}   e^{-iHt} \ket{\psi_{m_{0:{\tilde{p}}+1}} }   \notag \\
&\le \bra{\psi_{m_{0:{\tilde{p}}+1}}}   \br{ |\nb_{X_{0:p}}- N_p|  + c_{\tau,1} \hat{\mathcal{D}}_{X_{0:p}} + c_{\tau,2} s}^s  \ket{\psi_{m_{0:{\tilde{p}}+1}} }    \notag \\
&=\bra{\psi_{m_{0:{\tilde{p}}+1}}}  \br{ |m_{0:p}- N_p|  + c_{\tau,1} \hat{\mathcal{D}}_{X_{0:p}} + c_{\tau,2} s}^s  \ket{\psi_{m_{0:{\tilde{p}}+1}} }   \notag \\
&=\bra{\psi_{m_{0:{\tilde{p}}+1}}}  \br{ c_{\tau,1} \hat{\mathcal{D}}_{X_{0:p}} + c_{\tau,2} s}^s  \ket{\psi_{m_{0:{\tilde{p}}+1}} }   
\end{align}
with $m_{0:p} = \sum_{j=0}^{p} m_j$,
where we choose $N_p$ as 
$
N_p= m_{0:p} 
$
in the last equation. 
Note that the choice ensures $N_p\ge N$ because of $m_0\ge N$. 

From $\Pi_{\mathcal{S}(X[\ell],1/2, q_\ast)}\Pi_{m_{0:{\tilde{p}}+1}}\ket{\psi_{m_{0:{\tilde{p}}+1}} } =\ket{\psi_{m_{0:{\tilde{p}}+1}} } $, we then consider 
\begin{align}
\label{proof_start_for_boson_move_outside_4}
&\bra{\psi_{m_{0:{\tilde{p}}+1}}} \br{ c_{\tau,1} \hat{\mathcal{D}}_{X_{0:p}} + c_{\tau,2} s}^s  \ket{\psi_{m_{0:{\tilde{p}}+1}} }    \notag \\
&= \bra{\psi_{m_{0:{\tilde{p}}+1}}}  \Pi_{\mathcal{S}(X[\ell],1/2, q_\ast)} \Pi_{m_{0:{\tilde{p}}+1}}  \br{ c_{\tau,1} \hat{\mathcal{D}}_{X_{0:p}} + c_{\tau,2} s}^s  \Pi_{\mathcal{S}(X[\ell],1/2, q_\ast)} \Pi_{m_{0:{\tilde{p}}+1}} \ket{\psi_{m_{0:{\tilde{p}}+1}} }     \notag \\
&\le \norm{ \Pi_{\mathcal{S}(X[\ell],1/2, q_\ast)} \Pi_{m_{0:{\tilde{p}}+1}} \br{ c_{\tau,1} \hat{\mathcal{D}}_{X_{0:p}} + c_{\tau,2} s}^s \Pi_{\mathcal{S}(X[\ell],1/2, q_\ast)} \Pi_{m_{0:{\tilde{p}}+1}} }     \notag \\
&= \br{ c_{\tau,1} \norm{\Pi_{\mathcal{S}(X[\ell],1/2, q_\ast)} \Pi_{m_{0:{\tilde{p}}+1}}\hat{\mathcal{D}}_{X_{0:p}} \Pi_{m_{0:{\tilde{p}}+1}}\Pi_{\mathcal{S}(X[\ell],1/2, q_\ast)} }+ c_{\tau,2} s}^s  ,
\end{align}
where we use the fact that $\hat{\mathcal{D}}_{X_{0:p}}$ commutes with $\Pi_{\mathcal{S}(X[\ell],1/2, q_\ast)}$ and $\Pi_{m_{0:{\tilde{p}}+1}}$.
By applying the inequalities~\eqref{proof_start_for_boson_move_outside_2}, \eqref{proof_start_for_boson_move_outside_3}, \eqref{proof_start_for_boson_move_outside_4} to \eqref{proof_start_for_boson_move_outside}, we obtain
\begin{align}
\label{proof_start_for_boson_move_outside_4.5}
\norm{  \Pi_{X[\ell], \le N-\delta N} e^{-iHt} \ket{\psi_{m_{0:{\tilde{p}}+1}} }} 
&\le  \frac{\br{ c_{\tau,1} \norm{\Pi_{\mathcal{S}(X[\ell],1/2, q_\ast)} \Pi_{m_{0:{\tilde{p}}+1}}\hat{\mathcal{D}}_{X_{0:p}} \Pi_{m_{0:{\tilde{p}}+1}}\Pi_{\mathcal{S}(X[\ell],1/2, q_\ast)} }+ c_{\tau,2} s}^{s/2}}{|N_p - N+\delta N|^{s/2}} .
\end{align}

By using the same analyses as the proof of Lemma~\ref{lmma:tilde_N_X_p_upper_bound}, we prove the following lemma:
\begin{lemma} \label{lmma:tilde_N_X_p_upper_bound_proj}
Let us define 
\begin{align}
\label{delta_X_ell_q_ast_definition}
\delta_{X,\ell,q_\ast}: = e^{-\ell/8} q_\ast |X| \br{ \lambda_{1/2}   + \gamma \lambda_{1/4} \ell^{D-1}e^{-\ell/16}} ,
\end{align} 
We then obtain the upper bound of
\begin{align}
\label{proof_start_for_boson_move_outside_5}
\norm{\Pi_{\mathcal{S}(X[\ell],1/2, q_\ast)} \Pi_{m_{0:{\tilde{p}}+1}}\hat{\mathcal{D}}_{X_{0:p}} \Pi_{m_{0:{\tilde{p}}+1}}\Pi_{\mathcal{S}(X[\ell],1/2, q_\ast)} }\le \lambda_{1/2} \sum_{j=1}^{{\tilde{p}}} e^{-|p-j|/2} M_j  +\delta_{X,\ell,q_\ast}  \quad (1\le p\le {\tilde{p}}), 
\end{align}
with
\begin{align}
\label{ineq_lmma:tilde_N_X_p_upper_bound_case2}
M_j = \begin{cases} 
m_j& \for 1\le j \le {\tilde{p}}-1, \\
m_{\tilde{p}}+ m_{\tilde{p}+1}  &\for j={\tilde{p}}.
\end{cases}
\end{align} 
\end{lemma}

\noindent
\textit{Proof of Lemma~\ref{lmma:tilde_N_X_p_upper_bound_proj}}. From the inequality~\eqref{reduce_tilde_N_X_p_upp}, we can start from
\begin{align}
\label{reduce_tilde_N_X_p_upp_1st_ineq}
&\norm{\Pi_{\mathcal{S}(X[\ell],1/2, q_\ast)} \Pi_{m_{0:{\tilde{p}}+1}}\hat{\mathcal{D}}_{X_{0:p}} \Pi_{m_{0:{\tilde{p}}+1}}\Pi_{\mathcal{S}(X[\ell],1/2, q_\ast)} } \notag \\
&\le \lambda_{1/2} \sum_{j=1}^{{\tilde{p}}} e^{-|p-j|/2} M_j  +\sum_{i\in  \partial (X_{0:p})}  \sum_{j\in X_{1:{\tilde{p}}+1}^\co} e^{-\dist_{i,j}} 
\norm{\nb_j \Pi_{\mathcal{S}(X[\ell],1/2, q_\ast)}}  \notag \\
&\le \lambda_{1/2} \sum_{j=1}^{{\tilde{p}}} e^{-|p-j|/2} M_j  +q_\ast \sum_{i\in  \partial (X_{0:p})}  \sum_{j\in X_{1:{\tilde{p}}+1}^\co} 
e^{-\dist_{i,j} + \dist_{j,X[\ell]}/2}  
 \quad (1\le p\le {\tilde{p}}),
\end{align} 
where we use the definition~\eqref{mathcal_S_L_xi_N/def} in the last inequality.
We then decompose 
\begin{align}
\label{reduce_tilde_N_X_p_upp_second_ineq}
\sum_{i\in  \partial (X_{0:p})}  \sum_{j\in X_{1:{\tilde{p}}+1}^\co} e^{-\dist_{i,j} + \dist_{j,X[\ell]}/2}  
= \sum_{i\in  \partial (X_{0:p})}  \sum_{j\in X} e^{-\dist_{i,j}} 
+ \sum_{i\in  \partial (X_{0:p})}  \sum_{j\in X[\ell]^\co} e^{-\dist_{i,j} + \dist_{j,X[\ell]}/2} ,
\end{align} 
where we use $\dist_{j,X[\ell]}=0$ for $j\in X$.
For the first term, we obtain 
\begin{align}
\label{reduce_tilde_N_X_p_upp_2nd_term_2}
\sum_{i\in  \partial (X_{0:p})}  \sum_{j\in X}e^{-\dist_{i,j}} \le 
 \sum_{j \in X}  \sum_{i: \dist_{i,j}\ge \ell/4} e^{-\dist_{i,j}} \le 
 e^{-\ell/8}  |X|\sum_{i\in \Lambda} e^{-\dist_{i,j}/2} \le  \lambda_{1/2}e^{-\ell/8}  |X| ,
\end{align}
where, in the first inequality, we use $\dist_{i,j}\ge \ell/4$ for $i\in \partial (X_{0:p})$ and $j\in X$ (see Fig.~\ref{supp_fig_shell_choice_out.pdf}), and  in the third inequality, we apply the inequality~\eqref{oftern_used_inequality_lambda_c}.  
Also, for the second term in~\eqref{reduce_tilde_N_X_p_upp_second_ineq}, we use Eq.~\eqref{mathcal_S_L_xi_N/def} to obtain
\begin{align}
\label{reduce_tilde_N_X_p_upp_2nd_term_3}
 \sum_{i\in  \partial (X_{0:p})}  \sum_{j\in X[\ell]^\co} e^{-\dist_{i,j} + \dist_{j,X[\ell]}/2} 
 &\le  e^{-\ell/8}\sum_{i\in  \partial (X_{0:p})} \sum_{j\in X[\ell]^\co} e^{-\dist_{i,j}/2} \notag \\
 &\le  e^{-\ell/8}\sum_{i\in  \partial (X_{0:p})}\sum_{j: \dist_{i,j}\ge \ell/4} e^{-\dist_{i,j}/2}\notag \\
 &\le  \lambda_{1/4}  e^{-3\ell/16} |\partial (X_{0:p})| \le 
\gamma\lambda_{1/4}  |X| \ell^{D-1}e^{-3\ell/16},
\end{align}
where we use $\dist_{j,X[\ell]} \le \dist_{i,j} - \ell/4$ for $i\in  \partial (X_{0:p})$ and $j\in X[\ell]^\co$ in the first inequality, and in the last inequality, we use 
 $ |\partial (X_{0:p})| \le \abs{\partial X[\ell]}\le \gamma \ell^{D-1} |X|$ from~\eqref{supp_parameter_gamma_X}.
By applying the inequalities~\eqref{reduce_tilde_N_X_p_upp_2nd_term_2} and \eqref{reduce_tilde_N_X_p_upp_2nd_term_3} to~\eqref{reduce_tilde_N_X_p_upp_1st_ineq} and \eqref{reduce_tilde_N_X_p_upp_second_ineq}, we have 
\begin{align}
\label{reduce_tilde_N_X_p_upp_2nd_term_fin}
q_\ast \sum_{i\in  \partial (X_{0:p})}  \sum_{j\in X_{1:{\tilde{p}}+1}^\co} 
e^{-\dist_{i,j} + \dist_{j,X[\ell]}/2} 
&\le e^{-\ell/8} q_\ast |X| \br{ \lambda_{1/2}   + \gamma \lambda_{1/4} \ell^{D-1}e^{-\ell/16}} = \delta_{X,\ell,q_\ast},
\end{align} 
where the last equation is the definition~\eqref{delta_X_ell_q_ast_definition} for $\delta_{X,\ell,q_\ast}$.
By combining the inequalities~\eqref{reduce_tilde_N_X_p_upp_1st_ineq} and \eqref{reduce_tilde_N_X_p_upp_2nd_term_fin}, 
we prove the main inequality~\eqref{proof_start_for_boson_move_outside_5}. 
This completes the proof. $\square$


 {~}

\noindent\hrulefill{\bf [ End of Proof of Lemma~\ref{lmma:tilde_N_X_p_upper_bound_proj}]}

{~}

By combining the inequalities~\eqref{proof_start_for_boson_move_outside_4.5} and \eqref{proof_start_for_boson_move_outside_5},  we have
\begin{align}
\label{proof_start_for_boson_move_outside_fin}
\norm{  \Pi_{X[\ell], \le N-\delta N} e^{-iHt} \ket{\psi_{m_{0:{\tilde{p}}+1}} }} 
&\le \frac{1}{|m_{0:p} - N+\delta N|^{s/2}}  \br{c_{\tau,1}\lambda_{1/2} \sum_{j=1}^{{\tilde{p}}} e^{-|p-j|/2} M_j  +c_{\tau,1}\delta_{X,\ell,q_\ast}   + c_{\tau,2} s}^{s/2}   \notag \\
&=    \br{\frac{\tilde{M}_p +2 c_{\tau,2} s}{ M_{1:p} + m_0- N+\delta N }}^{s/2} 
\for \forall p\in [1, \tilde{p}-1],
\end{align}
where we define $M_{1:p} = \sum_{j=1}^p M_p$ and $\tilde{M}_p:= c_{\tau,1}\lambda_{1/2} \sum_{j=1}^{{\tilde{p}}} e^{-|p-j|/2} M_j $.   
Note that $c_{\tau,1}\delta_{X,\ell,q_\ast} \le c_{\tau,2} s$ from the condition~\eqref{delta_X_ell_q_ast_upp_choice}.  

Then, we aim to find the optimal $p$ which minimizes the RHS of the inequality~\eqref{proof_start_for_boson_move_outside_fin}.
We here prove the following lemma:
\begin{lemma}\label{Lema:upp_pi_X_ell_m_0_q+1}
If the length $\ell$ satisfies
\begin{align}
\label{condition_for_ell_prop_ge_g}
\ell \ge  \frac{2}{\tilde{f}_\tau} \log \br{\frac{5c_{\tau,1} \lambda_{1/2} e^{2\tilde{f}_\tau} \gamma (D/\tilde{f}_\tau)^D q_\ast  |X| }{2c_{\tau,2}} }
\end{align}
with
\begin{align}
\label{def_f_tau_ell_prop_dash}
\tilde{f}_\tau \coloneqq  \frac{1}{2} \log \br{1+\frac{1}{10e c_{\tau,1} \lambda_{1/2}  + 2} }, 
\end{align}
we reduce the inequality~\eqref{proof_start_for_boson_move_outside_fin} to the form of
\begin{align}
\label{proof_start_for_boson_move_outside_fin_cases1and2}
\norm{  \Pi_{X[\ell], \le N-\delta N} e^{-iHt} \ket{\psi_{m_{0:{\tilde{p}}+1}} }} 
&\le 2\exp\br{\frac{-m_0+ N-\delta N}{8e c_{\tau,2}}} 
\end{align}
by choosing $p\in [1,\tilde{p}-1]$ appropriately.
\end{lemma}

\subsubsection{Proof of Lemma~\ref{Lema:upp_pi_X_ell_m_0_q+1}}\label{sec:Proof of Lemma_Lema:upp_pi_X_ell_m_0_q+1}

We here consider the following two cases: 
\begin{align}
\label{boson_move_outside_case_two}
\min_{p\in [1:{\tilde{p}}-1]} \br{\frac{\tilde{M}_p}{M_{1:p} } }\begin{cases} 
\le \frac{1}{2e}  &\for \quad \textrm{Case 1} , \\ 
\ge \frac{1}{2e} & \for \quad \textrm{Case 2} .
\end{cases}
\end{align}

{\bf [Case~1]}  \\
In that case, we can find $p_\ast$ such that 
 \begin{align}
\frac{\tilde{M}_{p_\ast}}{M_{1:p_\ast}  } \le \frac{1}{2e}  .
\end{align}
By using the inequality of 
 \begin{align}
\frac{x'+y'}{x + y} = \frac{x'}{x+y}  + \frac{y'}{x+y} \le \frac{x'}{x} + \frac{y'}{y}  
\end{align}
for arbitrary positive $x,y,x',y'$, we have 
 \begin{align}
 \label{proof_start_for_boson_move_outside_fin_rrpre}
\frac{\tilde{M}_{p_\ast}+2 c_{\tau,2} s}{ M_{1:p_\ast} + m_0- N+\delta N } \le \frac{\tilde{M}_{p_\ast}}{M_{1:p_\ast} } + \frac{2 c_{\tau,2} s}{ m_0- N+\delta N }
\le \frac{1}{2e} + \frac{2 c_{\tau,2} s}{ m_0- N+\delta N } . 
\end{align}
Therefore, we choose $s$ as 
 \begin{align}
s= \floorf {\frac{m_0- N+\delta N}{4e c_{\tau,2}}  } ,
\end{align}
which reduces the inequality~\eqref{proof_start_for_boson_move_outside_fin_rrpre} to
 \begin{align}
 \label{proof_start_for_boson_move_outside_fin_rrpre_2}
\frac{\tilde{M}_{p_\ast}+2 c_{\tau,2} s}{ M_{1:p_\ast} + m_0- N+\delta N } \le \frac{1}{e} . 
\end{align}
By applying the above inequality to~\eqref{proof_start_for_boson_move_outside_fin}, we obtain the desired inequality of
\begin{align}
\label{proof_start_for_boson_move_outside_fin_case1}
\norm{  \Pi_{X[\ell], \le N-\delta N} e^{-iHt} \ket{\psi_{m_{0:{\tilde{p}}+1}} }} 
&\le 2\exp\br{\frac{-m_0+ N-\delta N}{8e c_{\tau,2}}} ,
\end{align}
where we use $e^{1/2} < 2$. 

{\bf [Case~2]}  \\
In this case, we can utilize Lemma~\ref{lemma:minimization_num_shell_frac} with $\Phi=2e$ to obtain 
\begin{align}
\label{upper_bound_tilde_M_1_}
\tilde{M}_1 
&\le \br{1+\frac{1}{2e \bar{\alpha}_{\infty} + \chi_\alpha}  }^{-{\tilde{p}}+2} \bar{\alpha}_\infty  M_{1:{\tilde{p}}-1}   \notag \\
&\le \br{1+\frac{1}{10e c_{\tau,1} \lambda_{1/2}  + 2}  }^{-{\tilde{p}}+2} 5c_{\tau,1} \lambda_{1/2}  M_{1:{\tilde{p}}-1}
= 5c_{\tau,1} \lambda_{1/2}  e^{-\tilde{f}_\tau (2\tilde{p}-4)}M_{1:{\tilde{p}}-1}
\end{align}
with $\tilde{f}_\tau $ defined in Eq.~\eqref{def_f_tau_ell_prop_dash}, 
where the parameters $\bar{\alpha}_{\infty}$ and $\chi_\alpha$ have been estimated as in Eqs.~\eqref{bar_alpha_infty_apply_upp} and \eqref{chi_alpha_infty_apply_upp}: 
$\bar{\alpha}_{\infty}= (4.0829\cdots)c_{\tau,1} \lambda_{1/2}$ and $\chi_\alpha = 1.5414\cdots$. 
From ${\tilde{p}}=\ell/2+1$ and the upper bound of 
\begin{align}
M_{1:{\tilde{p}}-1} \le M_{0:{\tilde{p}}}\le \norm{\Pi_{\mathcal{S}(X[\ell],1/2, q_\ast)} \nb_{X[\ell]}} \le q_\ast |X[\ell]|  \le  \gamma \ell^{D} q_\ast  |X| ,
\end{align}
we reduce the inequality~\eqref{upper_bound_tilde_M_1_} to
\begin{align}
\label{upper_bound_tilde_M_1_2}
\tilde{M}_1 
&\le 5c_{\tau,1} \lambda_{1/2} e^{2\tilde{f}_\tau} \gamma \ell^{D} q_\ast  |X|  e^{-\tilde{f}_\tau \ell} 
\le 5c_{\tau,1} \lambda_{1/2} e^{2\tilde{f}_\tau} \gamma (D/\tilde{f}_\tau)^D q_\ast  |X|  e^{-\tilde{f}_\tau \ell/2} ,
\end{align}
where we use the inequality of $x^z e^{-x} \le z^z e^{-x/2}$ for $z>0$. 
Therefore, by choosing 
\begin{align}
e^{\tilde{f}_\tau \ell/2}  \ge \frac{5c_{\tau,1} \lambda_{1/2} e^{2\tilde{f}_\tau} \gamma (D/\tilde{f}_\tau)^D q_\ast  |X| }{2c_{\tau,2}} 
\longrightarrow \ell \ge  \frac{2}{\tilde{f}_\tau} \log \br{\frac{5c_{\tau,1} \lambda_{1/2} e^{2\tilde{f}_\tau} \gamma (D/\tilde{f}_\tau)^D q_\ast  |X| }{2c_{\tau,2}} },
\end{align}
we have $\tilde{M}_1 \le 2c_{\tau,2} \le 2c_{\tau,2}  s$ from the inequality~\eqref{upper_bound_tilde_M_1_2}. 

Then, for $p=1$, the inequality~\eqref{proof_start_for_boson_move_outside_fin} reduces to
\begin{align}
\label{proof_start_for_boson_move_outside_fin_case2}
\norm{  \Pi_{X[\ell], \le N-\delta N} e^{-iHt} \ket{\psi_{m_{0:{\tilde{p}}+1}} }} 
&\le  \br{\frac{\tilde{M}_1+2 c_{\tau,2} s}{ M_1 + m_0- N+\delta N }}^{s/2} 
\le \br{\frac{4 c_{\tau,2} s}{m_0- N+\delta N }}^{s/2} .
\end{align}
By choosing 
\begin{align}
s= \floorf {\frac{m_0- N+\delta N}{4e c_{\tau,2}}  } ,
\end{align}
we reduce the inequality~\eqref{proof_start_for_boson_move_outside_fin_case2} to 
\begin{align}
\label{proof_start_for_boson_move_outside_fin_case1_2}
\norm{  \Pi_{X[\ell], \le N-\delta N} e^{-iHt} \ket{\psi_{m_{0:{\tilde{p}}+1}} }} 
&\le 2\exp\br{\frac{-m_0+ N-\delta N}{8e c_{\tau,2}}} .
\end{align}
From the inequalities~\eqref{proof_start_for_boson_move_outside_fin_case1} and \eqref{proof_start_for_boson_move_outside_fin_case1_2}, we prove the main inequality~\eqref{proof_start_for_boson_move_outside_fin_cases1and2}.
This completes the proof. $\square$

 {~}

\hrulefill{\bf [ End of Proof of Lemma~\ref{Lema:upp_pi_X_ell_m_0_q+1}] }

{~}

By combining the inequalities~\eqref{Pi_X_ell_N-delta N_psi_tau_upp} and \eqref{proof_start_for_boson_move_outside_fin_cases1and2}, we obtain
\begin{align}
\label{summation_from_m_0_m_q+1}
\norm{ \Pi_{X[\ell], \le N-\delta N} \ket{\psi(\tau)}}  
&\le\sum_{m_0,m_1,\ldots,m_{\tilde{p}+1}} 2\exp\br{\frac{-m_0+ N-\delta N}{8e c_{\tau,2}}} .
\end{align}
Because of $\Pi_{\mathcal{S}(X[\ell],1/2, q_\ast)} \ket{\psi}= \ket{\psi}$, the summation for $m_p$ is taken from $m_p=0$ to 
\begin{align}
m_p = \norm{ \Pi_{\mathcal{S}(X[\ell],1/2, q_\ast)} \nb_{X_p} } \le q_\ast |X_p| \le q_\ast \gamma \ell^{D} |X| ,
\end{align}
where we use $|X_p| \le |X[\ell]|\le \gamma \ell^D |X|$. 
We recall that $m_p$ is defined as the number of bosons in the region $X_p$ as in Eq.~\eqref{Pi_m_0:q+1_def}.
Also, the summation with respect to $m_0$ is taken from $m_0=N$, and hence, we have 
\begin{align}
\sum_{m_0=N}^\infty \exp\br{\frac{-m_0+ N-\delta N}{8e c_{\tau,2}}} \le \frac{1}{1-e^{-1/(8e c_{\tau,2})}}\exp\br{\frac{-\delta N}{8e c_{\tau,2}}} .
\end{align}
Therefore, the summation in~\eqref{summation_from_m_0_m_q+1} is upper-bounded by
\begin{align}
\label{summation_from_m_0_m_q+1__2}
\norm{ \Pi_{X[\ell], \le N-\delta N} \ket{\psi(\tau)}}  
&\le \frac{2 \br{q_\ast \gamma \ell^{D} |X|}^{{\tilde{p}}+1} }{1-e^{-1/(8e c_{\tau,2})}}\exp\br{\frac{-\delta N}{8e c_{\tau,2}}} \notag \\
&= \frac{2}{1-e^{-1/(8e c_{\tau,2})}}\exp\brr{\frac{-\delta N}{8e c_{\tau,2}} + \frac{\ell+4}{2} \log\br{q_\ast \gamma \ell^{D} |X|}} ,
\end{align}
where we use $\tilde{p}=\ell/2 +1$.
Under condition~\eqref{condition_for_ell_prop_ge_g}, we can choose $\ell$ such that 
\begin{align}
\ell = \ceilf{\frac{2}{\tilde{f}_\tau} \log \br{\frac{5c_{\tau,1} \lambda_{1/2} e^{2\tilde{f}_\tau} \gamma (D/\tilde{f}_\tau)^D q_\ast  |X| }{2c_{\tau,2}} }}= c'_\ell \log \br{q_\ast \gamma |X|} ,
\end{align}
where $c'_\ell $ is a constant of $\orderof{1}$. 
Then, we obtain an upper bound for  
\begin{align}
\label{summation_from_m_0_m_q+1__2_exp_term_up}
\frac{\ell+4}{2} \log\br{q_\ast \gamma \ell^{D} |X|} &=\frac{\ell+4}{2} \brr{ \log\br{q_\ast \gamma  |X|} + D \log\br{\ell} }  
= \frac{\ell+4}{2} \br{ \frac{\ell}{c_\ell'} +D  \log\br{\ell} } \notag \\
&\le 3\br{\frac{1}{c_\ell'} + D}\ell^2   
= 3c_\ell' \br{1 + c_\ell'  D }\log^2 \br{q_\ast \gamma |X|} ,
\end{align}
where in the inequality, we use $\ell\ge 1$, $\ell/2+2 \le 5\ell/2 \le 3\ell$, and $\log(\ell) \le \ell$. 
By combining the inequalities~\eqref{summation_from_m_0_m_q+1__2} and \eqref{summation_from_m_0_m_q+1__2_exp_term_up}, 
we obtain the main inequality~\eqref{main_ineq:prop:lower_probability}.
This completes the proof. $\square$

\section{Effective Hamiltonian theory}

In this section, we consider an effective Hamiltonian in the form of
\begin{align}
\bar{\Pi} H \bar{\Pi} , 
\end{align} 
where $\bar{\Pi} $ is an arbitrary projection such that $[\bar{\Pi}, \nb_i] = 0$ for $\forall i\in \Lambda$. 
Then, we often encounter a problem to estimate an error between the time evolutions $e^{-i\bar{\Pi} H \bar{\Pi}t}$ and $e^{-iHt}$.
Usually, we cannot ensure 
\begin{align}
\norm{ e^{-i\bar{\Pi} H \bar{\Pi}t}- e^{-iHt} } 
\end{align} 
in the level of the operator norm. 
If we consider specific quantum state $\ket{\psi}$ and operator $O$, we may be able to prove 
\begin{align}
\norm{ O(H,\tau) \ket{\psi} - O(\bar{\Pi} H \bar{\Pi}, \tau) \ket{\psi} } \approx 0 .
\end{align} 

In general, we can prove the following lemma which upper-bounds the error using the effective Hamiltonian. 
\begin{lemma}[Generalization of Lemma~13 in Ref.~\cite{PhysRevLett.127.070403}]  \label{effective Hamiltonian_accuracy}
Let $\ket{\psi}$ and $O$ be a quantum state and an operator of interest, respectively.  
Then, for arbitrary projection $\bar{\Pi}$ and $\tau$, we obtain 
\begin{align}
\label{error_time/evo_effectve_original_ineq_0}
\norm{ O(H,\tau) \ket{\psi} - O(\bar{\Pi} H \bar{\Pi}, \tau) \ket{\psi} } 
&\le  \norm{\bar{\Pi}^\co  e^{iH\tau} Oe^{-iH\tau} \ket{\psi} } 
+ \norm{\bar{\Pi}^\co Oe^{-iH\tau} \ket{\psi} } 
+  
\norm{\bar{\Pi}^\co   \ket{\psi} } 
+ \norm{\bar{\Pi}^\co e^{-iH\tau} \ket{\psi} } \notag \\
&+\int_0^\tau \norm{   \bar{\Pi}H_0 \bar{\Pi}^\co e^{iH (\tau - \tau_1)} O e^{-iH\tau}\ket{\psi} }   d\tau_1
+  \int_0^\tau \norm{   \bar{\Pi}H_0 \bar{\Pi}^\co e^{-iH\tau_1}\ket{\psi} }   d\tau_1  
\end{align} 
with $\bar{\Pi}^\co := 1- \bar{\Pi}$, 
where we set $\norm{O}=1$. 
\end{lemma}

We can easily extend the statement to a mixed state $\rho$:
\begin{corol}  \label{corol:effective Hamiltonian_accuracy}
Let $\rho$ be an arbitrary quantum state. Then, we generalize the inequality~\eqref{error_time/evo_effectve_original_ineq_0} to
\begin{align}
\label{error_time/evo_effectve_original_ineq_mix}
\norm{ O(H,\tau)  \rho - O(\bar{\Pi} H \bar{\Pi}, \tau)  \rho}_1 &\le  \norm{\bar{\Pi}^\co O(\tau) \sqrt{\rho} }_F+
\norm{\bar{\Pi}^\co O \sqrt{\rho(\tau)}}_F
+  \norm{\bar{\Pi}^\co \sqrt{\rho} }_F
+ \norm{\bar{\Pi}^\co \sqrt{\rho(\tau)} }_F  \notag \\
&+\int_0^\tau \norm{ \bar{\Pi}H_0 \bar{\Pi}^\co O(\tau-\tau_1) \sqrt{\rho(\tau_1) }  }_F  d\tau_1
+\int_0^\tau \norm{ \bar{\Pi}H_0 \bar{\Pi}^\co \sqrt{\rho(\tau_1)} }_F  d\tau_1 ,
\end{align}  
where $\norm{\cdots}_F$ is the Frobenius norm, i.e., $\norm{O}_F= \sqrt{\tr(O O^\dagger)}$.  
\end{corol}

\textit{Proof of Corollary~\ref{corol:effective Hamiltonian_accuracy}.}
Let us decompose the state $\rho$ as 
\begin{align}
\rho= \sum_i p_i \ket{\psi_i}\bra{\psi_i} .
\end{align}  
Then, for an arbitrary operator $\tilde{O}$, we have $$\norm{\tilde{O}\ket{\psi_i}\bra{\psi_i}}_1= \tr\br{ \sqrt{\tilde{O}\ket{\psi_i}\bra{\psi_i} \tilde{O}^\dagger}} = \sqrt{\bra{\psi_i} \tilde{O}^\dagger \tilde{O}\ket{\psi_i}} = \norm{\tilde{O}\ket{\psi_i}},$$
and hence
\begin{align}
\label{error_time/evo_effectve_original_ineq_corol}
\norm{ O(H,\tau)  \rho - O(\bar{\Pi} H \bar{\Pi}, \tau) \rho }_1 
&\le  \sum_i p_i  \biggl( \norm{\bar{\Pi}^\co  e^{iH\tau} Oe^{-iH\tau} \ket{\psi_i} } 
+ \norm{\bar{\Pi}^\co Oe^{-iH\tau} \ket{\psi_i} } 
+  
\norm{\bar{\Pi}^\co   \ket{\psi_i} } 
+ \norm{\bar{\Pi}^\co e^{-iH\tau} \ket{\psi_i} } \notag \\
&+\int_0^\tau \norm{   \bar{\Pi}H_0 \bar{\Pi}^\co e^{iH (\tau - \tau_1)} O e^{-iH\tau}\ket{\psi_i} }   d\tau_1
+  \int_0^\tau \norm{   \bar{\Pi}H_0 \bar{\Pi}^\co e^{-iH\tau_1}\ket{\psi_i} }   d\tau_1 \biggr) .
\end{align}  
Moreover, for an arbitrary operator $\tilde{O}$, we prove
\begin{align}
\sum_i p_i  \norm{\tilde{O} \ket{\psi_i} }= \sum_i p_i  \sqrt{\bra{\psi_i} \tilde{O}^\dagger \tilde{O} \ket{\psi_i} } 
\le  \sqrt{  \sum_i p_i  \bra{\psi_i} \tilde{O}^\dagger \tilde{O} \ket{\psi_i} } 
=  \sqrt{   \tr \br{\tilde{O}\rho \tilde{O}^\dagger }} = \norm{\tilde{O}\sqrt{\rho}}_F. 
\end{align}  
By combining the above two inequalities, we derive the desired inequality~\eqref{error_time/evo_effectve_original_ineq_mix}.
This completes the proof. $\square$

{~}\\

As another extension, we can also generalize the results to the time-dependent Hamiltonian $H_t$, which is utilized in proving Lemma~\ref{lemma:time_evo_highD_local_map} below.
Here, we utilize the notations in Eq.~\eqref{notation_time_dependent_operator}, i.e., 
$U_{A_t,\tau_1\to \tau_2} = \mathcal{T} e^{-i \int_{\tau_1}^{\tau_2} A_t   dt}$ and $O(A_t , \tau_1\to \tau_2) = U_{A_t, \tau_1\to \tau_2}^\dagger O U_{A_t, \tau_1\to \tau_2}$.
\begin{corol}  \label{corol:effective Hamiltonian_accuracy_depend_t}
Let us assume that $H_t$ satisfies $\norm{dH_t/dt} < \infty$.  
For simplicity, let us denote 
\begin{align}
U_{\tau'} =  U_{H_t, 0\to \tau'}
\end{align}  
for an arbitrary $\tau'$. 
We then obtain 
\begin{align}
\label{error_time/effective Hamiltonian_accuracy_depend_t}
&\norm{ O(H_t,0\to\tau)  \rho - O(\bar{\Pi} H_t \bar{\Pi}, 0\to\tau)  \rho}_1 \le  \norm{\bar{\Pi}^\co U_{\tau}^\dagger OU_{\tau} \sqrt{\rho} }_F+
\norm{\bar{\Pi}^\co O U_{\tau} \sqrt{\rho}}_F
+  \norm{\bar{\Pi}^\co \sqrt{\rho} }_F
+ \norm{\bar{\Pi}^\co U_{\tau} \sqrt{\rho} }_F  \notag \\
&\quad\quad\quad \quad\quad\quad \quad\quad\quad \quad\quad\quad +\int_0^\tau \norm{ \bar{\Pi}H_{\tau-\tau_1} \bar{\Pi}^\co U_{\tau_1}  U_{\tau}^\dagger O U_{\tau} \sqrt{\rho }  }_F  d\tau_1
+\int_0^\tau \norm{ \bar{\Pi}H_{\tau-\tau_1} \bar{\Pi}^\co   U_{\tau_1} \sqrt{\rho} }_F  d\tau_1 .
\end{align}  
The proof is exactly the same as for Lemma~\ref{effective Hamiltonian_accuracy}. 
\end{corol}

\subsubsection{Proof of Lemma~\ref{effective Hamiltonian_accuracy}} 
We take the same approach as Ref.~\cite[Supplementary Lemma~13]{PhysRevLett.127.070403}.
For the convenience of the readers, we show the full proof of this lemma.
Throughout the proof, we denote 
\begin{align}
\bar{\Pi}^\co =1- \bar{\Pi}, \quad \tilde{H}=\bar{\Pi} H \bar{\Pi}
\end{align}  
for simplicity. 
We start from the inequality as 
\begin{align}
\label{first_estimation_rho_0_error_0_2}
&\norm{ O(H,\tau) \ket{\psi} - O(\tilde{H}, \tau) \ket{\psi} }   \notag \\
&\le \left \| e^{iH\tau} O e^{-iH\tau} \ket{\psi} - e^{i\tilde{H}\tau} O e^{-iH\tau}  \ket{\psi} \right\| 
+ \left \| e^{i\tilde{H} \tau} O  ( e^{-iH\tau} - e^{-i\tilde{H}\tau}) \ket{\psi}\right \| \notag \\
&\le \left \| e^{iH\tau} O e^{-iH\tau} \ket{\psi} - e^{i\tilde{H}  \tau} O e^{-iH\tau} \ket{\psi} \right\|  
+ \left \| \ket{\psi} - e^{i\tilde{H}\tau} e^{-iH\tau}\ket{\psi}\right \|  ,
\end{align}  
where the first inequality is derived from the triangle inequality, and we use $\|O\| =1$ in the second inequality. 
In order to estimate the RHS in the inequality~\eqref{first_estimation_rho_0_error_0_2}, we need to upper-bound 
\begin{align}
\label{calculation_basic_effective_ham}
\norm{ e^{iH\tau} Oe^{-iH\tau} \ket{\psi}- e^{i\tilde{H} \tau} O e^{-iH\tau}\ket{\psi}  }
\end{align}  
for an arbitrary $O$. We note that by choosing $O=\hat{1}$ the second term in the RHS of \eqref{first_estimation_rho_0_error_0_2} is also given by the form of~\eqref{calculation_basic_effective_ham}. 
For the estimation of \eqref{calculation_basic_effective_ham}, we begin with the inequality of
\begin{align}
\label{calculation_basic_effective_ham_2}
&\norm{ e^{iH\tau} Oe^{-iH\tau} \ket{\psi}- e^{i\tilde{H} \tau} O e^{-iH\tau}\ket{\psi}  } \notag \\
\le&\norm{\bar{\Pi}^\co \br{ e^{iH\tau} Oe^{-iH\tau} \ket{\psi}- e^{i\tilde{H} \tau} O e^{-iH\tau}\ket{\psi}} }
+ \norm{ \bar{\Pi} \br{e^{iH\tau} Oe^{-iH\tau} \ket{\psi}- e^{i\tilde{H} \tau} O e^{-iH\tau}\ket{\psi}}} \notag \\
=&\norm{\bar{\Pi}^\co  e^{iH\tau} Oe^{-iH\tau} \ket{\psi} } 
+ \norm{\bar{\Pi}^\co Oe^{-iH\tau} \ket{\psi} } 
+ \norm{ \bar{\Pi} \br{ e^{iH\tau} O- e^{i\tilde{H} \tau} O} e^{-iH\tau}  \ket{\psi}} ,
\end{align}  
where, in the equation, we use 
\begin{align}
\label{calculation_basic_effective_ham_3}
\norm{ \bar{\Pi}^\co e^{i\tilde{H} \tau}  O e^{-iH\tau}\ket{\psi}} = 
\norm{\bar{\Pi}^\co  O e^{-iH\tau} \ket{\psi}} 
\end{align}  
because of $\bar{\Pi}^\co \tilde{H}=\bar{\Pi}^\co \bar{\Pi} H \bar{\Pi} = 0$. 

In order to upper-bound the third term in the RHS of \eqref{calculation_basic_effective_ham_2}, 
let us decompose the time to infinitely many pieces (i.e., $\tau=m_0 d\tau$ and $m_0\to \infty$).
We then estimate the norm $\norm{ \bar{\Pi} \br{ e^{iH\tau} O- e^{i\tilde{H} \tau} O} e^{-iH\tau}  \ket{\psi}}$ by using the following equation:
\begin{align}
\label{error_term_sum_small_dt}
e^{iH\tau} O - e^{i\tilde{H}\tau} O  &= \sum_{m=0}^{m_0-1} 
\br{ e^{i\tilde{H} md\tau} e^{iH (m_0-m)d\tau} O  - e^{i\tilde{H} (m+1)d\tau} e^{iH (m_0-1-m)d\tau} O   } \notag \\
&= 
\sum_{m=0}^{m_0-1} 
\br{ e^{i\tilde{H} m d\tau} e^{iH (\tau - m d\tau)} O  
- e^{i\tilde{H} (m d\tau +d\tau)} e^{iH (\tau - m d\tau  -d\tau)} O   }  \notag \\
&=- i  \int_0^\tau e^{i\tilde{H} \tau_1} \br{\tilde{H} -H  } e^{iH (\tau - \tau_1)} O  d\tau_1 
\end{align}  
in the limit of $m_0\to\infty$,
where, in the last equation, we use  
\begin{align}
& e^{i\tilde{H} m d\tau} e^{iH (\tau - m d\tau)} O  
- e^{i\tilde{H} (m d\tau+d\tau)} e^{iH (\tau -m d\tau -d\tau)} O    \notag \\
=&- i e^{i\tilde{H} m d\tau} \br{\tilde{H} -H  } e^{iH (\tau - m d\tau)} O d\tau  + \orderof{d\tau^2}.
\end{align}  
We consider the decomposition of 
\begin{align}
\bar{\Pi} \br{\tilde{H}-H  } = \bar{\Pi}H \bar{\Pi}  -  \bar{\Pi} H  \br{\bar{\Pi}+1-\bar{\Pi}} 
=- \bar{\Pi}H \bar{\Pi}^\co=-\bar{\Pi}H_0 \bar{\Pi}^\co  ,
\end{align}  
where we use $[\bar{\Pi},\tilde{H}]=0$ and $[V,\bar{\Pi}]=0$ which is derived from $[\nb_i, \bar{\Pi}]=0$ for $\forall i\in \Lambda$. 
We recall that $\bar{\Pi}^\co$ has been defined as $\bar{\Pi}^\co := 1- \bar{\Pi}$.
Therefore, from Eq.~\eqref{error_term_sum_small_dt}, we obtain 
\begin{align}
\label{error_term_sum_small_dt_reduce}
\norm{ \bar{\Pi}\br{ e^{iH\tau} O - e^{i\tilde{H}\tau} O} e^{-iH\tau}\ket{\psi}  } 
=&\norm{ i  \int_0^\tau    e^{i\tilde{H} \tau_1}  \bar{\Pi}H_0 \bar{\Pi}^\co  
 e^{iH (\tau - \tau_1)} d\tau_1  \cdot O e^{-iH\tau}\ket{\psi} } \notag \\
\le&\int_0^\tau \norm{   \bar{\Pi}H_0 \bar{\Pi}^\co  
e^{iH (\tau - \tau_1)} O e^{-iH\tau}\ket{\psi} }   d\tau_1 ,
\end{align}  
which reduces the inequality~\eqref{calculation_basic_effective_ham_2} to
\begin{align}
\label{calculation_basic_effective_ham_2_fini}
&\norm{ e^{iH\tau} Oe^{-iH\tau} \ket{\psi}- e^{i\tilde{H} \tau} O e^{-iH\tau}\ket{\psi}  } \notag \\
\le&\norm{\bar{\Pi}^\co  e^{iH\tau} Oe^{-iH\tau} \ket{\psi} } 
+ \norm{\bar{\Pi}^\co Oe^{-iH\tau} \ket{\psi} } 
+ \int_0^\tau \norm{   \bar{\Pi}H_0 \bar{\Pi}^\co  
e^{iH (\tau - \tau_1)} O e^{-iH\tau}\ket{\psi} }   d\tau_1.
\end{align}

In the same way, we can derive 
\begin{align}
\label{calculation_basic_effective_ham_fin_2}
\norm{  \ket{\psi}- e^{i\tilde{H} \tau} e^{-iH\tau}\ket{\psi}  } \le
\norm{\bar{\Pi}^\co   \ket{\psi} } 
+ \norm{\bar{\Pi}^\co e^{-iH\tau} \ket{\psi} } 
+  \int_0^\tau \norm{   \bar{\Pi}H_0 \bar{\Pi}^\co  
e^{-iH\tau_1}\ket{\psi} }   d\tau_1 
\end{align}  
by choosing $O=\hat{1}$ in the inequality~\eqref{calculation_basic_effective_ham_2_fini}. 
Therefore, by applying the inequalities~\eqref{calculation_basic_effective_ham_2_fini} and \eqref{calculation_basic_effective_ham_fin_2} to \eqref{first_estimation_rho_0_error_0_2}, we obtain the desired inequality~\eqref{error_time/evo_effectve_original_ineq_0}.
This completes the proof. $\square$

\subsection{Probability distribution for boson number after time evolution} \label{sec:Probability distribution for boson number after time evolution}

In estimating the errors in~\eqref{error_time/evo_effectve_original_ineq_0} for the effective Hamiltonian, we have to estimate the norms like
\begin{align}
 \norm{\bar{\Pi}^\co  e^{iH\tau_2} Oe^{-iH\tau_1} \ket{\psi} }  \AND 
  \norm{  \bar{\Pi}H_0 \bar{\Pi}^\co  e^{iH\tau_2} Oe^{-iH\tau_1} \ket{\psi} } .
\end{align}  
For this purpose, we let 
\begin{align}
O \to O_{X_0}\quad (X_0\subseteq i_0[r_0]) ,\quad  \ket{\psi} \to \rho_0 , \label{def_r_0X_0}
\end{align}
where $\rho_0$ satisfies the low-boson density condition in Assumption~\ref{only_the_assumption_initial} with $b_\rho=b_0/\gamma$.
We also assume that the operator $O_{X_0}$ satisfies the following condition, which is characterized by a parameter $q_0$:
\begin{align}
\label{O_X0_condition/_norm_lemma_operator_boson_pre}
\Pi_{X_0, > m+q_0} O_{X_0} \Pi_{X_0, m}=0 , \quad \Pi_{X_0, < m-q_0} O_{X_0} \Pi_{X_0, m} =0  \for \forall m >0.
\end{align}
Here, the constant $q_0$ is the maximum number of boson creation (or annihilation) by the operator $O_{X_0}$.
The parameter $q_0$ appears in the Lieb-Robinson bound (see Theorems~\ref{main_theorem_looser_Lieb_Robinson}, \ref{main_thm:boson_Lieb-Robinson_main} and \ref{main_thm:boson_Lieb-Robinson_main_1D}).
For example, if $O_{X_0}$ conserve the boson number in the region $X_0$, we have $q_0=0$. 
We also set the norm of $O_{X_0}$ as 
\begin{align}
\norm{O_{X_0}} = \zeta_0 
\end{align}
for simplicity. 

In the above setup, we consider a quantum state of\footnote{In the definition~\eqref{def:tilde_rho_tau_1tau2}, we consider a time evolution by a time-independent Hamiltonian. But, all the discussion can be straightforwardly generalized to 
\begin{align}
\label{def:tilde_rho_tau_1tau2_gene}
\tilde{\rho}_{\tau_1,\tau_2} := U_{H'_x,0\to \tau_2}^\dagger  O_{X_0} U_{H_x,0\to \tau_1}\rho_0 U_{H_x,0\to \tau_1}^\dagger O_{X_0}^\dagger U_{H'_x,0\to \tau_2} ,
\end{align}
where $H_x$ and $H_x'$ are time-dependent Hamiltonians and we use the notation~\eqref{notation_time_dependent_operator}.  
}
\begin{align}
\label{def:tilde_rho_tau_1tau2}
\tilde{\rho}_{\tau_1,\tau_2} := e^{iH\tau_2} O_{X_0} e^{-iH\tau_1} \rho_0 e^{iH\tau_1}  O_{X_0}^\dagger e^{-iH\tau_2}    \quad {\rm with} \quad  |\tau_1|,|\tau_2| \le t.  
\end{align}
Note that $\tilde{\rho}_{\tau_1,\tau_2}$ is simply equal to $\rho_0(t)$ by choosing $\tau_1=t$, $\tau_2=0$ and $O_{X_0}=\hat{1}$.
Our purpose in this section is to estimate the upper bound of the probability distribution
\begin{align}
\tr\brr{\Pi_{i,>q_i} \tilde{\rho}_{\tau_1,\tau_2}} \for i\in \Lambda.
\end{align}
As has been proved in Proposition~\ref{main_theorem_boson_concentration}, the distribution decays (sub)exponentially for $q_i \gtrsim [t\log (t)]^D$ if the site $i$ is sufficiently separated from the region $X_0$. 
We here prove the following lemma:
\begin{lemma}\label{lemm:boson_density_prob_gene_time_evo}
We define $\bar{\ell}_{t,R}$ as 
\begin{align}
\label{eq_bar_ell_t_r}
\bar{\ell}_{t,R} := \bar{c}_0t \log (Rq_0)   ,
\end{align}
where $\bar{c}_0$ is an appropriate constant and $R$ is a length scale such that $R\ge t$ and $R\ge r_0$ [see Eq.~\eqref{def_r_0X_0} for the definition of $r_0$]. 
Then, for a quantum state $\tilde{\rho}_{\tau_1,\tau_2}$ in Eq.~\eqref{def:tilde_rho_tau_1tau2},  we obtain
\begin{align}
\label{lemm:boson_density_prob_gene_time_evo/main01}
\tr \br{  \nb_{i[\ell_0]}^p \Pi_{i[\ell_0],> q_i}\tilde{\rho}_{\tau_1,\tau_2}  } 
\le \zeta_0^2 q_i^p e^{- \brr{q_i/(4 b_0 \ell_0^D) }^{1/\kappa}}   \for i[\ell_0] \subset X_0[\bar{\ell}_{t,R}]^\co  ,
\end{align}
and 
 \begin{align}
 \label{lemm:boson_density_prob_gene_time_evo/main02}
&\tr \br{\nb_{X_0[\bar{\ell}_{t,R}]}^p  \Pi_{X_0[\bar{\ell}_{t,R}] ,> q} \tilde{\rho}_{\tau_1,\tau_2}  } 
\le \zeta_0^2q^p  e^{- \brr{ ( q - 32e^2 q_0^3)/\xi_{t,R} }^{1/\kappa} }  ,
 \quad \xi_{t,R} := 16 b_0 \bar{\ell}_{t,R}^{2D}  |X_0| ,
\end{align}
where  $k\in \mathbb{N}$ and $\ell_0$ is assumed to be larger than $\bar{\ell}_{t,R}$ ($\ell_0\ge \bar{\ell}_{t,R}$). 
\end{lemma}

 \begin{figure}[tt]
\centering
\includegraphics[clip, scale=0.37]{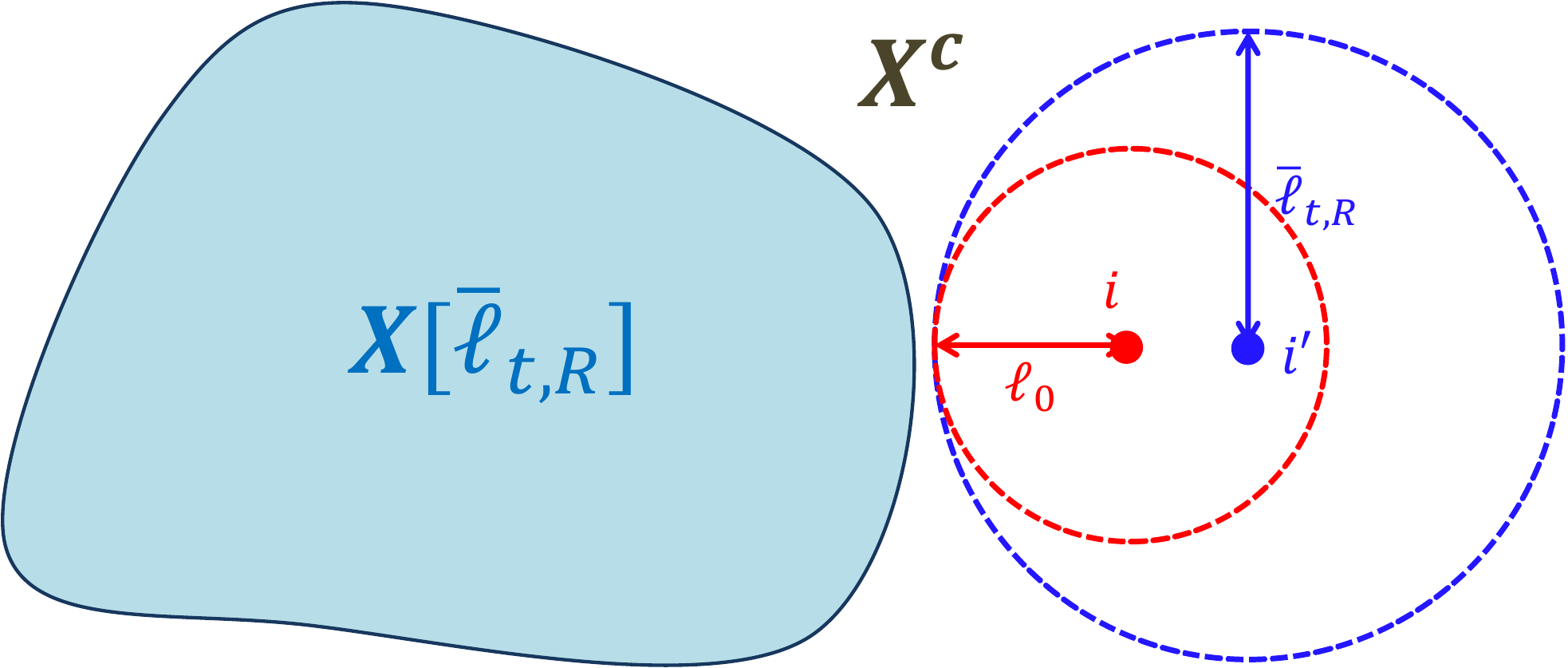}
\caption{For $\forall i[\ell_0] \in X[\bar{\ell}_{t,R}]^\co$ with $\ell_0< \bar{\ell}_{t,R}$, 
we can find another site $i'\in X^\co$ such that $i'[\bar{\ell}_{t,R}] \subset X[\bar{\ell}_{t,R}]^\co$ and $i'[\bar{\ell}_{t,R}] \supset i[\ell_0]$.
}
\label{supp_fig_site_i_i_dash_include.pdf}
\end{figure}

{\bf Remark.} 
In the inequality~\eqref{lemm:boson_density_prob_gene_time_evo/main01}, the condition $\ell_0\ge \bar{\ell}_{t,R}$ is assumed.
In the case of $\ell_0< \bar{\ell}_{t,R}$, we can still prove the following inequality: 
\begin{align}
\label{lemm:boson_density_prob_gene_time_evo/main1}
&\tr \br{\nb_{i[\ell_0]}^p  \Pi_{i[\ell_0],> q_i}\tilde{\rho}_{\tau_1,\tau_2} } 
\le \zeta_0^2q_i^p e^{- \brr{q_i/(4 b_0 \bar{\ell}_{t,R} ^D) }^{1/\kappa}}  \for i[\ell_0] \subset X_0[\bar{\ell}_{t,R}]^\co  .
\end{align}
The proof is given as follows.
First, for $i[\ell_0] \subset X_0[\bar{\ell}_{t,R}]^\co$, 
 there exists a site $i'$ such that $i'[\bar{\ell}_{t,R}]\supset i[\ell_0]$ and $i'[\bar{\ell}_{t,R}]\subset X_0[\bar{\ell}_{t,R}]^\co$ (see Fig.~\ref{supp_fig_site_i_i_dash_include.pdf}).
Hence, we obtain 
\begin{align}
\tr \br{\nb_{i[\ell_0]}^p  \Pi_{i[\ell_0],> q_i}\tilde{\rho}_{\tau_1,\tau_2} } 
&\le \tr \br{ \nb_{i'[\bar{\ell}_{t,R}]}^p \Pi_{i'[\bar{\ell}_{t,R}],> q_i} \tilde{\rho}_{\tau_1,\tau_2} },
\end{align}
which yields the inequality~\eqref{lemm:boson_density_prob_gene_time_evo/main1} from~\eqref{lemm:boson_density_prob_gene_time_evo/main01} by choosing $\ell_0=\bar{\ell}_{t,R}$.

%
%
%

\subsubsection{Proof of Lemma~\ref{lemm:boson_density_prob_gene_time_evo}}

The proof relies on the iterative uses of Proposition~\ref{main_theorem_boson_concentration}.
In the first step, we consider the moment function of 
\begin{align}
\tr \br{ \rho_0(\tau_1)  \nb_{i[\ell_0]}^s }  \for |\tau_1|\le t
\end{align}
for $\forall i \in \Lambda$. 
From Assumption~\ref{only_the_assumption_initial} with $b_\rho=b_0/\gamma$ in~\eqref{condition_for_moment_generating}, the initial state $\rho_0$ satisfies 
\begin{align}
\tr\br{\rho_0 \nb_{i[\ell_0]}^s} \le \frac{1}{e} \br{\frac{b_0}{e} s^\kappa \ell_0^D}^s    .
\end{align}
Hence, we obtain from Proposition~\ref{main_theorem_boson_concentration}
\begin{align}
\label{ineq_first_step_time_tau_1}
\tr \br{ \rho_0(\tau_1)  \nb_{i[\ell_0]}^s }   \le \frac{1}{e} \br{\frac{2b_0}{e} s^\kappa \ell_0^D}^s  \for \forall i\in \Lambda , \quad \ell_0\ge \bar{\ell}_1 ,
\end{align}
where $\bar{\ell}_1$ is determined by the condition~\eqref{cond_main_theorem_boson_concentration_X_0=empty}, i.e., $\bar{\ell}_1\propto t\log t$.

In the second step, we first consider the moment function of 
\begin{align}
\tr \br{ O_{X_0} \rho_0(\tau_1) O_{X_0}^\dagger  \nb_{i[\ell_0]}^s } =: \tr \br{ \tilde{\rho}_{\tau_1}  \nb_{i[\ell_0]}^s }  \for i [\ell_0]\subset X_0^\co .
\end{align}
From the inequality~\eqref{ineq_first_step_time_tau_1} and $\brr{O_{X_0} ,\nb_{i[\ell_0]}^s}=0$, we trivially obtain 
\begin{align}
\tr\br{\tilde{\rho}_{\tau_1}  \nb_{i[\ell_0]}^s} \le \frac{\zeta_0^2}{e} \br{\frac{2b_0}{e} s^\kappa \ell_0^D}^s    \for i[\ell_0] \in X_0^\co , \quad \ell_0\ge \bar{\ell}_1 ,
\label{assump_main_theorem_boson_concentration_lemma}
\end{align}
where we use $\norm{O_{X_0}}= \zeta_0$.
On the other hand, if we consider a boson number operator $\nb_Z$ such that $Z\cap X_0\neq	 \emptyset$,
we cannot obtain the same inequality as~\eqref{assump_main_theorem_boson_concentration_lemma} due to the boson creation by $O_{X_0}$.
We then consider a subset $X[r]$ for generic $r \ge \bar{\ell}_1$ and $X\subset \Lambda$. 
From the condition~\eqref{O_X0_condition/_norm_lemma_operator_boson_pre},
 the operator $O_{X_0}$ increases the boson number up to $q_0$ in the region $X_0$, and hence we utilize Lemma~\ref{lem:boson_operator_norm}
 with 
 $$O_q \to O_{X_0} ,\quad {\rm and} \quad  \br{\nu_i= 1 \for  i \in X[r], \quad \And \quad \nu_i= 0 \for  i \in X[r]^\co},$$ 
which yields
\begin{align}
\tr\br{\tilde{\rho}_{\tau_1} \nb_{X[r]}^s} =\tr \br{ O_{X_0} \rho_0(\tau_1) O_{X_0}^\dagger  \nb_{X[r]}^s }
&\le 4^s \zeta_0^2 \tr\brr{ \rho_0(\tau_1)(4q_0^3 +\nb_{X[r]} )^s} \notag \\
&\le 4^s\zeta_0^2  \brrr{ 4q_0^3 + \sum_{i\in X} \brr{ \tr\br{ \rho_0(\tau_1)\nb_{i[r]}^s}}^{1/s} }^s .
\label{assump_main_theorem_boson_concentration_inX_0_lemma}
\end{align}
where in the second inequality we use $\nb_{X[r]} \preceq  \sum_{i\in X} \nb_{i[r]}$ and 
$$\ave{\br{O_1+O_2+\cdots +O_m}^s} \le \br{\ave{O_1^s}^{1/s}+\ave{O_2^s}^{1/s}+\cdots +\ave{O_m^s}^{1/s}}^s$$
for arbitrary positive semi-definite Hermitian operators $\{O_1,O_2,\ldots,O_m\}$ as long as $\{O_j\}_{j=1}^m$ commute with each other. 
By applying the inequality~\eqref{ineq_first_step_time_tau_1} to~\eqref{assump_main_theorem_boson_concentration_inX_0_lemma}, we have 
\begin{align}
\tr(\tilde{\rho}_{\tau_1} \nb_{X[r]}^s)
&\le 4^s\zeta_0^2  \br{ 4q_0^3 + \frac{|X|}{e^{1/s}} \cdot \frac{2b_0}{e} s^\kappa  r^D} ^s 
\le \frac{\zeta_0^2}{e}  \br{ 16 e q_0^3 + |X| \cdot \frac{8b_0}{e} s^\kappa r^D  } ^s \for \forall r\ge \bar{\ell}_1 \notag \\
&\le \frac{\zeta_0^2}{e}  \br{ 16 e q_0^3 + |X| \cdot \frac{8b_0\bar{\ell}_1^D}{e} s^\kappa \max (r^D,1)  } ^s \for \forall r\ge 0 ,
\label{assump_main_theorem_boson_concentration_inX_0_lemma__2}
\end{align}
where we can choose $\forall X\subset \Lambda$.

In the final step, we consider the moment function for the quantum state
\begin{align}
\tilde{\rho}_{\tau_1,\tau_2}=e^{iH\tau_2} O_{X_0} e^{-iH\tau_1} \rho e^{iH\tau_1}  O_{X_0}^\dagger e^{-iH\tau_2}  
=  e^{iH\tau_2} \tilde{\rho}_{\tau_1} e^{-iH\tau_2} = \tilde{\rho}_{\tau_1}(-\tau_2) . 
\end{align}
Here, the initial state $\tilde{\rho}_{\tau_1}(0)$ satisfies the inequalities~\eqref{assump_main_theorem_boson_concentration_inX_0_lemma__2} and~\eqref{assump_main_theorem_boson_concentration_lemma},  
which give the initial conditions~\eqref{assump_main_theorem_boson_concentration_inX_0} and \eqref{assump_main_theorem_boson_concentration} for Proposition~\ref{main_theorem_boson_concentration}, respectively, and hence 
 we can let in Proposition~\ref{main_theorem_boson_concentration}
\begin{align}
b_0\to 2b_0,\quad  q_0 \to 16 e q_0^3, \quad q_1 \to  \frac{8b_0\bar{\ell}_1^D }{e} ,\quad \bar{\ell}_0 \to \bar{\ell}_1 \propto t\log(t) .
\end{align}
Then, from the inequalities~\eqref{main_ineq:main_theorem_boson_concentration} and \eqref{main_ineq:main_theorem_boson_concentration_X} we have
\begin{align}
\label{ineq_i_in_bar_C_0_t_r}
\tr \br{ \tilde{\rho}_{\tau_1}(-\tau_2) \nb_{i[\ell_0]}^s } \le  \frac{\zeta_0^2}{e}  \br{\frac{4b_0}{e} s^\kappa \ell_0^D}^s    
\for i[\ell_0] \subset X_0[\bar{\ell}_2]^\co ,
\end{align}
and 
\begin{align}
\label{ineq_i_in_bar_C_0_t_r_bigger_region}
\tr \br{ \tilde{\rho}_{\tau_1}(-\tau_2) \nb_{X_0[\ell_0]}^s } 
&\le \frac{\zeta_0^2}{e}   \br{32 e q_0^3 + \frac{16b_0\bar{\ell}_1^D\ell_0^{D}}{e} s^\kappa  |X_0|   } ^s ,
\end{align}
where $\ell_0 \ge \max(2D\bar{\ell_2},\bar{\ell}_1)$ and $\bar{\ell}_2$ is given by the condition~\eqref{cond_main_theorem_boson_concentration}, i.e.,
\begin{align}
\bar{\ell}_2= t\log [\Theta(t q_0 |X_0|) ],
\end{align}
which is also given in the form of Eq.~\eqref{eq_bar_ell_t_r} because of $X_0 = i_0[r_0] \subseteq i_0[R]$. 
In the following, we define $\bar{\ell}_{t,R}$ as $2D\bar{\ell}_2$, i.e., $\bar{\ell}_2 := \bar{\ell}_{t,R}/(2D)$.

By applying Lemma~\ref{lem:probability_dist_moment} [i.e., the inequality~\eqref{ineq:lem:probability_dist_moment_pre}] to the inequality~\eqref{ineq_i_in_bar_C_0_t_r}, we obtain 
\begin{align}
\tr \br{ \nb_{i[\ell_0]}^p  \Pi_{i[\ell_0],> q_i} \tilde{\rho}_{\tau_1,\tau_2}  } 
\le \zeta_0^2 q_i^p  e^{- \brr{q_i/(4 b_0 \ell_0^D) }^{1/\kappa}}   \for i[\ell_0] \subset X_0[\bar{\ell}_{t,R}/(2D)]^\co  ,
\end{align}
where we assume $\ell_0\ge \bar{\ell}_{t,R}$. 
Note that $i[\ell_0] \subset X_0[\bar{\ell}_{t,R}/(2D)]^\co $ is included in the condition $i[\ell_0] \subset X_0[\bar{\ell}_{t,R}]^\co$.
Moreover, we apply Lemma~\ref{lem:probability_dist_moment} to the inequality~\eqref{ineq_i_in_bar_C_0_t_r_bigger_region} with $\ell_0=\bar{\ell}_{t,R}$, which yields 
\begin{align}
\tr \br{ \nb_{X_0[\bar{\ell}_{t,R}]}^p \Pi_{X_0[\bar{\ell}_{t,R}] ,> q} \tilde{\rho}_{\tau_1,\tau_2}  } 
&\le \zeta_0^2 q^p \exp\brr{ - \br{\frac{ q - 32 e^2 q_0^3 }{16b_0\bar{\ell}_{t,R}^{2D} 
 |X_0| } }^{1/\kappa} } ,
\end{align}
where we use $\bar{\ell}_1 \le \bar{\ell}_{t,R}$. 
We thus prove the main inequalities~\eqref{lemm:boson_density_prob_gene_time_evo/main01} and \eqref{lemm:boson_density_prob_gene_time_evo/main02}.
This completes the proof of Lemma~\ref{lemm:boson_density_prob_gene_time_evo}. $\square$

\subsection{Effective Hamiltonian by the truncation of the boson number: first result}

By combining the obtained statements in Lemma~\ref{effective Hamiltonian_accuracy} (or Corollary~\ref{corol:effective Hamiltonian_accuracy}) and Lemma~\ref{lemm:boson_density_prob_gene_time_evo}, we aim to derive an explicit error bound for the effective Hamiltonian.
We here define the projection operator $\bar{\Pi}_{L,\bar{q}}$ as 
\begin{align}
\label{def_pi_bar_bar_q_e_L}
\bar{\Pi}_{L,\bar{q}}:= \prod_{i\in L} \Pi_{i, \le \bar{q}} , 
\end{align}
which uniformly truncates the boson number at each site in the region $L$ by $\bar{q}$. 
This type of projection will be considered in proving a non-optimal Lieb-Robinson bound in Theorem~\ref{main_theorem_looser_Lieb_Robinson} in Sec.~\ref{sec:Bosonic Lieb-Robinson bound: looser results}.

%

\begin{figure}[tt]
\centering
\includegraphics[clip, scale=0.4]{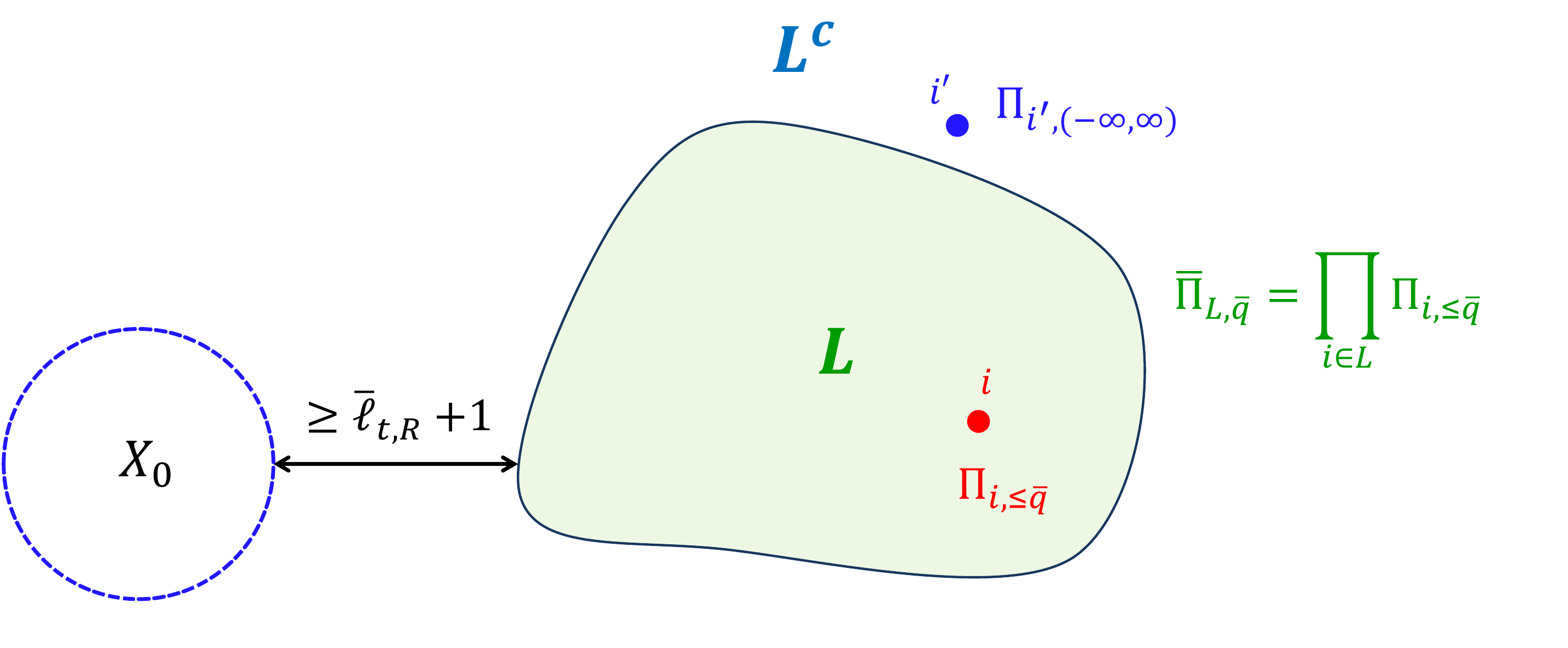}
\caption{Definition of the projection $\bar{\Pi}_{L,\bar{q}}$. In the projection, we truncate the boson number at each of the sites in the region $L$ up to $\bar{q}$, 
while there are no restrictions on the boson numbers in $L^\co$. For the projection, 
the approximation error for dynamics by the effective Hamiltonian $\bar{\Pi}_{L,\bar{q}} H \bar{\Pi}_{L,\bar{q}}$ is estimated in Proposition~\ref{prop:Effective Hamiltonian_local_region}.
}
\label{supp_fig_Projection_L_bar_q}
\end{figure}

We here approximate the dynamics $e^{-iHt}$ by using the effective Hamiltonian $\bar{\Pi}_{L,\bar{q}}H\bar{\Pi}_{L,\bar{q}}$, whose approximation error is obtained from the following proposition: 
\begin{prop} \label{prop:Effective Hamiltonian_local_region}
In the same setup as in Sec.~\ref{sec:Probability distribution for boson number after time evolution}, we obtain the approximation error as 
\begin{align}
\label{prop:Effective Hamiltonian_local_region_eq}
\norm{ O_{X_0}(H,t)  \rho_0 - O_{X_0}(\bar{\Pi}_{L,\bar{q}} H\bar{\Pi}_{L,\bar{q}}, t)  \rho_0 }_1 &\le  4 \zeta_0 \gamma |L| \br{1+ \bar{J}t\bar{q} } e^{- \brr{\bar{q}/(4 b_0 \bar{\ell}_{t,R} ^D) }^{1/\kappa}/2},
\end{align}  
where we assume $\dist_{X_0,L} \ge \bar{\ell}_{t,R}+1$ (see Fig.~\ref{supp_fig_Projection_L_bar_q}). 
\end{prop}

\subsubsection{Proof of Proposition~\ref{prop:Effective Hamiltonian_local_region}}

In applying Lemma~\ref{effective Hamiltonian_accuracy} (or Corollary~\ref{corol:effective Hamiltonian_accuracy}) to the estimation of the norm in~\eqref{prop:Effective Hamiltonian_local_region_eq}, we need to estimate the norms of
$
\norm{ \bar{\Pi}_{L,\bar{q}}^\co O_{X_0} (\tau_2) \sqrt{\rho_0(\tau_1)} }_F 
$
and 
$
\norm{ \bar{\Pi}_{L,\bar{q}}H_0 \bar{\Pi}_{L,\bar{q}}^\co O_{X_0} (\tau_2) \sqrt{\rho_0(\tau_1)} }_F   
$
for arbitrary $|\tau_1|,|\tau_2| \le t$. 
For the quantities, we aim to prove the following upper bounds:
\begin{align}
\label{ineq_O_X_bar_Pi_tau_2_1_upp_1st}
\norm{ \bar{\Pi}_{L,\bar{q}}^\co O_{X_0} (\tau_2) \sqrt{\rho_0(\tau_1)} }_F\le  \zeta_0  |L|^{1/2} e^{- \brr{\bar{q}/(4 b_0 \bar{\ell}_{t,R} ^D) }^{1/\kappa}/2},
 \end{align}
and 
\begin{align}
\label{ineq_O_X_bar_Pi_tau_2_2_upp_1st}
\norm{ \bar{\Pi}_{L,\bar{q}}H_0 \bar{\Pi}_{L,\bar{q}}^\co O_{X_0} (\tau_2) \sqrt{\rho_0(\tau_1)} }_F   
\le2\gamma \bar{J} \zeta_0 \bar{q} |L|  e^{- \brr{\bar{q}/(4 b_0 \bar{\ell}_{t,R} ^D) }^{1/\kappa}/2} .
\end{align}
By applying the above inequalities with appropriate $\tau_1$ and $\tau_2$ to the upper bounds in~\eqref{error_time/evo_effectve_original_ineq_mix}, we obtain 
\begin{align}
\norm{ O_{X_0}(H,t)  \rho_0 - O_{X_0}(\bar{\Pi}_{L,\bar{q}} H\bar{\Pi}_{L,\bar{q}}, t)  \rho_0 }_1 
&\le  4 \zeta_0  |L|^{1/2} e^{- \brr{\bar{q}/(4 b_0 \bar{\ell}_{t,R} ^D) }^{1/\kappa}/2}
+ 4t\gamma \bar{J} \zeta_0 \bar{q} |L| e^{- \brr{\bar{q}/(4 b_0 \bar{\ell}_{t,R} ^D) }^{1/\kappa}/2} ,
\end{align}  
which reduces to the main inequality~\eqref{prop:Effective Hamiltonian_local_region_eq} from $\gamma\ge 1$.
Therefore, the remaining tasks are proving the inequalities~\eqref{ineq_O_X_bar_Pi_tau_2_1_upp_1st} and \eqref{ineq_O_X_bar_Pi_tau_2_2_upp_1st}.

{~}\\

{\bf  [Proofs of the inequalities~\eqref{ineq_O_X_bar_Pi_tau_2_1_upp_1st} and \eqref{ineq_O_X_bar_Pi_tau_2_2_upp_1st}]}

By using the inequality~\eqref{lemm:boson_density_prob_gene_time_evo/main1} with $p=0$ in Lemma~\ref{lemm:boson_density_prob_gene_time_evo}, 
we immediately prove the first inequality~\eqref{ineq_O_X_bar_Pi_tau_2_1_upp_1st} as follows:
\begin{align}
\label{ineq_O_X_bar_Pi_tau_2_1_upp_derive}
\norm{ \bar{\Pi}_{L,\bar{q}}^\co O_{X_0} (\tau_2) \sqrt{\rho_0(\tau_1)} }_F^2=\tr \br{ \bar{\Pi}_{L,\bar{q}}^\co O_{X_0} (\tau_2) \rho_0(\tau_1)  O_{X_0} (\tau_2)^\dagger \bar{\Pi}_{L,\bar{q}}^\co }   
&\le \sum_{i\in L}\tr \br{  \Pi_{i, > \bar{q}} O_{X_0} (\tau_2) \rho_0(\tau_1)  O_{X_0} (\tau_2)^\dagger  }   \notag \\
&\le  \zeta_0^2 |L| e^{- \brr{\bar{q}/(4 b_0 \bar{\ell}_{t,R} ^D) }^{1/\kappa}} ,
\end{align}
where we use $i\in X_0[\bar{\ell}_{t,R}]^\co$ for $\forall i\in L$ from $\dist_{X_0,L}\ge \bar{\ell}_{t,R}+1 >\bar{\ell}_{t,R}$, which is assumed in the statement of Proposition~\ref{prop:Effective Hamiltonian_local_region}. 

Second, to prove the inequality~\eqref{ineq_O_X_bar_Pi_tau_2_2_upp_1st}, we first decompose
\begin{align}
\label{decomposition/pi_L_H_0_co}
\bar{\Pi}_{L,\bar{q}}H_0 \bar{\Pi}_{L,\bar{q}}^\co = \bar{\Pi}_{L,\bar{q}} \br{H_{0,L}  +  \partial h_L} \bar{\Pi}_{L,\bar{q}}^\co ,
\end{align}
where we use the notation~\eqref{def:Ham_surface} for $\partial h_L $ and $\bar{\Pi}_{L,\bar{q}} H_{0,L^\co }\bar{\Pi}_{L,\bar{q}}^\co =H_{0,L^\co } \bar{\Pi}_{L,\bar{q}} \bar{\Pi}_{L,\bar{q}}^\co =0$. 
Then, by using 
\begin{align}
\label{upper_bound_hopping_operator}
| b_i b_j^\dagger| = \sqrt{(b_i b_j^\dagger)^\dagger b_i b_j^\dagger} = \sqrt{\nb_i (\nb_j+1)} \preceq \nb_i+\nb_j ,
\end{align}
and 
\begin{align}
\label{definition_bar_pai_ij_bar_q}
&\bar{\Pi}_{L,\bar{q}} b_i b_j^\dagger \bar{\Pi}_{L,\bar{q}}^\co = \bar{\Pi}_{L,\bar{q}} b_i b_j^\dagger \bar{\Pi}_{i,j,\bar{q}}^\co ,\quad 
\bar{\Pi}_{i,j,\bar{q}}:=\begin{cases}
 \Pi_{i,\le \bar{q}}\Pi_{j,\le \bar{q}} &\for i,j\in L, \\
 \Pi_{i,\le \bar{q}} &\for i\in L , \quad j\in L^\co,
 \end{cases}
\end{align}
we obtain 
\begin{align}
\label{ineq_O_X_bar_Pi_tau_2_2_upp_0}
\norm{ \bar{\Pi}_{L,\bar{q}}H_0 \bar{\Pi}_{L,\bar{q}}^\co O_{X_0} (\tau_2) \sqrt{\rho_0(\tau_1)} }_F   
&\le \bar{J}\sum_{i\in L} \sum_{j: \dist_{i,j}=1}  \norm{ \bar{\Pi}_{L,\bar{q}} (b_i b_j^\dagger +{\rm h.c.} )\bar{\Pi}_{i,j,\bar{q}}^\co O_{X_0} (\tau_2) \sqrt{\rho_0(\tau_1)} }_F    \notag \\
&\le\bar{J}\sum_{i\in L} \sum_{j: \dist_{i,j}=1}  \norm{\br{|b_i b_j^\dagger| +|b_jb_i^\dagger|}\bar{\Pi}_{i,j,\bar{q}}^\co O_{X_0} (\tau_2) \sqrt{\rho_0(\tau_1)} }_F    \notag \\
&\le2 \bar{J}\sum_{i\in L} \sum_{j: \dist_{i,j}=1}  \norm{(\nb_i+\nb_j)\bar{\Pi}_{i,j,\bar{q}}^\co O_{X_0} (\tau_2) \sqrt{\rho_0(\tau_1)} }_F .
\end{align}
By applying the inequality~\eqref{lemm:boson_density_prob_gene_time_evo/main1} with $p=2$ to $\norm{(\nb_i+\nb_j)\bar{\Pi}_{i,j,\bar{q}}^\co O_{X_0} (\tau_2) \sqrt{\rho_0(\tau_1)} }_F$, we obtain 
\begin{align}
\label{ineq_O_X_bar_Pi_tau_2_2_upp_02}
\norm{(\nb_i+\nb_j)\bar{\Pi}_{i,j,\bar{q}}^\co O_{X_0} (\tau_2) \sqrt{\rho_0(\tau_1)} }_F^2  
&= \tr \brr{(\nb_i+\nb_j)^2 \bar{\Pi}_{i,j,\bar{q}}^\co O_{X_0} (\tau_2) \rho_0(\tau_1) O_{X_0} (\tau_2)^\dagger } \notag \\
&\le  \tr \brr{\nb_{i[1]}^2\Pi_{i[1],>\bar{q}} O_{X_0} (\tau_2) \rho_0(\tau_1) O_{X_0} (\tau_2)^\dagger }  \notag \\
&\le  \zeta_0^2 \bar{q}^2 e^{- \brr{\bar{q}/(4 b_0 \bar{\ell}_{t,R} ^D) }^{1/\kappa}} ,
\end{align}
where we use $\nb_i+\nb_j \preceq \nb_{i[1]}$ and $ \bar{\Pi}_{i,j,\bar{q}}^\co= \Pi_{i[1],>\bar{q}} \bar{\Pi}_{i,j,\bar{q}}^\co$ from $\dist_{i,j}=1$ in the first inequality, and in the second inequality, we ensure $i[1] \subset X_0[\bar{\ell}_{t,R}]^\co$ for $\forall i\in L$ from the condition $\dist_{X_0,L}\ge \bar{\ell}_{t,R}+1$. 

By combining the inequalities~\eqref{ineq_O_X_bar_Pi_tau_2_2_upp_0} and \eqref{ineq_O_X_bar_Pi_tau_2_2_upp_02}, we arrive at the inequality~\eqref{ineq_O_X_bar_Pi_tau_2_2_upp_1st} as follows: 
\begin{align}
\label{ineq_O_X_bar_Pi_tau_2_2_upp_0_fin}
\norm{ \bar{\Pi}_{L,\bar{q}}H_0 \bar{\Pi}_{L,\bar{q}}^\co O_{X_0} (\tau_2) \sqrt{\rho_0(\tau_1)} }_F   
&\le  2\bar{J}  \sum_{i\in L} \sum_{j: \dist_{i,j}=1} \zeta_0\bar{q} e^{- \brr{\bar{q}/(4 b_0 \bar{\ell}_{t,R} ^D) }^{1/\kappa}/2}   \notag \\
&\le 2 \gamma \bar{J} \zeta_0\bar{q} |L|e^{- \brr{\bar{q}/(4 b_0 \bar{\ell}_{t,R} ^D) }^{1/\kappa}/2} ,
\end{align}
where we use $\sum_{j: \dist_{i,j}=1} 1 \le |i[1]|\le \gamma$ for $\forall i\in \Lambda$.
This completes the proof. $\square$

\subsection{Effective Hamiltonian by the truncation of the boson number: second result}
\label{sec:Effective Hamiltonian by the truncation of the boson number: second result}

\begin{figure}[tt]
\centering
\includegraphics[clip, scale=0.4]{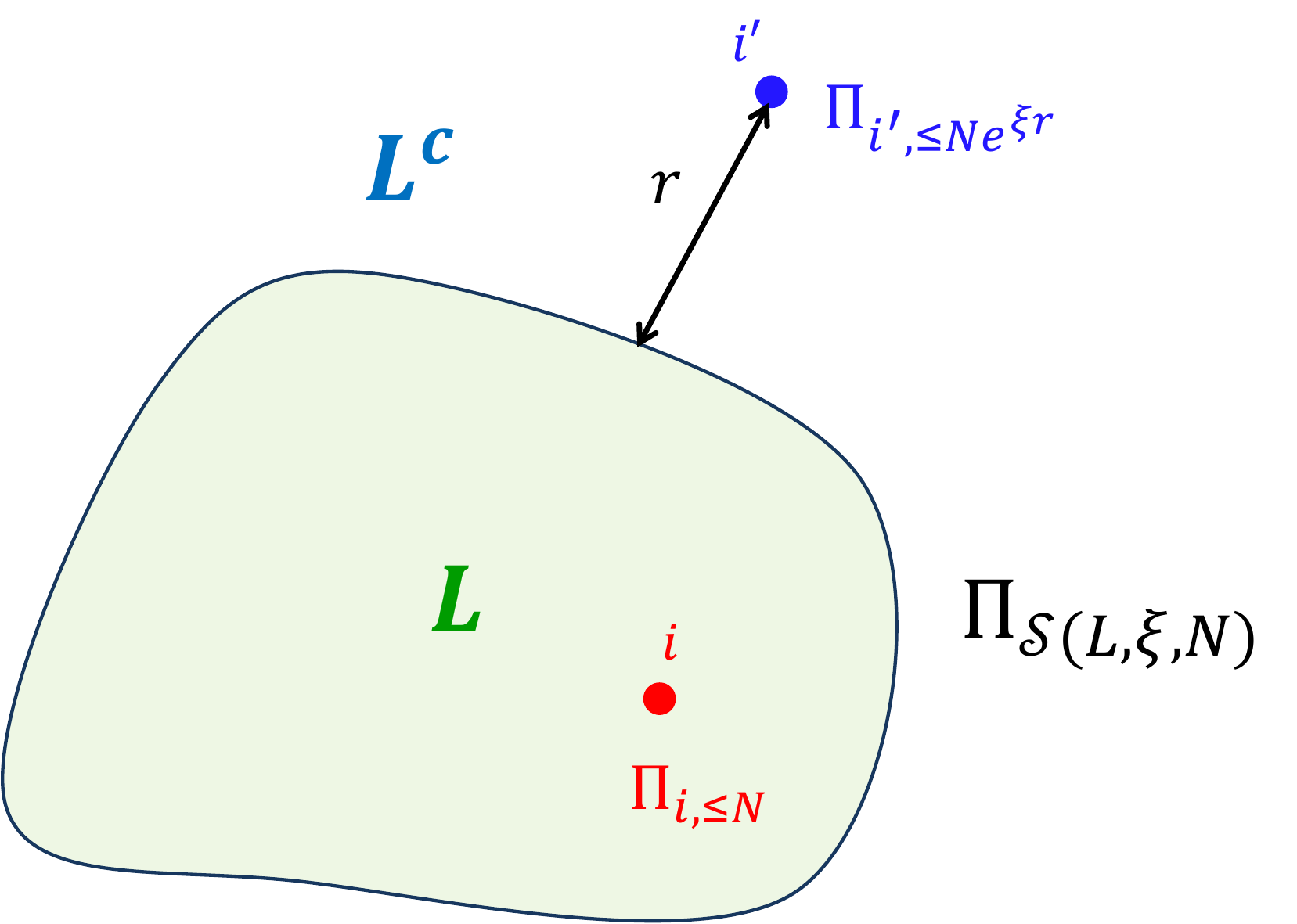}
\caption{Definition of the projection $\Pi_{\mathcal{S}(L,\xi,N)}$. In this projection, we truncate the boson number at each of the sites in the whole region up to $q_i$ ($i\in \Lambda$), 
where $q_i=N$ for $i\in L$ and $q_i=Ne^{\xi \dist_{i,L}}$ for $i\in L^\co$.
As the site $i$ is separated from the region $L$, the truncation number rapidly increases with the distance. 
The approximation error for dynamics by the effective Hamiltonian $\Pi_{\mathcal{S}(L,1/2,N)} H \Pi_{\mathcal{S}(L,1/2,N)}$ is given by
Proposition~\ref{prop:boson_number_truncation_1}. 
}
\label{supp_fig_Proj_S_L_xi_N}
\end{figure}

In Eq.~\eqref{mathcal_S_L_xi_N/def}, we have define the set $\mathcal{S}(L,\xi,N)$ as 
\begin{align}
\label{mathcal_S_L_xi_N/def_re}
\mathcal{S}(L,\xi,N) \coloneqq  \brrr{ \bold{N} | N_i \le N e^{\xi\dist_{i,L}} \for i\in  \Lambda} .
\end{align}
We consider the effective Hamiltonian, which is defined by the projection of (see Fig.~\ref{supp_fig_Proj_S_L_xi_N})
\begin{align}
\Pi_{\mathcal{S}(L,\xi,N)} := \prod_{i\in L} \Pi_{i,\le N}   \prod_{i\in L^\co}\Pi_{i,\le N e^{\xi\dist_{i,L}} } .
\end{align}
In our analyses, it is often pretty convenient to restrict the Hilbert space which is spanned by $\Pi_{\mathcal{S}(L,\xi,N)}$.

In the following, we choose $\xi=1/2$ and denote the projection
\begin{align}
\label{def_tilde_Pi_(L,N)}
\Pi_{\mathcal{S}(L,1/2,N)}=\tilde{\Pi}_{(L,N)} 
\end{align}
for simplicity.
We here consider the approximate time evolution by using the following effective Hamiltonian:
\begin{align}
\label{eff_Ham_mathcal_S_L_xi_N/def_re}
\tilde{\Pi}_{(L,N)} H\tilde{\Pi}_{(L,N)} =\tilde{H}_{(L,N)}.
\end{align}
We aim to prove the following proposition:
\begin{prop} \label{prop:boson_number_truncation_1}
Under the assumption of 
\begin{align}
\label{assumption_for_N_ineq}
 N \ge \max\brrr{ 64e^2 q_0^3 , 2 (8\kappa)^{\kappa} \xi_{t,R}},
\end{align}
the approximation error using the effective Hamiltonian $\tilde{H}_{(L,N)}$ is upper-bounded by
\begin{align}
\label{prop:effective_hamiltonian_error_t^d}
\norm{ \brr{ O_{X_0}(H,t) -  O_{X_0}(\tilde{H}_{(L,N)},t) }  \rho_0 }_1  \le 2^{2D+5}\zeta_0 \gamma |L|  D! \br{1+ 2 c_\kappa \gamma \bar{J} t \xi_{t,R}^2  }e^{- \brr{N/ (2\xi_{t,R})}^{1/\kappa}/(8\kappa)} ,
\end{align}
where $c_\kappa :=2^{4\kappa + 3}\kappa\Gamma(2\kappa)$. 
We also obtain 
\begin{align}
\label{effective_hamiltonian_error_t^d_2}
\norm{O_{X_0}(H,t) \rho_0  -  O_{X_0}(\tilde{H}_{(L,N)},t)  \tilde{\Pi}_{(L,N)} \rho_0 \tilde{\Pi}_{(L,N)} }_1  
\le 2^{2D+6}\zeta_0 \gamma |L|  D! \br{1+  c_\kappa \gamma \bar{J} t \xi_{t,R}^2  }e^{- \brr{N/ (2\xi_{t,R})}^{1/\kappa}/(8\kappa)}.
\end{align}
\end{prop}

\subsubsection{Proof of Proposition~\ref{prop:boson_number_truncation_1}}

The proof is similar to the one for Proposition~\ref{prop:Effective Hamiltonian_local_region}.
To apply Corollary~\ref{corol:effective Hamiltonian_accuracy} to the approximation error, we aim to estimate the norms of
$$
\norm{ \tilde{\Pi}_{(L,N)}^\co O_{X_0} (\tau_2) \sqrt{\rho_0(\tau_1) } }_F   
\quad 
{\rm and} 
\quad
\norm{ \tilde{\Pi}_{(L,N)} H_0 \tilde{\Pi}_{(L,N)}^\co O_{X_0} (\tau_2) \sqrt{\rho_0(\tau_1)}  }_F   
$$
for arbitrary $|\tau_1|,|\tau_2| \le t$. 
For the above two quantities, we prove the following inequalities for $N\ge  2 (8\kappa)^{\kappa} \xi_{t,R}$ (see below for the proofs):
\begin{align}
\label{ineq_O_X_bar_Pi_tau_2_1//upper_bound}
\norm{ \tilde{\Pi}_{(L,N)}^\co O_{X_0} (\tau_2) \sqrt{\rho_0(\tau_1) } }_F   
&\le 2^{2D+3}\zeta_0 \gamma |L|  D! e^{- \brr{N/(2\xi_{t,R})}^{1/\kappa}/(4\kappa)} ,
\end{align}
and 
\begin{align}
\label{ineq_O_X_bar_Pi_tau_2_2//upper_bound}
&\norm{ \tilde{\Pi}_{(L,N)} H_0 \tilde{\Pi}_{(L,N)}^\co O_{X_0} (\tau_2) \sqrt{\rho_0(\tau_1)} }_F  \le 
2^{2D+5} c_\kappa  \zeta_0\bar{J} \xi_{t,R}^2 \gamma^2 |L| D! e^{- \brr{N/ (2\xi_{t,R})}^{1/\kappa}/(8\kappa)}  ,
\end{align}
where $c_\kappa :=2^{4\kappa + 3}\kappa\Gamma(2\kappa)$.

By applying the inequalities~\eqref{ineq_O_X_bar_Pi_tau_2_1//upper_bound} and \eqref{ineq_O_X_bar_Pi_tau_2_2//upper_bound} with appropriate choices of $\tau_1$ and $\tau_2$ to the upper bounds in~\eqref{error_time/evo_effectve_original_ineq_mix}, we have 
\begin{align}
\label{prop:effective_hamiltonian_error_t^d_pre}
&\norm{ O_{X_0}(H,t)  \rho_0 - O_{X_0}(\tilde{H}_{(L,N)}, t)  \rho_0 }_1  \notag \\
&\le  2^{2D+5}\zeta_0 \gamma |L|  D! e^{- \brr{N/(2\xi_{t,R})}^{1/\kappa}/(4\kappa)}
+ 2^{2D+6} t c_\kappa  \zeta_0\bar{J} \xi_{t,R}^2 \gamma^2 |L| D! e^{- \brr{N/ (2\xi_{t,R})}^{1/\kappa}/(8\kappa)} \notag \\
&\le 2^{2D+5}\zeta_0 \gamma |L|  D! \br{1+  2c_\kappa \gamma \bar{J} t \xi_{t,R}^2  }e^{- \brr{N/ (2\xi_{t,R})}^{1/\kappa}/(8\kappa)} .
\end{align}  
We thus prove the first main inequality~\eqref{prop:effective_hamiltonian_error_t^d}.

In deriving the inequality of~\eqref{effective_hamiltonian_error_t^d_2}, 
because of 
\begin{align}
&\norm{O_{X_0}(H,t) \rho_0  -  O_{X_0}(\tilde{H}_{(L,N)},t)  \tilde{\Pi}_{(L,N)} \rho_0 \tilde{\Pi}_{(L,N)} }_1    \notag \\
&= \norm{O_{X_0}(H,t) \rho_0  - O_{X_0}(\tilde{H}_{(L,N)},t)  \rho_0 +O_{X_0}(\tilde{H}_{(L,N)},t)  \rho_0-  O_{X_0}(\tilde{H}_{(L,N)},t)  \tilde{\Pi}_{(L,N)} \rho_0 \tilde{\Pi}_{(L,N)} }_1   \notag \\
&\le  \norm{O_{X_0}(H,t) \rho_0  - O_{X_0}(\tilde{H}_{(L,N)},t)  \rho_0 }_1
+ \norm{ O_{X_0}(\tilde{H}_{(L,N)},t)  \br{ \rho_0-  \tilde{\Pi}_{(L,N)} \rho_0 \tilde{\Pi}_{(L,N)} }}_1 ,
\end{align}
we only have to estimate 
\begin{align}
\label{derive_effective_hamiltonian_error_t^d_2_fin}
\norm{ O_{X_0}(\tilde{H}_{(L,N)},t) \br{ \rho_0 - \tilde{\Pi}_{(L,N)} \rho_0 \tilde{\Pi}_{(L,N)}}  }_1 &\le
\zeta_0 \norm{\rho_0 - \tilde{\Pi}_{(L,N)} \rho_0 \tilde{\Pi}_{(L,N)}}_1 \notag \\
&\le 2\zeta_0  \norm{\rho_0 \tilde{\Pi}_{(L,N)}^\co }_1  
\le 2 \zeta_0 \norm{ \tilde{\Pi}_{(L,N)}^\co \sqrt{\rho_0}}_F ,
\end{align}
where we use 
\begin{align}
 \norm{\rho_0 - \tilde{\Pi}_{(L,N)} \rho_0 \tilde{\Pi}_{(L,N)}}_1 
 &=\norm{(1 - \tilde{\Pi}_{(L,N)}) \rho_0 \tilde{\Pi}_{(L,N)} + \rho_0 (1-\tilde{\Pi}_{(L,N)}) }_1 
 \le 2 \norm{\rho_0 \tilde{\Pi}_{(L,N)}^\co }_1  
\end{align}  
and~\footnote{For a general Schatten $p$ norm, one can prove the generalized H\"older inequality (see Prop. 2.5 in Ref.~\cite{sutter2018approximate}):
\begin{align}
\left \| \prod_{j=1}^s O_j \right \|_p \le \prod_{j=1}^s \|O_j \|_{p_j} , \label{sup_generalized_Holder_ineq}
\end{align}
where $\sum_{j=1}^s 1/p_j =1/p$.
The inequality immediately yields
\begin{align}
\left \| O_1 O_2 \right \|_1 \le \|O_1\|_2 \|O_2\|_2,
 \label{sup_generalized_Holder_ineq_operator_norm}
\end{align}
where we set $p=1$, $p_1=2$ and $p_2=2$ in \eqref{sup_generalized_Holder_ineq}.}
\begin{align}
\norm{\rho_0 \tilde{\Pi}_{(L,N)}^\co}_1 =\norm{\tilde{\Pi}_{(L,N)}^\co\rho_0 }_1  \le  \norm{\sqrt{\rho_0}}_2 \norm{\tilde{\Pi}_{(L,N)}^\co \sqrt{\rho_0} }_2
=  \norm{\tilde{\Pi}_{(L,N)}^\co \sqrt{\rho_0} }_F.
\end{align}  
Note that Schatten $2$ norm $\norm{\cdot}_2$ is equivalent to the Frobenius norm $\norm{\cdot}_F$.
By using the inequality~\eqref{ineq_O_X_bar_Pi_tau_2_1//upper_bound} with $O_{X_0}=\hat{1}$, we immediately obtain 
\begin{align}
\norm{ \tilde{\Pi}_{(L,N)}^\co \sqrt{\rho_0}}_F \le2^{2D+3}\zeta_0 \gamma |L|  D! e^{- \brr{N/(2\xi_{t,R})}^{1/\kappa}/(4\kappa)} .
\end{align}
Therefore, by combining the above inequality with~\eqref{prop:effective_hamiltonian_error_t^d_pre}, 
\begin{align}
\norm{ O_{X_0}(\tilde{H}_{(L,N)},t) \br{ \rho_0 - \tilde{\Pi}_{(L,N)} \rho_0 \tilde{\Pi}_{(L,N)}}  }_1 &\le
 2^{2D+5}\zeta_0 \gamma |L|  D! \br{1/2+1+  2c_\kappa \gamma \bar{J} t \xi_{t,R}^2  }e^{- \brr{N/ (2\xi_{t,R})}^{1/\kappa}/(8\kappa)} , \notag 
\end{align}
which yields the second main inequality~\eqref{effective_hamiltonian_error_t^d_2}. 
This completes the proof of Proposition~\ref{prop:boson_number_truncation_1}. 
$\square$

{~}\\

{\bf [Proof of the inequalities~\eqref{ineq_O_X_bar_Pi_tau_2_1//upper_bound} and \eqref{ineq_O_X_bar_Pi_tau_2_2//upper_bound}]}

From the inequalities~\eqref{lemm:boson_density_prob_gene_time_evo/main01} and \eqref{lemm:boson_density_prob_gene_time_evo/main02} in Lemma~\ref{lemm:boson_density_prob_gene_time_evo}, 
we can obtain the following looser bound which holds for all $i\in \Lambda$:
  \begin{align}
 \label{lemm:boson_density_prob_gene_time_evo/main0102}
&\tr \br{\nb_{i[\bar{\ell}_{t,R}]}^p  \Pi_{i[\bar{\ell}_{t,R}] ,> q_i} \tilde{\rho}_{\tau_1,\tau_2}  } 
\le \zeta_0^2q_i^p  e^{- \brr{ ( q_i - 32e^2 q_0^3)/\xi_{t,R} }^{1/\kappa} }  \for \forall i\in \Lambda.
\end{align}
We utilize the inequality~\eqref{lemm:boson_density_prob_gene_time_evo/main0102} with $p=0$ for the estimation of the norm $\norm{ \tilde{\Pi}_{(L,N)}^\co O_{X_0} (\tau_2) \sqrt{\rho_0(\tau_1) } }_F $.
For simplicity, we here adopt the notation of 
\begin{align}
\bar{N}_i := 
\begin{cases} 
N &\for i\in L, \\  
Ne^{\dist_{i,L}/2} &\for i \in L^\co ,
\end{cases} 
\end{align}
and obtain
\begin{align}
\label{ineq_O_X_bar_Pi_tau_2_1upp}
\norm{ \tilde{\Pi}_{(L,N)}^\co O_{X_0} (\tau_2) \sqrt{\rho_0(\tau_1) } }_F 
&\le \sum_{i\in \Lambda} \norm{\Pi_{i, >\bar{N}_i} O_{X_0} (\tau_2) \sqrt{\rho_0(\tau_1)}}_F \notag\\
&\le \zeta_0 \sum_{i\in \Lambda}  \exp\brr{ -\frac{1}{2} \br{\frac{\bar{N}_i - 32e^2 q_0^3 }{\xi_{t,R}} }^{1/\kappa} }   \notag \\
&\le \zeta_0 \sum_{i\in \Lambda} e^{- [\bar{N}_i /(2\xi_{t,R})]^{1/\kappa}/2}  ,
\end{align}
where in the last inequality we use the condition $N \ge 64e^2 q_0^3 $ in \eqref{assumption_for_N_ineq} to derive $32e^2 q_0^3  \le N/2 \le \bar{N}_i/2$. 
For the summation in the RHS of \eqref{ineq_O_X_bar_Pi_tau_2_1upp}, we have
 \begin{align}
\label{ineq_O_X_bar_Pi_tau_2_1upp_2}
&\sum_{i\in \Lambda} e^{- \brr{\bar{N}_i /(2\xi_{t,R})}^{1/\kappa}/2}  
= \sum_{i\in L} e^{- \brr{N /(2^{\kappa+1}\xi_{t,R})}^{1/\kappa} } + \sum_{i\in L^\co} e^{- [Ne^{\dist_{i,L}/2} /(2^{\kappa+1}\xi_{t,R})]^{1/\kappa} }   \notag \\
&\le |L| e^{- \brr{N /(2^{\kappa+1}\xi_{t,R})}^{1/\kappa} }  +  \gamma |L|   2^{2D+2} D! e^{- \brr{N /(2^{\kappa+1}\xi_{t,R})}^{1/\kappa}/(2\kappa)}  \notag \\
&\le   2^{2D+3} \gamma |L|  D! e^{- \brr{N/(2\xi_{t,R})}^{1/\kappa}/(4\kappa)} \for 
N\ge 2^{\kappa+1}(2\kappa)^{\kappa} \xi_{t,R}= 2 (4\kappa)^{\kappa} \xi_{t,R},
\end{align}
where we apply the inequality~\eqref{ineq:summation_set_S_L_xi_N} in Lemma~\ref{Lemm_Often used inequalities} to the second term in the first line by choosing
 \begin{align}
\kappa \to \kappa,\quad \Phi\to 2^{\kappa+1}\xi_{t,R},\quad q\to N, \quad \ell\to 0, \quad \xi\to 2 .
\end{align}
Note that $N\ge  2 (4\kappa)^{\kappa} \xi_{t,R}$ originates from the condition in~\eqref{ineq:summation_set_S_L_xi_N}, i.e., 
$q \ge (\xi \kappa)^\kappa \Phi$. 
By applying the inequality~\eqref{ineq_O_X_bar_Pi_tau_2_1upp_2} to~\eqref{ineq_O_X_bar_Pi_tau_2_1upp}, we prove the first inequality~\eqref{ineq_O_X_bar_Pi_tau_2_1//upper_bound}.

Second, we estimate the norm of $\norm{ \tilde{\Pi}_{(L,N)} H_0 \tilde{\Pi}_{(L,N)}^\co O_{X_0} (\tau_2) \sqrt{\rho_0(\tau_1)}  }_F $.  
By using the same inequality as~\eqref{ineq_O_X_bar_Pi_tau_2_2_upp_0}, we obtain
\begin{align}
\label{ineq_(L,N)_H_0_upper_nbound_1}
\norm{ \tilde{\Pi}_{(L,N)} H_0 \tilde{\Pi}_{(L,N)}^\co O_{X_0} (\tau_2) \sqrt{\rho_0(\tau_1)} }_F   
&\le \bar{J}\sum_{i\in \Lambda} \sum_{j: \dist_{i,j}=1}  \norm{ \tilde{\Pi}_{(L,N)} (b_i b_j^\dagger +{\rm h.c.} ) \br{\Pi_{i,\le \bar{N}_i}\Pi_{j,\le \bar{N}_j}}^{\co}  O_{X_0} (\tau_2) \sqrt{\rho_0(\tau_1)} }_F    \notag \\
&\le 2 \bar{J}\sum_{i\in \Lambda} \sum_{j: \dist_{i,j}=1}  \norm{(\nb_i+\nb_j)\br{\Pi_{i,\le \bar{N}_i}\Pi_{j,\le \bar{N}_j}}^{\co} O_{X_0} (\tau_2) \sqrt{\rho_0(\tau_1)} }_F .
\end{align}
Furthermore, from the same inequality as~\eqref{ineq_O_X_bar_Pi_tau_2_2_upp_02}, we obtain 
\begin{align}
\label{ineq_(L,N)_H_0_upper_nbound_2}
&\norm{(\nb_i+\nb_j)\br{\Pi_{i,\le \bar{N}_i}\Pi_{j,\le \bar{N}_j}}^{\co} O_{X_0} (\tau_2) \sqrt{\rho_0(\tau_1)} }_F^2  \notag \\
&= \tr \brr{(\nb_i+\nb_j)^2 \br{\Pi_{i,\le \bar{N}_i}\Pi_{j,\le \bar{N}_j}}^{\co} O_{X_0} (\tau_2) \rho_0(\tau_1) O_{X_0} (\tau_2)^\dagger } \notag \\
&\le  \tr \brr{\nb_{i[1]}^2\Pi_{i,>\bar{N}_i} O_{X_0} (\tau_2) \rho_0(\tau_1) O_{X_0} (\tau_2)^\dagger } 
+\tr \brr{\nb_{j[1]}^2 \Pi_{j,>\bar{N}_j}  O_{X_0} (\tau_2) \rho_0(\tau_1) O_{X_0} (\tau_2)^\dagger }  \notag \\
&\le \zeta_0^2 \br{ \bar{N}_i^2 e^{- \brr{(\bar{N}_i-32e^2 q_0^3 )/\xi_{t,R} }^{1/\kappa}} +\bar{N}_j^2 e^{- \brr{(\bar{N}_j-32e^2 q_0^3 )/\xi_{t,R} }^{1/\kappa}} },
\end{align}
where we use $\nb_i+\nb_j \preceq \nb_{i[1]}$ ($\nb_i+\nb_j \preceq \nb_{j[1]}$) from $\dist_{i,j}=1$ in the first inequality, and in the second inequality, we apply the inequality~\eqref{lemm:boson_density_prob_gene_time_evo/main0102} with $p=2$.
By combining the inequalities~\eqref{ineq_(L,N)_H_0_upper_nbound_1} and \eqref{ineq_(L,N)_H_0_upper_nbound_2}, we have 
\begin{align}
\label{ineq_(L,N)_H_0_upper_nbound_3}
&\norm{ \tilde{\Pi}_{(L,N)} H_0 \tilde{\Pi}_{(L,N)}^\co O_{X_0} (\tau_2) \sqrt{\rho_0(\tau_1)} }_F   \notag \\
&\le 
2\zeta_0\bar{J}\sum_{i\in \Lambda} \sum_{j: \dist_{i,j}=1} \br{ \bar{N}_i e^{- \brr{(\bar{N}_i-32e^2 q_0^3 )/\xi_{t,R} }^{1/\kappa}/2} +\bar{N}_j e^{- \brr{(\bar{N}_j-32e^2 q_0^3 )/\xi_{t,R} }^{1/\kappa}/2} } \notag \\
&\le 4\zeta_0 \gamma \bar{J}\sum_{i\in \Lambda} \bar{N}_i e^{- \brr{(\bar{N}_i-32e^2 q_0^3 )/\xi_{t,R} }^{1/\kappa}/2} 
\le  4\zeta_0 \gamma \bar{J}\sum_{i\in \Lambda} \bar{N}_i e^{- \brr{\bar{N}_i/ (2^{\kappa+1}\xi_{t,R})}^{1/\kappa}},
\end{align}
where we use $\sqrt{x^2+y^2} \le x+y$ in taking the square root of~\eqref{ineq_(L,N)_H_0_upper_nbound_2}, $\sum_{j: \dist_{i,j}=1} 1 \le |i[1]|\le \gamma$ for $\forall i\in \Lambda$, and  $32e^2 q_0^3  \le N_i/2$ from the condition
$N \ge 64e^2 q_0^3$. 
Finally, by employing the similar analyses to~\eqref{ineq_O_X_bar_Pi_tau_2_1upp_2}, we upper-bound the summation in~\eqref{ineq_(L,N)_H_0_upper_nbound_3}.
For the purpose, we first apply the inequality~\eqref{x^m_e^-x/Phi_upper_bound} to $\bar{N}_i e^{- \brr{\bar{N}_i/ (2^{\kappa+1}\xi_{t,R})}^{1/\kappa}}$ by letting
 \begin{align}
s\to 1, \quad \kappa \to \kappa,\quad \Phi\to  (2^{\kappa+1}\xi_{t,R}) ,
\end{align}
which yields 
 \begin{align}
\bar{N}_i e^{- \brr{\bar{N}_i/ (2^{\kappa+1}\xi_{t,R})}^{1/\kappa}} 
&\le 2(2^{\kappa} \cdot 2^{\kappa+1}\xi_{t,R})^{2} \kappa\Gamma(2\kappa) e^{- \brr{\bar{N}_i/ (2^{\kappa+1}\xi_{t,R})}^{1/\kappa}/2}  \notag \\
&=c_\kappa \xi_{t,R}^2 e^{- \brr{\bar{N}_i/ (2^{\kappa+1}\xi_{t,R})}^{1/\kappa}/2} \quad \textrm{with} \quad 
c_\kappa :=2^{4\kappa + 3}\kappa\Gamma(2\kappa) .
\end{align}
By using the above upper bound, we obtain  
 \begin{align}
\label{ineq_(L,N)_H_0_upper_nbound_4}
&\sum_{i\in \Lambda} \bar{N}_i e^{- \brr{\bar{N}_i/ (2^{\kappa+1}\xi_{t,R})}^{1/\kappa}} \le  c_\kappa \xi_{t,R}^2\sum_{i\in L} e^{- \brr{N/ (2^{\kappa+1}\xi_{t,R})}^{1/\kappa}/2} +c_\kappa \xi_{t,R}^2  \sum_{i\in L^\co} e^{- \brr{Ne^{\dist_{i,L}/2}/ (2^{\kappa+1}\xi_{t,R})}^{1/\kappa}/2} \notag \\
&= c_\kappa \xi_{t,R}^2|L| e^{- \brr{N/ (2^{\kappa+1}\xi_{t,R})}^{1/\kappa}/2}   + c_\kappa \xi_{t,R}^2  \sum_{i\in L^\co} e^{- \brr{Ne^{\dist_{i,L}/2}/ (2^{2\kappa+1}\xi_{t,R})}^{1/\kappa}}  \notag \\
&\le  c_\kappa \xi_{t,R}^2|L| e^{- \brr{N/ (2^{\kappa+1}\xi_{t,R})}^{1/\kappa}/2}   
+ c_\kappa \xi_{t,R}^2 \gamma |L| 2^{2D+2}D! e^{- \brr{N/ (2^{2\kappa+1}\xi_{t,R})}^{1/\kappa}/(2\kappa)} \notag \\
&\le 2^{2D+3} c_\kappa \xi_{t,R}^2 \gamma |L| D! e^{- \brr{N/ (2\xi_{t,R})}^{1/\kappa}/(8\kappa)} 
 \for N\ge 2^{2\kappa+1}(2\kappa)^{\kappa} \xi_{t,R}= 2 (8\kappa)^{\kappa} \xi_{t,R},
\end{align}
where in the second inequality, we use the inequality~\eqref{ineq:summation_set_S_L_xi_N} with 
 \begin{align}
\kappa \to \kappa,\quad \Phi\to  2^{2\kappa+1}\xi_{t,R} ,\quad q\to N, \quad \ell\to 0, \quad \xi\to 2 .
\end{align}
Here, the condition $N\ge 2 (8\kappa)^{\kappa} \xi_{t,R}$ in \eqref{ineq_(L,N)_H_0_upper_nbound_4} originates from the condition  
$q \ge (\xi \kappa)^\kappa \Phi$ in~\eqref{ineq:summation_set_S_L_xi_N}. 
By applying the inequality~\eqref{ineq_(L,N)_H_0_upper_nbound_4} to \eqref{ineq_(L,N)_H_0_upper_nbound_3}, we prove the desired inequality~\eqref{ineq_O_X_bar_Pi_tau_2_2//upper_bound}.

\subsection{Effective Hamiltonian by the truncation of the boson number: third result}\label{sec:Effective Hamiltonian: third result}

\begin{figure}[tt]
\centering
\includegraphics[clip, scale=0.5]{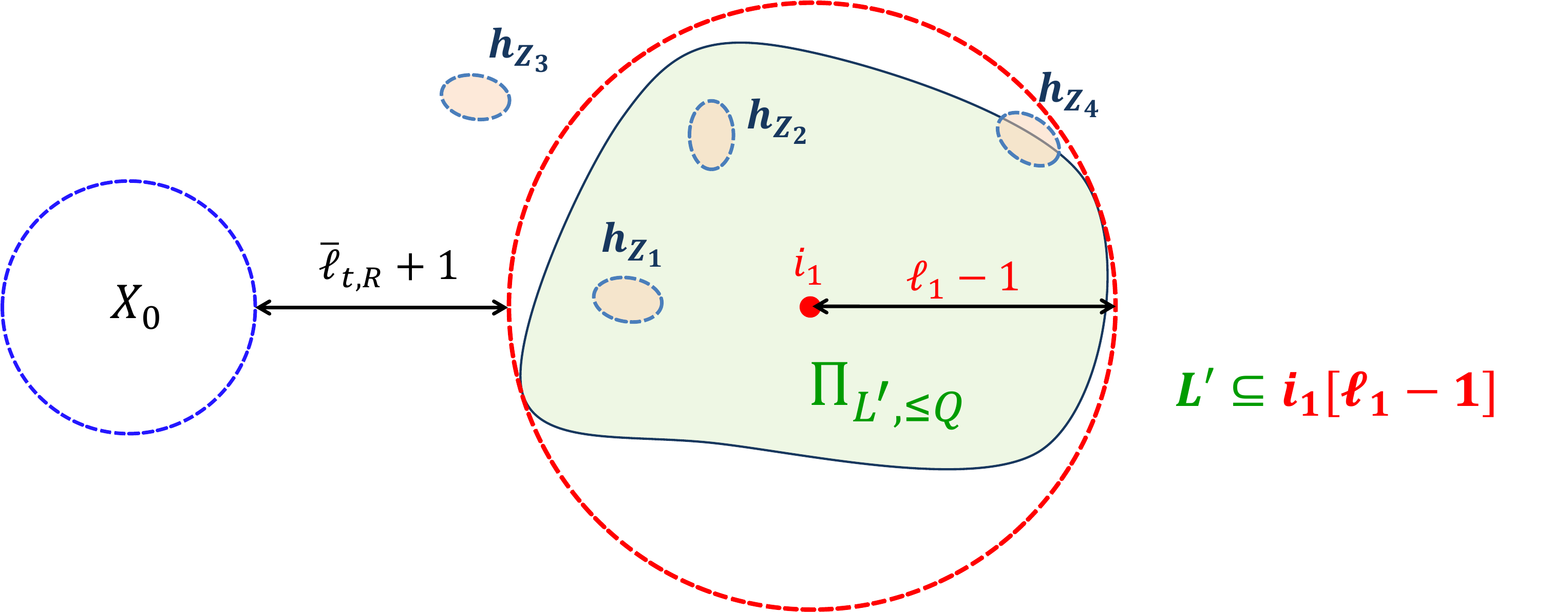}
\caption{Definition of the projection $\Pi_{L',\le Q}$.
In the projection, we truncate the boson number in the region $L'$ up to $Q$, i.e., $\norm{\nb_L \Pi_{L',\le Q}} = Q$,
where the region $i_1[\ell_1]$ ($\supset L'$) is separated from the region $X_0$ at least by the distance of $\bar{\ell}_{t,R}$.
Unlike the other projections $\bar{\Pi}_{L,\bar{q}}$ and $\tilde{\Pi}_{(L,N)}$ in Figs.~\ref{supp_fig_Projection_L_bar_q} and \ref{supp_fig_Proj_S_L_xi_N}, where the boson truncations are performed for each of the local sites, the boson truncation by $\Pi_{L',\le Q}$ is non-local in the sense that it cannot be decomposed into the product form like $\prod_{i\in L'}\Pi_{i,\le q_i}$.
This point breaks down the locality of the effective Hamiltonian around the boundary. 
In the figure, the interactions $h_{Z_1}$, $h_{Z_2}$ and $h_{Z_3}$ commute with each other by the projection $\Pi_{L',\le Q}$, 
while the boundary interaction $h_{Z_4}$ becomes non-local by the projection $\Pi_{L',\le Q}$. 
In Proposition~\ref{prop:boson_number_truncation_2}, we consider the projection $\Pi_{L', \le Q} \tilde{\Pi}_{(L,N)}$ and 
upper-bound the approximation error for dynamics by the effective Hamiltonian $\tilde{H}_{(L,N,L',Q)}$ in Eq.~\eqref{definition_tilde_L_N_L'_Q}.}
\label{supp_fig_proj_L_dash_leQ}
\end{figure}

In this section, we further consider the projection $\Pi_{L', \le Q}$ onto the region $L'$ such that (see Fig.~\ref{supp_fig_proj_L_dash_leQ})
\begin{align}
\label{cond_for_L_i_1_ell_1}
L'\subseteq i_1[\ell_1-1] \quad (\ell_1 \ge \bar{\ell}_{t,R}),\quad \dist_{i_1[\ell_1],X_0} \ge \bar{\ell}_{t,R} \to  i_1[\ell_1] \subset X_0[\bar{\ell}_{t,R}]^\co .
\end{align}
We then define the effective Hamiltonian $\tilde{H}_{(L,N,L',Q)}$ as 
\begin{align}
\label{definition_tilde_L_N_L'_Q}
\tilde{H}_{(L,N,L',Q)} =\Pi_{L', \le Q} \tilde{\Pi}_{(L,N)} H\tilde{\Pi}_{(L,N)} \Pi_{L', \le Q} ,
\end{align}
where the projection $\tilde{\Pi}_{(L,N)}$ has been defined in Eq.~\eqref{def_tilde_Pi_(L,N)}.

Here, we should note that, in general, the effective Hamiltonian $\tilde{H}_{(L,N,L',Q)}$ is no longer local in the sense that 
\begin{align}
\label{Non_commuteness_effective_interactions}
[\Pi_{L', \le Q} h_Z \Pi_{L', \le Q}, \Pi_{L', \le Q} h_{Z'} \Pi_{L', \le Q} ] \neq 0  \for Z\cap Z'\neq \emptyset ,
\end{align}
where $H=\sum_Z h_Z$ as in Eq.~\eqref{def:Ham}. 
To see the point, let us consider four interaction terms $h_{Z_1}$, $h_{Z_2}$, $h_{Z_3}$ and $h_{Z_4}$ such that (see Fig.~\ref{supp_fig_proj_L_dash_leQ})
$$Z_1,Z_2 \subset L' \quad (Z_1\cap Z_2=\emptyset), \quad Z_3 \subset L^{'\co},\quad (Z_4 \cap L') \neq \emptyset \And (Z_4 \cap L^{'\co})\neq \emptyset .$$
Here, the effective interaction terms $\Pi_{L', \le Q} h_{Z_1} \Pi_{L', \le Q} $, $\Pi_{L', \le Q}  h_{Z_2} \Pi_{L', \le Q} $ and $\Pi_{L', \le Q}  h_{Z_3} \Pi_{L', \le Q} $  commute with each other since 
\begin{align}
[\Pi_{L', \le Q}, h_{Z_i}  ]=0  \for i=1,2,3. 
\end{align}
It is trivially obtained from $[\nb_i, \Pi_{L', \le Q}]=0$ for $\forall i\in \Lambda$ and $[ b_i b_j^\dagger, \Pi_{L', \le Q}]=0$ as long as $i,j\in L'$. 
We thus obtain 
\begin{align}
\label{Commuteness_effective_interactions}
[\Pi_{L', \le Q}h_{Z_i} \Pi_{L', \le Q}, \Pi_{L', \le Q}h_{Z_j} \Pi_{L', \le Q} ]=
\Pi_{L', \le Q}[h_{Z_i} , h_{Z_j} ] \Pi_{L', \le Q} = 0
\end{align}
for $i,j=1,2,3$. 
On the other hand, we cannot ensure the above equation for $h_{Z_4}$ because of 
\begin{align}
[\Pi_{L', \le Q}, h_{Z_4}  ]\neq 0  .
\end{align}
This point leads to the breakdown of the locality of the interactions as in~\eqref{Non_commuteness_effective_interactions}.
We also discuss this point in Sec.~\ref{Sec:Generalization: small time evolution by effective Hamiltonian} when we consider the Lieb-Robinson bound for the effective Hamiltonian.

From Eq.~\eqref{Commuteness_effective_interactions}, we at least prove the following corollary:
\begin{corol} \label{corol:locality_effective_Hamiltonian}
If we consider a subset Hamiltonian $H_X$ such that $X\subseteq L'$ or $X\subseteq L^{'\co}$, the effective Hamiltonian $\tilde{H}_{X,(L,N,L',Q)}$ is still a local Hamiltonian in the sense that
\begin{align}
[\Pi_{L', \le Q} h_Z \Pi_{L', \le Q}, \Pi_{L', \le Q} h_{Z'} \Pi_{L', \le Q} ] = 0  \for Z\cap Z'= \emptyset ,
\end{align}
where $h_Z$ and $h'_Z$ are local interaction terms in $H_X$, i.e., $Z,Z'\subset X$. 
\end{corol}

On the approximation error for the effective Hamiltonian $\tilde{H}_{(L,N,L',Q)}$, 
we prove the following proposition:
\begin{prop} \label{prop:boson_number_truncation_2}
We adopt the same assumption as Eq.~\eqref{assumption_for_N_ineq} for $N$. 
Let us choose the subset $L'$ such that the condition~\eqref{cond_for_L_i_1_ell_1} is satisfied.
Then, the approximation error using the effective Hamiltonian $\tilde{H}_{(L,N)}$ is upper-bounded by
\begin{align}
\label{effective_hamiltonian_error_t^d___2}
&\norm{ \brr{ O_{X_0}(H,t) -  O_{X_0}(\tilde{H}_{(L,N,L',Q)},t) }  \rho_0 }_1  \notag \\
& \le  2^{2D+5}\zeta_0 \gamma |L|  D! \br{1+  2c_\kappa \gamma \bar{J} t \xi_{t,R}^2  }e^{- \brr{N/ (2\xi_{t,R})}^{1/\kappa}/(8\kappa)} 
+4\zeta_0 (1 + 2 t\gamma \bar{J} Q)  e^{- \brr{Q/(4 b_0 \ell_1^D) }^{1/\kappa}/2},
\end{align}
where $c_\kappa$ has been defined in Proposition~\ref{prop:boson_number_truncation_1}. 
Moreover, for $\tilde{\Pi}_{(L,N,L',Q)} \rho_0 \tilde{\Pi}_{(L,N,L',Q)}$, we obtain the upper bound of
\begin{align}
\label{effective_hamiltonian_error_t^d_2___2}
&\norm{ \brr{ O_{X_0}(H,t) -  O_{X_0}(\tilde{H}_{(L,N,L',Q)},t) }  \tilde{\Pi}_{(L,N,L',Q)} \rho_0 \tilde{\Pi}_{(L,N,L',Q)} }_1   \notag \\
&\le  2^{2D+6}\zeta_0 \gamma |L|  D! \br{1+  c_\kappa \gamma \bar{J} t \xi_{t,R}^2  }e^{- \brr{N/ (2\xi_{t,R})}^{1/\kappa}/(8\kappa)} 
+8\zeta_0 (1 +  t\gamma \bar{J} Q)  e^{- \brr{Q/(4 b_0 \ell_1^D) }^{1/\kappa}/2}. 
\end{align}
\end{prop}

\subsubsection{Proof of Proposition~\ref{prop:boson_number_truncation_2}}

In the proof, we denote
\begin{align}
\label{def_tilde_Pi_(L,N,L',Q)}
\tilde{\Pi}_{(L,N,L',Q)} =\Pi_{L', \le Q} \tilde{\Pi}_{(L,N)} .
\end{align}
In applying Corollary~\ref{corol:effective Hamiltonian_accuracy}, 
we aim to estimate the norms of
\begin{align}
\label{ineq_O_X_bar_Pi_tau_2_1//___2}
\norm{ \tilde{\Pi}_{(L,N,L',Q)}^\co O_{X_0} (\tau_2) \sqrt{\rho_0(\tau_1) } }_F   
\end{align}
and 
\begin{align}
\label{ineq_O_X_bar_Pi_tau_2_2//___2}
\norm{ \tilde{\Pi}_{(L,N,L',Q)} H_0 \tilde{\Pi}_{(L,N,L',Q)}^\co O_{X_0} (\tau_2) \sqrt{\rho_0(\tau_1)}  }_F   
\end{align}
for arbitrary $|\tau_1|,|\tau_2| \le t$. 

First, we have the decomposition of 
\begin{align}
\tilde{\Pi}_{(L,N,L',Q)}^\co 
=\Pi_{L', \le Q}^\co \tilde{\Pi}_{(L,N)} + \Pi_{L', \le Q} \tilde{\Pi}_{(L,N)}^\co + \Pi_{L', \le Q}^\co \tilde{\Pi}_{(L,N)}^\co 
= \Pi_{L', \le Q}^\co \tilde{\Pi}_{(L,N)} + \tilde{\Pi}_{(L,N)}^\co,
\end{align}
where we use $\Pi_{L', \le Q}+\Pi_{L', \le Q}^\co=1$ in the second equation.
By using the above, we upper-bound Eqs.~\eqref{ineq_O_X_bar_Pi_tau_2_1//___2} and \eqref{ineq_O_X_bar_Pi_tau_2_2//___2} 
as follows:
\begin{align}
\norm{ \tilde{\Pi}_{(L,N,L',Q)}^\co O_{X_0} (\tau_2) \sqrt{\rho_0(\tau_1) } }_F   
\le  \norm{ \tilde{\Pi}_{(L,N)}^\co O_{X_0} (\tau_2) \sqrt{\rho_0(\tau_1) } }_F  
+ \norm{ \Pi_{L', \le Q}^\co  O_{X_0} (\tau_2) \sqrt{\rho_0(\tau_1) } }_F    ,
\end{align}
and 
\begin{align}
&\norm{ \tilde{\Pi}_{(L,N,L',Q)} H_0 \tilde{\Pi}_{(L,N,L',Q)}^\co O_{X_0} (\tau_2) \sqrt{\rho_0(\tau_1)}  }_F    \notag \\
&\le \norm{ \tilde{\Pi}_{(L,N,L',Q)} H_0 \Pi_{L', \le Q}^\co \tilde{\Pi}_{(L,N)} O_{X_0} (\tau_2) \sqrt{\rho_0(\tau_1)}  }_F   
+ \norm{ \tilde{\Pi}_{(L,N,L',Q)} H_0 \tilde{\Pi}_{(L,N)}^\co  O_{X_0} (\tau_2) \sqrt{\rho_0(\tau_1)}  }_F     \notag \\
&\le \norm{\Pi_{L', \le Q} H_0 \Pi_{L', \le Q}^\co \tilde{\Pi}_{(L,N)} O_{X_0} (\tau_2) \sqrt{\rho_0(\tau_1)}  }_F   
+ \norm{ \tilde{\Pi}_{(L,N)}  H_0 \tilde{\Pi}_{(L,N)}^\co O_{X_0} (\tau_2) \sqrt{\rho_0(\tau_1)}  }_F .
\end{align}
We have already upper-bounded $\norm{ \tilde{\Pi}_{(L,N)}^\co O_{X_0} (\tau_2) \sqrt{\rho_0(\tau_1) } }_F$ and $\norm{ \tilde{\Pi}_{(L,N)}  H_0 \tilde{\Pi}_{(L,N)}^\co O_{X_0} (\tau_2) \sqrt{\rho_0(\tau_1)}  }_F$ in the inequalities~\eqref{ineq_O_X_bar_Pi_tau_2_1//upper_bound} and \eqref{ineq_O_X_bar_Pi_tau_2_2//upper_bound}, respectively.
Therefore, we need to estimate the norms of $\norm{ \Pi_{L', \le Q}^\co  O_{X_0} (\tau_2) \sqrt{\rho_0(\tau_1) } }_F$ and 
$\norm{\Pi_{L', \le Q} H_0 \Pi_{L', \le Q}^\co \tilde{\Pi}_{(L,N)} O_{X_0} (\tau_2) \sqrt{\rho_0(\tau_1)}  }_F$, which are upper-bounded by the following inequalities (see below for the proofs):
 \begin{align}
\label{ineq_lemma:Pi_L_leQ_norm_ineq/1}
\norm{ \Pi_{L', \le Q}^\co O_{X_0} (\tau_2) \sqrt{\rho_0(\tau_1)} }_F 
&\le  \zeta_0 e^{- \brr{Q/(4 b_0 \ell_1^D) }^{1/\kappa}/2},
\end{align}
and 
 \begin{align}
\label{ineq_lemma:Pi_L_leQ_norm_ineq/2}
\norm{\Pi_{L', \le Q} H_0 \Pi_{L', \le Q}^\co \tilde{\Pi}_{(L,N)} O_{X_0} (\tau_2) \sqrt{\rho_0(\tau_1)}  }_F
&\le 4 \zeta_0 \gamma \bar{J} Q  e^{- \brr{Q/(4 b_0 \ell_1^D) }^{1/\kappa}/2}.
\end{align}

By applying the upper bounds in the inequalities~\eqref{ineq_O_X_bar_Pi_tau_2_1//upper_bound}, \eqref{ineq_O_X_bar_Pi_tau_2_2//upper_bound}, \eqref{ineq_lemma:Pi_L_leQ_norm_ineq/1} and \eqref{ineq_lemma:Pi_L_leQ_norm_ineq/2} to 
to Corollary~\ref{corol:effective Hamiltonian_accuracy}, we obtain the upper bound of 
\begin{align}
&\norm{ \brr{ O_{X_0}(H,t) -  O_{X_0}(\tilde{H}_{(L,N,L',Q)},t) }  \rho_0 }_1  \notag \\
&\le 2^{2D+5}\zeta_0 \gamma |L|  D! \br{1+ 2 c_\kappa \gamma \bar{J} t \xi_{t,R}^2  }e^{- \brr{N/ (2\xi_{t,R})}^{1/\kappa}/(8\kappa)} 
+\zeta_0 (4 + 8t \gamma \bar{J} Q)  e^{- \brr{Q/(4 b_0 \ell_1^D) }^{1/\kappa}/2},
\label{effective_hamiltonian_error_t^d__pre_2}
\end{align}
where we use the inequality~\eqref{prop:effective_hamiltonian_error_t^d_pre} to estimate the contributions from \eqref{ineq_O_X_bar_Pi_tau_2_1//upper_bound} and \eqref{ineq_O_X_bar_Pi_tau_2_2//upper_bound}.
We thus prove the main inequality~\eqref{effective_hamiltonian_error_t^d___2}.

In deriving the inequality~\eqref{effective_hamiltonian_error_t^d_2___2}, we only have to estimate [see also~\eqref{derive_effective_hamiltonian_error_t^d_2_fin}]
\begin{align}
\norm{ O_{X_0}(\tilde{H}_{(L,N,L',Q)},t) \br{ \rho_0 - \tilde{\Pi}_{(L,N,L',Q)} \rho_0 \tilde{\Pi}_{(L,N,L',Q)}}  }_1 &\le
2\zeta_0 \norm{ \tilde{\Pi}_{(L,N,L',Q)}^\co \sqrt{\rho_0}}_F \notag \\
&\le 2\zeta_0 \br{ \norm{ \tilde{\Pi}_{(L,N)}^\co \sqrt{\rho_0}}_F + \norm{\Pi_{L', \le Q}^\co \sqrt{\rho_0}}_F} .
\end{align}
By applying the upper bounds~\eqref{ineq_O_X_bar_Pi_tau_2_1//upper_bound} and~\eqref{ineq_lemma:Pi_L_leQ_norm_ineq/1} to the above inequality, we obtain the upper bound of
\begin{align}
&\norm{ O_{X_0}(\tilde{H}_{(L,N,L',Q)},t) \br{ \rho_0 - \tilde{\Pi}_{(L,N,L',Q)} \rho_0 \tilde{\Pi}_{(L,N,L',Q)}}  }_1  \notag \\
&\le 2^{2D+4}\zeta_0 \gamma |L|  D! e^{- \brr{N/ (2\xi_{t,R})}^{1/\kappa}/(8\kappa)} 
+2\zeta_0   e^{- \brr{Q/(4 b_0 \ell_1^D) }^{1/\kappa}/2},
\label{effective_hamiltonian_error_t^d__pre_2_fin}
\end{align}
which yields the inequality~\eqref{effective_hamiltonian_error_t^d_2___2} by combining the inequality~\eqref{effective_hamiltonian_error_t^d__pre_2}. 
This completes the proof. $\square$

{~}\\

{\bf [Proof of the inequalities~\eqref{ineq_lemma:Pi_L_leQ_norm_ineq/1} and \eqref{ineq_lemma:Pi_L_leQ_norm_ineq/2}]}

We can adopt similar techniques to the proof of Proposition~\ref{prop:Effective Hamiltonian_local_region}. 
As in the inequality~\eqref{ineq_O_X_bar_Pi_tau_2_1_upp_derive}, we utilize the inequality~\eqref{lemm:boson_density_prob_gene_time_evo/main01} with $p=0$ in Lemma~\ref{lemm:boson_density_prob_gene_time_evo}, which yields 
\begin{align}
\label{ineq_O_X_bar_Pi_tau_2_1_upp_derive__re2}
\norm{ \Pi_{L', \le Q}^\co O_{X_0} (\tau_2) \sqrt{\rho_0(\tau_1)} }_F^2 
&=\tr \brr{\Pi_{L', > Q} O_{X_0} (\tau_2) \rho_0(\tau_1)  O_{X_0} (\tau_2)^\dagger }    \notag \\
&\le \tr \brr{\Pi_{i_1[\ell_1], > Q} O_{X_0} (\tau_2) \rho_0(\tau_1)  O_{X_0} (\tau_2)^\dagger } \le  \zeta_0^2 e^{- \brr{Q/(4 b_0 \ell_1^D) }^{1/\kappa}} ,
\end{align}
where in the first inequality we use $\Pi_{L',>Q}= \Pi_{L',>Q} \Pi_{i_1[\ell_1],> Q}$ because of $L'\subseteq i_1[\ell_1-1] \subset i_1[\ell_1] $,
and in the second inequality, we use $\dist_{i_1[\ell_1],X_0}\ge \bar{\ell}_{t,R}$ and $\ell_1 \ge \bar{\ell}_{t,R}$.
We thus prove the first inequality~\eqref{ineq_lemma:Pi_L_leQ_norm_ineq/1}.

Next, to prove the second inequality~\eqref{ineq_lemma:Pi_L_leQ_norm_ineq/2}, 
we first utilize the decomposition~\eqref{decomposition/pi_L_H_0_co} as 
\begin{align}
\label{decomposition/pi_L_H_0_co_re2}
\Pi_{L', \le Q} H_0 \Pi_{L', \le Q}^\co = \Pi_{L', \le Q} \br{H_{0,L'}  +  \partial h_{L'}} \Pi_{L', \le Q}^\co .
\end{align}
We then obtain
\begin{align}
\label{norm_upper_bound_L'_leQ_H_0_L,NX_0_1}
&\norm{\Pi_{L', \le Q} H_0 \Pi_{L', \le Q}^\co \tilde{\Pi}_{(L,N)} O_{X_0} (\tau_2) \sqrt{\rho_0(\tau_1)}  }_F^2 \notag \\
&= \tr \brr{ \br{H_{0,L'}  +  \partial h_{L'}}^\dagger \Pi_{L', \le Q}\br{H_{0,L'}  +  \partial h_{L'}} \Pi_{L', \le Q}^\co \tilde{\Pi}_{(L,N)} O_{X_0} (\tau_2) \rho_0(\tau_1) O_{X_0} (\tau_2) \tilde{\Pi}_{(L,N)} \Pi_{L', \le Q}^\co} .
\end{align}
To obtain an operator upper bound for $\br{H_{0,L'}  +  \partial h_{L'}}^\dagger \Pi_{L', \le Q}\br{H_{0,L'}  +  \partial h_{L'}}$, we first derive 
\begin{align}
\label{norm_upper_bound_L'_leQ_H_0_L,NX_0_1.5}
\abs{ H_{0,L'}  +  \partial h_{L'}}
\preceq \bar{J}  \sum_{i\in L'} \sum_{j: \dist_{i,j}=1} \br{ | b_i b_j^\dagger| +| b_i^\dagger b_j|  }  
\preceq 2\bar{J} \sum_{i\in L'} \sum_{j: \dist_{i,j}=1} \br{ \nb_i +\nb_j}  \preceq  4 \gamma \bar{J}\nb_{L'[1]},
\end{align}
where we use the operator inequality~\eqref{upper_bound_hopping_operator} in the second inequality, and the last inequality is derived from 
\begin{align}
&\sum_{i\in L'} \sum_{j: \dist_{i,j}=1} \nb_i \preceq \gamma \sum_{i\in L'} \nb_i  \preceq \gamma \nb_{L'[1]} , \notag \\
&\sum_{i\in L'} \sum_{j: \dist_{i,j}=1} \nb_j \preceq \sum_{j\in L'[1]} \sum_{i: \dist_{i,j}=1} \nb_j \preceq
\gamma \sum_{j\in L'[1]} \nb_j  \preceq \gamma \nb_{L'[1]}.
\end{align}
Note that $i\in L'$ and  $j: \dist_{i,j}=1$ implies $j\in L'[1]$. 
Because $[ | H_{0,L'}  +  \partial h_{L'}|, \nb_{L'[1]}]= 0$, we can obtain from~\eqref{norm_upper_bound_L'_leQ_H_0_L,NX_0_1.5}
\begin{align}
\label{norm_upper_bound_L'_leQ_H_0_L,NX_0_2}
\br{H_{0,L'}  +  \partial h_{L'}}^\dagger \Pi_{L', \le Q}\br{H_{0,L'}  +  \partial h_{L'}} 
&\preceq \abs{ H_{0,L'}  +  \partial h_{L'}}^2  \preceq  16 \gamma^2 \bar{J}^2\nb_{L'[1]}^2,
\end{align}
where we use the fact that for arbitrary Hermitian operators $A$ and $B$ such that $|A| \preceq |B|$, we have $|A|^s \preceq |B|^s$ ($s\in \mathbb{N}$) as long as $[|A|,|B|]=0$\footnote{In general, we cannot ensure $|A|^s \preceq |B|^s$ form $|A| \preceq |B|$ unless $[|A|,|B|]= 0$.
This point has made the proof of Subtheorem~\ref{Schuch_boson_extend} rather challenging [see the discussion around Eq.~\eqref{case_s=1_moment/general2}].}.

By combining the inequalities~\eqref{norm_upper_bound_L'_leQ_H_0_L,NX_0_1} and \eqref{norm_upper_bound_L'_leQ_H_0_L,NX_0_2}, we obtain
\begin{align}
&\norm{\Pi_{L', \le Q} H_0 \Pi_{L', \le Q}^\co \tilde{\Pi}_{(L,N)} O_{X_0} (\tau_2) \sqrt{\rho_0(\tau_1)}  }_F^2 \notag \\
&\le 16 \gamma^2 \bar{J}^2 \tr \brr{\nb_{L'[1]}^2 \Pi_{L', \le Q}^\co \tilde{\Pi}_{(L,N)} O_{X_0} (\tau_2) \rho_0(\tau_1) O_{X_0} (\tau_2) \tilde{\Pi}_{(L,N)} \Pi_{L', \le Q}^\co} \notag \\
&\le 16 \gamma^2 \bar{J}^2 \tr \brr{\nb_{L'[1]}^2 \Pi_{L'[1], > Q}  O_{X_0} (\tau_2) \rho_0(\tau_1) O_{X_0} (\tau_2)} \notag \\
&\le 16 \gamma^2 \bar{J}^2 \tr \brr{\nb_{i_1[\ell_1]}^2 \Pi_{i_1[\ell_1], > Q}  O_{X_0} (\tau_2) \rho_0(\tau_1) O_{X_0} (\tau_2)}  \le 
16\zeta_0^2 \gamma^2 \bar{J}^2 Q^2  e^{- \brr{Q/(4 b_0 \ell_1^D) }^{1/\kappa}},
\end{align}
where we use $L'[1] \subseteq i_1[\ell_1]$ from $L' \subseteq i_1[\ell_1-1]$ in the third inequality and 
apply the inequality~\eqref{lemm:boson_density_prob_gene_time_evo/main01} with $p=2$ in the last inequality. 
Note that we have assumed $\dist_{i_1[\ell_1],X_0} \ge \bar{\ell}_{t,R}$ (i.e., $i_1[\ell_1] \subset X_0[\bar{\ell}_{t,R}]^\co$).

\section{Bosonic Lieb-Robinson bound: looser results}\label{sec:Bosonic Lieb-Robinson bound: looser results}

Based on Proposition~\ref{main_theorem_boson_concentration}
 (or Proposition~\ref{prop:Effective Hamiltonian_local_region}), 
we can truncate the boson number on each of the sites up to $\orderof{\bar{\ell}_{t,R}^D}=\tO(t^D)$ [see Eq.~\eqref{eq_bar_ell_t_r} for the definition of $\bar{\ell}_{t,R}$].
By such a truncation, we can upper-bound the local energy on each site by $\tO(t^D)$. 
This energy upper bound allows us to derive the Lieb-Robinson bound with a velocity of $\tO(t^D)$. 
However, as we will prove below, the optimal Lieb-Robinson velocity is given by $\tO(t^{D-1})$ instead of $\tO(t^D)$.
Nevertheless, the following analyses are quite simple, and we can utilize them to develop a quantum algorithm to simulate the dynamics by quantum computer (see Result~3 in the main manuscript).



%

We here prove the following theorem:
\begin{theorem} \label{main_theorem_looser_Lieb_Robinson}
Let us consider an initial state $\rho_0$ that satisfies Assumption~\ref{only_the_assumption_initial} with $b_\rho=b_0/\gamma$ in~\eqref{condition_for_moment_generating}.
Also, we consider an arbitrary operator $O_{X_0}$ ($X_0\subseteq i_0[r_0]$\footnote{Roughly speaking, $r_0$ is the spatial size of the operator $O_{X_0}$.}, $\norm{O_{X_0}}=\zeta_0$) such that 
the condition~\eqref{O_X0_condition/_norm_lemma_operator_boson_pre} holds. 
Then, there exists a unitary operator $U_{X[R]}$ such that the following Lieb-Robinson bound holds: 
\begin{align}
\label{ineq:main_theorem_looser_Lieb_Robinson}
&\norm{ \brr{O_{X_0}(H,t)   - U_{i_0[R]}^\dagger O_{X_0} U_{i_0[R]}  } \rho_0}_1 \notag \\
&\le 4 \zeta_0 \gamma^2 R^D  \br{1+ R/c_1} \exp\brr{- \frac{1}{2}\br{\frac{R-r_0}{8 b_0c_1 \bar{J}t \bar{\ell}_{t,R} ^D} }^{1/\kappa}}
+\zeta_0 R^{D+1} e^{-(R-r_0)/(4k)}  ,
\end{align}  
where $c_1: = 2^{6-D} e^2k\gamma^3(2k)^{2D}$ and we assume $R\ge \max \brr{r_0 + 16k, r_0 +2( \bar{\ell}_{t,R}+1)}$. 
\end{theorem}

{\bf Remark.}
By using the $\Theta$ notation in Eq.~\eqref{Theta_notation_def}, we can express the inequality~\eqref{ineq:main_theorem_looser_Lieb_Robinson} as 
\begin{align}
\label{ineq:main_theorem_looser_Lieb_Robinson_simple}
\norm{ \brr{O_{X_0}(H,t)   - U_{i_0[R]}^\dagger O_{X_0} U_{i_0[R]}  } \rho_0}_1
&\le   \zeta_0  \Theta(R^{D+1}) 
\brrr{e^{- [\Theta(R-r_0)/ (t\bar{\ell}_{t,R} ^D)]^{1/\kappa}} +e^{-(R-r_0)/(4k)} } \notag \\
&\le \zeta_0 \exp\brr{- \Theta\br{\frac{R-r_0}{t\bar{\ell}_{t,R} ^D}}^{1/\kappa} +\Theta(\log(R)) }.
\end{align}  
Because of $\bar{\ell}_{t,R} \propto t \log (Rq_0)$, for $r_0=\orderof{1}$ and $q_0=\poly(R)$,
the approximation error becomes sufficiently small as long as 
\begin{align}
R \propto t^{D+1} \log^{D+1}(t) ,
\end{align}
which gives the Lieb-Robinson velocity of $\tO(t^D)$.

On the unitary operator $U_{i_0[R]}$, we show the explicit form in the inequality~\eqref{lem:Lieb-Robinson_effectve_ham_loose_main_ineq}, which is constructed by using the effective Hamiltonian. 
The construction only depends on the interaction terms within the region of $i_0[R]$.
 Hence, when we consider the subset Hamiltonian $H_{i_0[R]}$, 
we obtain the same inequality as \eqref{ineq:main_theorem_looser_Lieb_Robinson} for $\norm{ \brr{O_{X_0}(H_{i_0[R]},t)   - U_{i_0[R]}^\dagger O_{X_0} U_{i_0[R]}  } \rho_0}_1$ by using the same unitary operator $U_{i_0[R]}$, which yields 
\begin{align}
\label{ineq:approx_subset_Hamiltonian}
&\norm{ \brr{O_{X_0}(H,t)   -O_{X_0}(H_{i_0[R]},t)} \rho_0}_1 \notag \\
&\le \norm{ \brr{O_{X_0}(H,t)   - U_{i_0[R]}^\dagger O_{X_0} U_{i_0[R]} } \rho_0}_1  + \norm{ \brr{O_{X_0}(H_{i_0[R]},t)   - U_{i_0[R]}^\dagger O_{X_0} U_{i_0[R]} } \rho_0}_1 \notag \\
&\le 8 \zeta_0 \gamma^2 R^D  \br{1+ R/c_1} \exp\brr{- \frac{1}{2}\br{\frac{R-r_0}{8 b_0c_1 \bar{J}t \bar{\ell}_{t,R} ^D} }^{1/\kappa}}
+2\zeta_0 R^{D+1} e^{-(R-r_0)/(4k)}  .
\end{align}  
Therefore, we can obtain the approximation for $O_{X_0}(H,t)$ by simply considering the time evolution by the subset Hamiltonian $H_{i_0[R]}$ on the region $i_0[R]$.

\subsection{Proof of Theorem~\ref{main_theorem_looser_Lieb_Robinson}: basic strategy}

 \begin{figure}[tt]
\centering
\includegraphics[clip, scale=0.5]{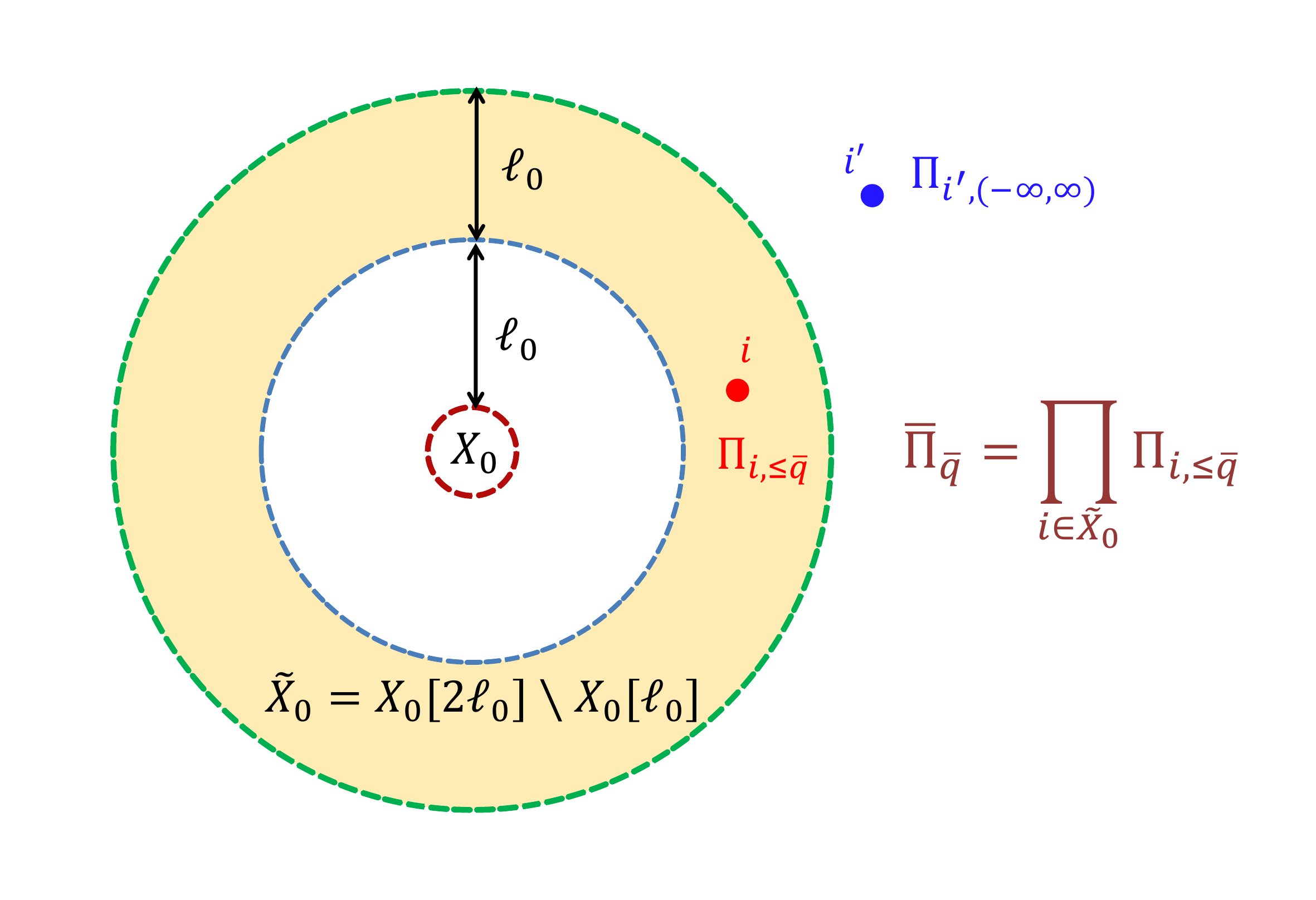}
\caption{Schematic picture of the region $\tilde{X}_0$.
In this region, we truncate the boson number up to $\bar{q}$ and define the projection as $\bar{\Pi}_{\bar{q}}$. 
We first approximate the original dynamics $e^{-iHt}$ by using the dynamics $e^{-i\bar{\Pi}_{\bar{q}}H\bar{\Pi}_{\bar{q}}t}$.
Then, because the effective Hamiltonian $\bar{\Pi}_{\bar{q}}H\bar{\Pi}_{\bar{q}}$ has a finite local energy in the region $\tilde{X}_0$, 
we can derive the Lieb-Robinson bound and approximate  $e^{-i\bar{\Pi}_{\bar{q}}H\bar{\Pi}_{\bar{q}}t}$ by $e^{-i\bar{\Pi}_{\bar{q}}H_{X_0[2\ell_0]}\bar{\Pi}_{\bar{q}}t}$. 
}
\label{supp_fig_tilde_X_0}
\end{figure}

For the proof, we consider the projection operator $\bar{\Pi}_{\tilde{X}_0,\bar{q}}$ (see Fig.~\ref{supp_fig_tilde_X_0}), where 
we use the notation in Eq.~\eqref{def_pi_bar_bar_q_e_L}, i.e., 
\begin{align}
\label{def_pi_bar_bar_q_e}
\bar{\Pi}_{\tilde{X}_0,\bar{q}}:= \prod_{i\in \tilde{X}_0} \Pi_{i, \le \bar{q}} , \quad \tilde{X}_0:= X_0[2\ell_0]\setminus X_0[\ell_0] \quad (\ell_0\ge 8k),
\end{align} 
where $X_0 \subseteq i_0[r_0]$ and
\begin{align}
\label{def_ell_0_R=2ell_0+r_0}
R=2\ell_0 + r_0  ,\quad   X_0[2\ell_0] \subseteq i_0[r_0+ 2\ell_0] = i_0[R] .
\end{align}
Note that the condition $R\ge r_0 + 16k$ implies $\ell_0\ge 8k$. 
In the following, we simply denote it as 
\begin{align}
\bar{\Pi}_{\tilde{X}_0,\bar{q}} \to \bar{\Pi}_{\bar{q}}
\end{align}
by omitting the region dependence $\tilde{X}_0$. 
We choose $\bar{q}$ ($\ge 2$) such that
\begin{align}
\label{choice_of_bar_q_t^d}
\bar{q}= \frac{\ell_0}{c_1\bar{J} t} , \quad c_1: = 2^{6-D} e^2k\gamma^3(2k)^{2D},
\end{align} 
where the choice is utilized in ensuring the condition~\eqref{uppper_bpund_1/c_3'_ineq} below\footnote{
Precisely speaking, in Eqs.~\eqref{def_ell_0_R=2ell_0+r_0} and $\eqref{choice_of_bar_q_t^d}$ for the definitions of $\ell_0$ and $\bar{q}$, 
we should define $\ell_0=\floorf{(R-r_0)/2}$ and $\bar{q}= \floorf{\ell_0/(c_1 t)}$, but for simplicity we assume they become integers without the floor function.}.

The basic strategy for the proof is similar to the one in Ref.~\cite{PhysRevLett.127.070403}, and 
we summarize the tasks as follows:
\begin{enumerate}
\item{} We approximate the dynamics $e^{-iHt}$ by using the effective Hamiltonian $\bar{\Pi}_{\bar{q}}H\bar{\Pi}_{\bar{q}}$; that is, 
our first task is to estimate the error of 
\begin{align}
\label{effective_hamiltonian_error_t^d}
\norm{ \brr{ O_{X_0}(H,t) -  O_{X_0}(\bar{\Pi}_{\bar{q}}H\bar{\Pi}_{\bar{q}},t) }  \rho_0 }_1  .
\end{align}
\item{} For the time evolution of $O_{X_0}(\bar{\Pi}_{\bar{q}}H\bar{\Pi}_{\bar{q}},t)$, 
we approximate the dynamics by using the Hamiltonian $\bar{\Pi}_{\bar{q}}H_{X_0[2\ell_0]}\bar{\Pi}_{\bar{q}}$ which is supported on $X_0[2\ell_0]$.
Here, in the Hamiltonian $\bar{\Pi}_{\bar{q}}H\bar{\Pi}_{\bar{q}}$,
the interaction terms are bounded in the region $\tilde{X}_0$ due to the truncation of the boson number.
However, in the region $\tilde{X}_0^\co$, we still encounter the unboundedness of the norm of the interactions. 
In Ref.~\cite{PhysRevLett.127.070403}, the authors have developed analyses to derive the
Lieb-Robinson bound for $e^{-i\bar{\Pi}_{\bar{q}}H\bar{\Pi}_{\bar{q}}t}$ only from the finite upper bound of the Hamiltonian norm in the region $\tilde{X}_0$.
\end{enumerate}

We start from the estimation of the norm~\eqref{effective_hamiltonian_error_t^d}.
The error is immediately obtained by using Proposition~\ref{prop:Effective Hamiltonian_local_region} 
with the choice of $L=\tilde{X}_0$ in~\eqref{prop:Effective Hamiltonian_local_region_eq}: 
\begin{align}
\label{effective_hamiltonian_error_t^d_upper_bound__0}
\norm{ \brr{ O_{X_0}(H,t) -  O_{X_0}(\bar{\Pi}_{\bar{q}}H\bar{\Pi}_{\bar{q}},t) }  \rho_0 }_1  
\le 4 \zeta_0 \gamma |\tilde{X}_0| \br{1+ \bar{J}t\bar{q} } e^{- \brr{\bar{q}/(4 b_0 \bar{\ell}_{t,R} ^D) }^{1/\kappa}/2},
\end{align}
where the following condition should be satisfied
\begin{align}
\label{cond_ell_0_first_1}
\ell_0=\dist_{X_0,\tilde{X}_0}  \ge \bar{\ell}_{t,R}+1.
\end{align} 
By using $|\tilde{X}_0| \le | X_0[2\ell_0] | \le |i_0[R]| \le \gamma R^D$, and $\bar{J}t\bar{q} = \ell_0/c_1$ ($\le R/c_1$) , we reduce the inequality~\eqref{effective_hamiltonian_error_t^d_upper_bound__0} to 
\begin{align}
\label{effective_hamiltonian_error_t^d_upper_bound}
\norm{ \brr{ O_{X_0}(H,t) -  O_{X_0}(\bar{\Pi}_{\bar{q}}H\bar{\Pi}_{\bar{q}},t) }  \rho_0 }_1
\le 4 \zeta_0 \gamma^2 R^D  \br{1+ R/c_1} \exp\brr{- \frac{1}{2}\br{\frac{R-r_0}{8  b_0c_1 \bar{J}t \bar{\ell}_{t,R} ^D} }^{1/\kappa}},
\end{align}
where we use $\ell_0=(R-r_0)/2$.
Remember that the length $\bar{\ell}_{t,R}$ has been defined in Eq.~\eqref{eq_bar_ell_t_r}. 
Therefore, the remaining task is to derive the Lieb-Robinson bound for the effective Hamiltonian $\bar{\Pi}_{\bar{q}} H \bar{\Pi}_{\bar{q}}$.
In detail, we aim to estimate the approximation error of 
 \begin{align}
 \label{lem:Lieb-Robinson_effectve_ham_loose_main_ineq}
\norm{ \Bigl[  O_{X_0} (\bar{\Pi}_{\bar{q}}  H\bar{\Pi}_{\bar{q}}  ,t) - O_{X_0} ( \bar{\Pi}_{\bar{q}} H_{X_0[\ell]} \bar{\Pi}_{\bar{q}} , t) \Bigr]  \rho_0 }_1 
\le \norm{  O_{X_0} (\bar{\Pi}_{\bar{q}}  H\bar{\Pi}_{\bar{q}}  ,t) - O_{X_0} ( \bar{\Pi}_{\bar{q}} H_{X_0[\ell]} \bar{\Pi}_{\bar{q}} , t)  }
\end{align} 
for a particular choice of $\ell$, 
where we use $\norm{O_1 O_2}_1 \le \norm{O_1}\cdot \norm{O_2}_1$ in the inequality. 
For this purpose, we prove the following lemma:
\begin{lemma} \label{lem:Lieb-Robinson_effectve_ham_loose}
Under the choice of $\bar{q}$ as in Eq.~\eqref{choice_of_bar_q_t^d}, we prove that 
 \begin{align}
 \label{lem:Lieb-Robinson_effectve_ham_loose_main_ineq}
\norm{  O_{X_0} (\bar{\Pi}_{\bar{q}}  H\bar{\Pi}_{\bar{q}}  ,t) - O_{X_0} ( \bar{\Pi}_{\bar{q}} H_{X_0[2\ell_0-2k]} \bar{\Pi}_{\bar{q}} , t) }
\le \zeta_0 R^{D+1} e^{-(R-r_0)/(4k)} 
\end{align} 
for $\ell_0 \ge 8k$. 
We show the proof in Sec.~\ref{sec:Derivation of the Lieb-Robinson bound}.
\end{lemma}

Finally, by combining the two inequalities~\eqref{effective_hamiltonian_error_t^d_upper_bound} and \eqref{lem:Lieb-Robinson_effectve_ham_loose_main_ineq}, we obtain the main inequality~\eqref{ineq:main_theorem_looser_Lieb_Robinson}.
The two conditions $\ell_0\ge \bar{\ell}_{t,R}+1$ in Eq.~\eqref{cond_ell_0_first_1} and $\ell_0\ge 8k$ in Lemma~\ref{lem:Lieb-Robinson_effectve_ham_loose} 
is integrated by the condition $R\ge \max \brr{r_0 + 16k, r_0 +2( \bar{\ell}_{t,R}+1)}$ in the main statement, where $\ell_0 = (R-r_0)/2$.   
This completes the proof of Theorem~\ref{main_theorem_looser_Lieb_Robinson}. $\square$

\subsection{Proof of Lemma~\ref{lem:Lieb-Robinson_effectve_ham_loose}: derivation of the Lieb-Robinson bound}\label{sec:Derivation of the Lieb-Robinson bound}

We here calculate the norm of 
 \begin{align}
\label{ineq:prop:error_time_evolution_effective_Ham_t^D}
&\norm{ O_{X_0}(\bar{\Pi}_{\bar{q}} H \bar{\Pi}_{\bar{q}}, t) - O_{X_0}(\bar{\Pi}_{\bar{q}} H_{X_0[2\ell_0]} \bar{\Pi}_{\bar{q}}, t)  } .
\end{align} 
Now, the effective Hamiltonian $\bar{\Pi}_{\bar{q}} H_{X_0[2\ell_0]} \bar{\Pi}_{\bar{q}}$ is supported on $X_0[2\ell_0]$ as $\bar{\Pi}_{\bar{q}}$ is supported on $\tilde{X}_0 \subset X_0[2\ell_0]$.
For the estimation of the norm~\eqref{ineq:prop:error_time_evolution_effective_Ham_t^D}, we can utilize the technique in Ref.~\cite[Supplementary Section S.V.C]{PhysRevLett.127.070403}. 
By using the interaction picture, we start from the decomposition of 
 \begin{align}
 \label{decomposition_of_tine_evolution_bar_H}
e^{-i \bar{\Pi}_{\bar{q}} H \bar{\Pi}_{\bar{q}} t} = e^{-i \bar{\Pi}_{\bar{q}} V \bar{\Pi}_{\bar{q}} t}\ \mathcal{T} \exp \br{    
-i\int_0^t e^{i \bar{\Pi}_{\bar{q}} V \bar{\Pi}_{\bar{q}}  \tau} \bar{\Pi}_{\bar{q}} H_0 \bar{\Pi}_{\bar{q}} e^{-i \bar{\Pi}_{\bar{q}} V \bar{\Pi}_{\bar{q}} \tau} d\tau} .
\end{align} 
The primary motivation to use the interaction picture lies in that we impose no assumptions on the norm of $V$ [see Def.~\ref{def_Short-range interactions_0} and Table~\ref{tab:necessary_assumption}]. 

For simplicity, we adopt the notations of
 \begin{align}
&\bar{H}:= \bar{\Pi}_{\bar{q}} H \bar{\Pi}_{\bar{q}} ,\quad \bar{H}_0:= \bar{\Pi}_{\bar{q}} H_0 \bar{\Pi}_{\bar{q}} ,\quad \bar{V}:= \bar{\Pi}_{\bar{q}} V \bar{\Pi}_{\bar{q}} ,\notag \\
&\bar{H}_\tau := e^{i \bar{V}  \tau}  \bar{H}_0 e^{- i \bar{V}  \tau},
\end{align} 
which simplify Eq.~\eqref{decomposition_of_tine_evolution_bar_H} as 
 \begin{align}
 \label{decomposition_of_tine_evolution_bar_H_simp}
e^{-i \bar{H} t} = e^{-i \bar{V} t}\ \mathcal{T} e^{    
-i\int_0^t\bar{H}_\tau d\tau} .
\end{align} 
We expand $\bar{H}_\tau$ as 
 \begin{align}
 \label{def:tilde_H_tau_bar}
&\bar{H}_\tau = \sum_{\langle i, j\rangle } J_{i,j}  e^{i \bar{V}  \tau} \bar{\Pi}_{\bar{q}} (b_i^\dagger b_j + {\rm h.c.}) \bar{\Pi}_{\bar{q}} e^{- i \bar{V}  \tau}   
=\sum_{\langle i, j\rangle } \bar{h}_{\mathcal{B}_{i,j},\tau} , \notag\\
&\bar{h}_{\mathcal{B}_{i,j},\tau}:=  \bar{\Pi}_{\bar{q}}  e^{i \bar{V}  \tau} (b_i^\dagger b_j + {\rm h.c.}) e^{- i \bar{V}  \tau}    \bar{\Pi}_{\bar{q}} ,\quad 
 \mathcal{B}_{i,j}= i[k-1] \cup j[k-1] , \quad \diam (\mathcal{B}_{i,j}) \le 2k ,
\end{align} 
where use the fact that the operator $e^{i \bar{V}  \tau}  b_i e^{-i \bar{V}  \tau} $ is supported on the subset $i[k-1]$ because $\bar{V}$ has interaction length of $k$ (see Def.~\ref{def_Short-range interactions_0}). 
We here obtain 
 \begin{align}
 \label{upper_bound_norm_tilde_h_Z_i_j_tau_bar}
\norm{ \bar{h}_{\mathcal{B}_{i,j},\tau}}  \le |J_{i,j}|  \cdot \norm{\bar{\Pi}_{\bar{q}} (b_i^\dagger b_j + {\rm h.c.}) \bar{\Pi}_{\bar{q}}} 
\le 4 \bar{J} \bar{q} \quad (i,j \in \tilde{X}_0),
\end{align}
 where we use the inequality~\eqref{upper_bound_hopping_operator} to derive
\begin{align}
\norm{\bar{\Pi}_{\bar{q}} b_i^\dagger b_j  \bar{\Pi}_{\bar{q}}} \le  \norm{\bar{\Pi}_{\bar{q}} \br{\nb_i+\nb_j } \bar{\Pi}_{\bar{q}}}\le 2\bar{q}.
\end{align}
Then, by using the inequality~\eqref{upper_bound_norm_tilde_h_Z_i_j_tau_bar} and the upper bound of
\begin{align}
\max_{i_0\in \Lambda}  \# \brrr{\langle i,j \rangle | \mathcal{B}_{i,j} \ni i_0} 
\le \gamma \max_{i_0 \in \Lambda} \br{  |i_0[k] | }\le \gamma^2 k^D ,
\end{align}
we obtain
 \begin{align}
 \label{def:tilde_H_tau_bar_extensive}
\max_{i_0\in \Lambda}\sum_{\substack{\langle i, j\rangle \\   \{i,j\} \subset \tilde{X}_0,\ \mathcal{B}_{i,j}\ni i_0}} \norm{\bar{h}_{\mathcal{B}_{i,j},\tau}} \le  
4 \bar{J} \bar{q} \cdot \gamma^2 k^D =c_3\ell_0,
\end{align} 
where $c_3$ is defined as 
 \begin{align}
 \label{g_extensive_ness_finite}
c_3 :=   \frac{4 \bar{J} \gamma^2 k^D \bar{q}}{\ell_0}.
\end{align} 

 \begin{figure}[tt]
\centering
\includegraphics[clip, scale=0.35]{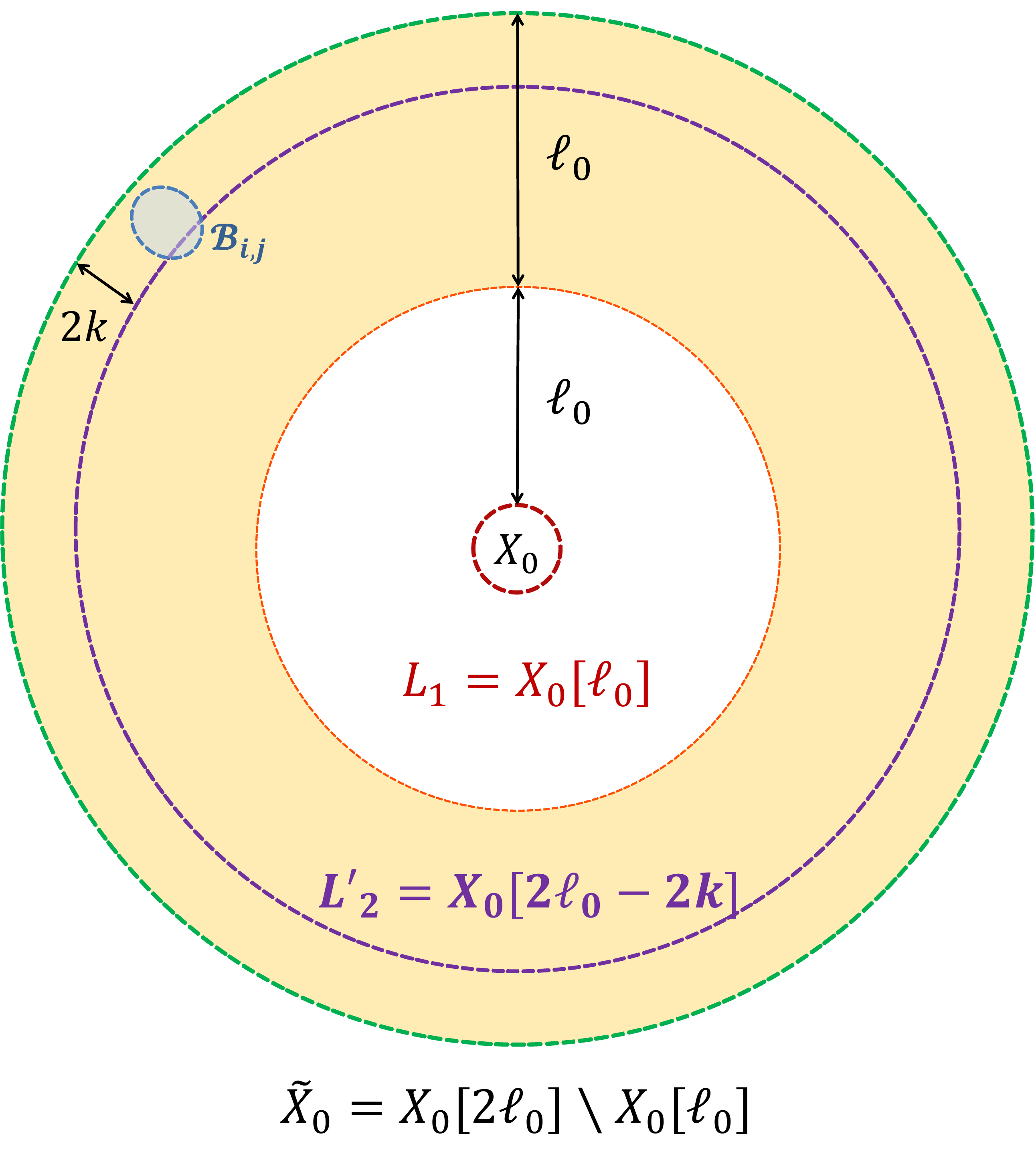}
\caption{Schematic picture for the positions of the subsets $X_0$, $\tilde{X_0}$, $L_1$ and $L_2'$. 
We here consider the time evolution of $O_{X_0}$ for the decomposed unitary operator~\eqref{decomposition_of_tine_evolution_bar_H_simp}.
Then, as in Eq.~\eqref{def_bar_O_L1_99}, the contribution from the interaction terms in $V$ is included in the subset $X_0[k] \subseteq X_0[\ell_0]=L_1$.
To treat the contribution from the interactions $H_\tau=\sum_{\langle i, j\rangle } \bar{h}_{\mathcal{B}_{i,j},\tau} $ in Eq.~\eqref{def:tilde_H_tau_bar}, we utilize Lemma~\ref{lem:Lieb--Robinson_start}.
From the lemma, the time evolution $\bar{O}_{L_1} (\bar{H}_\tau, 0\to t)$ is approximated by $\bar{O}_{L_1} \br{\bar{H}_{L_2',\tau}, 0\to t}$ with $L_2'=X_0[2\ell_0-2k]$, where we should consider the time evolution of the boundary interaction terms $\bar{h}_{\mathcal{B}_{i,j},\tau}$ ($\mathcal{B}_{i,j} \in \mathcal{S}_{L_2'}$) as shown in the inequality~\eqref{lem:Lieb--Robinson_start_apply/t^D}. 
From the choice of $L_2'$, we can always ensure that any interaction term $\bar{h}_{\mathcal{B}_{i,j},\tau}$ in $\mathcal{S}_{L_2'}$ is supported on the region 
$\tilde{X}_0$. 
}
\label{supp_fig_Lieb_Robinson_t_D}
\end{figure}

In the following, we utilize the following established lemma:
\begin{lemma}[Supplementary Lemma~14 in Ref.~\cite{PhysRevLett.127.070403}] \label{lem:Lieb--Robinson_start}
Let $H_\tau$ be an arbitrary time-dependent Hamiltonian in the form of
\begin{align}
\label{lemma_Ham__tauH}
H_\tau = \sum_{Z\subset \Lambda} h_{Z,\tau} . 
\end{align}
We also write a subset Hamiltonian on $L$ as follows:
\begin{align}
\label{subset_Hamiltonian_L_tau}
H_{L,\tau} = \sum_{Z:Z\subseteq L} h_{Z,\tau}.
\end{align}
Then, for an arbitrary subset $L$ such that $L \supseteq X$, we obtain 
\begin{align}
\label{lem:ineq:Lieb--Robinson_start}
\| O_X(H_\tau, 0\to t) - O_X(H_{L,\tau}, 0\to t) \|  \le \sum_{Z: Z \in \mathcal{S}_{L}} \int_0^t \left\| [h_{Z,x}(H_{L,\tau}, x\to 0), O_X] \right\| dx ,
\end{align}
where $\mathcal{S}_{L}$ is defined as a set of subsets $\{Z\}_{Z\subseteq \Lambda}$, which overlap the surface region of $L$:
\begin{align}
\label{set_mathcal_S_L}
\mathcal{S}_{L}:= \{ Z\subseteq \Lambda | Z \cap L \neq \emptyset, Z \cap L^\co \neq \emptyset\} .
\end{align}
Note that the notation $h_{Z,x}(H_{L,\tau}, x\to 0)$ has been defined in Eq.~\eqref{notation_time_dependent_operator}.
\end{lemma}

{~}\\

By using the decomposition~\eqref{decomposition_of_tine_evolution_bar_H_simp}, we now obtain (see Fig.~\ref{supp_fig_Lieb_Robinson_t_D})
 \begin{align}
 \label{O_X_0_bar_H_t_eq}
O_{X_0}(\bar{H},t) = \bar{O}_{L_1} (\bar{H}_\tau, 0\to t),  \quad L_1= X_0[\ell_0],
\end{align} 
where we define
 \begin{align}
 \label{def_bar_O_L1_99}
\bar{O}_{L_1} = e^{i \bar{\Pi}_{\bar{q}} V \bar{\Pi}_{\bar{q}} t} O_{X_0} e^{-i \bar{\Pi}_{\bar{q}} V \bar{\Pi}_{\bar{q}} t} 
=  e^{i \bar{\Pi}_{\bar{q}} V_{X_0[k]} \bar{\Pi}_{\bar{q}} t} O_{X_0} e^{-i \bar{\Pi}_{\bar{q}} V_{X_0[k]} \bar{\Pi}_{\bar{q}} t} ,
\end{align} 
where $V_{X_0[k]} $ picks up all the interaction terms $v_Z$ such that $Z\subset X_0[k]$. 
In deriving Eq.~\eqref{def_bar_O_L1_99}, we use $[\bar{\Pi}_{\bar{q}} v_Z\bar{\Pi}_{\bar{q}},\bar{\Pi}_{\bar{q}} v_{Z'}\bar{\Pi}_{\bar{q}}]=0 $ for arbitrary $Z,Z'$. Also, because of $\diam(Z)\le k$, we have $[\bar{\Pi}_{\bar{q}} v_Z\bar{\Pi}_{\bar{q}},O_{X_0}]\neq 0$ only if $Z$ is included in $X_0[k]$ (i.e., $Z\subseteq X_0[k]$).

We, in the following, aim to approximate $ \bar{O}_{L_1} (\bar{H}_\tau, 0\to t)$ by $ \bar{O}_{L_1} (\bar{H}_{L_2', \tau}, 0\to t)$ with 
 \begin{align}
L'_2= X_0[2\ell_0-2k] . 
\end{align} 
Lemma~\ref{lem:Lieb--Robinson_start} immediately gives the following upper bound on the approximation error:
 \begin{align}
 \label{lem:Lieb--Robinson_start_apply/t^D}
\norm{ \bar{O}_{L_1} (\bar{H}_\tau, 0\to t) -  \bar{O}_{L_1} (\bar{H}_{L_2', \tau}, 0\to t)} 
\le \sum_{i,j: \mathcal{B}_{i,j} \in \mathcal{S}_{L_2'}} \int_0^t \left\| [\bar{h}_{\mathcal{B}_{i,j} ,x}(\bar{H}_{L_2', \tau}, x\to 0), \bar{O}_{L_1}] \right\| dx . 
\end{align} 
For an arbitrary operator $O_Z$, we have already derived the following lemma:
\begin{lemma}[Supplementary Lemma~15 in Ref.~\cite{PhysRevLett.127.070403}] \label{lem:Lieb-Robinson_calculation}
Let us choose the subset $Z$ such that $Z\in \mathcal{S}_{L_2'}$ and $\diam(Z)\le 2k$.
Then, for an arbitrary operator $O_Z$, we have the upper bound of 
 \begin{align}
 \label{ineq_lem:Lieb-Robinson_calculation}
\left\| [ O_Z (\bar{H}_{L'_2,\tau}, x \to 0), \bar{O}_{L_1}] \right\| \le 2e^3 \norm{\bar{O}_{L_1}} \cdot \norm{O_Z} e^{-\ell_0/(2k)}
\end{align} 
under the conditions of 
 \begin{align}
 \label{cond_for_time_t_LR_short_time}
\ell_0\ge 8k, \quad x\le \frac{1}{e c_3'} , \quad c_3' := 16ekc_3 \gamma (2k)^D=2^{6-D} ek\bar{J}\gamma^3(2k)^{2D}  \frac{\bar{q}}{\ell_0},
\end{align} 
where $c_3$ has been defined in \eqref{g_extensive_ness_finite}\footnote{
The condition of $\ell_0\ge 8k$ is not explicitly assumed in the original statement of Ref.~\cite[Supplementary Lemma~15]{PhysRevLett.127.070403}, 
but assumed implicitly (see \cite[Supplementary (S.84)]{PhysRevLett.127.070403}).}. 
\end{lemma}

Under the choice of~\eqref{choice_of_bar_q_t^d} for $\bar{q}$, we have 
 \begin{align}
 \label{uppper_bpund_1/c_3'_ineq}
\frac{1}{e c_3'} =\frac{\ell_0}{\bar{q}} \cdot \frac{1}{2^{6-D} e^2k\bar{J}\gamma^3(2k)^{2D}} = t ,
\end{align} 
which allows us to utilize the inequality~\eqref{ineq_lem:Lieb-Robinson_calculation} for $\forall x\le t$. 
Hence, we apply Lemma~\ref{lem:Lieb-Robinson_calculation} to the inequality~\eqref{lem:Lieb--Robinson_start_apply/t^D} as 
 \begin{align}
 \label{lem:Lieb--Robinson_start_apply/t^D_2}
\norm{ \bar{O}_{L_1} (\bar{H}_\tau, 0\to t) -  \bar{O}_{L_1} (\bar{H}_{L_2', \tau}, 0\to t)} 
&\le 2e^3  \zeta_0 \int_0^t  \sum_{i,j: \mathcal{B}_{i,j} \in \mathcal{S}_{L_2'}} \norm{\bar{h}_{\mathcal{B}_{i,j} ,x}}  e^{-(R-r_0)/(4k)}  dx \notag \\
&\le 8 t e^3\zeta_0 \bar{J} \bar{q}\gamma R^D e^{-(R-r_0)/(4k)} ,
\end{align} 
where we use $\ell_0=(R-r_0)/2$, $\norm{\bar{O}_{L_1}}=\norm{O_{X_0}}=\zeta_0$ [see Eq.~\eqref{def_bar_O_L1_99}], and the norm inequality~\eqref{upper_bound_norm_tilde_h_Z_i_j_tau_bar} for $\norm{\bar{h}_{\mathcal{B}_{i,j} ,x}}$ to derive
 \begin{align}
 \label{sum_h_Bij_gamma^2}
\sum_{i,j: \mathcal{B}_{i,j} \in \mathcal{S}_{L_2'}}  \norm{\bar{h}_{\mathcal{B}_{i,j} ,x}}
 \le 4 \bar{J} \bar{q}\sum_{i,j: \mathcal{B}_{i,j} \in \mathcal{S}_{L_2'}} 1
 \le 4 \bar{J} \bar{q}\sum_{i\in X_0[2\ell_0]} \sum_{j:\dist_{i,j}=1} 1 \le 4 \bar{J} \bar{q} \gamma^2 R^D ,
\end{align} 
where we use $|X_0[2\ell_0]|\le |i_0[r_0+2\ell_0]| = |i_0[R]| \le \gamma R^D$ from $X_0\in i_0[r_0]$. 

Finally, from the definition~\eqref{choice_of_bar_q_t^d} for $\bar{q}$, we have 
 \begin{align}
 \label{uppper_bpund_1/c_3'_ineq_22}
\bar{J}\bar{q}t \le  \frac{\ell_0}{2^{6-D} e^2k\gamma^3(2k)^{2D}} 
= \frac{\ell_0}{8e^3 \gamma^2}  \cdot \frac{e}{2^{D+3} k^{2D+1} \gamma}
\le \frac{R}{8e^3 \gamma^2} ,
\end{align} 
which reduces the inequality~\eqref{lem:Lieb--Robinson_start_apply/t^D_2} to
 \begin{align}
 \label{lem:Lieb--Robinson_start_apply/t^D_3__0}
\norm{ \bar{O}_{L_1} (\bar{H}_\tau, 0\to t) -  \bar{O}_{L_1} (\bar{H}_{L_2', \tau}, 0\to t)} 
\le \zeta_0 R^{D+1} e^{-(R-r_0)/(4k)} .
\end{align} 
From the decomposition~\eqref{decomposition_of_tine_evolution_bar_H_simp} and definition~\eqref{def_bar_O_L1_99}, we can write
 \begin{align}
\bar{O}_{L_1} (\bar{H}_{L_2', \tau}, 0\to t) =\bar{O}_{L_1} (\bar{H}_{L_2'}, t) = O_{X_0} ( \bar{\Pi}_{\bar{q}} H_{L_2'} \bar{\Pi}_{\bar{q}} , t) , 
\end{align} 
which yields 
 \begin{align}
 \label{lem:Lieb--Robinson_start_apply/t^D_3}
\norm{ [ O_{X_0} (\bar{\Pi}_{\bar{q}}  H\bar{\Pi}_{\bar{q}}  ,t) - O_{X_0} ( \bar{\Pi}_{\bar{q}} H_{X_0[2\ell_0-2k]} \bar{\Pi}_{\bar{q}} , t) }
\le \zeta_0 R^{D+1} e^{-(R-r_0)/(4k)} ,
\end{align} 
where we use the explicit expression of $L_2'=X_0[2\ell_0-2k]$. 
This completes the proof of Lemma~\ref{lem:Lieb-Robinson_effectve_ham_loose}.
$\square$

\section{Main subtheorem on small time evolution}

As mentioned, the obtained bound in Theorem~\ref{main_theorem_looser_Lieb_Robinson} is not optimal. 
In the following sections, we aim to refine the upper bound to the optimal forms as in Theorems~\ref{main_thm:boson_Lieb-Robinson_main} and \ref{main_thm:boson_Lieb-Robinson_main_1D}.
For this purpose, we first derive an essential analytical ingredient for the refinements. 

\begin{figure}[tt]
\centering
\includegraphics[clip, scale=0.5]{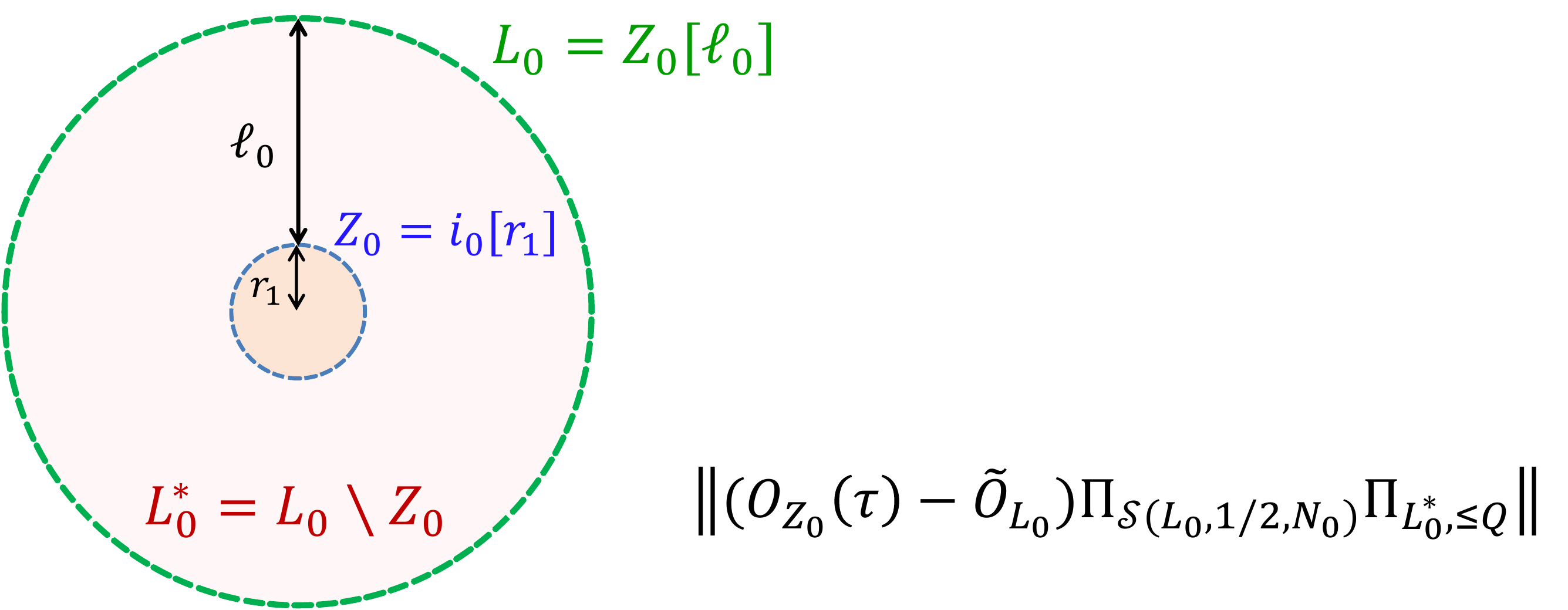}
\caption{Schematic picture of the setup of Subtheorem~\ref{subtheorem_small_time_evolution_general}. 
As the initial condition, we consider the projections $\Pi_{\mathcal{S}(L_0,1/2,N_0)}$ and $\Pi_{L_0^\ast, \le Q}$, where $L_0=Z_0[\ell_0]$ and $L_0^\ast = L_0\setminus Z_0$. 
The projection $\Pi_{\mathcal{S}(L_0,1/2,N_0)}$ implies that there are not too many bosons around the region $L_0$ [see Eq.~\eqref{mathcal_S_L_xi_N/def_re}].
On the projection $\Pi_{L_0^\ast, \le Q}$, the total number of bosons in the region $L_0^\ast$ is smaller than $Q$.
Under these two projections, we approximate the small time evolution $O_{Z_0} (H, \tau)$ in the better way as~\eqref {ineq:subtheorem_small_time_evolution_general}. 
This bound is the crucial ingredient for the optimal Lieb-Robinson bounds in Theorems~\ref{main_thm:boson_Lieb-Robinson_main} and \ref{main_thm:boson_Lieb-Robinson_main_1D}.
}
\label{fig:small_time_Lieb_Robinson.pdf}
\end{figure}

Here, we consider the following setup (see Fig.~\ref{fig:small_time_Lieb_Robinson.pdf}).
Let $O_{Z_0}$ be an arbitrary operator supported on a subset $Z_0 \subseteq i_0[r_1]$.  
We then consider an approximation of the time-evolved operator $O_{Z_0}(\tau)$ in the region $L_0= Z_0[\ell_0]$:
\begin{align}
\label{O_Z_0_tilde_O_L_0}
\norm { \br{ O_{Z_0} (\tau) -  \tilde{O}_{L_0} } \ket{\psi} } , \quad L_0:= Z_0[\ell_0] ,\quad L_0^\ast := L_0\setminus Z_0 ,
\end{align}
where $ \tilde{O}_{L_0}$ is an appropriate approximation of the operator $O_{Z_0} (\tau)$ [see the inequality~\eqref{ineq:subtheorem_small_time_evolution_general_re} for an example of the choice].
As a spectral constraint, we assume the equation of
\begin{align}
\label{Pi_S_L_0N_0_L_0_Q}
\Pi_{\mathcal{S}(L_0,1/2,N_0)} \Pi_{L_0^\ast, \le Q} \ket{\psi}  = \ket{\psi} ,
\end{align}
where $\mathcal{S}(L_0,1/2,N_0)$ has been defined by Eq.~\eqref{mathcal_S_L_xi_N/def_re}. 
The following subtheorem plays a central role in deriving the optimal Lieb-Robinson bounds:
\begin{subtheorem} \label{subtheorem_small_time_evolution_general}
Let the boson number $Q$ be larger than $|L_0^\ast|$, i.e., $Q\ge |L_0^\ast|$.
For an arbitrary operator $O_{Z_0}$ ($Z_0 \subseteq i_0[r_1]$, $\norm{O_{Z_0}}=\zeta_{Z_0}$) with the condition~\eqref{O_X_condition/_norm_lemma_operator} with $q=q_{Z_0}$, there exists an appropriate operator $\tilde{O}_{L_0}$ such that
\begin{align}
\label{ineq:subtheorem_small_time_evolution_general}
&\norm{\br{O_{Z_0} (\tau) - \tilde{O}_{L_0}} \Pi_{\mathcal{S}(L_0,1/2,N_0)} \Pi_{L_0^\ast, \le Q} }
\le\zeta_{Z_0} \Theta(Q^2) \brrr{ e^{-\Theta\br{\ell_0}} + e^{\Theta(\ell_0)}\brr{\Theta(\tau) +\Theta\br{\frac{\tau Q\log(Q)}{\ell_0^2}}  }^{\Theta(\ell_0)}} ,
\end{align}
where $\tau\le 1/(4\gamma\bar{J})$ and $\ell_0\ge\log[\Theta\brr{\log (q_{Z_0} N_0 )}]$ [see Eq.~\eqref{ell_0_conditionb_subtheorem}].
Note that the $\Theta$ notation has been defined in Eq.~\eqref{Theta_notation_def}. 
\end{subtheorem}

{\bf Remark.} 
In the inequality~\eqref{ineq:subtheorem_small_time_evolution_general}, the approximate operator $\tilde{O}_{L_0}$ is constructed as in Eq.~\eqref{def_tilde_O_L_subtheorem} and Eq.~\eqref{U_L_choice_subtheorem}. 
In the construction, we do not utilize information of the Hamiltonian outside the region $L_0$, and hence for the operator $O_{Z_0} (H_{L_0},\tau)$, we obtain the same inequality 
as~\eqref{ineq:subtheorem_small_time_evolution_general}. 
Therefore, by using a similar analysis to~\eqref{ineq:approx_subset_Hamiltonian}, we can obtain 
\begin{align}
\label{ineq:subtheorem_small_time_evolution_general_re}
&\norm{\brr{O_{Z_0} (H, \tau) - O_{Z_0} (H_{L_0}, \tau) } \Pi_{\mathcal{S}(L_0,1/2,N_0)} \Pi_{L_0^\ast, \le Q} } \notag \\
&\le\zeta_{Z_0} \Theta(Q^2) \brrr{ e^{-\Theta\br{\ell_0}} + e^{\Theta(\ell_0)}\brr{\Theta(\tau) +\Theta\br{\frac{\tau Q\log(Q)}{\ell_0^2}}  }^{\Theta(\ell_0)}}
\end{align}
by using the Hamiltonian $H_{L_0}$ supported on the subset $L_0$.

\subsection{Proof outline}


For the proof, we cannot use a simple approach that has been used for Theorem~\ref{main_theorem_looser_Lieb_Robinson}.
That is, if we simply truncate the boson number up to $Q$ at each site, we should construct the effective Hamiltonian as in Eq.~\eqref{effective_hamiltonian_error_t^d} with $\bar{q}=Q$.
By using similar analyses to Sec.~\ref{sec:Derivation of the Lieb-Robinson bound},  we can obtain the simple upper bound  to~\eqref{lem:Lieb-Robinson_effectve_ham_loose_main_ineq} in the form of 
\begin{align}
\norm {\br{ O_{Z_0} (\tau)-U_{L_0}^\dagger O_{Z_0} U_{L_0} }\Pi_{\mathcal{S}(L_0,1/2,N_0)} \Pi_{L_0^\ast, \le Q}  } \lesssim  
\zeta_{Z_0} e^{-\Theta(\ell_0)},
\label{looser_bound_to_be_improved}
\end{align}
where $U_{L_0}$ is an appropriate unitary operator and the boson number $Q$ satisfies $Q\lesssim \ell_0/(\bar{J}\tau)$ from Eq.~\eqref{choice_of_bar_q_t^d}.  
In the above bound, the norm exponentially decays with $\ell_0$ as long as $\tau \le \Theta \br{\ell_0/Q}$, which is qualitatively worse than the bound~\eqref{ineq:subtheorem_small_time_evolution_general} ensuring the decay up to $\tau \le \Theta \brr{\ell_0^2/(Q\log(Q))}$.
To refine the upper bound~\eqref{looser_bound_to_be_improved}, 
we need to utilize the fact that all the sites cannot have $Q$ bosons simultaneously.
For example, in the 1D case, in the subset from the site $i_0$ to $L_0^\co$, the number of bosons on one site is at most $\orderof{Q/\ell_0}$ \textit{in average}. 
We need to take this point into account for the proof of Subtheorem~\ref{subtheorem_small_time_evolution_general}. 

\begin{figure}[tt]
\centering
\includegraphics[clip, scale=0.5]{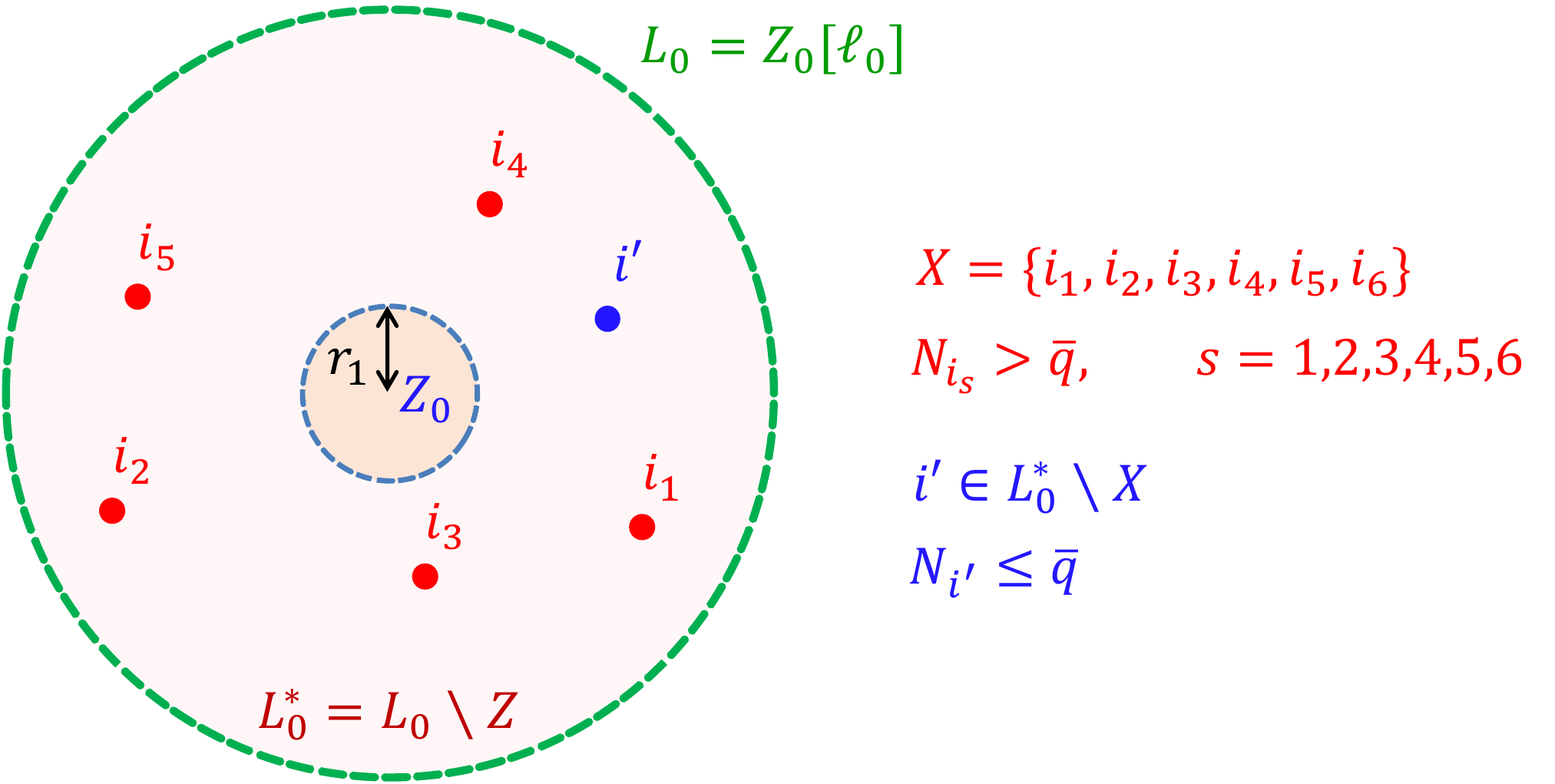}
\caption{Schematic picture of the subset $\tilde{\mathcal{S}}(L_0^\ast,\bold{N}_X,\bar{q})$.
In the picture, we consider the case $X=\{i_1,i_2,i_3,i_4,i_5,i_6\}$.
Then, if $\bold{N} \in \tilde{\mathcal{S}}(L_0^\ast,\bold{N}_X,\bar{q})$, we have  $N_{i_s} > \bar{q}$ for $s\in [6]$ and $N_{i'} \le \bar{q}$ for $i'\in L_0^\ast \setminus X$.
Here, the threshold $\bar{q}$ in Eq.~\eqref{def:bar_q_mathfrak_q} is appropriately chosen according to our purpose.
By combining $\Pi_{\tilde{\mathcal{S}}(L_0^\ast,\bold{N}_X,\bar{q})}$ with the projection $\Pi_{L_0^\ast, \le Q}$, we can ensure that the size of $X$ must be smaller than or equal to $Q/\bar{q}$, 
which leads to the expansion of Eq.~\eqref{Pi_L_ast_L_1/2_N0}. 
}
\label{fig:Set_tilde_S_dash.pdf}
\end{figure}

For this purpose, we first define a threshold boson number $\bar{q}$ as follows:
\begin{align}
\label{def:bar_q_mathfrak_q}
\bar{q} = \frac{\mathfrak{q}}{\ell_0} Q ,
\end{align}
where we choose $\bar{q}$ (or $\mathfrak{q}$) appropriately afterward [see Eq.~\eqref{eq:chose_bar_q_subtheorem} below].
We then construct a set $\tilde{\mathcal{S}}(L_0^\ast,\bold{N}_X,\bar{q})$ ($X\subseteq L_0^\ast$) as follows (see Fig.~\ref{fig:Set_tilde_S_dash.pdf}):
\begin{align}
\tilde{\mathcal{S}}(L_0^\ast,\bold{N}_X,\bar{q}) = \brrr{ \bold{N} | N_i > \bar{q} \for i\in X,\quad  
N_i \le \bar{q} \for i\in L_0^\ast\setminus X}  .
\end{align}
In this set, the boson numbers are fixed in $X$ and constrained only in $L_0^\ast \setminus X$. Here, $N_i$ is larger than $\bar{q}$ in the region $X$ but is less than or equal to $\bar{q}$ in the region $L_0^\ast\setminus X$. 
Then, we define the projection $\Pi_{\tilde{\mathcal{S}}(L_0^\ast,\bold{N}_X,\bar{q})}$ as 
\begin{align}
\Pi_{\tilde{\mathcal{S}}(L_0^\ast,\bold{N}_X,\bar{q})}:= \sum_{\bold{N}\in \tilde{\mathcal{S}}(L_0^\ast,\bold{N}_X,\bar{q})} \ket{\bold{N}} \bra{\bold{N}}
=\prod_{i\in X} \Pi_{i,N_i} \prod_{i\in L_0^\ast\setminus X} \Pi_{i, \le \bar{q}}  ,
\end{align}
where $\Pi_{\tilde{\mathcal{S}}(L_0^\ast,\bold{N}_X,\bar{q})}$ is supported on the subset $L_0^\ast$ and no restrictions are imposed for the boson numbers in $L_0^{\ast\co}$. 
We then obtain 
\begin{align}
\sum_{X: X\subseteq L_0^\ast}\sum_{\bold{N}_X : N_i>\bar{q} \ (i\in X)} \Pi_{\tilde{\mathcal{S}}(L_0^\ast,\bold{N}_X,\bar{q})} 
&= \sum_{X: X\subseteq L_0^\ast} \prod_{i\in X} \Pi_{i,>\bar{q}} \prod_{i\in L_0^\ast\setminus X} \Pi_{i, \le \bar{q}} \notag \\
&=\prod_{i\in L_0^\ast} (\Pi_{i,>\bar{q}}+ \Pi_{i, \le \bar{q}})   =  \hat{1}, 
\end{align}
where the summation $\sum_{\bold{N}_X : N_i>\bar{q} \ (i\in X)}$ takes all the patterns of $\{N_i\}_{i\in X}$ such that $N_i > \bar{q}$ ($i\in X$).
If the total number of bosons in the region $L_0^\ast$ is smaller than or equal to $Q$, we have  
\begin{align}
\bar{q} |X| \le Q  \longrightarrow |X| \le \frac{\ell_0}{\mathfrak{q}} .
\end{align}
Therefore, we can expand the projection $\Pi_{\mathcal{S}(L_0,1/2,N_0)} \Pi_{L_0^\ast, \le Q}$ as 
\begin{align}
\label{Pi_L_ast_L_1/2_N0}
\Pi_{\mathcal{S}(L_0,1/2,N_0)} \Pi_{L_0^\ast, \le Q} = \sum_{\substack{X: X\subset L_0^\ast \\ |X| \le \ell_0/\mathfrak{q}}} 
\sum_{\substack{\bold{N}_X : N_i>\bar{q} \ (i\in X) \\ N_X\le Q}} \Pi_{\tilde{\mathcal{S}}(L_0^\ast,\bold{N}_X,\bar{q})} \Pi_{\mathcal{S}(L_0,1/2,N_0)} \Pi_{L_0^\ast, \le Q}  ,
\end{align}
which yields 
\begin{align}
&O_{Z_0} (\tau) \Pi_{\mathcal{S}(L_0,1/2,N_0)} \Pi_{L_0^\ast, \le Q}  
=\sum_{\substack{X: X\subset L_0^\ast \\ |X| \le \ell_0/\mathfrak{q}}} 
\sum_{\substack{\bold{N}_X : N_i>\bar{q} \ (i\in X) \\ N_X\le Q}} 
 O_{Z_0} (\tau) \Pi_{\tilde{\mathcal{S}}(L_0^\ast,\bold{N}_X,\bar{q})} \Pi_{\mathcal{S}(L_0,1/2,N_0)} \Pi_{L_0^\ast, \le Q} .
\end{align}

We thus need to consider the following quantity:
\begin{align}
 O_{Z_0} (\tau) \Pi_{\tilde{\mathcal{S}}(L_0^\ast,\bold{N}_X,\bar{q})} \Pi_{\mathcal{S}(L_0,1/2,N_0)} \Pi_{L_0^\ast, \le Q} 
\end{align}
for each of the choices of $X\subset L_0^\ast$ and $\bold{N}_X$.
In the following, we aim to estimate the approximation error of $O_{Z_0}(\tau)$ under the above projections:
\begin{prop}\label{Prop:Lieb-Robinson_short_time_L_X_bar_q}
Let us take $\ell_0$ and $\delta \ell$ such that 
 \begin{align}
&\ell_0  \ge 4\delta \ell +12 k +4, \quad \delta \ell \ge 8 \log  \brr{\mu_0 \gamma N_0\br{\lambda_{1/4} +1 + 2^{4D+3} D!} + 4c_{\tau,1}(4q_{Z_0}^3+2N_0\lambda_{1/2}) } -8\log(\mu_1)   , \label{new_condition_ell_0_LR_prop}  \\
&\mu_0 :=4\br{ \lambda_{1/4} c_{\tau,1}^2+2 c_{\tau,1}} ,\quad 
\mu_1:=  4\brr{2 c_{\tau,2}+c_{\tau,1} c_{\tau,2}(1+ \lambda_1)} ,\label{def_mu_i_i_j_upp_re}
\end{align} 
where the two conditions for $\ell_0$ and $\delta \ell$ are introduced in \eqref{new_condition_ell_0_LR} and \eqref{condi_for_delta_ell}, respectively.
Then, for an arbitrary choices of $X$ and $\bar{q}$ in $\tilde{\mathcal{S}}(L_0^\ast,\bold{N}_X,\bar{q})$, we obtain the Lieb-Robinson bound for an appropriate choices of $U_{L_0}$ [see Eq.~\eqref{U_L_choice_subtheorem}]: 
\begin{align}
\label{main_ineqProp:Lieb-Robinson_short_time_L_X_bar_q}
\norm{ \br{ O_{Z_0} (\tau) -  U_{L_0}^\dagger O_{Z_0} U_{L_0} } \Pi_{\tilde{\mathcal{S}}(L_0^\ast,\bold{N}_X,\bar{q})} \Pi_{\mathcal{S}(L_0,1/2,N_0)} \Pi_{L_0^\ast, \le Q} } \le \delta_1 + \delta_2 ,
\end{align}
where $\delta_1$ and $\delta_2$ are defined as follows:
\begin{align}
\label{def_delta_1_proposition_O_Z_0}
&\delta_1 = 8 \zeta_{Z_0} |L_0^\ast|  e^{-\tilde{q}}+ 
56\tau  \gamma \zeta_{Z_0} \bar{J} e^{-\tilde{q}}\bar{N}_{\tilde{L}}  , \\
&\label{def_delta_2_proposition_O_Z_0}
\delta_2 = \frac{16\gamma\bar{J} \tau \zeta_{Z_0} }{2^D} \bar{N}_{\tilde{L}} 
\brr{\frac{\tilde{c}_1 \tau}{\ell_0} 
\br{\frac{\tilde{c}_2 Q}{\ell_0}+ \tilde{c}_{\tilde{q},\bar{q}}} }^{\ell_0/(4k)} ,
\end{align}
with
 \begin{align} 
&\tilde{c}_1:= 2^{D+6} e^2 \gamma^3 k^{2D+1} \bar{J} ,
\quad \tilde{c}_2:=2k  (4+\mu_0 \lambda_{3/4}) ,\quad 
\tilde{c}_{\tilde{q},\bar{q}}:=4\mu_1 \tilde{q} + 4\bar{q} + \mu_0 \lambda_{3/4} \bar{q} .\label{def_tilde_c_1_2_3_0}
\end{align} 
Here, $\tilde{q}$ is a free parameter which can be arbitrarily chosen [see Eq.~\eqref{choice_of_q_i_effective} below] and $\bar{N}_{\tilde{L}}$ is upper-bounded by the inequality~\eqref{ineq:lem:boson_number_ball_change_0}.
\end{prop}

From Proposition~\ref{Prop:Lieb-Robinson_short_time_L_X_bar_q}, by defining 
\begin{align}
\label{def_tilde_O_L_subtheorem}
\tilde{O}_{L_0} := \sum_{\substack{X: X\subset L_0^\ast \\ |X| \le \ell_0/\mathfrak{q}}} \sum_{\substack{\bold{N}_X : N_i>\bar{q} \ (i\in X) \\ N_X\le Q}} U_{L_0}(L_0^\ast,\bold{N}_X,\bar{q})^\dagger O_{Z_0} U_{L_0}(L_0^\ast,\bold{N}_X,\bar{q}) \Pi_{\tilde{\mathcal{S}}(L_0^\ast,\bold{N}_X,\bar{q})}  
\end{align}
with $U_{L_0}(L_0^\ast,\bold{N}_X,\bar{q})$ an appropriate unitary operator, 
we obtain 
\begin{align}
\label{main_ineq_subtheorem_delta_1_delta_2}
&\norm{\br{O_{Z_0} (\tau) - \tilde{O}_{L_0}} \Pi_{\mathcal{S}(L_0,1/2,N_0)} \Pi_{L_0^\ast, \le Q} }
\le  \sum_{\substack{X: X\subset L_0^\ast \\ |X| \le \ell_0/\mathfrak{q}}} \sum_{\substack{\bold{N}_X : N_i>\bar{q} \ (i\in X) \\ N_X\le Q}} (\delta_1+\delta_2)  \le \sum_{\substack{X: X\subset L_0^\ast \\ |X| \le \ell_0/\mathfrak{q}}}Q^{|X|} (\delta_1+\delta_2) 
\notag \\
&
\le 2(\delta_1+\delta_2) (Q|L_0^\ast|)^{\ell_0/ \mathfrak{q}} 
\le 2(\delta_1+\delta_2) Q^{2\ell_0/ \mathfrak{q}}= 2(\delta_1+\delta_2) \exp\brr{\frac{2Q}{\bar{q}} \log(Q) } ,
\end{align}
where in the third inequality, we use
\begin{align}
\label{sum_r_0_ell_0_mathfrak_q}
\sum_{r=0}^{\ell_0/\mathfrak{q}} \sum_{\substack{X: X\subset L_0^\ast \\ |X| =r}} 1 =  \sum_{r=0}^{\ell_0/\mathfrak{q}}  \binom{|L_0^\ast|}{r} \le 
2 |L_0^\ast|^{\ell_0/ \mathfrak{q}} ,
\end{align}
in the fourth inequality, we use  $Q\ge |L_0^\ast|$ from the assumption, and in the last equation, we use $Q=\bar{q}\ell_0 /\mathfrak{q}$.

\subsection{Completing the proof of Subtheorem~\ref{subtheorem_small_time_evolution_general}} 
\label{sec:Completing the proof of Subtheorem_small_time_evolution_general}

We have obtained all the ingredients to complete the proof.
In the following discussions, we fully utilize the $\Theta$ notation in Eq.~\eqref{Theta_notation_def} to avoid the complication of the notations.
We first choose $\delta \ell$ and $\ell_0$ such that the conditions in~\eqref{new_condition_ell_0_LR_prop} are satisfied:
\begin{align}
&\delta \ell  \ge 8 \log  \brr{\mu_0 \gamma N_0\br{\lambda_{1/4} +1 + 2^{4D+3} D!} + 4c_{\tau,1}(4q_{Z_0}^3+2N_0\lambda_{1/2}) } -8\log(\mu_1) \notag \\
&\longrightarrow \delta \ell =\Theta\brr{\log (q_{Z_0} N_0 )},
\end{align}
and 
 \begin{align}
 \label{ell_0_conditionb_subtheorem}
\ell_0  \ge 4\delta \ell +12 k +4 = \Theta\brr{\log (q_{Z_0} N_0 )}.
\end{align} 
We second choose the free parameter $\tilde{q}$ such that 
 \begin{align}
 \label{choice_of_tilde_q_proof}
\tilde{c}_{\tilde{q},\bar{q}}=4\mu_1 \tilde{q} + 4\bar{q} + \mu_0 \lambda_{3/4} \bar{q} \le 2\mu_0 \lambda_{3/4}\bar{q} \longrightarrow \tilde{q} =\Theta(\bar{q}),   
\end{align} 
where we have defined $\tilde{c}_{\tilde{q},\bar{q}}$ in Eq.~\eqref{def_tilde_c_1_2_3_0}.
Then, the quantity $\bar{N}_{\tilde{L}} $ is upper-bounded from~\eqref{ineq:lem:boson_number_ball_change_0} as follows:
\begin{align}
\label{upp_bar_N_tilde_L_Q^2}
\bar{N}_{\tilde{L}} \le  \Theta(Q+\bar{q}|L_0^\ast|) \le \Theta(\bar{q} Q ) ,
\end{align}
which yields
\begin{align}
\label{delta_1_theta_not}
\delta_1 =  8 \zeta_{Z_0} |L_0^\ast|  e^{-\tilde{q}}+ 
56\tau  \gamma \zeta_{Z_0} \bar{J} e^{-\tilde{q}}\bar{N}_{\tilde{L}} 
\le \zeta_{Z_0}  \Theta(\bar{q} Q )  e^{-\Theta(\bar{q})} ,
\end{align}
where we use $ |L_0^\ast| \le Q$. 
Moreover, we have 
\begin{align}
\label{delta_2_theta_not}
\delta_2 &= \frac{16\gamma\bar{J} \tau \zeta_{Z_0} }{2^D} \bar{N}_{\tilde{L}} 
\brr{\frac{\tilde{c}_1 \tau}{\ell_0} 
\br{\frac{\tilde{c}_2 Q}{\ell_0}+ \tilde{c}_{\tilde{q},\bar{q}}} }^{\ell_0/(4k)}  
= \zeta_{Z_0} \Theta(\bar{q}Q)  \brr{\Theta\br{\frac{\tau Q}{\ell_0^2}} + \Theta\br{\frac{\tau\bar{q}}{\ell_0} } }^{\Theta(\ell_0)} ,
\end{align}
where we use the fact that $\tilde{c}_1$ and $\tilde{c}_2$ are $\orderof{1}$ constants [see Eq.~\eqref{def_tilde_c_1_2_3_0}].

In the following, we choose $\bar{q}$ such that
 \begin{align}
 \label{eq:chose_bar_q_subtheorem}
 \bar{q} = \Theta\br{\frac{Q\log(Q)}{\ell_0}}+\Theta\br{\ell_0} ,\quad \textrm{which implies}\quad \exp\brr{\frac{2Q}{\bar{q}} \log(Q) }  \le 
  e^{\Theta(\ell_0)} .
\end{align}
By combining the above choice with Eqs~\eqref{delta_1_theta_not} and \eqref{delta_2_theta_not}, we can upper-bound $2(\delta_1+\delta_2) \exp\brr{\frac{2Q}{\bar{q}} \log(Q) }$ in the inequality~\eqref{main_ineq_subtheorem_delta_1_delta_2} in the form of 
\begin{align}
2(\delta_1+\delta_2) \exp\brr{\frac{2Q}{\bar{q}} \log(Q) }  \le\zeta_{Z_0} \Theta(Q^2) \brrr{ e^{-\Theta\br{\ell_0}} + e^{\Theta(\ell_0)}\brr{\Theta(\tau) +\Theta\br{\frac{\tau Q\log(Q)}{\ell_0^2}}  }^{\Theta(\ell_0)}},
\end{align}
where we use $\bar{q} \le \Theta(Q) +\Theta\br{\ell_0} \le  \Theta(Q) $ because of $Q\ge |L_0^\ast| = |Z_0[\ell_0]| - |Z_0| \ge  \Theta(\ell_0)$.
By applying the above inequality to~\eqref{main_ineq_subtheorem_delta_1_delta_2}, we obtain the main inequality~\eqref{ineq:subtheorem_small_time_evolution_general}.
This completes the proof of Subtheorem~\ref{subtheorem_small_time_evolution_general}. $\square$ 

 \subsection{Proof of Proposition~\ref{Prop:Lieb-Robinson_short_time_L_X_bar_q}: spectral constraint for construction of effective Hamiltonian}
  \label{sec:proof_prop_spectral constraint for construction of effective Hamiltonian}
 
We here consider an arbitrary quantum state $\ket{\psi_{\bold{N}}}$ such that
\begin{align}
\label{spectral_constraint_boson_number_psi_N}
 \Pi_{\tilde{\mathcal{S}}(L_0^\ast,\bold{N}_X,\bar{q})} \Pi_{\mathcal{S}(L_0,1/2,N_0)} \Pi_{L_0^\ast, \le Q} \ket{\psi_{\bold{N}}}= \ket{\psi_{\bold{N}}}.
\end{align}
Here, the quantum state $\ket{\psi_{\bold{N}}}$ is characterized by $\bold{N}$ as follows:
\begin{align}
\label{property_of_psi_vec_N}
&\Pi_{i,\le N_i} \ket{\psi_{\bold{N}}} =\ket{\psi_{\bold{N}}},\quad  N_i = N_0 e^{\dist_{i,L_0}/2} \for i\in L_0^{\ast\co}, \notag \\
&\Pi_{i,\le N_i} \ket{\psi_{\bold{N}}} =\ket{\psi_{\bold{N}}},\quad  N_i =\bar{q}  \for i\in  L_0^\ast \setminus X  , \notag \\
&\Pi_{i,N_i} \ket{\psi_{\bold{N}}} =\ket{\psi_{\bold{N}}}   , \quad {\rm i.e.} \quad 
\nb_i \ket{\psi_{\bold{N}}} =N_i \ket{\psi_{\bold{N}}},\quad  N_i > \bar{q} \for  i\in X ,
\end{align} 
with
\begin{align}
\label{property_of_psi_vec_N_N_i_le_Q}
\sum_{i\in X} N_i \le Q , 
\end{align}
where the condition for $i\in L_0^{\ast\co}$ originates from the projection $\Pi_{\mathcal{S}(L_0,1/2,N_0)}$, 
the conditions for $i\in L_0^\ast \setminus X$ and $i\in X$ originate from the projection $ \Pi_{\tilde{\mathcal{S}}(L_0^\ast,\bold{N}_X,\bar{q})}$, and the condition $\sum_{i\in X} N_i \le Q $ originates from the projection $\Pi_{L_0^\ast, \le Q}$. 
Note that any quantum state in the Hilbert space spanned by 
$\Pi_{\tilde{\mathcal{S}}(L_0^\ast,\bold{N}_X,\bar{q})} \Pi_{\mathcal{S}(L_0,1/2,N_0)} \Pi_{L_0^\ast, \le Q}$ is described by the form of $\ket{\psi_{\bold{N}}}$. 
Therefore, for an arbitrary operator $O$, we obtain 
\begin{align}
\norm{ O \Pi_{\tilde{\mathcal{S}}(L_0^\ast,\bold{N}_X,\bar{q})} \Pi_{\mathcal{S}(L_0,1/2,N_0)} \Pi_{L_0^\ast, \le Q} } 
=\sup_{\psi} \norm{ O \Pi_{\tilde{\mathcal{S}}(L_0^\ast,\bold{N}_X,\bar{q})} \Pi_{\mathcal{S}(L_0,1/2,N_0)} \Pi_{L_0^\ast, \le Q} \ket{\psi}} 
\le\sup_{\psi_{\bold{N}}} \norm{ O \ket{\psi_{\bold{N}}}}  . 
\end{align}

To prove the main inequality~\eqref{main_ineqProp:Lieb-Robinson_short_time_L_X_bar_q}, we start from the approximation by using the effective Hamiltonian as 
\begin{align}
\label{choice_of_effective_Ham_tilde_KL}
\tilde{H}[\tilde{L}',\bold{q}] :=\bar{\Pi}_{\tilde{L}',\bold{q}} H\bar{\Pi}_{\tilde{L}',\bold{q}} ,
\quad \bar{\Pi}_{\tilde{L}',\bold{q}}:=\prod_{i\in \tilde{L}'} \Pi_{i,\le q_i}
\end{align}
with
\begin{align}
\tilde{L}' := \tilde{L}[-1] , 
\end{align}
where the subset $\tilde{L}$ ($\subset L_0^\ast$) will be defined in Eq.~\eqref{definition_subset_tilde_L} (see also Fig.~\ref{fig:subset_tilde_L.pdf}), and $\bold{q}=\{q_i\}_{i\in \tilde{L}'}$ are appropriately chosen as in Eq.~\eqref{choice_of_q_i_effective}. 
In the present section and the subsequent section~\ref{sec:proof_prop_Error estimation for the effective Hamiltonian}, we aim to upper-bound
\begin{align}
\label{effective_upp_O_{Z_0}_tau}
\norm{ e^{iH \tau} O_{Z_0} e^{-iH \tau} \ket{\psi_{\bold{N}}}-  e^{i \tilde{H}[\tilde{L}',\bold{q}] \tau} O_{Z_0} e^{-i \tilde{H}[\tilde{L}',\bold{q}] \tau} \ket{\psi_{\bold{N}}} } ,
\end{align}
where the explicit upper bound for the above quantity will be given in Lemma~\ref{lem:effective Hamiltonian_accuracy_2}.


For the estimation of the norm~\eqref{effective_upp_O_{Z_0}_tau}, we utilize Lemma~\ref{effective Hamiltonian_accuracy},  
where it is critical to upper-bound the norm of
\begin{align}
\norm { \Pi_{i,> q_i} e^{iH\tau_2} O_{Z_0}e^{-iH\tau_1} \ket{\psi_{\bold{N}}}  } \for \forall i\in \tilde{L}.
\end{align}
In the present section, the final goal is to prove the statement in Corollary~\ref{corol:prob._upper_bound_time_evo}. 
We first prove the following lemma:
\begin{lemma} \label{lemm:prob._upper_bound_effective}
For arbitrary $\tau_1$ and $\tau_2$ such that $\tau_1,\tau_2\le \tau \le 1/(4\gamma\bar{J})$, we have 
\begin{align}
\label{main_ineq_lemm:prob._upper_bound_effective}
\norm { \nb_i^{s/2} e^{iH\tau_2} O_{Z_0}e^{-iH\tau_1} \ket{\psi_{\bold{N}}}  }^2
\le \zeta_{Z_0}^2 \brr{ 4N_{i} +\mu_0  \sum_{j \in \Lambda} e^{-3\dist_{i,j}/4}  N_j  + \mu_1 s +4c_{\tau,1} e^{-\dist_{i,Z_0}} (4q_{Z_0}^3+N^\ast_{Z_0})}^s   
\end{align}
for $\forall i\in L_0^\ast$, where we define $\mu_0$, $\mu_1$ and $N^\ast_{Z_0}$ as 
 \begin{align}
\label{def_mu_i_i_j_upp}
\mu_0 :=4\br{ \lambda_{1/4} c_{\tau,1}^2+2 c_{\tau,1}} ,\quad 
\mu_1:=  4\brr{2 c_{\tau,2}+c_{\tau,1} c_{\tau,2}(1+ \lambda_1)} ,\quad 
N^\ast_{Z_0}:=  N_{Z_0} + c_{\tau,1}\sum_{j\in Z_0}  \sum_{j' \in \Lambda} e^{-\dist_{j,j'}} N_{j'} .
\end{align}
\end{lemma}

\subsubsection{Proof of Lemma~\ref{lemm:prob._upper_bound_effective}} 

\begin{figure}[tt]
\centering
\includegraphics[clip, scale=0.35]{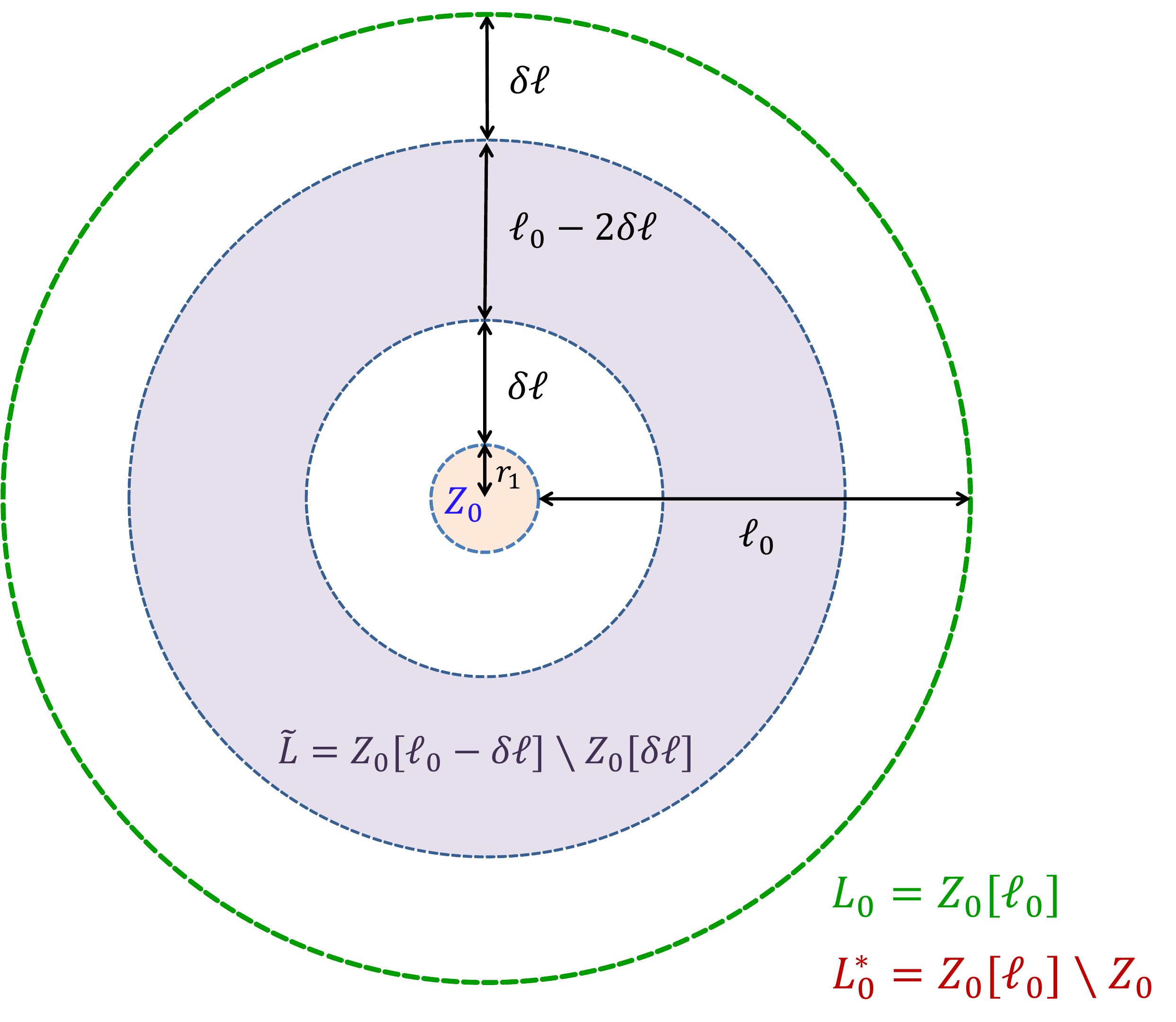}
\caption{Schematic picture of the region $\tilde{L}$. 
By defining the subset $\tilde{L}$ ($\subset L_0^\ast$) as in Eq.~\eqref{definition_subset_tilde_L}, 
we have $\dist_{\tilde{L}, Z_0} = \dist_{\tilde{L}, L_0^\co} = \delta \ell $. 
As shown in Eq.~\eqref{choice_of_effective_Ham_tilde_KL}, we are going to perform the truncations of boson number in the region $\tilde{L}[-1]$.
}
\label{fig:subset_tilde_L.pdf}
\end{figure}

The proof is similar to the one for Lemma~\ref{lemm:boson_density_prob_gene_time_evo}.
For the proof, we denote 
\begin{align}
\label{Def_time_evo_tau1_tau2_N}
&\rho_{\tau_1} = e^{-iH\tau_1} \ket{\psi_{\bold{N}}}   \bra{\psi_{\bold{N}}}  e^{iH\tau_1}, \notag \\
&\rho_{\tau_1,\tau_2} = e^{iH\tau_2} O_{Z_0}e^{-iH\tau_1} \ket{\psi_{\bold{N}}}   \bra{\psi_{\bold{N}}}  e^{iH\tau_1} O_{Z_0}^\dagger e^{-iH\tau_2} .
\end{align}
Then, by using Subtheorem~\ref{Schuch_boson_extend}, we obtain 
\begin{align}
\label{first_step_tau_2_time}
\norm { \nb_i^{s/2} e^{iH\tau_2} O_{Z_0}e^{-iH\tau_1} \ket{\psi_{\bold{N}}}  }^2=\tr \brr{ \nb_i^s \rho_{\tau_1,\tau_2}  } 
&\le \tr \brr{ \br{\nb_i + c_{\tau,1} \hat{\mathcal{D}}_{i} + c_{\tau,2}  s}^s O_{Z_0} \rho_{\tau_1}O_{Z_0}^\dagger   } ,
\end{align}
where $\hat{\mathcal{D}}_i$ is defined as $\hat{\mathcal{D}}_i=\sum_{j\in \Lambda} e^{-\dist_{i,j}} \nb_j$ using Eq.~\eqref{Eq:def_tilde_D_X}.   
We next apply Lemma~\ref{lem:boson_operator_norm}, where the condition~\eqref{O_X_condition/_norm_lemma_operator} is now given by  
\begin{align}
\Pi_{Z_0, > m+q_{Z_0}} O_{Z_0} \Pi_{Z_0, m} , \quad \Pi_{Z_0, < m-q_{Z_0}} O_{Z_0} \Pi_{Z_0, m} =0,
\end{align}
which is assumed in the main statement of Subtheorem~\ref{subtheorem_small_time_evolution_general}.
We then apply the inequality~\eqref{main_ineq_lem:boson_operator_norm} to~\eqref{first_step_tau_2_time} 
with $q\to q_{Z_0}$, $O_L\to O_{Z_0}$ and $\bar{\nu}\to c_{\tau,1} e^{-\dist_{i,Z_0}}$:  
\begin{align}
\label{upp_O_{Z_0}_nb_i_O_{Z_0}_upp}
O_{Z_0}^\dagger \br{\nb_i + c_{\tau,1} \hat{\mathcal{D}}_{i} + c_{\tau,2}  s}^s  O_{Z_0}  
\preceq 4^{s}\zeta_{Z_0}^2 \brr{\nb_i + c_{\tau,1} \hat{\mathcal{D}}_{i} + c_{\tau,2}  s +  c_{\tau,1} e^{-\dist_{i,Z_0}} \br{4q_{Z_0}^3 +\nb_{Z_0}} }^s .
\end{align}
Note that $i\notin Z_0$ from $i\in \tilde{L}$ and $\tilde{L} \cap Z_0=0$ (see Fig.~\ref{fig:subset_tilde_L.pdf}). 
By applying the inequality~\eqref{upp_O_{Z_0}_nb_i_O_{Z_0}_upp} to \eqref{first_step_tau_2_time}, we obtain
\begin{align}
\label{first_step_tau_2_time_final_form__0}
\tr \brr{ \nb_i^s \rho_{\tau_1,\tau_2}  } \le
 \zeta_{Z_0}^2 \tr \brrr{4^s  \brr{\nb_i + c_{\tau,1} \hat{\mathcal{D}}_{i} + c_{\tau,2}  s +  c_{\tau,1} e^{-\dist_{i,Z_0}} \br{4q_{Z_0}^3 +\nb_{Z_0}} }^s  \rho_{\tau_1}} .
\end{align}

Then, by using the H\"older inequality as in~\eqref{generalized Holder_ineq}, we have 
\begin{align}
\label{nb_i_1toi_s_expec}
\tr \brr{\nb_{i_1} \nb_{i_2} \cdots \nb_{i_s} \rho_{\tau_1}} \le \prod_{m=1}^s  \brrr{ \tr \brr{ \nb_{i_m}^s \rho_{\tau_1}}}^{1/s} .
\end{align}
For each of $\{\tr \brr{ \nb_{i_m}^s \rho_{\tau_1}}\}_{m=1}^s$, by applying Subtheorem~\ref{Schuch_boson_extend}, we have 
\begin{align}
\label{nb_i_1toi_s_expec_i_m}
 \brrr{ \tr \brr{ \nb_{i_m}^s \rho_{\tau_1}}}^{1/s}
&\le \brrr{ \tr \brr{ \br{ \nb_{i_m} + c_{\tau,1} \mathcal{D}_{i_m} + c_{\tau,2} s }^s  \ket{\psi_{\bold{N}}}   \bra{\psi_{\bold{N}}}  }}^{1/s} \notag \\
&\le N_{i_m} + c_{\tau,1} \mathcal{D}_{i_m}(\bold{N}) + c_{\tau,2} s 
\end{align}
with $\mathcal{D}_{i_m}(\bold{N})$ defined as 
\begin{align}
\mathcal{D}_{i_m} (\bold{N}) := \sum_{j \in \Lambda} e^{-\dist_{i_m,j}} N_j ,
\end{align}
where we use $\rho_{\tau_1=0}=\ket{\psi_{\bold{N}}}\bra{\psi_{\bold{N}}}$ and Eq.~\eqref{property_of_psi_vec_N}.
We thus reduce the inequality~\eqref{first_step_tau_2_time_final_form__0} to the form of
\begin{align}
\label{first_step_tau_2_time_final_form}
\tr \brr{ \nb_i^s \rho_{\tau_1,\tau_2}  } \le
4^s  \zeta_{Z_0}^2  \brr{N_i + c_{\tau,1} \mathcal{D}_{i}(\bold{N})+ c_{\tau,1} \tilde{\tilde{\mathcal{D}}}_i(\bold{N}) + 2 c_{\tau,2}  s +  c_{\tau,1} e^{-\dist_{i,Z_0}} \br{4q_{Z_0}^3 +N^\ast_{Z_0} +c_{\tau,2}s}}^s   ,
\end{align}
where we define $\tilde{\tilde{\mathcal{D}}}_i(\bold{N})$ and $N^\ast_{Z_0}$ as 
\begin{align}
\label{def:tilde_mathcal_D_i_X}
&\tilde{\tilde{\mathcal{D}}}_i(\bold{N})  = \sum_{j \in \Lambda} e^{-\dist_{i,j}} 
\br{N_j +  c_{\tau,1} \mathcal{D}_{j}(\bold{N}) + c_{\tau,2} s}  , \notag \\
&N^\ast_{Z_0} :=N_{Z_0} + c_{\tau,1} \sum_{j\in Z_0} \mathcal{D}_j (\bold{N}) = N_{Z_0} + c_{\tau,1}\sum_{j\in Z_0}  \sum_{j' \in \Lambda} e^{-\dist_{j,j'}} N_{j'} .
\end{align}

To estimate the RHS of~\eqref{first_step_tau_2_time_final_form}, we consider
\begin{align}
\label{first_step_tau_1+tau_2_time_final_form_RHS}
c_{\tau,1} \mathcal{D}_{i}(\bold{N})+ c_{\tau,1} \tilde{\tilde{\mathcal{D}}}_i(\bold{N})   
=&  c_{\tau,1}  \sum_{j \in \Lambda} e^{-\dist_{i,j}}  N_j  +c_{\tau,1}  \sum_{j \in \Lambda} e^{-\dist_{i,j}}  \br{N_j  + c_{\tau,2} s+  c_{\tau,1} \sum_{j'\in \Lambda}e^{-\dist_{j,j'}} N_{j'}}   \notag \\
=&2 c_{\tau,1}  \sum_{j \in \Lambda} e^{-\dist_{i,j}}  N_j  +c_{\tau,1} c_{\tau,2} s \sum_{j \in \Lambda} e^{-\dist_{i,j}} 
+c_{\tau,1}^2  \sum_{j \in \Lambda} \sum_{j'\in \Lambda}e^{-\dist_{i,j}-\dist_{j,j'}} N_{j'} .
\end{align}
By using the inequality~\eqref{oftern_used_inequality_lambda_c} in Lemma~\ref{Lemm_Often used inequalities}, we obtain 
\begin{align}
\label{sum_dist_i_j_j'_exp_0}
&c_{\tau,1} c_{\tau,2} s \sum_{j \in \Lambda} e^{-\dist_{i,j}}  \le c_{\tau,1} c_{\tau,2} \lambda_1 s ,
\end{align}
and
\begin{align}
\sum_{j'\in \Lambda} e^{-\dist_{i,j'}-\dist_{j',j}}  
&\le \sum_{j'\in \Lambda} e^{-(\dist_{i,j'}+\dist_{j',j})/4 - 3\dist_{i,j}/4}   \le e^{-3\dist_{i,j}/4} \sum_{j'\in \Lambda} e^{-\dist_{i,j'}/4} \le  e^{-3\dist_{i,j}/4} \lambda_{1/4}, 
\label{sum_dist_i_j_j'_exp_1}
\end{align}
where we use $\dist_{i,j'}+\dist_{j',j}\ge \dist_{i,j}$. 

By combining the inequalities~\eqref{sum_dist_i_j_j'_exp_0} and \eqref{sum_dist_i_j_j'_exp_1} with 
\eqref{first_step_tau_1+tau_2_time_final_form_RHS}, we have 
\begin{align}
\label{first_step_tau_1+tau_2_time_final_form_RHS_2}
&c_{\tau,1} \mathcal{D}_{i}(\bold{N})+ c_{\tau,1} \tilde{\tilde{\mathcal{D}}}_i(\bold{N})     
\le \br{\lambda_{1/4} c_{\tau,1}^2+2 c_{\tau,1} }  \sum_{j \in \Lambda} e^{-3\dist_{i,j}/4}  N_j  + c_{\tau,1} c_{\tau,2} \lambda_1 s .
\end{align}
By applying the inequality~\eqref{first_step_tau_1+tau_2_time_final_form_RHS_2} to~\eqref{first_step_tau_2_time_final_form}, 
we prove the main inequality~\eqref{main_ineq_lemm:prob._upper_bound_effective}.
This completes the proof. $\square$

 {~}

\hrulefill{\bf [ End of Proof of Lemma~\ref{lemm:prob._upper_bound_effective}] }

{~}

In Lemma~\ref{lemm:prob._upper_bound_effective}, 
the upper bound~\eqref{main_ineq_lemm:prob._upper_bound_effective} is still complicated since it includes the summation with respect to all the sites $j\in \Lambda$.
We focus on the case where $i\in \tilde{L}$ and further simplify the form of \eqref{main_ineq_lemm:prob._upper_bound_effective}, where we define the subset $\tilde{L}$ as follows (see Fig.~\ref{fig:subset_tilde_L.pdf}): 
\begin{align}
\label{definition_subset_tilde_L}
\tilde{L}  := Z_0[\ell_0-\delta \ell] \setminus  Z_0[\delta \ell] .
\end{align}
Note that the region $\tilde{L}$ is separated from $L_0^\co$ and $Z_0$ by a distance $\delta \ell$, i.e.,
 \begin{align}
 \label{definition_subset_tilde_L_length}
\dist_{\tilde{L}, Z_0} = \dist_{\tilde{L}, L_0^\co} = \delta \ell .  
\end{align}
In the following, we aim to prove the following lemma:

\begin{lemma}\label{lemm:prob._upper_bound_effective_simple}
If we choose $\delta \ell$ such that  
\begin{align}
\label{condi_for_delta_ell}
\delta \ell  &\ge 8 \log  \brr{\mu_0 \gamma N_0\br{\lambda_{1/4} +1 + 2^{4D+3} D!} + 4c_{\tau,1}(4q_{Z_0}^3+2N_0\lambda_{1/2}) } -8\log(\mu_1)  ,
\end{align}
we obtain 
\begin{align}
\label{main_ineq_lemm:prob._upper_bound_effective_simple}
\norm { \nb_i^{s/2} e^{iH\tau_2} O_{Z_0}e^{-iH\tau_1} \ket{\psi_{\bold{N}}}  }^2
\le  \zeta_{Z_0}^2 \br{ 4N_{i} +\mu_0  \sum_{j \in L_0^\ast} e^{-3\dist_{i,j}/4}  N_j  + 2\mu_1 s }^s  \for  \forall i\in \tilde{L} .
\end{align}
\end{lemma}

\subsubsection{Proof of Lemma~\ref{lemm:prob._upper_bound_effective_simple}}

Our purpose is to prove 
\begin{align}
\label{main_ineq_lemm:prob._upper_bound_effective}
4N_{i} +\mu_0  \sum_{j \in \Lambda} e^{-3\dist_{i,j}/4}  N_j  + \mu_1 s +4c_{\tau,1} e^{-\dist_{i,Z_0}} (4q_{Z_0}^3+N^\ast_{Z_0}) 
\le 4N_{i} +\mu_0  \sum_{j \in L_0^\ast} e^{-3\dist_{i,j}/4}  N_j  + 2\mu_1 s 
\end{align}
for $i\in \tilde{L}$, 
where the first term appears in~\eqref{main_ineq_lemm:prob._upper_bound_effective} of Lemma~\ref{lemm:prob._upper_bound_effective}. 
For the proof, we need to upper-bound the quantities of
\begin{align}
\label{N_j_e^dist_N_3Z3^ast}
\sum_{j \in Z_0} e^{-3\dist_{i,j}/4}  N_j + \sum_{j \in L_0^\co} e^{-3\dist_{i,j}/4}  N_j  \AND
N^\ast_{Z_0} = N_{Z_0}+\sum_{j\in Z_0}  \sum_{j' \in \Lambda} e^{-\dist_{j,j'}} N_{j'}  .
\end{align}
Note that $\Lambda\setminus L_0^\ast= Z_0\sqcup L_0^\co$ ($L_0^\ast=L_0\setminus Z_0$). 
We first consider $\sum_{j \in Z_0} e^{-3\dist_{i,j}/4}  N_j +\sum_{j \in L_0^\co} e^{-3\dist_{i,j}/4}  N_j$. 
From $\bold{N}\in \mathcal{S}(L_0,1/2,N_0)$, we obtain $N_j \le e^{\dist_{j,L_0}/2} N_0$  [see also Eq.~\eqref{property_of_psi_vec_N}], and hence 
\begin{align}
\label{lemm:prob._upper_bound_effective_simple_proof0.5_1}
\sum_{j \in Z_0} e^{-3\dist_{i,j}/4}  N_j  \le\sum_{j \in Z_0} e^{-3\dist_{i,j}/4} e^{\dist_{j,L_0}/2} N_0
 \le e^{-\dist_{i,Z_0}/2} N_0\sum_{j \in \Lambda} e^{-\dist_{i,j}/4}  \le \lambda_{1/4} e^{-\delta \ell/2} N_0 ,
 \end{align}
 and
\begin{align}
\label{lemm:prob._upper_bound_effective_simple_proof1}
\sum_{j \in L_0^\co} e^{-3\dist_{i,j}/4}  N_j  
&\le \sum_{j \in L_0^\co}  e^{-3\dist_{i,j}/4}  e^{\dist_{j,L_0}/2} N_0
\le  \sum_{r=\delta \ell} \sum_{j: \dist_{i,j}=r} e^{-\dist_{i,j}/4} N_0
\le  \sum_{r=\delta \ell} \gamma r^D e^{-r/4}  N_0 ,
\end{align}
where we use $\dist_{i,Z_0} \ge \delta \ell$ and $\dist_{i,j} \ge \delta \ell$ for $\forall j\in  L_0^{\co}$ as long as $i\in \tilde{L}$ (see Fig.~\ref{fig:subset_tilde_L.pdf}). 
By using the inequality~\eqref{summation_s_D_e_a_m}, we obtain 
\begin{align}
\label{lemm:prob._upper_bound_effective_simple_proof2}
\sum_{r=\delta \ell} r^D e^{-r/4} 
\le e^{-\delta \ell/8}  \sum_{r=1} r^D e^{-r/8} 
\le e^{-\delta \ell/8}  e^{-1/8} \br{1 + 2^D D! 8^{D+1}} 
\le e^{-\delta \ell/8} \br{\lambda_{1/4} +1 + 2^{4D+3} D!} .
\end{align}
By combining the inequalities~\eqref{lemm:prob._upper_bound_effective_simple_proof0.5_1}, \eqref{lemm:prob._upper_bound_effective_simple_proof1}, and \eqref{lemm:prob._upper_bound_effective_simple_proof2}, 
we arrive at the inequality of
\begin{align}
\label{lemm:prob._upper_bound_effective_simple_proof_1_fin}
\sum_{j \in Z_0} e^{-3\dist_{i,j}/4}  N_j  + \sum_{j \in L_0^\co} e^{-3\dist_{i,j}/4}  N_j  
&\le \lambda_{1/4} e^{-\delta \ell/2} N_0 + \gamma  N_0 e^{-\delta \ell/8} \br{\lambda_{1/4} +1 + 2^{4D+3} D!} \notag \\
&\le \gamma  N_0 e^{-\delta \ell/8} \br{\lambda_{1/4} +1 + 2^{4D+3} D!}
\end{align}

We second consider $N^\ast_{Z_0}$ in Eq.~\eqref{N_j_e^dist_N_3Z3^ast}, which is upper-bounded by using $\dist_{j',L_0} \le \dist_{j',Z_0} \le  \dist_{j,j'}$ for $\forall j\in Z_0$ and $\forall j' \in L_0^\co$ as follows:
\begin{align}
\label{lemm:prob._upper_bound_effective_simple_proof_2_fin}
N^\ast_{Z_0} 
&\le N_0+ N_0\sum_{j\in Z_0}  \br{ \sum_{j' \in L_0^\co} e^{-\dist_{j,j'}} \br{e^{\dist_{j',L_0}/2} } +
 \sum_{j' \in L_0} e^{-\dist_{j,j'}} }  \notag \\
 &\le N_0+ N_0\sum_{j\in Z_0}  \br{ \sum_{j' \in L_0^\co} e^{-\dist_{j,j'}/2 }+
 \sum_{j' \in L_0} e^{-\dist_{j,j'}}}  \notag \\
&\le N_0 \br{1+\sum_{j\in Z_0}\sum_{j' \in \Lambda}e^{-\dist_{j,j'}/2}}  \le 2N_0 \lambda_{1/2},
\end{align}
where in the fourth inequality we use the inequality~\eqref{oftern_used_inequality_lambda_c} with $\lambda_{1/2}\ge1$. 
By applying the inequalities~\eqref{lemm:prob._upper_bound_effective_simple_proof_1_fin} and \eqref{lemm:prob._upper_bound_effective_simple_proof_2_fin} 
to the original bound~\eqref{main_ineq_lemm:prob._upper_bound_effective}, 
the LHS of~\eqref{main_ineq_lemm:prob._upper_bound_effective} is bounded from above by
\begin{align}
\label{lemm:prob._upper_bound_effective_simple_proof2.5}
&4N_{i} +\mu_0  \sum_{j \in \Lambda} e^{-3\dist_{i,j}/4}  N_j  + \mu_1 s +4c_{\tau,1} e^{-\dist_{i,Z_0}} (4q_{Z_0}^3+N^\ast_{Z_0}) \notag \\
&\le 
4N_{i} +\mu_0  \sum_{j \in L_0} e^{-3\dist_{i,j}/4}  N_j 
+ \mu_1 s +  e^{-\delta \ell/8} \mu_0  \gamma N_0\br{\lambda_{1/4} +1 + 2^{4D+3} D!} 
+4c_{\tau,1} e^{-\delta \ell}  \br{4q_{Z_0}^3+2N_0 \lambda_{1/2}}  ,
\end{align}
where we use $\dist_{i,Z_0} \ge \delta \ell$ from $i\in \tilde{L}$ [see Eq.~\eqref{definition_subset_tilde_L_length}]. 
 
Finally, the condition~\eqref{condi_for_delta_ell} yields 
\begin{align}
\label{lemm:prob._upper_bound_effective_simple_proof3}
e^{-\delta \ell/8} \mu_0\gamma N_0\br{\lambda_{1/4} +1 + 2^{4D+3} D!} 
+4c_{\tau,1} e^{-\delta \ell}  (4q_{Z_0}^3+2N_0 \lambda_{1/2})   \le \mu_1 \le \mu_1 s .
\end{align}
By applying the above inequality to \eqref{lemm:prob._upper_bound_effective_simple_proof2.5}, we prove the main inequality~\eqref{main_ineq_lemm:prob._upper_bound_effective_simple}. 
This completes the proof. $\square$

 {~}

\hrulefill{\bf [ End of Proof of Lemma~\ref{lemm:prob._upper_bound_effective_simple}] }

{~}

By using Lemma~\ref{lemm:prob._upper_bound_effective_simple}, we immediately obtain the following corollary:
\begin{corol} \label{corol:prob._upper_bound_time_evo}
Let us define
\begin{align}
\label{notation_tilde_N_i}
\tilde{N}_i:=4N_{i} +\mu_0   \sum_{j \in L_0^\ast}   e^{-3\dist_{i,j}/4}N_j .
\end{align}
Then, for $Y\subset \tilde{L}$ and $q_i \ge  e\tilde{N}_i + 2e\mu_1 p$, we obtain
\begin{align}
\label{corol:ineq_prob._upper_bound_time_evo}
\norm {\nb_Y^{p/2} \Pi_{i,> q_i} e^{iH\tau_2} O_{Z_0}e^{-iH\tau_1} \ket{\psi_{\bold{N}}}  }^2 \le 
e\zeta_{Z_0}^2 \brr{e\tilde{N}_Y + |Y| (q_i-\tilde{N}_i)}^p \exp \br{\frac{-q_i +e \tilde{N}_i }{2e\mu_1}},
\end{align} 
where $\tilde{N}_Y=\sum_{i\in Y}\tilde{N}_i$. 
\end{corol}

\subsubsection{Proof of Corollary~\ref{corol:prob._upper_bound_time_evo}}

From the definition~\eqref{notation_tilde_N_i} and the inequality~\eqref{main_ineq_lemm:prob._upper_bound_effective_simple}, we have 
\begin{align}
\tr \brr{\nb_i^s \rho_{\tau_1,\tau_2}} 
\le  \zeta_{Z_0}^2 \br{ \tilde{N}_i  + 2\mu_1 s }^s  \for  \forall i\in \tilde{L} ,
\end{align}
where $\rho_{\tau_1,\tau_2}$ is the density-matrix expression of $e^{iH\tau_2} O_{Z_0}e^{-iH\tau_1} \ket{\psi_{\bold{N}}}$ as in Eq.~\eqref{Def_time_evo_tau1_tau2_N}.
In the following, we assume that $s$ is larger than or equal to $p$, i.e., $s\ge p$.
Then, by using a similar inequality to~\eqref{nb_i_1toi_s_expec}, we obtain 
\begin{align}
\label{ineq_for_nb_Y^p_nb_i_rho_tau_1_tau_2}
\tr \brr{\nb_Y^p \nb_i^{s-p} \rho_{\tau_1,\tau_2}} 
\le  \zeta_{Z_0}^2 \br{ \tilde{N}_Y  + 2 \mu_1|Y| s }^p \br{ \tilde{N}_i  + 2 \mu_1 s }^{s-p}  \for  \forall i\in \tilde{L} .
\end{align}

The proof of the inequality~\eqref{corol:ineq_prob._upper_bound_time_evo} is the same as that for Lemma~\ref{lem:probability_dist_moment}.
By using the Markov inequality, we obtain 
\begin{align}
\label{ineq_for_nb_Y^p_nb_i_rho_tau_1_tau_2_second}
\norm {\nb_Y^{p/2} \Pi_{i,> q_i} e^{iH\tau_2} O_{Z_0}e^{-iH\tau_1} \ket{\psi_{\bold{N}}}  }^2 
&\le  \frac{1}{q_i^{s-p}} \norm { \nb_Y^{p/2} \nb_i^{(s-p)/2} e^{iH\tau_2} O_{Z_0}e^{-iH\tau_1} \ket{\psi_{\bold{N}}}  }^2 \notag \\
&\le \zeta_{Z_0}^2 \br{\tilde{N}_Y + 2\mu_1 |Y|s}^p \br{\frac{\tilde{N}_i  + 2 \mu_1 s  }{q_i}}^{s-p} ,
\end{align} 
where the second inequality is derived from~\eqref{ineq_for_nb_Y^p_nb_i_rho_tau_1_tau_2}.  
Then, by choosing $s$ as 
\begin{align}
s= \floorf{\frac{q_i - e\tilde{N}_i}{2e\mu_1}},
\end{align}
we have $\tilde{N}_i  + 2 \mu_1 s \le q_i/e$ and $\tilde{N}_Y + 2\mu_1 |Y|s \le \tilde{N}_Y + |Y| (q_i-\tilde{N}_i)/e$.
Because of the initial condition for $q_i$, i.e., $q_i \ge  e\tilde{N}_i + 2e\mu_1 p$, we can ensure that the condition $s\ge p$ is satisfied for the above choice. 
Therefore, we prove the main inequality~\eqref{corol:ineq_prob._upper_bound_time_evo}. $\square$

 {~}

\hrulefill{\bf [ End of Proof of Corollary~\ref{corol:prob._upper_bound_time_evo}] }

{~}

\subsection{Proof of Proposition~\ref{Prop:Lieb-Robinson_short_time_L_X_bar_q}: Error estimation for the effective Hamiltonian}\label{sec:proof_prop_Error estimation for the effective Hamiltonian}

By using Corollary~\ref{corol:prob._upper_bound_time_evo}, we aim to upper-bound
the approximation error~\eqref{effective_upp_O_{Z_0}_tau}.
To estimate it, we utilize Lemma~\ref{effective Hamiltonian_accuracy} with
$$
\ket{\psi}\to \ket{\psi_{\bold{N}}}, \quad O\to O_{Z_0}, \AND  \bar{\Pi}\to \bar{\Pi}_{\tilde{L}',\bold{q}}=\prod_{i\in \tilde{L}'} \Pi_{i,\le q_i},
$$
where we need to consider the quantities of
\begin{align}
\label{effective Hamiltonian_accuracy_O_{Z_0}_sub_1}
\norm{\bar{\Pi}_{\tilde{L}',\bold{q}}^\co  e^{iH\tau_2} O_{Z_0}e^{-iH\tau_1} \ket{\psi_{\bold{N}}}}  
\end{align}
and
\begin{align}
\label{effective Hamiltonian_accuracy_O_{Z_0}_sub_2}
  \norm{\bar{\Pi}_{\tilde{L}',\bold{q}} H_0 \bar{\Pi}_{\tilde{L}',\bold{q}}^\co  e^{iH\tau_2} O_{Z_0}e^{-iH\tau_1} \ket{\psi_{\bold{N}}}}  
\end{align}
for appropriate choices of $\tau_1$, $\tau_2$ and $O_{Z_0}$.

In the above setup, we need to estimate each term that appears in the inequality~\eqref{error_time/evo_effectve_original_ineq_0}. 
We prove the following lemma: 
\begin{lemma}  \label{lem:effective Hamiltonian_accuracy_2}
Let $\tilde{q}$ be an arbitrary integer that will be appropriately chosen according to our purpose [see Eq.~\eqref{choice_of_tilde_q_proof}].
We here choose $\{q_i\}_{i\in \tilde{L}}$ as 
\begin{align}
\label{choice_of_q_i_effective}
q_i =  e\tilde{N}_i + 4e\mu_1 \tilde{q} \for i\in \tilde{L},
\end{align}
where $\tilde{N}_i$ has been defined in Eq.~\eqref{notation_tilde_N_i}\footnote{Strictly speaking, to ensure the condition in Corollary~\ref{corol:prob._upper_bound_time_evo}, i.e., $q_i \ge  e\tilde{N}_i + 2e\mu_1 p$, we have to add an additional condition $\tilde{q}\ge p/2$. 
Actually, we only utilize the cases of $p=0$ or $p=2$ in the proof, and hence $\tilde{q}\ge p/2=1$ is trivially satisfied.}.
Then, under the choices of $\ket{\psi}\to\ket{\psi_{\bold{N}}}$, $O\to O_{Z_0}$ and $\bar{\Pi}\to \bar{\Pi}_{\tilde{L}',\bold{q}}$, the error in  the inequality~\eqref{error_time/evo_effectve_original_ineq_0} reduces to the following form:
\begin{align}
\label{error_time/evo_effectve_original_ineq_lemma_acc}
&\norm{ e^{iH \tau} O_{Z_0} e^{-iH \tau} \ket{\psi_{\bold{N}}}-  e^{i \tilde{H}[\tilde{L}',\bold{q}] \tau} O_{Z_0} e^{-i \tilde{H}[\tilde{L}',\bold{q}] \tau} \ket{\psi_{\bold{N}}} }\le 8 \zeta_{Z_0} |L_0^\ast|  e^{-\tilde{q}}+ 
48\tau  \gamma \zeta_{Z_0} \bar{J} e^{-\tilde{q}}\bar{N}_{\tilde{L}} = \delta_1,
\end{align}  
where we use the definition~\eqref{def_delta_1_proposition_O_Z_0} for $\delta_1$, and we define 
\begin{align}
\label{def_bar_N_tilde_L_2}
 \bar{N}_{\tilde{L}} := \sum_{i\in \tilde{L}} q_i  .
\end{align}
\end{lemma}

{\bf Remark.} 
In the definition~\eqref{choice_of_q_i_effective}, the region where $\{q_i\}_{i\in \tilde{L}}$ have been defined is $\tilde{L}$ instead of $\tilde{L}'$. 
The definition of $q_i$ for $i\in \tilde{L}\setminus \tilde{L}'= \partial \tilde{L}$ is not necessary for the construction of the effective Hamiltonian, but only used for a technical reason in deriving the upper bound~\eqref{ineq_O_X_bar_Pi_tau_2_2_upp_02__re}.

\subsubsection{Proof of Lemma~\ref{lem:effective Hamiltonian_accuracy_2}}


The proof is similar to the one for Proposition~\ref{prop:Effective Hamiltonian_local_region}. 
First of all, from the definition $\bar{\Pi}_{\tilde{L}',\bold{q}}=\prod_{i\in \tilde{L}'} \Pi_{i,\le q_i}$, we can immediately derive 
\begin{align}
\norm {\bar{\Pi}^\co_{\tilde{L}',\bold{q}}  e^{iH\tau_2} O_{Z_0} e^{-iH\tau_1} \ket{\psi_{\bold{N}}}  }
& \le
  \sum_{i\in \tilde{L}'}\norm { \Pi_{i,> q_i}  e^{iH\tau_2} O_{Z_0} e^{-iH\tau_1}\ket{\psi_{\bold{N}}}  }  
  \le  \sqrt{e}  \zeta_{Z_0} \sum_{i\in \tilde{L}'}\exp \br{\frac{-q_i +e \tilde{N}_i }{4e\mu_1}} ,
\end{align}
where we apply Corollary~\ref{corol:prob._upper_bound_time_evo} with $p=0$.
Here, we have chosen $q_i$ as 
\begin{align}
q_i = e \tilde{N}_i + 4e\mu_1 \tilde{q}  \for i\in \tilde{L}, 
\end{align}
and hence we obtain 
\begin{align}
\label{projection_tile_L_2'_error}
\norm {\bar{\Pi}^\co_{\tilde{L}',\bold{q}}  e^{iH\tau_2} O_{Z_0} e^{-iH\tau_1} \ket{\psi_{\bold{N}}}  } 
& \le 2 \zeta_{Z_0} |\tilde{L}'|  e^{-\tilde{q}}.
\end{align}

The second task is to upper-bound 
\begin{align}
\label{effective Hamiltonian_accuracy_O_{Z_0}_sub_2_re}
  \norm{\bar{\Pi}_{\tilde{L}',\bold{q}} H_0 \bar{\Pi}_{\tilde{L}',\bold{q}}^\co  e^{iH\tau_2} O_{Z_0}e^{-iH\tau_1} \ket{\psi_{\bold{N}}}}  .
\end{align}
As in Eq.~\eqref{decomposition/pi_L_H_0_co}, we first obtain 
\begin{align}
&\bar{\Pi}_{\tilde{L}',\bold{q}} H_0 \bar{\Pi}_{\tilde{L}',\bold{q}}^\co=
\bar{\Pi}_{\tilde{L}',\bold{q}} \br{H_{0,\tilde{L}'}  +  \partial h_{\tilde{L}'}}  \bar{\Pi}_{\tilde{L}',\bold{q}}^\co \  .
\end{align}
We then obtain the same inequality as~\eqref{ineq_O_X_bar_Pi_tau_2_2_upp_0}: 
\begin{align}
\label{effective Hamiltonian_accuracy_O_{Z_0}_sub_2_re_2}
\norm{\bar{\Pi}_{\tilde{L}',\bold{q}} H_0 \bar{\Pi}_{\tilde{L}',\bold{q}}^\co  e^{iH\tau_2} O_{Z_0}e^{-iH\tau_1} \ket{\psi_{\bold{N}}}}  
\le 
2\bar{J}\sum_{i\in \tilde{L}'} \sum_{j: \dist_{i,j}=1}  \norm{(\nb_i+\nb_j)\bar{\Pi}_{i,j,q_i,q_j}^\co  e^{iH\tau_2} O_{Z_0}e^{-iH\tau_1} \ket{\psi_{\bold{N}}}}   ,
\end{align}
where we define $\bar{\Pi}_{i,j,q_i,q_j}^\co $ in the similar way to Eq.~\eqref{definition_bar_pai_ij_bar_q}, i.e., 
\begin{align}
\bar{\Pi}_{i,j,q_i,q_j}= \begin{cases}
 \Pi_{i,\le q_i}\Pi_{j,\le q_j} &\for i,j\in \tilde{L}', \\
 \Pi_{i,\le q_i} &\for i\in \tilde{L}' , \quad j\in \tilde{L}'^\co.
 \end{cases}
\end{align}
Then, as in the inequality~\eqref{ineq_O_X_bar_Pi_tau_2_2_upp_02}, we obtain
\begin{align}
\label{ineq_O_X_bar_Pi_tau_2_2_upp_02__re}
&\norm{(\nb_i+\nb_j)\bar{\Pi}_{i,j,q_i,q_j}^\co  e^{iH\tau_2} O_{Z_0}e^{-iH\tau_1} \ket{\psi_{\bold{N}}}} 
\le  \norm{\nb_{\{i,j\}} \br{ \Pi_{i,>q_i} + \Pi_{j,>q_j} }   e^{iH\tau_2} O_{Z_0}e^{-iH\tau_1} \ket{\psi_{\bold{N}}}}  \notag \\
&\le  \norm{\nb_{\{i,j\}}\Pi_{i,>q_i} e^{iH\tau_2} O_{Z_0}e^{-iH\tau_1} \ket{\psi_{\bold{N}}}} 
+  \norm{\nb_{\{i,j\}} \Pi_{j,>q_j}    e^{iH\tau_2} O_{Z_0}e^{-iH\tau_1} \ket{\psi_{\bold{N}}}} \notag \\
&\le  \sqrt{e}\zeta_{Z_0} \brr{e\tilde{N}_i+e\tilde{N}_j + 2 (q_i-\tilde{N}_i)}  \exp \br{\frac{-q_i +e \tilde{N}_i}{4e\mu_1}}
+\sqrt{e}\zeta_{Z_0} \brr{e\tilde{N}_i+e\tilde{N}_j + 2 (q_j-\tilde{N}_j)}  \exp \br{\frac{-q_j +e \tilde{N}_j}{4e\mu_1}} \notag \\
&\le  \sqrt{e}\zeta_{Z_0} \brr{e\tilde{N}_i+e\tilde{N}_j + 2 (q_i-\tilde{N}_i)}  e^{-\tilde{q}}
+\sqrt{e}\zeta_{Z_0}  \brr{e\tilde{N}_i+e\tilde{N}_j + 2 (q_j-\tilde{N}_j)}  e^{-\tilde{q}} \notag \\
&\le 2\sqrt{e}\zeta_{Z_0}\brr{q_i+q_j + (1-1/e)(e\tilde{N}_i+e\tilde{N}_j)}  e^{-\tilde{q}} 
 \le 2\sqrt{e} (2-1/e) \cdot  \zeta_{Z_0}\br{q_i+q_j }  e^{-\tilde{q}}
  \le 6 \zeta_{Z_0}\br{q_i+q_j }  e^{-\tilde{q}} ,
\end{align}
where we use Corollary~\ref{corol:prob._upper_bound_time_evo} with $p=2$, $Y=\{i,j\}$ and the definition of $q_i = e\tilde{N}_i + 4e\mu_1 \tilde{q}$ for $i\in \tilde{L}$. Note that for $i\in \tilde{L}'=\tilde{L}[-1]$ we have $j\in \tilde{L}$ as long as $\dist_{i,j}=1$. 

By combining the inequalities~\eqref{effective Hamiltonian_accuracy_O_{Z_0}_sub_2_re_2} and \eqref{ineq_O_X_bar_Pi_tau_2_2_upp_02__re}, we obtain
\begin{align}
\label{effective Hamiltonian_accuracy_O_{Z_0}_sub_2_re_3}
\norm{\bar{\Pi}_{\tilde{L}',\bold{q}} H_0 \bar{\Pi}_{\tilde{L}',\bold{q}}^\co  e^{iH\tau_2} O_{Z_0}e^{-iH\tau_1} \ket{\psi_{\bold{N}}}}  
&\le 
12\zeta_{Z_0} \bar{J} e^{-\tilde{q}}  \sum_{i\in \tilde{L}'} \sum_{j: \dist_{i,j}=1}\br{q_i+q_j }   \le 
24\gamma \zeta_{Z_0} \bar{J} e^{-\tilde{q}}\bar{N}_{\tilde{L}}   ,
\end{align}
where for the last inequality we use 
\begin{align}
&\sum_{i\in \tilde{L}'} \sum_{j: \dist_{i,j}=1} q_i\le \sum_{i\in \tilde{L}'}  \gamma q_i \le  \gamma  \sum_{i\in \tilde{L}} q_i 
=\gamma\bar{N}_{\tilde{L}}, \notag \\
&\sum_{i\in \tilde{L}'} \sum_{j: \dist_{i,j}=1} q_j \le 
\sum_{j\in \tilde{L}'[1]} \sum_{i: \dist_{i,j}=1} q_j \le  \gamma  \sum_{j\in \tilde{L}} q_j
=\gamma\bar{N}_{\tilde{L}} .
\end{align}
Note that $ \tilde{L}'[1]=\tilde{L}$ from the definition of $\tilde{L}'=\tilde{L}[-1]$.

Thus, we have obtained the upper bounds for the norms of~\eqref{effective Hamiltonian_accuracy_O_{Z_0}_sub_1} and \eqref{effective Hamiltonian_accuracy_O_{Z_0}_sub_2} in the forms of~\eqref{projection_tile_L_2'_error} and \eqref{effective Hamiltonian_accuracy_O_{Z_0}_sub_2_re_3}, respectively.
By applying these bounds to the inequality~\eqref{error_time/evo_effectve_original_ineq_0}, we have the error of 
\begin{align}
&\norm{ e^{iH \tau} O_{Z_0} e^{-iH \tau} \ket{\psi_{\bold{N}}}-  e^{i \tilde{H}[\tilde{L}',\bold{q}] \tau} O_{Z_0} e^{-i \tilde{H}[\tilde{L}',\bold{q}] \tau} \ket{\psi_{\bold{N}}} } \notag \\
&\le \norm{\bar{\Pi}_{\tilde{L}',\bold{q}}^\co  e^{iH\tau} O_{Z_0}e^{-iH\tau} \ket{\psi_{\bold{N}}} } 
+ \norm{\bar{\Pi}_{\tilde{L}',\bold{q}}^\co O_{Z_0}e^{-iH\tau} \ket{\psi_{\bold{N}}} } 
+  
\zeta_{Z_0} \norm{\bar{\Pi}_{\tilde{L}',\bold{q}}^\co   \ket{\psi_{\bold{N}}} } 
+\zeta_{Z_0}  \norm{\bar{\Pi}_{\tilde{L}',\bold{q}}^\co e^{-iH\tau} \ket{\psi_{\bold{N}}} } \notag \\
&\quad +\int_0^\tau\br{ \norm{   \bar{\Pi}_{\tilde{L}',\bold{q}} H_0 \bar{\Pi}_{\tilde{L}',\bold{q}}^\co e^{iH (\tau - \tau_1)} O_{Z_0} e^{-iH\tau}\ket{\psi_{\bold{N}}} }  
+\zeta_{Z_0} \norm{   \bar{\Pi}_{\tilde{L}',\bold{q}} H_0 \bar{\Pi}_{\tilde{L}',\bold{q}}^\co e^{-iH  \tau_1}\ket{\psi_{\bold{N}}} } }d\tau_1 \notag \\
&\le 8 \zeta_{Z_0} |\tilde{L}'|  e^{-\tilde{q}}+ 
48\tau  \gamma \zeta_{Z_0} \bar{J} e^{-\tilde{q}}\bar{N}_{\tilde{L}} .
\end{align}
We thus prove the main inequality~\eqref{error_time/evo_effectve_original_ineq_lemma_acc} by using $|\tilde{L}'|  \le |L_0^\ast|$. 
This completes the proof.  $\square$

 {~}

\hrulefill{\bf [ End of Proof of Lemma~\ref{lem:effective Hamiltonian_accuracy_2}] }

{~}

 
We here consider the number of bosons in a region $Y\subseteq \tilde{L}$ under the projection $\bar{\Pi}_{\tilde{L}',\bold{q}}$.
We are now interested in the upper bound of
\begin{align}
\sum_{i\in Y}  q_i   \for Y \subseteq \tilde{L} .
\end{align} 
Note that by choosing $Y=\tilde{L}$ the above quantity reduces to $\bar{N}_{\tilde{L}}$ in Eq.~\eqref{def_bar_N_tilde_L_2}. 
Originally, we imposed the condition~\eqref{spectral_constraint_boson_number_psi_N} for the boson numbers in the region $L_0$, 
where the boson number is smaller than or equal to $\bar{q}$ in $\tilde{L}\setminus X^\co$, but larger than $\bar{q}$ in $X$. 
Here we can prove the following lemma:
\begin{lemma} \label{lem:boson_number_ball_change}
Under the choice of \eqref{choice_of_q_i_effective}, we obtain the following upper bound:
\begin{align}
\label{ineq:lem:boson_number_ball_change_def_bar_Q}
\sum_{i\in Y}  q_i  \le  e (4+\mu_0 \lambda_{3/4}) Q +  e\br{4\mu_1 \tilde{q} + 4\bar{q} + \mu_0 \lambda_{3/4} \bar{q} } |Y|.
\end{align} 
In particular, by choosing $Y=\tilde{L}$, we obtain 
\begin{align}
&\bar{N}_{\tilde{L}} := \sum_{i\in \tilde{L}} q_i  
\le  e (4+\mu_0 \lambda_{3/4}) Q +  e\br{4\mu_1 \tilde{q} + 4\bar{q} + \mu_0 \lambda_{3/4} \bar{q} } |\tilde{L}| .
\label{ineq:lem:boson_number_ball_change_0}
\end{align} 
\end{lemma}

%

\subsubsection{Proof of Lemma~\ref{lem:boson_number_ball_change}}

By using the definitions of $q_i =  e\tilde{N}_i + 4e\mu_1 \tilde{q}$ and $\tilde{N}_i:=4N_{i} +\mu_0   \sum_{j \in L_0^\ast}   e^{-3\dist_{i,j}/4}N_j$, we first consider 
\begin{align}
\label{sum_q_i_in_L_upp}
\sum_{i\in Y} q_i  &= \br{\sum_{i\in (Y \cap X)} +   \sum_{i\in (Y \setminus X)} } \br{ e\tilde{N}_i + 4e\mu_1 \tilde{q}} \notag \\
&=4e\mu_1 \tilde{q}|Y| + e \br{\sum_{i\in (Y \cap X)} +   \sum_{i\in (Y \setminus X)} } \br{4N_{i} +\mu_0   \sum_{j \in L_0^\ast}   e^{-3\dist_{i,j}/4}N_j} .
\end{align} 
First, we have 
\begin{align}
\label{sum_q_i_in_L_upp_1}
\sum_{i\in (Y \cap X)}N_i +   \sum_{i\in (Y \setminus X)} N_i \le Q+ \bar{q}|Y|  , 
\end{align} 
where we use $\sum_{i\in X} N_i \le Q$ and $N_i\le \bar{q}$ for $i\in (Y \setminus X)$ from Eqs.~\eqref{property_of_psi_vec_N} and \eqref{property_of_psi_vec_N_N_i_le_Q}, respectively. 
Note that $Y\subseteq \tilde{L} \subset L_0^\ast$. 
Second, we aim to estimate
\begin{align}
\br{\sum_{i\in (Y \cap X)} +   \sum_{i\in (Y \setminus X)} }   \sum_{j \in L_0^\ast}   e^{-3\dist_{i,j}/4}N_j  
&= \sum_{i\in Y}  \sum_{j \in L_0^\ast} e^{-3\dist_{i,j}/4}N_j   \notag \\
&\le \sum_{i\in Y}  \sum_{j \in X} e^{-3\dist_{i,j}/4}N_j  +  \sum_{i\in Y}  \sum_{j \in L_0^\ast \setminus X} e^{-3\dist_{i,j}/4} \bar{q}. 
\end{align} 
By using
\begin{align}
\max_{i\in \Lambda}\br{ \sum_{j \in \Lambda}e^{-3\dist_{i,j}/4} } \le \lambda_{3/4} 
\end{align} 
from the inequality~\eqref{oftern_used_inequality_lambda_c}, 
we reduce the above inequality to 
\begin{align}
\sum_{i\in Y}  \sum_{j \in X} e^{-3\dist_{i,j}/4}N_j  +  \sum_{i\in Y}  \sum_{j \in L_0^\ast \setminus X} e^{-3\dist_{i,j}/4} \bar{q}
\le  \lambda_{3/4} \sum_{j \in X} N_j  +  \lambda_{3/4} \sum_{i\in Y} \bar{q} \le \lambda_{3/4} (Q + \bar{q}|Y|) . 
\end{align} 
Hence, we obtain
\begin{align}
\label{sum_q_i_in_L_upp_2}
\br{\sum_{i\in (Y \cap X)} +   \sum_{i\in (Y \setminus X)} }   \sum_{j \in L_0^\ast}   e^{-3\dist_{i,j}/4}N_j    \le \lambda_{3/4} \br{Q + \bar{q}|Y|}. 
\end{align} 
By combining the inequalities~\eqref{sum_q_i_in_L_upp_1} and \eqref{sum_q_i_in_L_upp_2}, we have 
\begin{align}
\sum_{i\in Y} q_i  &= 4e\mu_1 \tilde{q}|Y| + e \br{\sum_{i\in (Y \cap X)} +   \sum_{i\in (Y \setminus X)} } \br{4N_{i} +\mu_0   \sum_{j \in L_0^\ast}   e^{-3\dist_{i,j}/4}N_j} \notag \\
&\le 4e\mu_1 \tilde{q}|Y| + e (4+\mu_0 \lambda_{3/4}) \br{Q + \bar{q}|Y|} ,
\end{align} 
which yields the desired inequality~\eqref{ineq:lem:boson_number_ball_change_def_bar_Q}.
This completes the proof. $\square$

 {~}

\hrulefill{\bf [ End of Proof of Lemma~\ref{lem:boson_number_ball_change}] }

{~}

\subsection{Proof of Proposition~\ref{Prop:Lieb-Robinson_short_time_L_X_bar_q}: Lieb-Robinson bound for effective Hamiltonian}
\label{sec::short_Lieb-Robinson bound for effective Hamiltonian}
In this section, we adopt an additional condition for $\ell_0$ as follows:
 \begin{align}
 \label{new_condition_ell_0_LR}
\ell_0 - (2\delta \ell + 2 +6k)\ge \frac{\ell_0}{2} \longrightarrow  \ell_0  \ge 4\delta \ell +12 k +4,
\end{align} 
where $k$ has characterized the maximum interaction length in $V$ [see Eq.~\eqref{def:Ham}].
This condition will be used in~\eqref{proof_of_corollay_boson_LR_effective_Ham_condition_use} for the proof of Corollary~\ref{corol:boson_LR_effective_Ham}.

In this section, we partially follow the discussion in Sec.~\ref{sec:Derivation of the Lieb-Robinson bound}.
We here calculate the norm of 
 \begin{align}
\label{ineq:prop:error_time_evolution_effective_Ham}
&\norm{ O_{Z_0}\br{\tilde{H}[\tilde{L}',\bold{q}] ,\tau} - U_{L_0}^\dagger O_{Z_0} U_{L_0}  } ,\quad \tilde{H}[\tilde{L}',\bold{q}]=\bar{\Pi}_{\tilde{L}',\bold{q}} H \bar{\Pi}_{\tilde{L}',\bold{q}},
\end{align} 
where $U_{L_0}$ is an appropriate unitary operator [see Eq.~\eqref{U_L_choice_subtheorem}].
As in Eq.~\eqref{decomposition_of_tine_evolution_bar_H}, we consider the decomposition of 
 \begin{align}
 \label{decomposition_of_tine_evolution_tilde_H}
e^{-i \tilde{H}[\tilde{L}',\bold{q}] \tau} = e^{-i \tilde{V}[\tilde{L}',\bold{q}] \tau}\ \mathcal{T} \exp \br{    
-i\int_0^{\tau} e^{i \tilde{V}[\tilde{L}',\bold{q}] x} \tilde{H}_0[\tilde{L}',\bold{q}]  e^{-i \tilde{V}[\tilde{L}',\bold{q}] x} dx
},
\end{align} 
where $\tilde{V}[\tilde{L}',\bold{q}]=\bar{\Pi}_{\tilde{L}',\bold{q}} V \bar{\Pi}_{\tilde{L}',\bold{q}}$. 

We here denote the time evolution of $e^{i \tilde{V}[\tilde{L}',\bold{q}] x} \tilde{H}_0[\tilde{L}',\bold{q}]  e^{-i \tilde{V}[\tilde{L}',\bold{q}] x}$ by
 \begin{align}
 \label{def:tilde_H_tau}
\tilde{H}_x := e^{i \tilde{V}[\tilde{L}',\bold{q}] x } \tilde{H}_0[\tilde{L}',\bold{q}]   e^{-i \tilde{V}[\tilde{L}',\bold{q}] x} 
&=\bar{\Pi}_{\tilde{L}',\bold{q}}\sum_{\langle i, j\rangle } J_{i,j} e^{i \tilde{V}[\tilde{L}',\bold{q}] x}  (b_i^\dagger b_j + {\rm h.c.}) e^{-i \tilde{V}[\tilde{L}',\bold{q}] x}  \bar{\Pi}_{\tilde{L}',\bold{q}}   \notag\\
&=:\sum_{\langle i, j\rangle } \tilde{h}_{\mathcal{B}_{i,j},x} , 
\quad \mathcal{B}_{i,j}= i[k-1] \cup j[k-1] , \quad \diam (\mathcal{B}_{i,j}) \le 2k .
\end{align} 
By using the notation of $\tilde{H}_x$, we rewrite the decomposition~\eqref{decomposition_of_tine_evolution_tilde_H} as 
 \begin{align}
 \label{decomposition_of_tine_evolution_tilde_H_0_2}
e^{-i \tilde{H}[\tilde{L}',\bold{q}] \tau} = e^{-i \tilde{V}[\tilde{L}',\bold{q}] \tau}\ \mathcal{T}e^{-\int_0^\tau \tilde{H}_xdx}.
\end{align} 
We note that $\tilde{V}[\tilde{L}',\bold{q}]$ is a commuting Hamiltonian, i.e.,  
 \begin{align}
 \label{tilde_V_tilde_L_vec_q_extend}
\tilde{V}[\tilde{L}',\bold{q}] = \sum_{Z: \diam(Z)\le k}  \tilde{v}_{Z}[\tilde{L}',\bold{q}] ,\quad 
\brr{  \tilde{v}_{Z}[\tilde{L}',\bold{q}] ,  \tilde{v}_{Z'}[\tilde{L}',\bold{q}] } =0.
\end{align} 
Also, we have obtain a similar inequality to~\eqref{upper_bound_norm_tilde_h_Z_i_j_tau_bar}
 \begin{align}
 \label{upper_bound_norm_tilde_h_Z_i_j_tau}
\norm{ \tilde{h}_{\mathcal{B}_{i,j},x}}  \le |J_{i,j}|  \cdot \norm{\bar{\Pi}_{\tilde{L}',\bold{q}} (b_i^\dagger b_j + {\rm h.c.}) \bar{\Pi}_{\tilde{L}',\bold{q}}} 
\le 2\bar{J} (q_i + q_j ) \quad (i,j \in \tilde{L}').
\end{align}

We next consider the same equation as Eq.~\eqref{O_X_0_bar_H_t_eq}:
 \begin{align}
 \label{decomp_O_Z_0_time_evo_eff}
O_{Z_0}\br{\tilde{H}[\tilde{L}',\bold{q}] ,\tau} = \tilde{O}_{Z_0[k]} (\tilde{H}_x, 0\to \tau), 
\end{align} 
where we use the notation~\eqref{notation_time_dependent_operator} and $ \tilde{O}_{Z_0[k]}$ ($\|\tilde{O}_{Z_0[k]} \|=\| O_{Z_0}\|=\zeta_{Z_0})$ is defined as 
\begin{align}
\label{def_tilde_O_L_0:1}
\tilde{O}_{Z_0[k]}:= O_{Z_0}\br{\tilde{V}[\tilde{L}',\bold{q}] ,\tau} = e^{i \tilde{V}_{Z_0[k]}[\tilde{L}',\bold{q}] \tau} O_{Z_0} e^{-i \tilde{V}_{Z_0[k]}[\tilde{L}',\bold{q}] \tau}. 
\end{align} 
Here, the operator $\tilde{V}_{Z_0[k]}[\tilde{L}',\bold{q}]$ picks up all the interaction terms $\tilde{v}_{\tilde{Z}}[\tilde{L}',\bold{q}]$ in Eq.~\eqref{tilde_V_tilde_L_vec_q_extend} such that $\tilde{Z}\subset Z_0[k]$ [see also Eqs.~\eqref{def:Ham} and \eqref{def:Ham_subset}].

\begin{figure}[tt]
\centering
\includegraphics[clip, scale=0.35]{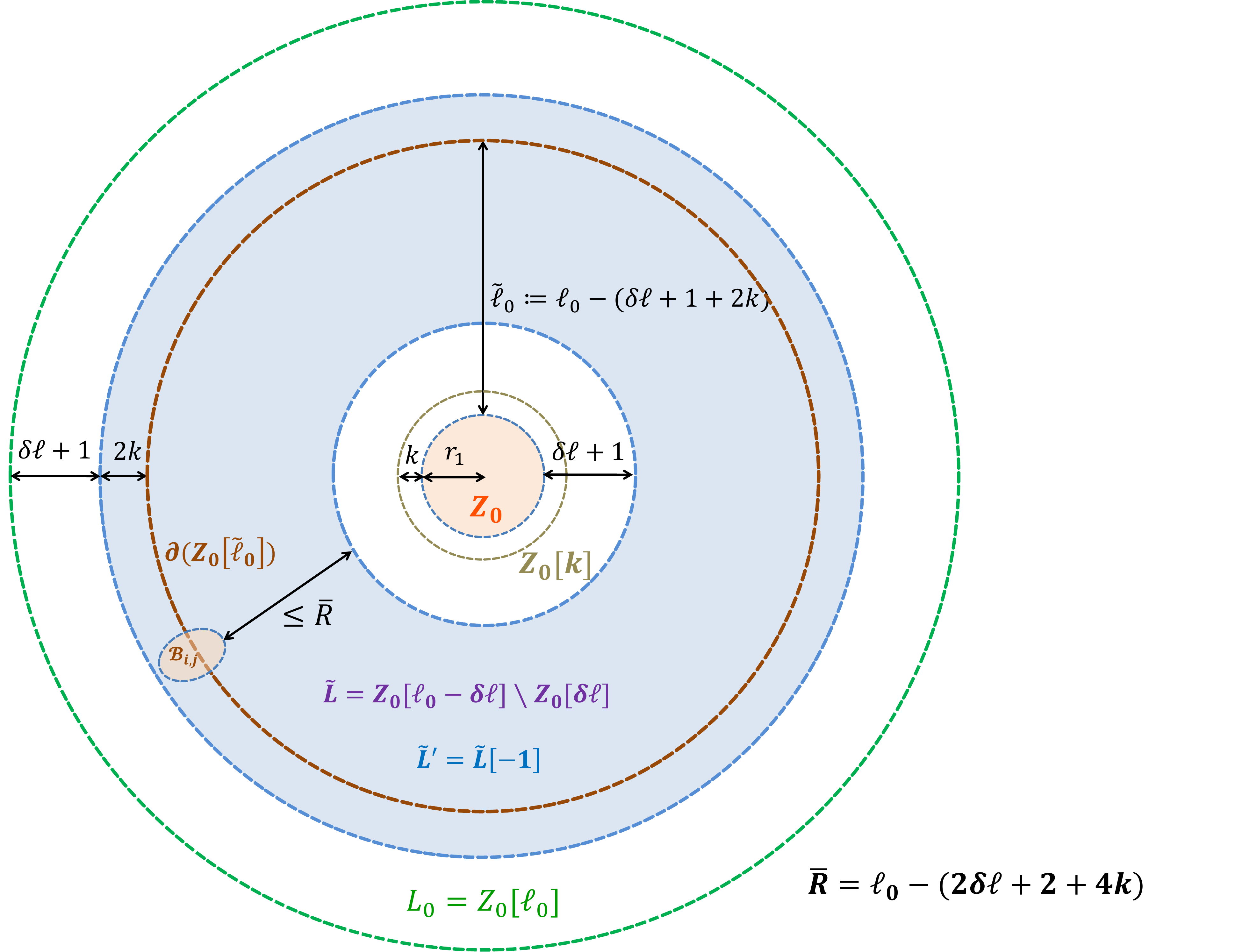}
\caption{Schematic pictures of the regions $L_0$, $\tilde{L}'$ and $Z_0[\tilde{\ell}_0]$.
The time evolution $O_{Z_0}\br{\tilde{H}[\tilde{L}',\bold{q}] ,\tau} $ is decomposed as in Eq.~\eqref{decomp_O_Z_0_time_evo_eff}, where the contribution from $\tilde{V}[\tilde{L}',\bold{q}]$ is included in $\tilde{O}_{Z_0[k]}$. 
To approximate the time evolution of $\tilde{O}_{Z_0[k]} (\tilde{H}_x, 0\to \tau)$, we use Lemma~\ref{lem:Lieb--Robinson_start} with the choice of the approximating region as 
$Z_0[\tilde{\ell}_0]$ (inside the region by brown dots). 
Then, from the inequality~\eqref{norm_calculation_LR_h_Z_tilde}, we have to estimate the time evolution of the boundary-interacting terms $\tilde{h}_{\mathcal{B}_{i,j},x}(\tilde{H}_{Z_0[\tilde{\ell}_0],\tau_1}, x\to 0)$. Here, from $\diam(\mathcal{B}_{i,j}) \le 2k$, we can ensure $\mathcal{B}_{i,j} \subset \tilde{L}'$ for 
$\mathcal{B}_{i,j} \in \mathcal{S}_{Z_0[\tilde{\ell}_0]}$. 
The length $\bar{R}$ is utilized in Lemma~\ref{lem:Lieb--Robinson_average_boson_prepare}, which is now given by the minimum distance between $\mathcal{B}_{i,j}$ and $Z_0[\delta \ell+1]$; that is, it is bounded from above by $\tilde{\ell}_0-2k-(\delta \ell+1)=\ell_0-(2\delta \ell+4k+2)$ [see Eq.~\eqref{replace_ment_LR3}].
}
\label{supp_fig_Region_setting_LR.pdf}
\end{figure}

In the following, we approximate $\tilde{O}_{Z_0[k]}  (\tilde{H}_x, 0\to \tau)$ by using the subset Hamiltonian $\tilde{H}_{L,x}$:
 \begin{align}
 \label{approximate_L_0:1_H_L_tau}
\tilde{O}_{Z_0[k]}  (\tilde{H}_x, 0\to \tau) \approx \tilde{O}_{Z_0[k]}  (\tilde{H}_{L_0,x} , 0\to \tau) , \quad 
\tilde{H}_{L,x}&= \sum_{\langle i,j\rangle : \mathcal{B}_{i,j}\subset L} \tilde{h}_{\mathcal{B}_{i,j},x} , 
\end{align} 
where we choose the subset $L$ appropriately afterward.
To estimate the approximation error for~\eqref{approximate_L_0:1_H_L_tau}, we utilize Lemma~\ref{lem:Lieb--Robinson_start}, 
in which we let $O_X \to \tilde{O}_{Z_0[k]} $ as in Eq.~\eqref{def_tilde_O_L_0:1} and choose the subset $L$ as $Z_0[\tilde{\ell}_0]$ (see also Fig.~\ref{supp_fig_Region_setting_LR.pdf}) with
\begin{align}
\label{def_tilde_ell_0_proof}
\tilde{\ell}_0 = \ell_0 - (\delta \ell +2 k  +1).
\end{align}

From the inequality~\eqref{lem:ineq:Lieb--Robinson_start} in Lemma~\ref{lem:Lieb--Robinson_start}, we obtain the error for the approximation~\eqref{approximate_L_0:1_H_L_tau} in the following form:
 \begin{align}
 \label{norm_calculation_LR_h_Z_tilde}
\norm{ \tilde{O}_{Z_0[k]}  (\tilde{H}_x, 0\to \tau) - \tilde{O}_{Z_0[k]}  (\tilde{H}_{Z_0[\tilde{\ell}_0],x} , 0\to \tau) }
\le \sum_{\langle i,j\rangle: \mathcal{B}_{i,j} \in \mathcal{S}_{Z_0[\tilde{\ell}_0]}} \int_0^{\tau} \left\| [ \tilde{h}_{\mathcal{B}_{i,j},x}(\tilde{H}_{Z_0[\tilde{\ell}_0],\tau_1}, x\to 0), \tilde{O}_{Z_0[k]} ] \right\| dx ,
\end{align} 
where $\mathcal{S}_{Z_0[\tilde{\ell}_0]}$ has been defined in Eq.~\eqref{set_mathcal_S_L}.
We note that as long as $\mathcal{B}_{i,j} \in \mathcal{S}_{Z_0[\tilde{\ell}_0]}$ we can ensure $\mathcal{B}_{i,j}\in \tilde{L}'$ because of $\diam(\mathcal{B}_{i,j})\le 2k$.
Then, we have to upper-bound the norm of the commutator like
 \begin{align}
\left\| [ \tilde{h}_{\mathcal{B}_{i,j},x}(\tilde{H}_{Z_0[\tilde{\ell}_0],\tau_1}, x\to 0), \tilde{O}_{Z_0[k]} ] \right\| \quad 
\br{\mathcal{B}_{i,j}\in \mathcal{S}_{Z_0[\tilde{\ell}_0]}} .
 \label{norm_calculation_LR}
\end{align} 
The difficulty here is that the norms of the local interaction terms~$\{\tilde{h}_{\mathcal{B}_{i,j},\tau}\}_{i,j}$ are not uniformly bounded; 
that is, the upper bound~\eqref{upper_bound_norm_tilde_h_Z_i_j_tau} implies that the norm is small on average but may be large for a particular choice of $i,j$.

 Before showing the Lieb-Robinson bound for~\eqref{norm_calculation_LR}, 
we prove the following convenient lemma:
\begin{lemma} \label{lem:Lieb--Robinson_average_boson_prepare}
Let $H$ be an arbitrary Hamiltonian such that 
\begin{align}
&H = \sum_{Z: \diam(Z)\le \bar{k}} h_Z , \quad \sum_{Z: Z \ni i} {\rm sign}( \norm{ h_{Z,\tau}} )  \le \bar{\kappa} ,\label{Ham_cond_diam_l}
\end{align}
with ${\rm sign}(x) = 1$ ($x>0$) and ${\rm sign}(0) = 0$.
Also, for a fixed region $X[\bar{R}]$, we assume
\begin{align}
\label{Ham_cond_local_norm}
\sum_{j=1}^m \norm{h_{Z_j}} \le \bar{g}_0 + \bar{g}_1 m  \quad {\rm s.t.} \quad 
Z_j \cap Z_{j'} = \emptyset,    \quad Z_j \subset X[\bar{R}]  \for \forall j,j'\in [1,m], 
\end{align}
where $\{Z_j\}_{j=1}^m$ are arbitrarily chosen.  
Then, the Hamiltonian $H$ satisfies the Lieb-Robinson bound in the form of
\begin{align}
\label{ineq:lem:Lieb--Robinson_average_boson_prepare}
\norm{ [O_X(H,t),O_Y] } \le  \frac{2 |\partial X|}{\gamma \bar{k}^D}\norm{O_X} \cdot \norm{O_Y} 
\brr{\frac{4e \bar{\kappa}  \gamma \bar{k}^{D+1} t }{R} \br{\frac{\bar{k}\bar{g}_0}{R}+ \bar{g}_1 } }^{R/\bar{k}} ,\quad 
R=\min (\dist_{X,Y}, \bar{R}-\bar{k}). 
\end{align}  
\end{lemma}

{\bf Remark.} 
We here consider a time-independent Hamiltonian, but the same proof technique is applied to arbitrary time-dependent Hamiltonian as long as $H(\tau)$ ($0\le \tau \le t$) satisfies the conditions~\eqref{Ham_cond_diam_l} and \eqref{Ham_cond_local_norm}\footnote{
For time-dependent Hamiltonian, we need to assume
\begin{align}
\sum_{j=1}^m \norm{h_{Z_j, \tau_j}} \le \bar{g}_0 + \bar{g}_1 m  
\end{align}
for arbitrary choices of $\{\tau_j\}_{j=1}^m$.}. 

The condition~\eqref{Ham_cond_local_norm} roughly implies that the average energy of $m$ interaction terms $\{h_{Z_j}\}_{j=1}^m$ is given by 
$\bar{g}_0/m + \bar{g}_1$. If we consider $\orderof{R}$ interaction terms (i.e., $m=\orderof{R}$), 
the average energy is $\orderof{R^{-1}}\bar{g}_0 + \bar{g}_1$. Thus, by denoting 
\begin{align}
\frac{\bar{g}_0}{(R/\bar{k})}+ \bar{g}_1 =\epsilon_{\rm ave} ,
\end{align}  
we can rewrite the inequality~\eqref{ineq:lem:Lieb--Robinson_average_boson_prepare} by
\begin{align}
\norm{ [O_X(H,t),O_Y] } \le  \frac{2 |\partial X|}{\gamma \bar{k}^D}\norm{O_X} \cdot \norm{O_Y} 
\br{\frac{C \epsilon_{\rm ave} t }{R} }^{R/\bar{k}} ,
\end{align}  
where $C$ is an $\orderof{1}$ constant. 
Therefore, this lemma refines the standard Lieb-Robinson bound so that the Lieb-Robinson velocity is proportional to the \textit{average local energy} instead of the upper bound on the local energy\footnote{On the point that the Lieb-Robinson velocity is generally proportional to the local energy, please see Ref.~\cite[Inequality. (2.4)] {Phd_kuwahara} for example.}.

\subsubsection{Proof of Lemma~\ref{lem:Lieb--Robinson_average_boson_prepare}}

For the proof, we refine the recursion method in the original proof~\cite{ref:Hastings2006-ExpDec,Nachtergaele2006}.
The key idea is to utilize the following slightly improved inequality:
\begin{align}
\label{start_LR_proof_refine}
\norm { [O_X(H,t) , O_Y]  } \le   2 \norm{O_X} \sum_{Z: Z\in \mathcal{S}(X) } \int_0^t  \norm{ [ h_Z (H_{X^\co}, x)  ,O_Y]}  dx 
\for [O_X,O_Y]=0 ,
\end{align}
where $\mathcal{S}(X)$ characterizes the set of subsets that overlap with the surface of $X$ as has been defined in Eq.~\eqref{set_mathcal_S_L}.
The derivation is given as follows. First of all, we adopt the decomposition of
\begin{align}
e^{-iHt} = e^{-i H_X t}  e^{-i H_{X^\co}t} \mathcal{T} e^{-i \int_0^t H_{\partial X} (H_X+H_{X^\co}, x) dx } ,
\end{align}
where we define $H_{\partial X}:=H-H_{X^\co}-H_X= \sum_{Z: Z\in \mathcal{S}(X) } h_Z$.
The above equation reduces the norm $\norm{ [O_X(H,t),O_Y] } $ to 
\begin{align}
\norm{ [O_X(H,t),O_Y] } 
&=
\norm{ \br{ \mathcal{T} e^{-i \int_0^t H_{\partial X} (H_X+H_{X^\co}, x) dx } }^\dagger  e^{iH_{X^\co}t} O_X(H_X,t) e^{-iH_{X^\co}t} \br{ \mathcal{T} e^{-i \int_0^t H_{\partial X} (H_X+H_{X^\co}, x) dx } } ,O_Y] } \notag \\
&\le 2 \norm{O_X}   \int_0^t  \norm{ [ H_{\partial X} (H_X+H_{X^\co}, x)  ,O_Y]}  dx 
=2 \norm{O_X}   \int_0^t  \norm{ [ H_{\partial X} (H_{X^\co}, x)  ,O_Y]}  dx ,
\end{align}
which yields the inequality~\eqref{start_LR_proof_refine}, where the first equation is derived from 
$[e^{iH_{X^\co}t} O_X(H_X,t) e^{-iH_{X^\co}t}, O_Y]= [O_X(H_X,t), O_Y]=0$, and we use $\norm{ [ H_{\partial X} (H_X+H_{X^\co}, x)  ,O_Y]}=
\norm{ [ H_{\partial X} (H_{X^\co} ,x) ,O_Y(H_X, x)]}=\norm{ [ H_{\partial X} (H_{X^\co} ,x) ,O_Y]}$ in the second equation.

\begin{figure}[tt]
\centering
\includegraphics[clip, scale=0.45]{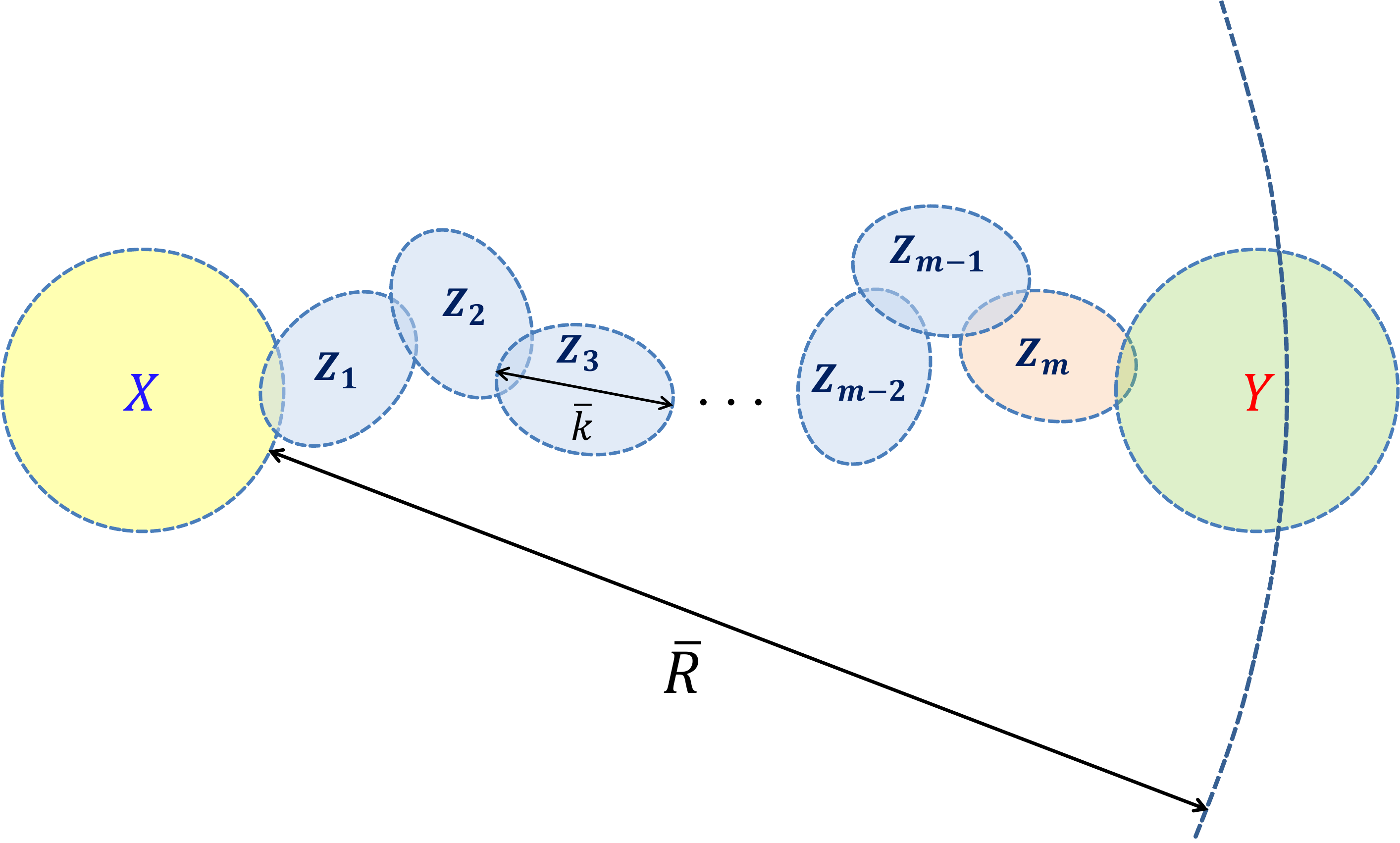}
\caption{Schematic picture of the subsets $\{Z_j\}_{j=1}^m$ that appear in the inequality~\eqref{start_LR_proof_refine_mth}.
In the first step, from the inequality~\eqref{start_LR_proof_refine}, the subset $Z_1$ should overlap with $X$.
In the second step, from the inequality~\eqref{start_LR_proof_refine_2}, the subset $Z_2$ has an overlap with $Z_1$ but no overlap with $X$.
In the third step, from the inequality~\eqref{start_LR_proof_refine_3}, the subset $Z_3$ has an overlap with $Z_2$, but no overlaps with $X$ and $Z_1$. By repeating the step, the subset $Z_{j_0}$ has overlaps with $Z_{j_0-1}$ and $Z_{j_0+1}$, but no overlaps with the other subsets $\{Z_{j}\}_{j\neq j_0-1,j_0,j_0+1}$. This point is critical in efficiently upper-bounding the norm $\norm{h_{Z_1}} \cdot \norm{h_{Z_2}} \cdots \norm{h_{Z_m}}$ by using the original condition~\eqref{Ham_cond_local_norm}. 
}
\label{supp_fig_LR_proof_refine.pdf}
\end{figure}

By using the inequality~\eqref{start_LR_proof_refine}, as long as $[h_{Z_1}, O_Y]=0$, we also obtain 
\begin{align}
\label{start_LR_proof_refine_2}
 \norm{ [ h_{Z_1} (H_{X^\co}, x_1)  ,O_Y]} \le 2 \norm{h_{Z_1}}  \sum_{\substack{ Z_2: Z_2\cap Z_1 \neq \emptyset \\  Z_2\cap X = \emptyset}}
  \int_0^{x_1}  \norm{ [ h_{Z_2} (H_{(X\cup Z_1)^\co}, x_2)  ,O_Y]}  dx_2 ,
\end{align}
where in the summation for $Z_2$ we pick up interactions in $H_{X^\co}$, which yields the constraint of $Z_2\cap X = \emptyset$. 
Furthermore, as long as $[h_{Z_2}, O_Y]=0$, we have 
\begin{align}
\label{start_LR_proof_refine_3}
\norm{ [ h_{Z_2} (H_{(X\cup Z_1)^\co}, x_2)  ,O_Y]} \le 2 \norm{h_{Z_2}}  \sum_{\substack{ Z_3: Z_3\cap Z_2 \neq \emptyset  \\  Z_3\cap X = \emptyset , Z_3\cap Z_1 = \emptyset} } \int_0^{x_2}  \norm{ [ h_{Z_3} (H_{(X\cup Z_1\cup Z_2)^\co}, x_3)  ,O_Y]}  dx_3  .
\end{align}
By iteratively applying the inequality~\eqref{start_LR_proof_refine}, we obtain  (see Fig.~\ref{supp_fig_LR_proof_refine.pdf})
\begin{align}
\label{start_LR_proof_refine_mth}
&\norm{ [O_X(H,t),O_Y] } \notag \\
&\le
2^m \norm{O_X} \sum_{Z_1: Z_1\in \mathcal{S}(X)}  \norm{h_{Z_1}} \sum_{\substack{ Z_2: Z_2\cap Z_1 \neq \emptyset \\  Z_2\cap X = \emptyset}} \norm{h_{Z_2}}
\sum_{\substack{ Z_3: Z_3\cap Z_2 \neq \emptyset  \\  Z_3\cap X = \emptyset , Z_3\cap Z_1 = \emptyset} } \norm{h_{Z_3}}
\cdots \sum_{\substack{ Z_m: Z_m\cap Z_{m-1} \neq \emptyset \\  Z_m \cap X = \emptyset, Z_m \cap Z_1 = \emptyset , \ldots,  Z_m \cap Z_{m-2} = \emptyset }} 
\notag \\
&\times \int_0^t   \int_0^{x_1}  \cdots \int_0^{x_m}  
\norm{ [ h_{Z_m} (H_{(X\cup Z_1\cup Z_2 \cup \cdots \cup Z_{m-1})^\co}, x_m)  ,O_Y] }  dx_1dx_2 \cdots dx_m , \notag \\
&\le \frac{2(2t)^m}{m!}\norm{O_X} \cdot \norm{O_Y} 
\sum_{Z_1: Z_1\in \mathcal{S}(X)}  \norm{h_{Z_1}} \sum_{\substack{ Z_2: Z_2\cap Z_1 \neq \emptyset \\  Z_2\cap X = \emptyset}} \norm{h_{Z_2}}
\cdots \sum_{\substack{ Z_m: Z_m\cap Z_{m-1} \neq \emptyset \\  Z_m \cap X = \emptyset, Z_m \cap Z_1 = \emptyset , \ldots,  Z_m \cap Z_{m-2} = \emptyset }}  \norm{h_{Z_m}} ,
\end{align} 
where we assume $[h_{Z_{m-1}},O_Y]=0$ and $Z_m\in Z[\bar{R}]$.  
Because of the constraints $[h_{Z_{m-1}},O_Y]=0$ and $Z_m\in Z[\bar{R}]$, we choose the parameter $m$ as 
\begin{align}
m =  \min\br{ \floorf{\frac{\dist_{X,Y}}{\bar{k}}}+1 , \floorf{\frac{\bar{R}}{\bar{k}}} } 
\ge  \frac{R}{\bar{k}} 
\label{condition_for_m_choosing}
\end{align}
with $R:=\min(\dist_{X,Y},\bar{R}-\bar{k})$ because of $\diam(Z_j) \le \bar{k}$ from the condition in Eq.~\eqref{Ham_cond_diam_l}.

We then aim to upper-bound the norm of 
\begin{align}
\label{product_h_Z_1...h_Zm}
\norm{h_{Z_1}} \cdot \norm{h_{Z_2}} \cdots \norm{h_{Z_m}} \le \br{\frac{\sum_{j=1}^m \norm{h_{Z_j}} }{m}}^m 
\end{align}
under the condition that 
$$Z_j\cap Z_{j-1} \neq \emptyset, \quad Z_j \cap X = \emptyset, \quad Z_j \cap Z_1 = \emptyset, \quad \ldots , \quad Z_j \cap Z_{j-2} = \emptyset  \for \forall  j \in [m],$$ 
where we use the inequality of arithmetic and geometric means. 
Note that from the above condition we have $Z_{j}\cap Z_{j'} =\emptyset$ as long as $|j-j'|\neq 1$.  
Therefore, 
by using the condition~\eqref{Ham_cond_local_norm}, we have 
\begin{align}
\label{product_h_Z_1...h_Z_upper_bound_1}
\sum_{j=1}^m \norm{h_{Z_j}} 
&=\sum_{j : {\rm odd}} \norm{h_{Z_j} } + \sum_{j : {\rm even}} \norm{h_{Z_j}} \notag \\
&\le \bar{g}_0 +  \sum_{j : {\rm odd}}\bar{g}_1+
\bar{g}_0 + \sum_{j : {\rm even}} \bar{g}_1\notag \\
&\le 2\bar{g}_0 +\bar{g}_1m .
\end{align}
By combining the inequalities~\eqref{product_h_Z_1...h_Zm} and \eqref{product_h_Z_1...h_Z_upper_bound_1}, we have 
\begin{align}
\label{product_h_Z_1...h_Zm_fin}
\norm{h_{Z_1}} \cdot \norm{h_{Z_2}} \cdots \norm{h_{Z_m}} \le \br{\frac{2\bar{g}_0}{m}+ \bar{g}_1 }^m .
\end{align}

Finally, for an arbitrary subset $Z \subset \Lambda$ with $\diam(Z)\le \bar{k}$, by using the condition~\eqref{Ham_cond_diam_l}, we have 
\begin{align}
\label{product_h_Z_1...h_Z_upper_bound__2}
\sum_{Z': Z'\cap Z \neq \emptyset}  {\rm sign} \br{\norm{h_{Z'}}}
\le \sum_{i\in Z}\sum_{Z': Z'\ni i}  {\rm sign} \br{\norm{h_{Z'}}} \le \bar{\kappa}  |Z| \le \bar{\kappa}  \gamma \bar{k}^D ,
\end{align}
where we use $|Z| \le  \gamma \bar{k}^D$ from $\diam(Z)\le \bar{k}$. 
In the same way, we obtain
\begin{align}
\label{product_h_Z_1...h_Z_upper_bound__2}
\sum_{Z: Z \in \mathcal{S}(X)}  {\rm sign} \br{\norm{h_{Z}}}
\le \sum_{i\in \partial X}\sum_{Z: Z\ni i}  {\rm sign} \br{\norm{h_{Z}}} \le \bar{\kappa}  |\partial X| .
\end{align}
Thus, by using the inequalities~\eqref{product_h_Z_1...h_Z_upper_bound_1} and \eqref{product_h_Z_1...h_Z_upper_bound__2}, we reduce the inequality~\eqref{start_LR_proof_refine_mth} to 
\begin{align}
\label{start_LR_proof_refine_mth__2}
\norm{ [O_X(H,t),O_Y] } 
&\le \frac{2(2t)^m}{m!} \norm{O_X} \cdot \norm{O_Y} \cdot \bar{\kappa}  |\partial X| \cdot  \br{ \bar{\kappa}  \gamma \bar{k}^D}^{m-1}
\br{\frac{2\bar{g}_0}{m}+ \bar{g}_1 }^m   \notag \\
&=  \frac{2 |\partial X|}{m! \cdot \gamma \bar{k}^D}\norm{O_X} \cdot \norm{O_Y} 
\brr{2 \bar{\kappa}  \gamma \bar{k}^D t  \br{\frac{2\bar{g}_0}{m}+ \bar{g}_1 } }^m\notag \\
&\le \frac{2 |\partial X|}{\gamma \bar{k}^D}\norm{O_X} \cdot \norm{O_Y} 
\brr{\frac{4e \bar{\kappa}  \gamma \bar{k}^D t }{m} \br{\frac{\bar{g}_0}{m}+ \bar{g}_1 } }^m,
\end{align} 
where we use $m!\le (m/e)^m$ in the last inequality.
Under the choice of~\eqref{condition_for_m_choosing}, we reduce the inequality~\eqref{start_LR_proof_refine_mth__2} to \eqref{ineq:lem:Lieb--Robinson_average_boson_prepare}. 
This completes the proof. $\square$

 {~}

\hrulefill{\bf [ End of Proof of Lemma~\ref{lem:Lieb--Robinson_average_boson_prepare}] }

{~}

The final task is to obtain all the parameters in Lemma~\ref{lem:Lieb--Robinson_average_boson_prepare} to apply them to 
our purpose [i.e., upper-bounding the norm~\eqref{norm_calculation_LR}].
First of all, from Eq.~\eqref{def:tilde_H_tau}, the interaction length of the Hamiltonian $\tilde{H}_\tau$ at most $(2k)$, i.e., 
\begin{align}
\label{replace_ment_LR1}
\bar{k} \to 2k.
\end{align} 
Also, the constant $\bar{\kappa}$ characterizes how many interaction terms $h_Z$ such that $Z\ni i$ exists in the Hamiltonian. 
For the case of $\tilde{H}_{\tau}$ in Eq.~\eqref{approximate_L_0:1_H_L_tau}, the $\bar{\kappa}$ is given by
\begin{align}
\bar{\kappa}  =\max_{i_0\in \Lambda}  \# \brrr{\langle i,j \rangle | \mathcal{B}_{i,j} \ni i_0} 
\le \max_{i_0\in \Lambda} \br{ \gamma |i_0[k] | }\le \gamma^2 k^D ,
\end{align}
because $\mathcal{B}_{i,j} = i[k-1] \cup j[k-1] \ni i_0$ ($\dist_{i,j}=1$) implies that the sites $i$ and $j$ should be included in $i_0[k]$.  
Therefore, we have 
\begin{align}
\label{replace_ment_LR2}
\bar{\kappa} \to \gamma^2 k^D. 
\end{align}

The remaining parameters are $\bar{g}_0$ and $\bar{g}_1$ in the condition~\eqref{Ham_cond_local_norm}. 
For this purpose, we need to calculate the upper bound of 
\begin{align}
\label{upper_bound_norm_Ham_many}
\sum_{s=1}^m \norm{h_{\mathcal{B}_{i_s,j_s}}} \le 2\bar{J}\sum_{s=1}^m (q_{i_s} + q_{j_s}) \quad (\mathcal{B}_{i_s,j_s}\subseteq \tilde{L}',\ 
\mathcal{B}_{i_s,j_s} \cap\mathcal{B}_{i_{s'},j_{s'}}=\emptyset),
\end{align}
where the inequality is derived from~\eqref{upper_bound_norm_tilde_h_Z_i_j_tau}.
By using Lemma~\ref{lem:boson_number_ball_change} with $Y=\{i_s,j_s\}_{s=1}^m$ ($|Y|=2m$), we upper-bound 
\begin{align}
\label{upper_bound_norm_Ham_many__2}
2\bar{J}\sum_{s=1}^m (q_{i_s} + q_{j_s}) 
\le 2 e \bar{J} (4+\mu_0 \lambda_{3/4}) Q  +  4e\bar{J} \br{4\mu_1 \tilde{q} + 4\bar{q} + \mu_0 \lambda_{3/4} \bar{q}} m,
\end{align}
which gives 
\begin{align}
\label{replace_ment_LR4}
\bar{g}_0  \to 2e \bar{J} (4+\mu_0 \lambda_{3/4})  Q, \quad   \bar{g}_1 \to 4e\bar{J} \br{4\mu_1 \tilde{q} + 4\bar{q} + \mu_0 \lambda_{3/4} \bar{q}} .
\end{align} 
Finally, the length $\bar{R}$ is defined by the region such that the condition~\eqref{Ham_cond_local_norm} is satisfied.
Hence, in our setup, we have to choose the length $\bar{R}$ so that we can ensure the condition of
\begin{align}
\mathcal{B}_{i,j}[\bar{R}] \subseteq \tilde{L}'  \for \forall \mathcal{B}_{i,j} \in \mathcal{S}_{Z_0[\tilde{\ell}_0]}  .
\end{align} 
Because of $\diam(\mathcal{B}_{i,j}) < 2k$ and $\tilde{\ell}_0=\ell_0 - (\delta \ell +2 k  +1)$ as in Eq.~\eqref{def_tilde_ell_0_proof},
we have to set $\bar{R}$ as $\ell_0 - (2\delta \ell +2 + 4k)$ (see Fig.~\ref{supp_fig_Region_setting_LR.pdf}), i.e., 
\begin{align}
\label{replace_ment_LR3}
\bar{R}\to \ell_0 - (2\delta \ell + 2 +4k). 
\end{align}


Based on the above parameter correspondences, we can prove the following corollary:
\begin{corol} \label{corol:boson_LR_effective_Ham}
For the operator of $\tilde{h}_{\mathcal{B}_{i,j},x}$ ($\mathcal{B}_{i,j} \in \mathcal{S}_{Z_0[\tilde{\ell}_0]}$), we obtain the Lieb-Robinson bound as follows:
 \begin{align}
 \label{ineq:corol:boson_LR_effective_Ham}
\left\| \brr{ \tilde{h}_{\mathcal{B}_{i,j},x}(\tilde{H}_{Z_0[\tilde{\ell}_0],x}, x \to 0), \tilde{O}_{Z_0[k]} } \right\| 
\le  \frac{2\zeta_{Z_0} |\mathcal{B}_{i,j}|}{\gamma (2k)^D}\norm{\tilde{h}_{\mathcal{B}_{i,j},x}} 
\brr{\frac{\tilde{c}_1 x}{\ell_0} 
\br{\frac{\tilde{c}_2 Q}{\ell_0}+ \tilde{c}_{\tilde{q},\bar{q}}} }^{\ell_0/(4k)} ,
\end{align} 
where $\tilde{c}_1$,  $\tilde{c}_2$ and  $\tilde{c}_3$ are defined as 
\begin{align}
\label{def_tilde_c_1_2_3}
\tilde{c}_1:= 2^{D+6} e^2 \gamma^3 k^{2D+1} \bar{J} ,
\quad \tilde{c}_2:=2k  (4+\mu_0 \lambda_{3/4}) ,\quad 
\tilde{c}_{\tilde{q},\bar{q}}:=4\mu_1 \tilde{q} + 4\bar{q} + \mu_0 \lambda_{3/4} \bar{q} .
\end{align} 
\end{corol}

\subsubsection{Proof of Corollary~\ref{corol:boson_LR_effective_Ham}}

We first notice that the parameter $R$ in the inequality~\eqref{ineq:lem:Lieb--Robinson_average_boson_prepare} satisfies 
$R=\min (\dist_{\mathcal{B}_{i,j},Z_0[k]}, \bar{R}-2k)=\bar{R}-2k$ because of $\mathcal{B}_{i,j} \in \mathcal{S}_{Z_0[\tilde{\ell}_0]}$, $\diam(\mathcal{B}_{i,j})\le 2k$ and $\bar{R}= \ell_0 - (2\delta \ell + 2 +4k)$ [see Eq.~\eqref{replace_ment_LR3}].
We then obtain 
 \begin{align}
 \label{proof_of_corollay_boson_LR_effective_Ham_condition_use}
\bar{R}-2k= \ell_0 - (2\delta \ell + 2 +6k) \ge \frac{\ell_0}{2} ,
\end{align} 
where the inequality results from the condition~\eqref{new_condition_ell_0_LR}. 
Hence, under the choice of~\eqref{replace_ment_LR3}, we replace $R=\min (\dist_{X,Y}, \bar{R}-\bar{k})$ in \eqref{ineq:lem:Lieb--Robinson_average_boson_prepare} as 
 \begin{align}
R \to  \frac{\ell_0}{2} .
\end{align} 
Then, by applying the replacements~\eqref{replace_ment_LR1},~\eqref{replace_ment_LR2},~\eqref{replace_ment_LR3}, and~\eqref{replace_ment_LR4} to 
the inequality~\eqref{ineq:lem:Lieb--Robinson_average_boson_prepare}, we have the inequality~\eqref{ineq:corol:boson_LR_effective_Ham} as follows:
\begin{align}
&\norm{ \brr{\tilde{h}_{\mathcal{B}_{i,j},x} (\tilde{H}_{Z_0[\tilde{\ell}_0],x}, x \to 0), \tilde{O}_{Z_0[k]} } } \notag\\
&\le \frac{2 |\mathcal{B}_{i,j}|}{\gamma (2k)^D}\norm{\tilde{h}_{\mathcal{B}_{i,j},x}} \cdot \norm{\tilde{O}_{Z_0[k]} } 
\brr{\frac{4e \br{\gamma^2 k^D}  \gamma  (2k)^{D+1} x}{(\ell_0/2)} 
\br{\frac{(2k\cdot 2e \bar{J} (4+\mu_0 \lambda_{3/4})  Q}{(\ell_0/2)}+ 4e\bar{J} \br{4\mu_1 \tilde{q} + 4\bar{q} + \mu_0 \lambda_{3/4} \bar{q}} } }^{(\ell_0/2)/ (2k)}  \notag \\
&= \frac{2\zeta_{Z_0} |\mathcal{B}_{i,j}|}{\gamma (2k)^D}\norm{\tilde{h}_{\mathcal{B}_{i,j},x}} 
\brr{\frac{2^{D+6} e^2 \gamma^3 k^{2D+1} \bar{J} x}{\ell_0} 
\br{\frac{2k  (4+\mu_0 \lambda_{3/4})  Q}{\ell_0}+4\mu_1 \tilde{q} + 4\bar{q} + \mu_0 \lambda_{3/4} \bar{q}  } }^{\ell_0/(4k)} ,
\end{align} 
where we use $\norm{\tilde{O}_{Z_0[k]} } =\zeta_{Z_0}$ from Eq.~\eqref{def_tilde_O_L_0:1}. 
Then, by using the definitions in~\eqref{def_tilde_c_1_2_3}, we reduce the above inequality to the main inequality~\eqref{ineq:corol:boson_LR_effective_Ham}.
This completes the proof. $\square$

 {~}

\hrulefill{\bf [ End of Proof of Corollary~\ref{corol:boson_LR_effective_Ham}] }

{~}

%
%

By applying Corollary~\ref{corol:boson_LR_effective_Ham} to the inequality~\eqref{norm_calculation_LR_h_Z_tilde},
we obtain 
 \begin{align}
&\norm{ \tilde{O}_{Z_0[k]}  (\tilde{H}_x, 0\to \tau) - \tilde{O}_{Z_0[k]}  (\tilde{H}_{Z_0[\tilde{\ell}_0],x} , 0\to \tau) }  \notag \\
&\le \sum_{\langle i,j\rangle: \mathcal{B}_{i,j} \in \mathcal{S}_{Z_0[\tilde{\ell}_0]}} \int_0^{\tau}  \frac{2\zeta_{Z_0} |\mathcal{B}_{i,j}|}{\gamma (2k)^D}\norm{\tilde{h}_{\mathcal{B}_{i,j},x}} 
\brr{\frac{\tilde{c}_1 x}{\ell_0} 
\br{\frac{\tilde{c}_2 Q}{\ell_0}+ \tilde{c}_{\tilde{q},\bar{q}}} }^{\ell_0/(4k)}   dx  \notag \\
&\le \sum_{\langle i,j\rangle: \mathcal{B}_{i,j} \in \mathcal{S}_{Z_0[\tilde{\ell}_0]}}  
\frac{2\tau \zeta_{Z_0} \cdot 2\gamma k^D}{\gamma (2k)^D} \cdot 2\bar{J} (q_i+q_j)
\brr{\frac{\tilde{c}_1 \tau}{\ell_0} 
\br{\frac{\tilde{c}_2 Q}{\ell_0}+ \tilde{c}_{\tilde{q},\bar{q}}} }^{\ell_0/(4k)}  \notag \\
&\le \frac{16\gamma\bar{J} \tau \zeta_{Z_0} }{2^D} \bar{N}_{\tilde{L}} 
\brr{\frac{\tilde{c}_1 \tau}{\ell_0} 
\br{\frac{\tilde{c}_2 Q}{\ell_0}+ \tilde{c}_{\tilde{q},\bar{q}}} }^{\ell_0/(4k)},
\label{effective_Ham_approx_local_1}
\end{align} 
where in the second inequality, we use $|\mathcal{B}_{i,j}|\le |i[k]| + |j[k]|\le 2\gamma k^D$ from $\mathcal{B}_{i,j}= i[k-1] \cup j[k-1]$ and $\norm{h_{\mathcal{B}_{i,j},\tau}}\le 2\bar{J}(q_i+q_j)$ as in~\eqref{upper_bound_norm_tilde_h_Z_i_j_tau}, and in the last inequality, we use 
 \begin{align}
 \label{summation_i_j_b_ij_tide_ell_0}
\sum_{\langle i,j\rangle: \mathcal{B}_{i,j} \in \mathcal{S}_{Z_0[\tilde{\ell}_0]}}  
 (q_i+q_j)
\le 2 \sum_{i \in \tilde{L}} \gamma q_i \le 2\gamma  \bar{N}_{\tilde{L}}
\end{align} 
with $\bar{N}_{\tilde{L}}$ defined in Eq.~\eqref{def_bar_N_tilde_L_2}.

Finally, we can estimate the norm~\eqref{ineq:prop:error_time_evolution_effective_Ham}.
Because the operators $\tilde{V}_{Z_0[k]}[\tilde{L}',\bold{q}]$ and $\tilde{H}_{Z_0[\tilde{\ell}_0],x}$ are supported on 
$Z_0[\tilde{\ell}_0]\subseteq L_0=Z_0[\ell_0]$, 
by choosing $U_{L_0}$ as 
 \begin{align}
 \label{U_L_choice_subtheorem}
U_{L_0} =  e^{-i \tilde{V}_{Z_0[k]}[\tilde{L}',\bold{q}] \tau}\ \mathcal{T} \exp \br{    
-i\int_0^{\tau} \tilde{H}_{Z_0[\tilde{\ell}_0],x} d\tau
},
\end{align} 
we have 
 \begin{align}
U_{L_0}^\dagger O_{Z_0} U_{L_0} =  \tilde{O}_{Z_0[k]}  (\tilde{H}_{Z_0[\tilde{\ell}_0],x} , 0\to \tau),
\end{align} 
where $\tilde{H}_{Z_0[\tilde{\ell}_0],x}$ has been defined by Eq.~\eqref{def:tilde_H_tau} and Eq.~\eqref{approximate_L_0:1_H_L_tau}. 
By using the above choice of $U_{L_0}$, we obtain 
 \begin{align}
\norm{ O_{Z_0}\br{\tilde{H}[\tilde{L}',\bold{q}] ,\tau} - U_{L_0}^\dagger O_{Z_0} U_{L_0}  } 
&\le  \frac{16\gamma\bar{J} \tau \zeta_{Z_0} }{2^D} \bar{N}_{\tilde{L}} 
\brr{\frac{\tilde{c}_1 \tau}{\ell_0} 
\br{\frac{\tilde{c}_2 Q}{\ell_0}+ \tilde{c}_{\tilde{q},\bar{q}}} }^{\ell_0/(4k)} = \delta_2,  
\label{LR:error_time_evolution_effective_Ham}
\end{align} 
where use the definition~\eqref{def_delta_2_proposition_O_Z_0}. 

By combining the inequalities~\eqref{error_time/evo_effectve_original_ineq_lemma_acc} and \eqref{LR:error_time_evolution_effective_Ham}, we prove the main inequality~\eqref{main_ineqProp:Lieb-Robinson_short_time_L_X_bar_q}.
This completes the proof of Proposition~\ref{Prop:Lieb-Robinson_short_time_L_X_bar_q}. $\square$

\subsection{Generalization: small time evolution by effective Hamiltonian} \label{Sec:Generalization: small time evolution by effective Hamiltonian}

 \begin{figure}[tt]
\centering
\includegraphics[clip, scale=0.5]{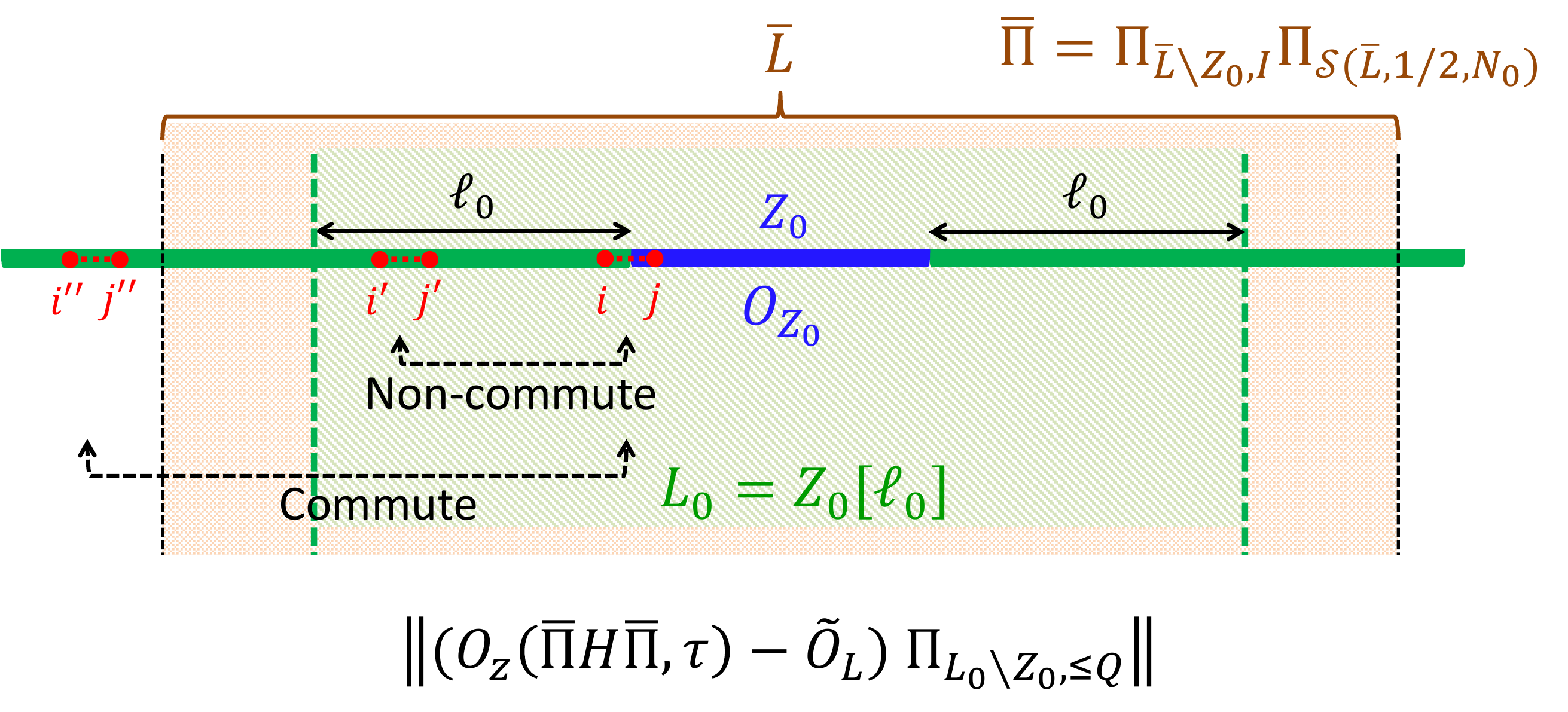}
\caption{Schematic picture of the setup of Proposition~\ref{subtheorem_small_time_evolution_general_eff}. 
Here, we consider the effective Hamiltonian~\eqref{eff_Ham_pi_bar_sub_theorem} that is spanned by the projection $\bar{\Pi}$.
The projection $\bar{\Pi}$ includes a non-local projection $\Pi_{\bar{L}\setminus Z_0, I}$ onto the region $\bar{L}\setminus Z_0$.
Such a setup is indeed used in proving the optimal Lieb-Robinson bound in one-dimensional systems (see also Fig.~\ref{supp_fic_3D_Set_vecN_def.pdf}).
The difficulty here is that the locality of the interaction no longer holds at the boundary of $\bar{L}$ as in Eq.~\eqref{non_locality_around_bar_L}. 
Even in this situation, we can prove the same statement as Subtheorem~\ref{subtheorem_small_time_evolution_general}.
}
\label{fig:small_time_Lieb_Robinson_eff_H.pdf}
\end{figure}

We here consider the same setup as Subtheorem~\ref{subtheorem_small_time_evolution_general} and generalize the statement to the effective Hamiltonian as 
\begin{align}
\label{eff_Ham_pi_bar_sub_theorem}
\bar{\Pi} H \bar{\Pi} ,\quad \bar{\Pi}= \Pi_{\bar{L}\setminus Z_0, I}\Pi_{\mathcal{S}(\bar{L},1/2,N_0)} \quad (I\subset \mathbb{N}) ,
\end{align}
where $\Pi_{\bar{L}\setminus Z_0, I}$ is a projection onto the eigenspace of $\nb_{\bar{L}\setminus Z_0}$ with eigenvalues in the range of $I \subset \mathbb{N}$.
Our purpose is to approximate the time evolution $O_{Z_0}(\bar{\Pi} H \bar{\Pi},\tau)$ onto the region $L_0= Z_0[\ell_0] \subseteq \bar{L}$ (see Fig.~\ref{fig:small_time_Lieb_Robinson_eff_H.pdf}). 
As has been discussed in Sec.~\ref{sec:Effective Hamiltonian: third result}, the interactions in the Hamiltonian $\bar{\Pi} H \bar{\Pi}$ are still local in the region of 
$\bar{L}\setminus Z_0$ and $(\bar{L}\setminus Z_0)^\co$ (see Corollary~\ref{corol:locality_effective_Hamiltonian}), whereas boundary interactions around $\partial (\bar{L}\setminus Z_0)$ is non-local. 

Then, we have to carefully treat the boundary interaction terms $\partial h_{\bar{L}} $ and $\partial h_{Z_0}$ [see Eq.~\eqref{def:Ham_surface} for their notations]. 
As shown in Fig.~\ref{fig:small_time_Lieb_Robinson_eff_H.pdf}, the hopping term 
\begin{align}
\bar{\Pi}  b_i b_j^\dagger \bar{\Pi}= \Pi_{\mathcal{S}(\bar{L},1/2,N_0)} \Pi_{\bar{L}\setminus Z_0, I}  b_i b_j^\dagger \Pi_{\bar{L}\setminus Z_0, I} \Pi_{\mathcal{S}(\bar{L},1/2,N_0)} \quad  \for i\in \bar{L}\setminus Z_0,\quad j\in (\bar{L}\setminus Z_0)^\co
\end{align}
is non-local in the region $\bar{L}\setminus Z_0$. That is, for the above pair of $i,j$, we have a commutation relation of
\begin{align}
\label{non_locality_around_bar_L}
\brr{ \bar{\Pi}  b_i b_j^\dagger \bar{\Pi} ,  \bar{\Pi}  b_{i'} b_{j'}^\dagger \bar{\Pi} }\neq 0 \for  i',j' \in \bar{L}\setminus Z_0 .
\end{align}
Note that we trivially obtain 
\begin{align}
\brr{ \bar{\Pi}  b_i b_j^\dagger \bar{\Pi} ,  \bar{\Pi}  b_{i''} b_{j''}^\dagger \bar{\Pi} }= 0 \for  i'',j'' \in (\bar{L}\setminus Z_0)^\co ,\quad  \{i'',j''\} \cap \{i,j\}=\emptyset.
\end{align}

In proving the optimal Lieb-Robinson bound in one-dimensional systems, we need to extend the statement of Subtheorem~\ref{subtheorem_small_time_evolution_general} to the effective Hamiltonian $\bar{\Pi} H \bar{\Pi}$ (see Sec.~\ref{sec:Proof of Lemma_upper_bound_Delta_O_vec_s}).
We here prove the following generalization:
\begin{prop} \label{subtheorem_small_time_evolution_general_eff}
Let us adopt the same setup as Subtheorem~\ref{subtheorem_small_time_evolution_general}. 
Then, for $\tau\le 1/(4\gamma\bar{J})$, we have the following approximation error by using the subset Hamiltonian $\bar{\Pi} H_{L_0} \bar{\Pi}$ as 
\begin{align}
\label{ineq:subtheorem_small_time_evolution_general_eff2_}
&\norm{\brr{O_{Z_0} (\bar{\Pi} H \bar{\Pi},\tau) -O_{Z_0} (\bar{\Pi} H_{L_0} \bar{\Pi},\tau) }  \Pi_{\tilde{L}, \le Q} } \notag \\
&\le\zeta_{Z_0} \Theta(Q^4 N_0 \bar{L}) \brrr{ e^{-\Theta\br{\ell_0}} + e^{\Theta(\ell_0)}\brr{\Theta(\tau) +\Theta\br{\frac{\tau Q\log(Q)}{\ell_0^2}}  }^{\Theta(\ell_0)}},
\end{align}
where $\ell_0\ge\log[\Theta\brr{\log (q_{Z_0} N_0 )}]$.
\end{prop}

\subsection{Proof of Proposition~\ref{subtheorem_small_time_evolution_general_eff}}
In the proof, we omit the projection $\bar{\Pi}$ and simply denote it as 
\begin{align}
\bar{\Pi} O \bar{\Pi} \to \grave{O} 
\end{align}
for an arbitrary operator $O$. 
For the proof, we need to prove Proposition~\ref{Prop:Lieb-Robinson_short_time_L_X_bar_q} for the effective Hamiltonian.
First of all, we have the same statement as Subtheorem~\ref{Schuch_boson_extend} for the effective Hamiltonian $\bar{\Pi} H \bar{\Pi}$ (see Corollary~\ref{Schuch_boson_extend_eff_Ham}).
Therefore, we can apply the same proof techniques in Secs.~\ref{sec:proof_prop_spectral constraint for construction of effective Hamiltonian} and \ref{sec:proof_prop_Error estimation for the effective Hamiltonian} to this case. 
In the following, we consider how the proof in Sec.~\ref{sec::short_Lieb-Robinson bound for effective Hamiltonian} should be modified.

We here adopt the same notations as those in Sec.~\ref{sec::short_Lieb-Robinson bound for effective Hamiltonian}. 
We separately treat the non-local interaction terms and start from the decomposition of 
\begin{align}
\grave{\tilde{H}}_0[\tilde{L}',\bold{q}] 
&= \bar{\Pi}_{\tilde{L}',\bold{q}} \grave{H}\bar{\Pi}_{\tilde{L}',\bold{q}} \notag \\
&=\grave{\tilde{H}}_{0,Z_0} [\tilde{L}',\bold{q}] + \grave{\tilde{H}}_{0,\bar{L}\setminus Z_0} [\tilde{L}',\bold{q}] + \grave{\tilde{H}}_{0,\bar{L}^\co} [\tilde{L}',\bold{q}] 
+ \partial \grave{\tilde{h}}_{\bar{L}} [\tilde{L}',\bold{q}] +\partial \grave{\tilde{h}}_{Z_0} [\tilde{L}',\bold{q}] ,
\end{align} 
where we have defined $\bar{\Pi}_{\tilde{L}',\bold{q}}:=\prod_{i\in \tilde{L}'} \Pi_{i,\le q_i}$ and $\tilde{L}'= \tilde{L}[-1]$. 
Note that the boundary interaction terms $\partial \grave{\tilde{h}}_{\bar{L}} [\tilde{L}',\bold{q}]$ and $\partial \grave{\tilde{h}}_{Z_0} [\tilde{L}',\bold{q}]$ are non-local in the region $\bar{L}\setminus Z_0$. 
As in Eq.~\eqref{def:tilde_H_tau}, we define $\grave{\tilde{H}}_x$ as 
 \begin{align}
 \label{tilde_H_x_decomp_tildeH'}
\grave{\tilde{H}}_x := e^{i \grave{\tilde{V}}[\tilde{L}',\bold{q}] x } \grave{\tilde{H}}_0[\tilde{L}',\bold{q}]   e^{-i \grave{\tilde{V}}[\tilde{L}',\bold{q}] x} 
&=\sum_{\langle i, j\rangle } \grave{\tilde{h}}_{\mathcal{B}_{i,j},x} , 
\quad \mathcal{B}_{i,j}= i[k-1] \cup j[k-1] , \quad \diam (\mathcal{B}_{i,j}) \le 2k \notag \\
&= \grave{\tilde{H}}'_x  + \partial \grave{\tilde{h}}_{\bar{L},x}+\partial \grave{\tilde{h}}_{Z_0,x},
\end{align} 
where in the last decomposition, we use the notations of
 \begin{align}
 \label{notation_tilde_H'_x}
\tilde{H}'_x 
&:= e^{i \tilde{V}[\tilde{L}',\bold{q}] x } \br{\tilde{H}_{0,Z_0} [\tilde{L}',\bold{q}] + \tilde{H}_{0,\bar{L}\setminus Z_0} [\tilde{L}',\bold{q}] + \tilde{H}_{0,\bar{L}^\co} [\tilde{L}',\bold{q}] }   e^{-i \tilde{V}[\tilde{L}',\bold{q}] x} \notag \\
&=\sum_{\langle i, j\rangle : \{i,j\} \subset Z_0} \tilde{h}_{\mathcal{B}_{i,j},x} 
+ \sum_{\langle i, j\rangle : \{i,j\} \subset \bar{L}\setminus Z_0} \tilde{h}_{\mathcal{B}_{i,j},x}
+ \sum_{\langle i, j\rangle : \{i,j\} \subset \bar{L}^\co} \tilde{h}_{\mathcal{B}_{i,j},x}  ,
\end{align} 
and 
 \begin{align}
\partial \tilde{h}_{\bar{L},x}:=  e^{i \tilde{V}[\tilde{L}',\bold{q}] x }\partial \tilde{h}_{\bar{L}} [\tilde{L}',\bold{q}] e^{-i \tilde{V}[\tilde{L}',\bold{q}] x} 
,\quad \partial \tilde{h}_{Z_0,x}:=  e^{i \tilde{V}[\tilde{L}',\bold{q}] x }\partial \tilde{h}_{Z_0} [\tilde{L}',\bold{q}] e^{-i \tilde{V}[\tilde{L}',\bold{q}] x} .
\end{align} 
Here, the Hamiltonian $\grave{\tilde{H}}'_x$ only includes local interactions in the sense that $\brr{\grave{\tilde{h}}_{\mathcal{B}_{i,j},x}, \grave{\tilde{h}}_{\mathcal{B}_{i',j'},x}}=0$ as long as $\{i,j\} \cap \{i',j'\} =\emptyset$.
Also, we notice that $\tilde{H}'_x$ commute with $\bar{\Pi}$ in Eq.~\eqref{eff_Ham_pi_bar_sub_theorem}, i.e., 
 \begin{align}
 \label{notation_tilde_H'_x_commute_with_bar_Pi}
[\tilde{H}'_x ,\bar{\Pi}] = 0  \longrightarrow   \grave{\tilde{H}}'_x = \tilde{H}'_x \bar{\Pi} = \bar{\Pi} \tilde{H}'_x 
\end{align} 

By using Eq.~\eqref{tilde_H_x_decomp_tildeH'}, we obtain a similar decomposition to Eq.~\eqref{decomposition_of_tine_evolution_tilde_H} as 
 \begin{align}
 \label{decomposition_of_tine_evolution_tilde_H_eff}
e^{-i \grave{\tilde{H}}[\tilde{L}',\bold{q}] \tau} 
& =e^{-i \grave{\tilde{V}}[\tilde{L}',\bold{q}] \tau} U_{\grave{\tilde{H}}_x, 0\to \tau} = e^{-i \grave{\tilde{V}}[\tilde{L}',\bold{q}] \tau}\ \mathcal{T} \exp \br{    
-i\int_0^{\tau} \grave{\tilde{H}}_x  dx}  \notag \\
&= e^{-i \grave{\tilde{V}}[\tilde{L}',\bold{q}] \tau} U_{\grave{\tilde{H}}'_x, 0\to \tau} 
\mathcal{T} \exp \brr{ -i\int_0^{\tau}\partial \grave{\tilde{h}}_{\bar{L},\tau'}\br{\grave{\tilde{H}}'_{x},\tau' \to \tau}   +\partial \grave{\tilde{h}}_{Z_0,\tau'}\br{\grave{\tilde{H}}'_{x},\tau'\to \tau}d\tau'} ,
\end{align} 
where we use the notations in Eq.~\eqref{notation_time_dependent_operator}, and the last equation is derived from
\begin{align}
\mathcal{T} e^{-i\int_0^\tau (A_{\tau'} + B_{\tau'}) d\tau'} 
&= U_{A_x,0\to \tau}  \mathcal{T} e^{-i \int_0^\tau B_{\tau'}(A_x,\tau'\to \tau) d\tau'}
\end{align} 
for arbitrary time-dependent operators $A_x$ and $B_s$.

By using the expression~\eqref{decomposition_of_tine_evolution_tilde_H_eff}, we aim to analyze the following time evolution:
 \begin{align}
 \label{O_Z_0_time_evo__effective}
O_{Z_0}\br{\grave{\tilde{H}}[\tilde{L}',\bold{q}] ,\tau} =U_{\partial}^\dagger \tilde{O}_{Z_0[k]} \br{\grave{\tilde{H}}'_x, 0\to \tau} U_{\partial}, 
\end{align} 
where $ \tilde{O}_{Z_0[k]}$ ($\|\tilde{O}_{Z_0[k]} \|=\| O_{Z_0}\|=\zeta_{Z_0})$ has been defined in Eq.~\eqref{def_tilde_O_L_0:1}, i.e.,
\begin{align}
\tilde{O}_{Z_0[k]}:= O_{Z_0}\br{\grave{\tilde{V}}[\tilde{L}',\bold{q}] ,\tau} = e^{i \grave{\tilde{V}}_{Z_0[k]}[\tilde{L}',\bold{q}] \tau} O_{Z_0} e^{-i \grave{\tilde{V}}_{Z_0[k]}[\tilde{L}',\bold{q}] \tau}, \quad [\tilde{O}_{Z_0[k]},\bar{\Pi}] = 0, 
\end{align} 
and $U_{\partial}$ is defined as 
 \begin{align}
 \label{def_U__partial}
U_{\partial}=
\mathcal{T} \exp \brr{ -i\int_0^{\tau} \partial \grave{\tilde{h}}_{\bar{L},\tau'} \br{\grave{\tilde{H}}'_{x},\tau' \to \tau}   +\partial \grave{\tilde{h}}_{Z_0,\tau'} \br{\grave{\tilde{H}}'_{x},\tau'\to \tau}d\tau'} .
\end{align} 
In Eq.~\eqref{O_Z_0_time_evo__effective}, we separately approximate the terms $U_{\partial}^\dagger$,  $\tilde{O}_{Z_0[k]}\br{\grave{\tilde{H}}'_x, 0\to \tau}$, and  $U_{\partial}$. 

In the following, we denote the approximation of $U_{\partial}$ by $\tilde{U}_{\partial}$, which will be given in Eq.~\eqref{def_U__partial_tilde} explicitly.
By using the notation of  $\tilde{U}_{\partial}$, we obtain 
 \begin{align}
&\norm{ O_{Z_0}\br{\grave{\tilde{H}}[\tilde{L}',\bold{q}] ,\tau} - \tilde{U}_{\partial}^\dagger \tilde{O}_{Z_0[k]} \br{\grave{\tilde{H}}'_{Z_0[\ell_1],x}, 0\to \tau} \tilde{U}_{\partial} } \notag \\
&= \norm{U_{\partial}^\dagger \tilde{O}_{Z_0[k]} \br{\grave{\tilde{H}}'_x, 0\to \tau} U_{\partial} - \tilde{U}_{\partial}^\dagger \tilde{O}_{Z_0[k]} \br{\grave{\tilde{H}}'_{Z_0[\ell_1],x}, 0\to \tau} \tilde{U}_{\partial} } 
\notag \\
&\le \norm{ \tilde{O}_{Z_0[k]}  \br{\grave{\tilde{H}}'_x, 0\to \tau} - \tilde{O}_{Z_0[k]}  \br{\grave{\tilde{H}}'_{Z_0[\ell_1],x} , 0\to \tau}}   \notag\\
&\quad +\norm{U_{\partial}^\dagger \tilde{O}_{Z_0[k]}  \br{\grave{\tilde{H}}'_{Z_0[\ell_1],x} , 0\to \tau} U_{\partial} - \tilde{U}_{\partial}^\dagger \tilde{O}_{Z_0[k]}  \br{\grave{\tilde{H}}'_{Z_0[\ell_1],x} , 0\to \tau} \tilde{U}_{\partial}}   \notag \\
&\le \norm{ \tilde{O}_{Z_0[k]}  \br{\grave{\tilde{H}}'_x, 0\to \tau} - \tilde{O}_{Z_0[k]}  \br{\grave{\tilde{H}}'_{Z_0[\ell_1],x} , 0\to \tau}}    
+ 2\zeta_{Z_0} \norm{U_{\partial}-\tilde{U}_{\partial}} ,
\label{approx_O_{Z_0}_Z_0[r_1]_refine}
\end{align} 
where we use $\norm{\tilde{O}_{Z_0[k]}  \br{\grave{\tilde{H}}'_{Z_0[\ell_1],x} , 0\to \tau}}=\norm{O_{Z_0}}=\zeta_{Z_0}$, 
and the length $\ell_1$ is appropriately chosen afterward.  
Also, for the last inequality in~\eqref{approx_O_{Z_0}_Z_0[r_1]_refine}, we used 
 \begin{align}
\norm{U_1^\dagger O U_1 - U_2^\dagger O U_2} 
= \norm{U_1^\dagger O U_1-U_2^\dagger O U_1+U_2^\dagger O U_1 - U_2^\dagger O U_2}  
&\le \norm{(U_1-U_2)^\dagger O U_1}+\norm{U_2^\dagger O (U_1 -U_2)}    \notag \\
&\le 2\norm{O} \cdot \norm{U_1 -U_2} . 
\end{align} 
Here, $\grave{\tilde{H}}'_{Z_0[\ell_1],x}$ is the subset Hamiltonian for $\grave{\tilde{H}}'_x$ in which we pick up the interaction terms in~\eqref{notation_tilde_H'_x} such that $\mathcal{B}_{i,j}\subset Z_0[\ell_1]$.

To estimate the RHS of the inequality~\eqref{approx_O_{Z_0}_Z_0[r_1]_refine}, we first aim to consider the approximation of $\tilde{O}_{Z_0[k]} \br{\grave{\tilde{H}}'_x, 0\to \tau} $ onto the region of $Z_0[\ell_1]$, where we assume that $\ell_1$ is smaller than 
$\tilde{\ell}_0$ which has been defined in Eq.~\eqref{def_tilde_ell_0_proof}, i.e., $\tilde{\ell}_0 = \ell_0 - (\delta \ell +2 k  +1)$.
Because the Hamiltonian $\grave{\tilde{H}}'_x$ does not include the non-local interactions, we can obtain the same approximation error as Eq.~\eqref{effective_Ham_approx_local_1}, which yields
 \begin{align}
&\norm{ \tilde{O}_{Z_0[k]}  \br{\grave{\tilde{H}}'_x, 0\to \tau} - \tilde{O}_{Z_0[k]}  \br{\grave{\tilde{H}}'_{Z_0[\ell_1],x} , 0\to \tau} } \le \zeta_{Z_0} \Theta(\tau) \bar{N}_{\tilde{L}} 
\brrr{\frac{\tau}{\ell_1} 
\brr{ \Theta(Q/\ell_1) +\Theta(\tilde{q}) +\Theta(\bar{q}) }}^{\Theta(\ell_1)} ,
\label{effective_Ham_approx_local_r_1}
\end{align} 
where $\ell_1\ge \Theta(\delta \ell)$ that originates from a similar condition to~\eqref{new_condition_ell_0_LR}.

 \begin{figure}[tt]
\centering
\includegraphics[clip, scale=0.45]{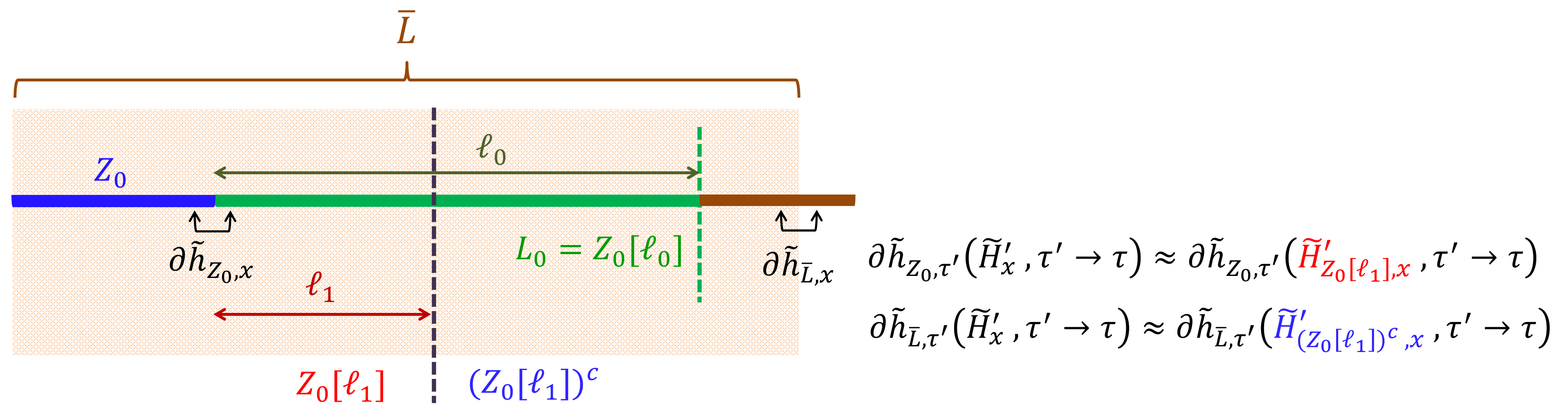}
\caption{Schematic picture of the approximations for $\partial \tilde{h}_{Z_0,\tau'}\br{\tilde{H}'_{x},\tau'\to \tau}$ and $\partial \tilde{h}_{\bar{L},\tau'} \br{\tilde{H}'_{x},\tau' \to \tau} $ in Eq.~\eqref{approx:small_time_Lieb_Robinson_eff_H_proof.pdf}, 
where $\partial \tilde{h}_{\bar{L},x}$ and $\partial \tilde{h}_{Z_0,x}$ are defined in Eq.~\eqref{tilde_H_x_decomp_tildeH'}.
The length $\ell_1$ is a free parameter that is introduced in the inequality~\eqref{approx_O_{Z_0}_Z_0[r_1]_refine}. 
}
\label{fig:small_time_Lieb_Robinson_eff_H_proof.pdf}
\end{figure}

We next estimate the error $\norm{U_{\partial}-\tilde{U}_{\partial}}$.
We first notice that because of Eq.~\eqref{notation_tilde_H'_x_commute_with_bar_Pi} 
 \begin{align}
\partial \grave{\tilde{h}}_{Z_0,\tau'}\br{\grave{\tilde{H}}'_{x},\tau'\to \tau}
=\bar{\Pi}  \tilde{h}_{Z_0,\tau'} \br{\tilde{H}'_x,\tau'\to \tau}\bar{\Pi}  .
\end{align} 
The same equation holds for $\partial \grave{\tilde{h}}_{\bar{L},\tau'}\br{\grave{\tilde{H}}'_{x},\tau'\to \tau}$.  
We then adopt the following approximations (see also Fig.~\ref{fig:small_time_Lieb_Robinson_eff_H_proof.pdf}): 
 \begin{align}
 \label{approx:small_time_Lieb_Robinson_eff_H_proof.pdf}
&\partial \tilde{h}_{Z_0,\tau'}\br{\tilde{H}'_{x},\tau'\to \tau} \approx \partial \tilde{h}_{Z_0,\tau'}\br{\tilde{H}'_{Z_0[\ell_1],x},\tau' \to \tau} ,\notag \\
&\partial \tilde{h}_{\bar{L},\tau'} \br{\tilde{H}'_{x},\tau' \to \tau} \approx  \partial \tilde{h}_{\bar{L},\tau'}\br{\tilde{H}'_{Z_0[\ell_1]^\co,x},\tau' \to \tau}  .
\end{align} 
By using the above approximated time evolution, we define
 \begin{align}
 \label{def_U__partial_tilde}
&\tilde{U}_{\partial}:=
\mathcal{T} \exp \brr{ -i\int_0^{\tau}\bar{\Pi} \partial \tilde{h}_{\bar{L},\tau'} \br{\tilde{H}'_{Z_0[\ell_1]^\co,x},\tau' \to \tau}\bar{\Pi}  +\bar{\Pi}\partial \tilde{h}_{Z_0,\tau'}\br{\tilde{H}'_{Z_0[\ell_1],x},\tau' \to \tau} \bar{\Pi} d\tau_1} \notag \\
&=\mathcal{T} \exp \brr{ -i\int_0^{\tau}\bar{\Pi}\partial \tilde{h}_{\bar{L},\tau'} \br{\tilde{H}'_{Z_0[\ell_1]^\co,x},\tau' \to \tau} \bar{\Pi}d\tau_1}
\otimes \mathcal{T} \exp \brr{ -i\int_0^{\tau}\bar{\Pi}\partial \tilde{h}_{Z_0,\tau'}\br{\tilde{H}'_{Z_0[\ell_1],x},\tau' \to \tau}\bar{\Pi} d\tau_1}.
\end{align} 
Under the above choice of $\tilde{U}_{\partial}$, we obtain 
 \begin{align}
  \label{def_U__partial_tilde_approx_inside}
 &\tilde{U}_{\partial}^\dagger \tilde{O}_{Z_0[k]} \br{\grave{\tilde{H}}'_{Z_0[\ell_1],x}, 0\to \tau} \tilde{U}_{\partial} 
 =\tilde{U}_{\partial,Z_0[\ell_1]}^\dagger   \tilde{O}_{Z_0[k]}\br{\grave{\tilde{H}}'_{Z_0[\ell_1],x}, 0\to \tau} \tilde{U}_{\partial,Z_0[\ell_1]} , \notag \\
 &\tilde{U}_{\partial,Z_0[\ell_1]}:=
\mathcal{T} \exp \brr{ -i\int_0^{\tau}\bar{\Pi}\partial \tilde{h}_{Z_0,\tau'}\br{\tilde{H}'_{Z_0[\ell_1],x},\tau' \to \tau} \bar{\Pi}d\tau_1} ,
\end{align} 
which only depends on the interaction terms of $H$ inside $Z_0[\ell_1]$.

From now, we aim to upper-bound the norm difference between $U_{\partial}$ and $\tilde{U}_{\partial}$.
We start from the inequality of
 \begin{align}
 \label{def_U__partial_norm_deff}
\norm{ U_{\partial}-\tilde{U}_{\partial}}\le &
\int_0^{\tau} \norm{\partial \tilde{h}_{Z_0,\tau'}\br{\tilde{H}'_{x},\tau'\to \tau} -\partial \tilde{h}_{Z_0,\tau'}\br{\tilde{H}'_{Z_0[\ell_1],x},\tau' \to \tau} }  dx \notag \\
&+ \int_0^{\tau} \norm{\partial \tilde{h}_{\bar{L},\tau'}\br{\tilde{H}'_{x},\tau' \to \tau} - \partial \tilde{h}_{\bar{L},\tau'}\br{\tilde{H}'_{Z_0[\ell_1]^\co,x},\tau' \to \tau} }dx .
\end{align}
Here, we use the inequality of 
 \begin{align}
\norm{\mathcal{T} e^{-i \int_0^\tau A_x dx} - \mathcal{T} e^{-i \int_0^\tau B_x dx} }
\le  \int_0^\tau \norm{A_x- B_x} dx
\end{align} 
for arbitrary time-dependent operators $A_x$ and $B_x$, 
which is derived from the decomposition of
 \begin{align}
\mathcal{T} e^{-i \int_0^\tau A_x dx} - \mathcal{T} e^{-i \int_0^\tau B_x dx} 
&=\sum_{m=0}^{M-1} \br{ \mathcal{T} e^{-i \int_{\tau_m}^\tau A_x dx}  \mathcal{T} e^{-i \int_0^{\tau_m} B_x dx}  - \mathcal{T} e^{-i \int_{\tau_{m+1}}^\tau A_x dx}  \mathcal{T} e^{-i \int_0^{\tau_{m+1}} B_x dx}} \notag \\
&=\sum_{m=0}^{M-1} \mathcal{T} e^{-i \int_{\tau_{m+1}}^\tau A_x dx} \br{-i A_{\tau_m}d\tau + i B_{\tau_{m+1}}d\tau}  \mathcal{T} e^{-i \int_0^{\tau_m} B_x dx} +\orderof{d\tau^2} ,
\end{align} 
where $\tau_m=m d\tau$ and $\tau_M=\tau$ ($d\tau \to +0$).  

For the first approximation in~\eqref{def_U__partial_norm_deff}, we can obtain the same approximation error as Eq.~\eqref{effective_Ham_approx_local_r_1}:
 \begin{align}
&\norm{\partial \tilde{h}_{Z_0,\tau'}\br{\tilde{H}'_{x},\tau' \to \tau} - \partial \tilde{h}_{Z_0,\tau'}\br{\tilde{H}'_{Z_0[\ell_1],x},\tau' \to \tau}}\notag \\
&\le \norm{\partial \tilde{h}_{Z_0,\tau'}} \Theta(\tau) \bar{N}_{\tilde{L}} 
\brrr{\frac{\tau}{\ell_1} 
\brr{ \Theta(Q/\ell_1) +\Theta(\tilde{q}) +\Theta(\bar{q}) }}^{\Theta(\ell_1)} ,
\label{effective_Ham_approx_local_r_1_2}
\end{align} 
Because of the projection $\Pi_{\mathcal{S}(\bar{L},1/2,N_0)}$, we have $\norm{\nb_i \Pi_{\mathcal{S}(\bar{L},1/2,N_0)}}\le N_0$ for $i\in \bar{L}$, and hence
 \begin{align}
\norm{\partial \tilde{h}_{Z_0,\tau'}} 
&\le \sum_{i\in \partial Z_0} \sum_{j\in Z_0^\co: \dist_{i,j}=1} \norm{\Pi_{\mathcal{S}(\bar{L},1/2,N_0)} J_{i,j}\br{ b_i b_j^\dagger +{\rm h.c.} } \Pi_{\mathcal{S}(\bar{L},1/2,N_0)}} \le  4 \gamma \bar{J} |Z_0| N_0, 
\label{norm_partial_tilde_h/Z}
\end{align} 
where we use the inequality of~\eqref{upper_bound_norm_tilde_h_Z_i_j_tau} and $|\partial Z_0| \le |Z_0|$.
Therefore, by applying the inequalities~\eqref{effective_Ham_approx_local_r_1_2} and \eqref{norm_partial_tilde_h/Z} to the first term of the RHS in~\eqref{def_U__partial_norm_deff}, we obtain
 \begin{align}
&\int_0^{\tau} \norm{\partial \tilde{h}_{Z_0,\tau'}\br{\tilde{H}'_{x},\tau'\to \tau} -\partial \tilde{h}_{Z_0,\tau'}\br{\tilde{H}'_{Z_0[\ell_1],x},\tau' \to \tau} }  dx  \notag \\
&\le 4\gamma \bar{J} |Z_0| N_0 \cdot \Theta(\tau^2) \bar{N}_{\tilde{L}} 
\brrr{\frac{\tau}{\ell_1} 
\brr{ \Theta(Q/\ell_1) +\Theta(\tilde{q}) +\Theta(\bar{q}) }}^{\Theta(\ell_1)}.
 \label{def_U__partial_norm_deff_first_term}
\end{align}

For the second approximation in~\eqref{def_U__partial_norm_deff}, we first utilize Lemma~\ref{lem:Lieb--Robinson_start} by choosing 
\begin{align}
O _X\to \partial \tilde{h}_{\bar{L},\tau'}, \quad H_\tau \to \tilde{H}'_{x},  \quad L \to Z_0[\ell_1]^\co
\end{align}
in the inequality~\eqref{lem:ineq:Lieb--Robinson_start}, which yields 
\begin{align}
&\norm{\partial \tilde{h}_{\bar{L},\tau'} \br{\tilde{H}'_{x},\tau' \to \tau} - \partial \tilde{h}_{\bar{L},\tau'}\br{\tilde{H}'_{Z_0[\ell_1]^\co,x},\tau' \to \tau} } \notag \\
\le& \sum_{\mathcal{B}_{i,j}: \mathcal{B}_{i,j} \in \mathcal{S}_{Z_0[\ell_1]}} \int_{\tau'}^\tau 
\left\| \brr{ \tilde{h}_{\mathcal{B}_{i,j},\tau''}\br{\tilde{H}'_{Z_0[\ell_1]^\co,x}, \tau''\to \tau'}, \partial \tilde{h}_{\bar{L},\tau'} } \right\| d\tau''.
\label{effective_Ham_approx_local_r_1_3_start}
\end{align}
For the time-evolved operator $\tilde{h}_{\mathcal{B}_{i,j},\tau''}\br{\tilde{H}'_{Z_0[\ell_1]^\co,x}, \tau''\to \tau'}$, we can apply the same inequality as~\eqref{effective_Ham_approx_local_r_1_2}, which gives
 \begin{align}
&\norm{\tilde{h}_{\mathcal{B}_{i,j},\tau''}\br{\tilde{H}'_{Z_0[\ell_1]^\co,x}, \tau''\to \tau'} 
-\tilde{h}_{\mathcal{B}_{i,j},\tau''}\br{\tilde{H}'_{Z_0[\ell_1+\ell_2] \setminus Z_0[\ell_1],x}, \tau''\to \tau'}}\notag \\
&\le \norm{\tilde{h}_{\mathcal{B}_{i,j},\tau''}} \Theta(\tau) \bar{N}_{\tilde{L}} 
\brrr{\frac{\tau}{\ell_2} 
\brr{ \Theta(Q/\ell_2) +\Theta(\tilde{q}) +\Theta(\bar{q}) }}^{\Theta(\ell_2)}  \for \mathcal{B}_{i,j} \in \mathcal{S}_{Z_0[\ell_1]} ,
\label{effective_Ham_approx_local_r_1_3}
\end{align} 
where we choose $\ell_2$ such that $\ell_1 + \ell_2 \le \tilde{\ell_0}$.
By using the same inequality as~\eqref{summation_i_j_b_ij_tide_ell_0} with \eqref{upper_bound_norm_tilde_h_Z_i_j_tau}, we obtain 
\begin{align}
\sum_{\mathcal{B}_{i,j}: \mathcal{B}_{i,j} \in \mathcal{S}_{Z_0[\ell_1]}} \norm{\tilde{h}_{\mathcal{B}_{i,j},\tau''}} 
\le \sum_{\mathcal{B}_{i,j}: \mathcal{B}_{i,j} \in \mathcal{S}_{Z_0[\tilde{\ell_0}]}} \norm{\tilde{h}_{\mathcal{B}_{i,j},\tau''}} 
\le 4\gamma \bar{J} \bar{N}_{\tilde{L}} .
\label{effective_Ham_approx_local_r_1_3_sum}
\end{align}
From the inequalities~\eqref{effective_Ham_approx_local_r_1_3} and~\eqref{effective_Ham_approx_local_r_1_3_sum}, we reduce the inequality~\eqref{effective_Ham_approx_local_r_1_3_start} to 
\begin{align}
&\norm{\partial \tilde{h}_{\bar{L},\tau'} \br{\tilde{H}'_{x},\tau' \to \tau} - \partial \tilde{h}_{\bar{L},\tau'}\br{\tilde{H}'_{Z_0[\ell_1]^\co,x},\tau' \to \tau} } \notag \\
\le& \sum_{\mathcal{B}_{i,j}: \mathcal{B}_{i,j} \in \mathcal{S}_{Z_0[\ell_1]}} \int_{\tau'}^\tau 
\left\| \brr{\tilde{h}_{\mathcal{B}_{i,j},\tau''}\br{\tilde{H}'_{Z_0[\ell_1+\ell_2] \setminus Z_0[\ell_1],x}, \tau''\to \tau'}, \partial \tilde{h}_{\bar{L},\tau'} } \right\| d\tau'' \notag \\
&+ \sum_{\mathcal{B}_{i,j}: \mathcal{B}_{i,j} \in \mathcal{S}_{Z_0[\ell_1]}} \int_{\tau'}^\tau 
\left\| \brr{\tilde{h}_{\mathcal{B}_{i,j},\tau''}\br{\tilde{H}'_{Z_0[\ell_1]^\co,x}, \tau''\to \tau'} 
-\tilde{h}_{\mathcal{B}_{i,j},\tau''}\br{\tilde{H}'_{Z_0[\ell_1+\ell_2] \setminus Z_0[\ell_1],x}, \tau''\to \tau'}, \partial \tilde{h}_{\bar{L},\tau'} } \right\| d\tau''\notag \\
\le & \sum_{\mathcal{B}_{i,j}: \mathcal{B}_{i,j} \in \mathcal{S}_{Z_0[\ell_1]}} \int_{\tau'}^\tau 
2 \norm{\partial \tilde{h}_{\bar{L},\tau'}} \cdot \norm{\tilde{h}_{\mathcal{B}_{i,j},\tau''}} \Theta(\tau) \bar{N}_{\tilde{L}} 
\brrr{\frac{\tau}{\ell_2} 
\brr{ \Theta(Q/\ell_2) +\Theta(\tilde{q}) +\Theta(\bar{q}) }}^{\Theta(\ell_2)}  d\tau''\notag \\
\le &2 \tau \norm{\partial \tilde{h}_{\bar{L},\tau'}} \cdot 4\gamma \bar{J} \bar{N}_{\tilde{L}}^2  \Theta(\tau) 
\brrr{\frac{\tau}{\ell_2} \brr{ \Theta(Q/\ell_2) +\Theta(\tilde{q}) +\Theta(\bar{q}) }}^{\Theta(\ell_2)} ,
\label{effective_Ham_approx_local_r_1_3_fin_pre}
\end{align}
where in the second inequality we use 
 \begin{align}
\norm{\brr{\tilde{h}_{\mathcal{B}_{i,j},\tau''}\br{\tilde{H}'_{Z_0[\ell_1+\ell_2] \setminus Z_0[\ell_1],x}, \tau''\to \tau'}, \partial \tilde{h}_{\bar{L},\tau'} }} 
= \norm{\brr{\tilde{h}_{\mathcal{B}_{i,j},\tau''}, \partial \tilde{h}_{\bar{L},\tau'} }} = 0, 
\end{align} 
because of $Z_0[\ell_1+\ell_2]\cap \partial \bar{L}[2k]=\emptyset$ for $\ell_1+\ell_2 \le \tilde{\ell}_0$. 
Finally, we can derive the upper bound for $\norm{\partial \tilde{h}_{\bar{L},\tau'} }$ as follows: 
 \begin{align}
\norm{\partial \tilde{h}_{\bar{L},\tau'}  } 
&\le \sum_{i\in \partial \bar{L}} \sum_{j\in \bar{L}^\co: \dist_{i,j}=1} \norm{\Pi_{\mathcal{S}(\bar{L},1/2,N_0)} J_{i,j}\br{ b_i b_j^\dagger +{\rm h.c.} } \Pi_{\mathcal{S}(\bar{L},1/2,N_0)}} \le  4 e^{1/2} \gamma \bar{J} |\bar{L}| N_0, 
\end{align} 
where we use~\eqref{upper_bound_hopping_operator} and $\norm{ \nb_j \Pi_{\mathcal{S}(\bar{L},1/2,N_0)}} \le e^{1/2}N_0$ for $j\in \bar{L}^\co$ with $\dist_{i,j}=1$. 
We thus reduce the inequality~\eqref{effective_Ham_approx_local_r_1_3_fin_pre} to 
\begin{align}
&\norm{\partial \tilde{h}_{\bar{L},\tau'} \br{\tilde{H}'_{x},\tau' \to \tau} - \partial \tilde{h}_{\bar{L},\tau'}\br{\tilde{H}'_{Z_0[\ell_1]^\co,x},\tau' \to \tau} } 
 \le\Theta\br{\tau^2 \bar{N}_{\tilde{L}}^2  N_0 |\bar{L}| }
\brrr{\frac{\tau}{\ell_2} \brr{ \Theta(Q/\ell_2) +\Theta(\tilde{q}) +\Theta(\bar{q}) }}^{\Theta(\ell_2)}   ,
\label{effective_Ham_approx_local_r_1_3_fin}
\end{align}
and hence we arrive at the inequality of 
 \begin{align}
&\int_0^{\tau} \norm{\partial \tilde{h}_{\bar{L},\tau'} \br{\tilde{H}'_{x},\tau' \to \tau} -\partial \tilde{h}_{\bar{L},\tau'}  \br{\tilde{H}'_{Z_0[\ell_1]^\co,x},\tau' \to \tau} }dx  \notag \\
&\le  \Theta\br{\tau^3 \bar{N}_{\tilde{L}}^2  N_0 |\bar{L}| }
\brrr{\frac{\tau}{\ell_2} \brr{ \Theta(Q/\ell_2) +\Theta(\tilde{q}) +\Theta(\bar{q}) }}^{\Theta(\ell_2)} .
 \label{def_U__partial_norm_deff_second_term}
\end{align}

By combining the upper bounds of~\eqref{def_U__partial_norm_deff_first_term} and \eqref{def_U__partial_norm_deff_second_term}, 
we can obtain the approximation error $\norm{ U_{\partial}-\tilde{U}_{\partial}}$ from~\eqref{def_U__partial_norm_deff}.
Therefore, by choosing $\ell_1=\ell_2 \propto \ell_0$ from $\ell_1 + \ell_2 \le \tilde{\ell}_0 = \ell_0 - (\delta \ell +2 k  +1)$, we obtain 
 \begin{align}
 \label{def_U__partial_norm_deff_fin}
\norm{ U_{\partial}-\tilde{U}_{\partial}}\le 
 \Theta\br{\bar{N}_{\tilde{L}}^2 N_0  |\bar{L}|} \brrr{\frac{\tau}{\ell_0} 
\brr{ \Theta(Q/\ell_0) +\Theta(\tilde{q}) +\Theta(\bar{q}) }}^{\Theta(\ell_0)}  ,
\end{align}
where we use $ \Theta\br{\tau^3 \bar{N}_{\tilde{L}}^2  N_0 |\bar{L}| } \le  \Theta\br{\bar{N}_{\tilde{L}}^2  N_0 |\bar{L}| }$ because of $\tau\le 1/(4\gamma\bar{J})=\orderof{1}$.
By using the above inequality and~\eqref{effective_Ham_approx_local_r_1}, we finally reduce the inequality~\eqref{approx_O_{Z_0}_Z_0[r_1]_refine} to the form of 
 \begin{align}
&\norm{ O_{Z_0}\br{\tilde{H}[\tilde{L}',\bold{q}] ,\tau} - \tilde{U}_{\partial}^\dagger \tilde{O}_{Z_0[k]} \br{\tilde{H}'_{Z_0[\ell_1],x}, 0\to \tau} \tilde{U}_{\partial} } 
\le \zeta_{Z_0} \Theta\br{\bar{q}^2Q^2 N_0  |\bar{L}|} \brrr{\frac{\tau}{\ell_0} 
\brr{ \Theta(Q/\ell_0) +\Theta(\tilde{q}) +\Theta(\bar{q}) }}^{\Theta(\ell_0)}   ,
\label{approx_O_{Z_0}_Z_0[r_1]_refine_fin}
\end{align} 
where we use $\bar{N}_{\tilde{L}} \le \bar{q}Q$ from the inequality~\eqref{upp_bar_N_tilde_L_Q^2}. 
From the definition of Eq.~\eqref{def_U__partial_tilde}, the approximate operator $\tilde{U}_{\partial}^\dagger \tilde{O}_{Z_0[k]} \br{\tilde{H}'_{Z_0[\ell_1],x}, 0\to \tau} \tilde{U}_{\partial}$ is supported on $Z_0[\ell_1] \subseteq Z_0[\ell_0] = L_0$. 
We thus prove the same statement as Proposition~\ref{Prop:Lieb-Robinson_short_time_L_X_bar_q} for the effective Hamiltonian.

Thus, by replacing the $\delta_2$ in Proposition~\ref{Prop:Lieb-Robinson_short_time_L_X_bar_q} with 
 \begin{align}
\tilde{\delta}_2 =  \zeta_{Z_0} \Theta\br{\bar{q}^2Q^2 N_0  |\bar{L}|} \brrr{\frac{\tau}{\ell_0} 
\brr{ \Theta(Q/\ell_0) +\Theta(\tilde{q}) +\Theta(\bar{q}) }}^{\Theta(\ell_0)}   ,
\end{align} 
the same analyses as in Sec.~\ref{sec:Completing the proof of Subtheorem_small_time_evolution_general} yields the main inequality~\eqref{ineq:subtheorem_small_time_evolution_general_eff2_}. 
Note that we have followed the same discussion as in~\eqref{ineq:subtheorem_small_time_evolution_general_re} in approximating $O_{Z_0} (\bar{\Pi} H_{L_0} \bar{\Pi},\tau)$ by $O_{Z_0} (\bar{\Pi} H_{L_0} \bar{\Pi},\tau)$ [see also Eq.~\eqref{def_U__partial_tilde_approx_inside}].
This completes the proof. $\square$

\section{Optimal Lieb-Robinson bound: high-dimensional cases} \label{sec:Optimal Lieb-Robinson bound: high-dimensional cases}

Based on Subtheorem~\ref{subtheorem_small_time_evolution_general}, we prove the optimal Lieb-Robinson bound with the Lieb-Robinson velocity as $v_{\rm LR} \propto t^{D-1}$.
Here, depending on the spatial dimension (i.e., $D\ge2$ or $D=1$), qualitatively different proofs are required. 
In the following sections~\ref{sec:Optimal Lieb-Robinson bound: high-dimensional cases} and \ref{sec:1D:boson_refined}, 
we derive the optimal Lieb-Robinson bounds for high dimensions and one dimension, respectively. 
For high-dimensional cases, we aim to prove the following theorem:

\begin{theorem} \label{main_thm:boson_Lieb-Robinson_main}
Let us consider the same setup as in Theorem~\ref{main_theorem_looser_Lieb_Robinson}. 
Then, under the condition
\begin{align}
\label{main_conditon_for_R_r_0}
R':=R-r_0 \ge \Theta\brr{t^D \log^{D+2} (q_0 r_0 t)}, 
\end{align} 
we obtain the Lieb-Robinson bound as 
\begin{align}
\label{ineq:main_thm:boson_Lieb-Robinson_main}
&\norm{ \brr{ O_{X_0}(t)- O_{X_0}\br{ H_{i_0[R]} ,t} }\rho_0  }_1 \le 
\norm{O_{X_0}}  \exp\brr{-\Theta\br{R'/t^D}^{1/{\kappa D}} + \Theta[\log(q_0 R)] }    .
\end{align} 
\end{theorem}

{\bf Remark.} From the condition~\eqref{main_conditon_for_R_r_0}, we can conclude that the Lieb-Robinson velocity $v_{\rm LR}$ depends on the time as 
\begin{align}
\label{main_conditon_for_R_r_0__2}
v_{\rm LR} \approx  
\tilde{\mathcal{O}}\br{t^{D-1}}  \for D\ge 2.  
\end{align} 
For $D\ge 2$, this estimation is qualitatively optimal up to a logarithmic correction, while for $D= 1$, we cannot apply the same analyses to prove.
In the 1D case, the choices in Eq.~\eqref{choices_of_Qand_delta_t_tau} give $\tau =\Theta \br{\ell_0/[\mathfrak{q}\log(\ell_0)]}$, but we have to make the parameter $\mathfrak{q} $ such that $\mathfrak{q} \ge \ell_0/\log(\ell_0)$ because of $\tau\le 1/(4\gamma \bar{J})$.
Therefore, in 1D, the condition $\tau\le 1/(4\gamma \bar{J})$ implies $\Delta r \ge \Theta(R'/t)$ ($R'=R-r_0$) from Eq.~\eqref{choice_of_Delta_r_highD}, and hence the condition~\eqref{condition_for_ell_0_prop_short_time_2D_fin_proof} is satisfied only for $R\ge \Theta(t^2)$, which is no longer better than Theorem~\ref{main_theorem_looser_Lieb_Robinson}.
This point is contrast to the high dimensional cases $(D\ge 2)$, where the choices in Eq.~\eqref{choices_of_Qand_delta_t_tau} does not influence the condition $\tau\le 1/(4\gamma \bar{J})$ as long as $\mathfrak{q} \ge \Theta(1)$; that is,  unlike the 1D case, we can freely control $\mathfrak{q}$.

In order to improve the Lieb-Robinson velocity to $\tilde{\mathcal{O}}\br{t^{0}}$ in 1D, we need more refined analyses which are summarized in Sec.~\ref{sec:1D:boson_refined}.


\subsection{Proof of Theorem~\ref{main_thm:boson_Lieb-Robinson_main}: preparation}
\label{sec:Proof of Theorem_main_thm:high_dim_preparation}

Throughout the proof, we consider the Hilbert space which is spanned by the projection $\Pi_{\mathcal{S}(i_0[R],1/2,N_0)}$ with $N_0= \poly(R)$, which was defined in Eq.~\eqref{mathcal_S_L_xi_N/def_proj}. 
We note that $i_0$ is chosen such that $X_0 \subseteq i_0[r_0]$. 
From Proposition~\ref{prop:boson_number_truncation_2} with $L\to i_0[R]$ and $N\to N_0$, 
we can achieve the approximation error of 
\begin{align}
&\norm{O_{X_0}(H,t) \rho_0  -  O_{X_0}(\tilde{H}_{(i_0[R],N_0)},t)   \tilde{\Pi}_{(i_0[R],N_0)}  \rho_0 \tilde{\Pi}_{(i_0[R],N_0)}  }_1  \notag \\
&\le  \zeta_0 \Theta\br{t\xi_{t,R}^2 |i_0[R]| } e^{- \brr{N_0/ (2\xi_{t,R})}^{1/\kappa}/(8\kappa)}  \notag \\
&\le  \zeta_0 e^{-R} ,\quad  \xi_{t,R} := 16 b_0 (2\bar{\ell}_{t,R})^D  |X_0|, 
\end{align} 
where we use~\eqref{effective_hamiltonian_error_t^d_2} in the first inequality, and for the second inequality, we choose $N_0$ appropriately as 
\begin{align}
\label{choice_N_0_upp_highD}
N_0 =64e^2 q_0^3 + \xi_{t,R} R^{\kappa} \polylog(R) \le \Theta\brr{q_0^3, R^{2D+\kappa} \polylog(R) } \le
\Theta\br{q_0^3, R^{2D+1+\kappa} }  .
\end{align}
Note that $64e^2 q_0^3$ in the above choice comes from the condition~\eqref{assumption_for_N_ineq}, 
and we use $t\le R$ and $|X_0|\le |i_0[R]|$ to derive $\xi_{t,R}\le \Theta[R^{2D} \log^D(q_0R)]$.
Under the above choice of $N_0$, in the following, we simply denote as 
\begin{align}
\label{notation_N_0_simplify}
\tilde{H}_{i_0[R] ,N_0} \to H  ,\quad \tilde{\Pi}_{(i_0[R],N_0)}  \rho_0 \tilde{\Pi}_{(i_0[R],N_0)}  \to \rho_0
\end{align}
without expressing the truncation of the boson numbers. 

\subsection{Proof of Theorem~\ref{main_thm:boson_Lieb-Robinson_main}} \label{sec:Proof_of_Theorem_main_thm:boson_Lieb-Robinson_main}

In the proof, we consider the length $R$ such that $R\le e^{\Theta(t)}$. 
In the asymptotic case as $R> e^{\Theta(t)}$,
the obtained upper bound~\eqref{ineq:main_theorem_looser_Lieb_Robinson} is much stronger than the inequality~\eqref{ineq:main_thm:boson_Lieb-Robinson_main}. 
Throughout the proof, we denote
$$\norm{O_{X_0}} = \zeta_0.$$ 

For the proof, as in Ref.~\cite{PhysRevLett.127.070403}, we utilize the connection of short-time unitary evolutions, which has also played critical roles in Refs.~\cite{Kuwahara_2016_njp,PhysRevLett.126.030604}. 
In Ref.~\cite{PhysRevLett.127.070403}, the time-independence of the initial state (i.e., $\rho_0(t)=\rho$) has been assumed, but in the present case, we generalize the method to the time-dependent initial state.
First, we introduce the time width $\tau \le  1/(4\gamma \bar{J})$ and decompose the total time $t$ to $ \bar{m}:=t/\tau$ [$=\orderof{t}$] pieces:
\begin{align}
\label{t_decomp_bar_m}
t := \bar{m} \tau .
\end{align}
For a fixed $R$, we define the subset $X_m$ as follows: 
\begin{align}
\label{Choice_Delta_r_X_m}
X_m:= i_0[r_0+ m\Delta r], \quad \Delta r = \left \lfloor \frac{R-r_0}{\bar{m}} \right \rfloor = \left \lfloor \frac{R'}{\bar{m}} \right \rfloor  \quad (R':=R-r_0),
\end{align}
where $X_{\bar{m}} \subseteq i_0[R]$.


We start from the approximation of 
\begin{align}
O_{X_0}(\tau) \xrightarrow{\rm approximate} O_{X_1}^{(1)}= O_{X_0}(H_{X_1}, \tau) ,
\end{align}
whose approximation error is obtained from
\begin{align}
\norm{\brr{ O_{X_0}(\tau) - O_{X_1}^{(1)} } \rho_0}_1.
\end{align} 
In the next step, we connect the two approximations of 
\begin{align}
O_{X_0}(\tau) \xrightarrow{\rm approximate} O_{X_1}^{(1)} ,\quad O_{X_1}^{(1)} (\tau) \xrightarrow{\rm approximate} O_{X_2}^{(2)}=O_{X_1}^{(1)}(H_{X_2}, \tau)  .
\end{align}
By using the decomposition of 
\begin{align}
O_{X_0}(2\tau) \rho_0 
&=e^{iH\tau} \brr{ O_{X_0}(\tau) - O_{X_1}^{(1)} } e^{-iH\tau} \rho_0  +e^{iH\tau}  O_{X_1}^{(1)} e^{-iH\tau}  \rho_0  \notag \\
&=e^{iH\tau} \brr{ O_{X_0}(\tau) - O_{X_1}^{(1)} } \rho_0(\tau ) e^{-iH\tau} +\brr{ O_{X_1}^{(1)}(\tau) - O_{X_2}^{(2)}}  \rho_0 
+ O_{X_2}^{(2)}  \rho_0 ,
\end{align} 
we obtain the approximation error of 
\begin{align}
\norm{ \brr{ O_{X_0}(2\tau)-O_{X_2}^{(2)} } \rho_0 }_1
\le \norm{ \brr{ O_{X_0}(\tau) - O_{X_1}^{(1)} } \rho_0(\tau ) }_1 + \norm{ \brr{ O_{X_1}^{(1)}(\tau) - O_{X_2}^{(2)}}  \rho_0 }_1 ,
\end{align} 
where we use the unitary invariance of the trace norm $\norm{\cdots}_1$.
By repeating the same process, we can obtain
\begin{align}
\label{Lieb_Robinson_connection_ineq}
&\norm{ \brr{ O_{X_0}(\bar{m}\tau)-O_{X_{\bar{m}}}^{(\bar{m})} } \rho_0 }_1
\le \sum_{j=1}^{\bar{m}} \norm{ \brr{ O_{X_{j-1}}^{(j-1)}(\tau) - O_{X_j}^{(j)} } \rho_0(\bar{m}\tau-j\tau) }_1 \notag \\
&\longrightarrow 
\norm{ \brr{ O_{X_0}(t)-O_{i_0[R]}^{(\bar{m})} } \rho_0 }_1
\le \sum_{j=1}^{\bar{m}} \norm{ \brr{ O_{X_{j-1}}^{(j-1)}(\tau) - O_{X_j}^{(j)} } \rho_0(\bar{m}\tau-j\tau) }_1 
\end{align} 
by setting $O_{i_0[R]}^{(\bar{m})}=O_{X_{\bar{m}}}^{(\bar{m})}$ from $X_{\bar{m}} \subseteq i_0[R]$, 
where each of $\brrr{O_{X_{j}}^{(j)} }_{j=1}^m$ is chosen as 
\begin{align}
\label{Choice_O_X_j_j_Delta/t}
O_{X_{j}}^{(j)} = O_{X_{j-1}}^{(j-1)} (H_{X_j}, \tau) \for j\in [\bar{m}] .
\end{align} 
We notice that each of the operators $\{O_{X_{j}}^{(j)}\}_{j=1}^m$ satisfies the same inequality as~\eqref{O_X0_condition/_norm_lemma_operator_boson_pre}, which originally holds for $O_{X_0}$\footnote{For an arbitrary operator $O_{X_0}$ such that the inequality~\eqref{O_X0_condition/_norm_lemma_operator_boson_pre} holds, the time evolved operator $O_{X_0}(H_{X_0[r]},t)$ also satisfies the same inequality by replacing $X_0\to X_0[r]$ in \eqref{O_X0_condition/_norm_lemma_operator_boson_pre} because of 
\begin{align}
\Pi_{X_0[r], > m+q_0} O_{X_0}(H_{X_0[r]},t) \Pi_{X_0[r], m} =e^{iH_{X_0[r]t}} \Pi_{X_0[r], > m+q_0} O_{X_0}\Pi_{X_0[r], m} e^{-iH_{X_0[r]t}} ,
\end{align} 
where we use $[H_{X_0[r]},\nb_{X_0[r]}]=0$.}.

To estimate the approximation errors which appear in~\eqref{Lieb_Robinson_connection_ineq}, we need to combine the following two points:
\begin{enumerate}
\item{} For the time-evolved density matrix $\rho_0(t -j\tau)$, the spectral constraint by Lemma~\ref{lemm:boson_density_prob_gene_time_evo} is imposed. Here, the initial low-boson-density condition~\eqref{only_the_assumption_initial} is loosened in the sense 
that only the average number of bosons in a ball region with radius $\bar{\ell}_{t,R}$ is small; that is, the boson number on one site can be as large as $\orderof{\bar{\ell}_{t,R}^D}=\tO(t^D)$.  
\item{} For short-time evolution, we have obtained a qualitatively optimal Lieb-Robinson bound as in Subtheorem~\ref{subtheorem_small_time_evolution_general} when the average boson number in a region is constrained as in Eq.~\eqref{Pi_S_L_0N_0_L_0_Q}. 
\end{enumerate}
By taking the above two points into account, we can prove the following proposition for short-time evolution:
\begin{prop} \label{prop:short_time_LR_2D-}
Let us set 
\begin{align}
\label{condition_for_ell_0_prop_short_time_2D}
\ell_0 \ge\log[\Theta(N_0)] \ge \bar{\ell}_{t,R} ,\quad \norm{O_X} = \zeta ,
\end{align}
where $N_0$ has been chosen as in Eq.~\eqref{choice_N_0_upp_highD}.
Also, we assume that boson creation by the operator $O_X$ ($X\subseteq i_0[R]$) is at most $q_0$; that is, the inequality~\eqref{O_X0_condition/_norm_lemma_operator_boson_pre} holds for $O_X$.
Then, for a short-time evolution of $e^{iH\tau}$, we obtain 
 \begin{align}
&\norm{ \brr{O_X(\tau) - O_X(H_{X[\ell]}, \tau)  } \rho_0(t_1)}_1 \notag \\
&\le  \zeta \Theta(N_0 R^{2D}Q^2) \brrr{ e^{- \Theta\br{Q/\ell_0^D}^{1/\kappa}}+ e^{-\Theta\br{\ell_0}} + e^{\Theta(\ell_0)}\brr{\Theta(\tau) +\Theta\br{\frac{\tau Q\log(Q)}{\ell_0^2}}  }^{\Theta(\ell_0)}} ,
 \label{main_ineq:prop:short_time_LR_2D-}  
\end{align}
for $\forall t_1\le t-\tau$ and $X\subset i_0[R-\ell]$, 
where $\ell=3\ell_0 + k$ [i.e., $\ell_0= (\ell-k)/3$].
\end{prop}

In the above proposition, we choose 
\begin{align}
\label{choices_of_Qand_delta_t_tau}
Q=\Theta \br{\mathfrak{q} \ell_0^D}\quad [\mathfrak{q} \le \Theta(\ell_0)] ,\quad  \tau =\Theta \br{ \frac{\ell_0^{-D+2}}{\mathfrak{q}\log(\ell_0)}  } ,\quad 
\Delta r = \ell = 3\ell_0 + k.
\end{align} 
We set the parameter $\mathfrak{q}$ appropriately afterward [see \eqref{choice of_mathfrak_q_1} below]. 
From the above choices, we can make the RHS of the inequality~\eqref{main_ineq:prop:short_time_LR_2D-} 
 \begin{align}
 \label{RHS_of_the prop_main_ineq_1}
&\Theta(N_0 R^{2D}Q^2)\brrr{ e^{- \Theta\br{Q/\ell_0^D}^{1/\kappa}}+ e^{-\Theta\br{\ell_0}} + e^{\Theta(\ell_0)}\brr{\Theta(\tau) +\Theta\br{\frac{\tau Q\log(Q)}{\ell_0^2}}  }^{\Theta(\ell_0)}}  \notag \\
&= \Theta \br{N_0 R^{2D}Q^2} \br{ e^{-\Theta(\mathfrak{q})^{1/\kappa}}  + e^{-\Theta(\ell_0)}} 
\le \Theta \br{N_0 R^{4D+2}} e^{-\Theta(\mathfrak{q})^{1/\kappa}}  ,
\end{align}
where in the last inequality, we use $Q \le \Theta \br{\ell_0^{D+1}}\le \Theta \br{R^{D+1}}$ and $e^{-\Theta(\mathfrak{q})^{1/\kappa}} \ge e^{-\Theta(\ell_0)}$ ($\kappa\ge1$) from $\mathfrak{q} \le \Theta(\ell_0)$. 

We then consider how to choose the parameter $\mathfrak{q}$. 
Here, from Eq.~\eqref{Choice_Delta_r_X_m}, we have $\Delta r = \floorf{\frac{R'}{\bar{m}}}=\floorf{\frac{R'}{(t/\tau)}}$, and hence 
the above choice~\eqref{choices_of_Qand_delta_t_tau} also yields
 \begin{align}
&\Delta r = \floorf{\frac{R'}{(t/\tau)}}  = \frac{R'}{t} \Theta \br{ \frac{\ell_0^{-D+2}}{\mathfrak{q}\log(\ell_0)}  } 
= \Theta \br{\frac{R'}{t} \cdot \frac{(\Delta r)^{-D+2}}{\mathfrak{q}\log(\Delta r)}  } \label{choice_of_Delta_r_highD} \\
&\longrightarrow (\Delta r)^{D-1} \log(\Delta r) =\Theta\br{\frac{R'}{\mathfrak{q}t} }\notag \\
&\longrightarrow \Delta r=\Theta\brr{\frac{R'}{\mathfrak{q}t\log(R'/t)}}^{1/{(D-1)}}=
\frac{t}{\log^{1/(D-1)}(R'/t)} \Theta\br{\frac{R'}{\mathfrak{q}t^D} }^{1/{(D-1)}}  \notag \\
&\longrightarrow \ell_0 = \frac{t}{\log^{1/(D-1)}(R'/t)} \Theta\br{\frac{R'}{\mathfrak{q}t^D} }^{1/{(D-1)}} ,
\label{Ineq_for_Delta/_r_with_t}
\end{align}
where we use $\ell_0 = (\Delta r-k)/3=\Theta(\Delta r)$ in the last equation. 
Because of $\mathfrak{q} \le \Theta(\ell_0)$, by choosing $\mathfrak{q}$ as
\begin{align}
\label{choice of_mathfrak_q_1}
\mathfrak{q} = \Theta\br{R'/t^D}^{1/D}  ,
\end{align} 
we reduces Eq.~\eqref{Ineq_for_Delta/_r_with_t} to 
 \begin{align}
\ell_0  =\frac{t}{\log^{1/(D-1)}(R'/t)} \Theta\br{R'/t^D}^{1/D} \propto  \frac{t}{\log^{1/(D-1)}(R'/t)} \mathfrak{q}  .
\label{Ineq_for_Delta/_r_with_t_2} 
\end{align}
Now, the condition $\mathfrak{q} \le \Theta(\ell_0)$ is ensured from the assumption of $R'\le R\le e^{\Theta(t)}$.
We thus adopt the choice of Eq.~\eqref{choice of_mathfrak_q_1} in the following.

Under the choices of Eqs.~\eqref{choices_of_Qand_delta_t_tau} and~\eqref{choice of_mathfrak_q_1}, by applying the inequalities~\eqref{main_ineq:prop:short_time_LR_2D-} and \eqref{RHS_of_the prop_main_ineq_1} to \eqref{Lieb_Robinson_connection_ineq}, we obtain
\begin{align}
\label{Lieb_Robinson_connection_ineq_bar_m}
\norm{ \brr{ O_{X_0}(t)-O_{X_{\bar{m}}}^{(\bar{m})} } \rho_0 }_1
&\le \bar{m} \zeta_0 \Theta \br{N_0R^{4D+2}} \exp\brr{-\Theta\br{R'/t^D}^{1/{\kappa D}}}\notag \\
&\le \zeta_0 \Theta \br{N_0R^{4D+3}} \exp\brr{-\Theta\br{R'/t^D}^{1/{\kappa D}}},
\end{align} 
where in the second inequality we use $\bar{m}= \floorf{R/\Delta r}\le R$ from Eq.~\eqref{Choice_Delta_r_X_m}.
From the choice of Eq.~\eqref{choice_N_0_upp_highD}, we reduce $\Theta \br{N_0R^{4D+3}}$ to 
\begin{align}
\Theta \br{N_0R^{4D+3}}\le  \Theta \br{q_0^3 R^{4D+3+ 2D+1+\kappa} }= \Theta \br{q_0^3 R^{6D+4+\kappa} } .
\end{align} 
By applying the above bound to the inequality~\eqref{Lieb_Robinson_connection_ineq_bar_m}, we obtain the desired upper bound for the approximation of $O_{X_0}(t)$. 
We prove the same upper bound for $\norm{ \brr{ O_{X_0}(H_{i_0[R]}, t)-O_{X_{\bar{m}}}^{(\bar{m})} } \rho_0 }_1$ under the choices as in Eq.~\eqref{Choice_O_X_j_j_Delta/t}, where $H_{i_0[R]}$ is the subset Hamiltonian on $i_0[R]$.
We thus prove the main inequality~\eqref{ineq:main_thm:boson_Lieb-Robinson_main} by letting $R'=R-r_0$.

Finally, we consider the condition $\ell_0 \ge \bar{\ell}_{t,R} = \Theta [t\log(q_0 R)]$ in Eq.~\eqref{condition_for_ell_0_prop_short_time_2D} is now given by
 \begin{align}
\label{condition_for_ell_0_prop_short_time_2D_fin_proof}
&\frac{t}{\log^{1/(D-1)}(R'/t)} \Theta\br{R'/t^D}^{1/D} \ge \Theta(\bar{\ell}_{t,R})
\longrightarrow  \br{R'/t^D}^{1/D} \ge \Theta(\log^{1+1/(D-1)}(q_0 R))  \notag \\
&\longrightarrow R-r_0 \ge  \Theta\brr{t^D \log^{D^2/(D-1)} (q_0 r_0 t)}
\longrightarrow R-r_0 \ge  \Theta\brr{t^D \log^{D+2} (q_0 r_0 t)},
\end{align}
where we use $D^2/(D-1) \le D+2$ for $D\ge 2$.
This completes the proof. $\square$

%
%
%

\subsection{Proof of Proposition~\ref{prop:short_time_LR_2D-}} \label{sec:Proof of Proposition_prop:short_time_LR_2D-}

For the proof, we utilize Subtheorem~\ref{subtheorem_small_time_evolution_general} appropriately to estimate the approximation error of
 \begin{align}
\norm{ \brr{O_X(\tau) - O_X(H_{X[\ell]}, \tau)  } \rho_0(t_1)}_1 
\end{align}
for $t_1\le t-\tau$. 
However, we cannot directly apply the subtheorem because we encounter the following difficulty. 
In high dimensional cases, when the radius $\ell$ becomes large, 
the subset size $|L_0^\ast|$ in Eq.~\eqref{O_Z_0_tilde_O_L_0} is roughly given by $|L_0^\ast|=|X[\ell] \setminus X|  = \Theta\brr{(\diam(X)+\ell)^D}$, 
and hence the number of bosons in the region $L_0^\ast$, denoted by $Q$, is proportional to $|L_0^\ast|$, i.e., $Q\propto  \Theta\brr{(\diam(X)+\ell)^D}$. 
Therefore, in applying the inequality~\eqref{ineq:subtheorem_small_time_evolution_general} to an operator $O_X$ supported on a large subset $X$ such that $\diam(X)\gtrsim \ell$, we have to  consider a significantly short time such that 
 \begin{align}
 \label{conditon_no_longer_helpful_Q}
 \frac{\tau Q\log(Q)}{\ell^2} \propto \tau \br{\frac{\diam(X)}{\ell}}^2 \diam(X)^{D-2} \log[\diam(X)] \lesssim \frac{1}{e} ,
\end{align}
which is necessary to make the inequality~\eqref{ineq:subtheorem_small_time_evolution_general} meaningful. 
Unfortunately, under the above condition, we can no longer derive the optimal Lieb-Robinson bound. 
To avoid the constraint~\eqref{conditon_no_longer_helpful_Q}, before applying the subtheorem to the derivation of the Lieb-Robinson bound, we first reduce the problem to the time evolution of local operators (see Fig.~\ref{fig:small_time_Lieb_Robinson_highD.pdf} below).

For this purpose, we first define the projection $\bar{\Pi}_{Y,r,\le Q}$ as follows:
\begin{align}
\label{definition_of_proj_bar_X_ell_leQ}
\bar{\Pi}_{Y,r,\le Q} := \prod_{i\in Y} \Pi_{i[r], \le Q} . 
\end{align}
From the projection, for any site $i\in Y$, the number of bosons in the ball region $i[r]$ is smaller than or equal to $Q$. 
We also define the following subsets (see Fig.~\ref{fig:small_time_Lieb_Robinson_highD.pdf}):  
\begin{align}
\label{set_X_2X_2_Tilde_X}
X_1:=X[2\ell_0] ,\quad X_2:=X[2\ell_0+k],\quad X_3:=X[3\ell_0 + k] ,\quad  \tilde{X} := X_2 \setminus X_1 = X[2\ell_0+k] \setminus  X[2\ell_0] ,
\end{align}
where $\ell_0$ has been defined in the main statement, i.e., $\ell_0=(\ell-k)/3$.
We then adopt the projection of 
\begin{align}
\label{def:bar_pi_ast_eq}
\bar{\Pi}^\ast := \bar{\Pi}_{\tilde{X},\ell_0,\le Q}= \prod_{i\in \tilde{X}} \Pi_{i[\ell_0], \le Q}  .
\end{align}

We next consider the decomposition of 
\begin{align}
H=H_1 + H_2 + \partial H_0 
\end{align}
with
\begin{align}
\label{defs:H1_H2_partialH_0}
&H_1 = H_{0,X_1} + V_{X_2} , \quad H_2 = H_{0,X_2^\co} +  V_{X_2^\co} + \sum_{Z \in \mathcal{S}_{X_2}} v_Z, \notag \\
& \partial H_0 = H-H_1-H_2 = H_{0, X_2\setminus X_1} + \partial h_{X_1} + \partial h_{X_2} = \sum_{\substack{\ave{i,j} \\ \{i,j\} \cap \tilde{X} \neq \emptyset}}J_{i,j} (b_i b_j^\dagger +{\rm h.c.}) ,
\end{align}
where we use the notations of Eq.~\eqref{def:Ham_surface}, and $\mathcal{S}_{X_2}$ has been defined in Eq.~\eqref{set_mathcal_S_L}, i.e., $\mathcal{S}_{X_2}:= \{ Z\subseteq \Lambda | Z \cap X_2 \neq \emptyset, Z \cap X_2^\co \neq \emptyset\}$.
Because the width of the region $X_2\setminus X_1$ is equal to $k$ (see Fig.~\ref{fig:small_time_Lieb_Robinson_highD.pdf}), 
we ensure that the support of $v_Z$ ($Z \in \mathcal{S}_{X_2}$) has no overlap with $X_1$, and hence 
\begin{align}
\label{eq_H_1_H_2_commute=0}
[H_1,H_2]= \sum_{Z \in \mathcal{S}_{X_2}} [H_{0,X_1},v_Z]=0 . 
\end{align}
By using the above notations, we consider 
\begin{align}
\label{The decomposition of_small time evolution}
e^{-iH\tau} = e^{-i \br{H_1 + H_2} \tau}   \mathcal{T} e^{-i\int_0^\tau  \partial H_{0,x} dx},
\end{align}
with 
\begin{align}
\label{def_h_partial_X_2,x}
\partial H_{0,x} := \partial H_0 \br{H_1 + H_2,x}=e^{i \br{H_1 + H_2} x}  \partial H_0 e^{-i \br{H_1 + H_2} x}.
\end{align}

Under the above setup, we will take the following strategy to prove Proposition~\ref{prop:short_time_LR_2D-}. 
From Eq.~\eqref{The decomposition of_small time evolution}, we reduce the target time evolution $O_X(\tau)$ to
\begin{align}
\label{O_X_tau_X_2_tau_partial_H_0x}
O_X(\tau)
&=\br{\mathcal{T} e^{-i\int_0^\tau  \partial H_{0,x} dx} }^\dagger O_X(H_1+H_2,\tau)\mathcal{T} e^{-i\int_0^\tau \partial H_{0,x} dx } \notag \\
&=: O_{X_2,\tau} (\partial H_{0,x} , 0\to \tau) , 
\end{align}
where we denote $O_X(H_1+H_2,\tau)=O_X(H_1,\tau) =: O_{X_2,\tau}$. 
Note that from Eq.~\eqref{eq_H_1_H_2_commute=0} and $[O_X,H_2]=0$, we have $$e^{i(H_1+H_2) \tau }O_Xe^{- i(H_1+H_2) \tau } = e^{iH_1\tau } e^{iH_2 \tau }O_Xe^{-iH_2\tau } e^{-iH_1\tau } = e^{iH_1\tau }O_Xe^{-iH_1\tau }.$$
We then aim to approximate the time evolution~\eqref{O_X_tau_X_2_tau_partial_H_0x} by using the effective Hamiltonian as 
\begin{align}
\label{def:partial_tilde_H_ast_0x}
\partial \tilde{H}^\ast_{0,x}  := \bar{\Pi}^\ast \partial H_{0,x} \bar{\Pi}^\ast = \bar{\Pi}^\ast e^{i \br{H_1 + H_2} x}  \partial H_0 e^{-i \br{H_1 + H_2} x}\bar{\Pi}^\ast  ,
\end{align}
where the second equation results from Eq.~\eqref{def_h_partial_X_2,x}. 
As a first task, we need to estimate the approximation error of
\begin{align}
\label{approx_partial_H_0_x_bar_Pi_ast}
O_{X_2,\tau} (\partial H_{0,x} , 0\to \tau)  \approx O_{X_2,\tau} (\bar{\Pi}^\ast \partial H_{0,x} \bar{\Pi}^\ast  , 0\to \tau), 
\end{align}
and then our second task is to approximate $\bar{\Pi}^\ast \partial H_{0,x} \bar{\Pi}^\ast$ onto $X_3$ ($=X[\ell]$). 
Because $\partial H_{0,x}$ is constructed from $b_ib_j^\dagger$ with $\{i,j\} \cap \tilde{X} \neq \emptyset$, we need to obtain the approximation of 
\begin{align}
\label{approx_partial_H_0_x_bar_Pi_ast_second}
\bar{\Pi}^\ast e^{i \br{H_1 + H_2} x} (b_ib_j^\dagger) e^{-i \br{H_1 + H_2} x}\bar{\Pi}^\ast 
\approx  \bar{\Pi}^\ast e^{i \br{H_{1,i[\ell_0]} + H_{2,i[\ell_0]}} x} (b_ib_j^\dagger) e^{-i \br{H_{1,i[\ell_0]} + H_{2,i[\ell_0]}} x}\bar{\Pi}^\ast   .
\end{align}
Because of $\{i,j\} \cap \tilde{X} \neq \emptyset$, the subset Hamiltonian $H_{1,i[\ell_0]} + H_{2,i[\ell_0]}$ is still supported on $X_3$. 
To the above approximation, we can utilize Subtheorem~\ref{subtheorem_small_time_evolution_general}. 
Combining the approximations~\eqref{approx_partial_H_0_x_bar_Pi_ast} and \eqref{approx_partial_H_0_x_bar_Pi_ast_second} leads to the main inequality~\eqref{main_ineq:prop:short_time_LR_2D-}.

 \begin{figure}[tt]
\centering
\includegraphics[clip, scale=0.25]{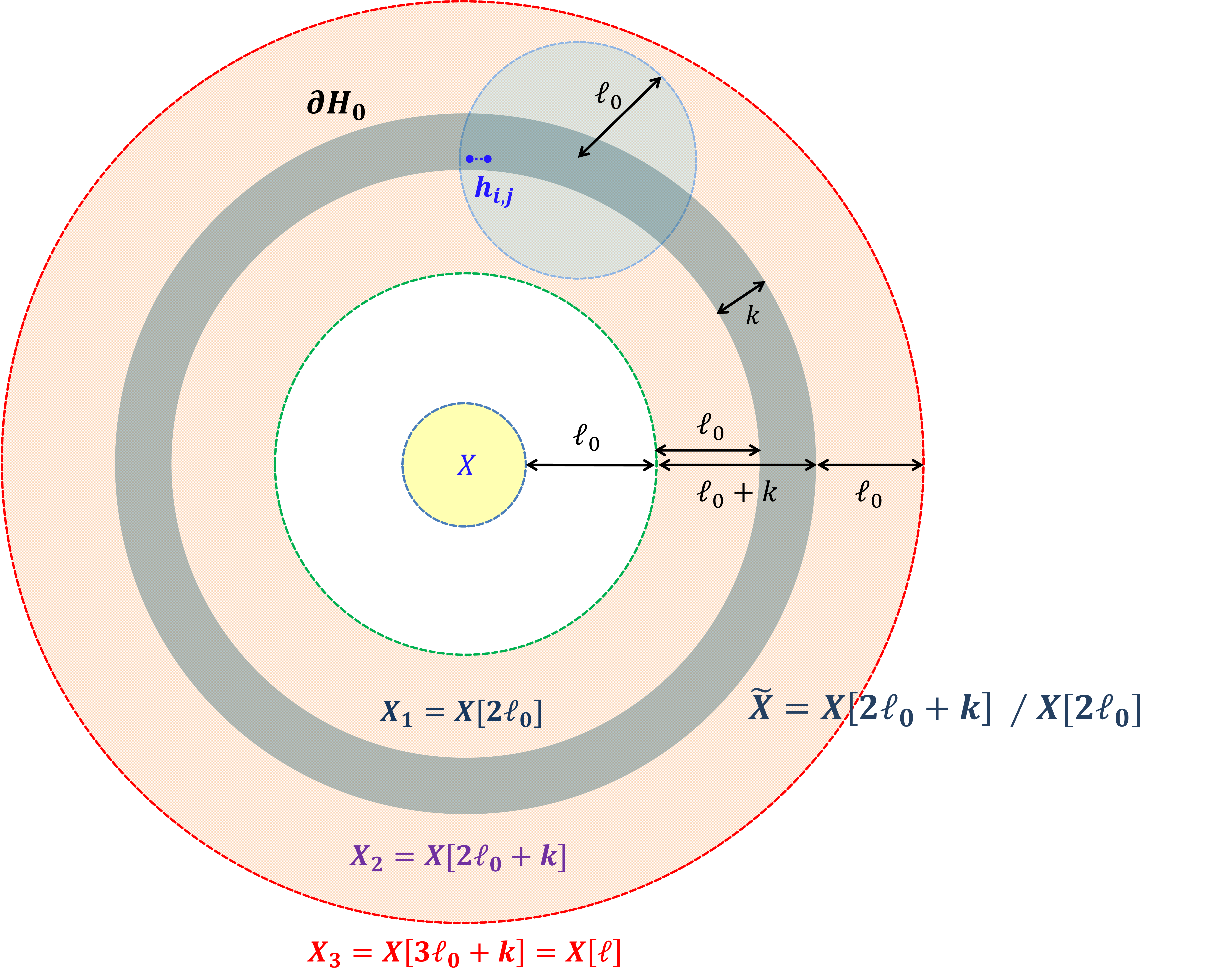}
\caption{Schematic picture of the setup for the proof. We first define the regions $X_1$, $X_2$ and $X_3$ as in Eq.~\eqref{set_X_2X_2_Tilde_X}. 
The shaded region $\tilde{X}$ is defined as $X_2\setminus X_1$. 
The Hamiltonian is decomposed to $H_1 + H_2 + \partial H_0$ [see Eq.~\eqref{defs:H1_H2_partialH_0}]. Roughly speaking, the Hamiltonians $H_1$ and $H_2$ acts on the region $X_1$ and $X_2^\co$, respectively, and the Hamiltonian $\partial H_0$ describes the interactions in the region $\tilde{X}$. By these choices, the Hamiltonians $H_1$ and $H_2$ are decoupled in the sense that $[H_1+H_2]=0$ as in Eq.~\eqref{eq_H_1_H_2_commute=0}. 
Then, the time evolution $e^{-iHt}$ is decomposed to $e^{-i \br{H_1 + H_2} \tau}$ and $\mathcal{T} e^{-i\int_0^\tau  \partial H_{0,x} dx}$ as in Eq.~\eqref{The decomposition of_small time evolution}, where $\partial H_{0,x}=\partial H_0 \br{H_1 + H_2,x}$. 
The time evolution $e^{-i \br{H_1 + H_2} \tau}=e^{-i H_1 \tau}e^{-i H_2\tau}$ keeps the operator $O_X$ in the region $X_2$, 
whereas the time evolution $\mathcal{T} e^{-i\int_0^\tau  \partial H_{0,x} dx}$ may induce the operator spreading outside the region $X_3$. 
In order to ensure the quasi-locality of the operator $\partial H_{0,x} = \partial H_{0} (H_1+H_2,x)$, we first approximate $\partial H_{0,x}$ by the effective Hamiltonian $\bar{\Pi}^\ast\partial H_{0,x}\bar{\Pi}^\ast$ (Lemma~\ref{lemma:time_evo_highD_local_map}), where $\bar{\Pi}^\ast$ is defined in Eq.~\eqref{def:bar_pi_ast_eq}.
Afterward, by using Subtheorem~\ref{subtheorem_small_time_evolution_general}, we approximate the effective Hamiltonian $\bar{\Pi}^\ast\partial H_{0,x}\bar{\Pi}^\ast$ onto the region $X_3$ [Corollary~\ref{corol::subtheorem_small_time_evolution_general} and the inequality~\eqref{time_evo_upp_tilde_h_ast_partial_X2_x}].
}
\label{fig:small_time_Lieb_Robinson_highD.pdf}
\end{figure}

First of all, for the approximation~\eqref{approx_partial_H_0_x_bar_Pi_ast}, we prove the following lemma:
\begin{lemma} \label{lemma:time_evo_highD_local_map}
For the time-evolved quantum state $\rho_0(t_1)$ with an arbitrary $t_1$ ($\le t-\tau$), we upper-bound the norm of
 \begin{align}
&\norm{ \brr{O_X(\tau) - O_{X_2,\tau} (\partial \tilde{H}^\ast_{0,x}, 0\to \tau)} \rho_0(t_1)}_1   \notag \\
&=\norm{ \brr{O_{X_2,\tau} (\partial H_{0,x}, 0\to \tau)- O_{X_2,\tau} (\partial \tilde{H}^\ast_{0,x}, 0\to \tau)} \rho_0(t_1)}_1 
\le \delta_{\ell_0,Q}^{(1)} ,
 \label{main_ineq:lemma:time_evo_highD_local_map}
\end{align}
where $ \delta_{\ell_0,Q}^{(1)} $ is defined as follows:
\begin{align}
\delta_{\ell_0,Q}^{(1)} :=4\gamma R^D(1 +2\tau \gamma^2 \bar{J}N_0 R^{D})\zeta e^{- \brr{Q/(4 b_0 \ell_0^D) }^{1/\kappa}/2}
=\Theta(N_0 R^{2D}) e^{- \Theta\br{Q/\ell_0^D}^{1/\kappa}} .
\end{align}
\end{lemma}

\subsubsection{Proof of Lemma~\ref{lemma:time_evo_highD_local_map}}

For the proof, we utilize Corollary~\ref{corol:effective Hamiltonian_accuracy_depend_t}. 
We here denote
\begin{align}
\label{def_U_tau_'_1}
U_{\tau'} =  U_{\partial H_{0,x}, 0\to \tau'} .
\end{align}  
From the inequality~\eqref{error_time/effective Hamiltonian_accuracy_depend_t}, in estimating the norm error in~\eqref{main_ineq:lemma:time_evo_highD_local_map}, we have to upper-bound
\begin{align}
\label{norm_estiamation_bar_Pi_ast_h_X1_1}
\norm{\bar{\Pi}^{\ast\co}U_{\tau_1} U_{\tau}^\dagger  O_{X_2,\tau} U_{\tau} \sqrt{\rho_0(t_1)} }_F ,
\end{align}
and
\begin{align}
\label{norm_estiamation_bar_Pi_ast_h_X1_2}
\norm{ \bar{\Pi}^\ast \partial H_{0,\tau-\tau_1} \bar{\Pi}^{\ast\co}  U_{\tau_1} U_{\tau}^\dagger O_{X_2,\tau} U_{\tau} \sqrt{\rho_0(t_1) }  }_F .
\end{align}
Note that all the terms in LHS of~\eqref{error_time/effective Hamiltonian_accuracy_depend_t} reduce to either of the above forms by letting $\tau_1 \to ( 0 \ {\rm or} \ \tau)$ and 
$O_{X_2} \to (\hat{1}\ {\rm or} \ O_{X_2})$ 


First, from the definition~\eqref{def_h_partial_X_2,x}, we aim to upper-bound 
\begin{align}
\norm{ \bar{\Pi}^\ast \partial H_{0,\tau-\tau_1} \bar{\Pi}^{\ast\co}} 
\le \sum_{\substack{\ave{i,j} \\ \{i,j\} \cap \tilde{X} \neq \emptyset}}\bar{J} \norm{ \bar{\Pi}^\ast\br{b_i b_j^\dagger +{\rm h.c.}}} 
\le\sum_{\substack{\ave{i,j} \\ \{i,j\} \cap \tilde{X} \neq \emptyset}}2\bar{J} \norm{ \bar{\Pi}^\ast \br{\nb_i+\nb_j}}   ,
\end{align}
where we use~\eqref{upper_bound_hopping_operator} in the last inequality.
Because the Hilbert space is restricted by $\tilde{\Pi}_{(i_0[R],N_0)}$ as in~\eqref{notation_N_0_simplify}, we can upper-bound as 
$\norm{\nb_i} \le N_0$ for $i\in i_0[R]$, which yields 
\begin{align}
\label{summation_h_Z_Z_in_ineq}
\norm{ \bar{\Pi}^\ast \partial H_{0,\tau-\tau_1} \bar{\Pi}^{\ast\co}}  
\le\sum_{\substack{\ave{i,j} \\ \{i,j\} \cap \tilde{X} \neq \emptyset}}4\bar{J} N_0 
\le  4\gamma \bar{J} N_0 |X_2\setminus X_1|
\le  4\gamma \bar{J} N_0 |X_2|  ,
\end{align}
where we use $\tilde{X}:=X_2\setminus X_1$ as in Eq.~\eqref{set_X_2X_2_Tilde_X}.
We hence upper-bound the norm~\eqref{norm_estiamation_bar_Pi_ast_h_X1_2} as 
\begin{align}
\label{norm_estiamation_bar_Pi_ast_h_X1_2_upp}
\norm{ \bar{\Pi}^\ast \partial H_{0,\tau-\tau_1} \bar{\Pi}^{\ast\co}U_{\tau_1} U_{\tau}^\dagger O_{X_2,\tau} U_{\tau}\sqrt{\rho_0(t_1) }  }_F
\le 4\gamma \bar{J} N_0 |X_2|  \norm{\bar{\Pi}^{\ast\co} U_{\tau_1} U_{\tau}^\dagger O_{X_2,\tau} U_{\tau} \sqrt{\rho_0(t_1)} }_F .
\end{align}
Thus, we only have to estimate the norm~\eqref{norm_estiamation_bar_Pi_ast_h_X1_1}.

We first reduce the norm~\eqref{norm_estiamation_bar_Pi_ast_h_X1_1} to 
\begin{align}
\label{norm_estiamation_bar_Pi_ast_h_X1_1_red}
\norm{\bar{\Pi}^{\ast\co}U_{\tau_1} U_{\tau}^\dagger O_{X_2,\tau} U_{\tau} \sqrt{\rho_0(t_1)} }_F 
=& \norm{\bar{\Pi}^{\ast\co} 
e^{i \br{H_1 + H_2}\tau_1}e^{-iH\tau_1}e^{iH\tau}e^{-i \br{H_1 + H_2}\tau}O_{X_2,\tau}e^{i \br{H_1 + H_2} \tau}e^{-iH\tau} \sqrt{\rho_0(t_1)} }_F \notag \\
=& \norm{\bar{\Pi}^{\ast\co}e^{i \br{H_1 + H_2}\tau_1}e^{iH(\tau-\tau_1)}O_X e^{-iH\tau} \sqrt{\rho_0(t_1)} }_F ,
\end{align}
where in the first equation, we use from Eqs.~\eqref{The decomposition of_small time evolution} and \eqref{def_U_tau_'_1}
\begin{align}
U_{\tau'}=\mathcal{T} e^{-i\int_0^{\tau'}  \partial H_0 \br{H_1 + H_2,x} dx}=e^{i \br{H_1 + H_2} \tau'} e^{-iH\tau'} , \quad H=H_1+H_2 +\partial H_0 ,
\end{align}
and the second equation is obtained from the definition of $O_X(H_1+H_2,\tau)=O_X(H_1,\tau) =: O_{X_2,\tau}$.
By applying the definition~\eqref{def:bar_pi_ast_eq} to Eq.~\eqref{norm_estiamation_bar_Pi_ast_h_X1_1_red}, we have
\begin{align}
\label{norm_estiamation_bar_Pi_ast_h_X1_1_red_2}
&\norm{\bar{\Pi}^{\ast\co}e^{i \br{H_1 + H_2}\tau_1}e^{iH(\tau-\tau_1)}  O_X  e^{-iH\tau} \sqrt{\rho_0(t_1)} }_F \notag \\
&\le \sum_{i\in \tilde{X}}\norm{\Pi_{i[\ell_0], > Q} e^{i \br{H_1 + H_2}\tau_1}   e^{iH(\tau-\tau_1)} O_X  e^{-iH\tau} \sqrt{\rho_0(t_1)} }_F.
\end{align}

For the estimation of the norm of
$$
\norm{\Pi_{i[\ell_0] , > Q}  e^{i \br{H_1 + H_2}\tau_1} e^{iH(\tau-\tau_1)} O_X  e^{-iH\tau} \sqrt{\rho_0(t_1)} }_F 
=\norm{\Pi_{i[\ell_0], > Q}  e^{i \br{H_1 + H_2}\tau_1} e^{iH(\tau-\tau_1)} O_X  e^{-iH\tau}e^{-iHt_1} \sqrt{\rho_0} }_F  ,
$$
we can apply the inequality~\eqref{lemm:boson_density_prob_gene_time_evo/main01} in Lemma~\ref{lemm:boson_density_prob_gene_time_evo} 
from $t_1+\tau \le t$.
In the lemma, the boson creation by the operator $O_{X_0}$ is assumed to be smaller than or equal to $q_0$ (see \eqref{O_X_condition/_norm_lemma_operator} for the formulation), which also holds for $O_X$ from the assumption of the main statement in Proposition~\ref{prop:short_time_LR_2D-}. 
Moreover, from the assumption of $\ell_0\ge \bar{\ell}_{t,R}$ and $\dist_{i,X} \ge 2\ell_0$ for $i\in \tilde{X} \ (=X[2\ell_0+k] \setminus X[2\ell_0])$, 
we have $i[\ell_0] \subset X[\ell_0]^\co \subseteq X[\bar{\ell}_{t,R}]^\co$. 
Therefore, by using the inequality~\eqref{lemm:boson_density_prob_gene_time_evo/main01} with $p=0$  obtain 
\begin{align}
\label{lemm:boson_density_prob_reappear}
\norm{\Pi_{i[\ell_0], > Q}  e^{i \br{H_1 + H_2}\tau_1} e^{iH(\tau-\tau_1)} O_X  e^{-iH\tau} \sqrt{\rho_0(t_1)} }_F^2
\le \zeta^2 e^{- \brr{Q/(4 b_0 \ell_0^D) }^{1/\kappa}}   .
\end{align}

By applying the inequality~\eqref{lemm:boson_density_prob_reappear} to~\eqref{norm_estiamation_bar_Pi_ast_h_X1_1_red} and \eqref{norm_estiamation_bar_Pi_ast_h_X1_1_red_2}, we obtain
\begin{align}
\label{norm_estiamation_bar_Pi_ast_h_X1_1_red_3}
&\norm{\bar{\Pi}^{\ast\co}U_{\tau_1} U_{\tau}^\dagger O_{X_2,\tau} U_{\tau} \sqrt{\rho_0(t_1)} }_F 
=\norm{\bar{\Pi}^{\ast\co}e^{i \br{H_1 + H_2}\tau_1} e^{iH(\tau-\tau_1)} O_X  e^{-iH\tau} \sqrt{\rho_0(t_1)} }_F 
\le |\tilde{X}| \zeta e^{- \brr{Q/(4 b_0 \ell_0^D) }^{1/\kappa}/2}   ,
\end{align}
which also yields from~\eqref{norm_estiamation_bar_Pi_ast_h_X1_2_upp}
\begin{align}
\label{norm_estiamation_bar_Pi_ast_h_X1_2_upp/2}
\norm{ \bar{\Pi}^\ast \partial H_{0,\tau_1} \bar{\Pi}^{\ast\co} U_{\tau-\tau_1}^\dagger O_{X_2,\tau} U_{\tau} \sqrt{\rho_0(t_1) }  }_F
\le 4\gamma \bar{J} N_0 |X_2|   \cdot |\tilde{X}| \zeta e^{- \brr{Q/(4 b_0 \ell_0^D) }^{1/\kappa}/2} .
\end{align}
By combining the above two inequalities with Corollary~\ref{corol:effective Hamiltonian_accuracy_depend_t}, we obtain 
\begin{align}
&\norm{ \brr{O_{X_2,\tau} (\partial H_{0,x}, 0\to \tau)- O_{X_2,\tau} (\partial \tilde{H}^\ast_{0,x}, 0\to \tau)} \rho_0(t_1)}_1  \notag \\
\le &4|\tilde{X}| \zeta e^{- \brr{Q/(4 b_0 \ell_0^D) }^{1/\kappa}/2} +8 \gamma \bar{J} N_0 |X_2|  \tau \cdot |\tilde{X}| \zeta e^{- \brr{Q/(4 b_0 \ell_0^D) }^{1/\kappa}/2}  \notag \\
\le & 4\gamma R^D(1 +2\tau \gamma^2 \bar{J}N_0 R^{D})\zeta e^{- \brr{Q/(4 b_0 \ell_0^D) }^{1/\kappa}/2}  ,
\end{align}
where we use $\tilde{X}, X_2 \subset i_0[R]$, which yields $|\tilde{X}|, |X_2| \le  |i_0[R]| \le \gamma R^D$. 
We thus prove the main inequality~\eqref{main_ineq:lemma:time_evo_highD_local_map}. 
This completes the proof. $\square$

 {~}

\hrulefill{\bf [ End of Proof of Lemma~\ref{lemma:time_evo_highD_local_map}] }

{~}

In the following, we aim to estimate the error to approximate $\partial \tilde{H}^\ast_{0,x}$ onto the subset $X_3$ ($=X[3\ell_0+k]$). 
We here denote
\begin{align}
h_{i,j} = J_{i,j} (b_ib_j^\dagger + {\rm h.c.}), \quad   \tilde{h}^\ast_{i,j}\br{H_1 + H_2,x}= \bar{\Pi}^\ast  h_{i,j} \br{H_1 + H_2 ,x }  \bar{\Pi}^\ast .
\end{align}
We then consider each of the interaction terms $\tilde{h}^\ast_{i,j}\br{H_1 + H_2,x}$ that is included in $\partial \tilde{H}^\ast_{0,x}$, i.e.,
\begin{align}
\partial \tilde{H}^\ast_{0,x}  = \bar{\Pi}^\ast e^{i \br{H_1 + H_2} x}  \partial H_0 e^{-i \br{H_1 + H_2} x}\bar{\Pi}^\ast
= \sum_{\ave{i,j}: \{i,j\} \cap \tilde{X} \neq \emptyset } \tilde{h}^\ast_{i,j}\br{H_1 + H_2,x},
\end{align}
where we use Eqs.~\eqref{def:partial_tilde_H_ast_0x} and \eqref{defs:H1_H2_partialH_0}. 
For each of $\tilde{h}^\ast_{i,j}\br{H_1 + H_2,x}$, we aim to estimate the approximation error of
\begin{align}
 \bar{\Pi}^\ast  h_{i,j} \br{H_1 + H_2 ,x }  \bar{\Pi}^\ast \approx  \bar{\Pi}^\ast h_{i,j \to X_3 ,x}  \bar{\Pi}^\ast = \tilde{h}^\ast_{i,j \to X_3 ,x} ,
\end{align}
where the operator $\tilde{h}_{i,j \to X_3,x}$ is an appropriate operator supported on $X_3$. 
Here, from the definitions in Eq.~\eqref{set_X_2X_2_Tilde_X}, we can ensure (see also Fig.~\ref{fig:small_time_Lieb_Robinson_highD.pdf})
$$
\{i,j\}[\ell_0-1] \subset X_3\for \{i,j\} \cap \tilde{X} \neq \emptyset \quad (\dist_{i,j}=1).
$$

By applying Subtheorem~\ref{subtheorem_small_time_evolution_general}, we immediately obtain the following corollary:
\begin{corol} \label{corol::subtheorem_small_time_evolution_general}
Under the constraints on the Hilbert space as in~\eqref{notation_N_0_simplify} and the projection~\eqref{def:bar_pi_ast_eq}, 
there exists an appropriate operator $\tilde{h}_{i,j \to X_3 ,x}$ such that  
\begin{align}
\label{main_ineq:corol:subtheorem_small_time_evolution_general}
\norm{ \brr{h_{i,j} \br{H_1 + H_2 ,x } - \tilde{h}^\ast_{i,j \to X_3 ,x}} \bar{\Pi}^\ast \Pi_{\mathcal{S}(i_0[R],1/2,N_0)} } \le \norm{h_{i,j}} \delta_{\ell_0,Q}^{(2)} \for x\le \tau, 
\end{align}
where $\ell_0\ge\log[\Theta(N_0)]$ and $\delta_{\ell_0,Q}$ is defined as 
\begin{align}
\delta_{\ell_0,Q}^{(2)}:= \Theta(Q^2) \brrr{ e^{-\Theta\br{\ell_0}} + e^{\Theta(\ell_0)}\brr{\Theta(\tau) +\Theta\br{\frac{\tau Q\log(Q)}{\ell_0^2}}  }^{\Theta(\ell_0)}}.
\end{align}
 The statement is proved by letting $O_{Z_0} \to h_{i,j} $ ($r_1=1,\ q_{Z_0}=1)$ in Subtheorem~\ref{subtheorem_small_time_evolution_general}.
\end{corol}

Therefore, by using the notation of 
\begin{align}
\partial \tilde{H}^\ast_{0,X_3,x}= \sum_{\substack{\ave{i,j} \\ \{i,j\} \cap \tilde{X} \neq \emptyset}} \tilde{h}^\ast_{i,j \to X_3 ,x} ,
\end{align}
we obtain 
\begin{align}
&\norm{ \mathcal{T} e^{-i\int_0^\tau    \partial \tilde{H}^\ast_{0,x} dx} - \mathcal{T} e^{-i\int_0^\tau  \partial \tilde{H}^\ast_{0,X_3,x} dx} }  \le\int_0^\tau  \norm{\partial \tilde{H}^\ast_{0,x}  - \partial \tilde{H}^\ast_{0,X_3,x} } dx  \notag \\
&\le \sum_{\substack{\ave{i,j} \\ \{i,j\} \cap \tilde{X} \neq \emptyset}} \int_0^\tau  \norm{    \tilde{h}^\ast_{i,j}\br{H_1+H_2,x} - \tilde{h}^\ast_{i,j \to X_3 ,x} } dx   \notag \\
&= \sum_{\substack{\ave{i,j} \\ \{i,j\} \cap \tilde{X} \neq \emptyset}} \int_0^\tau  \norm{   \bar{\Pi}^\ast \brr{ h_{i,j} \br{H_1 + H_2 ,x } - h_{i,j \to X_3 ,x}}  \bar{\Pi}^\ast \Pi_{\mathcal{S}(i_0[R],1/2,N_0)} } dx \notag \\
&\le \sum_{\substack{\ave{i,j} \\ \{i,j\} \cap \tilde{X} \neq \emptyset}} \tau \norm{h_{i,j}} \delta_{\ell_0,Q}^{(2)}  
\le 4\tau  \gamma \bar{J} N_0 |X_2| \delta_{\ell_0,Q}^{(2)} ,
\label{time_evo_upp_tilde_h_ast_partial_X2_x}
\end{align}
where the equation is a consequence from the constraints of the Hilbert space on $\Pi_{\mathcal{S}(i_0[R],1/2,N_0)}$ as in~\eqref{notation_N_0_simplify}, 
and we use the same analysis as~\eqref{summation_h_Z_Z_in_ineq} in the last inequality.

By using the above inequality, we obtain the approximation of
\begin{align}
\label{estimation_X_2_Xtime_evo_tilde_h_p_X}
 O_{X_2,\tau} (\partial \tilde{H}^\ast_{0,x}, 0\to \tau)  \approx O_{X_2,\tau} (\partial \tilde{H}^\ast_{0,X_3,x}, 0\to \tau)   .
\end{align}
From the inequality of
\begin{align}
\norm { U^\dagger O U - \tilde{U}^\dagger O\tilde{U} } = 
\norm{ U^\dagger O U - U^\dagger O \tilde{U} + U^\dagger O \tilde{U}- \tilde{U}^\dagger O\tilde{U}}
\le 2\norm{O}\cdot \norm{U - \tilde{U}} ,
\end{align}
we obtain the approximation error of \eqref{estimation_X_2_Xtime_evo_tilde_h_p_X} by using the inequality~\eqref{time_evo_upp_tilde_h_ast_partial_X2_x} as 
\begin{align}
\label{estimation_X_2_Xtime_evo_tilde_h_p_X_error}
\norm{ O_{X_2,\tau} (\partial \tilde{H}^\ast_{0,x}, 0\to \tau)  - O_{X_2,\tau} (\partial \tilde{H}^\ast_{0,X_3,x}, 0\to \tau)  } 
&\le 2 \norm{O_X} \cdot \norm{ \mathcal{T} e^{-i\int_0^\tau    \partial \tilde{H}^\ast_{0,x} dx} - \mathcal{T} e^{-i\int_0^\tau  \partial \tilde{H}^\ast_{0,X_3,x} dx} } \notag \\
&\le 8 \zeta\tau  \gamma \bar{J} N_0 |X_2| \delta_{\ell_0,Q}^{(2)} 
\le 8 \zeta\tau  \gamma^2 \bar{J} N_0 R^D \delta_{\ell_0,Q}^{(2)} ,
\end{align}
where we use $|X_2| \le i_0[R] \le \gamma R^D$. 

By combining the inequality~\eqref{main_ineq:lemma:time_evo_highD_local_map} in Lemma~\ref{lemma:time_evo_highD_local_map} with \eqref{estimation_X_2_Xtime_evo_tilde_h_p_X_error}, we finally arrive at the inequality of 
\begin{align}
&\norm{\brr{O_X(\tau)- O_{X_2,\tau} (\partial \tilde{H}^\ast_{0,X_3,x}, 0\to \tau)  }  \rho_0(t_1)}_1 \notag \\
&\le \delta_{\ell_0,Q}^{(1)} + 8 \zeta\tau  \gamma^2 \bar{J} N_0 R^D \delta_{\ell_0,Q}^{(2)}  \notag \\ 
&= \Theta(N_0 R^{2D}) e^{- \Theta\br{Q/\ell_0^D}^{1/\kappa}} + \zeta \Theta(N_0 R^D)  \cdot \Theta(Q^2) \brrr{ e^{-\Theta\br{\ell_0}} + e^{\Theta(\ell_0)}\brr{\Theta(\tau) +\Theta\br{\frac{\tau Q\log(Q)}{\ell_0^2}}  }^{\Theta(\ell_0)}} \notag \\
&= \zeta \Theta(N_0 R^{2D}Q^2)\brrr{ e^{- \Theta\br{Q/\ell_0^D}^{1/\kappa}}+ e^{-\Theta\br{\ell_0}} + e^{\Theta(\ell_0)}\brr{\Theta(\tau) +\Theta\br{\frac{\tau Q\log(Q)}{\ell_0^2}}  }^{\Theta(\ell_0)}} .
\end{align}
Therefore, we can choose the approximate operator for $O_X(\tau)$ as $O_{X_2,\tau} (\partial \tilde{H}^\ast_{0,X_3,x}, 0\to \tau) $. 
Here, the operator $O_{X_2,\tau} (\partial \tilde{H}^\ast_{0,X_3,x}, 0\to \tau) $ does not depend on the interaction terms outside $X_3$, 
and hence we can adopt the operator $O_{X_2,\tau} (\partial \tilde{H}^\ast_{0,X_3,x}, 0\to \tau) $ for the approximation of $O_X(H_{X_3},\tau)$.
Thus, by using the same analysis as~\eqref{ineq:approx_subset_Hamiltonian}, we prove the main inequality~\eqref{main_ineq:prop:short_time_LR_2D-}. 
This completes the proof of Proposition~\ref{prop:short_time_LR_2D-}. $\square$

%
%

\section{Lieb-Robinson bound in one dimensional systems} \label{sec:1D:boson_refined}

In this section, we prove the optimal Lieb-Robinson bound in which the Lieb-Robinson light cone is almost linear as $\tO(t)$.
The main statement is given as follows:

\begin{theorem} \label{main_thm:boson_Lieb-Robinson_main_1D}
We here adopt the same setups as in Theorems~\ref{main_theorem_looser_Lieb_Robinson} and \ref{main_thm:boson_Lieb-Robinson_main}. 
Then, under the condition of
\begin{align}
\label{main_conditon_for_R_r_0_1D}
R'':=R-r_0-\bar{\ell}_{t,R}-1\ge \Theta\brr{t \log^{2} (q_0 r_0 t)}, 
\end{align} 
we obtain the Lieb-Robinson bound as 
\begin{align}
\label{ineq:main_thm:boson_Lieb-Robinson_main_1D}
&\norm{ \brr{ O_{X_0}(t)- O_{X_0}\br{ H_{i_0[R]} ,t} }\rho_0  }_1 \le 
\norm{O_{X_0}} \exp\brrr{-\Theta\br{  \frac{R''}{t \log(R)} }^{1/\kappa} + \Theta[\log(R)]}  .
\end{align} 
\end{theorem}

{\bf Remark.} In the case where $r_0=\orderof{R}$ and $q_0=\orderof{R}$, the RHS of~\eqref{ineq:main_thm:boson_Lieb-Robinson_main_1D} decays with $R$ as long as 
\begin{align}
R \ge \Theta[ t \log^{1+\kappa} (t) ] .
\end{align} 
Therefore, the light cone becomes almost linear as $\tO(t)$ in one-dimensional systems. 

As a related result, Yin and Lucas have proved~\cite{PhysRevX.12.021039} that in one-dimensional systems
\begin{align}
\label{Yin_lucas_upp}
\tr \br{ \brr{O_{X}(t), O_Y} \rho_0  } \le  \br{\frac{\Theta(t)}{\dist_{X,Y}}}^{\Theta(\dist_{X,Y})} 
\end{align} 
under a similar assumption to Assumption~\ref{only_the_assumption_initial} on the initial state $\rho_0$. 
The considered quantity is rather weaker than $\norm{ \brr{O_{X}(t), O_Y} \rho_0}_1$\footnote{For example, we cannot discuss the gate complexity of the digital quantum simulation only from the bound~\eqref{Yin_lucas_upp}.}, 
but the upper bound is stronger in the sense that the commutator decays for $R \ge \Theta(t)$, which means the strictly linear light cone.
Moreover, the provided proof there is much simpler than ours.  
Thus, it leaves a hope to improve the present statement in Theorem~\ref{main_thm:boson_Lieb-Robinson_main_1D} to the better form~\eqref{Yin_lucas_upp} 
with a more straightforward proof.

 \begin{figure}[tt]
\centering
\includegraphics[clip, scale=0.4]{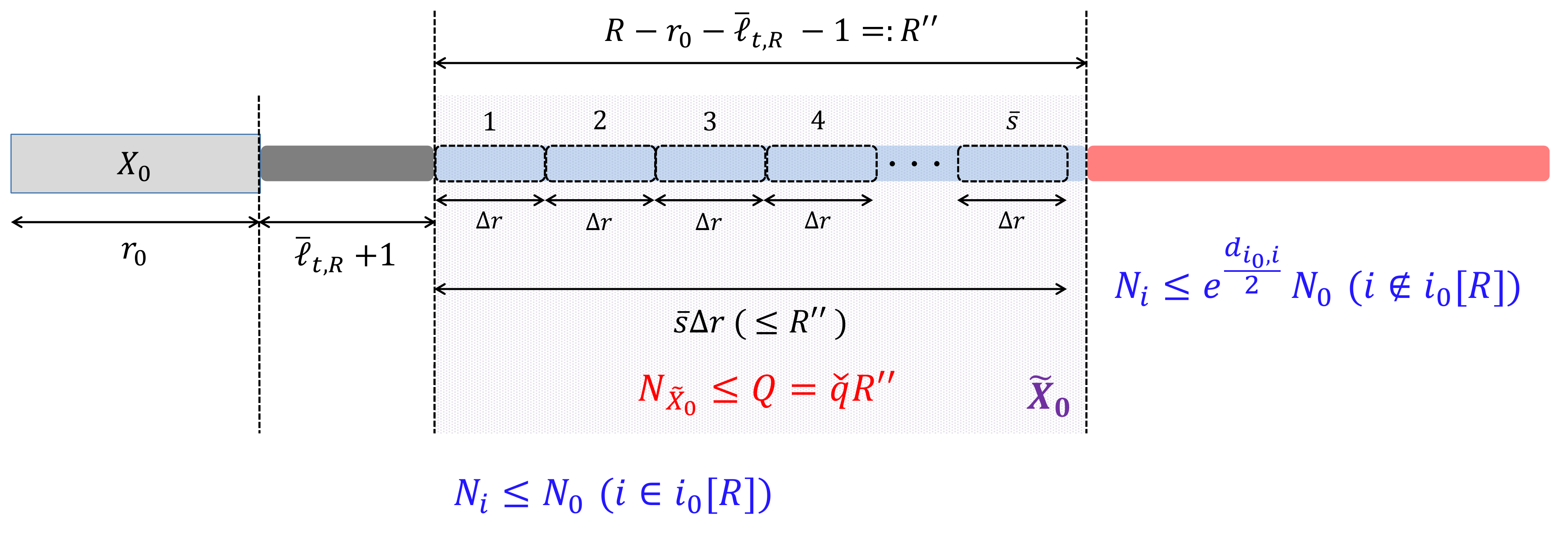}
\caption{Schematic picture of the setup for the proof of Theorem~\ref{main_thm:boson_Lieb-Robinson_main_1D}.
We adopt the projections $\Pi_{\mathcal{S}(i_0[R],1/2,N_0)}$ and $\Pi_{\tilde{X}_0, \le Q}$, where $\tilde{X}_0$ is defined as $\tilde{X}_0 := i_0[R] \setminus X_0[\bar{\ell}_{t,R}+1]$. We restrict ourselves to the Hilbert space spanned by these projections, and the error is estimated in the inequality~\eqref{effective_hamiltonian_error_t^d_2_1D}. 
In the main proof in Sec.~\ref{sec:1d_refined decomposition of the time evolution}, we decompose $\tilde{X}_0$ into $\bar{s}$ subspaces 
and refine the connection of short-time unitary evolutions in~\eqref{Lieb_Robinson_connection_ineq}.  
}
\label{supp_fic_3D_Set_vecN_def.pdf}
\end{figure}

\subsection{Preparation}

As in Sec.~\ref{sec:Proof of Theorem_main_thm:high_dim_preparation}, we restrict the Hilbert space which is spanned by the Fock state on $\mathcal{S}(i_0[R],1/2,N_0)$ ($X_0=i_0[r_0]$, see also Fig.~\ref{supp_fic_3D_Set_vecN_def.pdf}).
In addition, we adopt the projection $\Pi_{\tilde{X}_0, \le Q}$ onto the region $\tilde{X}_0$, where $\tilde{X}_0$ is defined as 
\begin{align}
\label{def_tilde_X_0}
\tilde{X}_0 := i_0[R] \setminus X_0[\bar{\ell}_{t,R}+1] .
\end{align}
In Proposition~\ref{prop:boson_number_truncation_2}, we set 
\begin{align}
\label{bar_q_def_R''_Q}
L \to i_0[R] ,\quad N\to N_0, \quad L'\to  \tilde{X}_0 ,\quad Q \to \check{q} R''  \quad (R''=R-r_0-\bar{\ell}_{t,R}-1) .
 \end{align}
Note that the definitition~\eqref{def_tilde_X_0} implies 
\begin{align}
\dist_{X_0,L'} = \bar{\ell}_{t,R}+1 ,
 \end{align}
and hence the condition~\eqref{cond_for_L_i_1_ell_1} is satisfied.
Here, we have $|\tilde{X}_0|=R''$, and we can replace $\ell_1$ in Eq.~\eqref{cond_for_L_i_1_ell_1} by $R''$, i.e., $\ell_1 \to R''$.
Then, from the inequality~\eqref{effective_hamiltonian_error_t^d___2}, we obtain
\begin{align}
\label{effective_hamiltonian_error_t^d_2_1D}
&\norm{ O_{X_0}(H,t) \rho_0 -  O_{X_0}\br{\tilde{H}_{(i_0[R],N_0,\tilde{X}_0,Q)},t}   \tilde{\Pi}_{(i_0[R],N_0,\tilde{X}_0,Q)} \rho_0 \tilde{\Pi}_{(i_0[R],N_0,\tilde{X}_0,Q)} }_1   \notag \\
&\le  \zeta_0 e^{-R}+ \zeta_0\Theta\br{tQ}e^{-\Theta\br{Q/R''}^{1/\kappa}} 
= \zeta_0 e^{-R}+ \zeta_0\Theta\br{tQ}e^{-\Theta\br{\check{q}}^{1/\kappa}} ,
\end{align}
where the effective Hamiltonian $\tilde{H}_{(i_0[R],N_0,\tilde{X}_0,Q)}$ has been defined in Eq.~\eqref{definition_tilde_L_N_L'_Q}, 
and we choose $N_0$ as in Eq.~\eqref{choice_N_0_upp_highD}, i.e., 
\begin{align}
N_0=\Theta\br{q_0^3, R^{2D+1+\kappa} }  =\Theta\br{q_0^3, R^{\kappa+3} } \quad \for D=1   .
\label{choice_N_0_order_R_1D}
 \end{align}
 We recall that the projection $\tilde{\Pi}_{(i_0[R],N_0,\tilde{X}_0,Q)} $ has been defined in Eq.~\eqref{def_tilde_Pi_(L,N,L',Q)}, i.e., 
 \begin{align}
\label{def_tilde_Pi_(L,N,L',Q)_recall}
\tilde{\Pi}_{(i_0[R],N_0,\tilde{X}_0,Q)} =\Pi_{\tilde{X}_0, \le Q} \tilde{\Pi}_{(i_0[R],N_0)} =
\Pi_{\tilde{X}_0, \le Q} \Pi_{\mathcal{S}(i_0[R],1/2,N_0)}  .
\end{align} 
Also, as in Eq.~\eqref{notation_N_0_simplify}, we adopt the simplified notation of 
\begin{align}
\label{notation_N_0_simplify_1D}
\tilde{H}_{(i_0[R],N_0,\tilde{X}_0,Q)} \to H  ,\quad \tilde{\Pi}_{(i_0[R],N_0,\tilde{X}_0,Q)} \rho_0 \tilde{\Pi}_{(i_0[R],N_0,\tilde{X}_0,Q)} \to \rho_0 . 
\end{align}

\subsection{Main proposition and the proof}

The proof of Theorem~\ref{main_thm:boson_Lieb-Robinson_main_1D} relies on the following proposition:
\begin{prop} \label{prop:1D_Lieb_Robinonsn_ingredient}
Let us choose $\check{q}$ as 
\begin{align}
\label{Choice_check_q_cond_prop}
\check{q} = \Theta\br{  \frac{R''}{t \log(R)} }.
\end{align}
Then, under the notations in Eq.~\eqref{notation_N_0_simplify_1D}, we can obtain the following approximation error
 \begin{align}
\label{main_ineq:prop:1D_Lieb_Robinonsn_ingredient}
\norm{ \left [O_{X_0}(t) -\bar{O}_{i_0[R]} \right] \rho_0 }_1 
&\le \zeta_0 e^{\Theta\brr{t\log(N_0)} - \Theta(R'') }  \for R''\ge \Theta[t\log^{1+\kappa}(N_0)] ,
\end{align}
where we can choose $\bar{O}_{i_0[R]}$ as $O_{X_0}(H_{i_0[R]},t)$ by following the same discussion as in Subtheorem~\ref{subtheorem_small_time_evolution_general} [see also \eqref{ineq:subtheorem_small_time_evolution_general_re}]. 
\end{prop}

By using Proposition~\ref{prop:1D_Lieb_Robinonsn_ingredient} with~\eqref{effective_hamiltonian_error_t^d_2_1D}, we obtain
 \begin{align}
\norm{ \left [O_{X_0}(t) -O_{X_0}(H_{i_0[R]},t)  \right] \rho_0 }_1
&\le\zeta_0 e^{-R}+ \zeta_0\Theta\br{tQ } \exp\brr{-\Theta\br{  \frac{R''}{t \log(R)} }^{1/\kappa}}  + \zeta_0 e^{\Theta\brr{t\log(R)} - \Theta(R'') }  .
\end{align}
Because the second term in the RHS is dominant, we prove the main inequality~\eqref{ineq:main_thm:boson_Lieb-Robinson_main_1D}.
Finally, because $N_0$ is given by Eq.~\eqref{choice_N_0_order_R_1D}, the condition $R''\ge \Theta[t\log^{1+\kappa}(N_0)]$ reduces to~\eqref{main_conditon_for_R_r_0_1D}.
This completes the proof of Theorem~\ref{main_thm:boson_Lieb-Robinson_main_1D}. $\square$

\subsection{Proof of Proposition~\ref{prop:1D_Lieb_Robinonsn_ingredient}: refined decomposition of the time evolution}
\label{sec:1d_refined decomposition of the time evolution}

In the case of one-dimensional systems,  if we use the same approach in Sec.~\ref{sec:Proof_of_Theorem_main_thm:boson_Lieb-Robinson_main}, we cannot obtain the optimal Lieb-Robinson light cone as $R=\tilde{\mathcal{O}}(t)$ (see also Remark in Theorem~\ref{main_thm:boson_Lieb-Robinson_main}).
To refine the bound, we take a similar but different approach.
In this section, we adopt the notations $\tau$, $X_m$ and $\bar{m}$ similar to Eqs.~\eqref{t_decomp_bar_m} and \eqref{Choice_Delta_r_X_m} (see also Fig.~\ref{supp_fic_3D_Set_vecN_def.pdf}):
\begin{align}
&t := \bar{m} \tau , \notag\\
&  X_m:= i_0[r_0+\bar{\ell}_{t,R}+1+ m\Delta r], \label{Choice_Delta_r_X_m_1D}\\
&\Delta r = \left \lfloor \frac{R-(r_0+\bar{\ell}_{t,R}+1)}{\bar{s}} \right \rfloor = \left \lfloor \frac{R''}{\bar{s}} \right \rfloor  \quad (\bar{s} \ge \bar{m}), \label{Choice_Delta_r_X_m_red}\\
&R'':= R-r_0-\bar{\ell}_{t,R}-1,
\end{align}
where $X_{\bar{s}} \subseteq i_0[R]$ and $\bar{s}$ is appropriately chosen [see~\eqref{upper_bound_X_0_m_delta_t_rho_0_proj_insert_fin2} below]. 

Here, we choose $\tau$ as an $\orderof{1}$ constant [$\tau\le1/(4\gamma \bar{J})$], which also implies 
\begin{align}
\bar{m} := t / \tau = \Theta(t) . 
\end{align}
On the other hand, we redefine the parameter $\Delta r$ as follows:
\begin{align}
\Delta r = \left \lfloor \frac{R''}{\bar{s}} \right \rfloor  \quad (\bar{s} \ge \bar{m}),
\end{align}
where 
By using $\Delta r$ and $R$, we choose $\check{q}$ such that
\begin{align}
\label{Choice_check_q_cond}
\check{q} = \Theta\br{  \frac{\Delta r}{\log(R)} },
\end{align}
which reduces to the form of Eq.~\eqref{Choice_check_q_cond_prop} from $\Delta r=\Theta(R''/\bar{s})$ and $\bar{s}=\Theta(t)$. 
From the inequality~\eqref{effective_hamiltonian_error_t^d_2_1D}, $\check{q}$ should be at least larger than $\log^{\kappa}(R)$, or else the upper bound in~\eqref{effective_hamiltonian_error_t^d_2_1D} is worse than the trivial bound. 
Thus, from $\kappa\ge 1$, we add the following additional condition to $\Delta r$ as 
\begin{align}
\label{cond_for_Delta_r_10}
\frac{\Delta r}{\log(R)}\ge \Theta[ \log^{\kappa}(R)] ,
\end{align}
which leads to the choice of $\Delta r$ as
\begin{align}
\label{cond_for_Delta_r_1}
 \Delta r = \Theta[\log^{1+\kappa}(N_0)]  \ge \Theta[\log^{1+\kappa}(R)] .
\end{align}
The above condition yields the condition $R''\ge \Theta[t\log^{1+\kappa}(N_0)]$ in the main inequality~\eqref{main_ineq:prop:1D_Lieb_Robinonsn_ingredient}.

\begin{figure}[tt]
\centering
\includegraphics[clip, scale=0.55]{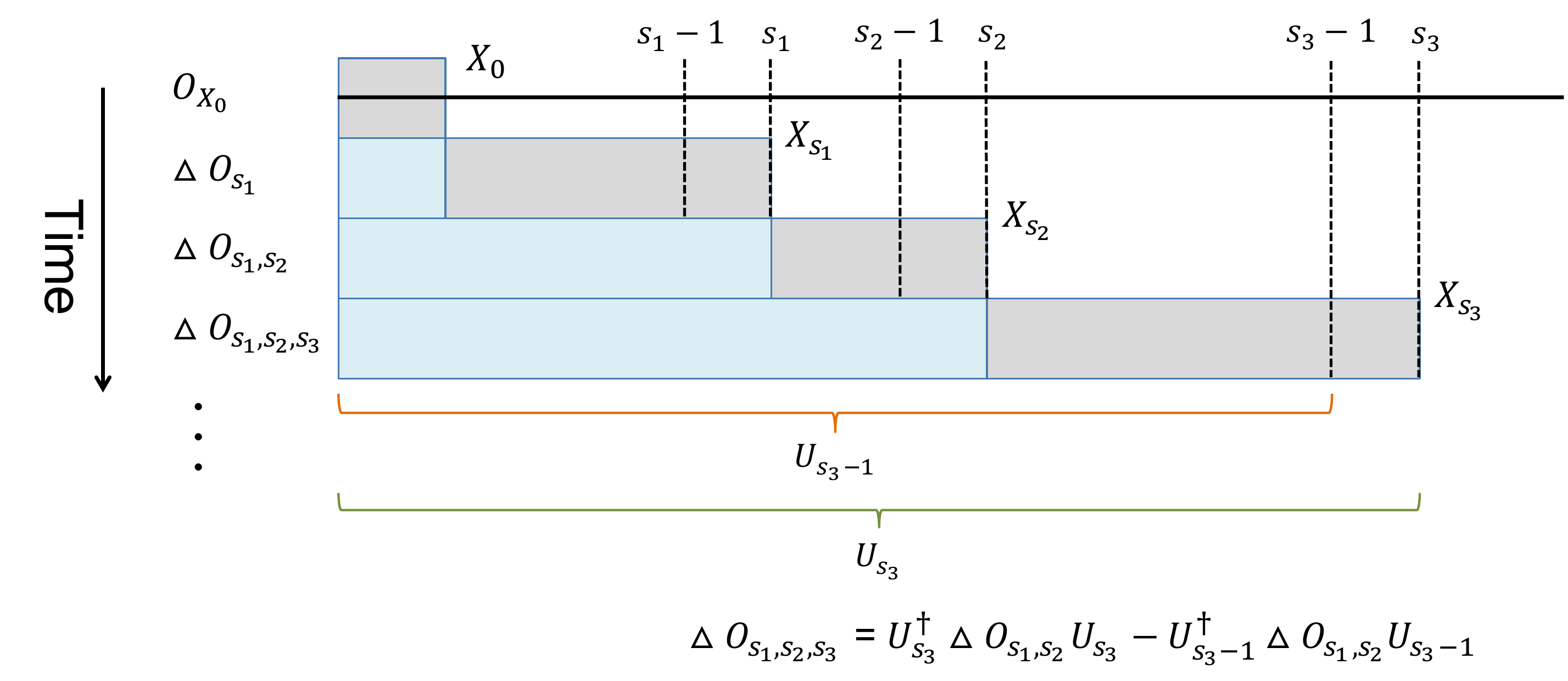}
\caption{Schematic picture of the new decomposition~\eqref{Eq_O_x_0_m_delta_t}.
We start from the time evolution of $O_{X_0}(\tau)$, which is decomposed to the summation of 
$\{\Delta O_{s_1}\}_{s_1=1}^{\bar{s}+1}$ as in Eq.~\eqref{connection_X_0_Delta_t}.
Here, $\Delta O_{s_1}$ is defined by using the unitary operator $U_{s_1}$ [see Eq.~\eqref{Def_U_s_H_Xs}] and is supported on the subset $X_{s_1}$.
In the next step, the time evolution of the decomposed operator, i.e., $\Delta O_{s_1}(\tau)$, is also decomposed to the summation of 
$\{\Delta O_{{s_1, s_2}}\}_{s_2=s_1+1}^{\bar{s}+1}$ as in Eq.~\eqref{connection_X_0_2Delta_t}. 
In the same way, the time evolution of $\Delta O_{{s_1, s_2}}(\tau)$ is further decomposed to $\{\Delta O_{{s_1, s_2,s_3}}\}_{s_3=s_2+1}^{\bar{s}+1}$. By connecting the decompositions, we obtain the full decomposition as in Eq.~\eqref{Eq_O_x_0_m_delta_t}. 
By truncating the summation for $s_1,s_2,\ldots,s_{\bar{m}}$ up to $\bar{s}$, all the decomposed operators $\Delta O_{\bold{s}_{\bar{m}}} $ are supported on the subset $X_{\bar{s}}\subseteq i_0[R]$, and hence we can obtain the refined approximation for the time evolution $O_{X_0}(t)$ [see Eq.~\eqref{Eq_O_x_0_m_delta_t_trn}].
}
\label{supp_fig_refined_decomp}
\end{figure}

In the following, we refine the previous decomposition, which has been given in Eq.~\eqref{Lieb_Robinson_connection_ineq} (see Fig.~\ref{supp_fig_refined_decomp} for the schematic picture of the present decomposition). 
Our purpose here is to derive the equation~\eqref{Eq_O_x_0_m_delta_t} and the inequality~\eqref{upper_bound_X_0_m_delta_t_rho_0}. 
We start from the decomposition of the time-evolved operator $O_{X_0}(\tau)$ as follows:
 \begin{align}
 \label{connection_X_0_Delta_t}
O_{X_0}(\tau) &= \sum_{s_1=1}^{\bar{s}+1}\Delta O_{s_1}  ,
\end{align}
where we define $\Delta O_{s_1}^{(1)}$ as
 \begin{align}
&\Delta O_{1} := U_{1}^{\dagger}O_{X_0} U_{1} , \notag \\
&\Delta O_{s_1}  :=U_{{s_1}}^{\dagger} O_{X_0} U_{{s_1}}  -U_{{s_1-1}}^{\dagger} O_{X_0} U_{{s_1-1}} \quad  (2\le s_1\le \bar{s}), \notag \\
&\Delta O_{\bar{s}+1}  := O_{X_0}(\tau)  -U_{{\bar{s}}}^{\dagger} O_{X_0} U_{{\bar{s}}},
\end{align}
where, for an arbitrary $s$, we denote $U_s$ as the time evolution in the region $X_s$, i.e., 
\begin{align}
\label{Def_U_s_H_Xs}
U_s := e^{-iH_{X_s} \tau}.
\end{align}
In the next step, we consider the decomposition of $\Delta O_{s_1}(\tau)$ as follows:
 \begin{align}
  \label{connection_X_0_2Delta_t}
 \Delta O_{s_1}(\tau) =\sum_{s_2=\min(\bar{s}+1,s_1+1)}^{\bar{s}+1} \Delta O_{s_1, s_2}  ,
\end{align}
where $\Delta O_{{s_1, s_2}}$ ($s_1\le \bar{s}$) is defined as 
 \begin{align}
 \label{def_delta_O_s_1_s_2}
&\Delta O_{s_1, s_1+1} :=U_{{s_1+1}}^{\dagger} \Delta O_{s_1} U_{{s_1+1}} , \notag \\
&\Delta O_{s_1,s_2}  :=U_{{s_2}}^{\dagger} \Delta O_{s_1} U_{{s_2}}
 -U_{{s_2-1}}^{\dagger} \Delta O_{s_1}  U_{{s_2-1}}  \quad  (s_1+2\le s_2\le \bar{s}), \notag \\
 &\Delta O_{s_1 , \bar{s}+1}  := \Delta O_{s_1} (\tau)  -U_{{\bar{s}}}^{\dagger} \Delta O_{s_1} U_{{\bar{s}}}.
\end{align}
Note that we adopt the same definition for $U_s$ as in Eq.~\eqref{Def_U_s_H_Xs}. 
In the case of $s_1=\bar{s}+1$, we define $\Delta O_{{s_1, s_2}}$ as 
 \begin{align}
\Delta O_{{\bar{s}+1, \bar{s}+1}}:= \Delta O_{\bar{s}+1} (\tau)  \for s_1=\bar{s}+1.
\end{align}
By combining Eqs.~\eqref{connection_X_0_Delta_t} and \eqref{connection_X_0_2Delta_t}, we obtain 
 \begin{align}
 \label{connection_X_0_Delta_t_2Delta_t}
O_{X_0}(2\tau) &=\sum_{s_1=1}^{\bar{s}+1}\sum_{s_2=\min(\bar{s}+1,s_1+1)}^{\bar{s}+1} \Delta O_{s_1, s_2}     .
\end{align}

By repeating this process, we recursively define $\Delta O_{{\bold{s}_m}}$ ($\bold{s}_m:=\{s_1,\ldots , s_{m-1},s_m\}$, $s_{m-1}\le \bar{s}$) as 
  \begin{align}
   \label{def_delta_O_vec_s_m_original}
&\Delta O_{\bold{s}_m} 
:= U_{{s_{m-1}+1}}^{\dagger} \Delta O_{\bold{s}_{m-1}}  U_{{s_{m-1}+1}}
\quad  (s_m=s_{m-1}+1), \notag \\
&\Delta O_{\bold{s}_m}  
:=U_{{s_m}}^{\dagger}\Delta O_{\bold{s}_{m-1}} U_{{s_m}}
 -U_{{s_m-1}}^{\dagger} \Delta O_{\bold{s}_{m-1}}  U_{{s_m-1}}  \quad  (s_{m-1}+2\le s_m\le \bar{s}), \notag \\
 &\Delta O_{\bold{s}_m}  := \Delta O_{\bold{s}_{m-1}} (\tau)  -U_{{\bar{s}}}^{\dagger}
 \Delta O_{\bold{s}_{m-1}}U_{{\bar{s}}}\quad  (s_m=\bar{s}+1) .
\end{align} 
In the case of $s_{m-1}=\bar{s}+1$, we define $\Delta O_{\bold{s}_m} $ as 
 \begin{align}
\Delta O_{\bold{s}_m}  := \Delta O_{\bold{s}_{m-1}} (\tau) \for s_{m-1}=s_m=\bar{s}+1. 
\end{align}
By using the above notations, we obtain 
 \begin{align}
  \label{connection_X_0_mDelta_t}
 \Delta O_{\bold{s}_{m-1}}(\tau) =\sum_{s_m=\min(\bar{s}+1,s_{m-1}+1)}^{\bar{s}+1} \Delta O_{\bold{s}_m}   .
\end{align}

By applying Eq.~\eqref{connection_X_0_mDelta_t} iteratively, we have the decomposition of $O_{X_0}(\bar{m} \tau)=O_{X_0}(t)$ as follows:
 \begin{align}
 \label{Eq_O_x_0_m_delta_t}
O_{X_0}(\bar{m} \tau)=O_{X_0}(t) &=\sum_{s_1=1}^{\bar{s}+1} \sum_{s_2=\min(\bar{s}+1,s_1+1)}^{\bar{s}+1} \cdots  \sum_{s_{\bar{m}}=\min(\bar{s}+1,s_{\bar{m}-1}+1)}^{\bar{s}+1} 
\Delta O_{\bold{s}_{\bar{m}}} .
\end{align}
Then, by truncating the summation for each of $\{s_m\}_{m=1}^{\bar{m}}$ up to $\bar{s}$, we define $\bar{O}_{X_{\bar{s}}}$ as 
 \begin{align}
  \label{Eq_O_x_0_m_delta_t_trn}
\bar{O}_{X_{\bar{s}}} :=   \sum_{s_1=1}^{\bar{s}} \sum_{s_2=s_1+1}^{\bar{s}} \cdots  \sum_{s_m=s_{m-1}+1}^{\bar{s}} 
\Delta O_{\bold{s}_{m}}  .
\end{align}
Here, the operator $\bar{O}_{X_{\bar{s}}}$ is supported on the subset $X_{\bar{s}} \subseteq  i_0[R]$ [see Eq.~\eqref{Choice_Delta_r_X_m_1D}]. 
Therefore, we need to upper-bound the norm of 
 \begin{align}
 \label{upper_bound_X_0_m_delta_t_rho_0}
\norm{ \left [O_{X_0}(t) -\bar{O}_{X_{\bar{s}}} \right] \rho_0 }_1
\le \sum_{\bold{s}_{\bar{m}} \in V_{\bar{s}+1}} \| \Delta O_{\bold{s}_{\bar{m}}} \rho_0 \|_1 ,
\end{align}
where $V_{\bar{s}+1}$ is a set of $\bold{s}_{\bar{m}}$ such that at least one $s_p \in \bold{s}_{\bar{m}}$ is equal to $\bar{s}+1$.

In general, for the estimation of $ \| \Delta O_{\bold{s}_{m}} \rho_0 \|_1$ ($1\le m\le \bar{m}$), we need to estimate the norms like 
  \begin{align}
\norm{ \br{U_{{s_m}}^{\dagger}\Delta O_{\bold{s}_{m-1}} U_{{s_m}}
 -U_{{s_m-1}}^{\dagger} \Delta O_{\bold{s}_{m-1}}  U_{{s_m-1}} }\rho_0 }_1 .
\end{align} 
However, we here encounter the following obstacle. 
By using the short-time Lieb-Robinson bound in Proposition~\ref{prop:short_time_LR_2D-}, we can only obtain an upper bound in the form of 
  \begin{align}
\norm{ \br{U_{{s_m}}^{\dagger}\Delta O_{\bold{s}_{m-1}} U_{{s_m}}
 -U_{{s_m-1}}^{\dagger} \Delta O_{\bold{s}_{m-1}}  U_{{s_m-1}} }\rho_0 }_1 \le \norm {\Delta O_{\bold{s}_{m-1}}} \varepsilon_{s_{m-1},s_{m}}  
\end{align}  
instead of 
  \begin{align}
  \label{connection_desire_short_time_LR}
\norm{ \br{U_{{s_m}}^{\dagger}\Delta O_{\bold{s}_{m-1}} U_{{s_m}}
 -U_{{s_m-1}}^{\dagger} \Delta O_{\bold{s}_{m-1}}  U_{{s_m-1}} }\rho_0 }_1 \le \norm {\Delta O_{\bold{s}_{m-1}}\rho_0}_1 \varepsilon_{s_{m-1},s_{m}}  ,
\end{align}
where $\varepsilon_{s_{m-1},s_{m}} $ is a small constant which depends on $s_{m-1}$ and $s_m$. 
If we could obtain the bound as \eqref{connection_desire_short_time_LR}, we can iteratively upper-bound $\norm {\Delta O_{\bold{s}_{m}}\rho_0}_1$ for $m\ge 1$, which may yield the desired inequality~\eqref{main_ineq:prop:1D_Lieb_Robinonsn_ingredient}.  
However, the inequality~\eqref{connection_desire_short_time_LR} does not seem to hold in general. 

In the following sections, we aim to refine the decomposition~\eqref{Eq_O_x_0_m_delta_t} furthermore. 

\subsection{Proof of Proposition~\ref{prop:1D_Lieb_Robinonsn_ingredient}: inserting projections}

 \begin{figure}[tt]
\centering
\includegraphics[clip, scale=0.55]{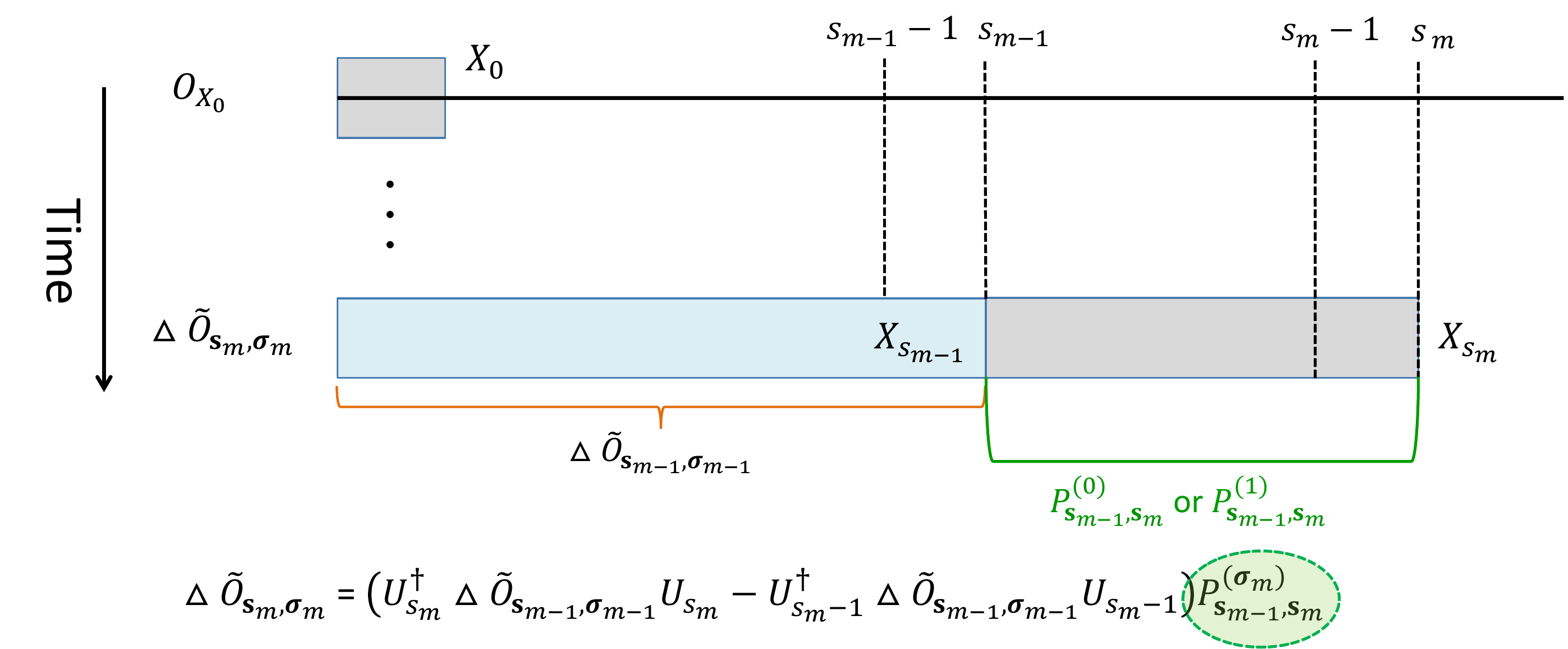}
\caption{New decomposed operator $\Delta \tilde{O}_{\bold{s}_m, \boldsymbol{\sigma}_m}$ in Eq.~\eqref{tilde_O_s_sigma_m}. 
We here adopt the projection operators $P^{(\sigma)}_{s,s'}$ with $0\le s\le s' \le \bar{s}+1$ and $\sigma=0,1$, which is supported on $X_{s'}\setminus X_s$.
The explicit definition of the projection will be given in Eq.~\eqref{def_projection_0_1_L_m_m_s_m-1}. 
In every step from the time $(m-1)\tau$ to $m\tau$, we insert the projection $P^{(0)}_{s_{m-1},s_m}$ or $P^{(1)}_{s_{m-1},s_m}$.
Regarding the projection indices $\boldsymbol{\sigma}_m=\{\sigma_j\}_{j=1}^m$, we redefine the new decomposed operators $\Delta \tilde{O}_{\bold{s}_m, \boldsymbol{\sigma}_m}$ as in Eq.~\eqref{def:tilde_O_Delta_O}. 
}
\label{supp_fig_Delta_tilde_O_def}
\end{figure}

To overcome the difficulty, we introduce a projection operator $P^{(\sigma)}_{s,s'}$ ($0\le s\le s'$, $\sigma=0,1$) such that $P^{(0)}_{s,s'}+P^{(1)}_{s,s'}=1$ which is supported on $X_{s'}\setminus X_s$ (see also Fig.~\ref{supp_fig_Delta_tilde_O_def}). 
By using the projection, we first decompose $\Delta O_{s_1} $ as 
  \begin{align}
\Delta O_{s_1} 
&= \Delta O_{s_1}P^{(0)}_{0,s_1} + \Delta O_{s_1} P^{(1)}_{0,s_1}  =: \Delta \tilde{O}_{s_1,0}+ \Delta \tilde{O}_{s_1,1} . 
\end{align}  
In the same way, we decompose $\Delta O_{s_1,s_2}$ in Eq.~\eqref{def_delta_O_s_1_s_2} as 
 \begin{align}
\Delta O_{s_1,s_2}  &= \sum_{\sigma_1=0,1}
\br{ U_{{s_2}}^{\dagger} \Delta \tilde{O}_{s_1,\sigma_1} U_{{s_2}} 
 -U_{{s_2-1}}^{\dagger} \Delta \tilde{O}_{s_1,\sigma_1} U_{{s_2-1}} } \br{ P^{(0)}_{s_1,s_2} + P^{(1)}_{s_1,s_2}} \notag \\
 &=: \sum_{\sigma_1=0,1}\sum_{\sigma_2=0,1}
\Delta \tilde{O}_{s_1,s_2,\sigma_1,\sigma_2},
\end{align}
where we consider the case of $s_1+2\le s_2\le \bar{s}$ in Eq.~\eqref{def_delta_O_s_1_s_2} for simplicity. 
The above decomposition reduces Eq.~\eqref{connection_X_0_Delta_t_2Delta_t} to
 \begin{align}
 \label{connection_X_0_Delta_t_2Delta_t_proj}
O_{X_0}(2\tau) &=\sum_{s_1=1}^{\bar{s}+1}\sum_{s_2=\min(\bar{s}+1,s_1+1)}^{\bar{s}+1} \sum_{\sigma_1=0,1}\sum_{\sigma_2=0,1} \Delta \tilde{O}_{s_1,s_2,\sigma_1,\sigma_2}   .
\end{align}

By repeating the same procedure, we decompose $\Delta \tilde{O}_{\bold{s}_m}$ as 
 \begin{align}
 \label{def:tilde_O_Delta_O}
&\Delta \tilde{O}_{\bold{s}_m} = \sum_{\boldsymbol{\sigma}_m \in \{0,1\}^{\otimes m}} \Delta \tilde{O}_{\bold{s}_m, \boldsymbol{\sigma}_m} ,\notag \\
&\Delta \tilde{O}_{\bold{s}_m, \boldsymbol{\sigma}_m} :=
\br{U_{{s_m}}^{\dagger}\Delta \tilde{O}_{\bold{s}_{m-1},\boldsymbol{\sigma}_{m-1}} U_{{s_m}}
 -U_{{s_m-1}}^{\dagger} \Delta \tilde{O}_{\bold{s}_{m-1},\boldsymbol{\sigma}_{m-1}}  U_{{s_m-1}} }
 P^{(\sigma_m)}_{s_{m-1},s_m}.
\end{align}
By using the notation of $\Delta \tilde{O}_{\bold{s}_m, \boldsymbol{\sigma}_m}$, we reduce Eqs.~\eqref{Eq_O_x_0_m_delta_t} and \eqref{Eq_O_x_0_m_delta_t_trn} to
 \begin{align}
 \label{tilde_O_s_sigma_m}
&O_{X_0}(t) =\sum_{\boldsymbol{\sigma}_{\bar{m}} \in \{0,1\}^{\otimes \bar{m}}} \sum_{s_1=1}^{\bar{s}+1} \sum_{s_2=\min(\bar{s}+1,s_1+1)}^{\bar{s}+1} \cdots  \sum_{s_{\bar{m}}=\min(\bar{s}+1,s_{\bar{m}-1}+1)}^{\bar{s}+1} 
\Delta \tilde{O}_{\bold{s}_{\bar{m}}, \boldsymbol{\sigma}_{\bar{m}}} , \notag \\
&\bar{O}_{X_{\bar{s}}} = \sum_{\boldsymbol{\sigma}_{\bar{m}} \in \{0,1\}^{\otimes \bar{m}}}   \sum_{s_1=1}^{\bar{s}} \sum_{s_2=s_1+1}^{\bar{s}} \cdots  \sum_{s_{\bar{m}}=s_{\bar{m}-1}+1}^{\bar{s}} 
\Delta \tilde{O}_{\bold{s}_{\bar{m}}, \boldsymbol{\sigma}_{\bar{m}}} .
\end{align}
By using the above equations, we have the approximation of 
 \begin{align}
 \label{upper_bound_X_0_m_delta_t_rho_0_proj_insert}
\norm{ \left [O_{X_0}(t) -\bar{O}_{X_{\bar{s}}} \right] \rho_0 }_1
\le \sum_{\boldsymbol{\sigma}_{\bar{m}} \in \{0,1\}^{\otimes \bar{m}}}  \sum_{\bold{s}_{\bar{m}} \in V_{\bar{s}+1}} \norm{ \Delta \tilde{O}_{\bold{s}_{\bar{m}}, \boldsymbol{\sigma}_{\bar{m}}} \rho_0}_1 .
\end{align}
Remember that $V_{\bar{s}+1}$ has been defined as the set of $\bold{s}_{\bar{m}}$ such that at least one $s_p \in \bold{s}_{\bar{m}}$ is equal to $\bar{s}+1$.
We aim to upper-bound the norm~$\norm{ \left [O_{X_0}(t) -\bar{O}_{X_{\bar{s}}} \right] \rho_0 }_1$ based on this decomposition by appropriately defining the projection $P^{(\sigma)}_{s,s'}$.

 \begin{figure}[tt]
\centering
\includegraphics[clip, scale=0.6]{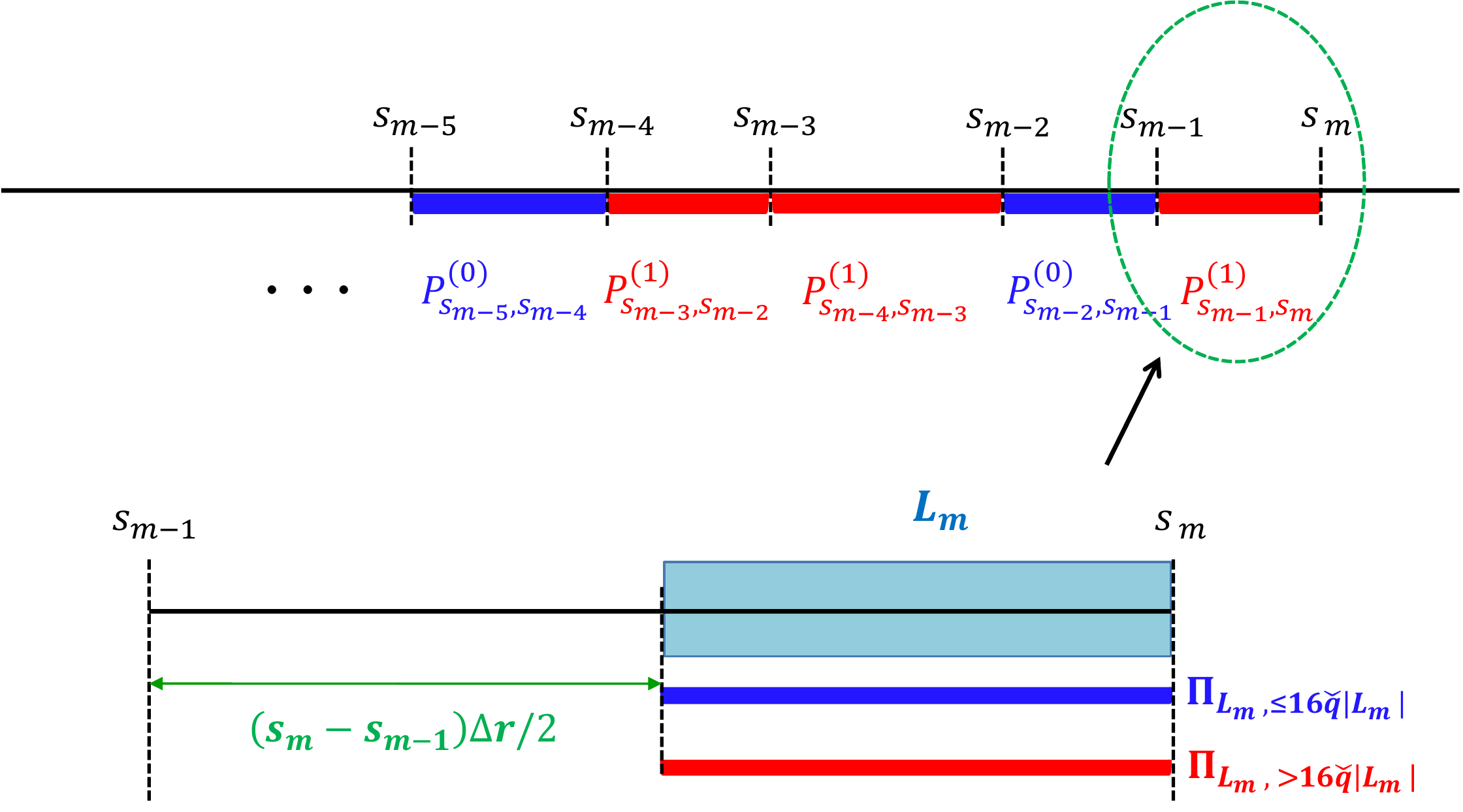}
\caption{Schematic picture of the definition~\eqref{def_projection_0_1_L_m_m_s_m-1} for the projections $P_{s_{m-1},s_m}^{(0)}$ and $P_{s_{m-1},s_m}^{(1)}$. Each of the projections characterizes the number of bosons in the region $L_m$, which is the half region between $X_{s_m}$ and $X_{s_{m-1}}$, i.e., $L_m=X_{s_{m}} \setminus X_{s_{m-1}} [(s_m-s_{m-1})\Delta r /2]$.
Here, the projection $P_{s_{m-1},s_m}^{(0)}$ implies a small boson number in the region $L_m$, while the projection $P_{s_{m-1},s_m}^{(1)}$ implies a large boson number. 
}
\label{supp_fig_projection_def.pdf}
\end{figure}

\subsection{Proof of Proposition~\ref{prop:1D_Lieb_Robinonsn_ingredient}: estimation of the decomposed operators $\Delta \tilde{O}_{\bold{s}_m, \boldsymbol{\sigma}_m}$}

In the following, we aim to upper-bound the norm of
 \begin{align}
\norm{ \Delta \tilde{O}_{\bold{s}_m, \boldsymbol{\sigma}_m} }
\end{align}
for arbitrary fixed choices of $\bold{s}_m$ ($\in V_{\bar{s}+1}$) and $\boldsymbol{\sigma}_m$. 
We here choose the projection operator $P_{s_{m-1},s_m}^{(\sigma_m)}$ as follows (see Fig.~\ref{supp_fig_projection_def.pdf}):
 \begin{align}
 \label{def_projection_0_1_L_m_m_s_m-1}
&P_{s_{m-1},s_m}^{(0)}= \Pi_{L_m , \le 16\check{q} |L_m|} ,\notag \\
&P_{s_{m-1},s_m}^{(1)}= \Pi_{L_m, >  16\check{q} |L_m|}, \notag \\
&L_m := X_{s_{m}} \setminus X_{s_{m-1}} [(s_m-s_{m-1})\Delta r /2],
\end{align}
where the constant $\check{q}$ is an appropriately defined parameter as in Eq.~\eqref{Choice_check_q_cond_prop}.
We recall that $\Pi_{L_m , \le 16\check{q} |L_m|}$ has been defined as a projector onto the eigenspace of $\nb_{L_m}$ with the eigenvalues smaller than $16\check{q} |L_m|$.
Here, $L_m$ is a half region from the point $s_{m-1}$ to $s_m$ (see Fig.~\ref{supp_fig_projection_def.pdf}). 
Roughly speaking, the projection $P_{s_{m-1},s_m}^{(0)}$ restricts the boson number in the region $L_m$, while the projection $P_{s_{m-1},s_m}^{(1)}$ permits unlimited number of bosons in the region $L_m$. 

\begin{figure}[tt]
\centering
\subfigure[Many low-boson-density projections]
{\includegraphics[clip, scale=0.55]{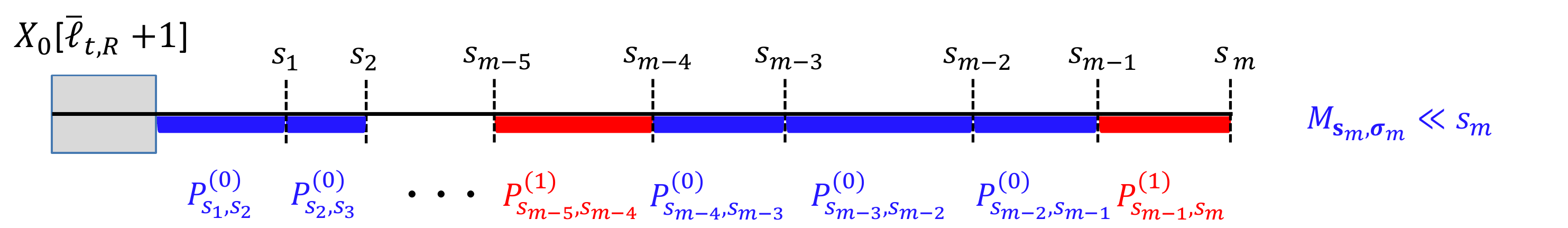}}
\subfigure[Many high-boson-density projections]
{\includegraphics[clip, scale=0.55]{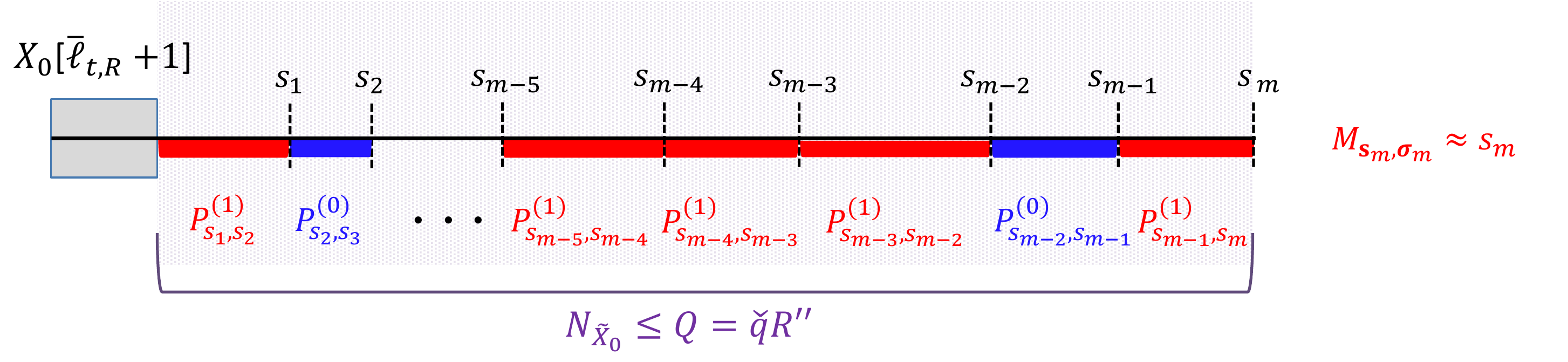}}
\caption{The two cases of the projection patterns $\{P_{s_{j-1},s_j}^{(\sigma_j)}\}_{j=1}^m$. 
Now, the parameter $M_{\bold{s}_m, \boldsymbol{\sigma}_m}= \sum_{j=1}^{m} \sigma_j  (s_j -s_{j-1})$ in Eq.~\eqref{Def_M_vec_sm} is used to characterize the ratio of the region with low (or high)-boson density. Note that $0\le M_{\bold{s}_m, \boldsymbol{\sigma}_m} \le s_m$.  
In the first case (a), the majority of the projections have the index $\sigma_j=0$; that is,  we have $M_{\bold{s}_m, \boldsymbol{\sigma}_m} \ll s_m$. 
We then apply the short-time Lieb-Robinson bound in Proposition~\ref{subtheorem_small_time_evolution_general_eff} to the time evolution with the projections $\{P_{s_{j-1},s_j}^{(\sigma_j)}\}_{j=1}^m$ such that $\sigma_j=0$, which yields Lemma~\ref{upper_bound_Delta_O_vec_s} and Corollary~\ref{corol:upper_bound_Delta_O_vec_s}. 
On the other hand, in the second case (b), the majority of the projections have the index $\sigma_j=1$, i.e., $M_{\bold{s}_m, \boldsymbol{\sigma}_m} \approx s_m$.
In this case, it means that so many bosons exist in the region $X_{s_m} \setminus X_0[\bar{\ell}_{t,R}+1]$. However, we have truncated the number of bosons in the region $\tilde{X}_0:=i_0[R]\setminus X_0[\bar{\ell}_{t,R}+1]$ by $Q$ as in Eq.~\eqref{notation_N_0_simplify_1D}. This allows us to efficiently upper-bound the norm $\norm{ \Delta \tilde{O}_{\bold{s}_m, \boldsymbol{\sigma}_m} }$ as in Lemma~\ref{lemma:Probability_upper_bound}.
}
\label{fig_low_high_boson_density}
\end{figure}

Based on the above definition, we take the following strategy (see also Fig.~\ref{fig_low_high_boson_density}):
\begin{enumerate}
\item{} The operator $\norm{ \Delta \tilde{O}_{\bold{s}_m, \boldsymbol{\sigma}_m} }$ includes $m$ projections $\{P_{s_{j-1},s_j}^{(\sigma_j)}\}_{j=1}^m$.
If the projection is given by $P_{s_{j-1},s_j}^{(0)}$, we can apply the Lieb-Robinson bound for the short-time evolution (i.e., Proposition~\ref{subtheorem_small_time_evolution_general_eff}) from time $(j-1)\tau$ to $j\tau$ in the region $X_{s_m}\setminus X_{s_{m-1}}$. 
On the other hand, if the projection is given by $P_{s_{j-1},s_j}^{(1)}$, we cannot obtain a meaningful Lieb-Robinson bound.
\item{} If many of the region have the indices $\sigma_j=0$ [Fig.~\ref{fig_low_high_boson_density} (a)], we can ensure that the norm $\norm{ \Delta \tilde{O}_{\bold{s}_m, \boldsymbol{\sigma}_m} }$ is sufficiently small by connecting the short-time Lieb-Robinson bound [see Lemma~\ref{upper_bound_Delta_O_vec_s} and Corollary~\ref{corol:upper_bound_Delta_O_vec_s}]. 
\item{} Conversely, if many of the region have the indices $\sigma_j=1$ [Fig.~\ref{fig_low_high_boson_density} (b)], we have to take another approach to upper-bound the norm $\norm{ \Delta \tilde{O}_{\bold{s}_m, \boldsymbol{\sigma}_m} }$. We have already upper-bounded the total number of bosons in the region $\tilde{X}_0$ by $Q$ as in Eq.~\eqref{notation_N_0_simplify_1D}. Hence, such a constraint allows us to upper-bound $\norm{ \Delta \tilde{O}_{\bold{s}_m, \boldsymbol{\sigma}_m} }$ when many of the region have large boson numbers [see Lemma~\ref{lemma:Probability_upper_bound}]. 
\end{enumerate}

We first consider the case with many low-boson-density projections [Fig.~\ref{fig_low_high_boson_density} (a)]. 
By using Proposition~\ref{subtheorem_small_time_evolution_general_eff}, we prove the following lemma:

\begin{lemma}\label{upper_bound_Delta_O_vec_s}
Let us choose $\check{q}$ as in~\eqref{Choice_check_q_cond}.
Then, we can upper-bound the norm $\norm{ \Delta \tilde{O}_{\bold{s}_m, \boldsymbol{\sigma}_m} } $ by using $\norm{ \Delta \tilde{O}_{\bold{s}_{m-1}, \boldsymbol{\sigma}_{m-1}} }$ as follows:
\begin{align}
  \label{upper_bound_norm_delta_O_vec_s_m}
&\norm{ \Delta \tilde{O}_{\bold{s}_m, \boldsymbol{\sigma}_m} } \le 2\norm{\Delta \tilde{O}_{\bold{s}_{m-1}, \boldsymbol{\sigma}_{m-1}}  }  \for \sigma_m=1 , \notag \\
&\norm{ \Delta \tilde{O}_{\bold{s}_m, \boldsymbol{\sigma}_m} } \le \lambda_Q \delta_{\Delta r}^{s_m -s_{m-1} -2} \norm{\Delta \tilde{O}_{\bold{s}_{m-1}, \boldsymbol{\sigma}_{m-1}}  } 
\for  \sigma_m=0,
\end{align} 
where $\delta_{\Delta r}$ and $\lambda_Q$ are constants such that 
 \begin{align}
 \label{def_delta_Delta_r_lambda_Q}
\delta_{\Delta r} = e^{-\Theta(\Delta r)} ,\quad \lambda_Q = \Theta(Q^4R^D N_0). 
\end{align}
\end{lemma}

\subsubsection{Proof of Lemma~\ref{upper_bound_Delta_O_vec_s}} \label{sec:Proof of Lemma_upper_bound_Delta_O_vec_s}

From the definition~\eqref{def:tilde_O_Delta_O}, the first inequality in~\eqref{upper_bound_norm_delta_O_vec_s_m} is trivial, and hence we need to prove the second inequality.
By using the expression~\eqref{def:tilde_O_Delta_O} and the triangle inequality, we first obtain
 \begin{align}
 \label{tilde_O_Delta_O_upp}
\norm{ \Delta \tilde{O}_{\bold{s}_m, \boldsymbol{\sigma}_m} }
 \le& 
\norm{
\brr{\Delta \tilde{O}_{\bold{s}_{m-1},\boldsymbol{\sigma}_{m-1}}(H, \tau) -  U_{{s_m}}^{\dagger}\Delta \tilde{O}_{\bold{s}_{m-1},\boldsymbol{\sigma}_{m-1}} U_{{s_m}}}
 P^{(\sigma_m)}_{s_{m-1},s_m}}  \notag \\
& +  \norm{
\brr{\Delta \tilde{O}_{\bold{s}_{m-1},\boldsymbol{\sigma}_{m-1}}(H, \tau)
 -U_{{s_m-1}}^{\dagger} \Delta \tilde{O}_{\bold{s}_{m-1},\boldsymbol{\sigma}_{m-1}}  U_{{s_m-1}} }
 P^{(\sigma_m)}_{s_{m-1},s_m}} \notag \\
=& \norm{
\brr{\Delta \tilde{O}_{\bold{s}_{m-1},\boldsymbol{\sigma}_{m-1}}(H, \tau) -  \Delta \tilde{O}_{\bold{s}_{m-1},\boldsymbol{\sigma}_{m-1}} (H_{X_{s_m}}, \tau) }
 P^{(\sigma_m)}_{s_{m-1},s_m}}  \notag \\
& +  \norm{
\brr{\Delta \tilde{O}_{\bold{s}_{m-1},\boldsymbol{\sigma}_{m-1}}(H, \tau)
 - \Delta \tilde{O}_{\bold{s}_{m-1},\boldsymbol{\sigma}_{m-1}}(H_{X_{s_m-1}}, \tau)   } 
 P^{(\sigma_m)}_{s_{m-1},s_m}} ,
\end{align}
where we use the definition~\eqref{Def_U_s_H_Xs} of $U_s$. 
From the abbreviation of the notation as in~\eqref{notation_N_0_simplify_1D}, $H$, $H_{X_{s_m}}$ and $H_{X_{s_m-1}}$ are given by the effective Hamiltonian under the projection $\tilde{\Pi}_{(i_0[R],N_0,\tilde{X}_0,Q)}$. 
For $\sigma_m=0$, we need to estimate 
 \begin{align}
&\norm{\brr{\Delta \tilde{O}_{\bold{s}_{m-1},\boldsymbol{\sigma}_{m-1}}(H, \tau) -  \Delta \tilde{O}_{\bold{s}_{m-1},\boldsymbol{\sigma}_{m-1}} (H_{X_{s_m}}, \tau)}
 P^{(0)}_{s_{m-1},s_m}} \notag \\
=&\norm{\brr{\Delta \tilde{O}_{\bold{s}_{m-1},\boldsymbol{\sigma}_{m-1}}(H,\tau) - \Delta \tilde{O}_{\bold{s}_{m-1},\boldsymbol{\sigma}_{m-1}} (H_{X_{s_m}}, \tau) }
\Pi_{L_m, \le  16\check{q} |L_m|} }   ,
\end{align}
where we use the definition~\eqref{def_projection_0_1_L_m_m_s_m-1} for $P^{(0)}_{s_{m-1},s_m}$.


Because of the abbreviated notation~\eqref{notation_N_0_simplify_1D}, we here consider the effective Hamiltonian which is constructed by the projections $\Pi_{\tilde{X}_0, \le Q}$ and $\Pi_{\mathcal{S}(i_0[R],1/2,N_0)}$.
We hence apply Proposition~\ref{subtheorem_small_time_evolution_general_eff} to $\Delta \tilde{O}_{\bold{s}_{m-1},\boldsymbol{\sigma}_{m-1}}(\tau)$ instead of Subtheorem~\ref{subtheorem_small_time_evolution_general}.
The operator $\Delta \tilde{O}_{\bold{s}_{m-1},\boldsymbol{\sigma}_{m-1}}$ is supported on $X_{s_{m-1}} \subseteq i_0[R]$, and hence the boson creation/annihilation by the operator is upper-bounded by 
\begin{align}
N_0 | X_{s_{m-1}}| \le \gamma R^D N_0  ,\quad {\rm i.e.,} \quad q_{Z_0}\to \gamma R^D N_0 ,
\end{align}
because of the constraint $\mathcal{S}(i_0[R],1/2,N_0)$. 
In the inequality~\eqref{ineq:subtheorem_small_time_evolution_general_eff2_} of Proposition~\ref{subtheorem_small_time_evolution_general_eff}, we set 
\begin{align}
O_{Z_0}\to \Delta \tilde{O}_{\bold{s}_{m-1},\boldsymbol{\sigma}_{m-1}},\quad   \zeta_{Z_0} \to \norm{\Delta \tilde{O}_{\bold{s}_{m-1},\boldsymbol{\sigma}_{m-1}}}, \quad \ell_0 \to  |L_m| ,\quad Q\to 16\check{q} |L_m| , \quad \tilde{L}\to  L_m, \quad \bar{L}=i_0[R] .
\end{align}
Here, the condition $\ell_0 \ge \Theta[\log (q_{Z_0}N_0)]$ in Proposition~\ref{subtheorem_small_time_evolution_general_eff} reduces to $|L_m| \ge \Theta[\log (N_0)]$ because of $q_{Z_0} \le \poly(N_0)$ from Eq.~\eqref{choice_N_0_order_R_1D}.  
We thus obtain
\begin{align}
 \label{applying the subtheorem_1D cases}
&\norm{\brr{\Delta \tilde{O}_{\bold{s}_{m-1},\boldsymbol{\sigma}_{m-1}}(H,\tau) -    \Delta \tilde{O}_{\bold{s}_{m-1},\boldsymbol{\sigma}_{m-1}} (H_{X_{s_m}}, \tau) }
 P^{(0)}_{s_{m-1},s_m}} \notag \\
&\le   \norm{\Delta \tilde{O}_{\bold{s}_{m-1},\boldsymbol{\sigma}_{m-1}}} \Theta(Q^4R^D N_0) \brrr{ e^{-\Theta(|L_m|)} + e^{\Theta(|L_m|)}\brr{\Theta(\tau) +\Theta\br{\frac{\check{q} \tau \log(\check{q} |L_m|)}{|L_m|}}  }^{\Theta(|L_m|)}} .
\end{align}

By using the condition~\eqref{Choice_check_q_cond}, i.e., $\check{q} \le  \Theta[\Delta r/\log(R)]$,  and $|L_m| \ge \Theta(\Delta r)$, we have 
 \begin{align}
\Theta(\tau) +\Theta\br{\frac{\check{q} \tau \log(\check{q} |L_m|)}{|L_m|}} =\Theta(\tau) ,
\end{align}  
which reduces the inequality~\eqref{applying the subtheorem_1D cases} to 
 \begin{align}
 \label{applying the subtheorem_1D cases_2}
&\norm{\brr{\Delta \tilde{O}_{\bold{s}_{m-1},\boldsymbol{\sigma}_{m-1}}(H,\tau) -    \Delta \tilde{O}_{\bold{s}_{m-1},\boldsymbol{\sigma}_{m-1}} (H_{X_{s_m}}, \tau) }
 P^{(0)}_{s_{m-1},s_m}} \notag \\
&\le   \norm{\Delta \tilde{O}_{\bold{s}_{m-1},\boldsymbol{\sigma}_{m-1}}} \Theta(Q^4R^D N_0) e^{-\Theta(|L_m|)} 
\end{align}
under an appropriate choice of $\tau$.
In the same way, we obtain 
 \begin{align}
  \label{applying the subtheorem_1D cases_3}
&\norm{\brr{\Delta \tilde{O}_{\bold{s}_{m-1},\boldsymbol{\sigma}_{m-1}}(H,\tau)
 - \Delta \tilde{O}_{\bold{s}_{m-1},\boldsymbol{\sigma}_{m-1}}(H_{X_{s_m-1}}, \tau)   }
 P^{(0)}_{s_{m-1},s_m}} \notag \\
& \le   \norm{\Delta \tilde{O}_{\bold{s}_{m-1},\boldsymbol{\sigma}_{m-1}}} \Theta(Q^4R^D N_0) e^{-\Theta(|L_m|-\Delta r)} ,
\end{align}
where we use $X_{s_m}=X_{s_m-1}[\Delta r]$.
By applying the inequalities~\eqref{applying the subtheorem_1D cases_2} and \eqref{applying the subtheorem_1D cases_3} to \eqref{tilde_O_Delta_O_upp}, we finally obtain 
 \begin{align}
 \label{tilde_O_Delta_O_upp_fini}
\norm{ \Delta \tilde{O}_{\bold{s}_m, \boldsymbol{\sigma}_m} } 
&\le \norm{\Delta \tilde{O}_{\bold{s}_{m-1},\boldsymbol{\sigma}_{m-1}}} \Theta(Q^4R^D N_0) e^{-\Theta(|L_m|-\Delta r)} \notag\\
&=\Theta(Q^4R^D N_0) e^{-\Theta\brr{(s_m-s_{m-1}-2)\Delta r/2 }}  \norm{\Delta \tilde{O}_{\bold{s}_{m-1},\boldsymbol{\sigma}_{m-1}}}, 
\end{align}
where we use the definition~\eqref{def_projection_0_1_L_m_m_s_m-1} of $L_m$, which implies $|L_m| = (s_m-s_{m-1})\Delta r /2$. 
We thus prove the inequality~\eqref{upper_bound_norm_delta_O_vec_s_m}. $\square$

 {~}

\hrulefill{\bf [ End of Proof of Lemma~\ref{upper_bound_Delta_O_vec_s}] }

{~}

By iteratively applying Lemma~\ref{upper_bound_Delta_O_vec_s} to $\norm{ \Delta \tilde{O}_{\bold{s}_m, \boldsymbol{\sigma}_m} }$, we can immediately prove the following corollary:
\begin{corol}\label{corol:upper_bound_Delta_O_vec_s}
Let us define $M_{\bold{s}_m, \boldsymbol{\sigma}_m}$ as 
 \begin{align}
 \label{Def_M_vec_sm}
M_{\bold{s}_m, \boldsymbol{\sigma}_m} := \sum_{j=1}^{m} \sigma_j  (s_j -s_{j-1}) ,
\end{align}
where we set $s_0=0$. 
We then obtain the upper bound of 
 \begin{align}
 \label{main_ineq:corol:upper_bound_Delta_O_vec_s}
\norm{ \Delta \tilde{O}_{\bold{s}_m, \boldsymbol{\sigma}_m} } \le \zeta_0 \br{2\lambda_Q\delta_{\Delta r}^{-2}}^{m} \delta_{\Delta r }^{s_m- M_{\bold{s}_m, \boldsymbol{\sigma}_m}} \norm{O_{X_0}} ,
\end{align}
where $\zeta_0$ has been defined as $\zeta_0=\norm{O_{X_0}}$.
\end{corol}

{~}\\

\textit{Proof of Corollary~\ref{corol:upper_bound_Delta_O_vec_s}.}
We here define $m_0$ and $m_1$ as the numbers of $\{\sigma_j\}_{j=1}^m$ such that $\sigma_j=0$ and $\sigma_j=1$, respectively. 
Note that $m_0+m_1=m$. 
Then, by recursively applying the inequality~\eqref{upper_bound_norm_delta_O_vec_s_m} to $\norm{ \Delta \tilde{O}_{\bold{s}_m, \boldsymbol{\sigma}_m} }$, we obtain \begin{align}
&\norm{ \Delta \tilde{O}_{\bold{s}_m, \boldsymbol{\sigma}_m} } 
\le 2^{m_1} \cdot  \br{\lambda_Q\delta_{\Delta r}^{-2}}^{m_0}  \delta_{\Delta r }^{M^\co_{\bold{s}_m, \boldsymbol{\sigma}_m}} 
\le  \br{2\lambda_Q\delta_{\Delta r}^{-2}}^{m}  \delta_{\Delta r }^{M^\co_{\bold{s}_m, \boldsymbol{\sigma}_m}} \norm{O_{X_0}}, 
\end{align} 
where we define $M^\co_{\bold{s}_m, \boldsymbol{\sigma}_m}$ as 
 \begin{align}
M^\co_{\bold{s}_m, \boldsymbol{\sigma}_m} := \sum_{j=1}^{m} (1-\sigma_j)  (s_j -s_{j-1}) .
\end{align}
By using $M_{\bold{s}_m, \boldsymbol{\sigma}_m}+ M^\co_{\bold{s}_m, \boldsymbol{\sigma}_m} = \sum_{j=1}^{m}  (s_j -s_{j-1})=s_m-s_0=s_m$ from $s_0=0$, we prove the inequality~\eqref{main_ineq:corol:upper_bound_Delta_O_vec_s}. 
This completes the proof. $\square$

 {~}

\hrulefill{\bf [ End of Proof of Corollary~\ref{corol:upper_bound_Delta_O_vec_s}] }

{~}

From the above corollary, we can ensure that the norm $\norm{ \Delta \tilde{O}_{\bold{s}_m, \boldsymbol{\sigma}_m} }$ decays exponentially as $M_{\bold{s}_m}$ decreases (i.e., $s_m-M_{\bold{s}_m}$ increases). Then, if $M_{\bold{s}_{\bar{m}}, \boldsymbol{\sigma}_{\bar{m}}}\ll s_m$, 
 the RHS of the inequality~\eqref{main_ineq:corol:upper_bound_Delta_O_vec_s} is sufficiently small as long as 
$$
\br{ s_m -M_{\bold{s}_{\bar{m}}, \boldsymbol{\sigma}_{\bar{m}}}} \Delta r  \gtrsim m \log  \br{\lambda_Q\delta_{\Delta r}^{-2}} .
$$
 On the other hand, for $M_{\bold{s}_{\bar{m}}, \boldsymbol{\sigma}_{\bar{m}}}\approx s_m$, the above condition cannot be satisfied unless we set $\Delta r$ sufficiently large beyond the condition $\Delta r =\Theta[\log^2(N_0)]$ as in~\eqref{cond_for_Delta_r_1}. 
To obtain meaningful bound without taking $\Delta r$ so large, we need another analysis to upper-bound $\norm{ \Delta \tilde{O}_{\bold{s}_m, \boldsymbol{\sigma}_m} } $. 
To see the point, let us expand $\Delta \tilde{O}_{\bold{s}_m, \boldsymbol{\sigma}_m}$ in Eq.~\eqref{def:tilde_O_Delta_O}. 
Then, the expanded terms, for example, include  
 \begin{align}
 \label{def:tilde_O_Delta_O_expansion_example}
\br{U_{s_1}   \cdots  U_{s_{m-1}}  U_{s_m} }^\dagger O_{X_0} 
\br{ U_{s_1} P^{(\sigma_{1})}_{0,s_{1}}  \cdots  U_{s_{m-1}} P^{(\sigma_{m-1})}_{s_{m-2},s_{m-1}} U_{s_m} P^{(\sigma_m)}_{s_{m-1},s_m}}.
\end{align}
In more detail, all the expanded terms are given in the form of
\begin{align}
 \label{def:tilde_O_Delta_O_expansion_all}
&\Delta \tilde{O}_{\bold{s}_m, \boldsymbol{\sigma}_m}  \rho_0
=\sum_{\zeta_1,\zeta_2,\ldots,\zeta_m=0,1}  
\br{U_{s_1-\zeta_1}   \cdots  U_{s_{m-1}-\zeta_{m-1}}  U_{s_m-\zeta_m} }^\dagger O_{X_0} \Phi(\bold{s}_m, \boldsymbol{\sigma}_m,\boldsymbol{\zeta}_m) \rho_0 \notag \\
&\Phi(\bold{s}_m, \boldsymbol{\sigma}_m,\boldsymbol{\zeta}_m)
= \prod_{j=1}^m (-1)^{\zeta_j} U_{{s_j - \zeta_j}}  P^{(\sigma_j)}_{s_{j-1},s_j},\quad 
\boldsymbol{\zeta}_m:=\{\zeta_1,\zeta_2 ,\ldots,\zeta_m \}
\end{align}
where the term~\eqref{def:tilde_O_Delta_O_expansion_example} corresponds to the case of $\zeta_j=0$ for $j=1,2,\ldots,m$. 
In the case where $M_{\bold{s}_{\bar{m}}, \boldsymbol{\sigma}_{\bar{m}}}\ll m$, 
the operator $\Phi(\bold{s}_m, \boldsymbol{\sigma}_m,\boldsymbol{\zeta}_m)$ roughly plays a role of projection onto quantum states with large boson number in the region 
$(X_{s_m} \setminus X_0[\bar{\ell_{t,R}+1}]) \subseteq \tilde{X}_0$ [see also Fig.~\ref{fig_low_high_boson_density} (b)].
Therefore, because the number of bosons is truncated up to $Q$ in the region $\tilde{X}_0$, we expect that
 \begin{align}
\norm{ \Phi(\bold{s}_m, \boldsymbol{\sigma}_m,\boldsymbol{\zeta}_m) \rho_0 } \ll 1 \for M_{\bold{s}_{\bar{m}}, \boldsymbol{\sigma}_{\bar{m}}}\approx s_m.
\end{align}
The difficulty stems from the unitary operator $U_{{s_j - \zeta_j}} $ that is inserted between $P^{(\sigma_j)}_{s_{j-1},s_j}$ and $P^{(\sigma_j)}_{s_{j-2},s_{j-1}}$ [see \eqref{upp_probability_upp_start} below].

In the following, we aim to derive a complementary statement to Corollary~\ref{corol:upper_bound_Delta_O_vec_s}, where 
the norm $\norm{ \Delta \tilde{O}_{\bold{s}_m, \boldsymbol{\sigma}_m} }$ decays exponentially as $M_{\bold{s}_m}$ increases.  
Indeed, we can prove the following lemma:
 
\begin{lemma}\label{lemma:Probability_upper_bound}
The norm of $\Delta O_{\bold{s}_{m},\boldsymbol{\sigma}_m}\rho_0$ is bounded from above by
 \begin{align}
\norm{ \Delta O_{\bold{s}_{m},\boldsymbol{\sigma}_m}  \rho_0 }_1 \le 2^{2m}\zeta_0  \exp\brr{-\Theta(1)  \br{\frac{M_{\bold{s}_m, \boldsymbol{\sigma}_m}\check{q}\Delta r }{2} -  \frac{Q}{8}  }},
\label{main_ineq_lemma:Probability_upper_bound}
\end{align}
where $M_{\bold{s}_m, \boldsymbol{\sigma}_m}$ has been defined in Eq.~\eqref{Def_M_vec_sm}. 
\end{lemma}

\subsubsection{Proof of Lemma~\ref{lemma:Probability_upper_bound}} \label{sec:Proof of Lemma_lemma:Probability_upper_bound}

For arbitrary choices of $\boldsymbol{\zeta}_m$, we aim to prove 
 \begin{align}
 \label{main_ineq_lemma:Probability_upper_bound_0}
\norm{ \Phi(\bold{s}_m, \boldsymbol{\sigma}_m,\boldsymbol{\zeta}_m) \rho_0 }_1 \le 2^m \exp\brr{-\Theta(1)  \br{\frac{M_{\bold{s}_m, \boldsymbol{\sigma}_m}\check{q}\Delta r }{2} -  \frac{Q}{8}  }  } ,
\end{align}
which immediately yields the desired inequality~\eqref{main_ineq_lemma:Probability_upper_bound} by counting all the patterns of $\{\boldsymbol{\zeta}_m\}$ in Eq.~\eqref{def:tilde_O_Delta_O_expansion_all}.

\begin{figure}[tt]
\centering
\includegraphics[clip, scale=0.6]{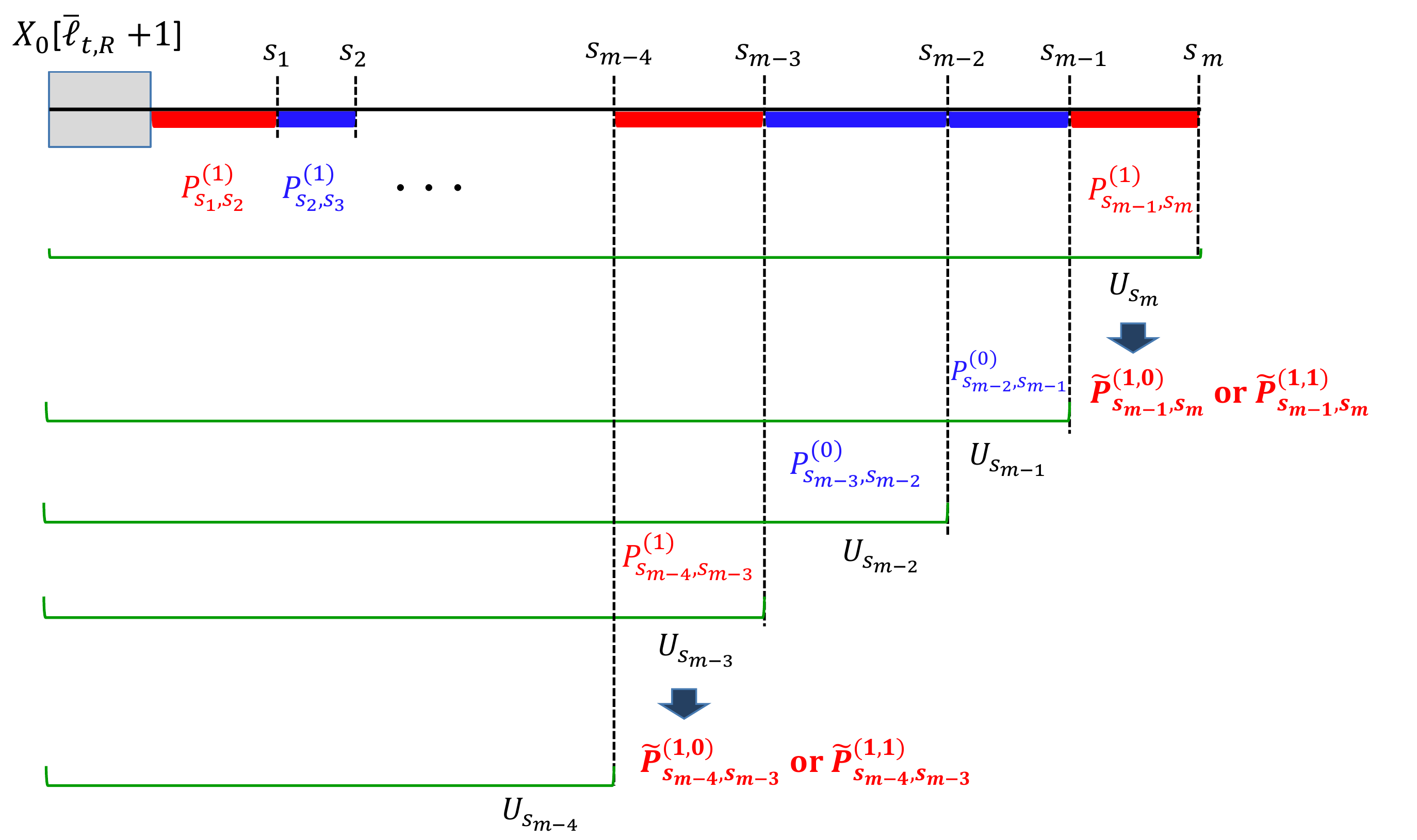}
\caption{Schematic picture of the decomposition of $U_{s_{j}} P^{(1)}_{s_{j-1},s_{j}}$. 
We have mentioned that the operator $\Phi(\bold{s}_m, \boldsymbol{\sigma}_m,\boldsymbol{\zeta}_m)$ in Eq.~\eqref{def:tilde_O_Delta_O_expansion_all} roughly plays a role in projection onto states with a large boson number. However, we cannot apply this intuitive discussion due to the outflow of bosons by the unitary operators $\{U_{s_j}\}_{j=1}^m$. To identify the influence of the unitary operators quantitatively, we introduce new projections $\tilde{P}_{s_{j-1},s_{j}}^{(1,0)}$ and $\tilde{P}_{s_{j-1},s_{j}}^{(1,1)}$ after $U_{s_{j}} P^{(1)}_{s_{j-1},s_{j}}$. 
The projection $\tilde{P}_{s_{j-1},s_{j}}^{(1,0)}$ characterizes the case that the over half of the bosons outflow from the region $X_{s_j} \setminus X_{s_{j-1}}$, while the projection $\tilde{P}_{s_{j-1},s_{j}}^{(1,1)}$ characterizes the case that many of the bosons still remains in the region $X_{s_j} \setminus X_{s_{j-1}}$. 
}
\label{supp_fig_high_boson_case_proof}
\end{figure}

We here consider the case of $\zeta_j=0$ for $j=1,2,\ldots,m$ for simplicity, but the same calculations are applied to the other cases.
We then need to consider the norm of
 \begin{align}
 \label{upp_probability_upp_start}
\norm{  U_{s_1} P^{(\sigma_{1})}_{0,s_{1}}  \cdots  U_{s_{m-1}} P^{(\sigma_{m-1})}_{s_{m-2},s_{m-1}} U_{s_m} P^{(\sigma_m)}_{s_{m-1},s_m}  \rho_0}_1.
\end{align}
If we simply consider the product of the projections $\{P^{(\sigma_j)}_{s_{j-1},s_{j}}\}_{j=1}^m$, it becomes a projection onto a region $X_m\setminus X_0[\bar{\ell}_{t,R}+1]$ with a large boson number.
However, due to the unitary operators $\{U_{s_j}\}_{j=1}^m$, such a simple discussion cannot be applied. 
For example, let us consider the case of $\sigma_m=1$ (see also Fig.~\ref{supp_fig_high_boson_case_proof}).
Then, by the projection $P^{(1)}_{s_{m-1},s_m}$, a quantum state $P^{(1)}_{s_{m-1},s_m}\ket{\psi}$ has a large boson number in the region $X_{s_m}\setminus X_{s_{m-1}}$. However, because of the unitary operator $U_{s_m}$, the bosons in $X_{s_m}\setminus X_{s_{m-1}}$ may outflow from the region, and hence the quantum state $U_{s_m} P^{(1)}_{s_{m-1},s_m}\ket{\psi}$ may no longer have many bosons in $X_{s_m}\setminus X_{s_{m-1}}$. 

To avoid the above difficulty, we decompose $U_{s_{j}} P^{(1)}_{s_{j-1},s_{j}}$ as follows:
 \begin{align}
U_{s_{j}} P^{(1)}_{s_{j-1},s_{j}} = \br{\tilde{P}_{s_{j-1},s_{j}}^{(1,0)} +\tilde{P}_{s_{j-1},s_{j}}^{(1,1)} }  U_{s_{j}} P^{(1)}_{s_{j-1},s_{j}} ,
\end{align}
where $\tilde{P}_{s_{j-1},s_{j}}^{(1,0)}$ and $\tilde{P}_{s_{j-1},s_{j}}^{(1,1)}$ are defined as follows (see Fig.~\ref{supp_fig_high_boson_case_proof}):
 \begin{align}
 \label{def_projection_0_1_L_m_m_s_m-1_10_11}
&\tilde{P}_{s_{j-1},s_{j}}^{(1,0)} = \Pi_{X_{s_j} \setminus X_{s_{j-1}} , \le  8\check{q} |L_j|} , \notag \\
&\tilde{P}_{s_{j-1},s_{j}}^{(1,1)}=\Pi_{X_{s_j} \setminus X_{s_{j-1}} , >  8\check{q} |L_j|}, \quad  8\check{q} |L_j|  = 4\check{q} \Delta r (s_j-s_{j-1})  ,
\end{align}
where $L_j$ has been defined in Eq.~\eqref{def_projection_0_1_L_m_m_s_m-1}, i.e., $L_j := X_{s_{j}} \setminus X_{s_{j-1}} [(s_j-s_{j-1})\Delta r /2]$. 
Note that original projection $P^{(1)}_{s_{j-1},s_{j}}$ ensures at least $16\check{q} |L_j|$ bosons in the region $L_j$ as in Eq.~\eqref{def_projection_0_1_L_m_m_s_m-1}. 
Hence, the projection $\tilde{P}_{s_{j-1},s_{j}}^{(1,0)}$ implies that over half of the bosons outflows from the region $X_{s_{j}} \setminus X_{s_{j-1}}$, 
while  the projection $\tilde{P}_{s_{j-1},s_{j}}^{(1,1)}$ means that many of the bosons still remains in the region $X_{s_{j}} \setminus X_{s_{j-1}}$. 

Because the projection $\tilde{P}_{s_{j-1},s_{j}}^{(1,0)}$ is supported on $X_{s_j} \setminus X_{s_{j-1}}$, it commutes with 
$U_{s_{j'}} P^{(\sigma_{j'})}_{s_{j'-1},s_{j'}}$ for $\forall j' \le j-1$ (see also Fig.~\ref{supp_fig_high_boson_case_proof}):
 \begin{align}
 \label{commutator_projection_j_j'01}
\brr{ \tilde{P}_{s_{j-1},s_{j}}^{(1,\theta_j)}, U_{s_{j'}} P^{(\sigma_{j'})}_{s_{j'-1},s_{j'}} }= 0 \for  \forall j'\le j-1,  \quad \theta_j=0,1.
\end{align}
For the fixed $\{\sigma_{j}\}_{j=1}^m$, we define the indices $\{j_1,j_2,\ldots, j_{m_1}\}$ such that 
 \begin{align}
 \label{def_j_1_j_m1}
\sigma_{j_1}= \sigma_{j_2} =\cdots = \sigma_{j_{m_1}} =1 .  
\end{align}
To see how to obtain an upper bound of $\norm{  U_{s_1} P^{(\sigma_{1})}_{0,s_{1}}  \cdots  U_{s_{m-1}} P^{(\sigma_{m-1})}_{s_{m-2},s_{m-1}} U_{s_m} P^{(\sigma_m)}_{s_{m-1},s_m}  \rho_0}_1$ using Eq.~\eqref{commutator_projection_j_j'01}, let us consider a simple case of $m=4$ and the norm $\norm{U_{s_1} P^{(0)}_{0,s_{1}}
U_{s_2} P^{(1)}_{s_{1},s_{2}}U_{s_3} P^{(0)}_{s_{2},s_{3}} U_{s_4} P^{(1)}_{s_{3},s_{4}}\rho_0}_1$ ($\sigma_1=\sigma_3=0$ and $\sigma_2=\sigma_4=1$). 
We first obtain
  \begin{align}
&\norm{ U_{s_1} P^{(0)}_{0,s_{1}}
U_{s_2} P^{(1)}_{s_{1},s_{2}}U_{s_3} P^{(0)}_{s_{2},s_{3}} U_{s_4} P^{(1)}_{s_{3},s_{4}}\rho_0}_1=  \norm{ \sum_{\theta_2,\theta_4}  U_{s_1} P^{(0)}_{0,s_{1}}
\tilde{P}_{s_{1},s_{2}}^{(1,\theta_{2})}  U_{s_2} P^{(1)}_{s_{1},s_{2}}U_{s_3} P^{(0)}_{s_{2},s_{3}} \tilde{P}_{s_{3},s_{4}}^{(1,\theta_{4})} U_{s_4}  P^{(1)}_{s_{3},s_{4}}  \rho_0}_1 .
\end{align}
Then, because the Hilbert space has been restricted by the projections $\Pi_{\mathcal{S}(i_0[R],1/2,N_0)}$ and $\Pi_{\tilde{X}_0, \le Q}$ [see Eq.~\eqref{notation_N_0_simplify_1D}], we have the following two inequalities:
  \begin{align}
& \norm{U_{s_1} P^{(0)}_{0,s_{1}}
\tilde{P}_{s_{1},s_{2}}^{(1,\theta_{2})}  U_{s_2} P^{(1)}_{s_{1},s_{2}}U_{s_3} P^{(0)}_{s_{2},s_{3}} \tilde{P}_{s_{3},s_{4}}^{(1,\theta_{4})} U_{s_4}  P^{(1)}_{s_{3},s_{4}}  \rho_0}_1  \notag \\
&= \norm{\Pi_{\tilde{X}_0, \le Q}  \tilde{P}_{s_{1},s_{2}}^{(1,\theta_{2})}  \tilde{P}_{s_{3},s_{4}}^{(1,\theta_{4})}  
U_{s_1} P^{(0)}_{0,s_{1}} P^{(1)}_{s_{1},s_{2}}  U_{s_2}U_{s_3} P^{(0)}_{s_{2},s_{3}} U_{s_4}  P^{(1)}_{s_{3},s_{4}}  \rho_0}_1  \notag \\
&\le  \norm{\Pi_{\tilde{X}_0, \le Q}  \tilde{P}_{s_{1},s_{2}}^{(1,\theta_{2})}  \tilde{P}_{s_{3},s_{4}}^{(1,\theta_{4})} } 
\cdot \norm{U_{s_1} P^{(0)}_{0,s_{1}}  P^{(1)}_{s_{1},s_{2}}  U_{s_2}U_{s_3} P^{(0)}_{s_{2},s_{3}} U_{s_4}  P^{(1)}_{s_{3},s_{4}}  \rho_0}_1  
\le \norm{\Pi_{\tilde{X}_0, \le Q}  \tilde{P}_{s_{1},s_{2}}^{(1,\theta_{2})}  \tilde{P}_{s_{3},s_{4}}^{(1,\theta_{4})} } 
\end{align}
and 
  \begin{align}
& \norm{U_{s_1} P^{(0)}_{0,s_{1}}
\tilde{P}_{s_{1},s_{2}}^{(1,\theta_{2})}  U_{s_2} P^{(1)}_{s_{1},s_{2}}U_{s_3} P^{(0)}_{s_{2},s_{3}} \tilde{P}_{s_{3},s_{4}}^{(1,\theta_{4})} U_{s_4}  P^{(1)}_{s_{3},s_{4}}  \rho_0}_1  \notag \\
&\le 
\norm{U_{s_1} P^{(0)}_{0,s_{1}}} 
\cdot \norm{\tilde{P}_{s_{1},s_{2}}^{(1,\theta_{2})}  U_{s_2} P^{(1)}_{s_{1},s_{2}} \Pi_{\mathcal{S}(i_0[R],1/2,N_0)}}
\cdot \norm{U_{s_3} P^{(0)}_{s_{2},s_{3}} } 
\cdot \norm{ \tilde{P}_{s_{3},s_{4}}^{(1,\theta_{4})} U_{s_4}  P^{(1)}_{s_{3},s_{4}}\Pi_{\mathcal{S}(i_0[R],1/2,N_0)}} \notag \\
&\le 
\norm{\tilde{P}_{s_{1},s_{2}}^{(1,\theta_{2})}  U_{s_2} P^{(1)}_{s_{1},s_{2}} \Pi_{\mathcal{S}(i_0[R],1/2,N_0)}}
\cdot  \norm{ \tilde{P}_{s_{3},s_{4}}^{(1,\theta_{4})} U_{s_4}  P^{(1)}_{s_{3},s_{4}}\Pi_{\mathcal{S}(i_0[R],1/2,N_0)}},
\end{align}
where we use the commutation relation~\eqref{commutator_projection_j_j'01} in the first inequality.
We then obtain 
  \begin{align}
&\norm{ U_{s_1} P^{(0)}_{0,s_{1}}
U_{s_2} P^{(1)}_{s_{1},s_{2}}U_{s_3} P^{(0)}_{s_{2},s_{3}} U_{s_4} P^{(1)}_{s_{3},s_{4}}\rho_0}_1 \notag \\
&\le  
\sum_{\theta_2,\theta_4}  \min\br{ \norm{\Pi_{\tilde{X}_0, \le Q}  \tilde{P}_{s_{1},s_{2}}^{(1,\theta_{2})}  \tilde{P}_{s_{3},s_{4}}^{(1,\theta_{4})} } , 
\norm{\tilde{P}_{s_{1},s_{2}}^{(1,\theta_{2})}  U_{s_2} P^{(1)}_{s_{1},s_{2}} \Pi_{\mathcal{S}(i_0[R],1/2,N_0)}}
\cdot  \norm{ \tilde{P}_{s_{3},s_{4}}^{(1,\theta_{4})} U_{s_4}  P^{(1)}_{s_{3},s_{4}}\Pi_{\mathcal{S}(i_0[R],1/2,N_0)}}} .\notag 
\end{align}

By generalizing the above inequality, we can derive 
 \begin{align}
 \label{upp_probability_upp_start_upp1}
&\norm{  U_{s_1} P^{(\sigma_{1})}_{0,s_{1}}  \cdots  U_{s_{m-1}} P^{(\sigma_{m-1})}_{s_{m-2},s_{m-1}} U_{s_m} P^{(\sigma_m)}_{s_{m-1},s_m}  \rho_0}_1  \notag \\
&\le \sum_{\theta_{j_1},\theta_{j_2}\cdots\theta_{j_{m_1}} =0,1}
\min\br{\norm{ \prod_{p=1}^{m_1} \tilde{P}_{s_{j_p-1},s_{j_p}}^{(1,\theta_{j_p})} \Pi_{\tilde{X}_0, \le Q}}, 
 \prod_{p=1}^{m_1} \norm{\tilde{P}_{s_{j_p-1},s_{j_p}}^{(1,\theta_{j_p})}  U_{s_{j_p}} P^{(1)}_{s_{j_p-1},s_{j_p}} \Pi_{\mathcal{S}(i_0[R],1/2,N_0)}}   }.
\end{align}
In the following, we consider the terms
 \begin{align}
 \label{seprately_norm_proj_1_and_2}
\norm{ \prod_{p=1}^{m_1} \tilde{P}_{s_{j_p-1},s_{j_p}}^{(1,\theta_{j_p})} \Pi_{\tilde{X}_0, \le Q}} 
\AND
 \prod_{p=1}^{m_1} \norm{\tilde{P}_{s_{j_p-1},s_{j_p}}^{(1,\theta_{j_p})}  U_{s_{j_p}} P^{(1)}_{s_{j_p-1},s_{j_p}} \Pi_{\mathcal{S}(i_0[R],1/2,N_0)}}   ,
\end{align}
separately.

We first consider the first term in~\eqref{seprately_norm_proj_1_and_2}. 
We here define $Q(\theta_1,\ldots, \theta_{m_1})$ as 
 \begin{align}
 \label{Q_theta_1_cdots_theta_m01}
Q(\theta_{j_1},\ldots, \theta_{j_{m_1}}) = \sum_{p=1}^{m_1} 8 \theta_{j_p} \check{q} |L_{j_p} | .
\end{align}
Then, from the definition~\eqref{def_projection_0_1_L_m_m_s_m-1_10_11}, the produce of the projections $\prod_{p=1}^{m_1} \tilde{P}_{s_{j_p-1},s_{j_p}}^{(1,\theta_{j_p})}$ ensures at least more than $Q(\theta_1,\ldots, \theta_{m_1})$ bosons in the region $X_{s_m} \setminus X_0[\bar{\ell}_{t,R}+1] \subseteq \tilde{X}_0$. 
Therefore, we have 
 \begin{align}
\prod_{p=1}^{m_1} \tilde{P}_{s_{j_p-1},s_{j_p}}^{(1,\theta_{j_p})} \Pi_{\tilde{X}_0, \le Q(\theta_1,\ldots, \theta_{m_1})}=0  ,
\end{align}
which yields the equation of
 \begin{align}
 \label{upp_probability_upp_start_upp1_first}
 \norm{ \prod_{p=1}^{m_1} \tilde{P}_{s_{j_p-1},s_{j_p}}^{(1,\theta_{j_p})} \Pi_{\tilde{X}_0, \le Q}}=0 \for  Q(\theta_1,\ldots, \theta_{m_1}) \ge Q .
\end{align}

 \begin{figure}[tt]
\centering
\includegraphics[clip, scale=0.6]{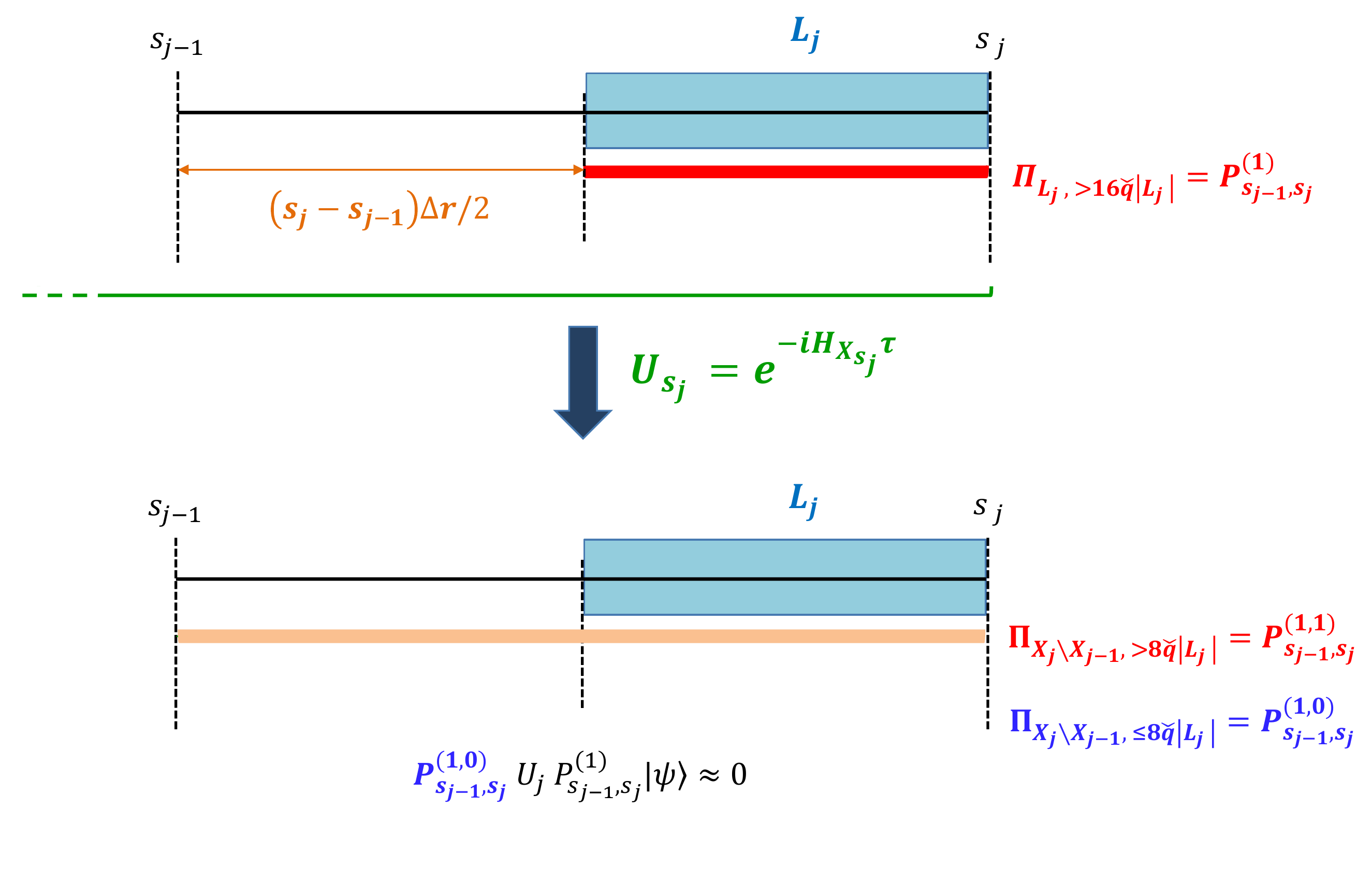}
\caption{Estimation of the norm~\eqref{norm_est_tilde_P_U_s_P_1}, i.e., $||\tilde{P}_{s_{j-1},s_{j}}^{(1,0)}  U_{s_{j}} P^{(1)}_{s_{j-1},s_{j}} \Pi_{\mathcal{S}(i_0[R],1/2,N_0)}||$. 
The setup is similar to the one in Proposition~\ref{prop:lower_probability}. 
We consider the setup that a quantum state initially has more than $16\check{q} |L_j|$ bosons in the region $L_j$. 
Then, the unitary operator $U_{s_j}$ induces the outflow of the bosons from the region $L_j$. 
Proposition~\ref{prop:lower_probability} provides the probability of the number of bosons that still remain in the region $X_{s_j} \setminus X_{s_{j-1}}$.
From the statement, we can ensure that the probability to observe less than $8\check{q} |L_j|$ bosons is smaller than $e^{-\Theta(\check{q} |L_j|)}$ as in the inequality~\eqref{main_ineq:prop:lower_probability_apply___0}. 
}
\label{supp_Apply_boson_flow.pdf}
\end{figure}

We next consider the second term in~\eqref{seprately_norm_proj_1_and_2}. 
In the estimation, we can apply Proposition~\ref{prop:lower_probability} to the norm of (see Fig.~\ref{supp_Apply_boson_flow.pdf})
\begin{align}
\label{norm_est_tilde_P_U_s_P_1}
\norm{\tilde{P}_{s_{j-1},s_{j}}^{(1,0)}  U_{s_{j}} P^{(1)}_{s_{j-1},s_{j}} \Pi_{\mathcal{S}(i_0[R],1/2,N_0)}}  .
\end{align}
For an arbitrary quantum state $\ket{\psi}$, the quantum state $\ket{\tilde{\psi}}=  P^{(1)}_{s_{j-1},s_{j}} \Pi_{\mathcal{S}(i_0[R],1/2,N_0)} \ket{\psi}$ satisfies 
\begin{align}
\Pi_{\mathcal{S}(i_0[R],1/2,N_0)}\ket{\tilde{\psi}}= \ket{\tilde{\psi}},\quad\Pi_{L_j, >  16\check{q} |L_j|}\ket{\tilde{\psi}} = \ket{\tilde{\psi}}.
\end{align}
By using the inequality~\eqref{main_ineq:prop:lower_probability} with 
$$N\to 16\check{q} |L_j|, \quad q_\ast \to N_0,\quad \delta N \to 8\check{q} |L_j|,\quad  X \to L_j ,\quad  \ell \to \frac{(s_j-s_{j-1})\Delta r}{2} , $$
we obtain for the time evolution $U_{s_j} = e^{-iH_{X_{s_j}}\tau}$
\begin{align}
\label{main_ineq:prop:lower_probability_apply___0}
\norm{ \Pi_{X_j\setminus X_{j-1}, \le 8\check{q} |L_j|}U_{s_j} \ket{\tilde{\psi}} }^2 \le \frac{2}{1-e^{-1/(8e c_{\tau,2})}}\exp\brr{\frac{-8\check{q} |L_j|}{8e c_{\tau,2}} + 3c_\ell' \br{1 + c_\ell'  D }\log^2 \br{N_0 \gamma |L_j|} }
\end{align}
as long as the following condition from Eq.~\eqref{main_ineq:prop:lower_probability_ell_cond} is satisfied:
\begin{align}
&\frac{(s_j-s_{j-1})\Delta r}{2} \ge  \ceilf{\frac{2}{\tilde{f}_\tau} \log \br{\frac{5c_{\tau,1} \lambda_{1/2} e^{2\tilde{f}_\tau} \gamma (D/\tilde{f}_\tau)^D N_0  |L_j| }{2c_{\tau,2}} }}= c'_\ell \log \br{N_0 \gamma |L_j|} , \notag \\
&\tilde{f}_\tau \coloneqq  \frac{1}{2} \log \br{1+\frac{1}{10e c_{\tau,1} \lambda_{1/2}  + 2} }. 
\label{cond_for_delta_r_apply}
\end{align}
Because of the condition~\eqref{cond_for_Delta_r_1}, i.e., $\Delta r = \Theta[\log^{1+\kappa}(N_0)]$, the condition~\eqref{cond_for_delta_r_apply} is satisfied.

By applying the inequality~\eqref{main_ineq:prop:lower_probability_apply___0} to the norm~\eqref{norm_est_tilde_P_U_s_P_1}, we have 
\begin{align}
\label{main_ineq:prop:lower_probability_apply}
\norm{\tilde{P}_{s_{j-1},s_{j}}^{(1,0)}  U_{s_{j}} P^{(1)}_{s_{j-1},s_{j}} \Pi_{\mathcal{S}(i_0[R],1/2,N_0)}}=
\sup_{\psi} \norm{ \Pi_{X_j\setminus X_{j-1}, \le 8\check{q} |L_j|}U_{s_j} \ket{\tilde{\psi}} } \le 
e^{- \Theta(1) \check{q}  |L_j| } .
\end{align}
On the other hand, for $\theta_j=1$, we only obtain the trivial inequality of
\begin{align}
\label{main_ineq:prop:lower_probability_apply_trivial}
\norm{\tilde{P}_{s_{j-1},s_{j}}^{(1,\theta_j)}  U_{s_{j}} P^{(1)}_{s_{j-1},s_{j}} \Pi_{\mathcal{S}(i_0[R],1/2,N_0)}}\le 1 \for \theta_j=1.
\end{align}
By applying the inequalities~\eqref{main_ineq:prop:lower_probability_apply} and \eqref{main_ineq:prop:lower_probability_apply_trivial} to the second term in~\eqref{seprately_norm_proj_1_and_2}, we obtain the following upper bound:
\begin{align}
 &\prod_{p=1}^{m_1}  \norm{\tilde{P}_{s_{j_p-1},s_{j_p}}^{(1,\theta_{j_p})}  U_{s_{j_p}} P^{(1)}_{s_{j_p-1},s_{j_p}} \Pi_{\mathcal{S}(i_0[R],1/2,N_0)}}   
 \le \exp\brr{-\Theta(1) \sum_{p=1}^{m_1} \br{1- \theta_{j_p}}\check{q} |L_{j_p}|  }\notag \\
& \le  \exp\brr{-\Theta(1)  \br{\frac{M_{\bold{s}_m, \boldsymbol{\sigma}_m}\check{q} \Delta r }{2} -  \frac{Q(\theta_1,\ldots, \theta_{m_1})}{8}  }  },
 \label{upp_probability_upp_start_upp1_second}
\end{align}
where we use the definition~\eqref{Q_theta_1_cdots_theta_m01} and the inequality of 
\begin{align}
 \sum_{p=1}^{m_1} |L_{j_p}|  = \frac{\Delta r}{2}\sum_{p=1}^{m_1}  (s_{j_p}-s_{j_p-1})
 = \frac{\Delta r}{2}\sum_{j=1}^m \sigma_j  (s_j-s_{j-1}) = \frac{M_{\bold{s}_m, \boldsymbol{\sigma}_m}\Delta r }{2} .
\end{align}
Note that we have defined as in Eqs.~\eqref{def_j_1_j_m1} and~\eqref{Def_M_vec_sm}.

By combining the inequalities~\eqref{upp_probability_upp_start_upp1_first} and \eqref{upp_probability_upp_start_upp1_second}, we reduce the inequality~\eqref{upp_probability_upp_start_upp1} to
 \begin{align}
 \label{upp_probability_upp_start_upp_fin}
&\norm{  U_{s_1} P^{(\sigma_{1})}_{0,s_{1}}  \cdots  U_{s_{m-1}} P^{(\sigma_{m-1})}_{s_{m-2},s_{m-1}} U_{s_m} P^{(\sigma_m)}_{s_{m-1},s_m}  \rho_0}_1  \notag \\
&\le \sum_{\theta_{j_1},\theta_{j_2}\cdots\theta_{j_{m_1}} =0,1}
 \exp\brr{-\Theta(1) \br{\frac{M_{\bold{s}_m, \boldsymbol{\sigma}_m}\check{q} \Delta r }{2} -  \frac{Q}{8}  } }\le 2^m  \exp\brr{-\Theta(1)  \br{\frac{M_{\bold{s}_m, \boldsymbol{\sigma}_m}\check{q}\Delta r }{2} -  \frac{Q}{8}  }} ,
\end{align}
where we use $\sum_{\theta_{j_1},\theta_{j_2}\cdots\theta_{j_{m_1}} =0,1} 1 \le 2^{m_1} \le 2^m$. 
We thus prove the inequality~\eqref{main_ineq_lemma:Probability_upper_bound_0}. 
This completes the proof. $\square$

%

 {~}

\hrulefill{\bf [ End of Proof of Lemma~\ref{lemma:Probability_upper_bound}] }

{~}

\subsection{Proof of Proposition~\ref{prop:1D_Lieb_Robinonsn_ingredient}: completing the proof}

We have finally obtained all the ingredients to prove Proposition~\ref{prop:1D_Lieb_Robinonsn_ingredient}.
By combining Corollary~\ref{corol:upper_bound_Delta_O_vec_s} and Lemma~\ref{lemma:Probability_upper_bound}, we obtain
 \begin{align}
 \label{combination_corol_lemm_delta_O_s_sigma}
\norm{ \Delta O_{\bold{s}_{m},\boldsymbol{\sigma}_m}  \rho_0 }_1 
&\le  \zeta_0  \min\br{2^{2m}\exp\brr{-\Theta(1)  \br{\frac{M_{\bold{s}_m, \boldsymbol{\sigma}_m}\check{q}\Delta r }{2} -  \frac{Q}{8}  }} ,\br{2\lambda_Q\delta_{\Delta r}^{-2}}^{m} \delta_{\Delta r }^{s_m- M_{\bold{s}_m, \boldsymbol{\sigma}_m}}  }   \notag \\
&\le \zeta_0  \max\br{ 2^{2m} \exp\brr{-\Theta(1)  \br{\frac{ \check{q}s_m \Delta r }{4} -  \frac{\check{q} R''}{8}  }} ,\br{2\lambda_Q\delta_{\Delta r}^{-2}}^{m} 
\delta_{\Delta r }^{s_m/2} } ,
\end{align}
where we use $Q=\check{q} R''$ as in Eq.~\eqref{bar_q_def_R''_Q}. 
Note that the condition $0\le  M_{\bold{s}_m, \boldsymbol{\sigma}_m} \le s_m$ from the definition~\eqref{Def_M_vec_sm} implies that either $M_{\bold{s}_m, \boldsymbol{\sigma}_m}$ or $(s_m- M_{\bold{s}_m, \boldsymbol{\sigma}_m})$ is greater than $s_m/2$.  

We now consider the case of $m=\bar{m}$ and $\bold{s}_{\bar{m}} \in V_{\bar{s}+1}$ [see~\eqref{upper_bound_X_0_m_delta_t_rho_0_proj_insert}].
Then, we can replace $s_m$ with $\bar{s}$ in the inequality~\eqref{combination_corol_lemm_delta_O_s_sigma}. 
Also,  because of the definition of $\Delta r$ in Eq.~\eqref{Choice_Delta_r_X_m_red}, we have 
\begin{align}
\Delta r \ge \frac{R''}{\bar{s}} -1 \longrightarrow \bar{s}\Delta r \ge R'' - \bar{s} .
\end{align}
Then, the above inequality reduces the inequality~\eqref{combination_corol_lemm_delta_O_s_sigma} to
 \begin{align}
\norm{ \Delta O_{\bold{s}_{\bar{m}},\boldsymbol{\sigma}_{\bar{m}}}  \rho_0 }_1 
&\le \zeta_0  \max\br{ 2^{2\bar{m}}\exp\brr{\Theta(1)  \br{\frac{ -\check{q} R'' + 2\check{q}\bar{s}  }{8} } } ,
\br{2\lambda_Q\delta_{\Delta r}^{-2}}^{\bar{m}} \delta_{\Delta r }^{\bar{s}/2}} \notag \\
&= \zeta_0  \br{2\lambda_Q\delta_{\Delta r}^{-2}}^{\bar{m}} \delta_{\Delta r }^{\bar{s}/2}
\end{align}
where $\br{2\lambda_Q\delta_{\Delta r}^{-2}}^{\bar{m}} \delta_{\Delta r }^{\bar{s}/2}$ is larger than $e^{-\Theta(\check{q} R'')}$ because of $\delta_{\Delta r }^{\bar{s}/2}= e^{-\Theta(R'')} $ from the definition~\eqref{def_delta_Delta_r_lambda_Q}, i.e., $\delta_{\Delta r }=e^{-\Theta(\Delta r)}$. 
By applying the above inequality to~\eqref{upper_bound_X_0_m_delta_t_rho_0_proj_insert}, we obtain 
 \begin{align}
 \label{upper_bound_X_0_m_delta_t_rho_0_proj_insert_fin}
\norm{ \left [O_{X_0}(t) -\bar{O}_{X_{\bar{s}}} \right] \rho_0 }_1
&\le \sum_{\boldsymbol{\sigma}_{\bar{m}}  \in \{0,1\}^{\otimes \bar{m}}}  \sum_{\bold{s}_{\bar{m}}  \in V_{\bar{s}+1}}\zeta_0  \br{2\lambda_Q\delta_{\Delta r}^{-2}}^{\bar{m}}  \delta_{\Delta r }^{\bar{s}/2}\notag \\
&\le \zeta_0  \brr{4(\bar{s}+1)\lambda_Q\delta_{\Delta r}^{-2}}^{\bar{m}} \delta_{\Delta r }^{\bar{s}/2}.
\end{align}
Finally, by choosing $\bar{s}=8\bar{m}$ in the inequality~\eqref{upper_bound_X_0_m_delta_t_rho_0_proj_insert_fin}, we arrive at the desired inequality of
 \begin{align}
 \label{upper_bound_X_0_m_delta_t_rho_0_proj_insert_fin2}
\norm{ \left [O_{X_0}(t) -\bar{O}_{X_{\bar{s}}} \right] \rho_0 }_1 
&\le \zeta_0 \exp\brrr{\bar{m}\log \brr{4(\bar{s}+1)\lambda_Q}- 2\bar{m} \log(\delta_{\Delta r}) } = \zeta_0 e^{\Theta\brr{t\log(N_0)} - \Theta(R'') }  ,
\end{align}
where we use $\log(\lambda_Q) =\Theta[\log(N_0)]$ from Eq.~\eqref{def_delta_Delta_r_lambda_Q} and 
$\bar{m} \log(\delta_{\Delta r}) = \bar{m} \Theta(\Delta r) = \Theta(R'')$.
This completes the proof of Proposition~\ref{prop:1D_Lieb_Robinonsn_ingredient}. $\square$

\section{Possible Scenario for Acceleration by Time-Independent Hamiltonian}

Finally, we discuss the possibility of realizing the acceleration of information propagation through more natural Hamiltonian dynamics. In the protocol of the main manuscript, we employed a highly artificial Hamiltonian by fine-tuning local interactions. Therefore, considering the standard Bose-Hubbard Hamiltonian with a simple initial state (e.g., the Mott state), it is natural to expect that the linear light cone should be established, and acceleration no longer occurs. Additionally, even if a 1D information path with high boson densities is realized, it should not persist for a long time under Hamiltonian dynamics with repulsive interactions. However, these simple intuitions are extremely challenging to make mathematically rigorous\footnote{Indeed, even for the homogeneous Bose-Hubbard model with the Mott state as the initial state, there have been no mathematical tools to treat information propagation in general.}.

\begin{figure}[tt]
    \centering
    \includegraphics[clip, scale=0.3]{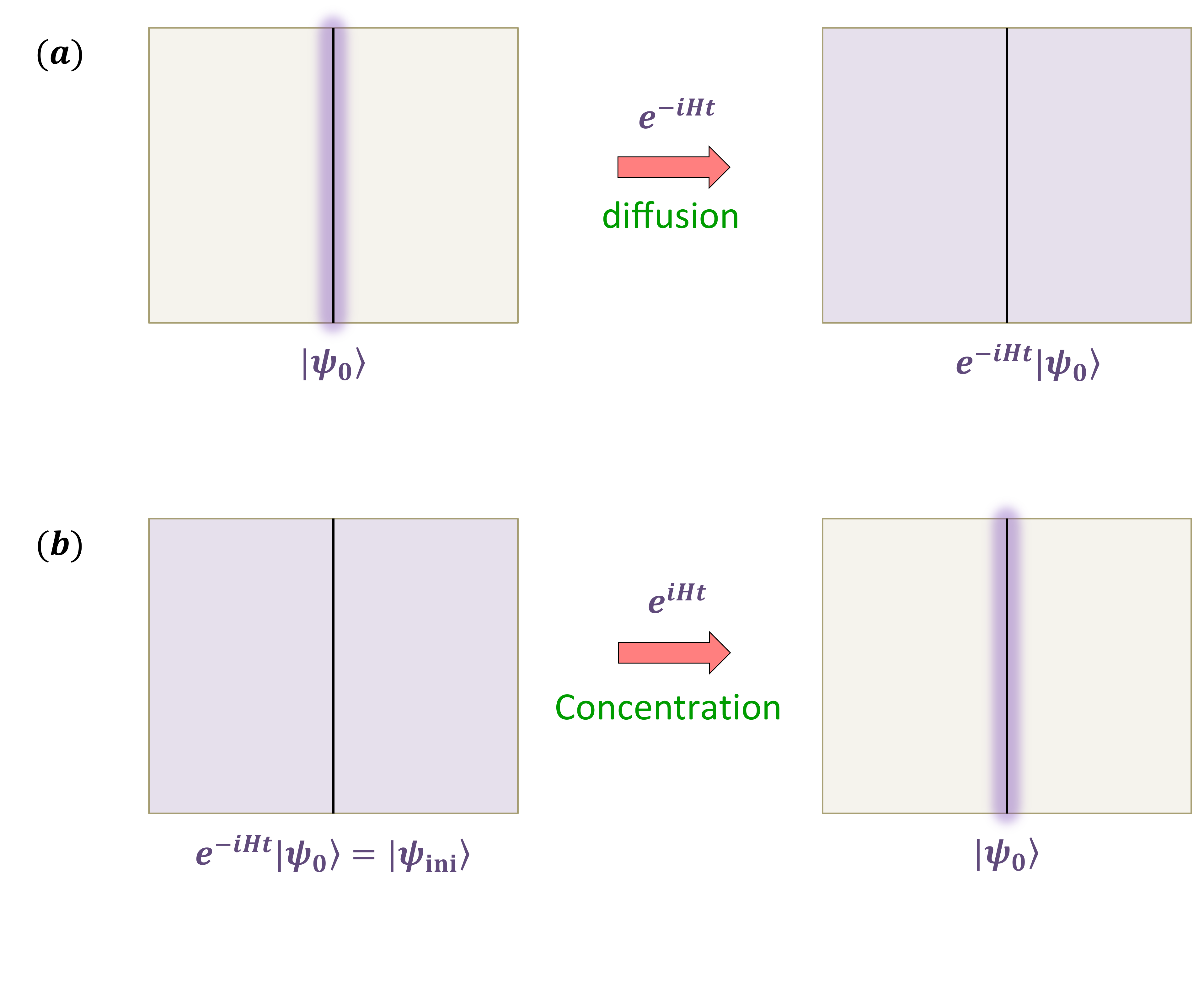}
    \caption{In process (a), the bosons diffuse, and the state transitions to a steady state with low-boson density. By reversing the time evolution, we induce boson concentration (b). During this process, the acceleration of information propagation can occur.}
    \label{Fig::pseudo_particle_transport}
\end{figure}

Here, we demonstrate that a special choice of the initial state can induce the acceleration of information propagation. To illustrate this, we first highlight the essence of acceleration, summarized by the following two points: i) concentrating bosons on a particular one-dimensional region, ii) the speed of information propagation is proportional to local boson numbers (see Ref.~\cite{PhysRevA.85.053625}). While the first step is unlikely in natural dynamics, using reverse time evolution allows us to consider a situation where bosons automatically concentrate on a one-dimensional region (Fig.~\ref{Fig::pseudo_particle_transport}):

\begin{enumerate}
    \item{} We first consider a quantum state $\ket{\psi_0}$ that has an information path with high boson densities on a 1D region.
    \item{} We then evolve the state by a translation-invariant Bose-Hubbard Hamiltonian, resulting in a steady-state $e^{-iHt}\ket{\psi_0}$ after a long time.    
    \item{} We choose the state $e^{-iHt}\ket{\psi_0}$ as our initial state of interest $\ket{\psi_{\rm ini}}$, which is expected to satisfy the low-boson density condition.
    \item{} Considering the Hamiltonian $-H$, the time evolution $e^{iHt}\ket{\psi_{\rm ini}}$ yields boson concentration onto the one-dimensional path.
\end{enumerate}

Importantly, the initial state $\ket{\psi_{\rm ini}}$ is quite rare in the whole phase space, but we always have to exclude the possibility that the initial state of interest is not included in this class of states. In the above situation, it might be possible to observe the acceleration of information propagation in the above setup.

At this stage, the computational cost for the numerical simulation of the above dynamics is quite substantial. However, it remains an intriguing open question to identify whether we can observe the acceleration of information propagation under the above dynamics or other natural situations.

%
%
%





%
%
%
%
%
%

\def\bibsection{\section*{Supplementary References}} 

\bibliography{LR_boson}